\documentclass[10pt]{amsart}

\usepackage{amsmath, latexsym, amsfonts, amssymb,amsthm,MnSymbol,stmaryrd}
\usepackage{graphicx, mathtools,xcolor,hyperref,dsfont,epsfig,caption,wrapfig,subfig,pifont,tikz}
\usetikzlibrary{patterns}
\usetikzlibrary{arrows}
\usetikzlibrary{calc}
\usepackage{datetime}
\usepackage{movie15}
\usepackage[mathscr]{eucal}
\hypersetup{
    linktoc=page,
    linkcolor=red,          
    citecolor=blue,        
    filecolor=blue,      
    urlcolor=cyan,
    colorlinks=true           
}

\newcommand{\nocontentsline}[3]{}
\newcommand{\tocless}[2]{\bgroup\let\addcontentsline=\nocontentsline#1{#2}\egroup}
\numberwithin{equation}{section}

\newtheorem{theorem}{Theorem}[section]
\newtheorem{lemma}[theorem]{Lemma}
\newtheorem{proposition}[theorem]{Proposition}
\newtheorem{corollary}[theorem]{Corollary}

\newtheorem{remark}[theorem]{Remark}
\newtheorem{definition}[theorem]{Definition}

\theoremstyle{definition}

\renewcommand{\tilde}{\widetilde}          
\DeclareMathSymbol{\leqslant}{\mathalpha}{AMSa}{"36} 
\DeclareMathSymbol{\geqslant}{\mathalpha}{AMSa}{"3E} 
\DeclareMathSymbol{\eset}{\mathalpha}{AMSb}{"3F}     
\renewcommand{\leq}{\;\leqslant\;}                   
\renewcommand{\geq}{\;\geqslant\;}                   
\newcommand{\dd}{\text{\rm d}}             

\newcommand{\C}{\mathbb{C}}
\renewcommand{\H}{\mathbb{H}}
\newcommand{\D}{\mathbb{D}}
\newcommand{\R}{\mathbb{R}}
\newcommand{\Z}{\mathbb{Z}}
\newcommand{\N}{\mathbb{N}}

\newcommand{\E}{\mathbb{E}}
\renewcommand{\P}{\mathds{P}}

\newcommand{\cjd}{\rangle}
\newcommand{\cjg}{\langle}

\newcommand{\hf}{\frac{_1}{^2}}

\newcommand{\pl}{\partial}
\newcommand{\bbar}{\overline}
\newcommand{\mc}{\mathcal}
\newcommand{\la}{\lambda}
\newcommand{\til}{\widetilde}
\newcommand{\tT}{ \mathrlap{\widetilde{\phantom{\mc T}}}\mc T}
\let\standardtilde\tilde
\let\standardwidetilde\widetilde
\renewcommand{\tilde}[1]{{\vphantom{#1}\smash{\standardtilde{#1}}}}
\renewcommand{\widetilde}[1]{{\vphantom{#1}\smash{\standardwidetilde{#1}}}}
\newcommand{\btau}{{\boldsymbol{\tau}}}
\newcommand{\btheta}{{\boldsymbol{\theta}}}

\usepackage{xparse}

\def\eps{\varepsilon}

\def\T{\mathbb{T}}

\def\bi{\begin{itemize}}
\def\ei{\end{itemize}}

\def\bnum{\begin{enumerate}}
\def\enum{\end{enumerate}}
\def\<#1{\langle #1 \rangle}

\newcommand{\caA}{{\mathcal A}}

\newcommand{\bL}{\mathbf{L}}

\newcommand{\bv}{\mathbf{v}}

\newcommand{\bH}{\mathbf{H}}

\renewcommand{\d}{\mathrm{d}}
\newcommand{\A}{\mathbb{A}}
\newcommand{\cA}{\mathcal{A}}
\newcommand{\bvarphi}{\boldsymbol{\varphi}}

\renewcommand{\Re}{\mathrm{Re}}

\newcommand{\bzeta}{\boldsymbol{\zeta}}
\newcommand{\bs}{\boldsymbol}

\usepackage{tikz-cd}

\usepackage[dvipsnames]{xcolor}
\usepackage{pdfcomment}

\newcommand{\tBiHol}{\widehat{\rm BiHol}}
\newcommand{\BDT}{\mc{DT}}
\newcommand{\BDTT}{\mc{D}\tT}
\newcommand{\piBD}{\pi_0\mc{BD}}
\newcommand{\mcG}{\mc{G}}

\def\epsilon{\eps}
\def\v{{\rm v}}

\author{Guillaume Baverez}
\address{Beijing International Center for Mathematical Research, Peking University}
\email{guillaume.baverez@bicmr.pku.edu.cn}

\author{Colin Guillarmou}
\address{Universit\'e Paris-Saclay, CNRS,  Laboratoire de math\'ematiques d'Orsay, 91405, Orsay, France.}
\email{colin.guillarmou@universite-paris-saclay.fr}

\author{Antti Kupiainen}
\address{University of Helsinki, Department of Mathematics and Statistics}
\email{antti.kupiainen@helsinki.fi}

\author{R\'emi Rhodes}
\address{Aix-Marseille University, Institut de Math\'ematiques de Marseille (I2M), and Institut Universitaire de France (IUF)}
\email{remi.rhodes@univ-amu.fr}

 \author{Yuxiao Xie}
\address{Universit\'e Paris-Saclay, CNRS,  Laboratoire de math\'ematiques d'Orsay, 91405, Orsay, France.}
\email{yuxiao.xie@universite-paris-saclay.fr}

\title{Virasoro Conformal Blocks and modular functor from Liouville CFT}

\begin{document}

 \begin{abstract} 
In this article, we give a global construction of the Virasoro conformal blocks on Teichm\"uller space of Riemann surfaces  of genus $g$ with $m$ marked points, when the central charge is $c>25$. We prove that they are global holomorphic sections of a holomorphic line bundle and they satisfy the Ward identity, which encodes their conformal invariance. For the sphere with $3$ points, the space of conformal blocks is one-dimensional. For a given marked Riemann surface and a marked pair of pants decomposition, it is in general an infinite dimensional Hilbert space, isomorphic to $L^2(\R_+^{3g-3+m})$ with respect to some measure involving the DOZZ structure constants. We show that it carries a projective unitary representation of the mapping class group. More generally, there are unitary isomorphisms for each Moore-Seiberg move, changing a marked pair of pants decomposition into another one.
  \end{abstract}

\maketitle
\tableofcontents

\vskip 2mm



\section{Introduction}

Conformal field theories (CFT) in dimension $2$ are quantum field theories enjoying conformal symmetries. In physics, they may arise as formal path integrals associated to classical actions on Riemann surfaces, which are conformally covariant. A CFT has a space of states $\mc H$, which  is attached to the unit circle $\T$,  in the Hilbert space setting. The state-field correspondence relates states to local fields.
 Local conformal symmetries are generated infinitesimally by
holomorphic and anti-holomorphic vector fields. In a local
coordinate, their Laurent modes give rise to two commuting
representations of the \emph{Virasoro algebra} on the Hilbert space $\mc{H}$. The algebra coming from holomorphic vector fields  has generators $({\bf L}_n)_{n\in \Z}$ and $c_{\rm L}\in \C$ satisfying 
\[ [{\bf L}_n,{\bf L}_m]=(n-m) {\bf L}_{m+n}+\frac{c_{\rm L}}{12}(n^3-n)\delta_{m+n,0}.\]
Here, $c_{\rm L}$ may be thought of as the operator $c_{\rm L}{\rm Id}$ on $\mc{H}$, and is called the \emph{central charge} of the CFT.
The anti-holomorphic vector fields generate another copy $(\tilde{{\bf L}}_n)_{n\in \Z}$ of the Virasoro algebra, commuting with each 
${\bf L}_m$. As formalised by Segal \cite{Segal87}, each Riemann surface $S=(\Sigma,J,{\bf x},\bs{\zeta})$ equipped with a Riemannian
metric $g$ compatible with the complex structure $J$, $m$ marked points ${\bf x}=(x_1,\dots,x_m)$ and $b$ boundary circles parametrised by a collection $\bs{\zeta}=(\zeta_1,\dots,\zeta_b)$ of parametrisations $\zeta_k:\T\to \pl_k\Sigma$, induces  an operator $\mc{A}_{S_g}:\mc{H}^{\otimes b^-}\to \mc{H}^{\otimes b^+}$. Here   $b^-$ and $b^+$ are respectively the number of negatively and positively 
oriented boundary circles, and the orientations are that of the parametrisations $\zeta_1,\dots,\zeta_b$. This operator is called the Segal amplitude of $S_g=(\Sigma,g,{\bf x},\bs{\zeta})$.
For example, a disk induces a state $\Psi\in \mc{H}$, an annulus induces an element of $\mc{H}^{\otimes 2}$, a pair of pants induces an element in $\mc{H}^{\otimes 3}$, if the boundary parametrisations are positively oriented. In this picture, gluing surfaces with boundary along boundary circles maps  to compositions of the corresponding Segal amplitudes.

The $m$-point correlation functions on a closed Riemann surface $(\Sigma,J)$, equipped with a compatible Riemannian metric $g$, are complex numbers attached to a collection of $m$ disjoint 
points ${\bf x}=(x_1,\dots,x_m)$ and weights $\bs{\alpha}=(\alpha_1,\dots,\alpha_m)\in A^m$, where $A$ is a set 
parametrising highest weight representations $\mc{V}_{\alpha_1},\dots,\mc{V}_{\alpha_m}$ of Virasoro algebra: $\mc{V}_{\alpha_j}$ is generated by a state $\Psi_{\alpha_j}\in \mc{V}_{\alpha_j}$ such that for some $\Delta_{\alpha_j}\in \C$, 
\[ {\bf L}_0 \Psi_{\alpha_j}=\Delta_{\alpha_j} \Psi_{\alpha_j} , \quad {\bf L}_n \Psi_{\alpha_j}=0 \textrm{ for all }n>0.\]
 Denoting $S:=(\Sigma,J,{\bf x})$ and $S_g:=(\Sigma,g,{\bf x})$, the correlation functions, denoted by $\cjg V_{\alpha_1}(x_1)\dots V_{\alpha_m}(x_m)\cjd_{S_g}$, 
 are the main objects of interest in physics to describe the CFT. They are invariant under diffeomorphisms acting simultaneously on
the surface, the marked points and the metric, and are covariant under conformal changes of $g$. Consequently, after choosing a compatible
metric in each conformal class, they can be viewed as functions\footnote{Intrinsically, they should rather be viewed as sections of the corresponding anomaly line bundle, see e.g. \cite{Maibach-Peltola}.} on the moduli space of $S$.
 The general physical principle of CFT is that, when viewed as functions of $S$,
  the correlation functions admit factorisations in terms of universal  holomorphic sections of a line bundle $\mc{L}(\bs{\alpha})$ over the Teichm\"uller space of $S$. These holomorphic sections are algebro-geometric objects related to the representation theory of Virasoro algebra. They are called \emph{conformal blocks}, since they are the algebraic building blocks to construct CFTs. For a class of CFTs called diagonal, the factorisation of correlation functions should have the form
\begin{equation}\label{eq:def-correl-via-bloc}
\cjg V_{\alpha_1}(x_1)\dots V_{\alpha_m}(x_m)\cjd_{S_g}= \int_{\mc{S}^{i_0}} \bs{\rho}(\bs{\alpha},{\bf p})\|\mc{F}_{\bs{\alpha}}(\mc{X},{\bf p})\|_g^2 d\mu({\bf p}),
\end{equation}
 where $i_0$ is the complex dimension of the Teichm\"uller space $\mc{T}(S^\bullet)$ of a fixed base surface $S^\bullet$ with $m$ marked points diffeomorphic to $S$, $\mc{S}\subset A$ is a set called the \textit{spectrum} of the CFT equipped with a (spectral) measure $\mu$, $\bs{\rho}(\bs{\alpha},{\bf p})$ is a  product of three-point correlation functions on the Riemann sphere with marked points $(0,1,\infty)$ and weights in $\{\alpha_1,\dots,\alpha_m,p_1,\dots,p_{i_0}\}$, and finally 
 $\mc{F}_{\bs{\alpha}}(\mc{X},{\bf p})$ denote the conformal blocks and $\|\!\cdot\!\|_g$ is a Hermitian norm on $\mc{L}(\bs{\alpha})$ induced by the choice of metric $g$. Here $\mc{X}$ denotes the Teichm\"uller class of $S$, i.e.\ a marking of the surface is needed. The conformal block $\mc{F}_{\bs{\alpha}}(\mc{X},{\bf p})$ and the line bundle $\mc{L}(\bs{\alpha})$ both depend on an isotopy class $[\bs{\Xi}^\bullet]$ of marked pair of pants decompositions of $S^\bullet$, as we explain below, and $\bs{\rho}(\bs{\alpha},{\bf p})$ depends on the combinatorial graph $\mc{G}$ representing the  pants decomposition, so that two isotopy classes $[\bs{\Xi}_1^\bullet]$ and $[\bs{\Xi}_2^\bullet]$ induce the same graph $\mc{G}$ if and only if they are related by a diffeomorphism of $S^\bullet$. We thus denote them by $\mc{F}^{[\bs{\Xi}^\bullet]}_{\bs{\alpha}}(\mc{X},{\bf p})$,  $\mc{L}^{[\bs{\Xi}^\bullet]}(\bs{\alpha})$ and $\bs{\rho}^{\mc{G}}(\bs{\alpha},{\bf p})$  to emphasise their dependence on $[\bs{\Xi}^\bullet]$ and $\mc{G}$.
For each ${\bf p}$, the map $\mc{X}\mapsto \mc{F}_{\bs{\alpha}}^{[\bs{\Xi}^\bullet]}(\mc{X},{\bf p})$ is a holomorphic section of a holomorphic line bundle $\mc{L}^{[\bs{\Xi}^\bullet]}(\bs{\alpha})$ over $\mc{T}(S^\bullet)$, and for each $\mc{X}$, as a function of ${\bf p}\in \mc{S}^{i_0}$, 
\[ \|\mc{F}^{{[\bs{\Xi}^\bullet]}}_{\bs{\alpha}}(\mc{X},\cdot)\|_g \in L^2(\mc{S}^{i_0},  \bs{\rho}^{\mc{G}}(\bs{\alpha},{\bf p})\dd \mu({\bf p})).\]

Rather than starting from a path integral, one may seek to construct
correlation functions  through the conformal block factorisation \eqref{eq:def-correl-via-bloc}.
The consistency of this construction imposes constraints on the data, especially  the spectrum and the three-point structure
constants as the conformal blocks are supposed to be known. Finding consistent data gives rise to the construction of a new CFT: this is the essence of the huge development of the conformal bootstrap in physics, introduced by Belavin-Polyakov-Zamolodchikov \cite{BPZ84}.

The constraints imposed by the conformal blocks are subtle. In contrast to correlation functions, the conformal blocks are not invariant under diffeomorphisms. This is reflected by the fact that they depend on $[\bs{\Xi}^\bullet]$ and on a marking diffeomorphism $F:S^\bullet\to S$, hence  are defined on Teichm\"uller space, while the correlation functions are functions on the moduli space. This means that for  two Teichm\"uller points $\mc{X}_1,\mc{X}_2\in \mc{T}(S^\bullet)$ representing the same Riemann surface $S$ with marked points, and for two isotopy classes $[\bs{\Xi}_1^\bullet],[\bs{\Xi}_2^\bullet]$ of marked pants decompositions of $S^\bullet$, with underlying graphs $\mc{G}_1,\mc{G}_2$, for the CFT to be well-defined, one should have 
\begin{equation}\label{crossing_symmetry}
\int_{\mc{S}^{i_0}} \bs{\rho}^{\mc{G}_1}(\bs{\alpha},{\bf p})\|\mc{F}^{[\bs{\Xi}_1^\bullet]}_{\bs{\alpha}}(\mc{X}_1,{\bf p})\|_g^2 d\mu({\bf p})=\int_{\mc{S}^{i_0}} \bs{\rho}^{\mc{G}_2}(\bs{\alpha},{\bf p})\|\mc{F}^{[\bs{\Xi}_2^\bullet]}_{\bs{\alpha}}(\mc{X}_2,{\bf p})\|_g^2 \dd\mu({\bf p}).
\end{equation}
This equation is called \textit{crossing symmetry}, which,  in some case, allows one to deduce the three-point functions appearing in $\rho^{\mc{G}_i}(\bs{\alpha},{\bf p})$. 

The mapping class group ${\rm Mod}(S^\bullet)$, i.e. the set of isotopy classes of diffeomorphisms on $S^\bullet$ fixing the marked points, acts on the Teichm\"uller space by biholomorphisms, but it also acts on the isotopy classes $[\bs{\Xi}^\bullet]$. Using the latter action, it is expected in general that the mapping class group acts on the space of conformal blocks (viewed as functions by trivialising $\mc{L}^{[\bs{\Xi}^\bullet]}(\bs{\alpha})$)
\[ \mc{H}^{\mc{G}}_{\bs{\alpha}}:=\overline{{\rm span}\{ \mc{F}^{[\bs{\Xi}^\bullet]}_{\bs{\alpha}}(\mc{X},{\bf p})\,|\, \mc{X}\in \mc{T}(S^\bullet)\}}\simeq L^2(\mc{S}^{i_0}, \bs{\rho}^{\mc{G}}(\bs{\alpha},{\bf p})\dd \mu({\bf p})).\]
More precisely, each CFT should produce a projective representation of ${\rm Mod}(S^\bullet)$ on $\mc{H}^{\mc{G}}_{\bs{\alpha}}$. And more generally, there is a groupoid introduced by Moore-Seiberg \cite{Moore-Seiberg} (see also  \cite{BakalovKirillov2001} for a mathematical treatment) that describes the changes, or moves, to go from one class $[\bs{\Xi}_1^\bullet]$ to another class $[\bs{\Xi}_2^\bullet]$. For each move $M: [\bs{\Xi}_1^\bullet]\to [\bs{\Xi}_2^\bullet]$, there is a bounded linear isomorphism (here $\mc{G}_i$ is the graph of $[\bs{\Xi}_i^\bullet]$ for $i=1,2$)
\[ K^M : \mc{H}^{\mc{G}_1}_{\bs{\alpha}}\to \mc{H}^{\mc{G}_2}_{\bs{\alpha}}\] 
and a nowhere vanishing holomorphic section
$s^M$ of ${\rm Hom}(\mc L^{[\bs\Xi_1^\bullet]}(\bs\alpha), \mc L^{[\bs\Xi_2^\bullet]}(\bs\alpha))$
such that, for each $\mc{X}\in \mc{T}(S^\bullet)$,
\[ \mc F^{[\bs\Xi_2^\bullet]}_{\bs\alpha}(\mc{X})= (K^M\otimes s^M(\mc X)) \mc F^{[\bs\Xi_1^\bullet]}_{\bs\alpha}(\mc X).\]
Mapping classes act on decomposition classes while preserving the gluing graph, and are therefore represented by transition
operators between the corresponding decompositions.
At the mathematical level, there are serious difficulties in carrying out this program.  First, the conformal blocks are hard to construct, since they often are described as formal series in local plumbing parameters $\bs{\tau}$ on Teichm\"uller space, and the convergence of the series is definitely not trivial. Second, assuming the conformal blocks are constructed, to obtain the correlation functions, one needs to verify all the crossing symmetry equations \eqref{crossing_symmetry}. For certain CFTs, such as rational CFTs, the space of conformal blocks can be defined as solutions of PDEs on moduli space, but
the finite-rank differential-equation approach does not directly extend to cases with an infinite dimensional space of conformal blocks.

On the one hand, the construction and study of the space of conformal blocks is a possible way to construct CFTs. On the other hand, the construction of a CFT can yield results on the space of conformal blocks. 
In our work, we rather use the second route: by giving a rigorous meaning to the path integral of Liouville CFT using probability, and  combining the probabilistic tools with  analytic results and the geometry of Teichm\"uller space, we construct the space of Virasoro conformal blocks and the modular functor for central charge $c_{\rm L}>25$.

Foundational results on rational CFTs have been obtained for the Wess-Zumino-Witten models (with compact groups) at positive integral level. 
Tsuchiya-Ueno-Yamada constructed bundles of conformal blocks equipped with projectively flat connections and factorisation isomorphisms \cite{Tsuchiya-Ueno-Yamada}.
Building on this work, Andersen-Ueno constructed the associated modular functors \cite{AndersenUeno2007} and, for $\mathfrak{sl}_N$,
identified them with those arising from Witten-Reshetikhin-Turaev theory \cite{AndersenUeno2015}.
The categorical framework relating modular tensor categories, modular functors and three-dimensional topological quantum field
theories is systematically developed in \cite{BakalovKirillov2001}. These constructions concern the chiral data of conformal field theory, i.e. conformal blocks and modular functors, rather than the correlation functions of the corresponding full CFT. 

Other rigorous constructions of conformal blocks and modular functors are available in the rational setting, notably for strongly rational vertex operator algebras. In this setting, the irreducible modules of the chiral algebra form a finite set, and the central charge is rational.
More precisely, for strongly rational vertex operator algebras, Huang proved that the category of modules is a modular tensor category \cite{Huang08}. Factorisation and local freeness of the corresponding sheaves of conformal blocks were established by Damiolini-Gibney-Tarasca \cite{DGT}, while convergence and sewing of conformal blocks were proved by Gui \cite{Gui24}. Huang \cite{Huang1,Huang2} also constructed the genus-zero and genus-one chiral conformal field theories 
in the strongly-rational case.
More recently, Damiolini-Woike constructed the modular functor directly from these conformal blocks \cite{DamioliniWoike25}. Beyond the rational setting, Huang \cite{Huang3,Huang4} constructed the genus-zero and genus-one chiral conformal field theories in the 
$C_2$-cofinite case, while  Gui and Zhang proved convergence of sewing
for conformal blocks of $C_2$-cofinite vertex operator algebras
\cite{GuiZhangSewing}. They subsequently established sewing-factorization
theorems without assuming rationality or self-duality
\cite{GuiZhangFactorization}.

For a generic central charge $c_{\rm L}>25$, conformal blocks form
infinite dimensional spaces, and a rigorous global construction of the
associated modular functor has remained unavailable. However, in the physics
literature, Teschner made important contributions on this question.
First, Ponsot-Teschner identified explicit fusion and braiding
kernels through the harmonic analysis of continuous series of
representations of the modular double \cite{PonsotTeschner1999,PonsotTeschner2001}. This yields the standard conformal field theory and bootstrap picture of the Virasoro modular functor
\cite{Teschner_Teich}, although its identification with conformal blocks
relies on analytic assumptions and does not provide a 
mathematical construction of the conformal blocks and their sewing operations.

A notion of quantisation of Teichm\"uller space using the representation of marked Riemann surfaces in terms of hyperbolic geometry has been developed by Kashaev \cite{Kashaev_98,Kashaev2001DehnSpectrum} and subsequently by Chekhov-Fock \cite{Fock_Chekhov}; see also the higher rank generalisation
of Fock-Goncharov \cite{FockGoncharov2009QuantumDil}. These constructions build on ideal triangulations of hyperbolic surfaces, and the quantisation is done by using particular coordinates on a decorated Teichm\"uller space induced by the triangulations. Inspired by Verlinde \cite{Verlinde}, and assuming that the Virasoro modular functor related to Liouville CFT can be constructed rigorously, Teschner \cite{Teschner04,Teschner_Teich} formulated  
a beautiful and striking connection between quantum Teichm\"uller theory and Virasoro modular functor. In particular, in \cite{TeschnerVartanov2015}, Teschner-Vartanov showed the validity of the Moore-Seiberg relations in higher genus within the context of quantum
Teichmüller theory, and computed the corresponding central extension of the Moore-Seiberg groupoid. This work further developed the proposed correspondence between the modular functor obtained from quantum Teichm\"uller theory and the Virasoro modular functor related to Liouville CFT. This correspondence can be regarded as a precise mathematical version of Verlinde's conjecture, once the Virasoro modular functor is constructed.
In Teschner's formulation of the Verlinde conjecture \cite{Verlinde}, the Hilbert
spaces of Liouville conformal blocks should be identified with those of
quantum Teichm\"uller theory, in a way compatible with their projective mapping
class group representations and factorisation under changes of pants
decompositions. The Liouville spectral parameters ${\bf p}$ correspond to a joint spectrum of the quantum geodesic length operators, while elementary changes of
decomposition are governed on both sides by the same fusion and
braiding kernels.\\

In this work, we build on the recent probabilistic and analytic construction of Liouville CFT in the sequence of works \cite{DKRV16, Huang2018,Guillarmou2019,KRV19_local,KRV_DOZZ,GKRV20_bootstrap,GKRV21_Segal,BGKR1,GRW2} to provide:
\begin{enumerate}
    \item a full and mathematically rigorous construction of the space of Virasoro conformal blocks
    for central charge $c_{\rm L}=1+6Q^2>25$, isomorphic to $L^2(\R_+^{3g-3+m},\mu)$ in genus $g$ with $m$ marked points, for some measure $\mu$ expressed as a product of DOZZ structure constants;
    \item a holomorphic line bundle, called the Schwarzian line bundle, over Teichm\"uller space, 
    whose holomorphic sections contain the conformal blocks;
    \item a projective unitary representation of the Moore-Seiberg groupoid, and of the mapping class group, into the space of conformal blocks.
\end{enumerate} 
Altogether, this can be viewed as a mathematical construction of the Virasoro modular functor, which is one side of Teschner's formulation of the Verlinde conjecture.\\

The correlation functions for the Liouville CFT on a Riemann surface $S=(\Sigma,J,{\bf x})$ with marked points ${\bf x}$ and Riemannian metric $g$ compatible with $J$ are represented, in physics, by the formal path integral
\[\cjg V_{\alpha_1}(x_1)\dots V_{\alpha_m}(x_m)\cjd_{S_g} = \int \prod_{k=1}^m e^{\alpha_k\phi(x_k)}e^{-\frac{1}{4\pi}\int_\Sigma (|\dd\phi|_g^2+QK_g\phi+e^{\gamma \phi}){\rm dv}_g}D\phi  \]
where $K_g$ is the scalar curvature of $g$, $\gamma\in (0,2)$, $Q=\gamma/2+2/\gamma$,  $\alpha_k\in \C$, $\phi:\Sigma\to \R$ is a real valued distribution and $D\phi$ stands for the ``uniform'' measure on the space of distributions. The probabilistic construction of the path integral on all surfaces using the Gaussian Free Field and the theory of Gaussian Multiplicative Chaos was carried out in \cite{DKRV16,Huang2018,Guillarmou2019}. In the subsequent work \cite{GKRV20_bootstrap}, the Hilbert space of the theory is 
\[\mc{H}= L^2(H^{-1}(\T),\mu_0), \quad \mu_0=\prod_{n=1}^\infty e^{-\frac{x_n^2}{2}-\frac{y_n^2}{2}}\frac{\dd x_n\dd y_n}{2\pi} \otimes \dd c, \]
where elements in the Sobolev space $H^{-1}(\T)$ of order $-1$ on the unit circle $\T$ are parametrised by $(c,(x_n,y_n)_{n\in \N^*})\in \R\times (\R^{\N^*})^2$ using Fourier series 
\[\tilde{\varphi}(e^{{\rm i}\theta})=c+\sum_{n=1}^\infty \Big(\frac{x_n+{\rm i}y_n}{2\sqrt{n}}e^{{\rm i}n\theta}+\frac{x_n-{\rm i}y_n}{2\sqrt{n}}e^{-{\rm i}n\theta}\Big).\]
Then the Hilbert space is decomposed as a direct integral of irreducible representations of Virasoro algebra (here $\widehat\otimes$ denotes the Hilbert completion of the tensor product)
\begin{equation}\label{e:decomposition_H_Verma}
\mc{H}= \int^{\oplus}_{\R_+} \mc{V}_{Q+{\rm i}p}\,\widehat\otimes \,\bbar{\mc{V}}_{Q+{\rm i}p} \,\dd p
\end{equation}
where $\mc{V}_{Q+{\rm i}p}$ are highest weight representations of Virasoro algebra with central charge $c_{\rm L}=1+6Q^2$: there is a family $\Psi_{Q+{\rm i}p}$ of highest weight vectors in $\bigcap_{\eps>0}e^{\eps|c|}\mc{H}$ such that 
\[  {\bf L}_0 \Psi_{Q+{\rm i}p}=\tilde{{\bf L}}_0\Psi_{Q+{\rm i}p}=\Delta_{Q+{\rm i}p} \Psi_{Q+{\rm i}p}, \quad {\bf L}_n  \Psi_{Q+{\rm i}p}=\tilde{{\bf L}}_n  \Psi_{Q+{\rm i}p}=0, \, \forall n>0,\] 
where $({\bf L}_n)_n,(\tilde{{\bf L}}_n)_n$ are two commuting Virasoro algebras acting as unbounded operators on $\mc{H}$. The Verma module $\mc{V}_{Q+{\rm i}p}$ is algebraically given by $\mc{V}_{Q+{\rm i}p}={\rm span}\{ {\bf L}_{-\nu(k)}\dots {\bf L}_{-\nu(1)}\Psi_{Q+{\rm i}p}\,|\, \nu=(\nu(1),\ldots,\nu(k))\in\mc{T}\}$
where $\mc{T}$ denotes the set of Young diagrams, and $\Delta_{\alpha}=\frac{\alpha}{2}(Q-\frac{\alpha}{2})$. 
Here we have used a Hilbert space structure (or completion) on each $\mc{V}_{Q+{\rm i}p}$ induced by the positive-definite Shapovalov forms, which identifies this completion with $L^2(\mc{T})$, after choosing the orthonormal descendant basis associated with the
Shapovalov form. The proof of  \eqref{e:decomposition_H_Verma} is done by diagonalising the self-adjoint (extension of the) operator ${\bf H}:={\bf L}_0+\tilde{{\bf L}}_0$, using methods from scattering theory.

To construct the conformal blocks, we need to introduce geometric data. We fix a Riemann surface $\tilde{S}^\bullet=(\Sigma^\bullet,J^\bullet,{\bf x}^\bullet,\bs{\zeta}^\bullet,{\bf Y}^\bullet)$ with $m$ marked points ${\bf x}^\bullet=(x_1^\bullet,\dots,x_m^\bullet)$, decorated with tangent vectors ${\bf Y}^\bullet=(Y_1^\bullet,\dots,Y_m^\bullet)$ in $T^{1,0}\Sigma^\bullet$ at the marked points, and with $b$ real analytic boundary parametrisations $\bs\zeta^\bullet=(\zeta_1^\bullet,\ldots,\zeta_b^\bullet)$ with each $\zeta_k^\bullet:\T\to \pl_k \Sigma^\bullet$. Here possibly $m=0$, $b=0$. An \textit{admissible decorated marked Riemann surface} 
is a collection $(F,\Sigma,J,{\bf x},\bs{\zeta},{\bf Y})$ where $F:\Sigma^\bullet\to \Sigma$ is a marking diffeomorphism sending marked points, decorating vectors, and boundary parametrisations of $S^\bullet$ to ${\bf x},{\bf Y},\bs{\zeta}$ (we also require that it is holomorphic near ${\bf x}^\bullet$ and $\pl \Sigma^\bullet$). Two such surfaces are said equivalent if there is a biholomorphism
preserving the marked points, decorating vectors and boundary parametrisations, and intertwining the markings up to the prescribed
relative isotopy. The space of equivalence classes of such decorated marked Riemann surfaces is the (infinite dimensional if $b>0$) decorated Teichm\"uller space $\tilde{\mc{T}}(S^\bullet)$. If we forget the decorating vectors and let $S^\bullet=(\Sigma^\bullet,J^\bullet,{\bf x}^\bullet,\bs{\zeta}^\bullet)$, the set of equivalence classes is the Teichm\"uller space $\mc{T}(S^\bullet)$. Such surfaces can be glued along boundary circles using the boundary parametrisations, and this descends to a gluing of their Teichm\"uller classes.
We call (decorated) \emph{complex building block} a surface as above but with genus $0$, $b\geq 1$ boundary circles and $m$ marked points (and decorating vectors) with $m+b=3$; there are three types: a pair of pants (of type $(m,b)=(0,3)$), an annulus with $1$ marked point (of type $(m,b)=(1,2)$) and a disk with $2$ marked points (of type $(2,1)$). 
A (decorated) \textit{marked complex building block decomposition} $\bs{\Xi}=(\Xi_1,\dots,\Xi_{j_0})$ of a surface $S$ of genus $g$ with $m$ marked points and $b$ boundary circles is a decomposition of $S$ into $j_0:=2g-2+m+b$ complex building blocks $\mc{P}_j$, each equipped with a marking diffeomorphism $\Xi_j:\Sigma_{m_j,b_j}^\bullet\to \mc{P}_j$ from a fixed canonical complex building block $\Sigma_{m_j,b_j}^\bullet$ to $\mc{P}_j$. 
Each diffeomorphism $\Xi_j$ maps marked points and decorating vectors of $\Sigma_{m_j,b_j}^\bullet$ to marked points and decorating vectors of $S$, and it also maps each parametrised boundary circle of $\Sigma_{m_j,b_j}^\bullet$ to either a parametrisation of a boundary circle in $\pl \Sigma$ or an interior closed embedded curve; the family of such interior curves is cutting the surface $S$ into the union of the subsurfaces $\mc{P}_j$. The number of cutting curves is denoted by $i_0:=3g-3+m+b$.
The corresponding equivalence classes (i.e.\ of admissible decorated marked Riemann surfaces equipped with a decorated marked complex building block decomposition), defined precisely in Definition~\ref{def:Decoratd_teich_DT} and denoted by $\tilde{\mc{X}}_{\bs{\Xi}}$, form the space $\mc D\tilde{\mc T}(\tilde S^\bullet)$. In the undecorated case, we write $\mc{D}\mc{T}(S^\bullet)$ and $\mc{X}_{\bs{\Xi}}$ instead of $\mc{D}\tilde{\mc{T}}(\tilde{S}^\bullet)$ and $\tilde{\mc{X}}_{\bs{\Xi}}$.
 An element $\tilde{\mc{X}}_{\bs{\Xi}}=[(F,\Sigma,J,{\bf x},\bs{\zeta},{\bf Y},\bs{\Xi})]$ is said to represent the isotopy 
 class $[\bs{\Xi}^\bullet]$ of decorated marked complex building blocks on the base surface $\tilde{S}^\bullet$ 
 if $F^{-1}\circ \bs{\Xi}$ belongs to the isotopy class  $[\bs{\Xi}^\bullet]$.
Notice also that there are natural projections
\[  \pi_{\mc{D}\tilde{\mc{T}}\to \tilde{\mc{T}}}: \mc{D}\tilde{\mc{T}}(\tilde{S}^\bullet)\to \tilde{\mc{T}}(\tilde{S}^\bullet), \qquad 
\pi_{\tilde{\mc{T}}\to \mc{T}}:\tilde{\mc{T}}(\tilde{S}^\bullet)\to \mc{T}(S^\bullet),\]
defined by forgetting the corresponding structures.

Our first result is the following:
\begin{theorem}\label{intro:th1}
Let $\tilde{S}^\bullet$ be a fixed decorated Riemann surface with genus $g$, $m\geq 0$ marked points and $b\geq 0$ real analytic parametrised boundary circles and let $[\bs{\Xi}^\bullet]=[(\Xi_1^\bullet,\dots,\Xi_{j_0}^\bullet)]$ be an isotopy class of decorated marked complex building block decompositions of $\tilde{S}^\bullet$, with $j_0=2g-2+m+b\geq 1$.
Let $\bs{\alpha}=(\alpha_1,\dots,\alpha_m)\in (-\infty,Q)^m$ be a collection of weights satisfying the Seiberg bounds with respect to the decomposition, i.e.\ for each $j$, if $(\alpha_{j1},\dots,\alpha_{jm_j})$ denotes the collection of weights attached to the marked points of the $j$-th complex building block of type $(m_j,b_j)$ induced by $\Xi_j^\bullet$, we have $$\sum_{k=1}^{m_j}\alpha_{jk}>(2-b_j)Q.$$
For each $\tilde{\mc{X}}_{\bs{\Xi}}\in \mc{D}\tilde{\mc{T}}(\tilde{S}^\bullet)$ in the isotopy class $[\bs{\Xi}^\bullet]$, there is a family 
\[\mc{B}_{\tilde{\mc{X}}_{\bs{\Xi}},\bs{\alpha}}({\bf p},{\bf p}') :\mc{V}_{Q+{\rm i}p'_1}\otimes \dots \otimes \mc{V}_{Q+{\rm i}p'_b}\to \C \]
of continuous linear functionals indexed by ${\bf p}\in \R_+^{i_0}$ and ${\bf p}'\in \R_+^{b}$, called \textbf{conformal blocks}, that satisfy the Ward identity of Theorem \ref{Ward_VOA_general}. Using the identification $\mc{V}_{Q+{\rm i}p'_k}\simeq L^2(\mc{T})$, the conformal blocks are defined for almost every $({\bf p},{\bf p}')\in \R_+^{i_0+b}$ and, for each compact interval $I\subset (0,\infty)$, they belong to the class
\[\mc{B}_{\tilde{\mc{X}}_{\bs{\Xi}},\bs{\alpha}} \in L^1(I^{i_0+b}; (L^2(\mc{T})^{\otimes b})^*).\]
In the special case with $m=0$ and $b\geq 1$ and where, for a representative $(F,\Sigma,J,\bs{\zeta},\bs{\Xi})$ of $\mc{X}_{\bs{\Xi}}$, the complex building blocks $\mc{P}_j$ are biholomorphic (as complex building blocks with boundary parametrisations) to $(\hat{\C}\setminus \cup_{k=1}^{b_j}\psi_{jk}(\D^\circ),J_{\C},(\psi_{j1}|_{\T},\ldots,\psi_{jb_j}|_{\T}))$ for some M\"obius transformations $\psi_{jk}$, the Ward identity becomes: 
for each holomorphic vector field $\v$ on $(\Sigma,J)$, one has
\begin{equation}\label{eintro:ward_identity}
\mc{B}_{\mc{X}_{\bs{\Xi}}}({\bf p},{\bf p}')\Big(\sum_{k=1}^{b}{\rm Id}^{\otimes(k-1)}\otimes{\bf L}_{\v_k}\otimes{\rm Id}^{\otimes(b-k)}\Big)=0,
\end{equation}
where ${\rm v}_k:=v_k(z)\pl_z=\zeta_k^*{\rm v}$, ${\bf L}_{{\rm v}_k}:=\sum_{n\in \Z}v_{kn}{\bf L}_n$ with $v_k(z)=-\sum_{n\in \Z}v_{kn}z^{n+1}$, and $\zeta_k:\T \to \pl\Sigma$ is the boundary parametrisation of the $k$-th boundary circle $\pl_k\Sigma$. \end{theorem}
The identity \eqref{eintro:ward_identity} is understood on the common domain of the Virasoro operators, as specified in Theorem~\ref{Ward_VOA_general}.
In the general case, the Ward identity contains the insertion terms and the Schwarzian terms associated with both the external
boundaries and the internal cutting curves, as stated in Theorem~\ref{Ward_VOA_general}. Theorem \ref{intro:th1}  also extends to complex weights $\bs{\alpha}\in\C^m$ satisfying the generalised Seiberg bounds \eqref{e:generalised_Seiberg_bounds_G} and the non-vanishing condition
\eqref{cdozznot=0}; in particular this applies to weights in $Q+{\rm i}\R^*$ (where $\R^*:=\R\setminus\{0\}$).

The spectral parameters ${\bf p}$ are called \textit{internal channel parameters} since they are attached to the cutting curves of the complex building block decomposition, while the parameters ${\bf p}'$ are called \textit{external channel parameters} and they describe the Verma modules $\mc{V}_{Q+{\rm i}p'_k}$ on which the conformal block acts.

One can glue two elements $\tilde{\mc{X}}^1_{\bs{\Xi}^1}\in \mc{D}\tilde{\mc{T}}(\tilde{S}_1^\bullet)$ and 
$\tilde{\mc{X}}^2_{\bs{\Xi}^2}\in \mc{D}\tilde{\mc{T}}(\tilde{S}_2^\bullet)$ along a boundary component: the result is denoted $\tilde{\mc{X}}_{\bs{\Xi}}\in \mc{D}\tilde{\mc{T}}(\tilde{S}^\bullet)$ where $\tilde{S}^\bullet=\tilde{S}_1^\bullet \# \tilde{S}_2^\bullet$ is the glued base surface, $\bs{\Xi}=(\bs{\Xi}_1,\bs{\Xi}_2)$, and $\tilde{\mc{X}}$ is the Teichm\"uller class induced by gluing $\tilde{\mc{X}}_1$ with $\tilde{\mc{X}}_2$. Under gluing of, say, $\pl_1\Sigma_1$ with $\pl_1\Sigma_2$, the conformal blocks satisfy
\[\mc{B}_{\tilde{\mc{X}}_{\bs{\Xi}}}({\bf p},{\bf p}')={\rm Tr}_{11}(\mc{B}_{\tilde{\mc{X}}^1_{\bs{\Xi}^1}}({\bf p}_1,{\bf p}'_1)\otimes\mc{B}_{\tilde{\mc{X}}^2_{\bs{\Xi}^2}}({\bf p}_2, {\bf p}_2'))|_{p'_{11}=p'_{21}=p},\] 
where ${\bf p}=({\bf p}_1,{\bf p}_2,p)$, ${\bf p}'=(p'_{12},\dots,p'_{1b_1},p'_{22},\dots,p'_{2b_2})$, and ${\rm Tr}_{11}$ is understood as  the following trace on the Verma modules: ${\rm Tr}:\mc{V}^*_{Q+{\rm i}p}\otimes \mc{V}^*_{Q+{\rm i}p}\to \C$  is defined by
\[{\rm Tr}(A\otimes B):=\sum_{\nu\in\mc T}A(e_\nu)B(e_\nu),
\]
where $(e_\nu)_{\nu\in\mc T}$ is the fixed orthonormal
descendant basis used to identify $\mc V_{Q+{\rm i}p}$ with $L^2(\mc T)$.
In fact, our conformal blocks are constructed from conformal blocks of marked complex building blocks through this gluing procedure. 

Next, we consider the case of closed Riemann surfaces, i.e. $S^\bullet=(\Sigma^\bullet,J^\bullet,{\bf x}^\bullet)$ is a fixed closed Riemann surface with $m$ marked points, and its decorated version is denoted $\tilde{S}^\bullet=(\Sigma^\bullet,J^\bullet,{\bf x}^\bullet,{\bf Y}^\bullet)$. 
The conformal block of $\tilde{\mc{X}}_{\bs{\Xi}}\in  \mc{D}\tilde{\mc{T}}(\tilde{S}^\bullet)$ is now a complex number $\mc{B}_{\tilde{\mc{X}}_{\bs{\Xi}},\bs{\alpha}}({\bf p})\in \C$, parametrised by ${\bf p}\in \R_+^{i_0}$, it can thus be considered as a function in $L^1(I^{i_0})$ for each compact interval $I\subset (0,\infty)$. For an isotopy class $[\bs{\Xi}^\bullet]$ of decorated marked complex building block decompositions of $\tilde{S}^\bullet$, there is a combinatorial graph $\mc{G}$, called \textit{gluing graph}, encoding the gluing rules for the boundary components of the complex building blocks to form $\tilde{S}^\bullet$ (see Definition \ref{def:gluing_graph}). Notice that the Seiberg bounds on $\bs{\alpha}\in(-\infty,Q)^m$ depend only on the graph $\mc{G}$ and not on the particular isotopy class $[\bs{\Xi}^\bullet]$ with gluing graph $\mc{G}$.

Our second result is the following:
\begin{theorem}\label{intro:th2}
Consider the setting of Theorem \ref{intro:th1} with $b=0$, i.e. with no boundaries, and let $[\bs{\Xi}^\bullet]$ be an isotopy class of decorated marked complex building block decompositions of $\tilde{S}^\bullet$, with gluing graph $\mc{G}$.\\
1) There exists a holomorphic line bundle $\mc{L}^{[\bs{\Xi}^\bullet]}(\bs{\alpha})$ over the finite dimensional Teichm\"uller space $\mc{T}(S^\bullet)$ such that, for almost all ${\bf p}\in \R_+^{i_0}$, the conformal block 
$\tilde{\mc{X}}_{\bs{\Xi}} \in \mc{D}\tilde{\mc{T}}(\tilde{S}^\bullet)\to \mc{B}_{\tilde{\mc{X}}_{\bs{\Xi}},\bs{\alpha}}({\bf p})\in \C$ descends as a holomorphic section of $\mc{L}^{[\bs{\Xi}^\bullet]}(\bs{\alpha})$  on $\mc{T}(S^\bullet)$, denoted 
\[(\mc{X}\mapsto \mc{F}_{\bs{\alpha}}^{[\bs{\Xi}^\bullet]}(\mc{X},{\bf p})) \,\, \in {\rm Hol}(\mc{T}(S^\bullet),\mc{L}^{[\bs{\Xi}^\bullet]}(\bs{\alpha})).\] 
2) A compatible Riemannian metric $g$ on a representative $S=(\Sigma,J,{\bf x})$ of $\mc X$ induces a Hermitian norm
on the fiber $\mc L_{\mc X}^{[\bs\Xi^\bullet]}(\bs\alpha)$, and we have
 \begin{equation}\label{intro:bootstrap}
\big\cjg \prod_{k=1}^mV_{\alpha_k}(x_k)\big\cjd_{\Sigma,g}=C_0 \int_{\R_+^{i_0}}\bs{\rho}^{\mc{G}}(\bs{\alpha},{\bf p})\|\mc{F}^{[\bs{\Xi}^\bullet]}_{\bs{\alpha}}(\mc{X},{\bf p})\|_{\mc{L}^{[\bs{\Xi}^\bullet]}(\bs{\alpha})}^2 \dd {\bf p},
\end{equation}
where $C_0>0$ is a constant depending only on the combinatorial
data of the decomposition, and 
$\bs{\rho}^{\mc{G}}(\bs{\alpha},{\bf p})$ is a product of $j_0$ DOZZ structure constants  \eqref{DOZZformula} with weights contained in $\{\alpha_1,\dots,\alpha_m,Q\pm{\rm i}p_1,\dots,Q\pm{\rm i}p_{i_0}\}$, see \eqref{rhoDOZZ}.\\
3) Define the Hilbert space 
\[\mc{H}^{\mc{G}}_{\bs{\alpha}}:= L^2(\R^{3g-3+m}_+, \bs{\rho}^{\mc{G}}(\bs{\alpha},{\bf p})\dd {\bf p})
\]
and the Hilbert bundle of conformal blocks over $\mc{T}(S^\bullet)$ associated to $[\bs{\Xi}^\bullet]$  by ${\rm CB}^{[\bs{\Xi}^\bullet]}_{\boldsymbol{\alpha}}:=\mc{H}^{\mc{G}}_{\bs{\alpha}}\otimes \mc{L}^{[\bs{\Xi}^\bullet]}(\bs{\alpha})$.
Fix a trivialisation $s:\mc{L}^{[\bs{\Xi}^\bullet]}(\bs{\alpha})\to \mc{T}(S^\bullet)\times \C$ of $\mc{L}^{[\bs{\Xi}^\bullet]}(\bs{\alpha})$, then the conformal blocks form a dense family in the Hilbert bundle, in the sense that
\[ \bbar{{\rm span}\{ s(\mc{F}^{[\bs{\Xi}^\bullet]}_{\bs{\alpha}}(\mc{X},\cdot)) \,|\, \mc{X}\in \mc{T}(S^\bullet)\}}=\mc{H}^{\mc{G}}_{\bs{\alpha}}.\]
\end{theorem}

The construction of the conformal blocks $\mc{F}^{[\bs{\Xi}^\bullet]}_{\bs{\alpha}}(\mc{X},\cdot)$ on a small neighbourhood of a point on Teichm\"uller space using  plumbing coordinates, as well as the conformal bootstrap formula \eqref{intro:bootstrap}, was achieved in \cite{GKRV21_Segal}. The novelty here is the global construction on Teichm\"uller space and the proof that this defines a proper holomorphic section of a line bundle.

The final result of this paper is the construction of the modular functor and the projective unitary representation of the mapping class group on $\mc{H}_{\bs{\alpha}}^{\mc{G}}$, for each combinatorial graph $\mc{G}$. We define the decorated mapping class group $\tilde{{\rm Mod}}(\tilde{S}^\bullet)$ (with $b=0$) to be the group of isotopy classes of diffeomorphisms of $S^\bullet$ fixing the marked points and decorating vectors (the isotopies are also required to fix these). 
There is a natural surjective homomorphism from $\tilde{\rm Mod}(\tilde{S}^\bullet)$ to the usual mapping class group ${\rm Mod}(S^\bullet)$, defined similarly but without the decorating vectors; 
its kernel is $\Z^m$, generated by the Dehn twists around the marked points.
\begin{theorem}\label{intro:RepMCG}
1) Consider two isotopy classes $[\bs{\Xi}_1^\bullet]$ and $[\bs{\Xi}_2^\bullet]$ of decorated marked complex building block decompositions with possibly different gluing graphs $\mcG_1,\mcG_2$, and let $\bs{\alpha}\in (-\infty,Q)^m$ be weights satisfying the Seiberg bounds with respect to both $\mcG_1$ and $\mcG_2$. Then there exists a unitary operator 
\[K_{\bs{\alpha}}^{[\bs{\Xi}_1^\bullet],[\bs{\Xi}_2^\bullet]}:\mc{H}_{\bs{\alpha}}^{\mcG_1}\to\mc{H}_{\bs{\alpha}}^{\mcG_2}\]
and a nowhere vanishing holomorphic section $s$ of ${\rm Hom}(\mc{L}^{[\bs{\Xi}_1^\bullet]}(\bs{\alpha}),\mc{L}^{[\bs{\Xi}_2^\bullet]}(\bs{\alpha}))
\simeq\mc{L}^{[\bs{\Xi}_2^\bullet]}(\bs{\alpha})\otimes(\mc{L}^{[\bs{\Xi}_1^\bullet]}(\bs{\alpha}))^*$ such that for all $\mc{X}\in \mc{T}(S^\bullet)$,
\[\mc{F}_{\bs{\alpha}}^{[\bs{\Xi}_2^\bullet]}(\mc{X})=(K_{\bs{\alpha}}^{[\bs{\Xi}_1^\bullet],[\bs{\Xi}_2^\bullet]}\otimes s(\mc{X}))(\mc{F}_{\bs{\alpha}}^{[\bs{\Xi}_1^\bullet]}(\mc{X})).\]
Moreover, such an operator $K_{\bs{\alpha}}^{[\bs{\Xi}_1^\bullet],[\bs{\Xi}_2^\bullet]}$ is unique up to multiplication by a complex constant of modulus $1$.\\
2) Fix a gluing graph $\mc{G}$ and an isotopy class $[\bs{\Xi}^\bullet]$ of decorated marked complex building block decompositions with gluing graph $\mc{G}$, and let $\bs{\alpha}$ be weights satisfying the Seiberg bounds with respect to $\mc{G}$.
Then the map 
\[  U^{\mc{G}}: [h]\in \tilde{{\rm Mod}}(\tilde{S}^\bullet) \to K_{\bs{\alpha}}^{[\bs{\Xi}^\bullet],[h^{-1}]\cdot[\bs{\Xi}^\bullet]} \in \mc{L}(\mc{H}_{\bs{\alpha}}^{\mc{G}}) \]
is a projective unitary representation of $\tilde{{\rm Mod}}(\tilde{S}^\bullet)$ on $\mc{H}_{\bs{\alpha}}^{\mc{G}}$. Moreover, it descends to a projective unitary 
representation of ${\rm Mod}(S^\bullet)$ on $\mc{H}_{\bs{\alpha}}^{\mc{G}}$. 
\end{theorem}
The construction of the projective representation above is done by decomposing a general change $[\bs{\Xi}_1^\bullet]\to[\bs{\Xi}_2^\bullet]$ into \textit{elementary moves}, as in the work of Moore-Seiberg \cite{Moore-Seiberg} and Bakalov-Kirillov  \cite{BakalovKirillov2001} for rational CFT. The moves are  of the following types: A-moves on a sphere with four disks removed, S-moves on a torus with one disk removed, T-moves induced by Dehn twists around boundary circles or marked points of a complex building block, and B-moves induced by braiding maps exchanging boundary circles or marked points of a complex building block. The T- and B-moves are represented by explicit phase multipliers, together with the corresponding relabelling of channel variables when needed. The A- and S-moves are represented by much more complicated operators involving the $6j$-symbols, as recently proved by Ghosal-Remy-Sun-Sun-Wu \cite{GRSSbloc,GRSSWBloc}, where they relate  explicitly the transition operators to Ponsot-Teschner's fusion kernels. In that work, the authors also prove analytic extension in $\C$ of the conformal blocks $\mc{F}_{\bs{\alpha}}^{[\bs{\Xi}^\bullet]}(\mc{X},{\bf p})$ for the 4-point sphere and the 1-point torus in the external parameters $\bs{\alpha}$ and the internal parameters ${\bf p}$. Their analysis is complementary to what we have developed here.

The proof of Theorem \ref{intro:th1} is based on the construction of Segal amplitudes for the Liouville CFT developed in the paper \cite{GKRV21_Segal}, and the recent paper \cite{BGKR1} on the deformation of these Segal amplitudes in terms of Virasoro generators. We need however to develop a general geometric formalism for Teichm\"uller spaces with boundary parametrisations. It is, in some way, related to 
works of Radnell-Schippers \cite{RadnellSchippers06}, but we concentrate here on real analytic parametrisations of the boundary, since we only need this case. 
In a related direction, Huang \cite{Huang2003FunctionalAnalyticII} constructed a locally convex completion of a vertex operator algebra carrying an action of a power of the determinant-line operad associated with genus-zero surfaces with analytically parametrised boundaries.
A fundamental step in our construction is to understand the proper definition of the conformal block of a single 
decorated marked complex building block; this is done in Section~\ref{definition_conformal_block_amplitudes}. Another crucial result used here, to prove the convergence of the conformal block of a glued surface, is the spectral decomposition theorem in \cite[Proposition 6.26]{GKRV20_bootstrap} for the (self-adjoint) Liouville Hamiltonian~${\bf H}$.

The line bundle in Theorem~\ref{intro:th2} arises from two covariance
properties. First, if $\bs\Xi_1$ and $\bs\Xi_2$ are two decorated
marked complex building block decompositions of the same surface, related
by an isotopy fixing the marked points and decorating vectors, then
(see Proposition~\ref{well-defined_blocks})
\[ \mc{B}_{\tilde{\mc{X}}_{\bs{\Xi}_2},\bs{\alpha}}({\bf p}) =e^{\frac{c_{\rm L}}{12} \Upsilon(\mc{X}_{\bs{\Xi}_1},\mc{X}_{\bs{\Xi}_2})}
\mc{B}_{\tilde{\mc{X}}_{\bs{\Xi}_1},\bs{\alpha}}({\bf p}),\]
where $\Upsilon(\mc{X}_{\bs{\Xi}_1},\mc{X}_{\bs{\Xi}_2})$ is a cocycle, independent of ${\bf p}$, involving Schwarzian derivatives of uniformising maps of an isotopy $t\in [1,2]\mapsto\bs{\Xi}_t$ of complex building blocks between $\bs{\Xi}_1$ and $\bs{\Xi}_2$. Replacing the decorating vectors $Y_k$ by $e^{t_k}Y_k$ (and deforming the $\Xi_j$ near $x_k$ accordingly) 
multiplies the conformal block by $e^{\sum_k\Delta_{\alpha_k}t_k}$. These two properties allow us to define the line bundle $\mc{L}^{[\bs{\Xi}^\bullet]}(\bs{\alpha})$ as a set of functions on decorated complex marked building block decompositions satisfying the two covariances just discussed.

The projective representation of the Moore-Seiberg moves and the mapping class group is proved using the construction above, together with the conformal bootstrap for Liouville CFT on surfaces with boundary \cite{GRW2}. The key fact is that the bootstrap in this case allows us to write correlations in terms of conformal blocks of the doubled surface along its boundary. For the A-move, using this boundary bootstrap formula for two different decompositions of the sphere with four disks removed, we obtain a linear relation between the conformal blocks for two different decompositions $\bs{\Xi}_1$ and $\bs{\Xi}_2$ obtained by cutting along the so-called s- and t-channels, as named in physics. The crucial observation in \cite{GRSSbloc,GRSSWBloc} is that, by choosing cleverly the cosmological constants in the boundary bootstrap, one can invert the linear relations between $\mc{B}_{\tilde{\mc{X}}_{\bs{\Xi}_1},\bs{\alpha}}$ and $\mc{B}_{\tilde{\mc{X}}_{\bs{\Xi}_2},\bs{\alpha}}$ and obtain an integral kernel in the ${\bf p}$ variable relating the two conformal blocks. This relation is only true on a subset of Teichm\"uller space. We prove that this subset contains an open part of a maximal totally real submanifold; holomorphic uniqueness then extends
the transition identity to the whole Teichm\"uller space.
The same argument works for the S-move on the torus with one disk removed. The T- and B-moves are obtained directly from the analysis of the conformal block of a single marked decorated complex building block. The unitarity of the operators is proved in the Appendices, using the density of conformal blocks and the bootstrap formula. We notice that we do not use the explicit formula for the integral kernels (called fusion kernels in physics) of the A-move or S-move operators. This construction gives for free that the fusion operators satisfy the Moore-Seiberg relations, including the pentagon relation, up to constant phases.
It is checked in \cite{GRSSWBloc} that the fusion kernels for the A- and S-moves are those proposed by Ponsot-Teschner \cite{PonsotTeschner1999,PonsotTeschner2001} and \cite{HadaszJaskolskiSuchanek2010,TeschnerVartanov2015}.

To summarise, the novelty of this paper is twofold. First, we construct Liouville conformal blocks as genuine holomorphic sections over Teichm\"uller space and establish the
convergence and gluing properties required for this construction. Second, we obtain projective unitary representations of the Moore-Seiberg groupoid and thus of the mapping class group. In particular, the projective relations between the fusion operators are derived from the geometry and the
uniqueness of the transition operators, rather than verified directly through identities on the special functions appearing in these kernels.
We also develop in Section~\ref{sec1} a fairly general complex-geometric framework, based on infinite dimensional Teichm\"uller spaces, which we hope should be useful for the study of other CFT models and geometric invariants. For example, our complex line bundle, called Schwarzian line bundle, has  properties very similar  to the Chern-Simons line bundle \cite{Guillarmou-Moroianu} related to the renormalised volume of hyperbolic convex co-compact 3-manifolds and their Chern-Simons invariants.\\

\textbf{Applications.} As mentioned above, this work is a missing fundamental piece to complete Teschner's program on the Verlinde conjecture, giving a correspondence between quantum Teichm\"uller theory and Virasoro modular functor. It
may serve as a starting point for extending geometric VOA methods to unitary theories with continuous spectrum. On the physics side, 
our construction provides a rigorous realisation of some of the conformal block ingredients appearing in the Virasoro TQFT proposed by Collier, Eberhardt and Zhang \cite{Collier_Eberhardt_Zhang}. 
In the semiclassical limit $\gamma\to 0$, conformal blocks are related to accessory parameters of associated differential equations; for the one-point torus, see the rigorous results of Desiraju, Ghosal and Prokhorov \cite{Desijaru_Ghosal_Prokhorov} on the Lamé equation.
Our constructions should provide the setting for the same analysis in higher genus.\\

\noindent\textbf{Acknowledgements.} C. Guillarmou, A. Kupiainen and R. Rhodes are supported by the Simons Collaboration grant ``Probabilistic Paths to Quantum Field Theory''. R. Rhodes is partially supported by the Institut Universitaire de France (IUF) and by the ANR-21-CE40-0003. A. Kupianen is partially supported by the Academy of Finland.
We thank J\"org Teschner for numerous discussions on the subject, David Radnell and Eric Schippers for explaining to us their work, Gwenael Massuyeau and Ramanujan Santaroubane for helps on topological questions, Xin Sun and Baojun Wu for discussions about their related ongoing project, Baptiste Cerclé for comments on the paper. This project took a long time; the mathematical content of the paper was finalised long before the rise of AI, 
but AI was used in the final stage for proofreading. At this stage, it also helped in the proof of Lemma \ref{orthogonality} and it suggested the pluriharmonicity argument to obtain 
the validity for the whole range of $\bs{\alpha}$ satisfying the Seiberg bounds (rather than for the weights $\alpha_k$ in a restricted range $(2Q/3,Q)$) in the proof of existence in Theorem \ref{th:kernelMS}.

\section{Riemann surfaces, moduli and Teichm\"uller spaces}\label{sec1}

In this section, we introduce the notions of Riemann surfaces of type $(g,m,b^+,b^-)$ and their moduli and Teichm\"uller spaces. Here $(\Sigma,J)$ will be a Riemann surface with $b=b^++b^-$ boundary connected components parametrised by real analytic maps $\zeta_j:\T\to \zeta_j(\T)=\pl_j\Sigma\subset\pl\Sigma$ ($\T\subset \C$ being the unit circle), so that $\zeta_j$ has positive orientation in $\Sigma$ if $j\leq b^+$ and  negative one if $j>b^+$, and
$g=g(\Sigma)$ will be the genus of the surface $\Sigma$ and $m$ the number of marked points.
Unless otherwise stated, we shall always assume that $g>0$ or $g=0$ and $m+b\geq3$.
These surfaces are the natural geometric objects on which we define the Segal amplitude for the Liouville CFT, as well as the conformal block amplitudes.

\subsection{Notations}

We call $J_{\C}$ the canonical complex structure on the Riemann sphere $\hat{\C}=\C\cup\{\infty\}$, associated to the complex coordinates $z=x+{\rm i}y$ on $\C$ and $1/z$ on $\hat{\C}\setminus \{0\}$. 
 If $D\subset \C$ open and $\Phi:D\to\C$ is holomorphic near a point $z$, we shall write $\Phi'(z)=\pl_z\Phi(z)$ for the complex derivative at $z$. Marked points on a surface $\Sigma$ are an ordered collection of distinct points ${\bf x}=(x_1,\dots,x_m)\in(\Sigma^\circ)^m$, and we shall always write $\{x\}:=\{x_1,\dots,x_m\}\subset \Sigma$ for the collection as a set.

We define the map
\begin{equation}\label{def_o}
o: \T\to \T, \quad o(e^{{\rm i}\theta})=e^{-{\rm i}\theta} ,\qquad o:\hat{\C}\to \hat{\C}, \quad o(z)=1/z,
\end{equation}
which is orientation reversing on $\T$. Let $\D$ be the closed unit disk in $\C$ and $\D^*:=\hat{\C}\setminus \D^\circ$, so that $o$ is a biholomorphic involution between $\D$ and $\D^*$. We define the closed annulus $\A_{\delta}:=\{z\in \C\, |\, |z|\in [\delta,1]\}$ if $\delta<1$ and more generally $\A_{\delta_1,\delta_2}:=\{z\,|\, |z|\in [\delta_1,\delta_2]\}$ if $\delta_1<\delta_2$, so that $\A_{\delta_1,\delta_2}=\delta_2\D\cap\delta_1\D^*$. We will use the following notations: if $\Sigma$ is a Riemann surface (for example $\Sigma=\hat{\C}$),
\begin{equation}\label{psi_in/out}  
\psi^{\rm out}:=\psi\circ o \textrm{ if }\psi:\D\to \psi(\D)\subset \Sigma \textrm{ is holomorphic}, 
\quad \psi^{\rm in}:=\psi\circ o \textrm{ if }\psi:\D^*\to \psi(\D^*)\subset \Sigma \textrm{ is holomorphic}.\end{equation}

For $\eps>0$ small, we define the spaces of biholomorphic maps
\begin{align*}
& {\rm BiHol}_\epsilon(\D):=\{\textrm{holomorphic embeddings }f:(1+\eps)\D\to\C\} , \\
& {\rm BiHol}_\epsilon(\D^*):=\{\textrm{holomorphic embeddings }f:(1+\eps)^{-1}\D^*\to\hat{\C}\setminus\{0\}\},\\
& {\rm BiHol}_\eps(\A):=\{\textrm{holomorphic embeddings }f:\A_{(1+\eps)^{-1},1+\eps}\to\C\setminus\{0\}\},
\end{align*} 
where by holomorphic embedding we mean a diffeomorphism (smooth up
to the boundary) onto its image which is holomorphic in the interior.
We equip these spaces with the Whitney $C^\infty$ topology. The restriction of an arbitrary element of
${\rm BiHol}_\eps(\D)$ or ${\rm BiHol}_\eps(\D^*)$ to
$\A_{(1+\eps)^{-1},1+\eps}$ does not necessarily belong to
${\rm BiHol}_\eps(\A)$, since it may take the value $0$ or $\infty$
on this annulus. On the other hand, restriction gives natural
inclusions
\[
\{f\in{\rm BiHol}_\eps(\D)\,|\,f(0)=0\} \subset {\rm BiHol}_\eps(\A)
, \quad  \{f\in{\rm BiHol}_\eps(\D^*)\,|\,f(\infty)=\infty\}
 \subset {\rm BiHol}_\eps(\A).
\]
We have a homeomorphic involution
\begin{equation}\label{defmcO}
\mc O:{\rm BiHol}_\eps(\A)\to
{\rm BiHol}_\eps(\A),\qquad
\mc O(f):=o\circ f\circ o.
\end{equation}
The same formula defines a homeomorphism
$\mc O:{\rm BiHol}_\eps(\D)\to {\rm BiHol}_\eps(\D^*),$
whose inverse is given by the same formula. It maps the subspace
$\{f\in{\rm BiHol}_\eps(\D)\,|\,f(0)=0\}$ onto
$\{f\in{\rm BiHol}_\eps(\D^*)\,|\,f(\infty)=\infty\}$, and is
compatible with the preceding restriction maps.
We identify the tangent spaces of these spaces at the identity map ${\rm Id}$ with function spaces which we describe next.

For $\eps>0$ small, we denote by ${\rm Hol}_\eps(\D)$, ${\rm Hol}_\eps(\D^*)$ and ${\rm Hol}_\eps(\A)$ the respective complex vector spaces of functions on $(1+\eps)\D$, $(1+\eps)^{-1}\D^*$ and $\A_{(1+\eps)^{-1},1+\eps}$ which are holomorphic in the interior and smooth up to the boundary, and in the case of $\D^*$ we allow a pole of order $\leq2$ at $\infty$. Notice that ${\rm Hol}_\eps(\D),{\rm Hol}_\eps(\D^*)\subset {\rm Hol}_\eps(\A)$ naturally. The space ${\rm Hol}_\eps(\D)$ is naturally identified with the tangent space $T_{\rm Id}{\rm BiHol}_\eps(\D)$ of ${\rm BiHol}_\eps(\D)$ in the sense that the tangent vector of a deformation $t\mapsto f_t\in{\rm BiHol}_\eps(\D)$ with $f_0={\rm Id}$ is defined to be $\partial_t|_{t=0}f_t\in{\rm Hol}_\eps(\D)$, and similarly for $\D^*$ and $\A$. (More properly, these should be viewed not as functions but as holomorphic vector fields, as we do below in Section \ref{sec:deformations}.) In the case of $\D^*$, to see that $\partial_t|_{t=0}f_t$ has at most a pole of order $\leq2$ at $\infty$, we use the bijection $\mc{O}:{\rm BiHol}_\eps(\D)\to{\rm BiHol}_\eps(\D^*)$ in \eqref{defmcO}. Indeed, for a deformation $t\mapsto f_t\in{\rm BiHol}_\eps(\D^*)$ with $f_0={\rm Id}$, if $\partial_t|_{t=0}(o\circ f_t\circ o)(z)=-\sum_{n\geq-1}v_nz^{n+1}\in{\rm Hol}_\eps(\D)$, we have 
\[\partial_t|_{t=0}f_t(z) = \partial_t|_{t=0}o\circ(o\circ f_t\circ o)(1/z) = -z^2 \partial_t|_{t=0}(o\circ f_t\circ o)(1/z) = \sum_{n\geq-1}v_nz^{-n+1}\in{\rm Hol}_\eps(\D^*).\] 
In other words, the differential $D\mc{O}({\rm Id}):{\rm Hol}_\eps(\D)\to{\rm Hol}_\eps(\D^*)$ of $\mc{O}$ at ${\rm Id}$ is the linear isomorphism $v\mapsto-z^2v(o(z))$, matching the transformation law of holomorphic vector fields.

Each  $f\in {\rm Hol}_\eps(\A)$ has a Laurent series expansion converging in $\A_{\delta,\delta^{-1}}$ with $\delta=(1+\eps)^{-1}$ 
\[f(z)=\sum_{n=-\infty}^\infty f_nz^{n+1},\]
providing this way  complex linear coordinates $f_n:{\rm Hol}_\eps (\A)\to\C$.  For $\epsilon>0$ fixed, we introduce the countable family of seminorms for $k\geq 0$
\begin{equation}\label{seminorm}
\|f\|^2_{\epsilon,k}:=\sum_{n\in \Z}(1+\epsilon)^{2|n|+2}(1+|n|)^{2k}|f_n|^2.
\end{equation}

We endow ${\rm Hol}_\eps(\A)$ with the family of norms
$(\|\cdot\|_{\eps,k})_{k\geq0}$, where $\|f\|_{\eps,k}$ is the
square root of the right-hand side of \eqref{seminorm}. This turns
${\rm Hol}_\eps(\A)$ into a Fr\'echet space, with the distance 
\begin{equation}\label{distance}
{\rm dist}_\eps (f_1,f_2):= \sum_{k\geq 0}2^{-k}\frac{\|f_1-f_2\|_{\epsilon,k}}{1+\|f_1-f_2\|_{\epsilon,k}},
\end{equation}
and this topology restricts to ${\rm Hol}_\eps(\D)$ and ${\rm Hol}_\eps(\D^*)$.

For fixed $\eps>0$, the norm $\|f\|_{\eps,k}$ is equivalent to
$ \|f(\delta\cdot)\|_{H^k(\T)}+\|f(\delta^{-1}\cdot)\|_{H^k(\T)}$ for 
$\delta=(1+\eps)^{-1}$. Consequently, the Fr\'echet topology defined by
\eqref{seminorm}--\eqref{distance} is equivalent to the Whitney $C^\infty$ topology on the space of maps from
$\A_{\delta,\delta^{-1}}$ to $\C$ which are holomorphic in the interior and smooth up to the boundary. In particular,
${\rm BiHol}_\eps(\A)$ is an open subset of ${\rm Hol}_\eps(\A)$. The analogous statement holds for the disk
and exterior-disk spaces, with the corresponding normalisation at $\infty$ in the latter case.
After identifying maps by restriction to smaller domains, we write
${\rm Hol}(S)=\bigcup_{\eps>0}{\rm Hol}_\eps(S)$ and ${\rm BiHol}(S)=\bigcup_{\eps>0}{\rm BiHol}_\eps(S)$, for
$S=\D,\D^*,\A$, only as sets of holomorphic germs. We do not equip these unions with a differential direct-limit structure: all
differentiability statements below are made on a space with $\eps>0$ fixed, possibly after restricting the maps to a slightly
smaller fixed domain.

\subsection{Riemann surfaces and complex building blocks}\label{complex_building_block}

\begin{figure}[h]
\centering
\begin{tikzpicture}[scale=0.7]
\draw[fill=gray!30,draw=black,very thick] (0,0) circle (2);
\draw[fill=white,draw=black,very thick] (0,0) circle (1);

\draw[->,>=stealth',black,very thick]
  (150:1.2cm) arc[start angle=150,end angle=210,radius=1.2cm]
  node[left] {\rm In};

\draw[->,>=stealth',black,very thick]
  (-30:2.2cm) arc[start angle=-30,end angle=30,radius=2.2cm]
  node[right] {\rm Out};

\node at (0.5,1.45) {$\mathbb A$};
\end{tikzpicture}
\caption{An annulus $\mathbb A$, which is a surface of type
$(0,0,1,1)$ with in/out boundaries.}
\label{pic_Annulus}
\end{figure}
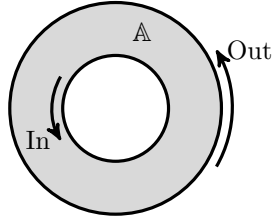

A compact Riemann surface, possibly with boundary, is a smooth oriented surface $\Sigma$ equipped with a set of smooth charts 
$\omega_j:U_j\to \C$, with $\bigcup_{j} U_j=\Sigma$, and $\omega_j(U_j)\subset \D$ with $\omega_j(U_j\cap \pl \Sigma)\subset \T$  if $U_j\cap \pl\Sigma\not=\emptyset$, with $\omega_j\circ \omega_i^{-1}:\omega_i(U_j\cap U_i)\to \omega_j(U_j\cap U_i)$ being holomorphic where it is defined, and analytic up to the boundary $\T$ if $U_j\cap U_i\cap \pl\Sigma\not=\emptyset$. 
A diffeomorphism $\zeta_k:\T\to \pl_k\Sigma$, called \textbf{parametrisation} of $\pl_k\Sigma$,  is said analytic if 
$\omega_i\circ \zeta_k: \zeta_k^{-1}(U_i)\to \T$ is real analytic for all $i$. We say that $\zeta_k$ is \textbf{outgoing} if $(\dot{\zeta}_k,\nu)$ is a positive basis of $T\Sigma$ at $\pl_k\Sigma$, where $\nu$ is a non-vanishing interior pointing vector at $\pl \Sigma$ and $\dot{\zeta}_k$ the tangent vector to the curve $\zeta_k(\T)$, while it is said to be \textbf{incoming} in the other case; see Figure \ref{pic_Annulus} for an example.
This also means that $\zeta_k$ extends holomorphically as a map $\zeta_k:\A_{\delta}\to V_k$ (resp.  $\zeta_k:\A_{1,\delta^{-1}}
\to V_k$) for some $\delta<1$ if $\zeta_k$ is outgoing (resp. incoming), where $V_k$ is a neighbourhood of $\pl_k\Sigma$ in $\Sigma$. 
The structure of Riemann surface induces an endomorphism $J:T\Sigma\to T\Sigma$ satisfying $J^2=-{\rm Id}$, obtained by setting
$J\pl_x:=\pl_y$ and $J\pl_y:=-\pl_x$ in the charts $U_j$ if $\omega_j=x+{\rm i}y$ ($x,y$ being real valued coordinates). The endomorphism is well-defined globally since the change of coordinates is holomorphic, and conversely the data of a $J$ satisfying $J^2=-{\rm Id}$ on $T\Sigma$ also induces a set of holomorphic charts. We then call $J$ the complex structure of $\Sigma$ and will denote $(\Sigma,J)$ the Riemann surface with complex structure $J$. The holomorphic tangent bundle 
$(T\Sigma)^{1,0}$ is defined as the subbundle $(T\Sigma)^{1,0}=\ker (J-{\rm i}\, {\rm Id})\subset \C T\Sigma$ inside the complexified tangent bundle. In local holomorphic coordinates $z$, local sections of $(T\Sigma)^{1,0}$ are $f(z)\pl_z$ for some smooth functions 
$f$ and a holomorphic vector field $\v$ is a smooth section of $(T\Sigma)^{1,0}$ such that in each holomorphic chart $z$, $\v=v(z)\pl_z$ with $v$ a holomorphic function.
 
We define the orientation sign $\sigma_k\in\{-1,+1\}$ with the convention 
\[\sigma_k=-1 \iff \zeta_k \textrm{ is incoming},\quad \sigma_k=+1 \iff \zeta_k \textrm{ is outgoing}.\]

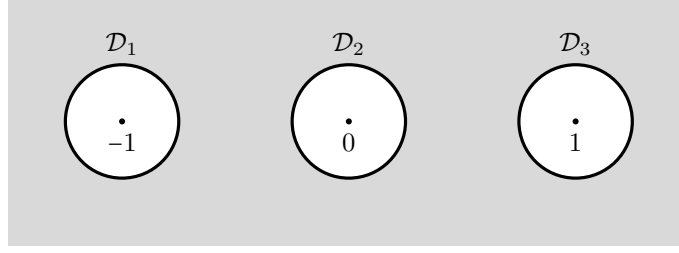
\begin{figure}[h]
\centering
\begin{tikzpicture}[scale=3]
\fill[gray!30] (-1.5,-0.55) rectangle (1.5,0.55);

\draw[fill=white,draw=black,very thick] (-1,0) circle (0.25);
\draw[fill=white,draw=black,very thick] (0,0) circle (0.25);
\draw[fill=white,draw=black,very thick] (1,0) circle (0.25);

\node[circle,fill,inner sep=0.8pt,label=below:$-1$] at (-1,0) {};
\node[circle,fill,inner sep=0.8pt,label=below:$0$] at (0,0) {};
\node[circle,fill,inner sep=0.8pt,label=below:$1$] at (1,0) {};

\node[anchor=south] at (-1,0.25) {$\mathcal{D}_1$};
\node[anchor=south] at (0,0.25) {$\mathcal{D}_2$};
\node[anchor=south] at (1,0.25) {$\mathcal{D}_3$};
\end{tikzpicture}
\caption{A pair of pants
$\hat{\C}\setminus \cup_{j=1}^3(\mc{D}_j^\bullet)^\circ$
with $\mc{D}_j^\bullet=D(z_j^\bullet,1/4)$,
$(z_1^\bullet,z_2^\bullet,z_3^\bullet)=(-1,0,1)$.}
\label{pic_pant}
\end{figure}

\begin{definition}[\textbf{Admissible Riemann surface}]
A connected Riemann surface $S=(\Sigma ,J,{\bf x},\bs{\zeta})$ with genus $g$ and with $m$ marked points ${\bf x}:=(x_{1},\dots,x_m)$ is said to be \textbf{admissible} if it has $b=b^++b^-$ analytic connected boundary components $\pl_1\Sigma,\dots,\pl_b\Sigma$ parametrised by analytic maps $\zeta_k:\T\to \pl_k \Sigma$ that are outgoing for $k\leq b^+$ and incoming for $k>b^+$. It is said to be of \textbf{type $(g,m,b^+,b^-)$}.
A Riemann surface of type $(0,m,b^+,b^-)$ with $m+b^++b^-=3$ is called a  \textbf{complex building block of type $(m,b^+,b^-)$}.

When $m\geq 1$, we also consider \textbf{decorated} Riemann surfaces  $\tilde{S}=(\Sigma,J,{\bf x},\bs{\zeta},{\bf Y})$ of type $(g,m,b^+,b^-)$, or decorated complex building blocks of type $(m,b^+,b^-)$,  
by adding to $(\Sigma,J,{\bf x},\bs{\zeta})$ a collection ${\bf Y}=(Y_1,\dots,Y_m)$ of non-zero tangent vectors $Y_1\in (T_{x_1}\Sigma)^{1,0},\dots,Y_m\in (T_{x_m}\Sigma)^{1,0}$ at the marked points. 
\end{definition}
A complex building block of type $(m,b^+,b^-)$ is diffeomorphic to a sphere  with $b=b^++b^-$ disjoint disks removed. When $b=3$ a complex building block is a pair of pants, when $b=2$ this is an annulus with $1$ marked point, when $b=1$ it is a disk with $2$ marked points. See Figure \ref{pic_pant}.

\begin{lemma}[\textbf{Filling surface, filling biholomorphisms}]\label{FillingSigma}
Let $S=(\Sigma,J,{\bf x},\bs{\zeta})$ be an admissible Riemann surface of type $(g, m,b^+,b^-)$. There is a closed Riemann surface 
$(\hat{\Sigma},\hat{J})$, called \textbf{filling} of $\Sigma$, of genus $g$, with $m+b$ marked points $\hat{{\bf x}}=({\bf x}', {\bf x})$ with ${\bf x}=(x_1,\dots,x_m)$ and ${\bf x}'=(x'_1,\dots,x'_b)$ for $b=b^++b^-$, and some biholomorphisms
\[ \forall k=1,\dots, b^+, \,\, \psi_k:\D^*\to \mc{D}_k\subset \hat{\Sigma},\quad  \forall k=b^++1,\dots, b, \,\, \psi_k:\D\to \mc{D}_k\subset \hat{\Sigma},\] 
called \textbf{filling biholomorphisms}, such that the disks $\mc{D}_k$ are pairwise disjoint, $\Sigma=\hat{\Sigma}\setminus \cup_{k=1}^b\mc{D}_k^\circ$, and $\zeta_k=\psi_k|_{\T}$ for all $k=1,\dots,b$. Moreover, there is $\eps>0$ small enough such that 
for $k=1,\dots,b^+$, $\psi_k^{\rm in}=\psi_k\circ o$ extends holomorphically to $(1+\eps)\D$ with  $x'_k=\psi_k^{\rm in}(0)=\psi_k(\infty)$, and for  $k=b^++1,\dots,b$,
 $\psi_k$ extends holomorphically to $(1+\eps)\D$ with $x'_k=\psi_k(0)$. 

 Conversely, for each closed Riemann surface 
$(\hat{\Sigma},\hat{J})$  of genus $g$ with $m+b$ marked points $\hat{{\bf x}}=({\bf x}',{\bf x})$ of the form 
 ${\bf x}=(x_1,\dots,x_m)$ and ${\bf x}'=(x'_1,\dots,x'_b)$, and with pairwise disjoint
local holomorphic charts $\psi_k:(1+\eps)\D\to \hat{\Sigma}$ such that $\psi_k(0)=x'_k$ and $\psi_k(\D)\cap\{{\bf x}\}=\varnothing$ for $k=1,\dots,b$, 
we obtain a Riemann surface $(\Sigma,J,{\bf x},\bs{\zeta})$ 
 of type $(g(\hat{\Sigma}),m,b^+,b^-)$ by setting $\Sigma=\hat{\Sigma}\setminus \cup_{k=1}^b\psi_k(\D)^\circ$, $\zeta_k:=\psi_k^{\rm out}|_{\T}$ for $k=1,\dots,b^+$ and $\zeta_k:=\psi_k|_{\T}$  for $k=b^++1,\dots,b$.
\end{lemma}
\begin{proof}
On a Riemann surface $\Sigma$ of type $(g,m,b^+,b^-)$, each incoming (resp. outgoing) parametrisation $\zeta_k$ for 
$k=b^++1,\dots,b$ (resp. $k=1,\dots,b^+$) extends holomorphically to a map $\zeta_k:  \A_{1,\delta^{-1}}\to V_k$  (resp. $\zeta_k: \A_{\delta}\to V_k$) where $V_k$ is an annulus containing $\pl_k\Sigma$ and $\delta<1$.
We glue $b^-$ disks $\mc{D}_{k}=\D$ for $k=b^++1,\dots,b$ to $\Sigma$ and $b^+$ disks $\mc{D}_{k}=\D^*=\hat{\C}\setminus \D^\circ $ 
for $k=1,\dots,b^+$ using the charts $\zeta_k$. This can be done as follows: to glue a disk $\D$ 
at an incoming boundary $\pl_{k}\Sigma:=\zeta_k(\T)$, we consider $\Sigma\sqcup \delta^{-1}\D/\sim$ where the equivalence relation is given by $z\in \A_{1,\delta^{-1}} \sim \zeta_k(z)\in V_k\cap \Sigma$, and similarly for an outgoing boundary we consider $\Sigma\sqcup \delta \D^*/\sim$ 
with the equivalence relation $z\in \A_\delta \sim \zeta_k(z)\in V_k\cap \Sigma$.
The identity map $\psi_{k}:\delta^{-1}\D\to \Sigma \sqcup \delta^{-1}\D$, the map $\zeta_k|_{\A_{1,\delta^{-1}}}$ for $k>b^+$, the map 
$\psi_{k}:\delta \D^* \to \Sigma \sqcup \delta \D^*$ given by $\psi_k(z)=z$ and the map $\zeta_k|_{\A_\delta}$ for $k\leq b^+$ 
give holomorphic charts, in a way that  we obtain a structure of closed Riemann surface denoted $(\hat{\Sigma},\hat{J})$  with $m+b$ marked points composed of ${\bf x}$ and ${\bf x}':=(x_1',\dots,x_b')$ with $x_k'=\psi_k(\infty)$ for $k=1,\dots,b^+$ and $x_k'=\psi_k(0)$ for $k=b^++1,\dots,b$. The converse is straightforward.
\end{proof}

There is a natural equivalence class of such Riemann surfaces, defined as follows.
\begin{definition}[\textbf{Moduli space}]\label{equivblocks}
Two admissible Riemann surfaces $(\Sigma^1,J^1,{\bf x}^1,\bs{\zeta}^1)$ and $(\Sigma^2,J^2,{\bf x}^2,\bs{\zeta}^2)$ of type $(g,m,b^+,b^-)$  are said equivalent if there exists $\Phi:(\Sigma^1,J^1)\to (\Sigma^2,J^2)$ biholomorphic, real analytic up to the boundary, such that 
$\zeta^2_k=\Phi\circ \zeta^1_k$ for $k=1,\dots,b=b^++b^-$ and $x_k^2=\Phi(x^1_k)$ for $k=1,\ldots,m$. The \textbf{moduli space} $\mc{M}_{g,m,b^+,b^-}$ is the set of equivalence classes of admissible Riemann surfaces of type $(g,m,b^+,b^-)$.
 
Two decorated Riemann surfaces  $(\Sigma^1,J^1,{\bf x}^1,\bs{\zeta}^1,{\bf Y}^1)$ and $(\Sigma^2,J^2,{\bf x}^2,\bs{\zeta}^2,{\bf Y}^2)$ of type $(g,m,b^+,b^-)$  are said equivalent if there exists $\Phi:(\Sigma^1,J^1)\to (\Sigma^2,J^2)$ biholomorphic, real analytic up to the boundary, such that 
$\zeta^2_k=\Phi\circ \zeta^1_k$ for $k=1,\dots,b=b^++b^-$ and $(x_k^2,Y_k^2)=(\Phi(x_k^1),D\Phi(x_k^1)Y^1_k)$ for $k=1,\ldots,m$. The \textbf{decorated moduli space} $\tilde{\mc{M}}_{g,m,b^+,b^-}$ 
 is the set of equivalence classes of decorated Riemann surfaces of type $(g,m,b^+,b^-)$. 
\end{definition}

We notice that the moduli space $\mc{M}_{g,m,b^+,b^-}$ is infinite dimensional when $b^++b^->0$
and that there is a natural projection 
\[ [ (\Sigma,J,{\bf x},\bs{\zeta},{\bf Y})]\in  \tilde{\mc{M}}_{g,m,b^+,b^-}\mapsto [ (\Sigma,J,{\bf x},\bs{\zeta})] \in \mc{M}_{g,m,b^+,b^-},\]
where the equivalence classes $[(\Sigma,J,{\bf x},\bs{\zeta},{\bf Y})]$ and $[(\Sigma,J,{\bf x},\bs{\zeta})]$ are respectively in the decorated case and the undecorated case.
It should not be confused with the finite dimensional moduli space of surfaces with boundary more commonly used \cite{Farb-Margalit} (sometimes called reduced moduli space), and defined without parametrisation of the boundary components and where the biholomorphisms used in the equivalence classes are not assumed to fix pointwise the boundary. 
In the case with no boundary, i.e.\ $b^+=b^-=0$, and in the stable range $2g-2+m>0$, the moduli space is denoted $\mc{M}_{g,m}$. It has
a natural structure of a complex orbifold of complex dimension $3g-3+m$ (see \cite{Imayoshi-Taniguchi}).

\subsection{Deformations of admissible Riemann surfaces}\label{sec:deformations}
In what follows, we shall need to differentiate functionals defined on sets of admissible Riemann surfaces of type $(g,m,b^+,b^-)$, we thus need 
a functional analytic setting and a topological space for such surfaces.

Throughout this paper, we let $\mc{U}_{\eps}$, $\mc{U}_{\eps}^-$ and $\mc{U}_{\eps}^+$ be arbitrarily small neighbourhoods of ${\rm Id}$ in ${\rm BiHol}_\eps(\A)$, $\{f\in{\rm BiHol}_\eps(\D)\,|\,f(0)=0\}$ and $\{f\in{\rm BiHol}_\eps(\D^*)\,|\,f(\infty)=\infty\}$, respectively. Notice that we can take $\mc{U}_{\eps}^+=\mc{O}(\mc{U}_{\eps}^-)$ (with $\mc{O}$ defined in \eqref{defmcO}) since $\mc{O}$ is a homeomorphism.

There is a right action of $\mc{U}_{\eps}^b=\mc{U}_{\eps}\times \dots \times \mc{U}_{\eps}$ on the set of admissible Riemann surfaces of type $(g,m,b^+,b^-)$ (with $b=b^++b^-$) defined as follows. Let $S=(\Sigma,J,{\bf x},\bs{\zeta})$ be an admissible Riemann surface of type $(g,m,b^+,b^-)$  and let $(\hat{\Sigma},\hat{J})$ be the (closed) filling surface from Lemma \ref{FillingSigma}, with $\bs{\psi}^0=(\psi_1^0,\dots,\psi_b^0)$ the filling biholomorphisms and ${\bf x}'=(x'_1,\dots,x_b')$ the points given by $x_k'=\psi_k^0(\infty)$ for $k\leq b^+$ and $x_k'=\psi_k^0(0)$ for $k>b^+$.
For $\bs{f}=(f_1,\dots,f_b)\in \mc{U}_{\eps}^b$ we define the new surface of type $(g,m,b^+,b^-)$ given by 
\begin{equation}\label{sigma^fzeta^f}
S\cdot {\bs{f}}:=(\Sigma^{\bs{f}},\hat{J},{\bf x},\bs{\zeta}^{\bs{f}})
\end{equation}
where $\bs{\zeta}^{\bs{f}}=(\zeta_1\circ f_1,\dots,\zeta_b\circ f_b)$ and $\Sigma^{\bs{f}}\subset \hat{\Sigma}$ is the connected surface 
which does not contain any of the points $x'_k$ and with boundary components $\pl_k \Sigma^{\bs{f}}=\zeta_k(f_k(\T))$ 
parametrised by $\zeta_k\circ f_k$ for $k=1,\dots,b$. Here if $\mc{U}_\eps$ is small enough, $\bs{\zeta}^{\bs{f}}$ is well-defined and 
$\Sigma^{\bs{f}}\subset \hat{\Sigma}$ is indeed a surface diffeomorphic to $\Sigma$ and with analytic boundary, 
which is a small deformation of $\Sigma$.
In the case where the genus $g=0$ and $m+b^++b^-=3$, for a point $\mc{X}=[S]\in \mc{M}_{0,m,b^+,b^-}$, we will see in the next section that for $\eps>0$ small enough, the set
\begin{equation}\label{Sigma(psi)} 
\mc{U}_{\eps}(\mc{X}) :=\{ [S\cdot \bs{f}]\, |\,  \bs{f}\in \mc{U}_\eps^b\}\subset\mc{M}_{0,m,b^+,b^-}
\end{equation}
provides a neighbourhood of $\mc{X}$ in a suitable $\eps$-subspace of $\mc{M}_{0,m,b^+,b^-}$. In other words, all small deformations of $\mc{X}$ are obtained in this way.

We will study functionals
$F:\mc{U}_\eps^b\to B$ with values in a Banach space $B$, using
the Fr\'echet topology described above. Differentiability is
understood in the following sense: $F$ is of class $C^1$ if
\[ D F(\bs f)\bs v :=\lim_{t\to0,\ t\in\R}
 \frac{F(\bs f+t\bs v)-F(\bs f)}{t}\]
exists for every $(\bs f,\bs v)$, is linear and continuous in
$\bs v$, and the map $(\bs f,\bs v)\mapsto D F(\bs f)\bs v$
is jointly continuous. In particular, for every $C^1$
curve $t\in (-\eps,\eps)\mapsto \bs{f}(t)$ with $\bs{f}(0)=\bs{f}$ and $\dot{\bs{f}}(0)=\bs{v}$, one has $
D F(\bs f)\bs v=\pl_tF(\bs{f}(t))|_{t=0}$. The functionals of interest will typically
be defined on the set $\mc{U}_\eps(\mc X)$ in
\eqref{Sigma(psi)}; we view them as functionals on
$\mc{U}_\eps^b$ satisfying the corresponding invariance under
biholomorphisms.


The tangent space $T_{\rm Id}\mc{U}_\eps^b$ of $\mc{U}_\eps^b$ at ${\rm Id}=({\rm Id},\dots,{\rm Id})$ is the space of holomorphic vector fields
\[T_{\rm Id}\mc{U}_\eps^b:=(T_{\rm Id}\mc{U}_\eps)^b:=\{ (v_1(z)\pl_z,\dots,v_b(z)\pl_z) \, |\,\forall k,v_k\in {\rm Hol}_\eps(\A)\}.\]
Similarly,
\[\begin{split}
T_{\rm Id}\mc{U}_\eps^- &:= \{ v(z)\pl_z \, |\,  v\in {\rm Hol}_\eps(\D), v(0)=0\},\label{Tangent_U_eps}\\
T_{\rm Id}\mc{U}_\eps^+ &:= \{ -z^{2}v(o(z))\pl_z\,|\,  v\in {\rm Hol}_\eps(\D), v(0)=0\}.
\end{split}\]

For ${\bf v}=(v_1\pl_z,\dots,v_b\pl_z)\in T_{\rm Id}\mc{U}_\eps^b$, we can consider the flow $f_k^t$ of $v_k\pl_z$ defined near $\T$ for some small time interval $t\in I$, and define $\bs{f}^t:=(f_1^t,\dots,f_b^t)\in \mc{U}_{\eps'}^b$  for $t\in I$ and some $\eps'>0$ small. For an admissible Riemann surface $S=(\Sigma,J,{\bf x}, \bs{\zeta})$
of type $(g,m,b^+,b^-)$, the family $S\cdot \bs{f}^t$ of surfaces arises as a deformation generated by ${\bf v}$. Such deformations are sometimes trivial when viewed in the moduli space $\mc{M}_{g,m,b^+,b^-}$. Indeed, consider a holomorphic vector field ${\rm v}$ on a neighbourhood $U\subset \hat{\Sigma}$ of $\Sigma$ (with $(\hat{\Sigma},\hat{J})$ the filling of $\Sigma$) and assume that ${\rm v}(x_k)=0$ for $k=1,\dots,m$. If $f^t$ denotes its flow for small time, we have 
\[ [(f^t(\Sigma),\hat{J},{\bf x},\bs{\zeta}^t)]=[S] \textrm{ in }\mc{M}_{g,m,b^+,b^-},\]
where $\bs{\zeta}^t= (f^t\circ \zeta_1,\dots,f^t\circ \zeta_b)$. This deformation can also be written as 
\[(f^t(\Sigma),\hat{J},{\bf x},\bs{\zeta}^t)=S\cdot \bs{f}^t\]
where $\bs{f}^t:=(\zeta_1^{-1}\circ f^t\circ \zeta_1,\dots,\zeta_b^{-1}\circ f^t\circ \zeta_b)$; here $\zeta_k:\T\to \pl_k\Sigma$ 
is extended as a holomorphic map $\A_{\delta,\delta^{-1}}\to\hat{\Sigma}$ for some $\delta<1$, so that $\zeta_k^{-1}\circ f^t\circ \zeta_k\in {\rm BiHol}_{\eps''}(\A)$ for some $\eps''>0$ if $t$ is small enough. The family ${\bf v}\in T_{\rm Id}\mc{U}_\eps^b$ of vector fields arising from this deformation is 
\begin{equation}\label{global_holo_vf}
{\bf v}=(v_1(z)\pl_z,\dots,v_b(z)\pl_z) ,\quad v_k(z)\pl_z:=\pl_t(\zeta_k^{-1}\circ f^t\circ \zeta_k(z))|_{t=0}\pl_z=\zeta_k^*{\rm v}.
\end{equation}
Such deformations will be called \textbf{trivial deformations}, but they will play an important role to uncover the local conformal symmetry, namely the Ward identities associated to the Virasoro algebra. 
If $B$ is a Banach space and $F: \mc{M}_{g,m,b^+,b^-} \to B$ is a differentiable map, viewed as a functional on the space of surfaces of type $(g,m,b^+,b^-)$, and if $(\Sigma,J,{\bf x},\bs{\zeta})$ is a representative of an element $\mc{X}\in \mc{M}_{g,m,b^+,b^-}$, then 
we have $F(S\cdot \bs{f}^t)=F(S)$ for small $t$, hence $DF(S)\bv=0$ for $\bv$ given by \eqref{global_holo_vf}.

%

\subsection{Model form of complex building blocks}\label{sec:model_forms}
We will work with special representatives of complex building blocks that we now define. The main difference in this case is that we can identify the filling surface from Lemma \ref{FillingSigma} with the Riemann sphere $\hat{\C}$, and the representative can be made unique by requiring that the marked points and the centers of the removed disks are located at the prescribed points $-1,0,1$. 

We shall need the following space which is slightly larger than ${\rm BiHol}_\eps(\D)$:
\[\tBiHol_\eps(\D) := \{\textrm{holomorphic embeddings }\psi:(1+\eps)\D\to\hat{\C}\textrm{ with }\psi(0)\neq\infty\}.\]
We equip this space with the Whitney $C^\infty$ topology, so that the subspace topology on ${\rm BiHol}_\eps(\D)\subset \tBiHol_\eps(\D)$ coincides with the Fr\'echet topology defined before by the distance ${\rm dist}_\eps (\cdot,\cdot)$ in \eqref{distance}.

\begin{definition}[\textbf{Model complex building blocks}]\label{model_building_block}
A \textbf{model complex building block} of type $(m,b^+,b^-)$ is an admissible Riemann surface with $m\leq 2$ marked points and $b=b^++b^-=3-m$ 
parametrised boundary components of the form
\[(\hat{\C}\setminus \cup_{k=1}^b\mc{D}^\circ_k,J_{\C},{\bf x},(\psi_1|_{\T},\dots,\psi_{b}|_{\T}))\]
where $J_\C$ is the canonical complex structure on $\hat{\C}$, and for some $\eps>0$, $\psi_k\circ o\in\tBiHol_\eps(\D)$ and $\mc{D}_k=\psi_k(\D^*)$ for $k=1,\dots,b^+$, $\psi_k\in\tBiHol_\eps(\D)$ and $\mc{D}_k=\psi_k(\D)$ for $k=b^++1,\dots,b$, and we require that the disks $\mc{D}_k$ are pairwise disjoint and ${\bf x}=(x_{b+1},\dots,x_3)\subset (\hat{\C}\setminus \cup_{k=1}^b\mc{D}_k)^m$ are distinct marked points.
If $x'_k:=\psi_k^{\rm in}(0)=\psi_k(\infty)$ for $k=1,\dots,b^+$ and $x_k':=\psi_k(0)$ for $k=b^++1,\dots,b$, we say that the model complex building block is \textbf{centered} if 
\[(x'_1,\dots,x'_b,x_{b+1},\dots,x_3)=(-1,0,1).\]
\end{definition}
 
Let us show that a complex building block  is equivalent to a unique centered  model complex building block that we call its \textbf{model form}. 
\begin{lemma}[\textbf{Model form}]\label{modelform}
Let  $S=(\mc{P},J,{\bf x},\bs{\zeta})$ be a complex building block of type $(m,b^+,b^-)$. There exists a biholomorphism $\Phi:\mc{P}\to \Phi(\mc{P})\subset \hat{\C}$ with $\hat{\C}\setminus \Phi(\mc{P})=\cup_{k=1}^b\mc{D}^\circ_k$ with the $\mc{D}_k$ pairwise disjoint, and there exist biholomorphisms $\psi_k:\D^*\to \mc{D}_k$ for $k=1,\dots,b^+$  admitting a holomorphic extension to a neighbourhood of $\D^*$ and biholomorphism $\psi_k:\D\to \mc{D}_k$ for $k=b^++1,\dots,b$ admitting a holomorphic extension to a neighbourhood of $\D$, such that $\Phi\circ \zeta_k=\psi_k|_{\T}$ for $k=1,\dots,b$.
In other words, each complex building block is equivalent to a model complex building block
\[(\hat{\C}\setminus\cup_{k=1}^b\mc{D}^\circ_k,J_{\C},{\bf z},( \psi_{1}|_{\T},\dots,\psi_{b}|_{\T}))\]
with ${\bf z}=\Phi({\bf x})\in (\hat{\C}\setminus \cup_{k=1}^b\mc{D}_k)^m$.  
The \textbf{uniformising biholomorphism} $\Phi$ is unique up to left composition by a M\"obius transformation $M\in {\rm PSL}_2(\C)$. Two complex building blocks $(\mc{P}^1,J^1,{\bf x}^1,\bs{\zeta}^1)$ and $(\mc{P}^2,J^2,{\bf x}^2,\bs{\zeta}^2)$ are equivalent if and only if they have the same model complex building block up to a M\"obius transformation. In particular, the map $\Phi$ is unique if we require that the   model complex building block it maps to is centered.
\end{lemma}
\begin{proof} We give the proof in the case with incoming boundaries, the general case is essentially the same.
First we consider the closed  filling Riemann surface $(\hat{\mc{P}},\hat{J})$ of Lemma \ref{FillingSigma} obtained by gluing $b$ disks $\mc{D}_1,\dots,\mc{D}_b$ to $\mc{P}$ using the boundary parametrisations. The glued surface is a sphere 
with $3$ marked points   $\hat{{\bf x}}=({\bf x}',\bf x)$, ${\bf x}'$ being the set of centers of the glued disks $\mc{D}_k$. 
By the uniformisation theorem, there is a  biholomorphic map 
$\hat{\Phi}: (\hat{\mc{P}},\hat{J})\to (\hat{\C},J_\C)$, and two such biholomorphisms are necessarily related by an element $M\in {\rm PSL}_2(\C)$. Now we set 
$\psi_k:=\hat{\Phi}\circ \zeta_k:\T\to \psi_k(\T)$ which extend holomorphically in some disk $(1+\eps)\D$. 
This produces a model block as desired. 
It is direct to check that a biholomorphism $\phi:\mc{P}^1\to \mc{P}^2$ mapping a model complex building block to another model block (thus with $\phi\circ \zeta^1_k=\zeta_k^2$) must extend holomorphically to $\hat{\C}$ and thus is an element in ${\rm PSL}_2(\C)$. 

Finally, let $S^1=(\mc{P}^1,J^1,{\bf x}^1,\bs{\zeta}^1)$ and $S^2=(\mc{P}^2,J^2,{\bf x}^2,\bs{\zeta}^2)$ be two complex building blocks represented as above with, say, $\mc{D}^1_k\subset \hat{\C}$ and $\mc{D}^2_k\subset \hat{\C}$ the disks as above  corresponding to $\mc{P}^1$ and $\mc{P}^2$. Assume they are equivalent by a map $\Psi:\mc{P}^2\to \mc{P}^1$. 
Let $\Phi^1$ be a  uniformising map putting $\mc{P}^1$ under model form 
$(\psi_1^1,\dots,\psi_b^1)$ with marked points ${\bf z}^1=(z^1_1,\dots z^1_m)$. Then $\Phi^2:=\Phi^1\circ \Psi: \mc{P}^2 \to \hat{\C}\setminus\cup_{k=1}^b\psi^1_k(\D^\circ)$ and $\Phi^2\circ \zeta^2_k:\T\to \pl \mc{D}^1_k$ is equal to $\Phi^1\circ \zeta^1_k=\psi^1_k$, which is holomorphic near 
$\D$. This shows that $\Phi^2$ is a map putting $S^2$ under model form, which in turn is the model form of $S^1$.
\end{proof}

The (decorated) moduli space of complex building blocks is the set of equivalence classes of (decorated) complex building blocks in the sense of Definition \ref{equivblocks}:
\[ \mc{P}_{m,b^+,b^-} :=\mc{M}_{0,m,b^+,b^-}, \quad   \tilde{\mc{P}}_{m,b^+,b^-} :=\tilde{\mc{M}}_{0,m,b^+,b^-}, \qquad\textrm{with }m+b^++b^-=3.\]
By convention, when $b^+=0$ we write $b=b^-$, $\mc{P}_{m,b}:=\mc{P}_{m,0,b}$ and  $\tilde{\mc{P}}_{m,b}:=\tilde{\mc{P}}_{m,0,b}$. Throughout the paper, we consider the fixed triplet of points in $\hat{\C}$
\begin{equation}\label{defz_0}
{\bf z}^\bullet:=(z^{\bullet}_1,z^{\bullet}_2,z^{\bullet}_3):=(-1,0,1).
\end{equation} 
Using Lemma \ref{modelform}, we can identify  
\[\mc{P}_{m,b}\simeq \bigcup_{\eps>0}\mc{P}^\eps_{m,b},\quad\mc{P}_{m,b^+,b^-}\simeq \bigcup_{\eps>0}\mc{P}^\eps_{m,b^+,b^-},\]
where (recall the notation $\psi^{\rm out}$ from \eqref{psi_in/out})
\begin{equation}\label{defPmb}
\begin{gathered}
\mc{P}^{\eps}_{m,b}:= \{ \bs{\psi}=(\psi_1,\dots,\psi_b)\in \tBiHol_{\eps}(\D)^{b}\,|\, \forall j\not=k,  \psi_j(\D)\cap \psi_k(\D)=\emptyset=\psi_j(\D)\cap\{z^{\bullet}_{\ell}\}_{\ell=b+1}^3, \psi_j(0)=z^{\bullet}_j\},\\
\mc{P}^{\eps}_{m,b^+,b^-}:=\{ ( \psi^{\rm out}_{1},\dots,\psi_{b^+}^{\rm out}, \psi_{b^++1},\dots,\psi_{b}) \, | \, (\psi_1,\dots,\psi_b)\in \mc{P}_{m,b}^\eps\}.
\end{gathered}
\end{equation}
The canonical complex building block associated to an element $\bs{\psi}\in \mc{P}^\eps_{m,b^+,b^-}$ is the surface 
$(\mc{P},J_\C, {\bf x}^\bullet,\bs{\zeta})$ with 
\begin{equation}\label{building_block_from_mf}
\mc{P}=\hat{\C}\setminus \Big(\bigcup_{k=1}^{b^+}\psi_k(\D^*)^\circ\cup \bigcup_{k=b^++1}^b\psi_k(\D)^\circ\Big),\quad \bs{\zeta}=\bs{\psi}|_{\T}, \quad {\bf x}^\bullet=(z^\bullet_{b+1},\dots,z^\bullet_3).
\end{equation}
Similarly for the decorated case, we have identifications $\tilde{\mc{P}}_{m,b}\simeq\bigcup_{\eps>0}\tilde{\mc{P}}_{m,b}^\eps$, $\tilde{\mc{P}}_{m,b^+,b^-}\simeq\bigcup_{\eps>0}\tilde{\mc{P}}_{m,b^+,b^-}^\eps$, with 
\begin{equation}
\tilde{\mc{P}}^{\eps}_{m,b}:= \{ (\bs{\psi},\bs{\lambda})\in \mc{P}^{\eps}_{m,b}\times (\C\setminus\{0\})^m\},
\end{equation}
where $\bs{\lambda}=(\lambda_1,\ldots,\lambda_m)$ with each $\la_j\in \C\setminus\{0\}\simeq(T_{z^{\bullet}_{b+j}}\C)^{1,0}\setminus\{0\}$ viewed as a non-zero holomorphic tangent vector at $z_{b+j}^\bullet$, and $\tilde{\mc{P}}_{m,b^+,b^-}^\eps$ is defined similarly.
For each $b\in \{1,2,3\}$, there is a canonical projection
\begin{equation}\label{defpi3tob} 
 \pi_{3\to b}: \mc{P}^{\eps}_{0,3} \to \tilde{\mc{P}}^{\eps}_{m,b}, \qquad (\psi_1,\psi_2,\psi_3)\mapsto 
(\psi_1,\dots,\psi_b, \psi'_{b+1}(0),\dots,\psi_3'(0)).
\end{equation}

The space $\mc{P}^{\eps}_{m,b}$ inherits its topology as a subspace of $\tBiHol_{\eps}(\D)^{b}$. We thus equip the moduli space $\mc{P}_{m,b}$ with the direct limit topology on its identified space $\bigcup_{\eps>0}\mc{P}^\eps_{m,b}$. For $\bs{\psi}^0=(\psi^0_1,\dots,\psi^0_b)\in\mc{P}^{\eps}_{m,b}\subset \mc{P}_{m,b}$, the right action of $(\mc{U}_{\eps}^-)^b$ produces a set
\[\bs{\psi}^0\cdot (\mc{U}_{\eps}^-)^b:=\{ \bs{\psi}^0\cdot\bs{f}:=(\psi^0_1\circ f_1,\dots, \psi^0_b\circ f_b)\,|\, \bs{f}=(f_1,\ldots,f_b)\in(\mc{U}_{\eps}^-)^b\}.\]
Assume that $\bs{\psi}^0\in\mc{P}^{\eps'}_{m,b}$ for some $\eps'>\eps$. Shrinking the neighbourhood $\mc{U}_{\eps}^-$ if necessary, we can assume $f((1+\eps)\D)\subset(1+\eps')\D$ for all $f\in\mc{U}_{\eps}^-$, so that $\bs{\psi}^0\cdot (\mc{U}_\eps^-)^b\subset\mc{P}_{m,b}^{\eps}$. We claim that it is a neighbourhood of $\bs{\psi}^0$ in $\mc{P}_{m,b}^{\eps}$. For each $k$, since $\psi_k^0((1+\eps)\D)\subset\psi_k^0((1+\eps')\D)^\circ$, there exists a neighbourhood $V_k$ of $\psi_k^0$ in $\{\psi\in\tBiHol_{\eps}(\D)\,|\,\psi(0)=\psi_k^0(0)\}$ such that $\psi_k((1+\eps)\D)\subset\psi_k^0((1+\eps')\D)$ for all $\psi_k\in V_k$. Then $(\psi_k^0)^{-1}\circ\psi_k:(1+\eps)\D\to(1+\eps')\D$ is well-defined and $(\psi_k^0)^{-1}\circ\psi_k\to{\rm Id}$ in ${\rm BiHol}_{\eps}(\D)$ as $\psi_k\to\psi_k^0$ in $\tBiHol_{\eps}(\D)$. Thus, shrinking $V_k$ if necessary, we can assume $(\psi_k^0)^{-1}\circ\psi_k\in\mc{U}_{\eps}^-$ for $\psi_k\in V_k$. Since $\psi_k=\psi_k^0\circ((\psi_k^0)^{-1}\circ\psi_k)$, we have $\bs{\psi}^0\cdot (\mc{U}_{\eps}^-)^b\supset\prod_kV_k$, which is an open set in $\mc{P}_{m,b}^{\eps}$ containing $\bs{\psi}^0$. Thus the claim is proved. Moreover, the map
\[\bs{f}\in(\mc{U}_\eps^-)^b\mapsto\bs{\psi}^0\cdot\bs{f}\in\mc{P}_{m,b}^\eps\]
is clearly a continuous embedding. To recapitulate, given a point $\bs{\psi}^0\in\bigcup_{\eps'>\eps}\mc{P}_{m,b}^{\eps'}\subset\mc{P}_{m,b}^\eps$, the above procedure produces a continuous parametrisation of a neighbourhood of $\bs{\psi}^0$ in $\mc{P}_{m,b}^\eps$ by $(\mc{U}_\eps^-)^b$. This allows us to make the identification
\begin{equation}\label{Tangent_space}
T_{\bs{\psi}^0}\mc{P}^{\eps}_{m,b}\simeq(T_{\rm Id}\mc{U}_\eps^-)^b=\{ (v_1(z)\pl_z,\dots,v_b(z)\pl_z) \, |\, \forall j, v_j\in {\rm Hol}_\eps(\D), v_j(0)=0\}.
\end{equation}
In fact, since $\bigcup_{\eps'>\eps}\mc{P}_{m,b}^{\eps'}$ is dense in $\mc{P}_{m,b}^\eps$, we shall make this identification formally for all $\bs{\psi}^0\in\mc{P}_{m,b}^\eps$ (keeping in mind that no actual differential structure on $\mc{P}_{m,b}^\eps$ has been defined). The space $\mc{P}^{\eps}_{m,b^+,b^-}$ can be treated similarly using $(\mc{U}_\eps^+)^{b^+}\times(\mc{U}_\eps^-)^{b^-}$.

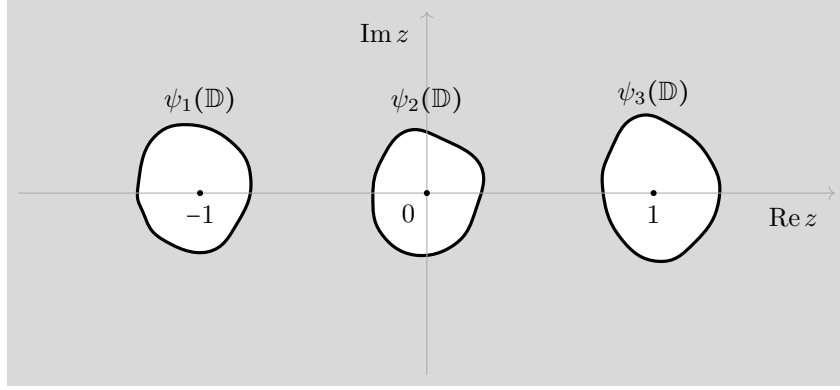
\begin{figure}[ht]
\centering
\begin{tikzpicture}[
  scale=3,
  line cap=round,
  line join=round
]

\fill[gray!30] (-1.85,-0.85) rectangle (1.85,0.85);

\draw[fill=white,draw=black,very thick]
  plot[smooth cycle,tension=0.75] coordinates {
    (-1.273, 0.035)
    (-1.210, 0.238)
    (-1.042, 0.301)
    (-0.853, 0.217)
    (-0.776, 0.056)
    (-0.832,-0.126)
    (-0.965,-0.259)
    (-1.168,-0.203)
    (-1.252,-0.084)
  };

\draw[fill=white,draw=black,very thick]
  plot[smooth cycle,tension=0.72] coordinates {
    (-0.238,-0.014)
    (-0.203, 0.140)
    (-0.091, 0.273)
    ( 0.070, 0.238)
    ( 0.238, 0.119)
    ( 0.217,-0.070)
    ( 0.126,-0.224)
    (-0.056,-0.273)
    (-0.203,-0.175)
  };

\draw[fill=white,draw=black,very thick]
  plot[smooth cycle,tension=0.78] coordinates {
    (0.776, 0.014)
    (0.811, 0.189)
    (0.923, 0.336)
    (1.091, 0.294)
    (1.245, 0.154)
    (1.287,-0.035)
    (1.182,-0.217)
    (1.021,-0.301)
    (0.853,-0.196)
  };

\draw[->,gray!65] (-1.8,0) -- (1.8,0)
  node[anchor=north east,xshift=-3pt,yshift=-2pt,black] {$\operatorname{Re}z$};

\draw[->,gray!65] (0,-0.8) -- (0,0.8)
  node[anchor=north east,xshift=-3pt,yshift=-1pt,black] {$\operatorname{Im}z$};

\node[anchor=south] at (-1,0.31) {$\psi_1(\D)$};
\node[anchor=south] at ( 0,0.31) {$\psi_2(\D)$};
\node[anchor=south] at ( 1,0.34) {$\psi_3(\D)$};

\node[circle,fill,inner sep=0.8pt,label=below:$-1$]
  at (-1,0) {};
\node[circle,fill,inner sep=0.8pt,label=below left:$0$]
  at (0,0) {};
\node[circle,fill,inner sep=0.8pt,label=below:$1$]
  at (1,0) {};

\end{tikzpicture}
\caption{The $3$ uniformising maps
$\bs{\psi}=(\psi_1,\psi_2,\psi_3)$ representing a model complex
building block with $3$ incoming boundaries.}
\label{fig:three-biholomorphic-disks}
\end{figure}

Let us now make an important remark. For $\bs{\psi}^0\in \mc{P}^{\eps}_{m,b^+,b^-}$ and $S=(\mc{P}, J,{\bf x},\bs{\zeta})$ the associated 
complex building block defined in \eqref{building_block_from_mf}, the action of $\mc{U}_\eps^b$ produces a set of complex building blocks $S\cdot \bs{f}$ for $\bs{f}\in  \mc{U}_\eps^b$
as in \eqref{sigma^fzeta^f}. They are \textit{not} under model form because the maps $\psi_k^0\circ f_k$ (or $\psi_k^0\circ f_k\circ o$ in the outgoing case) where $f_k\in\mc{U}_\eps \subset {\rm BiHol}_\eps(\A)$ are not elements in $\tBiHol_\epsilon(\D)$, i.e.\ they may not extend holomorphically to $\D$. Hence they cannot directly be considered as elements in $\mc{P}^{\eps}_{m,b^+,b^-}$ as defined in \eqref{defPmb}. 
However, Lemma \ref{modelform} implies that each $S\cdot \bs{f}$ can be uniformised into a model form, i.e.\ it is equivalent to an element in $\mc{P}^{\eps'}_{m,b^+,b^-}$ for some $\eps'>0$. 
The next lemma shows that this can be done in a  $C^1$ fashion.
\begin{lemma}[\textbf{Regularity of model form uniformisation}]\label{PhiC1}
Let $\bs{\psi}^0\in \mc{P}^{\eps}_{m,b^+,b^-}$ and $S=(\mc{P}, J,{\bf x},\bs{\zeta})$ be the associated 
complex building block defined in \eqref{building_block_from_mf}.
For $\bs{f}\in \mc{U}_\eps^b$ (with $b=b^++b^-$), the uniformising map $\Phi_{\bs{f}}$ of $(\mc{P}, J,{\bf x},\bs{\zeta})\cdot \bs{f}$ 
onto the centered model form, defined by Lemma \ref{modelform}, is continuous with respect to $\bs{f}$, and the mapping into the model form
\[
\bs{f} \in \mc{U}_\eps^b\mapsto  (\Phi_{\bs{f}}\circ \psi_1^0\circ f_1, \dots,\Phi_{\bs{f}}\circ \psi_b^0\circ f_b)\in \mc{P}^{\eps'}_{m,b^+,b^-}
\]
is $C^1$ for any $0<\eps'<\eps$ (we may need to shrink $\mc{U}_\eps$ depending on $\eps'$), where each $\Phi_{\bs{f}}\circ \psi_k^0\circ f_k$ implicitly extends to a holomorphic map on $(1+\eps')\D$ for $k>b^+$ and to $(1+\eps')^{-1}\D^*$ for $k\leq b^+$.
Moreover, this mapping is holomorphic in the sense that if $\bs{\tau}\mapsto \bs{f}(\bs{\tau})$ is a holomorphic map from an open neighbourhood of $0\in \C^\ell$ to $\mc{U}_\eps^b$ with $\bs{f}(0)={\rm Id}$, then 
\[\bs{\tau}\mapsto (\Phi_{\bs{f}(\bs{\tau})}\circ \psi_1^0\circ f_1(\bs{\tau}), \dots,\Phi_{\bs{f}(\bs{\tau})}\circ \psi_b^0\circ f_b(\bs{\tau}))\in \mc{P}^{\eps'}_{m,b^+,b^-}\]
is holomorphic on this open set.
\end{lemma}
\begin{proof} Let us deal with the case where the boundaries are all incoming; the other cases are proved in the same way. 
For this, we need to recall the explicit construction of $\Phi_{\bs{f}}$ using the Beltrami equation in $\hat{\C}$. 
Let $\mc{P}^{\bs{f}}$ be defined as in \eqref{sigma^fzeta^f} (with $\Sigma=\mc{P}$). 
It is a subset of the Riemann sphere $\mc{P}^{\bs{f}}=\hat{\C}\setminus \cup_{k=1}^b(\mc{D}^{\bs{f}}_k)^\circ$ 
where $\mc{D}^{\bs{f}}_k\subset \hat{\C}$ is the (topological) disk with analytic boundary containing $z^\bullet_k$.
The conformal structure $J$ of $\mc{P}^{\bs{f}}$ is given by $J_\C$. Shrinking $\mc{U}_\eps$ if necessary, we may assume that $f(\A_{(1+\eps')^{-1},1+\eps'})\subset\A_{(1+\eps)^{-1},1+\eps}^\circ$ for all $f\in\mc{U}_\eps$, so that the parametrisation $\psi_k:=\psi_k^0\circ f_k:\T\to \pl \mc{D}^{\bs{f}}_k$ extends holomorphically to $\A_{(1+\eps')^{-1},1+\eps'}$. We then extend $f_k$ to $(1+\eps')\D$ by the map $\hat{f}_k: z\mapsto q(z)f_k(z)+(1-q(z))z$ where $q\in C^\infty(\C,[0,1])$ is equal to $1$ in $\A_{(1+\eps'/2)^{-1},1+\eps'}$ and $0$ in $(1+\eps')^{-1}\D$. Then, if $\mc{U}_\eps^b$ is a small enough neighbourhood of the identity, the induced extension $\hat{\psi}_k=\psi_k^0\circ \hat{f}_k$ of $\psi_k$ from $(1+\eps')\D$ to $U_k$ is a smooth diffeomorphism with 
$U_k$ a neighbourhood of $\mc{D}^{\bs{f}}_k$ and $\hat{\psi}_k(\D)=\mc{D}_k^{\bs{f}}$.
Then consider the Beltrami coefficient $\mu_{\psi_k^0\circ \hat{f}_k^{-1}\circ (\psi_k^0)^{-1}}$ of $\psi_k^0\circ \hat{f}_k^{-1}\circ (\psi_k^0)^{-1}$ on $\hat\psi_k((1+\eps')\D)$, which is smooth and vanishes outside $\hat\psi_k((1+\eps'/2)^{-1}\D)$, and let $\mu_{\bs{f}}$ be the smooth Beltrami coefficient that extends $\sum_{k=1}^b \mu_{\psi_k^0\circ \hat{f}_k^{-1}\circ (\psi_k^0)^{-1}}$ by $0$ on $\hat{\C}$. 
The desired map $\Phi_{\bs{f}}$ is the restriction to $\mc{P}^{\bs{f}}$ of the quasiconformal map $\Phi_{\bs{f}}$ on $\hat{\C}$ that is the unique solution of the Beltrami equation 
\[\pl_{\bar{z}}\Phi_{\bs{f}}=\mu_{\bs{f}} \pl_{z}\Phi_{\bs{f}}, \quad 
\Phi_{\bs{f}}(z^\bullet_k)=z^\bullet_k \textrm{ for }k=1,2,3.\]
Indeed, $\Phi_{\bs{f}}|_{\mc{P}^{\bs{f}}}$ is a biholomorphism onto its image,
and $\Phi_{\bs{f}}\circ (\psi_k^0\circ \hat{f}_k \circ (\psi_k^0)^{-1})$ is a biholomorphism on $\psi_k^0((1+\eps')\D)$, thus 
$\Phi_{\bs{f}}\circ \psi_k^0\circ \hat{f}_k$ is the holomorphic extension to $(1+\eps')\D$ of $\Phi_{\bs{f}}\circ \psi_k^0\circ f_k$.
The function $\mu_{\bs{f}}\in C_c^\infty(\hat{\C})$  depends in a $C^1$ fashion on $\bs{f}$, with support contained in $\bigcup_{k=1}^b\hat\psi_k((1+\eps'/2)^{-1}\D)$. The map $\mu \mapsto \Phi^\mu|_{B(0,R)}$ solving $\pl_{\bar{z}}\Phi^\mu=\mu \pl_z\Phi^\mu$ on $\hat{\C}$ and which fixes $z^\bullet_k$ for $k=1,2,3$ is $C^1$, and in fact complex analytic, as a map
\[\{\mu \in L^\infty(R\D)\,|\, |\mu|\leq K<1\}\to C^0(R\D)\]
 for $R>0$ and $K<1$  fixed (see \cite{Ahlfors_Bers,NagBook}). This implies that 
  \[ \bs{f}\in \mc{U}_\eps^b \mapsto (\Phi_{\bs{f}}\circ \psi_1,\dots,\Phi_{\bs{f}}\circ \psi_b)\in \mc{P}^{\eps'}_{m,b}\] 
is $C^1$ if one uses the $C^0((1+\eps')\D)$ norm in the target space.
Since the seminorms $\|\cdot\|_{\eps'/2,k}$ of \eqref{seminorm} are all controlled by the $C^0((1+\eps')\D)$ norm thanks to holomorphy, we see that the mapping is $C^1$.

To prove the holomorphic dependence, we slightly modify the preceding extension procedure. Let \(U\subset\mathbb C^\ell\) be an open neighbourhood of $0$, and assume that $(\bs{\tau},z)\mapsto f_k(\bs{\tau},z)$
is jointly holomorphic on a fixed annulus. After shrinking \(U\) and the annulus if necessary, the inverse
$h_k(\bs{\tau}):=f_k(\bs{\tau})^{-1}$
is defined on a fixed annulus and depends holomorphically on $(\boldsymbol\tau,z)$, by the holomorphic implicit function theorem.
Choose $q\in C^\infty(\mathbb C,[0,1])$ as before, which is equal to $1$ on a fixed annular neighbourhood of the curves
$f_k(\boldsymbol\tau)(\mathbb T)$, for all $\bs{\tau}\in U$, and which is equal to $0$ on a slightly smaller disk. We define
$\hat h_k(\boldsymbol\tau,z):=q(z)h_k(\bs{\tau},z)+(1-q(z))z.$
Shrinking $U$ if necessary, this is a smooth orientation-preserving diffeomorphism. Moreover,
\[\partial_{\bar z}\hat h_k= (\partial_{\bar z}q)(h_k-z), \qquad
\partial_z\hat h_k = q\pl_zh_k+(1-q)+(\partial_zq)(h_k-z).
\]
It follows that the Beltrami coefficient $\bs{\tau}\mapsto\mu_{\hat h_k(\bs{\tau})} = \partial_{\bar z}\hat h_k(\bs{\tau})/\partial_z\hat h_k(\bs{\tau})$
is holomorphic with values in $L^\infty$. Set $H_{k,\bs{\tau}}:=\psi_k^0\circ\hat h_k(\bs{\tau})\circ(\psi_k^0)^{-1}$, 
and let \(\mu_{\bs{\tau}}\) coincide with $\mu_{H_{k,\boldsymbol\tau}}$ on the corresponding disks and vanish elsewhere. The supports may be chosen inside fixed compact subsets of the filling disks, and $\bs{\tau}\mapsto\mu_{\bs{\tau}}$
is holomorphic with values in $L^\infty(\widehat{\mathbb C})$. Let \(\Phi_{\boldsymbol\tau}\) be the normalised solution of
\[ \partial_{\bar z}\Phi_{\boldsymbol\tau} =\mu_{\bs{\tau}}\partial_z\Phi_{\bs{\tau}},
\qquad \Phi_{\bs{\tau}}(z_r^\bullet)=z_r^\bullet, \quad r=1,2,3.\]
By \cite[Theorem 11]{Ahlfors_Bers}, $\boldsymbol\tau\mapsto\Phi_{\bs{\tau}}$ is holomorphic. Since
$\mu_{\bs{\tau}}$ vanishes on the deformed building block, $\Phi_{\bs{\tau}}$ is conformal there. Furthermore,
$\Phi_{\bs{\tau}}$ and \(H_{k,\boldsymbol\tau}\) have the same Beltrami coefficient, and hence $\Phi_{\boldsymbol\tau}\circ H_{k,\boldsymbol\tau}^{-1}$ is holomorphic by \cite[Theorem 5(vii)]{Ahlfors_Bers}. On a neighbourhood of the boundary, $H_{k,\boldsymbol\tau}^{-1}\circ\psi_k^0 =
\psi_k^0\circ f_k(\boldsymbol\tau),$ and therefore $\Phi_{\boldsymbol\tau}\circ\psi_k^0\circ f_k(\boldsymbol\tau)$
extends holomorphically to the filling disk. Finally, choose a fixed circle $\{|\zeta|=r\}$ contained in the common annulus of definition. The holomorphic extension satisfies 
\[\bigl(\Phi_{\bs{\tau}}\circ\psi_k^0 \circ f_k(\bs{\tau})\bigr)(z)
 = \frac{1}{2\pi{\rm i}} \oint_{|\zeta|=r} \frac{ \Phi_{\bs{\tau}}
 \bigl(\psi_k^0(f_k(\bs{\tau},\zeta))\bigr)}
 {\zeta-z}\,\mathrm d\zeta, \qquad |z|<r.
\]
The integrand depends holomorphically on $\bs{\tau}$. Cauchy estimates on every slightly smaller disk therefore imply the desired holomorphic dependence with values in $\mc{P}^{\epsilon'}_{m,b^+,b^-}$.
\end{proof}

\subsection{Teichm\"uller space and mapping class group}\label{sub:Teichmuller}

To understand the conformal blocks, we will need to work with Riemann surfaces equipped with a marking. 

\begin{definition}[\textbf{Teichm\"uller space}]\label{def:Teichmuller}
1) Let $S^{\bullet}=(\Sigma^{\bullet},J^\bullet,{\bf x}^\bullet,\bs{\zeta}^\bullet)$ be a fixed base Riemann surface of type $(g,m,0,b)$. For $b^+,b^-$ with $b^++b^-=b$, the \textbf{Teichm\"uller space} $\mc{T}_{g,m,b^+,b^-}=\mc{T}_{g,m,b^+,b^-}(S^\bullet)$ is the set of equivalence classes  $[(F, \Sigma,J,{\bf x},{\bs \zeta})]$, where $(\Sigma,J,{\bf x},{\bs \zeta})$ is a Riemann surface of type $(g,m,b^+,b^-)$ and $F$ is an orientation preserving smooth diffeomorphism  $F: \Sigma^{\bullet}\to \Sigma$ that is holomorphic near $\{{\bf x}^\bullet\}\cup\pl\Sigma^{\bullet}$ and such that $F\circ \zeta^\bullet_j=\zeta_j\circ o$ for $j=1,\dots,b^+$, $F\circ \zeta^\bullet_j=\zeta_j$ for $j=b^++1,\ldots,b$, $F(x_k^\bullet)=x_k$ for $k=1,\ldots,m$,
and $(F^1,\Sigma^1,J^1,{\bf x}^1,{\bs \zeta}^1)$ and $(F^2,\Sigma^2,J^2,{\bf x}^2,{\bs \zeta}^2)$ are said equivalent if they are equivalent as Riemann surfaces of type $(g,m,b^+,b^-)$ by some biholomorphism $\Phi: (\Sigma^1,J^1)\to (\Sigma^2,J^2)$ such that $(F^2)^{-1}\circ \Phi\circ F^1$ is isotopic to the identity on $\Sigma^{\bullet}$ via an isotopy equal to the identity on the marked points ${\bf x}^\bullet$ and the boundary $\pl \Sigma^{\bullet}$; we say that the isotopy is relative to $\{{\bf x}^\bullet\}\cup\pl\Sigma^{\bullet}$. We 
also denote $\mc{T}_{g,m,b}:=\mc{T}_{g,m,0,b}$ when all boundary components are incoming, i.e. $b^+=0$ and $b=b^-$.

2) Let $\tilde{S}^\bullet=(\Sigma^{\bullet},J^\bullet,{\bf x}^\bullet,\bs{\zeta}^\bullet,{\bf Y}^\bullet)$ where $(\Sigma^{\bullet},J^\bullet,{\bf x}^\bullet,\bs{\zeta}^\bullet)$ is as above and ${\bf Y}^\bullet=(Y^\bullet_1,\dots,Y^\bullet_m)$ with non-zero $Y_j^\bullet\in(T_{x^\bullet_j}\Sigma^\bullet)^{1,0}$. For $b^+,b^-$ with $b^++b^-=b$, the \textbf{decorated Teichm\"uller space} $\tT_{g,m,b^+,b^-}=\tT_{g,m,b^+,b^-}(\tilde{S}^\bullet)$ is the set of equivalence classes $[(F, \Sigma,J,{\bf x},{\bs \zeta},{\bf Y})]$, where $(\Sigma,J,{\bf x},{\bs \zeta},{\bf Y})$ is a decorated Riemann surface of type $(g,m,b^+,b^-)$ and $F$ is as above and such that $DF(x_k^\bullet)Y_k^\bullet=Y_k$ for $k=1,\ldots,m$, and $(F^1,\Sigma^1,J^1,{\bf x}^1,{\bs \zeta}^1,{\bf Y}^1)$ and $(F^2,\Sigma^2,J^2,{\bf x}^2,{\bs \zeta}^2,{\bf Y}^2)$ are said equivalent if they are equivalent as decorated Riemann surfaces of type $(g,m,b^+,b^-)$ by some biholomorphism $\Phi: (\Sigma^1,J^1)\to (\Sigma^2,J^2)$ such that $(F^2)^{-1}\circ \Phi\circ F^1$ is isotopic to the identity on $\Sigma^{\bullet}$ via an isotopy relative to $\{{\bf x}^\bullet\}\cup\pl\Sigma^{\bullet}$ and preserving ${\bf Y}^\bullet$.
We also denote $\tT_{g,m,b}:=\tT_{g,m,0,b}$ when all boundary components are incoming.
\end{definition}
We notice that the Teichm\"uller space depends on the choice of a base surface $S^\bullet=(\Sigma^{\bullet},J^\bullet,{\bf x}^\bullet,\bs{\zeta}^\bullet)$ of type $(g,m,0,b)$, but for two different choices of such base surfaces with the same type, there is an isomorphism between the Teichm\"uller spaces. Similarly, the decorated Teichm\"uller space does not depend on the choice of $\tilde{S}^\bullet=(\Sigma^{\bullet},J^\bullet,{\bf x}^\bullet,\bs{\zeta}^\bullet,{\bf Y}^\bullet)$ in this sense.

For sets $U\subset\Sigma^\bullet$ and $Y\subset T\Sigma^{\bullet}$ with $\pi(Y)\subset U$ (where $\pi:T\Sigma^\bullet\to\Sigma^\bullet$ is the canonical projection), we denote by ${\rm Diff}(\Sigma^\bullet ,U)$ the group of orientation preserving smooth diffeomorphisms of $\Sigma^{\bullet}$ that are biholomorphic with respect to $J^\bullet$ in a neighbourhood of $U$ and equal to the identity on $U$, and by ${\rm Diff}(\Sigma^\bullet,U;Y)$ the subgroup of $F\in {\rm Diff}(\Sigma^\bullet,U)$ whose differential $DF$ is equal to the identity on $Y$. We also denote the subgroup of ${\rm Diff}(\Sigma^\bullet,U)$ (resp. ${\rm Diff}(\Sigma^\bullet,U;Y)$) of elements that are isotopic to the identity inside ${\rm Diff}(\Sigma^\bullet,U)$ (resp. ${\rm Diff}(\Sigma^\bullet,U;Y)$) by ${\rm Diff}_0(\Sigma^\bullet ,U)$ (resp. ${\rm Diff}_0(\Sigma^\bullet,U;Y)$).
We equip these spaces with the Whitney $C^\infty$ topology.

The \textbf{mapping class group} of $(\Sigma^{\bullet},\{{\bf x}^\bullet\}\cup \pl\Sigma^{\bullet})$ is defined by
\[ {\rm Mod}(\Sigma^{\bullet},\{{\bf x}^\bullet\} \cup \pl \Sigma^{\bullet}):={\rm Diff}(\Sigma^{\bullet},\{{\bf x}^\bullet\} \cup \pl \Sigma^{\bullet})/{\rm Diff}_0(\Sigma^{\bullet},\{{\bf x}^\bullet\} \cup \pl \Sigma^{\bullet}).\]
When taking the quotient, the condition of being holomorphic near the marked points and the boundary does not change the mapping
class group. Indeed, by a local straightening isotopy in disjoint coordinate neighbourhoods, every diffeomorphism fixing
$\{{\bf x}^\bullet\}\cup\pl\Sigma^\bullet$ pointwise is isotopic, relative to this set, to one which is holomorphic in a neighbourhood
of it.

Notice that for $[(F,\Sigma,J,{\bf x},\bs{\zeta})]\in\mc{T}_{g,m,b^+,b^-}$ (with the same choice of base surface $(\Sigma^{\bullet},J^\bullet,{\bf x}^\bullet,\bs{\zeta}^\bullet)$) and $h_0\in{\rm Diff}_0(\Sigma^{\bullet},\{{\bf x}^\bullet\} \cup \pl \Sigma^{\bullet})$, we have by definition
\[[(F\circ h_0,\Sigma,J,{\bf x},\bs{\zeta})]=[(F,\Sigma,J,{\bf x},\bs{\zeta})]\text{ in }\mc{T}_{g,m,b^+,b^-}.\]
Thus there is a well-defined right action of  ${\rm Mod}(\Sigma^{\bullet},\{{\bf x}^\bullet\} \cup \pl \Sigma^{\bullet})$ on 
$\mc{T}_{g,m,b^+,b^-}$ defined by ($[h]$ denotes the image of $h\in{\rm Diff}(\Sigma^{\bullet},\{{\bf x}^\bullet\} \cup \pl \Sigma^{\bullet})$ in ${\rm Mod}(\Sigma^{\bullet},\{{\bf x}^\bullet\} \cup \pl \Sigma^{\bullet})$)
\[[(F,\Sigma,J,{\bf x},\bs{\zeta})]\cdot[h]:=[(F\circ h,\Sigma,J,{\bf x},\bs{\zeta})].\]
Similarly, 
the \textbf{framed mapping class group} of $(\Sigma^{\bullet},\{{\bf x}^\bullet\}\cup \pl\Sigma^{\bullet})$ is defined by
\[ \tilde{{\rm Mod}}(\Sigma^{\bullet},\{{\bf x}^\bullet\} \cup \pl \Sigma^{\bullet}):={\rm Diff}(\Sigma^{\bullet},\{{\bf x}^\bullet\} \cup \pl \Sigma^{\bullet};{\bf Y}^\bullet)/{\rm Diff}_0(\Sigma^{\bullet},\{{\bf x}^\bullet\}\cup \pl \Sigma^{\bullet};{\bf Y}^\bullet),\]
where ${\bf Y}^\bullet=(Y^\bullet_1,\dots,Y^\bullet_m)$ with non-zero $Y_j^\bullet\in T_{x^\bullet_j}\Sigma^\bullet$.
Since diffeomorphisms in ${\rm Diff}(\Sigma^{\bullet},\{{\bf x}^\bullet\} \cup \pl \Sigma^{\bullet};{\bf Y}^\bullet)$ are assumed to be holomorphic at the marked points ${\bf x}^\bullet$, all groups here are actually independent of the choice of ${\bf Y}^\bullet$. As in \cite[Lemma 2.1]{AndersenBorotOrantin_modular}, for ${\bf x}^\bullet=(x^\bullet_1,\dots,x^\bullet_m)$, there is an exact sequence
\[ 0\to \Z^m\to \tilde{{\rm Mod}}(\Sigma^{\bullet},\{{\bf x}^\bullet\} \cup \pl \Sigma^{\bullet})\to {\rm Mod}(\Sigma^{\bullet},\{{\bf x}^\bullet\} \cup \pl \Sigma^{\bullet})\to 0 \]
where the $j$-th copy of $\Z$  corresponds to the cyclic group generated by the Dehn twist around the  boundary of an  embedded disk 
$\mc{D}_j^\bullet$ centered at $x_j^\bullet$ which is small enough so as not to contain any other marked point. 
We have
\[\tilde{{\rm Mod}}(\Sigma^{\bullet},\{{\bf x}^\bullet\} \cup \pl \Sigma^{\bullet})\simeq {\rm Mod}\Big(\Sigma^{\bullet}\setminus \bigcup_{j=1}^m\mc{D}^\bullet_{j}, \pl(\Sigma^{\bullet}\setminus \bigcup_{j=1}^m\mc{D}_{j}^\bullet)\Big).\]
There is a right action of $\tilde{{\rm Mod}}(\Sigma^{\bullet},\{{\bf x}^\bullet\} \cup \pl \Sigma^{\bullet})$ on 
$\tT_{g,m,b^+,b^-}$ by $[(F,\Sigma,J,{\bf x},\bs{\zeta},{\bf Y})]\cdot[h]:=[(F\circ h,\Sigma,J,{\bf x},\bs{\zeta},{\bf Y})]$.
 
There are natural projections 
\[ \pi_{\mc{T}\to \mc{M}}:\mc{T}_{g,m,b^+,b^-}\to \mc{M}_{g,m,b^+,b^-}, \quad \pi_{\tT\to\mc{T}}:\tT_{g,m,b^+,b^-}\to \mc{T}_{g,m,b^+,b^-},\]
corresponding to ${\rm Diff}_0(\Sigma^{\bullet},\{{\bf x}\}^\bullet \cup \pl \Sigma^{\bullet})\subset {\rm Diff}(\Sigma^{\bullet},\{{\bf x}^\bullet\} \cup \pl \Sigma^{\bullet})$ in the first case and ${\rm Diff}_0(\Sigma^{\bullet},\{{\bf x}^\bullet\} \cup \pl \Sigma^{\bullet};{\bf Y}^\bullet)\subset {\rm Diff}_0(\Sigma^{\bullet},\{{\bf x}^\bullet\}\cup \pl \Sigma^{\bullet})$ in the second case.

\subsection{Teichm\"uller space of complex building blocks}
For a complex building block  with $m$ marked points and $b=b^++b^-=3-m$ boundary components, 
we choose our base surface $\Sigma^{\bullet}$ of Definition \ref{def:Teichmuller} to be 
\begin{equation}\label{sigma0mb}
\Sigma^{\bullet}=\Sigma^{\bullet}_{m,b}:=\hat{\C}\setminus \bigcup_{k=1}^b (\mc{D}_k^\bullet)^\circ, \quad \mc{D}^\bullet_k:=D(z^{\bullet}_k,1/4), 
\end{equation}  
with marked points ${\bf x}^\bullet=(x_1^\bullet,\ldots,x_m^\bullet):=(z^{\bullet}_{b+1},\dots,z^\bullet_3)$, where $z^{\bullet}_k$ are defined in \eqref{defz_0}. Here $\Sigma^{\bullet}$ is equipped with the canonical complex structure $J_\C$ and with boundary parametrisations $\bs{\zeta}^\bullet=(\zeta^\bullet_1,\dots,\zeta_{b}^\bullet)$ where
\begin{equation}\label{param_base_model_block}
\zeta^\bullet_k(e^{{\rm i}\theta})=z^{\bullet}_k+\frac{1}{4}e^{{\rm i}\theta} .
\end{equation} 
We also choose ${\bf Y}^\bullet=(Y^\bullet_1,\dots,Y^\bullet_m)$ with $Y^\bullet_j=\pl_{z}\in (T_{x^\bullet_j}\hat{\C})^{1,0}$ in the complex coordinate $z$ on $\C$. We begin by defining two important types of diffeomorphisms on this base surface. 

For $k\in \{1,2,3\}$ and $\alpha\in \R$  and some $\eps>0$ small, consider 
a diffeomorphism 
\begin{equation}\label{Tkalpha}
T_{k,\alpha}: \hat{\C}\to \hat{\C}
\end{equation}
which is equal to $z\mapsto z_k^{\bullet}+e^{{\rm i}\alpha\chi(|z-z^{\bullet}_k|)}(z-z_k^{\bullet})$ on $D(z_k^{\bullet},1/4+2\eps)$ and to the identity outside, 
where $\chi\in C_c^\infty(\R)$ is a cutoff function equal to $1$ on $[0,1/4+\eps]$ and $0$ on $[1/4+2\eps,\infty)$. Notice that $T_{k,0}={\rm Id}$, and $T_{k,2\pi n}={\rm Id}$ near $\pl D(z^\bullet_k,1/4)$ for $n\in \Z$.
We define the \textbf{Dehn twist} around $z_k^\bullet$ (we also say around $\partial_k\Sigma_{m,b}^\bullet$ if $k\leq b$) to be $T_k:=T_{k,2\pi}$ (see \cite[Chapter 3]{Farb-Margalit}), which restricts to a diffeomorphism on $\Sigma^\bullet_{m,b}$, in the class ${\rm Diff}(\Sigma^\bullet_{m,b},\{{\bf x}^\bullet\} \cup \pl \Sigma^{\bullet}_{m,b};{\bf Y}^\bullet)$.  
The mapping class group of this surface is (see \cite{Birman})
\begin{equation}\label{MCG_of_pant} 
{\rm Mod}(\Sigma^{\bullet}_{m,b},\{{\bf x}^\bullet\}\cup\pl \Sigma_{m,b}^{\bullet})\simeq \Z^b, \quad \til{{\rm Mod}}(\Sigma^{\bullet}_{m,b}, \{{\bf x}^\bullet\}\cup\pl \Sigma_{m,b}^{\bullet};{\bf Y}^\bullet)\simeq\Z^3,
\end{equation}
where the $k$-th copy of $\Z$ corresponds to the cyclic group generated by the Dehn twist $T_k$ around $z_k^\bullet$. 

Let $\sigma$ be a transposition of $\{j,k,\ell\}=\{1,2,3\}$ ($j<k$) with $(\sigma(j),\sigma(k),\sigma(\ell))=(k,j,\ell)$. We define a diffeomorphism, called \textbf{braiding},
\begin{equation}\label{braiding}
B_{jk}: \hat{\C}\to \hat{\C}
\end{equation}
such that $B_{jk}(z_{i}^\bullet+z/4)=z_{\sigma(i)}^\bullet+z/4$ for $z\in\D$ and $i=1,2,3$, so that in particular,
\[B_{jk}\circ\zeta_i^\bullet=\zeta_{\sigma(i)}^\bullet,\quad B_{jk}(z_i^\bullet)=z_{\sigma(i)}^\bullet,\quad DB_{jk}(z_i^\bullet)Y_i^\bullet=Y_{\sigma(i)}^\bullet,\quad\text{ for }i=1,2,3.\]
It is defined to be
\begin{equation}\label{defBjk}
B_{jk}:=T_{k,\pi}T_{j,\pi}R_{jk,\pi},
\end{equation}
where $T_{i,\alpha}$ is the twist defined in \eqref{Tkalpha} and $R_{jk,\pi}$ is a rotation defined as follows. For $\alpha\in\R$ and $\eps>0$ small, we define a diffeomorphism
\[R_{jk,\alpha}:\hat{\C}\to\hat{\C}\]
which is equal to $z\mapsto y+e^{{\rm i}\alpha\chi(|z-y|)}(z-y)$ in $D(y,r)$ and to the identity outside, where $y=(z_j^\bullet+z_k^\bullet)/2$, $r=|z_j^\bullet-z_k^\bullet|/2+1/4+2\eps$, and $\chi\in C_c^\infty(\R)$ is a cutoff function equal to $1$ on $[|z_j^\bullet-z_k^\bullet|/2-1/4-\eps,|z_j^\bullet-z_k^\bullet|/2+1/4+\eps]$ and $0$ outside $(|z_j^\bullet-z_k^\bullet|/2-1/4-2\eps,|z_j^\bullet-z_k^\bullet|/2+1/4+2\eps)$.
In practice, we assume that either $j<k\leq b$ or $b<j<k$, so that $B_{jk}$ restricts to a diffeomorphism $\Sigma_{m,b}^\bullet\to\Sigma_{m,b}^\bullet$. If $b=3$ the possible cases are $B_{12},B_{13},B_{23}$, if $b=2$ only $B_{12}$ is possible and if $b=1$ only $B_{23}$ is possible. Notice that $B_{jk}$ is not in ${\rm Diff}(\Sigma^\bullet_{m,b},\{{\bf x}^\bullet\} \cup \pl \Sigma^{\bullet}_{m,b})$ since it does not fix $\{{\bf x}^\bullet\} \cup \pl \Sigma^{\bullet}_{m,b}$ pointwise. It is straightforward to check that (see also Lemma \ref{Lambda_lemma} below)
\[ B_{jk}^2=T_kT_jT_\ell^{(-1)^{\ell+1}} \]
up to isotopy relative to $\{{\bf x}^\bullet\}\cup\pl \Sigma_{m,b}^{\bullet}$ (or $\bigcup_{i=1}^3\partial\mc{D}_i^\bullet$), where $T_i=T_{i,2\pi}$ are the Dehn twists defined above.

We now explain how to represent in a more concrete way the Teichm\"uller and decorated Teichm\"uller spaces for the special case of complex building blocks, defined in terms of the fixed base surface $\Sigma_{m,b}^\bullet$.

\begin{lemma}\label{another_desc_of_Teich}
1) There is a bijective map between $\mc{T}_{0,m,b}$  and the set of equivalence classes 
$[\mu]$ of Beltrami differentials $\mu\in C_c^\infty((\Sigma_{m,b}^\bullet)^\circ \setminus \{{\bf x}^\bullet\})$ satisfying $\|\mu\|_{L^\infty}<1$, where  $\mu_1$ is said equivalent to $\mu_2$ if there is $f\in {\rm Diff}_0(\Sigma_{m,b}^{\bullet}, \{{\bf x}^\bullet\}\cup\pl\Sigma_{m,b}^{\bullet})$ such that $f^*\mu_1=\mu_2$. The inverse map is given by 
\[[\mu]\mapsto [(\Phi^\mu,\Phi^\mu(\Sigma_{m,b}^{\bullet}),J_\C,{\bf x}^\bullet,\bs{\zeta}^{\mu})]\]
where $\Phi^\mu$ solves the Beltrami equation $\pl_{\bar{z}}\Phi^\mu=\mu\pl_z\Phi^\mu$ in $\hat{\C}$ with $\Phi^\mu(z^{\bullet}_k)=z^{\bullet}_k$  for $k=1,2,3$ and $\bs{\zeta}^{\mu}=\Phi^\mu\circ\bs{\zeta}^\bullet$, i.e.\ $\zeta^\mu_k(e^{{\rm i}\theta})=\Phi^\mu(z^{\bullet}_k+e^{{\rm i}\theta}/4)$ for $k=1,\ldots,b$.\\
2) There is a bijective map between $\tT_{0,m,b}=\tT_{0,m,0,b}$  and the set of equivalence classes 
$[\mu]$ of Beltrami differentials $\mu\in C_c^\infty((\Sigma_{m,b}^\bullet)^\circ \setminus \{{\bf x}^\bullet\})$ satisfying $\|\mu\|_{L^\infty}<1$, where $\mu_1$ is said equivalent to $\mu_2$ if there is $f\in {\rm Diff}_0(\Sigma_{m,b}^{\bullet}, \{{\bf x}^\bullet\}\cup\pl\Sigma_{m,b}^{\bullet};{\bf Y}^\bullet)$  such that $f^*\mu_1=\mu_2$. The inverse map is given by 
\[[\mu]\mapsto [(\Phi^\mu,\Phi^\mu(\Sigma_{m,b}^{\bullet}),J_\C,{\bf x}^\bullet,\bs{\zeta}^{\mu},{\bf Y}^\mu)]\]
where $\Phi^\mu,\bs{\zeta}^{\mu}$ are as in 1) and ${\bf Y}^\mu=D\Phi^\mu({\bf Y}^\bullet)$, i.e.\ $Y^\mu_k=D\Phi^\mu(x^\bullet_k)Y^\bullet_k$ for $k=1,\ldots,m$.
\end{lemma}
\begin{proof} 1) Let $(F,\Sigma,J,{\bf x},\bs{\zeta})$ be a representative of an element in  $\mc{T}_{0,m,b}$. To the diffeomorphism $F:\Sigma_{m,b}^{\bullet}\to \Sigma$, we can associate a Beltrami differential $\mu=\mu_{F}$ on $\Sigma^{\bullet}_{m,b}$ with $\|\mu\|_{L^\infty}<1$ (see \cite[Section 1.4]{Imayoshi-Taniguchi}) which vanishes near $\{{\bf x}^\bullet\}\cup\pl\Sigma_{m,b}^{\bullet}$ since $F$ is holomorphic from $J_\C$ to $J$ near $\{{\bf x}^\bullet\}\cup\pl\Sigma_{m,b}^{\bullet}$; the Beltrami differential $\mu$ is obtained by comparing $F^*J$ with $J_\C$.
By the same argument as \cite[Theorem 1.6]{Imayoshi-Taniguchi}, if $F^1:\Sigma_{m,b}^{\bullet}\to \Sigma^1$ is another such diffeomorphism and $\mu^1$ the associated Beltrami differential, then $[(F,\Sigma,J,{\bf x},\bs{\zeta})]=[(F^1,\Sigma^1,J^1,{\bf x}^1,\bs{\zeta}^1)]$ if and only if there is a diffeomorphism $f:\Sigma_{m,b}^{\bullet}\to \Sigma_{m,b}^{\bullet}$ holomorphic near $\{{\bf x}^\bullet\}\cup\pl\Sigma_{m,b}^{\bullet}$ such that $f^*\mu=\mu^1$ and $f$ is isotopic to the identity relative to $\pl\Sigma_{m,b}^{\bullet}\cup \{{\bf x}^\bullet\}$  (in fact $f(z)=z$ near $\pl \Sigma_{m,b}^{\bullet}$). Conversely, for $\mu$ representing an equivalence class, we can extend $\mu$ by $0$ in $\hat{\C}\setminus \Sigma^{\bullet}$ and solve the Beltrami equation $\pl_{\bar{z}}\Phi^\mu=\mu\pl_z\Phi^\mu$ with $\Phi^\mu(z^\bullet_k)=z^\bullet_k$ for $k=1,2,3$. This induces $b$ biholomorphisms 
$\psi_k^\mu(z)=\Phi^\mu(z_k^\bullet+z/4)$ from $\D$ to $\Phi^\mu(\mc{D}_k^\bullet)$ for $k=1,\dots,b$, and if $\bs{\psi}^\mu=(\psi^\mu_1,\dots,\psi_b^\mu)$ and $\bs{\zeta}^\mu=\bs{\psi}^\mu|_{\T}$, we obtain a map $[\mu]\mapsto [(\Phi^\mu,\Phi^\mu(\Sigma_{m,b}^{\bullet}),J_\C,{\bf x}^\bullet,\bs{\zeta}^{\mu})]$ inverting the map above.\\
2) The proof is the same as above except that the diffeomorphism $f$ must have $Df(x_k^\bullet)$ fixing $Y_k^\bullet$ for each $k$, that is, $f'(x^\bullet_k)=1$.
\end{proof}

We thus equip the spaces $\mc{T}_{0,m,b},\tT_{0,m,b}$ with the topologies induced by the $C_c^\infty$ topology on the Beltrami differentials. More precisely, we use the direct limit topology as $C_c^\infty((\Sigma_{m,b}^\bullet)^\circ \setminus \{{\bf x}^\bullet\})=\bigcup_KC_c^\infty(K)$ for $K$ ranging over all compact subsets in $(\Sigma_{m,b}^\bullet)^\circ \setminus \{{\bf x}^\bullet\}$, and on each $C_c^\infty(K)$ we put the Whitney $C^\infty$ topology. Since $\mc{T}_{0,m,b},\tT_{0,m,b}$ consist of equivalence classes of Beltrami differentials, the fact that these topologies are Hausdorff is not a priori obvious and will follow from Proposition \ref{identification_Teichmuller} below when we show that they are homeomorphic to other spaces with more concrete topologies.

In the following definition, we consider a covering space over the space $\mc{P}^\eps_{m,b}$ (recall \eqref{defPmb}), and we shall see below that this is equivalent to the Teichm\"uller space of complex building blocks. 
\begin{definition}[\textbf{Space of model marked complex building blocks}]\label{teichmullerpant} For $\eps>0$, we define the space 
\[\mc{T}^{\eps}_{m,b}:= \Big\{(\bs{\psi},\btheta)\in \mc{P}^{\eps}_{m,b}\times \R^b \,\Big |\, \forall j=1,\dots,b, \, \frac{\psi_j'(0)}{|\psi_j'(0)|}=e^{{\rm i}\theta_j}\Big\},\]
where $\bs{\psi}=(\psi_1,\dots,\psi_b)$, $\btheta=(\theta_1,\dots,\theta_b)$. 
The natural projection
\begin{equation}\label{covering_map} 
 \pi_{\mc{P}}: \mc{T}^{\eps}_{m,b} \to \mc{P}^{\eps}_{m,b}, \quad 
(\bs{\psi}, \btheta)\mapsto \bs{\psi}
\end{equation}
is a covering map with discrete fiber $\Z^b$.
We also define the space of decorated elements of $\mc{T}^{\eps}_{m,b}$ to be
\[ \tT^{\eps}_{m,b}:=\mc{T}^{\eps}_{m,b}\times \C^m.\]
The map (with $e^{\bs{\lambda}}=(e^{\lambda_1},\ldots,e^{\lambda_m})$ for $\bs{\lambda}=(\lambda_1,\ldots,\lambda_m)$)
\begin{equation}\label{covering_map_decorated} 
\tilde{\pi}_{\mc{P}} := (\pi_{\mc{P}}, \exp): \tT^{\eps}_{m,b} \to \tilde{\mc{P}}^{\eps}_{m,b}, \quad 
(\bs{\psi}, \btheta,\bs{\lambda})\mapsto (\bs{\psi},e^{\bs{\lambda}})
\end{equation}
is a covering map with discrete fiber $\Z^3$.
\end{definition}

There is a canonical projection
\[ \tilde{\pi}_{3\to b}: \mc{T}^\eps_{0,3} \to \tT^\eps_{m,b}, \quad (\bs{\psi},\btheta)\mapsto ((\psi_1,\dots,\psi_b,\theta_1,\dots,\theta_b), (\log |\psi_{b+1}'(0)|+{\rm i}\theta_{b+1},
\dots,\log |\psi_{3}'(0)|+{\rm i}\theta_{3})) \]
which lifts the map $\pi_{3\to b}$ of \eqref{defpi3tob} in the sense that
\[\pi_{3\to b} \circ \pi_{\mc{P}} = \tilde{\pi}_{\mc{P}} \circ \tilde{\pi}_{3\to b}.\]

Our next goal is to establish a bijection between $\bigcup_{\eps>0}\mc{T}_{m,b}^\eps$ and the Teichm\"uller space $\mc{T}_{0,m,b}$, and likewise between $\bigcup_{\eps>0}\tT_{m,b}^\eps$ and the decorated Teichm\"uller space $\tT_{0,m,b}$. It is convenient to work with the following description of Teichm\"uller space:

\begin{lemma}\label{another_desc_of_Teich2}
There is a bijection between $\mc{T}_{0,m,b}$ (resp.\ $\tT_{0,m,b}$) and the set of equivalence classes $[F]$ of $F\in{\rm Diff}(\hat{\C} ,{\bf z}^\bullet)$ such that $F$ is holomorphic near $\{{\bf x}^\bullet\}\cup\bigcup_{k=1}^b\mc{D}_k^\bullet$, the equivalence relation being isotopy relative to $\{{\bf x}^\bullet\}\cup\bigcup_{k=1}^b\mc{D}_k^\bullet$ (fixing the value of $DF=F'$ at ${\bf x}^\bullet$ in the decorated case).
\end{lemma}
\begin{proof} The proof is contained in the proof of Lemma \ref{another_desc_of_Teich} where 
$F=\Phi^\mu$ with $[\mu]$ the equivalence class of Beltrami differentials of that lemma.
\end{proof}

Notice that for $[F]\in\mc{T}_{0,m,b}$ and $k=1,\ldots,b$, we have $F(z^{\bullet}_k+\cdot /4)\in\tBiHol_\eps(\D)$ for some $\eps>0$, and the natural projection to the moduli space (in model form) is just
\begin{equation}\label{normalisationF}
\mc{T}_{0,m,b} \to \bigcup_{\eps>0}\mc{P}^\eps_{m,b}, \quad [F]\mapsto (F(z^{\bullet}_k+\cdot /4))_{k=1}^b.
\end{equation}
To lift this map to $\bigcup_{\eps>0}\mc{T}^\eps_{m,b}$, we still need to determine the $\theta_k$, as well as the $\C^m$ part in the decorated case.
Here and below, ${\rm Diff}(\hat{\C},{\bf z}^\bullet)$ denotes the group of
orientation-preserving smooth diffeomorphisms of $\hat{\C}$ which
fix each $z_k^\bullet$ and are holomorphic in a neighbourhood of
these three points. It is equipped with the Whitney $C^\infty$
topology.

Consider the continuous group homomorphism
\begin{equation}\label{lambda}
\lambda=(\lambda_1,\lambda_2,\lambda_3): {\rm Diff}(\hat{\C} ,{\bf z}^\bullet) \to (\C\setminus \{0\})^3, \quad F\mapsto(F'(z_1^{\bullet}),\dots,F'(z_3^{\bullet})),
\end{equation}
where $(\C\setminus \{0\})^3$ is endowed with component-wise multiplication. 
The standard parametric local-straightening argument in three
disjoint coordinate disks shows that the inclusion of
${\rm Diff}(\hat{\C},{\bf z}^\bullet)$ into the group of all
orientation-preserving smooth diffeomorphisms of $\hat{\C}$ fixing
$z_1^\bullet,z_2^\bullet,z_3^\bullet$ induces isomorphisms on
$\pi_0$ and $\pi_1$. The latter group is contractible by
\cite[Theorem~1(a)]{EarleEells1967}. Consequently,
${\rm Diff}(\hat{\C},{\bf z}^\bullet)$ is connected and simply
connected. Since $\exp:\C^3\to(\C\setminus\{0\})^3$
is the universal covering, $\lambda$ admits a unique continuous
lifting $\Lambda$ satisfying $\Lambda({\rm Id})=0$.
The lift is a continuous group homomorphism
\begin{equation}\label{Lambda}
\Lambda=(\Lambda_1,\Lambda_2,\Lambda_3):{\rm Diff}(\hat{\C} ,{\bf z}^\bullet) \to \C^3, \quad F \mapsto (\log F'(z_1^\bullet), \log F'(z_2^\bullet), \log F'(z_3^\bullet)).
\end{equation}
By uniqueness of the lift and the identity
$\lambda(F\circ G)=\lambda(F)\lambda(G)$, one also has
\[
\Lambda(F\circ G)=\Lambda(F)+\Lambda(G),
\]
so that $\Lambda$ is a group homomorphism.
In other words, this uniquely determines the branch of $\Lambda_k:F\mapsto\log F'(z_k^\bullet)$ on ${\rm Diff}(\hat{\C} ,{\bf z}^\bullet)$.

\begin{lemma}\label{Lambda_lemma}
Recall the diffeomorphisms $T_{k,\alpha},B_{jk}:\hat{\C}\to\hat{\C}$ from \eqref{Tkalpha}, \eqref{braiding}.\\
1) We have $\Lambda_j(T_{k,\alpha})={\rm i}\alpha\delta_{jk}$, where $\delta_{jk}$ is the Kronecker delta.\\
2) Let $M_{jk}$ be the unique M\"obius transformation transposing $z_j^\bullet,z_k^\bullet$ and fixing the other point $z_\ell^\bullet$, so that $M_{jk}\circ B_{jk}\in{\rm Diff}(\hat{\C} ,{\bf z}^\bullet)$. We have 
\[ \begin{gathered}
\Lambda(M_{12}\circ B_{12})=(\log4+{\rm i}\pi, -\log4+{\rm i}\pi, {\rm i}\pi ) ,\quad \Lambda(M_{23}\circ B_{23})=({\rm i}\pi,-\log4+{\rm i}\pi,\log4+{\rm i}\pi),\\
\Lambda(M_{13}\circ B_{13})=({\rm i}\pi,-{\rm i}\pi,{\rm i}\pi).
\end{gathered}\]
\end{lemma}
\begin{proof}
1) First we have  $\lambda_j(T_{k,\alpha})=e^{{\rm i}\alpha\delta_{jk}}$, thus $\Lambda_j(T_{k,\alpha})={\rm i}\alpha\delta_{jk}$ by continuity of the family $\alpha\mapsto T_{k,\alpha}$.

2) First we prove a useful property of $\Lambda$. Let $G_1,G_2$ be orientation preserving diffeomorphisms of $\hat{\C}$ which are holomorphic near $z_1^\bullet,z_2^\bullet,z_3^\bullet$ such that 
\[(G_1(z_1^\bullet),G_1(z_2^\bullet),G_1(z_3^\bullet))=(G_2(z_1^\bullet),G_2(z_2^\bullet),G_2(z_3^\bullet))=(z_{\sigma(1)}^\bullet,z_{\sigma(2)}^\bullet,z_{\sigma(3)}^\bullet)\] 
for some permutation $\sigma$ of $\{1,2,3\}$. Then for $F\in{\rm Diff}(\hat{\C},{\bf z}^\bullet)$, we have
\begin{equation}\label{Lambda_property}
\Lambda(G_1^{-1}\circ F\circ G_2)=\Lambda(G_1^{-1}\circ G_2)+\sigma\Lambda(F),
\end{equation}
where $\sigma\Lambda=(\Lambda_{\sigma(1)},\Lambda_{\sigma(2)},\Lambda_{\sigma(3)})$. Indeed, since $\Lambda$ is a group homomorphism, $\Lambda(G_1^{-1}\circ F\circ G_2)=\Lambda(G_1^{-1}\circ F\circ G_1)+\Lambda(G_1^{-1}\circ G_2)$. We have 
\[\lambda_j(G_1^{-1}\circ F\circ G_1)=(G_1^{-1}\circ F\circ G_1)'(z_j^\bullet)=(G_1^{-1})'(F(G_1(z_j^\bullet)))F'(G_1(z_j^\bullet))G_1'(z_j^\bullet)=F'(z_{\sigma(j)}^\bullet)=\lambda_{\sigma(j)}(F),\] 
so the map $F\mapsto\Lambda(G_1^{-1}\circ F\circ G_1)$ is a group homomorphism lifting $\sigma\lambda=(\lambda_{\sigma(1)},\lambda_{\sigma(2)},\lambda_{\sigma(3)})$. By the uniqueness of the liftings $\Lambda_i$, we get $\Lambda(G_1^{-1}\circ F\circ G_1)=\sigma\Lambda(F)$.

Applying this to $M_{jk}\circ B_{jk}=M_{jk}^{-1}\circ(T_{k,\pi}\circ T_{j,\pi})\circ R_{jk,\pi}$ (recall the definition \eqref{defBjk}) and using 1), it remains to show that 
\[\Lambda_j(M_{jk}^{-1}\circ R_{jk,\pi})=-\log|M'_{jk}(z_j^\bullet)|, \quad 
\Lambda_k(M_{jk}^{-1}\circ R_{jk,\pi})=-\log|M'_{jk}(z_k^\bullet)|,\quad \Lambda_\ell(M_{jk}^{-1}\circ R_{jk,\pi})=(-1)^{\ell+1}{\rm i}\pi.\] 
To simplify the notation, let us write $R_\alpha$ for $R_{jk,\alpha}$. For $\alpha\in\R$, let $M_\alpha$ be the unique M\"obius transformation with $(M_\alpha(z_j^\bullet),M_\alpha(z_k^\bullet),M_\alpha(z_\ell^\bullet))=(R_\alpha(z_j^\bullet),R_\alpha(z_k^\bullet),z_\ell^\bullet)$ (note that $R_\alpha(z_\ell^\bullet)=z_\ell^\bullet$). In particular, $M_\pi=M_{jk}$. Then $\alpha\in[0,\pi]\mapsto M_\alpha^{-1}\circ R_\alpha$ is a continuous family in ${\rm Diff}(\hat{\C},{\bf z}^\bullet)$ from ${\rm Id}$ to $M_{jk}^{-1}\circ R_{jk,\pi}$. Now $\lambda_i(M_\alpha^{-1}\circ R_\alpha)=R_\alpha'(z_i^\bullet)/M_\alpha'(z_i^\bullet)$, and by definition $R_\alpha'(z_j^\bullet)=R_\alpha'(z_k^\bullet)=e^{{\rm i}\alpha}$, $R_\alpha'(z_\ell^\bullet)=1$. To compute the monodromy of $M_\alpha'$, using the property $(M(z_1)-M(z_2))^2=(z_1-z_2)^2M'(z_1)M'(z_2)$ for any M\"obius transformation $M$ and $z_1,z_2,M(z_1),M(z_2)\in\C$, we get (compare the proof of Lemma \ref{Gammaterm_lemma} below) that
\[M_\alpha'(z_i^\bullet)(M_\alpha(z_{i_1}^\bullet)-M_\alpha(z_{i_2}^\bullet))(M_\alpha(z_i^\bullet)-M_\alpha(z_{i_1}^\bullet))^{-1}(M_\alpha(z_i^\bullet)-M_\alpha(z_{i_2}^\bullet))^{-1}\]
is constant for each $i=1,2,3$, where $\{i,i_1,i_2\}=\{1,2,3\}$. We thus get that 
\[\log M_\pi'(z_j^\bullet)= \log|M'_{jk}(z_j^\bullet)|+{\rm i}\pi, \quad 
\log M_\pi'(z_k^\bullet)=\log|M'_{jk}(z_k^\bullet)|+{\rm i}\pi,\quad  \log M_\pi'(z_\ell^\bullet)=\log|M'_{jk}(z_\ell^\bullet)|+(-1)^{\ell}{\rm i}\pi.\] 
Note that in the case $\ell=2$, there is an additional monodromy of $2\pi$ coming from 
\[(M_\alpha(z_\ell^\bullet)-M_\alpha(z_j^\bullet))(M_\alpha(z_\ell^\bullet)-M_\alpha(z_k^\bullet))=(0-e^{{\rm i}\alpha})(0+e^{{\rm i}\alpha}).\] 
A direct computation gives 
$\log|M'_{jk}(z_i^\bullet)|=0$ except in the following cases:
\[  (j,k)=(1,2) : \quad |M'_{12}(z_1^\bullet)|=\frac{1}{4}, \quad  |M'_{12}(z_2^\bullet)|=4, \quad (j,k)=(2,3)\, :  \quad 
|M'_{23}(z_2^\bullet)|=4, \quad |M'_{23}(z_3^\bullet)|=\frac{1}{4}.\qedhere\]
\end{proof}

We are ready to describe the model form isomorphisms.

\begin{proposition}[\textbf{Model form}]\label{identification_Teichmuller}
Let $b$ and $m$ be such that $m+b=3$, $b\geq1$. There are homeomorphisms, defined in terms of the description of Teichm\"uller space in Lemma \ref{another_desc_of_Teich2},
\begin{align*}
&\Theta_{m,b}: \mc{T}_{0,m,b} \to \bigcup_{\eps>0}\mc{T}^{\eps}_{m,b}, \quad [F] \mapsto \big((F(z^{\bullet}_k+\cdot /4))_{k=1}^b,({\rm Im}\,\Lambda_k(F))_{k=1}^b\big),\\
&\tilde{\Theta}_{m,b}: \tT_{0,m,b} \to \bigcup_{\eps>0}\tT^{\eps}_{m,b}, \quad [F]\mapsto\big((F(z^{\bullet}_k+\cdot /4))_{k=1}^b,({\rm Im}\,\Lambda_k(F))_{k=1}^b,(\Lambda_k(F))_{k=b+1}^3\big),
\end{align*}
which are Gateaux differentiable with continuous derivative, when viewed as defined on the space of Beltrami differentials. Moreover, they obey the following transformation rules under Dehn twists and braidings. Suppose $\Theta_{m,b}([(F,\Sigma,J,{\bf x},\bs{\zeta})])=(\bs{\psi},\bs{\theta})$ and $\tilde{\Theta}_{m,b}([(F,\Sigma,J,{\bf x},\bs{\zeta},{\bf Y})])=(\bs{\psi},\bs{\theta},\bs{\la})$ if $b<3$.\\ 
1) \textbf{Action by Dehn twist}: Let  $T_k$ be 
 the Dehn twist around $\pl_k\Sigma^\bullet_{m,b}$ if $k=1,\dots,b$ or around $z_k^\bullet$ if $k=b+1,\ldots,3$. Then
\begin{equation}\label{Dehn_Twist_under_Theta} 
\begin{aligned}
&\Theta_{m,b}([(F\circ T_k,\Sigma,J,{\bf x},\bs{\zeta})])=(\bs{\psi},\bs{\theta}+2\pi\Pi_be_k), \\
&\tilde{\Theta}_{m,b}([(F\circ T_k,\Sigma,J,{\bf x},\bs{\zeta},{\bf Y})])=(\bs{\psi},(\bs{\theta},\bs{\la})+2\pi e_k),
\end{aligned}\end{equation}
where $e_i=(\delta_{ik})_{k=1}^3$ is the canonical basis of $\C^3$ if $i\leq b$, $e_i={\rm i}(\delta_{ik})_{k=1}^3$ if $i>b$, and $\Pi_b:(a_k)_{k=1}^3\mapsto(a_k)_{k=1}^b$.\\ 
2) \textbf{Action by braiding}: Let
$B_{jk}$ be the braiding map of \eqref{braiding} with $\{j,k,\ell\}=\{1,2,3\}$, $j<k$. For notational convenience, we add the subscript $_\tau$ to denote the result of swapping the components with indices $j,k$. Then if $(b,m,j,k)\not=(1,2,2,3)$, we have 
\begin{equation}\label{Braiding_under_Theta} 
\begin{aligned}
&\Theta_{m,b}([(F\circ B_{jk},\Sigma,J,{\bf x}_\tau,\bs{\zeta}_\tau)])=(M_{jk}\circ\bs{\psi}_\tau,\bs{\theta}_\tau+\pi\Pi_b(e_j+e_k+(-1)^{\ell+1}e_\ell)),\\
&\tilde{\Theta}_{m,b}([(F\circ B_{jk},\Sigma,J,{\bf x}_\tau,\bs{\zeta}_\tau,{\bf Y}_\tau)])=(M_{jk}\circ\bs{\psi}_\tau,(\bs{\theta}_\tau,\bs{\lambda}_\tau)+\pi(e_j+e_k+(-1)^{\ell+1}e_\ell)),
\end{aligned}\end{equation}
where $e_i,\Pi_b$ are as in 1) and $M_{jk}\in {\rm PSL}_2(\C)$ is the unique M\"obius transformation mapping ${\bf z}^\bullet$ to ${\bf z}^\bullet_\tau$, i.e.\ 
$(M_{jk}(z^\bullet_j),M_{jk}(z^\bullet_k),M_{jk}(z^\bullet_\ell))=(z^\bullet_k,z^\bullet_j,z^\bullet_\ell)$. In the case $(b,m,j,k)=(1,2,2,3)$, we have instead
 \begin{equation}\label{Braiding_under_Theta23}
 \tilde{\Theta}_{2,1}([(F\circ B_{jk},\Sigma,J,{\bf x}_\tau,\bs{\zeta}_\tau,{\bf Y}_\tau)])=(M_{23}\circ \psi_1,\theta_1+\pi,\la_3-\log4+{\rm i}\pi,\la_2+\log4+{\rm i}\pi).
 \end{equation}
\end{proposition}
Here we use the indices $\bs{\lambda}=(\lambda_{b+1},\ldots,\lambda_3)\in\C^m$ and ${\bf Y}=(Y_{b+1},\ldots,Y_3)$. To clarify the notation, let us write down $(\bs{\theta}_\tau,\bs{\lambda}_\tau)+\pi(e_j+e_k+(-1)^{\ell+1}e_\ell)$ explicitly. If $b=3$, there is no $\bs{\lambda}$ and this is $(\theta_{\tau(1)}+(-1)^{s(1)}\pi,\theta_{\tau(2)}+(-1)^{s(2)}\pi,\theta_{\tau(3)}+(-1)^{s(3)}\pi)$, where $\tau$ is the permutation of $\{1,2,3\}$ with $(\tau(j),\tau(k),\tau(\ell))=(k,j,\ell)$ and $s(j)=s(k)=0$, $s(\ell)=\ell+1$. If $b=2$, necessarily $(j,k,\ell)=(1,2,3)$ and this is $(\theta_2+\pi,\theta_1+\pi,\lambda_3+{\rm i}\pi)$. If $b=1$, necessarily $(j,k,\ell)=(2,3,1)$ and we are in the case \eqref{Braiding_under_Theta23}. 
\begin{proof}
In the decorated case, $\lambda$ is constant on an equivalence class $[F]\in\tT_{0,m,b}$, so $\Lambda$ descends to $\tT_{0,m,b}$ and $\tilde{\Theta}_{m,b}$ is well-defined. In the undecorated case, $\Pi_b\circ\lambda=(\lambda_k)_{k=1}^b$ is constant on an equivalence class $[F]\in\mc{T}_{0,m,b}$, so $\Pi_b\circ\Lambda=(\Lambda_k)_{k=1}^b$ and hence $\Theta_{m,b}$ are well-defined.
Continuity and differentiability follow from the corresponding properties of the map $\mu\mapsto\Phi^\mu$ in Lemma \ref{another_desc_of_Teich}. Moreover, these maps are clearly surjective. Since the relevant mapping class groups are generated by Dehn twists, 1) shows that they are injective. To see that they are homeomorphisms, it suffices to construct continuous local inverses of the map $\mu\mapsto(\Phi^\mu(z^{\bullet}_k+\cdot /4))_{k=1}^b$. Suppose $\mu_0\in C_c^\infty((\Sigma_{m,b}^\bullet)^\circ \setminus \{{\bf x}^\bullet\})$ and $\psi_k^0=\Phi^{\mu_0}(z^{\bullet}_k+\cdot /4)\in\tBiHol_\eps(\D)$ for all $k$ for some $\eps>0$. For $\psi_k$ close to $\psi_k^0$ in $\tBiHol_\eps(\D)$, we can find $F\in{\rm Diff}(\hat{\C},{\bf z}^\bullet)$ with $F(z^{\bullet}_k+\cdot /4)=\psi_k$ on $(1+\eps)\D$, $F=\Phi^{\mu_0}$ on $\hat{\C}\setminus\bigcup_{k=1}^b(z_k^\bullet+\frac{1+2\eps}{4}\D)$, such that the map $(\psi_k)_{k=1}^b\mapsto F_{\bar{z}}/F_z\in C_c^\infty(\hat{\C}\setminus\bigcup_{k=1}^b(z_k^\bullet+\frac{1+\eps}{4}\D))$ is continuous (where the latter space is equipped with the Whitney $C^\infty$ topology). This is the desired local inverse. In the decorated case, we also need to adjust $F'(z_k^\bullet)$ for $k>b$, but this can be easily done in a continuous fashion.

Property 1) follows immediately from 1) of Lemma \ref{Lambda_lemma} and the fact that $\Lambda$ is a group homomorphism.

Let us prove 2). By definition, $[(F\circ B_{jk},\Sigma,J,{\bf x}_\tau,\bs{\zeta}_\tau)]=[(M_{jk}\circ F\circ B_{jk},M_{jk}(\Sigma),J,{\bf x},M_{jk}\circ\bs{\zeta}_\tau)]$, and likewise for the decorated case, so braiding has the effect of left composition by $M_{jk}$ on the $\psi_i$. The change on the $\theta_i,\lambda_i$ follows from 2) of Lemma \ref{Lambda_lemma} and the identity $\Lambda(M_{jk}\circ F\circ B_{jk})=\Lambda(M_{jk}\circ B_{jk})+\Lambda_\tau(F)$ which follows from \eqref{Lambda_property}.
\end{proof}

The following commutative diagram summarises the most important model forms we have established for the moduli and Teichm\"uller spaces in the case of complex building blocks:

\[
\begin{tikzcd}
\tT_{0,m,b}
  \arrow[d,"\pi_{\tT\to\mc{T}}"]
  \arrow[r,"\tilde{\Theta}_{m,b}","\sim"']
& \bigcup_{\eps>0}\tT_{m,b}^\eps
  \arrow[d]
&[-2em] (\bs{\psi},\bs{\theta},\bs{\lambda})
  \arrow[d,mapsto]
\\
\mc{T}_{0,m,b}
  \arrow[d,"\pi_{\mc{T}\to\mc{M}}"]
  \arrow[r,"\Theta_{m,b}","\sim"']
& \bigcup_{\eps>0}\mc{T}_{m,b}^\eps
  \arrow[d]
& (\bs{\psi},\bs{\theta})
  \arrow[d,mapsto]
\\
\mc{P}_{m,b}
  \arrow[r,swap,"\sim"]
& \bigcup_{\eps>0}\mc{P}_{m,b}^\eps
& \bs{\psi}
\end{tikzcd}
\]

Recall the convention \eqref{Tangent_space} for the tangent spaces of the moduli space $\mc{P}_{m,b}^\eps$ in model form. Since the projection $\mc{T}_{m,b}^\eps\to\mc{P}_{m,b}^\eps$ is a covering map, we define for $(\bs{\psi}^0,\bs{\theta}^0)\in\mc{T}_{m,b}^\eps$
\begin{equation}\label{Tangent_space_Teich}
T_{(\bs{\psi}^0,\bs{\theta}^0)}\mc{T}^{\eps}_{m,b}:= T_{\bs{\psi}^0}\mc{P}^{\eps}_{m,b}=(T_{\rm Id}\mc{U}_\eps^-)^b=\{ (v_1(z)\pl_z,\dots,v_b(z)\pl_z) \, |\, \forall j, v_j\in {\rm Hol}_\eps(\D), v_j(0)=0\},
\end{equation}
and similarly $T_{(\bs{\psi}^0,\bs{\theta}^0,\bs{\lambda}^0)}\tT^{\eps}_{m,b}:=T_{(\bs{\psi}^0,\bs{\theta}^0)}\mc{T}^{\eps}_{m,b}\times\C^m$ for $\bs{\lambda}^0\in\C^m$.

Up to this point, we have mostly been focusing on the case $b^+=0$, i.e.\ when all boundaries are incoming. However, the general case can be easily reduced to this case via the map $o:z\mapsto 1/z$. More precisely, for $b=b^++b^-$, there is a natural identification between $\mc{T}_{g,m,b^+,b^-}$ and $\mc{T}_{g,m,b}$ given by 
\begin{equation}\label{teich_to_teich_in}
\mc{X}=[(F,\Sigma,J,{\bf x},\bs{\zeta})]\in \mc{T}_{g,m,b^+,b^-} \mapsto\mc{X}^{\rm in}:= [(F,\Sigma,J,{\bf x},\bs{\zeta}^{\rm in})]\in \mc{T}_{g,m,b},
\end{equation}
where $\bs{\zeta}^{\rm in}=(\zeta_1^{\rm in},\dots,\zeta_b^{\rm in})$ is defined, for $\sigma_k$ the orientation sign of $\zeta_k$, by
\[ \zeta_k^{\rm in}:= \zeta_k \textrm{ if }\sigma_k=-1, \quad  \zeta_k^{\rm in}:= \zeta_k\circ o \textrm{ if }\sigma_k=+1.\]
Conversely, if $\bs{\sigma}=(\sigma_1,\dots,\sigma_b)\in \{\pm1\}^b$ is a set of orientation signs with $\sigma_k=+1$ for $k=1,\ldots,b^+$, $\sigma_k=-1$ for $k=b^++1,\ldots,b$, we have the inverse map
\begin{equation}\label{teich_to_teich_sigma}
\mc{X}=[(F,\Sigma,J,{\bf x},\bs{\zeta})]\in \mc{T}_{g,m,b} \mapsto\mc{X}^{\bs{\sigma}}:= [(F,\Sigma,J,{\bf x},\bs{\zeta}^{\bs{\sigma}})]\in \mc{T}_{g,m,b^+,b^-},
\end{equation}
where $\bs{\zeta}^{\bs{\sigma}}=(\zeta_1^{\bs{\sigma}},\dots,\zeta_b^{\bs{\sigma}})$ is defined by 
\[ \zeta_k^{\bs{\sigma}}:=\zeta_k \, \textrm{ if }\sigma_k=-1, \quad \zeta_k^{\bs{\sigma}}:=\zeta_k\circ o \, \textrm{ if }\sigma_k=+1.\]
The same observation applies for the moduli spaces and in the decorated case. In particular, for the decorated Teichm\"uller space we define 
\begin{equation}\label{decor_teich_to_teich_sigma}
\tilde{\mc{X}}=[(F,\Sigma,J,{\bf x},\bs{\zeta},{\bf Y})]\in \tT_{g,m,b} \mapsto\tilde{\mc{X}}^{\bs{\sigma}}:= [(F,\Sigma,J,{\bf x},\bs{\zeta}^{\bs{\sigma}},{\bf Y})]\in \tT_{g,m,b^+,b^-}.
\end{equation}
In the case of complex building blocks, the model form of an element $\mc{X}=[(F,\Sigma,J,{\bf x},\bs{\zeta})]\in \mc{T}_{0,m,b^+,b^-}$ with $m+b^++b^-=3$ can be obtained by the model form of $\mc{X}^{\rm in}$ as follows. If $\Theta_{m,b}(\mc{X}^{\rm in})=(\bs{\psi},\bs{\theta})$, then $\bs{\psi}^{\rm out}:=(\psi^{\rm out}_{1},\dots,\psi_{b^+}^{\rm out}, \psi_{b^++1},\dots,\psi_{b})$ (recall \eqref{psi_in/out}) is a model form of the class $\pi_{\mc{T}\to\mc{M}}(\mc{X})=[(\Sigma,J,{\bf x},\bs{\zeta})]$ in the moduli space $\mc{P}_{m,b^+,b^-}^\eps$, and indeed one can define $\Theta_{m,b^+,b^-}(\mc{X}):=(\bs{\psi}^{\rm out},\bs{\theta})$. However, to simplify the exposition, we shall stick with the maps $\Theta_{m,b},\tilde{\Theta}_{m,b}$ and write out $o$ explicitly. In other words, the model forms $\bs{\psi}$ we consider in later sections will always have $\psi_k\in\tBiHol_\eps(\D)$.

As in the case of surfaces of type $(g,m,b^+,b^-)$, there is an action of $\mc{U}_\eps^b$ on the set of (possibly decorated) marked surfaces of type $(g,m,b^+,b^-)$, and thus on $\mc{T}_{g,m,b^+,b^-}$ (resp.\ $\tT_{g,m,b^+,b^-}$), lifting the action on the moduli space $\mc{M}_{g,m,b^+,b^-}$ (resp.\ $\tilde{\mc{M}}_{g,m,b^+,b^-}$) introduced in \eqref{sigma^fzeta^f}. 
For a marked Riemann surface $(F,\Sigma,J,{\bf x},\bs{\zeta})$ or a decorated one $(F,\Sigma,J,{\bf x},\bs{\zeta},{\bf Y})$, if $(\hat{\Sigma},\hat{J})$ denotes the filling of $(\Sigma,J,{\bf x},\bs{\zeta})$ from Lemma \ref{FillingSigma}, consider for $\bs{f}=(f_1,\dots,f_b)\in \mc{U}_\eps^b$ the (decorated) marked surfaces
\begin{equation}
(F,\Sigma,J,{\bf x},\bs{\zeta})\cdot \bs{f}:=(F^{\bs{f}},\Sigma^{\bs{f}},\hat{J},{\bf x},\bs{\zeta}^{\bs{f}}),\quad (F,\Sigma,J,{\bf x},\bs{\zeta},{\bf Y})\cdot \bs{f}:=(F^{\bs{f}},\Sigma^{\bs{f}},\hat{J},{\bf x},\bs{\zeta}^{\bs{f}},{\bf Y}),
\end{equation}
defined as follows. We let $(\Sigma^{\bs{f}},\hat{J},{\bf x},\bs{\zeta}^{\bs{f}})$ be defined by \eqref{sigma^fzeta^f} and choose the marking diffeomorphism $F^{\bs{f}}:\Sigma^\bullet\to \Sigma^{\bs{f}}$ in a small neighbourhood of $F$ (in the $C^\infty$ topology on maps from $\Sigma^\bullet$ to $\hat{\Sigma}$). 
We can also assume that $F^{\bs{f}}=F$ outside a small collar neighbourhood of $\pl \Sigma^\bullet_{m,b}$. Then $F^{\bs{f}}$ is unique up to isotopy, and we have well-defined actions of $\mc{U}_\eps^b$ on $\mc{T}_{g,m,b^+,b^-}$ and $\tT_{g,m,b^+,b^-}$ given for $\mc{X}=[(F,\Sigma,J,{\bf x},\bs{\zeta})]\in\mc{T}_{g,m,b^+,b^-}$ and $\tilde{\mc{X}}=[(F,\Sigma,J,{\bf x},\bs{\zeta},{\bf Y})]\in\tT_{g,m,b^+,b^-}$ by
\begin{equation}\label{action_bsf_tildeX} 
\mc{X}\cdot \bs{f}:=[(F^{\bs{f}},\Sigma^{\bs{f}},\hat{J},{\bf x},\bs{\zeta}^{\bs{f}})],\quad \tilde{\mc{X}}\cdot \bs{f}:=[(F^{\bs{f}},\Sigma^{\bs{f}},\hat{J},{\bf x},\bs{\zeta}^{\bs{f}},{\bf Y})].
\end{equation}
In the case of complex building blocks with incoming boundaries (i.e. with genus $g=0$, $m+b=3$, $(b^+,b^-)=(0,b)$), the model forms 
$(\bs{\psi}^{\bs{f}},\bs{\theta}^{\bs{f}}):=\Theta_{m,b}(\mc{X}\cdot \bs{f})$ and $(\bs{\psi}^{\bs{f}},\bs{\theta}^{\bs{f}},\bs{\lambda}^{\bs{f}}):=\tilde{\Theta}_{m,b}(\tilde{\mc{X}}\cdot \bs{f})$ depend in a $C^1$ fashion on $\bs{f}$ by Lemma \ref{PhiC1}. Moreover, we have the following formula for $\bs{\lambda}^{\bs{f}}$ in the decorated case. Without loss of generality, we assume that $\tilde{\mc{X}}=[(F,\Sigma,J,{\bf x},\bs{\zeta},{\bf Y})]$ with $\hat{\Sigma}=\hat{\C}$, so that $\Sigma\subset \hat{\C}$. Let $(\bs{\psi},\bs{\theta},\bs{\la})=\tilde{\Theta}_{m,b}(\tilde{\mc{X}})$ be the model form. We observe that
\[ [(F^{\bs{f}}, \Sigma^{\bs{f}},J_{\C},{\bf x},\bs{\zeta}^{\bs{f}},{\bf Y})]= [(\Phi_{\bs{f}}\circ F^{\bs{f}},  \Phi_{\bs{f}}(\Sigma^{\bs{f}}),J_\C,{\bf x},\Phi_{\bs{f}}\circ \bs{\zeta}^{\bs{f}},{\bf Y}^{\bs{f}})]\text{ in }\tT_{0,m,b}, \]
where $\Phi_{\bs{f}}$ is the uniformising map of Lemma \ref{PhiC1} (viewed as a diffeomorphism from a subset of $\hat{\C}$ to another subset of $\hat{\C}$) and ${\bf Y}^{\bs{f}}= (D\Phi_{\bs{f}}(x_{1})Y_1,\dots,D\Phi_{\bs{f}}(x_{m})Y_m)$ for ${\bf Y}=(Y_1,\dots,Y_m)$. Thus
\begin{equation}\label{extension_block_def2} 
\begin{split}
\tilde{\Theta}_{m,b}(\tilde{\mc{X}}\cdot\bs{f})&=\tilde{\Theta}_{m,b}([(F^{\bs{f}}, \Sigma^{\bs{f}},J_{\C},{\bf x},\bs{\zeta}^{\bs{f}},{\bf Y})])\\
&= \tilde{\Theta}_{m,b}( [(\Phi_{\bs{f}}\circ F^{\bs{f}},  \Phi_{\bs{f}}(\Sigma^{\bs{f}}),J_\C,{\bf x},\Phi_{\bs{f}}\circ \bs{\zeta}^{\bs{f}},{\bf Y}^{\bs{f}})])\\
&= ( \Theta_{m,b}(\mc{X}\cdot \bs{f}),\bs{\la}+(\log\Phi_{\bs{f}}'(x_1),\dots,\log\Phi_{\bs{f}}'(x_m))),
\end{split}
\end{equation}
where $\log\Phi_{\bs{f}}'(x_k)$ for $k=1,\dots,m$ can be uniquely defined since $\Phi_{\bs{f}}$ is close to ${\rm Id}$.

\subsection{Marked complex building block decompositions}

The conformal blocks will be defined as functions on Teichm\"uller space depending on a decomposition of the marked surface into 
marked complex building blocks. This leads us to formulate the following definition: 
\begin{definition}[\textbf{Marked complex building block decomposition}]\label{def:cbb_marking}
Let $S=(\Sigma,J,{\bf x},\bs{\zeta})$ be an admissible Riemann surface of type $(g,m,0,b)$ and let
$\tilde{S}=(\Sigma,J,{\bf x},\bs{\zeta},{\bf Y})$ be the same surface with a collection ${\bf Y}$ of decorating vectors at the marked points.
 We assume that $2g-2+m+b>0$. If $3g-3+m+b=0$, equivalently if $g=0$ and $m+b=3$, and $b\geq1$,
the surface is itself a complex building block and its marked complex
building block decomposition consists of a single marked complex
building block, with no internal edge. Otherwise,
$3g-3+m+b>0$ and a \textbf{marked complex building block
decomposition} of $S$ is a collection $\bs{\Xi}=(\Xi_j)_{j=1}^{j_0}$ of $j_0=2g-2+m+b$ orientation preserving smooth maps $\Xi_j:\mc{P}_j^\bullet\to\Sigma$ mapping a canonical complex building block $\mc{P}_j^\bullet=\Sigma^\bullet_{m_j,b_j}$ (with $m_j+b_j=3$, $b_j\geq1$) to the Riemann surface with boundary $\Xi_j(\mc{P}_j^\bullet)$ embedded in $\Sigma$, satisfying the following properties:\\
1) For each $j$, $\Xi_j|_{(\mc{P}_j^\bullet)^\circ}$ is diffeomorphic onto the image but may be non-injective on the boundary $\pl\mc{P}_j^\bullet$, $\Xi_j$ is holomorphic near $\{{\bf x}^\bullet_j\}\cup\pl\mc{P}_j^\bullet$, and $\Xi_j(\{{\bf x}^\bullet_j\})\subset \{{\bf x}\}$, where ${\bf x}^\bullet_j=(x_1^\bullet,\ldots,x_{m_j}^\bullet)$ are the marked points of $\mc{P}_j^\bullet$.
The \textbf{boundary} of $\bs{\Xi}$ is the set $\bigcup_{j=1}^{j_0}\Xi_j(\pl\mc{P}_j^\bullet)\subset\Sigma$.\\
2) One has $\Sigma=\bigcup_{j=1}^{j_0}\Xi_j(\mc{P}_j^\bullet)$, and $(\Xi_j(\mc{P}_j^\bullet))^\circ\cap(\Xi_{j'}(\mc{P}_{j'}^\bullet))^\circ=\emptyset$ for $j\not=j'$. \\
3) For each $(j,k)$ with $j\in \{1,\dots,j_0\}$ and $k\in \{1,\dots,b_j\}$ such that $\Xi_{j}\circ \zeta^\bullet_k(\T)\not\subset \pl\Sigma$, there exist $j'\in \{1,\dots,j_0\}$ and $k'\in \{1,\dots,b_{j'}\}$ with $(j,k)\not=(j',k')$ such that
\begin{equation}\label{gluing_rule}
\Xi_j\circ \zeta^\bullet_k=\Xi_{j'}\circ \zeta^\bullet_{k'}\circ o,
\end{equation}
where $o:e^{{\rm i}\theta}\mapsto e^{-{\rm i}\theta}$ on $\T$ and $\zeta_k^\bullet$ is the canonical parametrisation of \eqref{param_base_model_block} for $\pl_k\mc{P}_j^\bullet=\pl\mc{D}_k^\bullet$. In the case $j=j'$, we say that $\Xi_j(\mc{P}_j^\bullet)$ is self-glued on the curve $\Xi_j\circ \zeta^\bullet_k(\T)=\Xi_{j'}\circ \zeta^\bullet_{k'}(\T)\subset(\Xi_j(\mc{P}_j^\bullet))^\circ$. \\
4) For each parametrisation $\zeta_\ell$ in $\bs{\zeta}$ parametrising the boundary of $\Sigma$, there exist $j\in \{1,\dots,j_0\}$ and $k\in \{1,\dots,b_j\}$ such that
\begin{equation}\label{boundary_compatibility}
\zeta_\ell=\Xi_j\circ \zeta^\bullet_k.
\end{equation}
\noindent With ${\bf Y}^\bullet_j=(Y_{1}^\bullet,\dots,Y_{m_j}^\bullet)$ being the fixed collection of decorating vectors $Y_{k}^\bullet=\pl_z\in(T_{x_{k}^\bullet}\hat{\C})^{1,0}$ on $\mc{P}_j^\bullet=\Sigma^\bullet_{m_j,b_j}$ for each $j$, we say that a marked complex building block decomposition $\bs{\Xi}=(\Xi_j)_{j=1}^{j_0}$ of $S$ is a marked complex building block decomposition of $\tilde{S}$ if $D\Xi_j(x_{k}^\bullet)Y_{k}^\bullet=Y_{jk}$ for $k=1,\ldots,m_j$, where $Y_{jk}$ is the decorating vector attached to $\Xi_{j}(x_k^\bullet)$ in the collection ${\bf Y}$.
We denote by $\mc{BD}(S)$, resp. $\tilde{\mc{BD}}(\tilde{S})$, the set of all marked complex building block decompositions of $S$, resp. $\tilde{S}$.
\end{definition}

\begin{remark}
For each $j$, the open set $\Xi_j((\mc{P}_j^\bullet)^\circ)\subset\Sigma$ with the complex structure $J$ can be compactified into a Riemann surface $(\mc{P}_j,J)$ (with the same $J$ by abuse of notation) diffeomorphic to $\mc{P}_j^\bullet$. More precisely, there is a corresponding surjective gluing map $\pi_j:\mc{P}_j\to\Xi_j(\mc{P}_j^\bullet)=\overline{\Xi_j((\mc{P}_j^\bullet)^\circ)}\subset\Sigma$, and the diffeomorphic embedding $\Xi_j|_{(\mc{P}_j^\bullet)^\circ}$ extends to a diffeomorphism $\Xi^{\rm ex}_j:\mc{P}_j^\bullet\to\mc{P}_j$ such that $\Xi_j=\pi_j\circ \Xi^{\rm ex}_j$. Moreover, $\mc{P}_j$ inherits a structure of complex building block $(\mc{P}_j,J,{\bf x}_j,\bs{\zeta}_j)$ where ${\bf x}_j=\Xi_j({\bf x}_j^\bullet)$ and $\bs{\zeta}_j$ is such that $\pi_j\circ\bs{\zeta}_j=(\Xi_j\circ \zeta_{1}^\bullet,\dots,\Xi_j\circ \zeta_{b_j}^\bullet)$.
Notice that $\pi_j$ is a diffeomorphism if and only if $\Xi_j(\mc{P}_j^\bullet)$ has no self-gluing. In case $\Xi_j(\mc{P}_j^\bullet)$ is self-glued on the curve parametrised by $\Xi_j\circ \zeta^\bullet_k=\Xi_j\circ \zeta^\bullet_{k'}\circ o$, the Riemann surface $\Xi_j(\mc{P}_j^\bullet)$ necessarily has genus $1$ and is obtained from the complex building block $\mc{P}_j$ by gluing $\pl_k\mc{P}_j$ with $\pl_{k'}\mc{P}_j$ through the identification $\zeta_{jk}\sim\zeta_{jk'}\circ o$. 
In practice however, we shall largely ignore the distinction between $\mc{P}_j$ and $\Xi_j(\mc{P}_j^\bullet)$ and think of $\Xi_j$ as an embedded complex building block in $\Sigma$. This slight abuse of notation will be harmless since the precise meaning is always clear from context.
\end{remark}

\begin{remark}
We emphasise that our convention \eqref{gluing_rule} for gluing the boundaries is different from what one would use in the theory of modular functors. More precisely, there one would require that $\Xi_j\circ \zeta^\bullet_k=\Xi_{j'}\circ \zeta^\bullet_{k'}$ give the same parametrisation of the cutting curve $\Xi_j\circ \zeta^\bullet_k(\T)$ (viewed as an object in the cobordism category), so that $\pl_k\mc{P}_j$ is outgoing and $\pl_{k'}\mc{P}_{j'}$ is incoming, or vice versa. Our convention requires instead that the two parametrisations differ by $o:e^{{\rm i}\theta}\mapsto e^{-{\rm i}\theta}$, so that we can assume that both $\pl_k\mc{P}_j$ (parametrised by $\Xi_j\circ\zeta_k^\bullet$) and $\pl_{k'}\mc{P}_{j'}$ (parametrised by $\Xi_{j'}\circ\zeta_{k'}^\bullet$) are incoming, which allows us to avoid problems of duality when defining the conformal blocks later.
\end{remark}

Given a marked complex building block decomposition $\bs{\Xi}\in\mc{BD}(S)$, resp. $\bs{\Xi}\in\tilde{\mc{BD}}(\tilde{S})$, we obtain for each $j$ an element
\begin{equation}\label{element_in_subTeich} 
\mc{X}_{j}:=[ (\Xi_j, \mc{P}_j, J, {\bf x}_j,\bs{\zeta}_j)]\in \mc{T}_{0,m_j,b_j}, \quad \textrm{resp. }\tilde{\mc{X}}_{j}:=[ (\Xi_j, \mc{P}_j, J, {\bf x}_j,\bs{\zeta}_j,{\bf Y}_j)]\in \tT_{0,m_j,b_j},
\end{equation}
where ${\bf Y}_j=(Y_{j1},\ldots,Y_{jm_j})$ is a subcollection of ${\bf Y}$ as in Definition \ref{def:cbb_marking}.
Combining this with the maps $\Theta_{m,b},\tilde\Theta_{m,b}$ from Proposition \ref{identification_Teichmuller}, we obtain maps $\mc{MF}$ on $\mc{BD}(S)$ and $\tilde{\mc{MF}}$ on $\tilde{\mc{BD}}(\tilde{S})$, called the \textbf{model forms}, defined by
\begin{equation}\label{definition_MFXi}
\mc{MF}(\bs{\Xi}):=(\Theta_{m_j,b_j}(\mc{X}_j))_{j=1}^{j_0}, \quad \tilde{\mc{MF}}(\bs{\Xi}):=(\tilde{\Theta}_{m_j,b_j}(\tilde{\mc{X}}_j))_{j=1}^{j_0}.
\end{equation}
The conformal blocks will be defined via these model forms. Intuitively, we think of the map $\mc{MF}$ and $\tilde{\mc{MF}}$ as defined on the union ``$\bigcup_{S}\mc{BD}(S)$'' and 
 ``$\bigcup_{S}\tilde{\mc{BD}}(\tilde{S})$'' of all marked complex building block decompositions of all (decorated) Riemann surfaces of type $(g,m,0,b)$. The latter spaces can be defined rigorously as fiber bundles over the Teichm\"uller spaces $\mc{T}_{g,m,b}$ and $\tT_{g,m,b}$ whose fibers are the corresponding spaces of decompositions modulo the action of certain diffeomorphisms, as defined below.

\begin{definition}[\textbf{Teichm\"uller space decorated with marked complex building block decomposition}]\label{def:Decoratd_teich_DT}
Let $S^\bullet=(\Sigma^{\bullet},J^\bullet,{\bf x}^\bullet,\bs{\zeta}^\bullet)$ be a fixed base Riemann surface of type $(g,m,0,b)$ and let
$\tilde{S}^\bullet=(\Sigma^{\bullet},J^\bullet,{\bf x}^\bullet,\bs{\zeta}^\bullet,{\bf Y}^\bullet)$ be the same surface with a fixed collection of decorating vectors ${\bf Y}^\bullet$ at the marked points, as in Definition \ref{def:Teichmuller}. 
We define the following:\\
1) The space $\BDT_{g,m,b}=\BDT(S^\bullet)$ is the set of equivalence classes $[(F,\Sigma,J,{\bf x},\bs{\zeta},\bs{\Xi})]$, where $(F,\Sigma,J,{\bf x},\bs{\zeta})$ is a marked Riemann surface defining a point $[(F,\Sigma,J,{\bf x},\bs{\zeta})]\in\mc{T}_{g,m,b}=\mc{T}_{g,m,b}(S^\bullet)$ and $\bs{\Xi}\in\mc{BD}(S)$ is a marked complex building block decomposition of $S=(\Sigma,J,{\bf x},\bs{\zeta})$, and we say that $(F^1,\Sigma^1,J^1,{\bf x}^1,\bs{\zeta}^1,\bs{\Xi}^1)$ and $(F^2,\Sigma^2,J^2,{\bf x}^2,\bs{\zeta}^2,\bs{\Xi}^2)$ are equivalent if there exists a biholomorphism $\Phi:(\Sigma^1,J^1)\to(\Sigma^2,J^2)$ such that $(F^2)^{-1}\circ\Phi\circ F^1\in{\rm Diff}_0(\Sigma^\bullet,\{{\bf x}^\bullet\}\cup \pl\Sigma^\bullet)$ and, 
for each $j=1,\ldots,j_0$, there exists $h_j\in{\rm Diff}_0(\mc P_j^\bullet,\{{\bf x}_j^\bullet\}\cup\pl\mc P_j^\bullet)$
such that
\[
\Phi\circ\Xi_j^1=\Xi_j^2\circ h_j.
\]
2) The space $\BDTT_{g,m,b}=\BDTT(\tilde{S}^\bullet)$ is the set of equivalence classes $[(F,\Sigma,J,{\bf x},\bs{\zeta},{\bf Y},\bs{\Xi})]$, where $(F,\Sigma,J,{\bf x},\bs{\zeta},{\bf Y})$ is a decorated marked Riemann surface defining a point $[(F,\Sigma,J,{\bf x},\bs{\zeta},{\bf Y})]\in\tT_{g,m,b}=\tT_{g,m,b}(\tilde{S}^\bullet)$ and $\bs{\Xi}\in\tilde{\mc{BD}}(\tilde{S})$ is a marked complex building block decomposition of $\tilde{S}=(\Sigma,J,{\bf x},\bs{\zeta},{\bf Y})$. Moreover, we say that $(F^1,\Sigma^1,J^1,{\bf x}^1,\bs{\zeta}^1,{\bf Y}^1,\bs{\Xi}^1)$ and $(F^2,\Sigma^2,J^2,{\bf x}^2,\bs{\zeta}^2,{\bf Y}^2,\bs{\Xi}^2)$ are equivalent if there exists a biholomorphism $\Phi:(\Sigma^1,J^1)\to(\Sigma^2,J^2)$ such that $(F^2)^{-1}\circ\Phi\circ F^1\in{\rm Diff}_0(\Sigma^\bullet,\{{\bf x}^\bullet\}\cup \pl\Sigma^\bullet; {\bf Y}^\bullet)$ and,
for each $j=1,\ldots,j_0$, there exists $h_j\in{\rm Diff}_0( \mc P_j^\bullet,\{{\bf x}_j^\bullet\}\cup\pl\mc P_j^\bullet;{\bf Y}_j^\bullet)$
such that
\[
\Phi\circ\Xi_j^1=\Xi_j^2\circ h_j.
\]
Notice that the biholomorphism $\Phi$, if it exists, is uniquely
determined. Indeed, since the diffeomorphisms $h_j$ are equal to the
identity near the boundary, the above relations determine $\Phi$ on
an open collar of the cutting curves, and hence on the whole connected
Riemann surface by analytic continuation.
\end{definition}

Notice that for $[(F,\Sigma,J,{\bf x},\bs{\zeta},\bs{\Xi})]\in\BDT_{g,m,b}$ and $h_0\in{\rm Diff}_0(\Sigma^\bullet,\{{\bf x}^\bullet\}\cup \pl\Sigma^\bullet)$, we have by definition
\[[(F\circ h_0,\Sigma,J,{\bf x},\bs{\zeta},\bs{\Xi})]=[(F,\Sigma,J,{\bf x},\bs{\zeta},\bs{\Xi})]\text{ in }\BDT_{g,m,b}.\]
More generally, there is a right action of the mapping class group 
\[{\rm Mod}_{g,m,b}={\rm Mod}(\Sigma^\bullet,\{{\bf x}^\bullet\}\cup \pl\Sigma^\bullet
)\]
on $\BDT_{g,m,b}$ given by 
\[ [(F,\Sigma,J,{\bf x},\bs{\zeta},\bs{\Xi})]\cdot [h]:=[(F\circ h,\Sigma,J,{\bf x},\bs{\zeta},\bs{\Xi})],\]
which lifts the right action of ${\rm Mod}_{g,m,b}$ on $\mc{T}_{g,m,b}$ discussed in Section \ref{sub:Teichmuller}. There is a similar action of the framed mapping class group $\tilde{{\rm Mod}}_{g,m,b}=\tilde{{\rm Mod}}(\Sigma^\bullet,\{{\bf x}^\bullet\}\cup \pl\Sigma^\bullet)$ on $\BDTT_{g,m,b}$.

There are natural projections
\begin{equation}\label{BD_Teich_fibration}
\begin{alignedat}{2}
\pi_{\BDT\to\mc{T}}
&:\BDT_{g,m,b}\to\mc{T}_{g,m,b},
&\qquad [(F,\Sigma,J,{\bf x},\bs{\zeta},\bs{\Xi})]
&\mapsto [(F,\Sigma,J,{\bf x},\bs{\zeta})],\\
\pi_{\BDTT\to\tT}
&:\BDTT_{g,m,b}\to\tT_{g,m,b},
&\qquad [(F,\Sigma,J,{\bf x},\bs{\zeta},{\bf Y},\bs{\Xi})]
&\mapsto [(F,\Sigma,J,{\bf x},\bs{\zeta},{\bf Y})],\\
\pi_{\BDTT\to\BDT}
&:\BDTT_{g,m,b}\to \BDT_{g,m,b},
&\qquad [(F,\Sigma,J,{\bf x},\bs{\zeta},{\bf Y},\bs{\Xi})]
&\mapsto [(F,\Sigma,J,{\bf x},\bs{\zeta},\bs{\Xi})],
\end{alignedat}
\end{equation}
and we have the commutative diagram
\[
\begin{tikzcd}
\BDTT_{g,m,b}
  \arrow[d,"\pi_{\BDTT\to\tT}"]
  \arrow[r,"\pi_{\BDTT\to\BDT}"]
& \BDT_{g,m,b}
\arrow[d,"\pi_{\BDT\to \mc{T}}"]
\\
\tT_{g,m,b}
  \arrow[r,"\pi_{\tT\to\mc{T}}"]
& \mc{T}_{g,m,b}
\end{tikzcd}
\]

We now discuss the fibers of the projections $\pi_{\BDT\to\mc T}$ and $\pi_{\BDTT\to\tT}$. Let
\begin{equation}\label{BDred}
\mc{BD}_{\rm red}(S):=\mc{BD}(S)/{\sim},
\end{equation}
where
$\bs\Xi^1\sim\bs\Xi^2$ if for every $j$, there exists  $h_j\in{\rm Diff}_0(\mc P_j^\bullet,\{{\bf x}_j^\bullet\}\cup\pl\mc P_j^\bullet)$ 
such that $\Xi_j^1=\Xi_j^2\circ h_j$. We define $\tilde{\mc{BD}}_{\rm red}(\tilde S)$ similarly, using
the relative identity components preserving ${\bf Y}_j^\bullet$.
The fiber of $\BDT_{g,m,b}$ over a
point $\mc{X}=[(F,\Sigma,J,{\bf x},\bs{\zeta})]\in\mc{T}_{g,m,b}$
is naturally isomorphic to $\mc{BD}_{\rm red}(S)$, where $S=(\Sigma,J,{\bf x},\bs{\zeta})$. Indeed,  for $[(F',\Sigma',J',{\bf x}',\bs{\zeta}',\bs{\Xi}')]\in\pi_{\BDT\to\mc{T}}^{-1}(\mc{X})$, we have $[(F',\Sigma',J',{\bf x}',\bs{\zeta}')]=[(F,\Sigma,J,{\bf x},\bs{\zeta})]$ in $\mc{T}_{g,m,b}$, so there exist $h_0\in{\rm Diff}_0(\Sigma^\bullet,\{{\bf x}^\bullet\}\cup \pl\Sigma^\bullet)$ and a biholomorphism $\Phi:(\Sigma,J)\to(\Sigma',J')$ such that $F'=\Phi\circ F\circ h_0$.   Then 
\[[(F',\Sigma',J',{\bf x}',\bs{\zeta}',\bs{\Xi}')]=[(F\circ h_0,\Sigma,J,{\bf x},\bs{\zeta},\Phi^{-1}\circ\bs{\Xi}')]=[(F,\Sigma,J,{\bf x},\bs{\zeta},\Phi^{-1}\circ\bs{\Xi}')]\] 
in $\BDT_{g,m,b}$ with $\Phi^{-1}\circ\bs{\Xi}'\in\mc{BD}(S)$. 
Both equalities here are instances of the defining equivalence with all interior
reparametrisations $h_j$ equal to the identity.
Consequently,
\[\pi_{\BDT\to\mc{T}}^{-1}(\mc{X})=
\left\{[(F,\Sigma,J,{\bf x},\bs{\zeta},\bs{\Xi})] \,|\,[\bs{\Xi}]_{\rm red}\in\mc{BD}_{\rm red}(S)\right\}
\]
and the map
\[\pi_{\BDT\to\mc{T}}^{-1}(\mc{X}) \to \mc{BD}_{\rm red}(S),
\qquad [(F,\Sigma,J,{\bf x},\bs{\zeta},\bs{\Xi})] \mapsto[\bs{\Xi}]_{\rm red}
\]
is a well-defined bijection, where $[\cdot]_{\rm red}$ denotes the equivalence class in \eqref{BDred}.  Notice that this map depends on the choice of a representative marked Riemann surface $(F,\Sigma,J,{\bf x},\bs{\zeta})$ in $\mc{X}=[(F,\Sigma,J,{\bf x},\bs{\zeta})]\in\mc{T}_{g,m,b}$. 
However, the composition of this map with the map $\mc{MF}$ defined above on $\mc{BD}(S)$ is well-defined and gives a map, also called the \textbf{model form},
\begin{equation}\label{def_MF}
\mc{MF}:\BDT_{g,m,b}\to\Big(\bigsqcup_{b=1}^3\bigcup_{\eps>0}\mc{T}_{3-b,b}^\eps\Big)^{j_0},\quad [(F,\Sigma,J,{\bf x},\bs{\zeta},\bs{\Xi})]\mapsto \mc{MF}(\bs{\Xi})=(\Theta_{m_j,b_j}(\mc{X}_j))_{j=1}^{j_0},
\end{equation}
where $\mc{X}_j$ is defined by \eqref{element_in_subTeich} and $m_j,b_j$ (with $m_j+b_j=3$, $b_j\geq1$) depend on the type of $\Xi_j$. Similarly, for the projection $\pi_{\BDTT\to\tT}$, the fiber of $\BDTT_{g,m,b}$ over a point $\tilde{\mc{X}}=[(F,\Sigma,J,{\bf x},\bs{\zeta},{\bf Y})]\in\tT_{g,m,b}$ is isomorphic to $\tilde{\mc{BD}}_{\rm red}(\tilde{S})$ with $\tilde{S}=(\Sigma,J,{\bf x},\bs{\zeta},{\bf Y})$, and one has a well-defined model form map
\begin{equation}\label{def_MFtilde}
\tilde{\mc{MF}}:\BDTT_{g,m,b}\to\Big(\bigsqcup_{b=1}^3\bigcup_{\eps>0}\tT_{3-b,b}^\eps\Big)^{j_0},\quad [(F,\Sigma,J,{\bf x},\bs{\zeta},{\bf Y},\bs{\Xi})]\mapsto \tilde{\mc{MF}}(\bs{\Xi})=(\tilde{\Theta}_{m_j,b_j}(\tilde{\mc{X}}_j))_{j=1}^{j_0},
\end{equation}
with $\tilde{\mc{X}}_j$ again defined by \eqref{element_in_subTeich}. Notice that these maps are not injective since they forget the marking $F$. In fact, their fibers are discrete, as we shall explain now. 

\begin{definition}\label{def_pi_0_BD}
Let $S^\bullet=(\Sigma^{\bullet},J^\bullet,{\bf x}^\bullet,\bs{\zeta}^\bullet)$ be a fixed base Riemann surface of type $(g,m,0,b)$ and let
$\tilde{S}^\bullet=(\Sigma^{\bullet},J^\bullet,{\bf x}^\bullet,\bs{\zeta}^\bullet,{\bf Y}^\bullet)$ be the same surface with a fixed collection of decorating vectors ${\bf Y}^\bullet$ at the marked points, as in Definition \ref{def:Teichmuller}. We write $\mc{BD}_{g,m,b}=\mc{BD}(S^\bullet)$, resp. $\tilde{\mc{BD}}_{g,m,b}=\tilde{\mc{BD}}(\tilde{S}^\bullet)$, and denote by $\piBD_{g,m,b}=\piBD(S^\bullet)$, resp. $\pi_0\tilde{\mc{BD}}_{g,m,b}=\pi_0\tilde{\mc{BD}}(\tilde{S}^\bullet)$, the sets of isotopy classes of marked complex building decompositions of $S^\bullet$, resp. $\tilde{S}^\bullet$. More precisely, an element of $\piBD_{g,m,b}$ is a set of the form
\[[\bs{\Xi}^\bullet]:=\{h_0\circ\bs{\Xi}^\bullet\in\mc{BD}_{g,m,b} \,|\,h_0\in{\rm Diff}_0(\Sigma^\bullet,\{{\bf x}^\bullet\}\cup \pl\Sigma^\bullet)\}\]
for some $\bs{\Xi}^\bullet\in\mc{BD}_{g,m,b}$, and an element of $\pi_0\tilde{\mc{BD}}_{g,m,b}$ is a set of the form
\[[\bs{\Xi}^\bullet]:=\{h_0\circ\bs{\Xi}^\bullet\in\tilde{\mc{BD}}_{g,m,b} \,|\,h_0\in{\rm Diff}_0(\Sigma^\bullet,\{{\bf x}^\bullet\}\cup \pl\Sigma^\bullet;{\bf Y}^\bullet)\}\]
for some $\bs{\Xi}^\bullet\in\tilde{\mc{BD}}_{g,m,b}$.
\end{definition}
The  diffeomorphisms $h_j$ used in Definition \ref{def:Decoratd_teich_DT} do not change these isotopy
classes. 
Notice that there are free left actions of ${\rm Mod}_{g,m,b}$ on $\piBD_{g,m,b}$ and resp.\ of $\tilde{{\rm Mod}}_{g,m,b}$ on $\pi_0\tilde{\mc{BD}}_{g,m,b}$, defined by
\begin{equation}\label{action_Mod_pi0DD}
[h]\cdot[\bs{\Xi}^\bullet]:=[h\circ\bs{\Xi}^\bullet], 
\end{equation}
where we choose the representative $h$ such that it is analytic near the boundary of $\bs{\Xi}^\bullet$.\\

We can now define the following lifting of the map $\mc{MF}$ of \eqref{def_MF}:
\begin{equation}\label{def_MFhat}
\mc{MF}^{\rm aug}:\BDT_{g,m,b}\to \mc{MF}(\BDT_{g,m,b})\times\piBD_{g,m,b}, \quad [(F,\Sigma,J,{\bf x},\bs{\zeta},\bs{\Xi})]\mapsto (\mc{MF}(\bs{\Xi}),[F^{-1}\circ\bs{\Xi}]).
\end{equation}
Here we can choose $F$ such that $F^{-1}$ is analytic near the boundary of $\bs{\Xi}$, so that $F^{-1}\circ\bs{\Xi}\in\mc{BD}_{g,m,b}$. We shall see in a moment that $\mc{MF}^{\rm aug}$ is injective. Notice also that for $[h]\in {\rm Mod}_{g,m,b}$,
\[\mc{MF}^{\rm aug}([(F,\Sigma,J,{\bf x},\bs{\zeta},\bs{\Xi})]\cdot [h])=[h^{-1}]\cdot
\mc{MF}^{\rm aug}([(F,\Sigma,J,{\bf x},\bs{\zeta},\bs{\Xi})]),\]
where the left action of ${\rm Mod}_{g,m,b}$ on $\mc{MF}(\BDT_{g,m,b})\times\piBD_{g,m,b}$ is 
the extension of \eqref{action_Mod_pi0DD} acting trivially on $\mc{MF}(\BDT_{g,m,b})$, i.e.
\[[h]\cdot (\mc{MF}(\bs{\Xi}),[\bs{\Xi}^\bullet]):=(\mc{MF}(\bs{\Xi}),[h]\cdot [\bs{\Xi}^\bullet]).\]

Next, we shall describe the reverse procedure of gluing a collection of marked complex building blocks into a marked Riemann surface. This will give the inverse of the map $\mc{MF}^{\rm aug}$, and in particular right inverses of the map $\mc{MF}$. First, we introduce a combinatorial structure to encode the way in which the complex building blocks are glued together in a complex building block decomposition.

\begin{definition}[\textbf{Gluing graph}]\label{def:gluing_graph}
Let $S=(\Sigma,J,{\bf x},\bs{\zeta})$ be an admissible Riemann surface of type $(g,m,0,b)$ and $\bs{\Xi}=(\Xi_j)_{j=1}^{j_0}\in\mc{BD}(S)$ a marked complex building block decomposition of $S$. The \textbf{gluing graph} of $\bs{\Xi}$ is the (connected) non-oriented multi-graph $\mcG=(V,E)$ defined as follows:
\begin{itemize}
\item The set of vertices is $V=\{{\rm ve}_1,\dots,{\rm ve}_{j_0},{\rm ve}_1^{{\rm m}},\ldots,{\rm ve}_m^{{\rm m}},{\rm ve}^{\rm b}_1,\ldots,{\rm ve}^{\rm b}_b\}$, where ${\rm ve}_j$ denote \textbf{internal vertices} and are of the form ${\rm ve}_j=\{(j,1),(j,2),(j,3)\}$, 
${\rm ve}^{\rm m}_j=\{(j,\bullet)\}$ are external vertices associated to marked points, ${\rm ve}^{\rm b}_j=\{(j,\circ )\}$ are external vertices associated to boundary components. 
\item The set of edges $E=\{{\rm e}_1,\dots,{\rm e}_{i_0},{\rm e}_1^{{\rm m}},
\dots,e_m^{\rm m},{\rm e}_1^{{\rm b}},\dots,{\rm e}_{b}^{\rm b}\}$ consists of three types of edges:
\begin{itemize}
    \item For $j,k,j',k'$ as in \eqref{gluing_rule}, there is an edge ${\rm e}_i=\{(j,k),(j',k')\}$ between ${\rm ve}_j$ and ${\rm ve}_{j'}$. These edges are called the \textbf{internal edges}. Here we have fixed an enumeration of these edges, which is merely for notational convenience and will not be important in what follows.
    \item For $\ell,j,k$ such that $x_\ell=\Xi_j(z_k^\bullet)$ with $k>b_j$, there is an edge ${\rm e}^{\rm m}_\ell=\{(\ell,\bullet),(j,k)\}$ between ${\rm ve}^{\rm m}_\ell$ and ${\rm ve}_j$. These edges are \textbf{external edges associated to marked points}.
    \item For $\ell,j,k$ as in \eqref{boundary_compatibility} with $k\leq b_j$, there is an edge ${\rm e}^{\rm b}_\ell=\{(\ell,\circ ),(j,k)\}$ between ${\rm ve}^{{\rm b}}_\ell$ and 
    ${\rm ve}_j$. These edges are \textbf{external edges associated to boundary circles}.
\end{itemize}
\end{itemize}
The \textbf{internal subgraph} of $\mcG$ is the subgraph consisting of internal vertices and internal edges. Notice that necessarily $j_0=2g-2+m+b$, $i_0=3g-3+m+b$.
\end{definition}

It will sometimes be convenient later to pick orientations of the internal edges ${\rm e}_i$; in that case we denote an oriented edge by $\vec{\rm e}_i=((j,k),(j',k'))$, viewing $(j,k)$ as the starting vertex and $(j',k')$ the ending vertex. The oriented graph will then be denoted $\vec{\mc{G}}$.

For a gluing graph $\mcG$, we denote by $\mc{BD}^{\mcG}(S)$ the subset of elements in $\mc{BD}(S)$ whose gluing graph coincides with $\mcG$, and similarly by $\BDT_{g,m,b}^{\mcG}$ the subset of elements $[(F,\Sigma,J,{\bf x},\bs{\zeta},\bs{\Xi})]\in\BDT_{g,m,b}$ where $\bs{\Xi}\in\mc{BD}^{\mc{G}}(S)$ with $S=(\Sigma,J,{\bf x},\bs{\zeta})$, and by $\piBD_{g,m,b}^{\mcG}\subset\piBD_{g,m,b}$ the subset of isotopy classes in $\mc{BD}^{\mcG}_{g,m,b}=\mc{BD}^{\mcG}(S^\bullet)$. Notice that two marked decompositions $\bs{\Xi},\bs{\Xi}'\in\mc{BD}(S)$ (with $S=(\Sigma,J,{\bf x},\bs{\zeta})$) have the same gluing graph if and only if there exists a diffeomorphism $h\in{\rm Diff}(\Sigma,\{{\bf x}\}\cup\pl\Sigma)$ such that $\bs{\Xi}'=h\circ\bs{\Xi}$. The map $\mc{MF}$ of \eqref{def_MF} restricts to a surjective map
\[\mc{MF}|_{\BDT_{g,m,b}^{\mcG}}:\BDT_{g,m,b}^{\mcG}\to\mc{T}_{g,m,b}^{\mc{G}}:=\bigcup_{\eps>0}\mc{T}_{m_1,b_1}^\eps\times\cdots\times\bigcup_{\eps>0}\mc{T}_{m_{j_0},b_{j_0}}^\eps,\]
so that
\begin{equation}\label{hatMF}
\mc{MF}^{\rm aug}|_{\BDT_{g,m,b}^{\mcG}}:
\BDT_{g,m,b}^{\mcG}\to\mc{T}_{g,m,b}^{\mcG}\times\piBD_{g,m,b}^{\mcG}.
\end{equation}
Conversely, we can define a \textbf{gluing map}
\begin{equation}\label{gluing_map}
{\rm Gl}^{\mcG}:\mc{T}_{g,m,b}^{\mc{G}}\times\piBD_{g,m,b}^{\mcG}\to\BDT_{g,m,b}^{\mcG}
\end{equation}
as follows. By abuse of notation, we shall freely identify $\mc{T}_{g,m,b}^{\mc{G}}$ with $\mc{T}_{0,m_1,b_1}\times\cdots\times\mc{T}_{0,m_{j_0},b_{j_0}}$ using the model form map in Proposition \ref{identification_Teichmuller}, i.e. we think of ${\rm Gl}^{\mcG}$ as gluing either marked complex building blocks or their model forms.

Given a collection $\bs{\mc{X}}=(\mc{X}_j)_{j=1}^{j_0}\in\mc{T}_{g,m,b}^{\mcG}$ of marked complex building blocks with $\mc{X}_j=[(\Xi_j,\mc{P}_j,J_j,{\bf x}_j,\bs{\zeta}_j)]\in\mc{T}_{0,m_j,b_j}$, we construct a Riemann surface $S^{\bs{\mc{X}}}=(\Sigma^{\bs{\mc{X}}},J^{\bs{\mc{X}}},{\bf x}^{\bs{\mc{X}}},\bs{\zeta}^{\bs{\mc{X}}})$ of type $(g,m,0,b)$ by setting $\Sigma^{\bs{\mc{X}}}:=\bigsqcup_{j=1}^{j_0}\mc{P}_j/\!\sim$, where the equivalence relation $\sim$ is defined by $\Xi_j\circ\zeta_k^\bullet(e^{{\rm i}\theta})\in\mc{P}_j\sim\Xi_{j'}\circ\zeta_{k'}^\bullet(e^{-{\rm i}\theta})\in\mc{P}_{j'}$ ($\theta\in\R$) for each internal edge $\{(j,k),(j',k')\}$ in $\mcG$, $J^{\bs{\mc{X}}}$ is the unique complex structure on $\Sigma^{\bs{\mc{X}}}$ such that all inclusion maps $\iota^{\bs{\mc{X}}}_j:\mc{P}_j\to\Sigma^{\bs{\mc{X}}}$ are holomorphic, ${\bf x}^{\bs{\mc{X}}}=(x_1^{\bs{\mc{X}}},\ldots,x_m^{\bs{\mc{X}}})$ where $x_\ell^{\bs{\mc{X}}}:=\iota^{\bs{\mc{X}}}_j(x_{jk})$ (with ${\bf x}_j=(x_{j1},\ldots,x_{jm_j})$ the marked points of $\mc{P}_j$) if ${\rm e}^{\rm m}_\ell =\{(\ell,\bullet),(j,b_j+k)\}$ is an edge in $\mcG$, and finally $\bs{\zeta}^{\bs{\mc{X}}}=(\zeta_1^{\bs{\mc{X}}},\ldots,\zeta_b^{\bs{\mc{X}}})$ where $\zeta_\ell^{\bs{\mc{X}}}:=\iota^{\bs{\mc{X}}}_j\circ\zeta_{jk}$ (with $\bs{\zeta}_j=(\zeta_{j1},\ldots,\zeta_{jb_j})$ the boundary parametrisations of $\mc{P}_j$) if $\{(\ell,\circ),(j,k)\}$ is an edge in $\mcG$. In particular, applying this with $\mc{X}_j=[({\rm Id},\mc{P}_j^\bullet=\Sigma_{m_j,b_j}^\bullet,J_{\C},{\bf x}^\bullet,\bs{\zeta}^\bullet)]$ yields a surface $S^{\mcG}=(\Sigma^{\mcG},J^{\mcG},{\bf x}^{\mcG},\bs{\zeta}^{\mcG})$ depending only on the graph $\mcG$. We will call this surface the \textbf{surface realisation} of $\mcG$.

For a general point $\bs{\mc{X}}=(\mc{X}_j)_{j=1}^{j_0}\in\mc{T}_{g,m,b}^{\mcG}$ and fixed representatives
$(\Xi_j,\mc P_j,J_j,{\bf x}_j,\bs\zeta_j)$ of the points $\mc X_j$,	 there is a unique diffeomorphism $F^{\bs{\mc{X}}}:\Sigma^{\mcG}\to\Sigma^{\bs{\mc{X}}}$ determined by the relation $\Xi_j^{\bs{\mc{X}}}:=\iota_j^{\bs{\mc{X}}}\circ\Xi_j = F^{\bs{\mc{X}}}\circ\iota_j^{\mcG}:\mc{P}_j^\bullet\to\Sigma^{\bs{\mc{X}}}$, where $\iota_j^{\bs{\mc{X}}}:\mc{P}_j\to\Sigma^{\bs{\mc{X}}}$ and $\iota_j^{\mcG}:\mc{P}_j^\bullet\to\Sigma^{\mcG}$ are the inclusion maps coming from the construction. Then $\bs{\Xi}^{\bs{\mc{X}}}=(\Xi_j^{\bs{\mc{X}}})_{j=1}^{j_0}\in\mc{BD}^{\mcG}(S^{\bs{\mc{X}}})$ is a marked complex building block decomposition of $S^{\bs{\mc{X}}}$ with gluing graph $\mcG$. It remains to determine the marking of $S^{\bs{\mc{X}}}$, which amounts to choosing a marking $\Sigma^\bullet\to\Sigma^{\mcG}$. This is where we use the component $\piBD_{g,m,b}^{\mcG}$. Given an isotopy class $[\bs{\Xi}^\bullet]\in\piBD_{g,m,b}^{\mcG}$ with $\bs{\Xi}^\bullet\in\mc{BD}(S^\bullet)$, we choose a diffeomorphism $I:\Sigma^\bullet\to\Sigma^{\mcG}$ such that $(I^{-1}\circ\iota_j^{\mcG})_{j=1}^{j_0}$ is in the class $[\bs{\Xi}^\bullet]$. Such a diffeomorphism is unique up to isotopy, so we can define
\[{\rm Gl}^{\mcG}(\bs{\mc{X}},[\bs{\Xi}^\bullet]):=[(F^{\bs{\mc{X}}}\circ I,\Sigma^{\bs{\mc{X}}},J^{\bs{\mc{X}}},{\bf x}^{\bs{\mc{X}}},\bs{\zeta}^{\bs{\mc{X}}},\bs{\Xi}^{\bs{\mc{X}}})]\in\BDT_{g,m,b}^{\mcG}.\]
It is straightforward to check that ${\rm Gl}^{\mcG}$ is the inverse of $\mc{MF}^{\rm aug}|_{\BDT_{g,m,b}^{\mcG}}$. In particular, for a fixed $[\bs{\Xi}^\bullet]\in\piBD_{g,m,b}^{\mcG}$, the map ${\rm Gl}^{\mcG}(\cdot,[\bs{\Xi}^\bullet])$ is a right inverse of $\mc{MF}|_{\BDT_{g,m,b}^{\mcG}}$. To recapitulate,  we have identified $\BDT_{g,m,b}^{\mc G}$ with
$\mc T_{g,m,b}^{\mc G}\times\piBD_{g,m,b}^{\mc G}$
as follows:
\[
\begin{tikzcd}[column sep=1em,row sep=2.5em]
\BDT_{g,m,b}^{\mcG}
  \arrow[dr,swap,"\pi_{\BDT\to\mc{T}}"]
  \arrow[rr,shift left=1.5pt,harpoon,"\mc{MF}^{\rm aug}"]
&& \mc{T}_{g,m,b}^{\mcG} \times \piBD_{g,m,b}^{\mcG}
  \arrow[dl,"\pi_{\BDT\to\mc{T}}\circ{\rm Gl}^{\mcG}"]
  \arrow[ll,shift left=1.5pt,harpoon,"{\rm Gl}^{\mcG}"]
\\
& \mc{T}_{g,m,b} &
\end{tikzcd}
\]

We can define a topology on $\BDT_{g,m,b}$ such that each $\BDT_{g,m,b}^{\mcG}$ is closed and open and homeomorphic to $\mc{T}_{g,m,b}^{\mcG} \times \piBD_{g,m,b}^{\mcG}$ via the maps $\mc{MF}^{\rm aug},{\rm Gl}^{\mcG}$, where $\piBD_{g,m,b}^{\mcG}$ is equipped with the discrete topology. Then the connected components of $\BDT_{g,m,b}$ are in one-to-one correspondence with the set $\bigsqcup_{\mcG}\piBD_{g,m,b}^{\mcG}$ since $\mc{T}_{g,m,b}^{\mcG}$ is connected. For an isotopy class $[\bs{\Xi}^\bullet]\in\piBD_{g,m,b}^{\mcG}$ for some graph $\mcG$, we denote the connected component corresponding to $[\bs{\Xi}^\bullet]$ by
\begin{equation}
\BDT_{g,m,b}^{[\bs{\Xi}^\bullet]} := \{[(F,\Sigma,J,{\bf x},\bs{\zeta},\bs{\Xi})] \,|\,F^{-1}\circ\bs{\Xi}\in[\bs{\Xi}^\bullet]\} = {\rm Gl}^{\mcG}(\mc{T}_{g,m,b}^{\mcG},[\bs{\Xi}^\bullet]).
\end{equation}
For any $[\bs{\Xi}^\bullet]\in\piBD_{g,m,b}^{\mcG}$, we have
\[\BDT_{g,m,b}^{\mcG} = \bigsqcup_{[h]\in{\rm Mod}_{g,m,b}}\BDT_{g,m,b}^{[h]\cdot[\bs{\Xi}^\bullet]}.\]
For a fixed representative $(F,S)$ of $\mc X$, we have an identification
\[ \big\{[\bs\Xi]_{\rm red}\in\mc{BD}_{\rm red}(S)
\,|\,
[F^{-1}\circ\bs\Xi]=[\bs\Xi^\bullet]\big\} \xrightarrow{\sim}\pi_{\BDT\to\mc T}^{-1}(\mc X)
\cap\BDT_{g,m,b}^{[\bs\Xi^\bullet]}.
\]
Families of points in this fiber will be
described using families of representatives $\bs\Xi$.

Finally, we introduce a useful abuse of language. If a point $\mc{X}_{\bs{\Xi}}=[(F,\Sigma,J,{\bf x},\bs{\zeta},\bs{\Xi})]$ is in $\mc{D}\mc{T}_{g,m,b}^{[\bs{\Xi}^\bullet]}$ for some $[\bs{\Xi}^\bullet]\in\piBD_{g,m,b}$, we shall say that the marked complex building block decomposition $\bs{\Xi}$ of $(\Sigma,J,{\bf x},\bs{\zeta})$ is in the \textbf{isotopy class} $[\bs{\Xi}^\bullet]$. Recall that this means $F^{-1}\circ\bs{\Xi}\in[\bs{\Xi}^\bullet]$, so this condition implicitly depends on the marking $F$. 

The discussion above applies also to the decorated case. In particular, we have an injective map
\begin{equation}\label{def_MFtildehat}
\tilde{\mc{MF}}^{\rm aug}:\BDTT_{g,m,b}\to \tilde{\mc{MF}}(\BDTT_{g,m,b})\times\pi_0\tilde{\mc{BD}}_{g,m,b}, \quad [(F,\Sigma,J,{\bf x},\bs{\zeta},{\bf Y},\bs{\Xi})]\mapsto (\tilde{\mc{MF}}(\bs{\Xi}),[F^{-1}\circ\bs{\Xi}]),
\end{equation}
which, for each gluing graph $\mc{G}$, restricts to a bijection
\[\tilde{\mc{MF}}^{\rm aug}|_{\BDTT^{\mc{G}}_{g,m,b}}:\BDTT^{\mc{G}}_{g,m,b} \xrightarrow{\sim} \tT^{\mc{G}}_{g,m,b}\times \pi_0\tilde{\mc{BD}}_{g,m,b}^{\mc{G}},\]
where $\tT^{\mc{G}}_{g,m,b}:=\bigcup_{\eps>0}\tT_{m_1,b_1}^\eps\times\cdots\times\bigcup_{\eps>0}\tT_{m_{j_0},b_{j_0}}^\eps$ and $\tilde{\mc{BD}}_{g,m,b}^{\mc{G}},\pi_0\tilde{\mc{BD}}_{g,m,b}^{\mc{G}},\BDTT_{g,m,b}^{\mcG}$ are the subsets with gluing graph $\mc{G}$, defined in the obvious way. The inverse of this map is a gluing map 
\[\tilde{\rm Gl}^{\mc{G}}: \tT^{\mc{G}}_{g,m,b}\times \pi_0\tilde{\mc{BD}}_{g,m,b}^{\mc{G}} \xrightarrow{\sim} \BDTT^{\mc{G}}_{g,m,b},\] 
defined in a similar manner.\\ 

\noindent \textbf{The plumbing map.}
There is a natural way to construct finite dimensional holomorphic families of marked Riemann surfaces with boundary using the so-called plumbing construction, which we now define.
\begin{definition}[\textbf{Plumbing}]\label{def:plumbing}
Let $\mc{X}_{\bs{\Xi}}=[(F,\Sigma,J,{\bf x},\bs{\zeta},\bs{\Xi})]\in \mc{D}\mc{T}_{g,m,b}$ and let 
$\bs{\mc{X}}=(\mc{X}_1,\dots,\mc{X}_{j_0})$ be the marked complex building blocks with model forms $\mc{M}\mc{F}(\mc{X}_{\bs{\Xi}})=(\Theta_{m_1,b_1}(\mc{X}_1),\ldots,\Theta_{m_{j_0},b_{j_0}}(\mc{X}_{j_0}))\in \mc{T}^\eps_{m_1,b_1}\times \dots \times \mc{T}^\eps_{m_{j_0},b_{j_0}}$ for some $\eps>0$, so that
${\rm Gl}^{\mcG}(\bs{\mc{X}},[\bs{\Xi}^\bullet])=\mc{X}_{\bs{\Xi}}$ if $\mc{G}$ is the gluing graph of $\bs{\Xi}$ and $[\bs{\Xi}^\bullet]=[F^{-1}\circ\bs{\Xi}]\in\piBD_{g,m,b}^{\mcG}$ is the isotopy class of $\bs{\Xi}$.
For each internal edge 
${\rm e}_i=\{(j,k),(j',k')\}$ of $\mcG$, we pick an orientation $\vec{\rm e}_i=((j,k),(j',k'))$ and choose $\tau_i\in \C$ with ${\rm Im}(\tau_i)> -\log(1+\eps)$.
For each $j=1,\dots,j_0$ and $k=1,\dots,b_j$ let 
\begin{equation}\label{def_taujk}
\tau_{jk}:=\begin{cases} 
\tau_i & \textrm{ if }\vec{\rm e}_i=((j,k),(j',k')), \\
0 & \textrm{ otherwise.}
\end{cases}
\end{equation}
The plumbed marked Riemann surface $\mc{X}_{\bs{\Xi}}(\bs{\tau})\in \mc{D}\mc{T}_{g,m,b}$  
for $\bs{\tau}=(\tau_1,\dots,\tau_{i_0})$ is defined by 
\[\mc{X}_{\bs{\Xi}}(\bs{\tau}):={\rm Gl}^{\mcG}(\bs{\mc{X}}(\bs{\tau}),[\bs{\Xi}^\bullet])\]
where the collection $\bs{\mc{X}}(\bs{\tau}):=(\mc{X}_1(\bs{\tau}),\dots,\mc{X}_{j_0}(\bs{\tau}))$ of marked complex building blocks is such 
that for each $j$, if $\Theta_{m_j,b_j}(\mc{X}_j)=(\bs{\psi}_j,\bs{\theta}_j)$, then
\[\Theta_{m_j,b_j}(\mc{X}_j(\bs{\tau}))=(\psi_{j1}(e^{{\rm i}\tau_{j1}}\cdot),\dots,\psi_{jb_j}(e^{{\rm i}\tau_{jb_j}}\cdot), \theta_{j1}+{\rm Re}(\tau_{j1}),\dots,\theta_{jb_j}+{\rm Re}(\tau_{jb_j})).\]
Similarly, in the case of a decorated $\tilde{\mc{X}}_{\bs{\Xi}}=[(F,\Sigma,J,{\bf x},\bs{\zeta},{\bf Y},\bs{\Xi})]\in \mc{D}\tT_{g,m,b}$, with $\tilde{\bs{\mc{X}}}=(\tilde{\mc{X}}_1,\dots,\tilde{\mc{X}}_{j_0})$ the associated decorated marked complex building blocks having model forms $\tilde{\Theta}_{m_j,b_j}(\tilde{\mc{X}}_j)=(\bs{\psi}_j,\bs{\theta}_j,\bs{\la}_j)$, 
we define $\tilde{\mc{X}}_{\bs{\Xi}}(\bs{\tau})\in \mc{D}\tT_{g,m,b}$ through its model forms
\[\tilde{\Theta}_{m_j,b_j}(\tilde{\mc{X}}_j(\bs{\tau}))=(\psi_{j1}(e^{{\rm i}\tau_{j1}}\cdot),\dots,\psi_{jb_j}(e^{{\rm i}\tau_{jb_j}}\cdot), \theta_{j1}+{\rm Re}(\tau_{j1}),\dots,\theta_{jb_j}+{\rm Re}(\tau_{jb_j}), \la_{j1},\dots,\la_{jm_j}).\]
In particular, $\pi_{\mc{D}\tT\to \mc{DT}}(\tilde{\mc{X}}_{\bs{\Xi}}(\bs{\tau}))=\mc{X}_{\bs{\Xi}}(\bs{\tau})$.
\end{definition}

In the case $b=0$, the Teichm\"uller space $\mc{T}_{g,m}:=\mc{T}_{g,m,0}$ of surfaces without boundary is a finite dimensional complex manifold of complex dimension $3g-3+m$. If 
$\mc{X}=[(F,\Sigma,J,{\bf x})]\in \mc{T}_{g,m}$ 
and $[\bs{\Xi}^\bullet]\in \piBD_{g,m}^{\mcG}$ (omitting the subscript $b=0$), consider the unique hyperbolic metric $g^{\rm h}$ on $\Sigma$ with finite volume compatible with $J$ and 
choose a marked complex building block decomposition $\bs{\Xi}^{\rm h}=(\Xi^{\rm h}_1,\dots,\Xi^{\rm h}_{j_0})$ of $S=(\Sigma,J,{\bf x})$ 
in the isotopy class 
$[\bs{\Xi}^\bullet]$ (i.e. with $[F^{-1}\circ\bs{\Xi}^{\rm h}]=[\bs{\Xi}^\bullet]$) such that for each $(j,k)$, 
$\Xi^{\rm h}_j\circ \zeta^\bullet_{jk}:\T\to\Sigma$ is a constant speed parametrisation of a hyperbolic geodesic.
These $i_0=3g-3+m$ closed geodesics form a maximal family of cutting curves and their lengths $\ell_1,\dots,\ell_{i_0}$ are part of the Fenchel-Nielsen coordinates for $\mc{T}_{g,m}$. We call $\bs{\Xi}^{\rm h}$ a \textbf{hyperbolic} marked complex building block decomposition associated to $\mc{X}$, with isotopy class $[\bs{\Xi}^\bullet]$. Similarly, in the decorated case, we say that $\bs{\Xi}^{\rm h}\in \tilde{\mc{BD}}(\tilde{S})$ (with $\tilde{S}=(\Sigma,J,{\bf x},{\bf Y})$) is a hyperbolic decorated marked complex building block decomposition associated to $\tilde{\mc{X}}\in \tT_{g,m}$ with isotopy class $[\bs{\Xi}^\bullet]\in \pi_0\tilde{\mc{BD}}_{g,m}$ if the boundary parametrisations are constant speed parametrisations of hyperbolic geodesics. 

The following results are proved by Hinich-Vaintrob \cite{Hinich-Vaintrob} (for the hyperbolic plumbing) and \cite{Hinich} (for the plumbing with 
${\rm Im}(\tau_i)>N$ large enough):
\begin{lemma}\label{lem:plumbing}
For $\bs{\tau}=(\tau_1,\dots,\tau_{i_0})\in\C^{i_0}$ close to $0$, the plumbing map $\bs{\tau}\mapsto \mc{X}(\bs{\tau}):=\pi_{\mc{D}\mc{T}\to \mc{T}}(\mc{X}_{\bs{\Xi}^{\rm h}}(\bs{\tau}))\in \mc{T}_{g,m}$ is a local complex embedding near $0$. In other words, the plumbing coordinates $\bs{\tau}$ are local complex coordinates in a neighbourhood of $\mc{X}$ in $\mc{T}_{g,m}$. 
Moreover, for any $\mc{X}_{\bs{\Xi}}$, not necessarily hyperbolic, there is $N>0$ large enough such that the map $\bs{\tau}\mapsto  \mc{X}(\bs{\tau}):=\pi_{\mc{D}\mc{T}\to \mc{T}}(\mc{X}_{\bs{\Xi}}(\bs{\tau}))\in \mc{T}_{g,m}$ is a holomorphic embedding on the set 
\begin{equation}\label{U_N}
U_N:=\{\bs{\tau} \in \C^{i_0} \,|\, \forall i=1,\dots,i_0,\, {\rm Im}(\tau_i)>N\}.
\end{equation}
\end{lemma}

\subsection{Maximal families of cutting curves and elementary moves}

Given a Riemann surface $S=(\Sigma,J,{\bf x},\bs{\zeta})$ of type $(g,m,0,b)$ and a marked complex building block decomposition $\bs{\Xi}=(\Xi_j)_{j=1}^{j_0}\in\mc{BD}(S)$, there is a family of $i_0=3g-3+m+b$ embedded analytic parametrised curves $\xi_i:\T\to \mc{C}_i\subset \Sigma^\circ\setminus\{{\bf x}\}$ given by 
$\xi_i=\Xi_j\circ \zeta_{k}^\bullet=\Xi_{j'}\circ \zeta_{k'}^\bullet\circ o$ for each internal edge ${\rm e}_i=\{(j,k),(j',k')\}$ of the gluing graph $\mc{G}$ of $\bs{\Xi}$, where $\zeta_{k}^\bullet$ is the canonical parametrisation of \eqref{param_base_model_block}. We call these curves the cutting curves of the marked decomposition, defined up to composition by $o:e^{{\rm i}\theta}\mapsto e^{-{\rm i}\theta}$ on $\T$.
This leads us to the following definition:

\begin{definition}[\textbf{Maximal family of cutting curves}]\label{def:maximal_family_cutting_curves}
Let $S=(\Sigma,J,{\bf x},\bs{\zeta})$ be an admissible Riemann surface of type $(g,m,0,b)$. A \textbf{(parametrised) cutting curve} on $S$ is an equivalence class $[\xi]:=\{\xi,\xi\circ o\}$ where $\xi:\T\to\Sigma^\circ\setminus\{{\bf x}\}$ is a real analytic embedding.
A \textbf{maximal family of cutting curves} on $S$ is an unordered collection $\bs{\xi}=\{[\xi_1],\dots,[\xi_{i_0}]\}$ of disjoint cutting curves such that each connected component of 
$\Sigma^\circ \setminus \bigcup_{i=1}^{i_0}\xi_i(\T)$ (equipped with the marked points in ${\bf x}$ that are contained in it) is homeomorphic to the interior of a complex building block. Notice that necessarily $i_0=3g-3+m+b$. We denote by $\mc{CF}(S)$ the set of all maximal families of cutting curves on $S$. Two maximal families $\bs{\xi}^1=\{[\xi_1^1],\dots,[\xi^1_{i_0}]\}$ and $\bs{\xi}^2=\{[\xi^2_1],\dots,[\xi^2_{i_0}]\}$ of cutting curves are said \textbf{equivalent} if $\{c_1^1,\ldots,c_{i_0}^1\}=\{c_1^2,\ldots,c_{i_0}^2\}$ as sets, where for $\ell=1,2$, $i=1,\ldots,i_0$, $c_i^\ell$ is the free homotopy class of the curve $\xi_i^\ell(\T)$ on $\Sigma^\circ\setminus\{{\bf x}\}$. 
\end{definition}

As mentioned above, for each Riemann surface $S$ we have a natural map
\[\pl:\mc{BD}^{\mc{G}}(S)\to\mc{CF}(S),\quad (\Xi_j)_{j=1}^{j_0}\mapsto\{[\Xi_{j(i)}\circ\zeta_{k(i)}^\bullet]\}_{i=1}^{i_0}=\{[\Xi_{j'(i)}\circ\zeta_{k'(i)}^\bullet]\}_{i=1}^{i_0},\]
where the internal edges of the gluing graph $\mc{G}$ are ${\rm e}_i=\{(j(i),k(i)),(j'(i),k'(i))\}$ for $i=1,\ldots,i_0$. Here the enumeration of the internal edges is not important. Notice that isotopic marked complex building block decompositions give equivalent maximal families of cutting curves. Conversely, we have the following straightforward result:

\begin{lemma}\label{lem:equivalent_curves}
Two marked complex building block decompositions $\bs{\Xi}^1$ and $\bs{\Xi}^2$ of an admissible Riemann surface $S=(\Sigma,J,{\bf x},\bs{\zeta})$ have equivalent maximal families of cutting curves if and only if:\\
- there exists a diffeomorphism $h_0\in {\rm Diff}_0(\Sigma,\{{\bf x}\}\cup \pl\Sigma)$ isotopic to ${\rm Id}$, and \\
- there exists a permutation $\sigma$ of $\{1,\dots,j_0\}$ and, for each $j=1,\dots,j_0$, a diffeomorphism
$h_j\in  {\rm Diff}(\Sigma_{m_j,b_j}^\bullet,\{{\bf x}_j^\bullet\}\cup \pl\Sigma_{m_j,b_j}^\bullet)$ (with ${\bf x}_j^\bullet$ the marked points of $\Sigma_{m_j,b_j}^\bullet$) such that for $\Xi_j^1$ in $\bs{\Xi}^1$ and $\Xi^2_{\sigma(j)}$ in $\bs{\Xi}^2$, we have
\[ h_0\circ\Xi^1_j=\Xi^2_{\sigma(j)} \circ B_j\circ h_j.\]
where $B_j$ is a finite product of braiding maps on $\Sigma_{m_j,b_j}^\bullet$ defined in  \eqref{braiding}.
The analogous statement holds for decorated marked complex building block decompositions of a decorated surface $\tilde S=(\Sigma,J,{\bf x},\bs{\zeta},{\bf Y})$, where now $h_0\in {\rm Diff}_{0}(\Sigma,\{{\bf x}\}\cup\pl\Sigma;{\bf Y})$ and each $h_j\in {\rm Diff}(\Sigma_{m_j,b_j}^\bullet,\{{\bf x}_j^\bullet\}\cup \pl\Sigma_{m_j,b_j}^\bullet;{\rm Y}_j^\bullet)$ (with ${\bf Y}_j^\bullet$ the decorating vectors of $\Sigma_{m_j,b_j}^\bullet$).
\end{lemma}
\begin{proof} We only prove the undecorated case, the decorated case being similar. One implication is clear. Conversely assume that the maximal families $\{[\xi_i^1]\}_{i=1}^{i_0},\{[\xi_i^2]\}_{i=1}^{i_0}$ of cutting curves of $\bs{\Xi}^1,\bs{\Xi}^2$, respectively, are equivalent. Up to renumbering and reversing the orientations of the cutting curves, there is a diffeomorphism $h_0\in{\rm Diff}_0(\Sigma,\{{\bf x}\}\cup \pl\Sigma)$ isotopic to ${\rm Id}$ such that 
$h_0\circ \xi^1_i=\xi^2_i$ for each $i=1,\dots,i_0$. Up to renumbering the collection $\bs{\Xi}^2=(\Xi^{2}_j)_j$, we can assume that for each $j$, 
$f_j:=(\Xi_j^2)^{-1}\circ h_0\circ \Xi^1_j$ is a diffeomorphism of the canonical complex building block $\mc{P}_j^\bullet=\Sigma^\bullet_{m_j,b_j}$ sending marked points to marked points and such that $f_j\circ \zeta_k^\bullet=\zeta^\bullet_{\tau(k)}$ for $k\in \{1,\dots,b_j\}$ with $\tau$ a permutation of this set determined by the relation $h_0\circ \Xi^1_j\circ\zeta_k^\bullet=\Xi^2_j\circ\zeta_{\tau(k)}^\bullet$. The diffeomorphism $f_j$ preserves each boundary circle $\pl_k\mc{P}_j^\bullet$ if and only if $\tau={\rm Id}$. If, in addition, there are two marked points, one has either  $f_j(z^\bullet_2)=z^\bullet_2$ or $f_j(z^\bullet_2)=z^\bullet_3$. Choose a product $B_j$ of the braiding maps defined in \eqref{braiding} that induces the same permutation of boundary components and marked points as $f_j$, and set $h_j:=B_j^{-1}\circ f_j$. Then $h_j={\rm Id}$ on $\{{\bf x}_j^\bullet\}\cup\pl\mc{P}_j^{\bullet}$, and hence $h_0\circ\Xi_j^1=\Xi_j^2\circ B_j\circ h_j$, as claimed.
\end{proof}

\begin{figure}[h]
\begin{tikzpicture}
\draw[thick] (-4.95,5) to[out=-20,in=200] (0,5) ;
\draw[thick] (-5,2) to[out=20,in=160] (0,3) ;
\draw[thick] (-5,4) to[out=20,in=160] (-3.5,3.8) ;
\draw[thick] (-5,3) to[out=-10,in=210] (-3.3,3.94) ;
\draw[thick] (-5,5) to[out=180,in=180] (-5,4) ;
\draw[thick,dashed] (-5,5) to[out=0,in=0] (-5,4) ;
\draw[thick] (-5,3) to[out=180,in=180] (-5,2) ;
\draw[thick,dashed] (-5,3) to[out=0,in=0] (-5,2) ;
\draw[thick] (0,4.4) to[out=190,in=30] (-1.5,4) ;
\draw[thick] (0,3.8) to[out=190,in=-30] (-1.7,4.1) ;
\draw[thick] (0,3.8) to[out=200,in=160] (0,3)  to[out=20,in=-20] (0,3.8) ;
\draw[thick] (0,5) to[out=200,in=160] (0,4.4)  to[out=20,in=-20] (0,5) ;
\draw[thick,blue] (-3.75,3.6) to[out=-30,in=190] (-1.2,3.85) ;
\draw[thick,blue,dashed] (-3.75,3.6) to[out=0,in=170] (-1.2,3.85) ;
\draw[thick,red] (-2.4,4.5) to[out=-100,in=100] (-2,3.12) ;
\draw[thick,red,dashed] (-2.4,4.5) to[out=-70,in=70] (-2,3.12) ;
\draw[thick] (2,5) to[out=-30,in=120] (7,5)  to[out=-60,in=60] (6.7,3)  to[out=-120,in=20] (2,3.5) ;
\draw[thick,dashed] (2,5) to[out=-20,in=20] (2,3.5) ;
\draw[thick]  (2,3.5)  to[out=160,in=200] (2,5) ;
\draw[thick]  (3.8,4)  to[out=-20,in=210] (6.3,4) ;
\draw[thick]  (4,3.94)  to[out=40,in=130] (6.2,3.95) ;
\draw[thick,blue]  (5.6,5.54)  to[out=250,in=120] (5,4.4) ;
\draw[thick,blue,dashed]  (5.6,5.54)  to[out=-10,in=5] (5,4.4) ;
\draw[thick,red]  (6.5,4)  to[out=90,in=20] (5,5) to[out=200,in=130] (3,4) to[out=-50,in=150] (5.5,3) to[out=-30,in=-90]   (6.5,4);
\end{tikzpicture}
\caption{A-move (left) and S-move (right): the blue curve is replaced by the red curve.}
\label{pic:moves}
\end{figure}
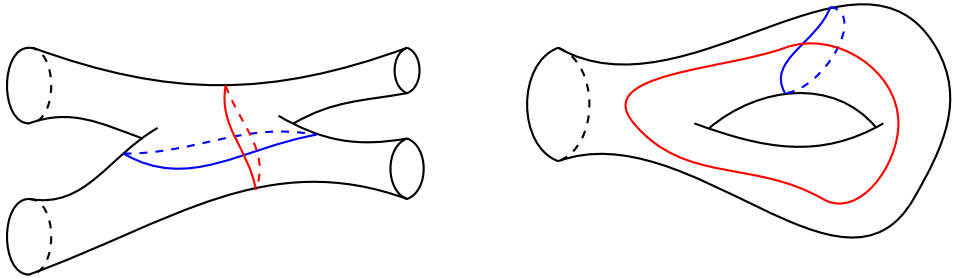

Hatcher-Lochak-Schneps \cite{Hatcher_Lochak_Schneps} proved the following result. Let $\bs{\xi}^1$ and $\bs{\xi}^2$ be two maximal families of cutting curves that are not equivalent. Then there exists a finite sequence of elementary moves between maximal families of cutting curves that transforms  $\bs{\xi}^1$ into $\bs{\xi}^2$. Each elementary move sends a maximal family of cutting curves to another non-equivalent maximal family.
Moreover, in constructing intermediate families connecting $\bs{\xi}^1$ to $\bs{\xi}^2$, only the equivalence classes of the families of cutting curves are relevant. In other words, the elementary moves act on equivalence classes of maximal families of cutting curves.
The elementary moves are of two types, denoted A or S, and are defined as follows (see Figure \ref{pic:moves}):\\

$\bullet$ \textbf{A-move on curves}. Take a complex building block decomposition $(\mc{P}^1_j)_j$ of $\Sigma$, and a cutting curve $\xi^1_i:\T\to\mc{C}^1_i$ bounding two distinct building blocks, say $\mc{P}_{j}^1$ and $\mc{P}_{j'}^1$. The gluing $\mc{P}^1_j\# \mc{P}^1_{j'}$ of these  building blocks   along $\mc{C}^1_i$ is topologically a sphere with $4$ exterior objects, each of which is either a removed disk or a marked point. Then we replace $\mc{C}^1_i$ by another non-equivalent cutting curve  $\xi^2_i:\T\to \mc{C}^{2}_i$, embedded in the interior of $\mc{P}^1_j\# \mc{P}^1_{j'}$, that intersects $\mc{C}^1_i$ transversely in exactly two points and that splits $\mc{P}^1_j\# \mc{P}^1_{j'}$ into two new building blocks $\mc{P}^2_j$ and $\mc{P}^2_{j'}$ (see Figure \ref{pic:moves}). Setting $\xi^2_k=\xi^1_k$ for $k\not=i$, the change from $\bs{\xi}^1=\{[\xi^1_1],\dots,[\xi^1_{i_0}]\}$ to $\bs{\xi}^2=\{[\xi^2_1],\dots,[\xi^2_{i_0}]\}$ is called an A-move. 
 
$\bullet$ \textbf{S-move on curves}. Take a complex building block decomposition $(\mc{P}^1_j)_j$ of $\Sigma$, and a cutting curve $\xi^1_i:\T\to\mc{C}^1_i$ corresponding to a self-gluing of some $\mc{P}^1_j$. Then the self-glued building block along $\mc{C}^1_i$, denoted $\#\mc{P}^1_j$, is topologically a torus with one exterior object, which is either a removed disk or a marked point. Then we replace $\mc{C}^{1}_i$ by another non-equivalent cutting curve  $\xi^2_i:\T\to \mc{C}^{2}_i$, embedded in the interior of $\#\mc{P}^1_j$, that intersects $\mc{C}^1_i$ transversely in exactly one point and that produces (by cutting $\#\mc{P}^1_j$ along that new curve) a new building block  $\mc{P}^{2}_j$ (see Figure \ref{pic:moves}). Setting $\xi^2_k=\xi^1_k$ for $k\not=i$, the change from $\bs{\xi}^1=\{[\xi^1_1],\dots,[\xi^1_{i_0}]\}$ to $\bs{\xi}^2=\{[\xi^2_1],\dots,[\xi^2_{i_0}]\}$ is called an S-move.\\

Notice that in either case, there are infinitely many non-homotopic choices of $\mc{C}_i^2$. However, if the complex building blocks come with markings, it is then possible to uniquely determine a choice of $\mc{C}_i^2$, called the \textbf{dual curve} of $\mc{C}_i^1$, by fixing once and for all a choice of $\mc{C}_i^2$ for the base surfaces $\Sigma^\bullet_{m,b}$ and pushing forward this choice via the markings. We formulate the precise definitions next.

\subsection{Elementary moves on the set of marked complex building block decompositions}\label{sub:elementary_moves}
Let $S=(\Sigma,J,{\bf x},\bs{\zeta})$ be a Riemann surface of type $(g,m,0,b)$. We now introduce the notion of elementary moves on the set $\mc{BD}(S)$ of marked complex building block decompositions of $S$. These moves are defined so that they can be used to realise any elementary move on curves described in the previous subsection. In particular, using the result of Hatcher-Lochak-Schneps \cite{Hatcher_Lochak_Schneps} and Lemma \ref{lem:equivalent_curves}, we will show that any two marked complex building block decompositions are related by a sequence of elementary moves.\\

$\bullet$ \textbf{A-move.} Consider a marked complex building block decomposition $\bs{\Xi}^1=(\Xi^1_j)_{j=1}^{j_0}$ of $S$ with $\Xi^1_j:\mc{P}_j^\bullet=\Sigma^\bullet_{m_j,b_j} \to \mc{P}_j\subset \Sigma$ for $j=1,\dots,j_0$.
Suppose $j\not=j'$ are such that $\{(j,k_j),(j',k_{j'})\}$ is an internal edge of the gluing graph of $\bs{\Xi}^1$, with $1\leq k_j\leq b_j$ and $1\leq k_{j'}\leq b_{j'}$, i.e.\ $\pl_{k_j}\mc{P}_j$ is glued to $\pl_{k_{j'}}\mc{P}_{j'}$ on $\Sigma$. 
We call $\mc{C}\subset\Sigma^\circ\setminus\{{\bf x}\}$ the cutting curve, parametrised by $[\Xi_j^1\circ\zeta_{k_j}^\bullet]=[\Xi_{j'}^1\circ\zeta_{k_{j'}}^\bullet]$ (recall Definition \ref{def:maximal_family_cutting_curves}), and $\Sigma_{\rm A}\subset \Sigma$ the surface obtained by gluing $\mc{P}_j$ and $\mc{P}_{j'}$ along $\mc{C}$. Relabelling the $\Xi_j$ if necessary, we may assume that $j=1,j'=2$ to simplify the notation. 
Denote $\hat{\H}:=\H\cup\{\infty\}$ with $\H:=\{{\rm Im}\,z\geq0\}$ and $\hat{\R}:=\R\cup\{\infty\}$ and $\tau(z):=\bar{z}$ the complex conjugation. 
The following terminology will be useful. Given a map $\varphi:U\subset\hat{\C}\to\hat{\C}$ defined on a set $U$ satisfying $\tau(U)=U$, we say $\varphi$ is \textbf{$\R$-symmetric} if $\varphi(\bar{z})=\overline{\varphi(z)}$ for all $z\in U$, $\varphi(U\cap\H)\subset\H$, and $\varphi|_{U\cap\hat{\R}}:U\cap\hat{\R}\to\hat{\R}$ preserves the orientation on $\hat{\R}$ (i.e.\ if $a,b,c\in U\cap\hat{\R}$ are positively cyclically ordered, then so are $\varphi(a),\varphi(b),\varphi(c)\in\hat{\R}$).

We now glue the canonical complex building blocks $\mc{P}_1^\bullet ,\mc{P}_{2}^\bullet$ along $\pl_{k_1}\mc{P}_1^\bullet\sim\pl_{k_2}\mc{P}_{2}^\bullet$ and realise the resulting surface as a subset of $\hat{\C}$, taking into account the complex involution $\tau(z)=\bar{z}$ as follows. Let $b_{\rm A}=b_1+b_2-2$ and $m_{\rm A}=m_1+m_2$. Consider $\Sigma_{\rm A}^\bullet= \hat{\C}\setminus \cup_{k=1}^{b_{\rm A}}(D_k^\bullet)^\circ$ where $D_1^\bullet,\ldots,D_{b_{\rm A}}^\bullet\subset\C$ are non-overlapping round disks which are symmetric under $\tau$ and whose intersections with $\R$ are the intervals $I_1,\dots, I_{b_{\rm A}}$. Let ${\bf x}_{\rm A}^\bullet$ be a collection of $m_{\rm A}$ marked points contained in $\R\setminus \cup_{k=1}^{b_{\rm A}}I_k$. If $M_1,\ldots,M_{b_{\rm A}}$ are $\R$-symmetric M\"obius transformations with $M_k(\D)=D_k^\bullet$ (e.g.\ $M_k(z)=z_k+r_kz$ for $D_k^\bullet=D(z_k,r_k)$ with $r_k>0$), then $S_{\rm A}^\bullet=(\Sigma_{\rm A}^\bullet,J_\C,{\bf x}_{\rm A}^\bullet,(M_1|_\T,\ldots,M_{b_{\rm A}}|_\T))$ is a Riemann surface of type $(0,m_{\rm A},b_{\rm A})$. We denote by $\mc{O}_1^\bullet,\dots,\mc{O}_4^\bullet$ the collection of disks $D^\bullet_k$ and marked points in ${\bf x}_{\rm A}^\bullet$, positively ordered along $\hat\R$, i.e.\ $y<y'$ if $y\in \mc{O}^\bullet_k\cap\R$ and $y'\in \mc{O}_{k'}^\bullet\cap\R$ with $k<k'$.

We fix an $\R$-symmetric real analytic embedding $\xi^{\rm s}:\T\to(\Sigma_{\rm A}^\bullet)^\circ\setminus\{{\bf x}_{\rm A}^\bullet\}$, called the \textbf{s-channel cut}, such that the bounded domain $\mc{D}^{\rm s}$ bounded by $\xi^{\rm s}(\T)$ contains $\mc{O}^\bullet_1\cup \mc{O}^\bullet_2$ in its interior, while $\hat{\C}\setminus \mc{D}^{\rm s}$ contains $\mc{O}^\bullet_3\cup \mc{O}^\bullet_4$. Similarly, we fix another $\R$-symmetric real analytic embedding $\xi^{\rm t}:\T\to(\Sigma_{\rm A}^\bullet)^\circ\setminus\{{\bf x}_{\rm A}^\bullet\}$, called the \textbf{t-channel cut}, intersecting $\xi^{\rm s}(\T)$ transversely at exactly two points and such that the bounded domain $\mc{D}^{\rm t}$ bounded by $\xi^{\rm t}(\T)$ contains $\mc{O}^\bullet_2\cup \mc{O}^\bullet_3$ in its interior and $\hat{\C}\setminus \mc{D}^{\rm t}$ contains $\mc{O}^\bullet_1\cup \mc{O}^\bullet_4$. 
Let $\mc{P}_1^{s}=\Sigma_{\rm A}^\bullet \setminus (\mc{D}^{\rm s})^\circ$ and $\mc{P}_2^{s}=\Sigma_{\rm A}^\bullet \cap \mc{D}^{\rm s}$ be the two complex building blocks associated to the s-channel cut, and similarly $\mc{P}_1^{t}=\Sigma_{\rm A}^\bullet \setminus (\mc{D}^{\rm t})^\circ$ and $\mc{P}_2^{t}=\Sigma_{\rm A}^\bullet \cap \mc{D}^{\rm t}$ for the t-channel cut, each equipped with the marked points in ${\bf x}_{\rm A}^\bullet$ contained in it. The type of the complex building block $\mc{P}_\ell^{\rm t}$ is denoted  $(m_\ell^*,b_\ell^*)$ for $\ell=1,2$.

\begin{remark}
We need to explain a subtlety concerning the ordering of the boundary parametrisations of these complex building blocks. Let us call a \emph{marking diffeomorphism} a diffeomorphism between admissible surfaces that maps marked points to marked points and boundary circles to boundary circles (compatible with the respective parametrisations). For a complex building block $\mc{P}$ embedded in $\hat{\C}$ of type $(m,b)$, it is \emph{not} always possible to choose an $\R$-symmetric marking diffeomorphism $\Sigma_{m,b}^\bullet\to\mc{P}$, depending on the ordering of the boundary parametrisations of $\mc{P}$. Indeed, let $\mc{O}_1,\mc{O}_2,\mc{O}_3$ denote the collection of boundary circles and marked points of $\mc{P}$, positively ordered along $\hat\R$, and suppose $\zeta_k(\T)=\mc{O}_{i_k}$ for $k=1,\ldots,b$ where $\bs{\zeta}=(\zeta_1,\ldots,\zeta_b)$ is the ordered collection of boundary parametrisations of $\mc{P}$. Since boundary circles are ordered before marked points in the canonical complex building block $\Sigma_{m,b}^\bullet$, the same must be true for $\mc{O}_1,\mc{O}_2,\mc{O}_3$ for an $\R$-symmetric marking diffeomorphism $\Sigma_{m,b}^\bullet\to\mc{P}$ to exist. Thus only the following cases are allowed:
\begin{align*}
&b=3:\qquad(i_1,i_2,i_3)\in\{(1,2,3),(2,3,1),(3,1,2)\};\\
&b=2:\qquad(i_1,i_2)\in\{(1,2),(2,3),(3,1)\};\\
&b=1:\qquad i_1\in\{1,2,3\}\text{ (no constraint)}.
\end{align*}
In particular, in the case $b=2$, prescribing $i_k=1$ for some $k$ (i.e.\ $\mc{O}_1=\zeta_k(\T)$) determines which of $\mc{O}_2,\mc{O}_3$ is a marked point and which is a boundary circle.
\end{remark}

We choose the ordering of the marked points and disks in $\mc{O}_1^\bullet,\dots,\mc{O}_4^\bullet$ such that for $\ell=1,2$, there exists an $\R$-symmetric marking diffeomorphism $\mc{P}_\ell^{\rm s}\to\Sigma_{m_\ell,b_\ell}^\bullet$ mapping the cutting curve $\xi^{\rm s}(\T)=\pl \mc{D}^{\rm s}$ to $\pl_{k_\ell}\Sigma_{m_\ell,b_\ell}^\bullet$. In the case $b_\ell=2$, this uniquely determines which of $\mc{O}_1,\mc{O}_2$ if $\ell=2$ (or $\mc{O}_3,\mc{O}_4$ if $\ell=1$) is the marked point, by the remark. On the other hand, for $\ell=1,2$, we choose the smallest $1\leq k_\ell^{*}\leq b_\ell^*$ such that there exists an $\R$-symmetric marking diffeomorphism $\mc{P}_\ell^{\rm t}\to\Sigma_{m_\ell^*,b_\ell^*}^\bullet$ mapping the cutting curve $\xi^{\rm t}(\T)=\pl \mc{D}^{\rm t}$ to $\pl_{k_\ell^*}\Sigma_{m_\ell^*,b_\ell^*}^\bullet$. Again by the remark, it is not always possible to choose $k_\ell^*=1$. These choices ensure that the marking diffeomorphisms in the next paragraph exist.

Choose two $\R$-symmetric marking diffeomorphisms
\[\Omega^{\rm s}_1 : \Sigma_{m_1,b_1}^\bullet \to \mc{P}^{\rm s}_1, \quad \Omega^{\rm s}_2 : \Sigma_{m_2,b_2}^\bullet \to \mc{P}^{\rm s}_2,\]
with $\Omega^{\rm s}_\ell$ biholomorphic near $\pl \Sigma_{m_\ell,b_\ell}^\bullet$ and near $\{{\bf z}^\bullet\}\cap \Sigma_{m_\ell,b_\ell}^\bullet$ for $\ell=1,2$, and satisfying 
\[ \Omega^{\rm s}_1\circ \zeta_{k_1}^\bullet (e^{{\rm i}\theta})=\xi^{\rm s}(e^{{\rm i}\theta}), \quad \Omega^{\rm s}_2\circ \zeta_{k_2}^\bullet (e^{{\rm i}\theta})=\xi^{\rm s}(e^{{\rm i}(\pi-\theta)}).\]
Similarly, choose two $\R$-symmetric marking diffeomorphisms
\[\Omega^{\rm t}_1 : \Sigma_{m_1^*,b_1^*}^\bullet \to \mc{P}^{\rm t}_1, \quad \Omega^{\rm t}_2 : \Sigma_{m_2^*,b_2^*}^\bullet \to \mc{P}^{\rm t}_2,\]
with $\Omega^{\rm t}_\ell$ biholomorphic near $\pl \Sigma_{m_\ell^*,b_\ell^*}^\bullet$ and near $\{{\bf z}^\bullet\}\cap \Sigma_{m_\ell^*,b_\ell^*}^\bullet$ for $\ell=1,2$, and satisfying 
\[ \Omega^{\rm t}_1\circ \zeta_{k_1^*}^\bullet (e^{{\rm i}\theta})=\xi^{\rm t}(e^{{\rm i}\theta}), \quad \Omega^{\rm t}_2\circ \zeta_{k_2^*}^\bullet (e^{{\rm i}\theta})=\xi^{\rm t}(e^{{\rm i}(\pi-\theta)}).\]
For practical convenience involving the fusion kernel in Section \ref{sec:transition_operator}, we further assume that 
\begin{equation}\label{particular_choice} 
\begin{gathered}
\xi^{\rm s}(e^{{\rm i}\theta})=q_{11}^{\rm s}e^{{\rm i}\theta}, 
\quad \xi^{\rm t}(e^{{\rm i}\theta})=z^{\rm t}_{11}+q_{11}^{\rm t}e^{{\rm i}\theta},\\
\Omega_j^{\rm s}\circ \zeta_{k}^\bullet(e^{{\rm i}\theta})=M_{jk}^{\rm s}(e^{{\rm i}\theta}), \textrm{ for }k\in \{1,\dots,b_j\}\setminus \{k_j\}, \quad \Omega_j^{\rm s}(z_{k}^\bullet)=
z^{\rm s}_{jk} , \textrm{ for }k=b_j+1,\dots,3,\\
\Omega_j^{\rm t}\circ \zeta_{k}^\bullet(e^{{\rm i}\theta})=M_{jk}^{\rm t}(e^{{\rm i}\theta}), \textrm{ for }k\in \{1,\dots,b_j^*\}\setminus \{k_j^*\}, \quad \Omega_j^{\rm t}(z_{k}^\bullet)=z^{\rm t}_{jk} , \textrm{ for }k=b_j^*+1,\dots,3,
\end{gathered}
\end{equation}
for some $q_{11}^{\rm s},q_{11}^{\rm t}>0$, $z^{\rm s}_{jk},z^{\rm t}_{jk}\in \R$, and some $\R$-symmetric M\"obius transformations $M_{jk}^{\rm s},M_{jk}^{\rm t}$. 
Here we assume also that 
\begin{gather*}
\{M_{jk}^{\rm s}\,|\,j=1,2,1\leq k\leq b_j, k\not=k_j\}=\{M_{jk}^{\rm t}\,|\,j=1,2,1\leq k\leq b_j^*, k\not=k_j^*\}=\{M_1,\ldots,M_{b_{\rm A}}\},\\
\{z_{jk}^{\rm s}\,|\,j=1,2,k=b_j+1,\ldots,3\}=\{z_{jk}^{\rm t}\,|\,j=1,2,k=b_j^*+1,\ldots,3\}=\{{\bf x}_{\rm A}^\bullet\}.
\end{gather*}
Notice that since $T_{k_2,\pi}\circ \zeta_{k_2}^\bullet(e^{{\rm i}\theta})=\zeta_{k_2}^\bullet(e^{{\rm i}(\pi+\theta)})$ (recall \eqref{Tkalpha}), we get $\Omega^{\rm s}_2\circ T_{k_2,\pi}\circ \zeta_{k_2}^\bullet=\xi^{\rm s}\circ o$ and similarly $\Omega^{\rm t}_2\circ T_{k_2^*,\pi}\circ \zeta_{k_2^*}^\bullet=\xi^{\rm t}\circ o$, 
so $\bs\Xi^{{\rm s},\bullet}_{\rm A}=(\Omega^{\rm s}_1,\Omega^{\rm s}_2\circ T_{k_2,\pi})$ and $\bs\Xi^{{\rm t},\bullet}_{\rm A}=(\Omega^{\rm t}_1,\Omega^{\rm t}_2\circ T_{k_2^*,\pi})$ are marked complex building block decompositions of $S_{\rm A}^\bullet$ with internal edges $\{(1,k_1),(2,k_2)\}$ and $\{(1,k_1^*),(2,k_2^*)\}$, respectively. 

The marked complex building block decomposition $\bs{\Xi}_{\rm A}^1=(\Xi^1_1,\Xi^1_2)$ of $\Sigma_{\rm A}\subset\Sigma$ (with the appropriate marked points and boundary parametrisations) induces a smooth orientation preserving diffeomorphism
\begin{equation}\label{markingOmegaA} 
\Omega_{\rm A}: \Sigma_{\rm A}^\bullet\to \Sigma_{\rm A}
\end{equation}
such that $\Omega_{\rm A}|_{\mc{P}_1^{\rm s}}=\Xi_1^1\circ (\Omega^{\rm s}_1)^{-1}$ and $\Omega_{\rm A}|_{\mc{P}_2^{\rm s}}=\Xi_2^1\circ T_{k_2,-\pi}\circ (\Omega^{\rm s}_2)^{-1}$.
The cutting curve is $\mc{C}=\Omega_{\rm A}(\xi^{\rm s}(\T))$. 
Next, we pull back the complex structure $J$ on $\Sigma$ to $\Sigma_{\rm A}^\bullet$ by $\Omega_{\rm A}$ and choose a curve $\hat{\xi}^{\rm t} :\T \to(\Sigma_{\rm A}^\bullet)^\circ\setminus\{{\bf x}_{\rm A}^\bullet\}$ 
which is a real analytic embedding with respect to $\Omega_{\rm A}^*J$ and is isotopic to $\xi^{\rm t}$
in $(\Sigma_{\rm A}^\bullet)^\circ\setminus\{{\bf x}_{\rm A}^\bullet\}$. We extend this isotopy to obtain an orientation preserving diffeomorphism $H_{J}:\Sigma_{\rm A}^\bullet\to \Sigma_{\rm A}^\bullet$ isotopic to the identity relative to $\cup_{k=1}^4\pl\mc{O}_k^\bullet$ such that $H_{J}\circ \xi^{\rm t}=\hat{\xi}^{\rm t}$, and $H_J$ is chosen to be biholomorphic from $J_\C$ to $\Omega_{\rm A}^*J$ near $\xi^{\rm t}(\T)$ (this is possible since $\xi^{\rm t}$ and ${\hat{\xi}}^{\rm t}$ are real analytic for their respective complex structures).
Cutting $\Sigma_{\rm A}^\bullet$ along $\hat{\xi}^{\rm t}$ produces two complex building blocks 
$\hat{\mc{P}}_{1}^{\rm t}$ and $\hat{\mc{P}}_2^{\rm t}$ as before, and we define the markings 
\begin{align*}
& F_{\rm A}: \Sigma^\bullet_{m_1^*,b_1^*}\to \hat{\mc{P}}_{1}^{\rm t}, \quad 
F_{\rm A}=H_J\circ \Omega^{\rm t}_1, \\
& G_{\rm A}:\Sigma^\bullet_{m_2^*,b_2^*}\to \hat{\mc{P}}_2^{\rm t}, \quad G_{\rm A}=H_J\circ \Omega^{\rm t}_2\circ T_{k_2^*,\pi}.
\end{align*}
This allows us to define a new marked complex building block decomposition $\bs{\Xi}^2=(\Xi^2_k)_{k=1}^{j_0}$ of $S$ by  
\begin{equation}\label{def_Xi^2_1_moveA} 
\Xi^2_1:=\Omega_{\rm A}\circ F_{\rm A}, \quad \Xi^2_2:=\Omega_{\rm A}\circ G_{\rm A}, \quad \Xi^2_k:=\Xi^1_k \textrm{ for }k>2,
\end{equation}
with the new complex building blocks $\mc{P}_1^*:=\Omega_{\rm A}(\hat{\mc{P}}_{1}^{\rm t})$ and
$\mc{P}_2^*:=\Omega_{\rm A}(\hat{\mc{P}}_{2}^{\rm t})$.
We will call the new cutting curve $\mc{C}^*:=\Omega_{{\rm A}}(\hat{\xi}^{\rm t}(\T))$ on $\Sigma_{{\rm A}}\subset\Sigma$ (parametrised by $[\Omega_{\rm A}\circ\hat{\xi}^{\rm t}]$) the curve \textbf{dual} to $\mc{C}$. The change $\bs{\Xi}^1\to \bs{\Xi}^2$ is called the \textbf{A-move of the first type} associated to $\bs{\Xi}^1$ with respect to the internal edge $\{(j,k_1),(j',k_2)\}$ (with $j=1$, $j'=2$) of the gluing graph of $\bs{\Xi}^1$. It replaces this edge by $\{(j,k_1^*),(j',k_2^*)\}$, which is uniquely determined by $k_1,k_2$ and the types $(m_j,b_j),(m_{j'},b_{j'})$ of $\Xi_j,\Xi_{j'}$. 
This induces an A-move on maximal families of cutting curves, with $\mc{C}$ replaced by $\mc{C}^*$. Notice that the A-move associated to a marked complex building block decomposition depends on the choices of $\Sigma_{\rm A}^\bullet$, $F_{\rm A}$, and $G_{\rm A}$, but the isotopy class of $\bs{\Xi}^2$ (in the sense of Definition \ref{def_pi_0_BD}) is independent of these choices. 

Similarly, one can do this for a fixed real analytic simple closed curve $\xi^{\rm t}$ intersecting $\xi^{\rm s}(\T)$ transversely at exactly two points and separating $\mc{O}_1^\bullet\cup\mc{O}_3^\bullet$ from $\mc{O}_2^\bullet\cup\mc{O}_4^\bullet$, i.e.\ they are contained in different connected components of $\hat{\C}\setminus\xi^{\rm t}(\T)$.
The resulting change of marked complex building block decompositions is called an \textbf{A-move of the second type}. Notice that $\xi^{\rm t}$ (and thus $\Omega_1^{\rm t},\Omega_2^{\rm t}$) cannot be $\R$-symmetric in this case. In particular, contrary to the previous case, there is no natural choice of $k_1^*,k_2^*$, and it suffices to fix one once and for all. \\ 

\begin{figure}[h]
\begin{tikzpicture}[scale=2.5]
\fill[gray!30] (-2.6,-1.25) rectangle (1.65,1.25);

\draw[fill=white,draw=black,very thick] (0.25,0) circle (0.3);
\node[anchor=south] at (0.25,0.3) {$\mathcal{O}_{3}$};

\draw[fill=white,draw=black,very thick] (1,0) circle (0.3);
\node[anchor=south] at (1,0.3) {$\mathcal{O}_{4}$};

\draw[fill=white,draw=black,very thick] (-0.45,0) circle (0.2);
\node[anchor=south] at (-0.45,0.2) {$\mc{O}_{2}$};

\draw[fill=white,draw=black,very thick] (-1.05,0) circle (0.18);
\node[anchor=south] at (-1.05,0.18) {$\mc{O}_{1}$};

\draw[draw=blue,very thick] (-1.2,0) circle (1);
\node[anchor=south,text=blue] at (-2,0.9) {$\xi^{\rm s}(\T)=\Omega_{\rm A}^{-1}(\mathcal{C})$};

\draw[draw=red,very thick] (-0.06,0) circle (0.68);
\node[anchor=south,text=red] at (0.45,-0.95) {$\hat{\xi}^{\rm t}(\T)=\Omega_{\rm A}^{-1}(\mathcal{C}^*)$};

\node[anchor=south] at (0.5,0.6) {$\mathcal{P}_{1}^{\rm s}$};
\node[anchor=south] at (-1.6,-0.5) {$\mathcal{P}_{2}^{\rm s}$};

\draw[->,gray!65] (-2.5,0) -- (1.6,0);
\node[anchor=north] at (1.45,0) {$\R$};
\end{tikzpicture}
\caption{A-move of the first type on $\Sigma_{\rm A}^\bullet=\mc{P}_1^{\rm s}\cup\mc{P}_{2}^{\rm s}$ with $b_1=b_{2}=3$}
\label{fig:A-move}
\end{figure}
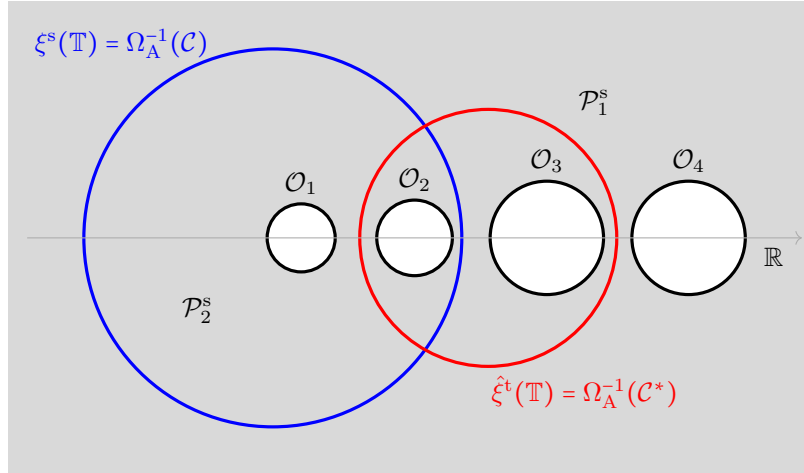

\noindent $\bullet$ \textbf{S-move.}
Consider a marked complex building block decomposition $\bs{\Xi}^1=(\Xi^1_j)_{j=1}^{j_0}$ of
$S=(\Sigma,J,{\bf x},\bs{\zeta})$ with $\Xi^1_j:\mc{P}_j^\bullet=\Sigma^\bullet_{m_j,b_j} 
\to \mc{P}_j\subset\Sigma$ for $j=1,\dots,j_0$. Suppose $j$ is such that $\{(j,1),(j,2)\}$ is an internal edge
of the gluing graph of $\bs{\Xi}^1$, i.e.\ $\pl_1\mc{P}_j$ is self-glued to $\pl_2\mc{P}_j$ on $\Sigma$.
We call $\mc C\subset\Sigma^\circ\setminus\{{\bf x}\}$ the corresponding cutting curve, parametrised by $[\Xi_j^1\circ\zeta_1^\bullet] =[\Xi_j^1\circ\zeta_2^\bullet]$.
We denote by $\#\mc{P}_j^\bullet$ the torus with a marked point $z_3^\bullet$ if $b_j=2$, or with a disk removed if $b_j=3$, obtained
by gluing $\pl_1\mc{P}_j^\bullet$ to $\pl_2\mc{P}_j^\bullet$ through the identification $\zeta_1^\bullet(z)\sim\zeta_2^\bullet(o(z))$ for $z\in\T$.
The map $\Xi_j^1$ descends to a diffeomorphism $\Xi_j^\#:\#\mc{P}_j^\bullet\to\mc{P}_j\subset\Sigma
$ onto the image $\mc{P}_j=\Xi_j^1(\mc{P}_j^\bullet)$. We equip $\#\mc{P}_j^\bullet$ with the pulled-back complex structure $J_j^\#:=(\Xi_j^\#)^*J$.
In particular, $\Xi_j^\#$ is biholomorphic for this complex structure. We fix once and for all a relative isotopy class of simple closed curves in $(\#\mc{P}_j^\bullet)^\circ$, disjoint from the marked point or boundary, and intersecting the glued curve $\zeta_1^\bullet(\T)\sim\zeta_2^\bullet(\T)$ transversely at exactly one point. This will be the relative isotopy class represented by the vertical cutting curve in the rectangular model of Section~\ref{sec:local_studyS}. We choose a $J_j^\#$-real-analytic parametrised representative $\xi:\T\to \xi(\T)\subset\#\mc{P}_j^\bullet$ in this class. For compatibility with the lifts used in that model, we choose $\xi$ so that the intersection point is
\[\xi(-1)=[\zeta_1^\bullet(-1)]=[\zeta_2^\bullet(-1)],\]
where $[x]$ denotes the projection of $x\in\mc{P}_j^\bullet$ on $\#\mc{P}_j^\bullet$.
Cutting $\#\mc{P}_j^\bullet$ along $\xi(\T)$ yields a new complex building block $\mc{P}_j^*$ of the same type $(0,m_j,b_j)$ as
$\mc{P}_j^\bullet$. Let $\pi:\mc{P}_j^*\to\#\mc{P}_j^\bullet$ be the canonical projection. If $b_j=3$, we equip $\mc{P}_j^*$ with
the incoming boundary parametrisations $\bs{\xi}=(\xi_1,\xi_2,\xi_3)$ satisfying $\pi\circ\xi_1=\xi$, 
$\pi\circ\xi_2=\xi\circ o$ and $\pi\circ\xi_3=\zeta_3^\bullet$. If $b_j=2$, we equip $\mc{P}_j^*$ with the incoming boundary parametrisations $\bs{\xi}=(\xi_1,\xi_2)$ defined in the same way and with the marked point $\pi^{-1}(z_3^\bullet)$.

We further fix a relative isotopy class of markings $F^{\rm S}:\Sigma_{m_j,b_j}^\bullet\to\mc{P}_j^*$
and choose a representative which is analytic near the marked point and the boundary and satisfies
$F^{\rm S}\circ\zeta_k^\bullet=\xi_k$, for $k=1,\ldots,b_j$, as well as  $F^{\rm S}(z_3^\bullet)=\pi^{-1}(z_3^\bullet)$
if $b_j=2$. We then define a new marked complex building block
decomposition $\bs{\Xi}^2=(\Xi_{j'}^2)_{j'=1}^{j_0}$ of $S$ by
\[
\Xi_j^2:=\Xi_j^\#\circ\pi\circ F^{\rm S},
\quad
\Xi_{j'}^2:=\Xi_{j'}^1\text{ for }j'\neq j.
\]
We call $\mc C^*:=\Xi_j^\#(\xi(\T))\subset\mc{P}_j\subset\Sigma$ (parametrised by $[\Xi_j^\#\circ\xi]$) the curve \textbf{dual} to
$\mc C$. The change $\bs{\Xi}^1\to\bs{\Xi}^2$
is called the \textbf{S-move} associated to $\bs{\Xi}^1$. It induces an S-move on maximal families of cutting curves, in which $\mc C$ is replaced by $\mc C^*$. Notice that the S-move associated to a marked complex building block decomposition is uniquely defined once the representatives $\xi$ and $F^{\rm S}$ are fixed.

\begin{remark}
For the A- and S-moves, the dual curves $\hat{\xi}^{\rm t},\xi$ and the markings $F^{\rm A},G^{\rm A},F^{\rm S}$ will be chosen in a way best suited for computations with the conformal blocks. These choices affect the value of the fusion kernels for the conformal blocks (Sections \ref{sec:local_study} and \ref{sec:local_studyS}), but not the final projective representation of the mapping class group (Theorem \ref{th:kernelMS}).
\end{remark}

$\bullet$ \textbf{B-move.}   Consider a marked complex building block decomposition $\bs{\Xi}^1=(\Xi^1_j)_{j=1}^{j_0}$ of $S=(\Sigma,J,{\bf x},\bs{\zeta})$ with $\Xi^1_j:\mc{P}_j^\bullet=\Sigma^\bullet_{m_j,b_j} \to \mc{P}_j\subset \Sigma$ for $j=1,\dots,j_0$. Take $j\in\{1,\ldots,j_0\}$ and $k,k'\in \{1,2,3\}$ such that either $k<k'\leq b_j$ or $b_j<k<k'$. Define $\Xi_j^2:=\Xi^1_j\circ B_{kk'}$ where $B_{kk'}$ is the braiding map defined in \eqref{braiding}, and let $\Xi_{j'}^2:=\Xi^1_{j'}$ for $j'\neq j$. Then $\bs{\Xi}^2=(\Xi_j^2)_{j=1}^{j_0}$ is another marked complex building block decomposition of $S$. A change of the form $\bs{\Xi}^1\to \bs{\Xi}^2$ is called a \textbf{B-move}. Notice that it does not change the maximal family of cutting curves.\\

$\bullet$ \textbf{T-move.}  Consider a marked complex building block decomposition $\bs{\Xi}^1=(\Xi^1_j)_{j=1}^{j_0}$ of $S=(\Sigma,J,{\bf x},\bs{\zeta})$ with $\Xi^1_j:\mc{P}_j^\bullet=\Sigma^\bullet_{m_j,b_j} \to \mc{P}_j\subset \Sigma$ for $j=1,\dots,j_0$. 
Take $j\in\{1,\ldots,j_0\}$, $k\in \{1,2,3\}$, and $n\in\Z$. Define $\Xi_j^2:=\Xi^1_j\circ T_{k}^n$, where $T_k=T_{k,2\pi}$ is the Dehn twist defined in \eqref{Tkalpha}, around $\pl_k\mc{P}_j^\bullet$ if $k\leq b_j$ or around the marked point $z_k^\bullet$ if $k>b_j$. Let $\Xi_{j'}^2:=\Xi^1_{j'}$ for $j'\neq j$. Then $\bs{\Xi}^2=(\Xi_j^2)_{j=1}^{j_0}$ is another marked complex building block decomposition of $S$. A change of the form $\bs{\Xi}^1\to \bs{\Xi}^2$ is called a \textbf{T-move}.
Notice that it does not change the maximal family of cutting curves, nor the gluing graph.\\

So far, we have defined elementary moves on the set $\mc{BD}(S)$ of marked complex building block decompositions of an admissible Riemann surface $S=(\Sigma,J,{\bf x},\bs{\zeta})$. Taking $S$ to be the fixed base Riemann surface $S^\bullet=(\Sigma^{\bullet},J^\bullet,{\bf x}^\bullet,\bs{\zeta}^\bullet)$ of type $(g,m,b)$ and descending to isotopy classes, they give corresponding elementary moves on the discrete set $\piBD_{g,m,b}$. Later in Section \ref{sec:projective_representation}, we shall consider elementary moves on $\BDT_{g,m,b}$, i.e.\ if $[(F,S,\bs{\Xi}^\ell)]\in\BDT_{g,m,b}$ with $F^{-1}\circ\bs{\Xi}^\ell\in[\bs{\Xi}_\ell^\bullet]\in\piBD_{g,m,b}$ for $\ell=1,2$, then the move $\bs{\Xi}^1\to\bs{\Xi}^2$ on $\mc{BD}(S)$ and the move $[\bs{\Xi}_1^\bullet]\to[\bs{\Xi}_2^\bullet]$ on $\piBD_{g,m,b}$ are moves of the same type.

\begin{remark}[Undecorated and decorated moves]\label{rem:moves-forgetful}
For a decorated surface $\tilde{S}=(S,{\bf Y})$, forgetting the decorating vectors ${\bf Y}$ induces maps
\begin{equation}\label{forgetfull_proj}
\tilde{\mc{BD}}(\tilde{S})\to\mc{BD}(S),\quad\pi_0\tilde{\mc{BD}}_{g,m,b}\to \piBD_{g,m,b}.
\end{equation}
They are equivariant with respect to the projection in the short exact sequence (recall Section \ref{sub:Teichmuller})
\[
0\to\Z^m\to\tilde{{\rm Mod}}_{g,m,b}
\to {\rm Mod}_{g,m,b}\to0.
\]
We say that a move on $\tilde{\mc{BD}}(\tilde{S})$ is an A/S/B/T-move if its projection to $\mc{BD}(S)$ is an A/S/B/T-move. In the case of the A- and S-moves, we require that the markings $\Omega_{\rm A},F^{\rm A},G^{\rm A},F^{\rm S}$ map the standard vector $\pl_z$ to the corresponding decorated vectors at the marked points on $\Sigma$. Notice that the map $\pi_0\tilde{\mc{BD}}_{g,m,b}\to \piBD_{g,m,b}$ is a surjective projection, and two points in $\pi_0\tilde{\mc{BD}}_{g,m,b}$ lie in the same fiber if and only if they differ only by T-moves around marked points. Indeed, a T-move is always non-trivial on $\pi_0\tilde{\mc{BD}}_{g,m,b}$, and it is trivial on $\piBD_{g,m,b}$ if and only if it is around a marked point. Such T-moves correspond to the kernel $\Z^m$ above.
On the other hand, a B-move exchanging two marked points generally changes both the undecorated and decorated classes. 
\end{remark}

The main result of this section is the fact that these elementary moves generate all possible changes of marked complex building block decompositions:
\begin{proposition}[\textbf{Changes of marked complex building block decompositions}]\label{prop_elementary_moves}
For any two marked complex building block decompositions  $\bs{\Xi}^1,\bs{\Xi}^2$ of $S=(\Sigma,J,{\bf x},\bs{\zeta})$, there is a finite sequence of elementary moves of type A, S, B, T as defined above that brings $\bs{\Xi}^1$ to $h_0\circ \bs{\Xi}^2$ for some diffeomorphism $h_0\in {\rm Diff}_{0}(\Sigma,\{{\bf x}\} \cup \pl\Sigma)$ isotopic to ${\rm Id}$, up to relabelling\footnote{Here relabelling means only reindexing the collection $(\Xi_j)_j$ of complex building block markings. We do not allow permutation of the marked points or boundary components.} of the markings in $\bs{\Xi}^2$. The analogous statement holds for two decorated marked complex building block decompositions of a decorated surface $\tilde S=(\Sigma,J,{\bf x},\bs{\zeta},{\bf Y})$, where $h_0$ now belongs to ${\rm Diff}_{0}(\Sigma,\{{\bf x}\}\cup\pl\Sigma;{\bf Y})$. In other words, all elements of $\piBD(S)$ and $\pi_0\tilde{\mc{BD}}(\tilde{S})$ are related by a finite sequence of elementary moves.
\end{proposition}
\begin{proof}
First consider the undecorated case and assume that $\bs{\Xi}^1$ and $\bs{\Xi}^2$ have equivalent maximal families of cutting curves. By Lemma \ref{lem:equivalent_curves}, after applying a global isotopy to $\bs{\Xi}^1$, relabelling the complex building blocks, and applying B-moves, the two decompositions differ by $\Xi_j^2=\Xi_j^1\circ h_j$ for some $h_j\in{\rm Diff}(\Sigma_{m_j,b_j}^\bullet,\{{\bf x}^\bullet\}\cup\pl\Sigma_{m_j,b_j}^\bullet)$ for each $j$. By \eqref{MCG_of_pant}, the undecorated mapping class group of a model complex building block ${\rm Mod}(\Sigma_{m_j,b_j}^\bullet,\{{\bf x}^\bullet\}\cup\pl\Sigma_{m_j,b_j}^\bullet)$
 is generated by the Dehn twists $T_k$ around its $b_j$ boundary components. We can thus assume $h_j\in{\rm Diff}_0(\Sigma_{m_j,b_j}^\bullet,\{{\bf x}^\bullet\}\cup\pl\Sigma_{m_j,b_j}^\bullet)$ is isotopic to ${\rm Id}$, up to applying T-moves on $\bs{\Xi}^1$.
Then the diffeomorphism $h_0$ of $\Sigma$ defined by $h_0=\Xi_j^1\circ h_j\circ(\Xi_j^1)^{-1}$ on each $\Xi_j^1(\Sigma_{m_j,b_j}^\bullet)=\Xi_j^2(\Sigma_{m_j,b_j}^\bullet)$ belongs to ${\rm Diff}_0(\Sigma,\{{\bf x}\}\cup\pl\Sigma)$ and satisfies $\bs{\Xi}^2=h_0\circ\bs{\Xi}^1$. Notice in particular that no T-move around a marked point is needed in the undecorated case.

It remains to show that there is a sequence of elementary moves of type A, S, B, T that brings the maximal family of cutting curves of $\bs{\Xi}^1$ to a family equivalent to that of $\bs{\Xi}^2$. By the result of Hatcher-Lochak-Schneps \cite{Hatcher_Lochak_Schneps}, it suffices to show that any A- or S-move on curves (as defined in the previous subsection) can be realised by a sequence of elementary moves on marked complex building block decompositions.

Consider an A-move on curves. Let $\mc{C}$ be a cutting curve of $\bs{\Xi}^1$ bounding two distinct complex building blocks $\mc{P}_j,\mc{P}_{j'}$ marked by $\Xi_j^1,\Xi_{j'}^1$, so that $\mc{C}=\Xi_j^1\circ\zeta_k^\bullet(\T)=\Xi_{j'}^1\circ\zeta_{k'}^\bullet(\T)$ for some $k,k'$. Up to applying B-moves, we may assume $k_1=k$ and $k_2=k'$ in the definition of the A-move. An A-move (of the first or second type) replaces $\mc{C}$ with a dual curve $\mc{C}^*\subset\mc{P}_j\cup\mc{P}_{j'}$. Recall the marking $\Omega_{\rm A}:\Sigma_{\rm A}^\bullet\to\mc{P}_j\cup\mc{P}_{j'}$ and the notation $\mc{O}_1^\bullet,\ldots,\mc{O}_4^\bullet$ from the construction.
Let $\mc{C}'$ be a cutting curve in $(\mc{P}_j\cup\mc{P}_{j'})^\circ\setminus \{{\bf x}\}$ that intersects $\mc{C}$ at exactly two points and splits $\mc{P}_j\cup\mc{P}_{j'}$ into two new complex building blocks. Then $\Omega_{\rm A}^{-1}(\mc{C}')$ splits $\hat{\C}$ into two connected components, each containing two of the $\mc{O}_k^\bullet$. Suppose that $\mc{O}_1^\bullet$ and $\mc{O}_i^\bullet$ lie in the same component. If $i=2$, then $\mc{C}'$ is homotopically equivalent to $\mc{C}$. If $i=4$ or $i=3$, we choose the A-move $\bs{\Xi}^1\to\bs{\Xi}^2$ to be respectively of the first or second type. If $\mc C^*$ denotes its prescribed dual curve, then there exists $n\in\Z$ such that $\mc{C}'$ is homotopically equivalent to $T_{\mc C}^n(\mc C^*)$, where $T_{\mc C}$ is the Dehn twist around $\mc C$. Applying to $\bs\Xi^1$ the boundary T-move around $\mc C$ corresponding to $T_{\mc C}^n$, the prescribed dual curve is therefore homotopically equivalent to $\mc{C}'$.

The case of S-moves on curves is proved similarly, using the fact that, if $\mc{C}',\mc{C}''$ are two curves on $\#\mc{P}_j^\bullet$, each intersecting $\zeta_1^\bullet(\T)\sim\zeta_2^\bullet(\T)$ at exactly one point and such that cutting $\#\mc{P}_j^\bullet$ along either curve produces a complex building block, then there exists $n\in\Z$ such that $\mc{C}'$ is homotopically equivalent to $T^n(\mc{C}'')$, where $T$ is the Dehn twist around $\zeta_1^\bullet(\T)\sim\zeta_2^\bullet(\T)$.

The decorated case is proved similarly using the decorated version of Lemma \ref{lem:equivalent_curves}.
\end{proof}

For the fusion kernel of the A-move in Section \ref{sec:local_study}, it will be important to assume $\R$-symmetry, which is unfortunately not possible for A-moves of the second type. The following lemma shows that it suffices to consider A-moves of the first type.

\begin{lemma}\label{lem:A-second-type}
An A-move of the second type can be obtained as a composition of A-moves of the first type, B-moves, and T-moves, up to isotopy.
\end{lemma}
\begin{proof}
It suffices to consider the case where there are only two complex building blocks. As in $\Sigma_{\rm A}^\bullet$ in the definition of the A-move, the surface has $4$ exterior objects $\mc{O}_1,\mc{O}_2,\mc{O}_3,\mc{O}_4$, each of which is either a marked point or a removed disk. 

First we have the following observation. Suppose two decompositions $\bs{\Xi}^1,\bs{\Xi}^2$ have the same gluing graph and induce the same partition of $\mc{O}_1,\mc{O}_2,\mc{O}_3,\mc{O}_4$. Let $\mc{C}_1,\mc{C}_2$ be their respective cutting curves, and $\mc{C}_1^*,\mc{C}_2^*$ the corresponding (uniquely defined) dual curves in the A-move of the first type. We assume in addition that the intersection number of $\mc{C}_1$ and $\mc{C}_2^*$ is two. After applying a T-move around $\mc{C}_1$ to $\bs{\Xi}^1$ if necessary, we can make $\mc{C}_1^*$ isotopic to $\mc{C}_2^*$. Thus $\bs{\Xi}^1,\bs{\Xi}^2$ are related by two A-moves of the first type and possibly B-moves and T-moves. This shows in particular that the choice of the dual curve for the A-move of the second type is irrelevant for this lemma. Moreover, notice that applying the A-move of the first type twice gives back the original decomposition, up to relabelling the marked complex building blocks: thus a composition of B-moves exchanging the exterior objects (since we required $k_1^*,k_2^*$ to be minimal in the definition), T-moves and an isotopy, gives back the original marked building block, showing that the inverse of an A-move of the first type is the composition of an A-move of a first type with possibly B- and T-moves.

An A-move of the first type changes the partition from $\{\mc{O}_1,\mc{O}_2\}\mid\{\mc{O}_3,\mc{O}_4\}$ to $\{\mc{O}_1,\mc{O}_4\}\mid\{\mc{O}_2,\mc{O}_3\}$, while an A-move of the second type changes it to $\{\mc{O}_1,\mc{O}_3\}\mid\{\mc{O}_2,\mc{O}_4\}$. If $\mc{O}_1,\mc{O}_2$ are both marked points or both disks and $\mc{C}$ the cutting curve separating $\{\mc{O}_1,\mc{O}_2\}\mid\{\mc{O}_3,\mc{O}_4\}$, then the B-move exchanging $\mc{O}_1,\mc{O}_2$ followed by the A-move of the first type gives the same partition as an A-move of the second type, with the cutting curves having intersection number with $\mc{C}$ equal to $2$ in both cases, and we are done by the discussion above. The same applies for $\mc{O}_3,\mc{O}_4$. We are left with the case where exactly one of $\mc{O}_1,\mc{O}_2$ is a marked point and also exactly one of $\mc{O}_3,\mc{O}_4$ is a marked point. Assume $\mc{O}_1,\mc{O}_4$ have the same type and $\mc{O}_2,\mc{O}_3$ have the same other type, and let $\mc{C}^*$ be the
cutting curve prescribed by the A-move of the second type, while $\mc{C}$ is the starting cutting curve associated to the partition $\{\mc{O}_1,\mc{O}_2\}\mid\{\mc{O}_3,\mc{O}_4\}$. Choose a curve
$\mc{C}'$ inducing the partition
$\{\mc O_1,\mc O_4\}\mid\{\mc O_2,\mc O_3\}$ with intersection number $2$ with both $\mc{C}$ and $\mc{C}^*$ (it is classical topological fact that it exists).  We can apply the A-move of the first type to get the partition $\{\mc{O}_1,\mc{O}_4\}\mid\{\mc{O}_2,\mc{O}_3\}$, and up to applying a suitable T-move about $\mc{C}$ before this A-move, the new cutting curve can be assumed to be isotopic to $\mc{C}'$. Then we apply the B-move exchanging $\mc{O}_2,\mc{O}_3$. Another A-move of the first type gives the partition $\{\mc{O}_1,\mc{O}_3\}\mid\{\mc{O}_2,\mc{O}_4\}$ and, up to applying a suitable T-move about $\mc{C}'$ before this second A-move, the new curve splitting $\{\mc{O}_1,\mc{O}_3\}\mid\{\mc{O}_2,\mc{O}_4\}$ can be assumed to be isotopic to $\mc{C}^*$. By the discussion above, we have thus decomposed our A-move of the second type into a composition of A-moves of the first type, B- and T-moves.
The remaining case is when $\mc{O}_1,\mc{O}_3$ have the same type and $\mc{O}_2,\mc{O}_4$ have the same other type. Let $\bs{\Xi}_1$ with cutting curve $\mc{C}$ splitting $\{\mc{O}_1,\mc{O}_2\}\mid\{\mc{O}_3,\mc{O}_4\}$ and $\bs{\Xi}_2$ its image by an A-move of the second type, with cutting curve $\mc{C}^*$ splitting $\{\mc{O}_1,\mc{O}_3\}\mid\{\mc{O}_2,\mc{O}_4\}$. A B-move exchanging $\mc{O}_1,\mc{O}_3$, followed by an A-move of the first type produces a cutting curve $\mc{C}'$ with partition
$\{\mc{O}_1,\mc{O}_2\}\mid\{\mc{O}_3,\mc{O}_4\}$ and with intersection number $2$ with 
$\mc{C}^*$. Since $\mc{C}$ has the same properties, a
suitable T-move about $\mc{C}^*$, applied before the
A-move of the first type, makes $\mc{C}'$ isotopic to $\mc{C}$. The new decomposition differs from $\bs{\Xi}_1$ only by B- and T-moves, up to isotopy. Since the inverse of an A-move of the first type is an A-move of the first type possibly composed with B- and T-moves, we have thus obtained $\bs{\Xi}_2$ as a composition of B-, T-, and A-moves of the first type.
\end{proof}

 \subsection{Admissible Riemannian metrics}
 
 Let $S=(\Sigma,J,{\bf x},\bs{\zeta})$ be an admissible Riemann surface and $\bs{\psi}=(\psi_1,\dots,\psi_b)$ be the associated family of filling biholomorphisms. We say that a Riemannian metric $g$ on $S$ is \textbf{admissible} if
 it is compatible with the complex structure $J$ (i.e. $J$ is a rotation of an angle $+\pi/2$ in $T\Sigma$ when measured with respect to $g$) and, near each boundary circle $\pl_j\Sigma$, one has
 \[\psi_j^*g=g_\A,\quad g_\A:= \frac{|dz|^2}{|z|^2}.\]
 
 Since we will work with Segal amplitudes, which depend on a choice of admissible Riemannian metric, it is convenient to associate to each 
deformation $S\cdot \bs{f}=(\Sigma^{\bs{f}},\hat{J},{\bf x},\bs{\zeta}^{\bs{f}})$ defined by
\eqref{sigma^fzeta^f} for $\bs{f}\in \mc{U}_\eps^{b}$, 
an admissible metric $g^{\bs{f}}$ on $S\cdot \bs{f}$ 
that depends in a $C^1$ fashion on $\bs{f}$. Choose an admissible metric $g$ and $\delta<1$ close
enough to $1$ so that $((\psi_j^{0})^{\rm in})^*g=g_{\A}$ on $\A_{1,\delta^{-1}}$ for all $j$.
If $\zeta_j$ is incoming, then for $f_j$ close enough to ${\rm Id}$, let $\A_{\delta f_j}$ be the annulus bounded by $\T$ and
$\delta f_j(\T)$, and let $g_{f_j}$ be a smooth metric on
$\A_{\delta f_j}$, depending in a $C^1$ fashion on $f_j$,
such that $g_{f_j}=g_{\A}$ near $\T$, such that
$g_{f_j}$ is admissible near $\delta f_j(\T)$ with
respect to the parametrisation $\delta f_j|_{\T}$, and such that
$g_{\rm Id}=g_{\A}$. The construction of $g_{f_j}$ can be made
explicit; see \cite[Section 3.4]{BGKR1}.
Set
\[\mc{C}_j^{f_j}:=(\psi_j^0\circ R_{\delta^{-1}})(\A_{\delta f_j})\subset\Sigma^{\bs f}, \quad \textrm{ with }R_{\delta^{-1}}(z)=\delta^{-1}z.\]
Similarly, for the outgoing $\zeta_j$, we replace $\psi_j^0$ and $f_j$ by $\psi_j^0\circ o$ and $o\circ f_j\circ o$ and 
define $g_{f_j}$ and $\mc{C}_j^{f_j}$ as above.
After shrinking the neighbourhood $\mc U_\eps$ if necessary, these
collars are pairwise disjoint, and we define
\begin{equation}\label{defgpsi}
g^{\bs f}:=\begin{cases}
g &\text{on } \Sigma^{\bs f}\setminus\cup_{j=1}^b\mc C_j^{f_j},\\[2mm]
(\psi_j^0)_*(R_{\delta^{-1}})_*g_{f_j}
&\text{on }\mc C_j^{f_j}.
\end{cases}
\end{equation}
The two expressions agree near the outer boundary
$\psi_j^0(\delta^{-1}\T)$ of $\mc C_j^{f_j}$ and  $g^{\bs f}$ is admissible on
$(\Sigma^{\bs f},\hat J,{\bf x},\bs\zeta^{\bs f})$. The family
$\bs f\mapsto g^{\bs f}$ is $C^1$ and satisfies
$g^{\rm Id}=g$.


We complete this section with a definition, which will be useful later. 
\begin{definition}\label{LiouvillePant}
Let $S=(\Sigma,J,{\bf x},\bs{\zeta})$ be a complex building block of type $(m,b^+,b^-)$ and $g$ be an admissible Riemannian metric on $S$. For $b=b^++b^-$, let
$\bs{\psi}=(\psi_1,\dots,\psi_b)\in \mc{P}^\eps_{m,b^+,b^-}$ be the model form of $S$, 
with uniformising map $\Phi_{\mc{P}}:\Sigma\to \mc{P}$,  where 
\[\mc{P}=\hat{\C}\setminus \bigg(\Big(\bigcup_{k=1}^{b^+}\psi_k^{\rm in}(\D^\circ)\Big)\cup \Big(\bigcup_{k=b^++1}^{b}\psi_k(\D^\circ)\Big)\bigg)\] 
is the corresponding model complex building block with boundary parametrised by $\bs{\zeta}=(\psi_1|_{\T},\dots,\psi_b|_{\T})$ and marked points ${\bf x}=(z_{{b+1}}^\bullet,\dots,z_3^\bullet)$.  We view $(\hat{\C},J_{\C},{\bf z}^\bullet)$ as the filling of $(\mc{P},J_\C,{\bf x},\bs{\zeta})$ and define $\hat{g}$ to be the continuous, piecewise smooth Riemannian metric on $\hat{\C}$ given by 
\[ \hat{g}=\begin{cases}
(\Phi_{\mc{P}})_*g & \textrm{ on }\mc{P},\\
 (\psi_k^{\rm in})_*(|\dd z|^2) & \textrm{ on }\psi_k^{\rm in}(\D), \, k=1,\dots,b^+,\\
 (\psi_k)_*(|\dd z|^2) & \textrm{ on }\psi_k(\D), \, k=b^++1,\dots,b.\\
\end{cases}\]
We  call $\hat{g}$ the \textbf{filling metric} of $g$ on  $\hat{\C}$.
\end{definition}

\section{Spectral resolution of the Liouville Hamiltonian and holomorphic Hilbert space}\label{spectral_resolution}

In the probabilistic construction of Liouville CFT developed in \cite{DKRV16,Guillarmou2019}, the correlation functions and partition functions of the theory are rigorously defined. Associated with this construction is a Hilbert space $\mathcal{H}$, which can be identified (essentially) with the $L^2$-space of fields on the unit circle $\T$, viewed as embedded in the Riemann sphere $\hat{\C}$. This Hilbert space is equipped with a natural Hamiltonian $\mathbf{H}$.

The conformal blocks arise from the spectral resolution of this Hamiltonian, established in \cite{GKRV20_bootstrap}, combined with the disintegration properties of the path integral representation of correlation functions into products of elements of the Hilbert space $\mathcal{H}$. These properties correspond to the Segal axioms, which were proved in \cite{GKRV21_Segal}.

In this section, we recall and expand the discussion from \cite[Section 6.4]{GKRV21_Segal} about the spectral resolution of the Liouville Hamiltonian ${\bf H}$.\\

\textbf{Notation}: from now on, we consider tensor products of separable Hilbert spaces $V$, all being of the form $V=L^2(Z,\mu)$ for some measured space $(Z,\mu)$. We always denote $V^{\otimes n}$ the completion of the $n$-th tensor product, which is a Hilbert space that can be identified with $L^2(Z^n,\mu^{\otimes n})$. Similarly, since $V^*$ identifies with $V$ via the inner product, $(V^*)^{\otimes n}$ identifies with its dual $V^{\otimes n}\simeq L^2(Z^n,\mu^{\otimes n})$.

\subsection{Liouville Hilbert space and Hamiltonian}
We denote by 
\[H^s(\T):=\{ \varphi(\theta)=\sum_{n\in \Z}\varphi_ne^{{\rm i}n\theta}\,|\, \sum_{n}(1+|n|)^{2s}|\varphi_n|^2<\infty\}\] 
the Sobolev space of order $s\in \R$ on the unit circle $\T$. Consider  
a random variable $\varphi\in H^{-s}(\T)$ ($s>0$)
\[ \varphi(\theta)=\sum_{n\geq 1}\frac{x_n+{\rm i}y_n}{2\sqrt{n}}e^{{\rm i}n\theta}+\frac{x_{n}-{\rm i}y_n}{2\sqrt{n}}e^{-{\rm i}n\theta}\]
where $x_n,y_n$ are i.i.d centered real Gaussians with variance $1$, on a probability space 
$(\Omega_\T,\mc{F}_\T,\P_\T)$. Concretely, $\Omega_\T=(\R^2)^{\N}$, 
\[ \P_\T= \prod_{n=1}^\infty e^{-\frac{x_n^2}{2}-\frac{y_n^2}{2}}\frac{\d x_n \d y_n}{2\pi},\]
and $\mc{F}_\T$ is the cylinder  Borel sigma-algebra.
We let $\mu_0$ be the pushforward of the measure $\d c\otimes \P_\T$ from $\R\times \Omega_\T$ 
to $H^{-s}(\T)$ by the random variable $c+\varphi$ (here $c$ is viewed as a constant function in $H^{-s}(\T)$) 
and define the Hilbert space
\[\mc{H}:=L^2(H^{-s}(\T),d\mu_0)\simeq L^2(\R\times \Omega_{\T},\d c\otimes \P_\T).\] 
We shall view functions in $\mc{H}$  as functions of $(c,\varphi)$, which in turn are functions of $c,(x_n,y_n)_n$.
The \textit{Liouville Hamiltonian} $\bH$ is an unbounded self-adjoint operator on $\mc{H}$, 
introduced in \cite{GKRV20_bootstrap} and formally of the form 
\[ \bH=-\frac{1}{2}\pl_c^2+{\bf P}+\frac{Q^2}{2}+e^{\gamma c}V\]
where $Q=\gamma/2+2/\gamma$ with $\gamma \in (0,2)$, 
\[{\bf P}:=\sum_{n\geq 1}n(\pl_{x_n}^*\pl_{x_n}+ \pl_{y_n}^*\pl_{y_n})\]
and $V$ is a non-negative unbounded operator which, when $\gamma<\sqrt{2}$, is the multiplication by a positive 
$L^p(\Omega_\T)$ function for any $p<2/\gamma^2$. As it is not necessary here, we do not recall the precise definition of $V$ and refer to  \cite[Section 5]{GKRV20_bootstrap} for details. Concretely, the operator $\bH$ is the Friedrichs extension of a quadratic form 
\[ \mc{Q}(F)=\int_{\R} \Big( \frac{1}{2}\|\pl_cF\|_{L^2(\Omega_\T)}^2+\frac{Q^2}{2}\|F\|_{L^2(\Omega_\T)}^2+\cjg {\bf P}F,F\cjd_{L^2(\Omega_\T)}+e^{\gamma c}\cjg VF,F\cjd_{L^2(\Omega_\T)}\Big)\d c.
\]
We denote by $\mc{D}(\mc{Q})=\{ F\in \mc{H}\,|\, \mc{Q}(F)<\infty\}$ 
the domain of this quadratic form and $\mc{D}'(\mc{Q})$ its dual space. The operator $\bH:\mc{D}(\mc{Q})\to \mc{D}'(\mc{Q})$ is bounded.

\subsection{Spectral resolution and Virasoro operators}\label{sectiion:spectral}

We recall the spectral resolution of the Hamiltonian $\bH$ and the role of the Virasoro operators in this diagonalisation, following \cite{GKRV20_bootstrap,GKRV21_Segal,BGKRV}.

Let $\mc{T}$ be the set of Young diagrams, that is the set of finite non-increasing sequences of positive integers
$\nu=\nu(1)\geq \dots \nu(k)>0$, and let $\mc{T}_n:=\{ \nu\in \mc{T}\, |\, |\nu|=n\}$ where the size  of a diagram is $|\nu|:=\sum_{j}\nu(j)$. 
By convention $\mc{T}$ also contains the empty diagram, denoted $0$, and whose size is $0$. 
We denote by  $d_n=\#\mc{T}_n$, the cardinality of $\mc{T}_n$. Let $\mc{H}_{\mc{T}}:=\bigoplus_{n=0}^\infty \C^{d_n}$ equipped with the standard Hermitian product $\cjg v,v'\cjd_{\mc{T}}=\sum_n \cjg v_n,v'_n\cjd_{\C^{d_n}}$ if $v=\sum_n v_n$ with $v_n\in \C^{d_n}$ and similarly for $v'$. We will denote by $(e_{\nu})_{\nu\in \mc{T}}$ an orthonormal basis of $\mc{H}_{\mc{T}}$. 
We can identify $\mc{H}_{\mc{T}}$ to $\ell^2(\mc{T})$ equipped with the standard counting measure on $\mc{T}$ (i.e. each $\nu$ has mass $1$).

We summarize the results of  \cite[Section 6]{GKRV21_Segal}, in particular \cite[Corollary 6.5]{GKRV21_Segal}. The natural spectral parameter for $\bH$ is 
\[\Delta_{\alpha}:=\frac{\alpha}{2}(Q-\frac{\alpha}{2}),\quad  \textrm{ for } \alpha\in \C.\] 
We will also use a function $\rho:\R\to \R$ which satisfies $\rho(c)=c$ on $\R_-$ and $\rho(c)=0$ in $c\geq 0$. $\rho$ can also be seen as  a function on $H^{-s}(\T)$ (depending only on the $0$-th Fourier mode $c$).

\begin{lemma}\cite[Section 6]{GKRV21_Segal}\label{l:diagonalisation}
There exists a family $(H_{Q+{\rm i}p,\nu,\tilde{\nu}})_{(p,\nu,\tilde{\nu})\in \R_+\times \mc{T}^2}\subset \bigcap_{\delta>0}e^{-\delta \rho}\mc{D}(\mc{Q})$ of 
eigenfunctions of ${\bf H}$:   
\[ ({\bf H}-2\Delta_{Q+{\rm i}p}-|\nu|-|\tilde{\nu}|)H_{Q+{\rm i}p,\nu,\tilde{\nu}}=0, \] 
spanning Verma modules  
\begin{equation}\label{Verma_modules} 
\mc{V}^\circ_{Q+{\rm i}p}:={\rm span}\{H_{Q+{\rm i}p,\nu,0}\,|\, \nu \in \mc{T}\}, \quad \bbar{\mc{V}}^\circ_{Q+{\rm i}p}:={\rm span}\{H_{Q+{\rm i}p,0,\tilde{\nu}}\,|\, \tilde{\nu}\in \mc{T}\}\end{equation}
and providing a spectral resolution of $\bH$ in the following sense: there is an isometry 
\begin{align}
& \Pi_{\mc{V}\otimes \bbar{\mc{V}}}: \mc{H}\to L^2(\R_+;L^2(\mc{T}^2))=L^2(\R_+\times \mc{T}^2)\nonumber \\ 
& (\Pi_{\mc{V}\otimes \bbar{\mc{V}}}u)(p;\nu,\tilde{\nu}):=\frac{1}{\sqrt{2\pi}}\cjg u, H_{Q+{\rm i}p,\nu,\tilde{\nu}}\cjd_{\mc{H}} \label{Pi2V}.
\end{align}
The right-hand side of \eqref{Pi2V} makes sense  for all $p\in \R_+$ if  $u\in \bigcup_{\delta>0}e^{\delta\rho}\mc{H}$, and extends to $\mc{H}$ by density. 
\end{lemma}
The functions $H_{Q+{\rm i}p,\nu,\tilde{\nu}}$ for $(\nu,\tilde{\nu})\not=(0,0)$ can be described in two ways:
\begin{enumerate}
\item[1)] First, they are descendant states that can be expressed in terms of the action of two commuting Virasoro algebras as follows. For each $n\in \Z$, there are bounded operators 
\begin{equation}\label{Virasoro_generators} 
{\bf L}_n: e^{\delta \rho}\mc{D}(\mc{Q})\to e^{\delta \rho}\mc{D}'(\mc{Q}), \quad \tilde{{\bf L}}_n: e^{\delta\rho}\mc{D}(\mc{Q})\to e^{\delta\rho}\mc{D}'(\mc{Q}), \quad \forall \delta\in \R
\end{equation}
called \textit{Virasoro generators}. They preserve ${\rm span}\{ H_{Q+{\rm i}p,\nu,\tilde{\nu}}\,|\, (\nu,\tilde{\nu}) \in \mc{T}^2\}$ and form two commuting Virasoro algebras with central charge $c_{\rm L}=1+6Q^2$ when acting on this linear space. These Virasoro representations are unitary in the sense that
 ${\bf L}_n^*={\bf L}_{-n}$, $\tilde{\bf L}_n^*=\tilde{\bf L}_{-n}$ for all $n\in\Z$; more precisely, $\cjg {\bf L}_nu,u'\cjd_{\mc{H}}=\cjg u,{\bf L}_{-n}u'\cjd_{\mc{H}}$ for $u,u'\in \mc{D}(\mc{Q})$, and similarly for $\tilde{\bf L}_n$. 

 Let us  define the \textit{primary state} 
\[\Psi_{Q+{\rm i}p}:=H_{Q+{\rm i}p,0,0},\] 
and, for each $\nu=(\nu(1),\dots,\nu(k))\in \mc{T}$ and $\tilde{\nu}=(\tilde{\nu}(1),\dots,\tilde{\nu}(\ell))\in \mc{T}$, the \textit{descendant state} 
\begin{equation}\label{primary}
\Psi_{Q+{\rm i}p,\nu,\tilde\nu}:={\bf L}_{-\nu(k)}\dots {\bf L}_{-\nu(1)}\tilde{\bf L}_{-\tilde{\nu}(\ell)}\dots \tilde{\bf L}_{-\tilde{\nu}(1)}\Psi_{Q+{\rm i}p},
\end{equation}
which are  proven, in \cite[Section 3.2.4]{BGKRV},  to be  well-defined as elements in $\bigcap_{\delta>0} e^{-\delta \rho}\mc{D}(\mc{Q})$. They satisfy 
\[ \forall n>0, \,\, {\bf L}_n\Psi_{Q+{\rm i}p}=0 ,\quad \forall \nu\in \mc{T},\,\, 
\tilde{{\bf L}}_n\Psi_{Q+{\rm i}p,\nu,0}=0.\] 
Also, there are  ($p$-dependent) positive definite real valued matrices   $(F_{Q+{\rm i}p}(\nu,\nu'))_{\nu,\nu'\in \mc{T},|\nu|=|\nu'|=n}$, called \textit{Shapovalov matrices}, defined for each $n>0$ on the set $\mc{T}_n$ of Young Diagrams of size $n$, such that 
\begin{equation}\label{def_of_H_Q+ip}
H_{Q+{\rm i}p,\nu,\tilde{\nu}}=\sum_{|\nu'|=|\nu|,|\tilde{\nu}'|=|\tilde{\nu}|}F_{Q+{\rm i}p}^{-1/2}(\nu,\nu')F_{Q+{\rm i}p}^{-1/2}(\tilde{\nu},\tilde{\nu}')\Psi_{Q+{\rm i}p,\nu',\tilde{\nu}'}.
\end{equation}
Here the $F_{Q+{\rm i}p}^{-1/2}(\nu,\nu')$ notation means the $(\nu,\nu')$ coefficient of the square root of the inverse of the Shapovalov matrices of size $|\nu|=|\nu'|$.  Furthermore, their dependence in $p$ is only through the quantity $\Delta_{Q+{\rm i}p}$ in such a way that $  F_{Q+{\rm i}p}(\nu,\nu')=F_{Q-{\rm i}p}(\nu,\nu')$. 
We refer to \cite[Proposition 4.9]{GKRV20_bootstrap} for more details on the definition of the Shapovalov matrices. Finally, by \cite[Theorem 1.2]{BGKRV}, there is a holomorphic extension $\alpha\mapsto \Psi_{\alpha,\nu,\tilde{\nu}}$ in $\alpha\in \C$ with values in $e^{-\delta \rho}\mc{D}(\mc{Q})$ for $|{\rm Re}(\alpha)-Q|<\delta$
\item[2)] Second, they are scattering states (\cite[Theorem 4.2]{BGKRV}): 
there is a constant $\mc{S}(Q+{\rm i}p)\in \C$ called reflection (or scattering) coefficient, satisfying
\begin{equation}\label{eq_fonct_scattering} 
\mc{S}(Q+{\rm i}p)\mc{S}(Q-{\rm i}p)=1,
\end{equation}
such that for each $\nu,\tilde{\nu}\in \mc{T}^2$ with $|\nu|+|\tilde{\nu}|=n$, there exists  $h_{Q\pm {\rm i}p,\nu,\tilde{\nu}}\in \ker ({\bf P}-n)$  of $L^2(\Omega_\T)$ norm $1$ such that for $\chi\in C^\infty$ equal to $1$ near $-\infty$ and $0$ in $(0,\infty)$,
\begin{equation}\label{expansion H}
H_{Q+{\rm i}p,\nu,\tilde{\nu}}-\chi(c)(e^{{\rm i}pc}h_{Q+{\rm i}p,\nu,\tilde{\nu}}+e^{-{\rm i}pc}\mc{S}(Q+{\rm i}p)h_{Q-{\rm i}p,\nu,\tilde{\nu}})\in \mc{H}.
\end{equation} 
Moreover for each $p>0$ fixed, the vectors $h_{Q\pm {\rm i}p,\nu,\tilde{\nu}}$ with $|\nu|+|\tilde{\nu}|=n$ form an orthonormal basis of $\ker ({\bf P}-n)$.
Furthermore, $H_{Q+{\rm i}p,\nu,\tilde{\nu}}$ is the unique solution in $\cap_{\delta>0}e^{-\delta\rho}\mc{D}(\mc{Q})$ of
$({\bf H}-2\Delta_{Q+{\rm i}p}-|\nu|-|\tilde{\nu}|)u=0$ with this asymptotic, 
for $h_{Q+{\rm i}p,\nu,\tilde{\nu}}$ fixed. Finally, by \cite[Theorem 1.2]{BGKRV},
\begin{equation}\label{symmetry_psi_Q+ip}
\Psi_{Q+{\rm i}p,\nu,\tilde\nu}=\mc{S}(Q+{\rm i}p)\Psi_{Q-{\rm i}p,\nu,\tilde\nu}.
\end{equation}
\end{enumerate}

\medskip
\noindent \textbf{Verma modules.} The Verma module \eqref{Verma_modules} $\mc{V}^\circ_{Q+{\rm i}p}$ (resp.  $\bbar{\mc{V}}^\circ_{Q+{\rm i}p}$) consists of finite linear combinations of basis elements $H_{Q+{\rm i}p,\nu,0}$ (resp. $H_{Q+{\rm i}p,0,\tilde{\nu}}$). 
We can equip $\mc{V}^\circ_{Q+{\rm i}p}$ with the Hilbert space structure of $L^2(\mc{T})$ by setting
\begin{equation}\label{Verma_Hilbert} 
\mc{V}_{Q+{\rm i}p}:=\{  \sum_{\nu\in \mc{T}}f(\nu)H_{Q+{\rm i}p,\nu,0} \,|\, f\in L^2(\mc{T})\}
\end{equation}
equipped with the natural complex vector space structure and the Hermitian product 
\begin{equation}\label{Hilbert_st_verma} 
\cjg \sum_{\nu\in \mc{T}}f(\nu)H_{Q+{\rm i}p,\nu,0},  \sum_{\nu\in \mc{T}}f'(\nu)H_{Q+{\rm i}p,\nu,0}\cjd_{\mc{V}_{Q+{\rm i}p}}:=\cjg f,f'\cjd_{L^2(\mc{T})}=\sum_{\nu \in \mc{T}}f(\nu)\bbar{f'}(\nu).
\end{equation}
Here we view \eqref{Verma_Hilbert} as an abstract completion of $\mc{V}_{Q+{\rm i}p}^\circ$. Finite combinations of $H_{Q+{\rm i}p,\nu,0}$ can be represented as functions belonging to $\bigcap_{\eps>0}e^{-\eps\rho}\mc{H}$ (and similarly for $ \bbar{\mc{V}}_{Q+{\rm i}p}$), 
but they are not contained in the Hilbert space $\mc{H}$. However, their direct integral in $p\in \R_+$ inherits a Hilbert space structure from $\mc{H}$ as we shall explain.
First, define the bounded map 
\begin{align}
& \Pi_{\mc{V}}: \mc{H}\to L^2(\R_+;L^2(\mc{T}))=L^2(\R_+\times \mc{T})\nonumber \\ 
& (\Pi_{\mc{V}}u)(p;\nu)=\frac{1}{\sqrt{2\pi}}
\cjg u, H_{Q+{\rm i}p,\nu,0}\cjd_{\mc{H}}=(\Pi_{\mc{V}\otimes \bbar{\mc{V}}}u)(p;\nu,0) \label{PiV}
\end{align}
which satisfies $\|\Pi_{\mc{V}}u\|_{L^2(\R_+\times \mc{T})}\leq \|u\|_{L^2}$; here we used the convention that $F_{Q+{\rm i}p}^{-1/2}(0,0)=1$. We notice that $\Pi_{\mc{V}}$ can be viewed as a part of the spectral resolution associated to ${\bf H}$: for a function $u\in \mc{H}$, 
\[ \frac{1}{\sqrt{2\pi}}\sum_{\nu\in \mc{T}}\int_{0}^\infty (\Pi_{\mc{V}}u)(p;\nu)H_{Q+{\rm i}p,\nu,0}\dd p\]
projects on a subspace of the continuous spectrum.
\begin{lemma}\label{surjectivity}
The map $\Pi_{\mc{V}\otimes \bbar{\mc{V}}}:\mc{H}\to L^2(\R_+\times \mc{T}^2)$  is surjective, and therefore a unitary map.   
\end{lemma}
\begin{proof} 
Consider for  $\la>Q^2/2$ the space 
$E_\la:=\{(\nu,\tilde{\nu})\in \mc{T}^2\,|\, |\nu|+|\tilde{\nu}|<\la-Q^2/2\}$ and for $\delta>0$ 
\[ W_\la: u\in e^{\delta \rho}\mc{H}\mapsto \big((\nu,\tilde{\nu})\mapsto \frac{1}{\sqrt{2\pi}}\cjg u,H_{Q+{\rm i}p(\la,\nu,\tilde{\nu}),\nu,\tilde{\nu}}\cjd_{\mc{H}}\big)\in L^2(E_\la),\]
where $p(\la,\nu,\tilde{\nu})>0$ is defined by $2\Delta_{Q+{\rm i}p(\la,\nu,\tilde{\nu})}+|\nu|+|\tilde{\nu}|=\la$.
Let $\chi\in C^\infty(\R)$ so 
that ${\rm supp}(\chi)\subset \R_-$ and $\chi(c)=1$ if $c\leq -1$. 
Set $v_{p,\nu,\tilde{\nu}}:=({\bf H}-2\Delta_{Q+{\rm i}p}-|\nu|-|\tilde{\nu}|)u_{p,\nu,\tilde{\nu}}$ 
with $u_{p,\nu,\tilde{\nu}}:=\chi(c)e^{{\rm i}pc}h_{Q+{\rm i}p,\nu,\tilde{\nu}}$. We have 
\[ v_{p,\nu,\tilde{\nu}}=-\frac{1}{2}[\pl_c^2,\chi]e^{{\rm i}pc}h_{Q+{\rm i}p,\nu,\tilde{\nu}}+\chi(c)e^{(\gamma+{\rm i}p) c}Vh_{Q+{\rm i}p,\nu,\tilde{\nu}} \in e^{\delta \rho}\mc{D}'(\mc{Q})\]
for some $\delta>0$, where we used \cite[Lemma 6.1]{GKRV20_bootstrap} to deal with the second term in the right-hand side. 
Using \cite[Lemma 6.18]{GKRV20_bootstrap} and the expansion \eqref{expansion H}, we see that if $2\Delta_{Q+{\rm i}p'}+|\nu'|+|\tilde{\nu}'|=
2\Delta_{Q+{\rm i}p}+|\nu|+|\tilde{\nu}|=\la$,  then 
\[
\cjg v_{p,\nu,\tilde{\nu}},H_{Q+{\rm i}p',\nu',\tilde{\nu}'}\cjd_{\mc{H}}= {\rm i}p\,\delta_{p}(p')\delta_{(\nu,\tilde{\nu})}(\nu',\tilde{\nu}'),
\]
where $\delta_{(\nu,\tilde{\nu})}$ and $\delta_{p}$ are respectively the characteristic functions of the singletons  $\{(\nu,\tilde{\nu})\}$ and $\{p\}$.
As a consequence, if $2\Delta_{Q+{\rm i}p'}+|\nu'|+|\tilde{\nu}'|=
2\Delta_{Q+{\rm i}p}+|\nu|+|\tilde{\nu}|$, we have
\begin{equation}\label{Wpunu} 
\begin{split} 
\cjg {\bf H}^{-1}v_{p,\nu,\tilde{\nu}},H_{Q+{\rm i}p',\nu',\tilde{\nu}'}\cjd_{\mc{H}}&=\cjg v_{p,\nu,\tilde{\nu}},{\bf H}^{-1}H_{Q+{\rm i}p',\nu',\tilde{\nu}'}\cjd_{\mc{H}}\\
& =\frac{{\rm i}p}{\la(p,|\nu|+|\tilde{\nu}|)}\delta_{p}(p')\delta_{(\nu,\tilde{\nu})}(\nu',\tilde{\nu}'),
\end{split}\end{equation}
where $\la(p,N):=2\Delta_{Q+{\rm i}p}+N$. Here, we have used \cite[Lemma 6.5]{GKRV20_bootstrap}, which shows that 
${\bf H}^{-1}:e^{\pm \eps\rho}\mc{D}'(\mc{Q})\to e^{\pm \eps \rho}\mc{D}(\mc{Q})$ is bounded for all $\eps\in (0,Q)$ small enough, and the fact that ${\bf H}^{-1}H_{Q+{\rm i}p',\nu',\tilde{\nu}'}=\la(p,N)^{-1}H_{Q+{\rm i}p',\nu',\tilde{\nu}'}$ since $\ker {\bf H}\cap e^{-\eps \rho}\mc{D}(\mc{Q})=0$ for $\eps>0$ small.
Now we claim that this implies the density of the range of $\Pi_{\mc{V}\otimes \bbar{\mc{V}}}$. Notice that the range of $\Pi_{\mc{V}\otimes \bbar{\mc{V}}}$ is closed since it is an isometry.
Let $G\in L^2(\R_+\times \mc{T}^2)$ be orthogonal to 
$\Pi_{\mc{V}\otimes \bbar{\mc{V}}}(\mc{H})$. By the spectral theorem, we have for each interval $I\subset (0,\infty)$ and each $u \in \mc{H}$
\begin{equation}\label{orthogonality_surj}
0=\cjg \Pi_{\mc{V}\otimes \bbar{\mc{V}}}({\bf 1}_{I}({\bf H})u),G\cjd_{L^2(\R_+\times \mc{T}^2)}=\int_{\R_+}\sum_{\nu,\tilde{\nu}\in \mc{T}}{\bf 1}_{I}(\la(p,|\nu|+|\tilde{\nu}|))W_{\la(p,|\nu|+|\tilde{\nu}|)}(u) (\nu,\tilde{\nu})\bbar{G(p,\nu,\tilde{\nu})}\dd p.
\end{equation}
For $p_0>0$ and $\nu_0,\tilde{\nu}_0\in\mc{T}$ fixed, write $N_0:=|\nu_0|+|\tilde{\nu}_0|$. Assume that $\la(p_0,N_0)\notin Q^2/2+\N$. Take
$I=(\la(p_0-\eps_1,N_0),\la(p_0+\eps_2,N_0))$ with $\eps_1,\eps_2>0$ arbitrarily small. The $p$-integral above splits into finitely many integrals on the intervals 
$I_N:=\{p>0\;|\; p^2+2N\in ((p_0-\eps_1)^2+2N_0,(p_0+\eps_2)^2+2N_0)\}$ for $N<N_0+(p_0+\eps_2)^2/2$, corresponding to Young diagrams $\nu,\tilde{\nu}$ satisfying $|\nu|+|\tilde{\nu}|=N$. Using the change of variables $p_N=\sqrt{p^2+2(N-N_0)}$ for $p\in I_N$ (so that $p_N\in (p_0-\eps_1,p_0+\eps_2)$), \eqref{orthogonality_surj} becomes
\begin{equation}\label{orthogonality_surj2} 
0=\int_{p_0-\eps_1}^{p_0+\eps_2}\sum_{\nu,\tilde{\nu}\in A(p,N_0)} W_{\la(p,N_0)} (u)(\nu,\tilde{\nu})\,p\,\frac{\bbar{G(\sqrt{p^2+2(N_0-|\nu|-|\tilde{\nu}|)},\nu,\tilde{\nu})}}{\sqrt{p^2+2(N_0-|\nu|-|\tilde{\nu}|)}}\dd p,
\end{equation}
where $A(p,N_0)=\{ (\nu,\tilde{\nu})\in \mc{T}^2\,|\, p^2+2(N_0-|\nu|-|\tilde{\nu}|)>0\}$.
Since $\eps_1,\eps_2>0$ are arbitrary, using \eqref{orthogonality_surj} with $u={\bf H}^{-1}v_{p_0,\nu_0,\tilde{\nu}_0}$ implies that as a function of $p\in \R_+$, we have almost everywhere in $p$ in a neighbourhood of $p_0$
\[0=\sum_{\nu,\tilde{\nu}\in A(p,N_0)} W_{\la(p,N_0)} ({\bf H}^{-1}v_{p_0,\nu_0,\tilde{\nu}_0})(\nu,\tilde{\nu})\,p\,\frac{\bbar{G(\sqrt{p^2+2(N_0-|\nu|-|\tilde{\nu}|)},\nu,\tilde{\nu})}}{\sqrt{p^2+2(N_0-|\nu|-|\tilde{\nu}|)}}.\]
Using \eqref{Wpunu} and that 
$\la \mapsto W_\la({\bf H}^{-1}v_{p_0,\nu_0,\tilde{\nu}_0})(\nu,\tilde{\nu})$ is $C^1$ on $(Q^2/2+|\nu|+|\tilde{\nu}|,\infty)$ for each $\nu,\tilde{\nu}\in \mc{T}$, we deduce that there exist $\eps>0$ and $C>0$ such that for $|p-p_0|<\eps$, we have
\[ |G(p,\nu_0,\tilde{\nu}_0)|\leq C\eps \Big(\sum_{\substack{(\nu,\tilde{\nu})\in A(p,N_0)\\ (\nu,\tilde{\nu})\not=(\nu_0,\tilde{\nu}_0)}} \frac{|G(\sqrt{p^2+2(N_0-|\nu|-|\tilde{\nu}|)},\nu,\tilde{\nu})|}{\sqrt{p^2+2(N_0-|\nu|-|\tilde{\nu}|)}}\Big).\]
Integrating the square of this inequality on $(p_0-\eps,p_0+\eps)$, we have for some $C'>0$ 
\[\int_{p_0-\eps}^{p_0+\eps}|G(p,\nu_0,\tilde{\nu}_0)|^2\dd p\leq C'^2\eps^2 \int_{p_0-\eps}^{p_0+\eps}\sum_{\substack{(\nu,\tilde{\nu})\in A(p,N_0)\\ (\nu,\tilde{\nu})\not=(\nu_0,\tilde{\nu}_0)}} |G(\sqrt{p^2+2(N_0-|\nu|-|\tilde{\nu}|)},\nu,\tilde{\nu})|^2\dd p.
\]
We see that 
\[\lim_{\eps\to 0} \frac{1}{\eps}\int_{p_0-\eps}^{p_0+\eps}|G(p,\nu_0,\tilde{\nu}_0)|^2\dd p=0, \]
thus, using the Lebesgue differentiation theorem, we deduce that $G(p_0,\nu_0,\tilde{\nu}_0)=0$ almost everywhere in $p_0$. Since $\nu_0,\tilde{\nu}_0$ are also arbitrary, we have thus proved that $G(p,\nu,\tilde{\nu})=0$ for almost all $p>0$ and all $\nu,\tilde{\nu}\in \mc{T}$. 
\end{proof}

Let us define the \textbf{holomorphic Hilbert space} 
\begin{equation}\label{defH^10}
\begin{split}
& \mc{H}^{1,0}:= \Pi_{\mc{V}\otimes \bbar{\mc{V}}}^{-1}(L^2_{1,0}(\R_+\times \mc{T}^2)) \quad \textrm{ with }\\
& L^2_{1,0}(\R_+\times \mc{T}^2):=\{ u\in L^2(\R_+\times \mc{T}^2)\,|\, \forall  \nu\in \mc{T},\, \forall \tilde{\nu}\in \mc{T}\setminus\{0\}, 
u(p,\nu,\tilde{\nu})=0\}.
\end{split}
\end{equation}
Since the vector space $L^2_{1,0}(\R_+\times \mc{T}^2)$ is closed in $L^2(\R_+\times \mc{T}^2)$, we deduce that $\mc{H}^{1,0}\subset \mc{H}$ is a Hilbert subspace of $\mc{H}$.
Define the isomorphism
\begin{equation}\label{iota10} 
\iota_{1,0}: L^2(\R_+\times \mc{T})\to  L_{1,0}^2(\R_+\times \mc{T}^2) ,\quad (\iota_{1,0}u)(p,\nu,\tilde{\nu}):=u(p,\nu)\delta_0(\tilde{\nu}).
\end{equation}
\begin{lemma}\label{PiV^*PiV}
The operator $\Pi_{\mc{V}}$ of \eqref{PiV} satisfies 
\begin{equation}\label{defPi10}
 \Pi^{1,0}= \Pi_{\mc{V}}^*\Pi_{\mc{V}}, \qquad \Pi_{\mc{V}}\Pi_{\mc{V}}^*={\rm Id}
\end{equation}
where $\Pi^{1,0}:\mc{H}\to \mc{H}^{1,0}$ is the orthogonal projection. 
The operator $\Pi_{\mc{V}}^*$ is thus an isomorphism 
\[\Pi_{\mc{V}}^*: L^2(\R_+\times \mc{T})\to \mc{H}^{1,0}\] 
and  has the following form: 
\begin{equation}\label{PiV^*}
\forall  f\in L^2(\R_+\times \mc{T}),\quad \Pi_{\mc{V}}^*f=\frac{1}{\sqrt{2\pi}}\int_{\R_+}\sum_{\nu\in \mc{T}}f(p,\nu)H_{Q+{\rm i}p,\nu,0}\,\d p \, \in \mc{H}.
\end{equation}
\end{lemma}
\begin{proof} First, we have $\Pi^{1,0}=\Pi_{\mc{V}\otimes \bbar{\mc{V}}}^{-1} \circ \iota_{1,0} \circ \Pi_{\mc{V}}$.
Next we observe that $\Pi_{\mc{V}}= \iota_{1,0}^*\circ \Pi_{\mc{V}\otimes \bbar{\mc{V}}}$ (viewing $\iota_{1,0}$ as mapping $L^2(\R_+\times \mc{T})\to L^2(\R_+\times \mc{T}^2)$), which proves \eqref{defPi10}.
We can compute the form of $\Pi_{\mc{V}}^*$: for $f\in L^2(\R_+\times \mc{T})$ and $u\in e^{\delta \rho}\mc{H}$
\[  \cjg \Pi_{\mc{V}}^*f,u\cjd_{\mc{H}}= \cjg f,\Pi_{\mc{V}}u\cjd_{L^2(\R_+\times \mc{T})}=\frac{1}{\sqrt{2\pi}}\int_{\R_+}\sum_{\nu\in \mc{T}}f(p,\nu) \cjg H_{Q+{\rm i}p,\nu,0},u\cjd_{\mc{H}}\,\d p=
 \frac{1}{\sqrt{2\pi}}\Big\cjg \int_{\R_+}\sum_{\nu\in \mc{T}}f(p,\nu)H_{Q+{\rm i}p,\nu,0}\,dp,u\Big\cjd_{\mc{H}}\]
so that, by density of $e^{\delta \rho}\mc{H}$ in $\mc{H}$, we get \eqref{PiV^*}.
\end{proof}
Lemma \ref{PiV^*PiV} implies that the direct integral $\mc{V}$,  over $p\in \R_+$, of the Verma modules $\mc{V}_{Q+{\rm i}p}$ can be represented under the form 
\begin{equation}\label{VQ+ip}
\mc{V}:= \Big\{ \Pi_{\mc{V}}^*f= \frac{1}{\sqrt{2\pi}}\int_{\R_+}\sum_{\nu\in \mc{T}} f(p,\nu)H_{Q+{\rm i}p,\nu,0}\dd p  \,|\, f\in L^2(\R_+\times \mc{T})\Big\}={\rm Ran}(\Pi_{\mc{V}}^*)=\mc{H}^{1,0},
\end{equation}
namely as the holomorphic Hilbert subspace of $\mc{H}$, and $\Pi_{\mc{V}}^*$ produces a unitary isomorphism of this space with $L^2(\R_+\times \mc{T})$.
This will thus be convenient to represent the collection of Verma modules $\mc{V}_{Q+{\rm i}p}$ in terms of this $L^2(\R_+\times \mc{T})$ space. 
In particular, since conformal blocks will be operators acting on tensor products of Verma modules  $\mc{V}_{Q+{\rm i}p_1}\otimes\dots\otimes \mc{V}_{Q+{\rm i}p_k}$, it is convenient, for functional analytic purpose, to view the conformal blocks as families in the variables $(p_1,\dots,p_k)$, and in fact as elements 
acting on tensor products of $\mc{V}=\mc{H}^{1,0}\simeq L^2(\R_+\times \mc{T})$.

We now define  the positive function (viewed as a weight) on $\R_+\times \mc{T}^2$
\begin{equation}\label{def_h_0}
h_0(p,\nu,\tilde{\nu}):=(2\Delta_{Q+{\rm i}p}+|\nu|+|\tilde{\nu}|).
\end{equation} 
The domain of $\bH$ is 
\[ \mc{D}(\bH)=\{ u\in \mc{H}\,|\, h_0 \Pi_{\mc{V}\otimes \bbar{\mc{V}}}u \in L^2(\R_+\times \mc{T}^2)\}\]
and $\Pi_{\mc{V}\otimes \bbar{\mc{V}}}\bH= h_0\Pi_{\mc{V}\otimes \bbar{\mc{V}}}$ on $\mc{D}(\bH)$. Clearly
\[  \Pi_{\mc{V}\otimes \bbar{\mc{V}}} : \mc{D}(\mc{Q}) \to  h_0^{-\frac{1}{2}}L^2(\R_+\times \mc{T}^2)\]
is  a unitary map and, by duality, it can  be extended as a unitary map 
\[ \Pi_{\mc{V}\otimes \bbar{\mc{V}}} : \mc{D}'(\mc{Q}) \to h_0^{\frac{1}{2}}L^2(\R_+\times \mc{T}^2).\]
This implies that $\Pi_{\mc{V}}:\mc{D}'(\mc{Q}) \to h_0^{\frac{1}{2}}L^2(\R_+\times \mc{T})$ is also bounded if we use the 
notational convention $h_{0}(p,\nu):=h_0(p,\nu,0)=(2\Delta_{Q+{\rm i}p}+|\nu|)$. Similarly we have for each $t\geq 0$
\[ \Pi_{\mc{V}\otimes \bbar{\mc{V}}}e^{-t\bH}= e^{-th_0}\Pi_{\mc{V}\otimes \bbar{\mc{V}}}.\]

We will also consider the subspace 
\begin{equation}\label{defQ10} 
\mc{D}(\mc{Q})^{1,0}:=\Pi_{\mc{V}\otimes \bbar{\mc{V}}}^{-1}(h_0^{-1/2}L^2_{1,0}(\R_+\times \mc{T}^2))=\mc{D}(\mc{Q})\cap \mc{H}^{1,0}
\end{equation}
which is a Hilbert space when equipped with the $\|\cdot\|_{\mc{Q}}$ norm.

\subsection{Virasoro action}
The Virasoro operators ${\bf L}_n$ and $\tilde{{\bf L}}_n$ introduced in \eqref{Virasoro_generators} act respectively on the Verma modules 
$\mc{V}_{Q+{\rm i}p}$ and $\bbar{\mc{V}}_{Q+{\rm i}p}$, as we now describe.
By \cite[Proposition 3.6]{BGKRV}, there are coefficients $\bs{\ell}_n^{Q+{\rm i}p}(\nu,\nu')$ for all $p\in \R_+$ such that 
\[{\bf L}_{n}H_{Q+{\rm i}p,\nu,0}= \sum_{\nu' \in \mc{T}}\bs{\ell}_{n}^{\, Q+{\rm i}p}(\nu',\nu)H_{Q+{\rm i}p,\nu',0}\] 
with $\bs{\ell}_{n}^{Q+{\rm i}p}(\nu',\nu)=0$ if $n>|\nu|$ or if $|\nu'|\not=|\nu|-n$. These coefficients can be computed  using the commutation relations of the Virasoro algebra, using \eqref{primary} and \eqref{def_of_H_Q+ip}. They only depend on $\Delta_{Q+{\rm i}p}$, $\nu,\nu'$ and the central charge $c_{\rm L}=1+6Q^2$. In particular, we have the same coefficients $\bs{\ell}_n^{Q+{\rm i}p}$ for $\bbar{\mc{V}}_{Q+{\rm i}p}$, i.e.\
\[\tilde{\bf L}_{n}H_{Q+{\rm i}p,0,\tilde{\nu}}= \sum_{\tilde{\nu}' \in \mc{T}}\bs{\ell}_{n}^{\, Q+{\rm i}p}(\tilde{\nu}',\tilde{\nu})H_{Q+{\rm i}p,0,\tilde{\nu}'}.\]
Using that the Shapovalov matrices $F^{1/2}_{Q+{\rm i}p}(\nu,\nu')$ are real valued and depend only on $\Delta_{Q+{\rm i}p}=\Delta_{Q-{\rm i}p}$, we deduce that 
\[ \bbar{\bs{\ell}_{n}^{\, Q+{\rm i}p}(\nu',\nu)}=\bs{\ell}_{n}^{\, Q+{\rm i}p}(\nu',\nu)=\bs{\ell}_{n}^{\, Q-{\rm i}p}(\nu',\nu).\]
Recall that ${\bf L}_n,\tilde{\bf L}_n:\mc{D}(\mc{Q})\to \mc{D}'(\mc{Q})$, for $n\in \Z$, are bounded (they are not bounded on $\mc{H}$) and 
${\bf L}_n^*={\bf L}_{-n}$ in the sense $\cjg {\bf L}_nu,u'\cjd_{\mc{H}}=\cjg u,{\bf L}_{-n}u'\cjd_{\mc{H}}$ for  $u,u'\in \mc{D}(\mc{Q})$. Therefore the following holds in $h_0^{\frac{1}{2}}L^2(\R_+\times \mc{T}^2)$: for each $u\in e^{\eps \rho}\mc{D}(\mc{Q})$ for $\eps>0$,
\begin{equation}\label{Lnln}
\begin{split} 
(\Pi_{\mc{V}\otimes \bbar{\mc{V}}} {\bf L}_n u)(p,\nu,\tilde{\nu})=&  \frac{1}{\sqrt{2\pi}}\cjg {\bf L}_nu, H_{Q+{\rm i}p,\nu,\tilde{\nu}}\cjd_{\mc{H}}= \frac{1}{\sqrt{2\pi}}\cjg u, {\bf L}_{-n}H_{Q+{\rm i}p,\nu,\tilde{\nu}}\cjd_{\mc{H}}\\
=& \sum_{\nu' \in \mc{T}}\bbar{\bs{\ell}_{-n}^{\, Q+{\rm i}p}(\nu',\nu)}(\Pi_{\mc{V}\otimes \bbar{\mc{V}}}u)(p,\nu',\tilde{\nu}),
\end{split}
\end{equation}
and similarly $(\Pi_{\mc{V}\otimes \bbar{\mc{V}}} \tilde{{\bf L}}_n u)(p,\nu,\tilde{\nu})=
\sum_{\tilde{\nu}' \in \mc{T}}\bbar{\bs{\ell}_{-n}^{\, Q+{\rm i}p}(\tilde{\nu}',\tilde{\nu})}(\Pi_{\mc{V}\otimes \bbar{\mc{V}}}u)(p,\nu,\tilde{\nu}')$.
The matrix
\begin{equation}\label{matrixln}
\forall \nu\in \mc{T}_{m-n}, \, \nu'\in\mc{T}_m , \quad [\bs{\ell}_{n}^{Q+{\rm i}p}]_{\nu,\nu'}:=\bbar{\bs{\ell}_{-n}^{Q+{\rm i}p}(\nu',\nu)} =\bs{\ell}_{-n}^{Q+{\rm i}p}(\nu',\nu)
\end{equation} 
defines for each $m\geq 0$ a family of bounded operators (matrices) for all $p>0$ 
\[ \bs{\ell}_n^{Q+{\rm i}p}: L^2(\mc{T}_m)=\C^{d_m}\to  L^2(\mc{T}_{m-n})=\C^{d_{m-n}}.\]
The formula \eqref{Lnln}  tells us that $\bs{\ell}^{Q+{\rm i}p}_n$ is the matrix representation of ${\bf L}_n$ in the basis $H_{Q+{\rm i}p,\nu,\tilde{\nu}}$.
Although the definition of $ [\bs{\ell}_{n}^{Q+{\rm i}p}]_{\nu,\nu'}$ may look counter-intuitive, we shall see in Lemma \ref{smallell_n} that $[\bs{\ell}_{n}^{Q+{\rm i}p}]_{\nu,\nu'}=\bs{\ell}_n^{Q+{\rm i}p}(\nu,\nu')$
by using that ${\bf L}_n^*={\bf L}_{-n}$ on $\mc{D}(\mc{Q})$, but, at this stage, we need to use \eqref{matrixln}.
We finally define the linear operator $\bs{\ell}_n: C_c^\infty(\R_+)\otimes \bigoplus_{m=0}^\infty \C^{d_m}\to C_c^\infty(\R_+)\otimes \bigoplus_{m=0}^\infty \C^{d_m}$ by (here $v$ is a finite sum) 
\[ \bs{\ell}_n( f(p) \otimes v):=f(p)\otimes (\bs{\ell}_{n}^{Q+{\rm i}p}v).\]

\begin{lemma}\label{smallell_n}
The operator $\bs{\ell}_n$ extends to a bounded operator  
\[ \bs{\ell}_n: h_{0}^{-\frac{1}{2}}L^2(\R_+\times \mc{T})\to h_{0}^{\frac{1}{2}}L^2(\R_+\times \mc{T})\] 
where we use the notational convention $h_{0}(p,\nu):=h_0(p,\nu,0)=(2\Delta_{Q+{\rm i}p}+|\nu|)$:  for all $u\in L^2(\R_+\times \mc{T})$
\begin{equation}\label{boundednessell_n} 
 \|h_{0}^{-\frac{1}{2}}\bs{\ell}_nh_{0}^{-\frac{1}{2}}u\|_{L^2(\R_+\times \mc{T})}\leq \|{\bf L}_n\|_{\mc{D}(\mc{Q})\to \mc{D}'(\mc{Q})} \|u\|_{L^2(\R_+\times \mc{T})}.
 \end{equation}
The following formula holds true on $\mc{D}(\mc{Q})$
\begin{equation}\label{LnPiV} 
\Pi_{\mc{V}}\bL_n =\bs{\ell}_n \Pi_{\mc{V}}.
\end{equation}
Moreover, we have
$\bs{\ell}_n^*=\bs{\ell}_{-n}=\bs{\ell}_n^\top$ in the sense that for all $u_1,u_2\in h_{0}^{-\frac{1}{2}}L^2(\R_+\times \mc{T})$, 
\[ \cjg \bs{\ell}_nu_1,u_2\cjd_{L^2(\R_+\times \mc{T})}=\cjg u_1,\bs{\ell}_{-n}u_2\cjd_{L^2(\R_+\times \mc{T})}=\cjg u_1,\bbar{\bs{\ell}_{-n}\bbar{u_2}}\cjd_{L^2(\R_+\times \mc{T})}.\]
\end{lemma}
\begin{proof}To prove this bound, it suffices to take $f\in \mc{H}$ so that $\Pi_{\mc{V}\otimes \bbar{\mc{V}}}f(p,\nu,\tilde{\nu})=u(p,\nu)\delta_{0}(\tilde{\nu})$ with $\delta_0$ the characteristic function of $\{0\}\subset \mc{T}$,  
and $\|f\|_{\mc{H}}=\|\delta_{0}(\tilde{\nu})u\|_{L^2(\R_+\times \mc{T}^2)}=\|u\|_{L^2(\R_+\times \mc{T})}$ by Lemma \ref{surjectivity}, and write 
\[\begin{split}
\|h_0^{-\frac{1}{2}}\bs{\ell}_nh_0^{-\frac{1}{2}}u\|_{L^2(\R_+\times \mc{T})}& =\|\delta_0(\tilde{\nu})\Pi_{\mc{V}\otimes \bbar{\mc{V}}}\bH^{-1/2} {\bf L}_n {\bH}^{-1/2}f \|_{L^2(\R_+\times \mc{T}^2)}\\
& \leq \|\bH^{-1/2} {\bf L}_n {\bH}^{-1/2}f\|_{\mc{H}}\\
& \leq  \|{\bf L}_n\|_{\mc{D}(\mc{Q})\to \mc{D}'(\mc{Q})}\|f\|_{\mc{H}}\leq \|{\bf L}_n\|_{\mc{D}(\mc{Q})\to \mc{D}'(\mc{Q})}\|u\|_{L^2(\R_+\times \mc{T})}
\end{split}\] 
and using that $\bH^{-1/2}:\mc{D}'(\mc{Q})\to \mc{H}$ and $\bH^{-1/2}:\mc{H}\to \mc{D}(\mc{Q})$ are unitary maps.
Finally, if $u_j\in h_0^{-1/2}L^2(\R_+\times \mc{T})$ for $j=1,2$, using Lemma \ref{surjectivity} there are $f_j\in \mc{D}(\mc{Q})$ so that  
$(\Pi_{\mc{V}\otimes \bbar{\mc{V}}}f_j)(p,\nu,\tilde{\nu})=\delta_0(\tilde{\nu})u_j(p,\nu)$ and $\|f_j\|_{\mc{H}}=\|u_j\|_{L^2(\R_+\times \mc{T})}$, thus 
\[\begin{split}
\cjg \bs{\ell}_nu_1,u_2\cjd_{L^2(\R_+\times \mc{T})}= & \cjg \delta_{0}(\tilde{\nu})\bs{\ell}_nu_1,\delta_{0}(\tilde{\nu})u_2\cjd_{L^2(\R_+\times \mc{T}^2)}
= \cjg {\bf L}_nf_1,f_2\cjd_{\mc{H}}\\
=& \cjg f_1,{\bf L}_{-n}f_2\cjd_{\mc{H}}=\cjg \delta_{0}(\tilde{\nu})u_1,\delta_{0}(\tilde{\nu})\bs{\ell}_{-n}u_2\cjd_{L^2(\R_+\times \mc{T}^2)}\\
=& \cjg u_1,\bs{\ell}_{-n}u_2\cjd_{L^2(\R_+\times \mc{T})}
\end{split}\]
where we used \eqref{Lnln} in the second and fourth identities and the fact that ${\bf L}_n^*={\bf L}_{-n}$ on $\mc{D}(\mc{Q})$ in the third identity.
This ends the proof.
\end{proof}
Notice that this result also implies that $\bs{\ell}_n^{Q+{\rm i}p}(\nu,\nu')=\bbar{\bs{\ell}_{-n}^{Q+{\rm i}p}(\nu',\nu)}$. The operators $\bs{\ell}_n$ also act naturally in two ways as maps 
$h_{0}^{-\frac{1}{2}}L^2(\R_+\times \mc{T}^2)\to h_{0}^{\frac{1}{2}}L^2(\R_+\times \mc{T}^2)$ by extending trivially the action on $\R_+\times \mc{T}$ to $\R_+\times \mc{T}^2$ by 
\[ (\bs{\ell}_n F)(p,\nu,\tilde{\nu}):=\sum_{\nu'} \bs{\ell}_n^{Q+{\rm i}p}(\nu,\nu')F(p,\nu',\tilde{\nu}),\quad (\tilde{\ell}_n F)(p,\nu,\tilde{\nu}):=\sum_{\tilde{\nu}'} \bs{\ell}_n^{Q+{\rm i}p}(\tilde{\nu},\tilde{\nu}')F(p,\nu,\tilde{\nu}')\]
so that 
\[\Pi_{\mc{V}\otimes \bbar{\mc{V}}} {\bf L}_n=\bs{\ell}_n \Pi_{\mc{V}\otimes \bbar{\mc{V}}},\qquad \Pi_{\mc{V}\otimes \bbar{\mc{V}}} \tilde{{\bf L}}_n=
\tilde{\ell}_n \Pi_{\mc{V}\otimes \bbar{\mc{V}}}.\]
Using this discussion, we can show 
\begin{lemma}\label{DQ10}
The Hilbert space $\mc{D}(\mc{Q})^{1,0}$ defined by \eqref{defQ10} is also equal to 
\[ \mc{D}(\mc{Q})^{1,0}=\{ u\in \mc{D}(\mc{Q})\,|\, \forall n>0, \, \tilde{\bf L}_{n}u=0\}.\]
\end{lemma}
\begin{proof}
First, let us show that $u\in \mc{D}(\mc{Q})^{1,0}$ is killed by $\tilde{\bf L}_n$ for all $n>0$. Indeed, 
since $\Pi_{\mc{V}\otimes \bbar{\mc{V}}}\tilde {\bf L}_nu=\tilde{\bs{\ell}}_n\Pi_{\mc{V}\otimes \bbar{\mc{V}}}u$, 
and since $(\Pi_{\mc{V}\otimes \bbar{\mc{V}}}u)(p,\nu,\tilde{\nu})=f(p,\nu)\delta_0(\tilde{\nu})$ for some $f$,  it suffices to use that $\bs{\ell}_n^{Q+{\rm i}p}\delta_{0}=0$  if $n>0$. Conversely, let $u\in \mc{D}(\mc{Q})$ such that $\tilde{\bf L}_{n}u=0$ for all $n>0$, and let $F:=\Pi_{\mc{V}\otimes\bar{\mc{V}}}u$. We decompose $F=\sum_{m=0}^\infty F_m$ where 
$F_m(p,\nu,\tilde{\nu}):=1_{|\tilde{\nu}|=m}F(p,\nu,\tilde{\nu})$ 
are level $m$ descendants. Since $\tilde{\bs{\ell}}_nF=0$ for all $n>0$ and $\tilde{\bs{\ell}}_n$ decreases the level by $n$, then $\tilde{\bs{\ell}}_nF_m=0$ for all $1\leq n\leq m$. Fix $m\geq 1$.
Using that $\tilde{\bs{\ell}}_n^*=\tilde{\bs{\ell}}_{-n}$ for all $n\in \N$, we obtain for almost all $(p,\nu)$ and all $(n_1,\dots,n_k)\in (\N^*)^k$ 
with $n_1+\dots+n_k=m$,
\[
\left\langle F_m(p,\nu,\cdot),
\tilde{\bs{\ell}}_{-n_1}^{Q+{\rm i}p}\cdots
\tilde{\bs{\ell}}_{-n_k}^{Q+{\rm i}p}\delta_0\right\rangle
=
\left\langle \tilde{\bs{\ell}}_{n_1}^{Q+{\rm i}p}F_m(p,\nu,\cdot),
\tilde{\bs{\ell}}_{-n_2}^{Q+{\rm i}p}\cdots
\tilde{\bs{\ell}}_{-n_k}^{Q+{\rm i}p}\delta_0\right\rangle
=0.
\]
Thus $F_m(p,\nu,\cdot)$ is orthogonal to the entire level $m$ space,
to which it belongs, and hence $F_m(p,\nu,\cdot)=0$ for all $m\geq 1$ and almost all $(p,\nu)$. 
This implies that $F=F_0$ and $u\in \mc{D}(\mc{Q})^{1,0}$.
\end{proof}

\subsection{Quantization of holomorphic vector fields}
Consider a holomorphic vector field ${\rm v}=v(z)\pl_z$ near $\T$, with $v(z)=-\sum_{n\in \Z}v_n z^{n+1}\in {\rm Hol}_{\eps}(\A)$ for some $\eps>0$. By \cite[Lemma 4.18]{BGKR1} (see the proof of \cite[Corollary 3.4]{BGKRV} for the extension to the weighted space $e^{-\eps \rho}\mc{D}(\mc{Q})$) the operators 
\begin{equation}\label{def_Hv}
\bH_{{\rm v}}={\bf L}_{{\rm v}} +\tilde {\bf L}_{{\rm v}} , \quad {\bf L}_{{\rm v}}:= \sum_{n\in \Z} v_n\bL_n, \quad 
\tilde\bL_{{\rm v}}=\sum_{n\in \Z} \bbar{v}_n\, \tilde\bL_n
\end{equation}
are bounded as map $e^{-\eps \rho}\mc{D}(\mc{Q})\to e^{-\eps \rho}\mc{D}'(\mc{Q})$ for each $\eps\geq 0$.
 The operator ${\bf H}_{{\rm v}}$ acts by 
 \[ {\bf H}_{{\rm v}}H_{Q+{\rm i}p,\nu,\tilde{\nu}}= \sum_{n}v_n \sum_{\nu'\in\mc{T},
 |\nu'|=|\nu|-n}\bs{\ell}_{n}^{Q+{\rm i}p}(\nu',\nu)H_{Q+{\rm i}p,\nu',\tilde{\nu}}+\sum_{n}\bbar{v}_n 
 \sum_{\tilde{\nu}'\in\mc{T},
 |\tilde{\nu}'|=|\tilde{\nu}|-n}\bs{\ell}_{n}^{Q+{\rm i}p}(\tilde{\nu}',\tilde{\nu})H_{Q+{\rm i}p,\nu,\tilde{\nu}'}.\]
as an element in $e^{-\eps \rho}\mc{D}'(\mc{Q})$ for each $\eps>0$.
Thus we  have for $u\in \mc{D}(\mc{Q})$
 \begin{equation}\label{def:ell_bv}
\begin{gathered}
 (\Pi_{\mc{V}\otimes \bbar{\mc{V}}} \bH_{{\rm v}} u)(p,\nu,\tilde{\nu})=(\bs{\ell}_{{\rm v}}+\tilde{\bs{\ell}}_{{\rm v}})(\Pi_{\mc{V}\otimes \bbar{\mc{V}}} u)(p,\nu,\tilde{\nu}), \quad \textrm{ with }\\
  \bs{\ell}_{{\rm v}}:h_{0}^{-\frac{1}{2}}L^2(\R_+\times \mc{T}^2)\to h_{0}^{\frac{1}{2}}L^2(\R_+\times \mc{T}^2) , \quad \bs{\ell}_{{\rm v}}=\sum_{n}v_n\bs{\ell}_n, \\
\tilde{\bs{\ell}}_{{\rm v}}:h_{0}^{-\frac{1}{2}}L^2(\R_+\times \mc{T}^2)\to h_{0}^{\frac{1}{2}}L^2(\R_+\times \mc{T}^2) , \quad \tilde{\bs{\ell}}_{{\rm v}}=\sum_{n}\bbar{v}_n\tilde{\bs{\ell}}_n.
\end{gathered}
\end{equation} 
If $v(z)=-\sum_{n\geq 0}v_nz^{n+1}$ only has positive powers of $z$, for each $p>0$ we define the following operator: for all $m\geq 0$
\begin{equation}\label{lvfixedp} 
\bs{\ell}_{{\rm v}}^{Q+{\rm i}p}=\sum_{n\geq 0}v_n\bs{\ell}_n^{Q+{\rm i}p}: L^2(\mc{T}_m)\to L^2(\bigcup\nolimits_{n\leq m}\!\mc{T}_n), \quad \tilde{\bs{\ell}}_{{\rm v}}^{ Q+{\rm i}p}=\sum_{n\geq 0}\bbar{v}_n\tilde{\ell}_n^{\, Q+{\rm i}p}: L^2(\mc{T}_m)\to L^2(\bigcup\nolimits_{n\leq m}\!\mc{T}_n).
\end{equation}
In that case, we can view $\bs{\ell}_{{\rm v}}^{Q+{\rm i}p},\tilde{\bs{\ell}}_{{\rm v}}^{\, Q+{\rm i}p}$ as well-defined  operators 
\[\bs{\ell}_{{\rm v}}^{Q+{\rm i}p}, \tilde{\bs{\ell}}_{{\rm v}}^{\, Q+{\rm i}p}: C_{c}(\mc{T})\to C_{c}(\mc{T}),\]
where $C_{c}(\mc{T})$ is the set of functions on $\mc{T}$ with finite support. If $u\in h_0^{-1/2}L^2(\R_+\times\mc{T})$, for ${\rm v}=-\sum_n v_nz^{n+1}\pl_z$ as above, we have $\bs{\ell}_{\rm v}u\in h_0^{1/2}L^2(\R_+\times\mc{T})$ where $\bs{\ell}_{{\rm v}}=\sum_n v_n\bs{\ell}_n$ 
acts by 
\begin{equation}\label{ellvQ+ip} 
(\bs{\ell}_{{\rm v}}u)(p,\nu)=(\bs{\ell}_{{\rm v}}^{Q+{\rm i}p}u(p,\cdot))(\nu).
\end{equation}
We shall often consider $(\bs{\ell}_{{\rm v}}u)(p,\cdot)$ as a family of functions in $(1+|\nu|)^{1/2}L^2(\mc{T})$ depending on $p$ and defined for almost every $p$. 

\section{Liouville amplitudes of complex building blocks and their variations}

In this section, we recall the notion of Liouville amplitudes associated with Riemann surfaces with marked points and parametrised boundary 
equipped with admissible metrics, following our previous works \cite{GKRV21_Segal,BGKR1}. 
These probabilistic constructions fit into the geometric axioms for conformal field theory introduced by Segal \cite{Segal87}.
The probabilistic construction of the amplitudes will be used here as a black-box. Accordingly, we only recall the minimal set of properties required for the study of conformal blocks. We refer to \cite{GKRV21_Segal} for a detailed presentation of the material employed in this section. In this section, since the genus of surfaces does not play an important role, we use $g$ to denote Riemannian metrics.

\subsection{Liouville amplitudes}

We consider a Riemann surface $S=(\Sigma,J,{\bf x},\bs{\zeta})$ with marked points and boundary parametrisation, with $g$ an admissible metric on $S$. The scalar curvature will be denoted by $K_{g}$.
To define the Liouville amplitude, we first introduce the parameters 
\begin{equation}\label{Liouville_parameters}
\gamma \in (0,2), \quad Q=\frac{2}{\gamma}+\frac{\gamma}{2}, \quad \mu>0, \quad c_{\rm L}=1+6Q^2.
\end{equation}
We recall from \cite[Sections 2.1 \& 2.2]{GKRV21_Segal} the following facts:
\begin{itemize}
\item  There is a probability space $(\Omega,\mc{F},\P)$ and a (distributional valued) random variable $X_{S,{\rm D}}\in H^{-s}(\Sigma)$ (with $s\in (0,1/2)$ any fixed number), called \textit{Gaussian Free Field}, with vanishing expectation and covariance kernel:  
\[ \E[ \cjg X_{S,{\rm D}},f_1\cjd \cjg X_{S,{\rm D}},f_2\cjd]=  \int_{\Sigma\times \Sigma}
G_{g,{\rm D}}(x,x')f_1(x)f_2(x'){\rm dv}_{g}(x){\rm dv}_{g}(x')\]
for any $f_1,f_2\in C_c^\infty(\Sigma^\circ)$, where $G_{g,{\rm D}}(x,x')=G_{S,{\rm D}}(x,x')$ is the Green function inverting $(2\pi)^{-1}\Delta_{g,{\rm D}}$ on $S$ with Dirichlet condition on $\pl \Sigma$, where $\Delta_{g,{\rm D}}$ is the non-negative Riemannian Laplacian. Viewing distributions as the dual of volume densities (or $2$-forms) and since $G_{g,{\rm D}}(x,x')$ depends only on $S$ but not on $g$, $X_{S,{\rm D}}$ is independent of $g$.
In the case where $\pl \Sigma=\emptyset$, the Gaussian Free Field is the random distribution $X_{g}$ orthogonal to constants, with vanishing expectation and covariance kernel 
\[\E[\cjg X_{g},f_1\cjd \cjg X_{g},f_2\cjd]= \int_{\Sigma\times \Sigma}
G_{g}(x,x')f_1(x)f_2(x'){\rm dv}_{g}(x){\rm dv}_{g}(x')\]
where $G_{g}$ is the integral kernel of the self-adjoint operator $R_g:L^2(\Sigma,{\rm v}_{g})\to L^2(\Sigma,{\rm v}_{g})$ such that $\Delta_{g}R_{g}=2\pi({\rm Id}-\Pi_0)$, $R_g\Pi_0=\Pi_0R_g=0$, where $\Pi_0$ is the orthogonal projection on the constants in $L^2(\Sigma,{\rm v}_{g})$. We write $\phi_g=X_g+c$ for the field, with $c$ the constant mode sampled on Lebesgue measure.
\item For $\tilde{\bs{\varphi}}=(\tilde{\varphi}_1,\dots,\tilde{\varphi}_b)\in H^{-s}(\T)^b$, let $\phi_{g}=\phi_S:=X_{S,{\rm D}}+P_S\tilde{\bs{\varphi}}$, where $P_S\tilde{\bs{\varphi}}$ is the harmonic function on $S$ with boundary value (in the weak sense) $P_S\tilde{\bs{\varphi}}|_{\pl_j\Sigma}=\tilde{\varphi}_j\circ \zeta_j^{-1}$; it only depends on $S$. In the case where $\pl \Sigma=\emptyset$, we let $\Phi_{g}:=X_{g}+c$ for some constant $c\in\R$ sampled with the Lebesgue measure $\dd c$.
 There is a random variable $M_\gamma^{g}$ taking values in the space of Radon measures on $\Sigma$, called \textit{Gaussian Multiplicative Chaos}, obtained as the limit 
 \[
 \lim_{\eps\to 0}\eps^{\gamma^2/2}e^{\gamma \phi^\eps_{g}}{\rm dv}_{g}.
 \]
 Here
$\phi^\eps_{g}\in C^0(\Sigma)$ is a Gaussian random variable obtained by regularization of $\phi_{g}$ at scale $\eps>0$:  $\phi_{g}^\eps(x)$ 
is the average on the geodesic circle (in the metric $g$) of radius $\eps>0$ centered at $x$.  
\item The Dirichlet-to-Neumann (DN) operator ${\bf D}_S:C^\infty(\T)^b\to  C^\infty(\T)^b$ of $\Sigma$ is defined by 
\[ {\bf D}_S(\tilde{\varphi}_1,\dots,\tilde{\varphi}_b)=(-\pl_\nu P_S\tilde{\bs{\varphi}}|_{\pl_1\Sigma}\circ \zeta_1,\dots,-\pl_\nu P_S\tilde{\bs{\varphi}}|_{\pl_b\Sigma}\circ \zeta_b)\]
where $\pl_\nu$ is the unit interior pointing normal vector at $\pl \Sigma$. It satisfies that  
$\mathbf{D}_S-{\bf D}: C^{-\infty}(\T)^b\to C^{\infty}(\T)^b$ is smoothing (see \cite[Section 4.2]{GKRV21_Segal}), where ${\bf D}=(D_\D,\dots,D_\D)$ is the product of DN operators on the unit disk $\D$, given by 
\[D_\D \big(\sum_{n}f_ne^{{\rm i}n\theta}\big)=\sum_{n}|n|f_ne^{{\rm i}n\theta}.\]
\end{itemize}

\begin{definition}{ \bf{(Amplitudes)}}\label{def:amp}
Let $S=(\Sigma,J, {\bf x},\bs{\zeta})$ be an admissible Riemann surface, possibly without boundary, and let $g$ be an admissible metric on $S$. 
Let $\boldsymbol{\alpha}=(\alpha_1,\dots,\alpha_m)$ be weights satisfying $\alpha_k<Q$ for all $k$.\\
\noindent {\bf (A)}
If $\partial\Sigma=\emptyset$, the amplitude of $S_g=(\Sigma,g,{\bf x})$ with weights $\bs{\alpha}$, also called  \textit{correlation function},  is defined  by  
\begin{equation}\label{def_amp_closed} 
\mathcal{A}_{S_g,\bs{\alpha}}:= \big(\frac{{\rm v}_{g}(\Sigma)}{{\det}(\Delta_{g})}\big)^\hf  \lim_{\eps\to 0}\int_\R  \mathbb{E}\Big[ 
\prod_{j=1}^m\eps^{\alpha_j^2/2}e^{\alpha_j\phi_{g}^{\eps}(x_j)}e^{-\frac{Q}{4\pi}\int_{\Sigma}K_{g}\phi_{g}\,{\rm dv}_{g} - \mu M^{g}_\gamma(\Sigma)} \Big]\,\dd c.
\end{equation}
where $\det(\Delta_{g})$ is the  determinant of the Laplacian, defined by zeta-regularization after projecting off $\ker \Delta_g=\R$, and ${\rm v}_{g}(\Sigma)$ is  the volume of $\Sigma$. 
This is positive but can be infinite, depending on the values of $\bs{\alpha}$, see below.
 \vskip 3mm
 
\noindent {\bf (B)}  If  $\partial\Sigma$ has $b>0$ boundary components, the amplitude of $S_g=(\Sigma,g,{\bf x},\bs{\zeta})$ with weights $\bs{\alpha}$ is
  a function $\tilde{\bs{\varphi}}\mapsto \caA_{S_g,\bs{\alpha} }(\tilde{\bs{\varphi}})$ of  the boundary fields $\tilde{\bs{\varphi}}:=(\tilde\varphi_1,\dots,\tilde\varphi_b)\in (H^{-s}(\T))^b$  and,  for $s\in (0,1/2)$, it is  defined by 
\begin{align}\label{amplitude}
 \mc{A}_{S_g,\bs{\alpha}}(\tilde{\bs{\varphi}}) :=\lim_{\eps\to 0}\det (\Delta_{g,{\rm D}})^{-\hf}\caA^0_{S}(\tilde{\bs{\varphi}})
 \E \big[\prod_{j=1}^m \eps^{\alpha_j^2/2}e^{\alpha_j\phi_g^\eps(x_j)}e^{-\frac{Q}{4\pi}\int_\Sigma K_g\phi_g\dd {\rm v}_g -\mu M_\gamma^g (\Sigma)}\big]
\end{align}
where $\phi_g=X_{S,{\rm D}}+P_S\tilde{\bs{\varphi}}$ as above, the expectation $\E$ is over the Dirichlet GFF $X_{S,{\rm D}}$, and 
 $\caA^0_{S}(\tilde{\bs{\varphi}})$ is the free field amplitude defined by  
\begin{align}\label{amplifree}
\caA^0_{S}(\tilde{\bs{\varphi}}):=e^{-\frac{1}{2}( \tilde{\bs{\varphi}}, (\mathbf{D}_S-\mathbf{D})  \tilde{\bs{\varphi}})_2}.
\end{align}
Here $(u_1,u_2)_2=\frac{1}{2\pi}\int_0^{2\pi}u_1(\theta)\bar{u}_2(\theta)\d\theta$ denotes the normalised $L^2$ product on $\T$. 
\end{definition}

We recall the main properties of amplitudes: points 1) to 5), except 2) are proved in \cite{GKRV21_Segal} while 2) is proved in \cite{BGKR1}:
\begin{proposition}\label{p:amplitudes_properties}
Let $S=(\Sigma,J,{\bf x},\bs{\zeta})$ be an admissible Riemann surface and let $g$ be an admissible metric on $S$, and set $S_g:=(\Sigma,g,{\bf x},\bs{\zeta})$. 
Let $\bs{\alpha}=(\alpha_1,\dots,\alpha_m)$ be  weights  satisfying $\alpha_j<Q$, attached to the marked points $x_j$.\\
1) \textbf{Boundedness:}  Let $\chi(\Sigma)$ denote the Euler characteristic of $\Sigma$  and set $b$ the number of boundary connected components and $m$ the number of marked points. Then 
\begin{equation}\label{Seiberg_bd}
\sum_{j=1}^m \alpha_j-Q\chi(\Sigma)>0\Longrightarrow  \mc{A}_{S_g,\bs{\alpha}}\in L^2(H^{-s}(\T)^b,\mu_0^{\otimes b})\simeq \mc{H}^{\otimes b}.
\end{equation}
2) \textbf{Annuli:} If $S=(\Sigma,J,\bs{\zeta})$ is an admissible annulus, i.e. an admissible Riemann surface of type $(0,0,1,1)$, equipped with an admissible metric, then
$\mc{A}_{S_g}(\tilde{\varphi}_1,\tilde{\varphi}_2)$ is the integral kernel of a bounded operator 
\[\mc{A}_{S_g}: \mc{H}\to \mc{H}.\]  
3) \textbf{Diffeomorphism invariance:} If $(\Sigma',J',{\bf x}',\bs{\zeta}')$ is  equivalent to $(\Sigma,J,{\bf x},\bs{\zeta})$  via a biholomorphism $\Phi:\Sigma\to \Sigma'$ in the sense of Definition  \ref{equivblocks}, 
then  
\[ \mc{A}_{\Sigma,g,{\bf x},\bs{\zeta},\bs{\alpha}}=\mc{A}_{\Sigma',\Phi_*g,{\bf x}',\bs{\zeta}',\bs{\alpha}}.\]
4) \textbf{Weyl anomaly:} If $g'=e^{\omega}g$ with $\omega|_{\pl \Sigma}=0$, then
\begin{equation}\label{weyl_anomaly_formula_amp}
\mc{A}_{\Sigma,g',{\bf x},\bs{\zeta},\bs{\alpha}}=e^{c_{\rm L}S_{\rm L}^0(\Sigma,g,g') -\sum_{j=1}^m\Delta_{\alpha_j}\omega(x_j)}\mc{A}_{\Sigma,g,{\bf x},\bs{\zeta},\bs{\alpha}},
\end{equation}
where $c_{\rm L}$ is the \textbf{central charge} defined in \eqref{Liouville_parameters}, and $S_{\rm L}^0$ is the Liouville anomaly functional defined by 
\begin{equation}\label{SL0}
S_{\rm L}^0(\Sigma,g,g'):= \frac{1}{96\pi}\int_{\Sigma}(|\dd\omega|^2_g+2K_g\omega) {\rm dv}_g.
\end{equation}
5) \textbf{Cutting/gluing:} The amplitudes satisfy the gluing/cutting property. Suppose  that $S_g=(\Sigma,g,{\bf x},\bs{\zeta})$ is cut along a parametrised embedded analytic circle $\zeta:\T \to \mc{C}$, producing two connected surfaces
 $S^1_{g_1}=(\Sigma_1,g_1,{\bf x}_1,(\bs{\zeta}_1,\zeta))$ and $S^2_{g_2}=(\Sigma_2,g_2,{\bf x}_2,(\zeta,\bs{\zeta}_2))$ with respectively $b_1$ and $b_2$ boundary circles,
 with $\zeta$ incoming for $\Sigma^1$ and outgoing for $\Sigma^2$. Let $\bs{\alpha}$ be weights associated to ${\bf x}$ satisfying the Seiberg bounds and let $\bs{\alpha}_1$ and $\bs{\alpha}_2$ be the corresponding weights for ${\bf x}_1$ and ${\bf x}_2$.
 Then, for $\mu_0^{\otimes b}$-a.e. $\tilde{\bvarphi}=(\tilde{\bvarphi}_1,\tilde{\bvarphi}_2)\in (H^{-s}(\T))^{b_1+b_2-2}$,    
\[\cA_{S_g,\bs{\alpha}}(\tilde{\bvarphi})=C\int_{H^{-s}(\T)}\cA_{S^1_{g_1},\bs{\alpha}_1}(\tilde{\bvarphi}_1,\tilde{\varphi})\cA_{S^2_{g_2},\bs{\alpha}_2}(\tilde{\varphi},\tilde{\bvarphi}_2)\d\mu_0(\tilde{\varphi})\]
where 
$$C= \frac{1}{\sqrt{2} \pi} \text{ if }\partial\Sigma \not =\emptyset,\qquad  \text{ and }C= \sqrt{2}  \text{  if }\partial\Sigma =\emptyset.$$ 
If the curve   $\mc{C}$ does not separate $\Sigma$, cutting along $\zeta$ produces a new surface $S'_{g'}=(\Sigma',g',{\bf x},\bs{\zeta}')$. Then, for  
$\mu_0^{\otimes b}$-a.e. $\tilde{\bvarphi}\in(H^{-s}(\T))^{b}$,  
\[\cA_{S_g,\bs{\alpha}}(\tilde{\bvarphi})=C\int\cA_{S'_{g'},\bs{\alpha}}(\tilde{\bvarphi},\tilde{\varphi},\tilde{\varphi})\d\mu_0 (\tilde{\varphi})\]
with the same constant $C$ as above.
\end{proposition}
\begin{proof} 1) is proved in  \cite[Theorem 4.4]{GKRV21_Segal}, 3) and 4) in  \cite[Proposition 4.7]{GKRV21_Segal}, 5) in \cite[Proposition 5.1 \& 5.2]{GKRV21_Segal},  while 2) follows from \cite[Lemma 4.17]{BGKR1}.
\end{proof}

Just as in the case of annuli, where the amplitude can be viewed as a bounded operator on $\mc{H}$,  the amplitude of an admissible Riemann surface with $m$ marked points and $b$ parametrised boundary circles, and with an admissible metric is an element in $\mc{H}^{\otimes b}$, provided that the parameters $\bs{\alpha}$ satisfy the so-called \textbf{Seiberg bounds}
\begin{equation}\label{Seiberg_bounds} 
\alpha_j<Q , \quad \sum_{j=1}^m \alpha_j-Q\chi(\Sigma)>0.
\end{equation}

Assume that  $S=(\Sigma,J,{\bf x},\bs{\zeta})$ is an admissible Riemann surface with $m$ marked points and $b$ boundary components, such that $m-\chi(\Sigma)>0$. 
Consider some biholomorphism $\psi_\ell:\D\to \psi_\ell(\D)\subset \Sigma$ (with respect to $J$)  with disjoint images for $\ell=1,\dots,m$, such that 
 $\psi_\ell(0)=x_\ell$ is a marked point. Assume that there is an admissible 
metric $g$ on $S=(\Sigma,J,{\bf x},\bs{\zeta})$ such that $\psi_\ell^*g=|\dd z|^2$ for each $\ell$.
Let $S'=(\Sigma',J,\bs{\zeta}')$ where $\Sigma'=\Sigma\setminus \cup_{\ell=1}^m \psi_\ell(\D^\circ)$ and $\bs{\zeta}'=(\zeta_1,\dots,\zeta_b,\psi_1|_{\T},\dots,\psi_m|_\T)$. Note that $S'$ has $b'=b+m$ boundary components and $\chi(\Sigma')=\chi(\Sigma)-m$.
By \cite[Theorem 4.4]{GKRV21_Segal}, since $m-\chi(\Sigma)>0$, then $\mc{A}_{S'_g}\in \mc{H}^{\otimes b'}$, and more precisely we have the bound on the amplitude (for $\tilde{\varphi}_k=c_k+\varphi_k$ where $c_k\in \R$ is the zero mode of $\tilde{\varphi}_k$)
\begin{equation}\label{boundamp}
|\mc{A}_{S'_g}(\tilde{\varphi}_1,\dots,\tilde{\varphi}_{b'})|\leq Ce^{\frac{s_0}{b'}\rho(\sum_{k=1}^{b'}c_k)-a\sum_{j=1}^{b'}(c_j-\frac{1}{b'}\sum_{k=1}^{b'}c_k)^2+R\rho(-\sum_{k=1}^{b'}c_k)}A(\varphi_1,\dots,\varphi_{b'})
\end{equation}
for $s_0=-Q\chi(\Sigma')=Q(m-\chi(\Sigma))$, some $C>0,a>0$ and $R\gg 1$ arbitrarily large,
and where $\rho$ is defined above Lemma \ref{l:diagonalisation} and $A\in L^2(\Omega_\T^{b'},\mathbb{P}_{\T}^{\otimes b'})$.
Combining this bound with \cite[Proposition 5.1]{GKRV21_Segal} (see also \cite[Proposition 2.6]{BGKR1} since the metric is not assumed to be admissible along the cutting curves $\psi_\ell(\T)$), we have for $\bs{\alpha}$ satisfying the Seiberg bounds  
\[\mc{A}_{S_g,\bs{\alpha}}(\tilde{\varphi}_1,\dots,\tilde{\varphi}_b)=C_{m,b}\int \mc{A}_{S'_g}(\tilde{\varphi}_1,\dots,\tilde{\varphi}_b,\tilde{\varphi}_1',\dots,\tilde{\varphi}_m')\Psi_{\alpha_1}(\tilde{\varphi}'_1)\dots \Psi_{\alpha_m}(\tilde{\varphi}'_m)\dd \mu_0(\tilde{\varphi}'_1)\dots\dd\mu_0(\tilde{\varphi}'_m),\]
with $C_{m,b}=Z_{\D,g_\D}^m(\sqrt{2}\pi)^{-m}$ if $b>0$ or $C_{m,b}=Z_{\D,g_\D}^m\sqrt{2}(\sqrt{2}\pi)^{-m+1}$ if $b=0$, where $Z_{\D,g_\D}$ (with $g_\D=|dz|^2$) is the normalisation constant defined in \cite[Section 4.3]{GKRV21_Segal}.
Using the bounds \eqref{boundamp}, the holomorphy of $\alpha\mapsto \Psi_\alpha$ with values in $e^{-\delta \rho}\mc{H}$ for $|{\rm Re}(\alpha)-Q|<\delta$, and $\Psi_{\bar{\alpha}}=\bbar{\Psi}_{\alpha}$, we obtain directly the following (a related result is also contained in \cite[Proposition 11.13]{GKRV21_Segal}):
\begin{lemma}\label{extension_holomorphe_amp}
The amplitude 
\begin{equation}\label{anlayticity_wrt_alpha}
\bs{\alpha}\mapsto \mc{A}_{S_g,\bs{\alpha}}
\end{equation} 
extends holomorphically in $\{\bs{\alpha}\in \C^m\, |\, \sum_{\ell=1}^m |{\rm Re}(\alpha_\ell)-Q|<Q(m-\chi(\Sigma))\}$ 
with values in $\mc{H}^{\otimes b}$, and the amplitude satisfies $\bbar{\mc{A}_{S_g,\bs{\alpha}}}=\mc{A}_{S_g,\bbar{\bs{\alpha}}}$.
\end{lemma}
We say that $\bs{\alpha}\in \C^m$ satisfy the \textbf{generalised Seiberg bound} if 
\begin{equation}\label{generalisedSB}
\sum_{\ell=1}^m |{\rm Re}(\alpha_\ell)-Q|<Q(m-\chi(\Sigma)).
\end{equation}

If it has $b^-$ incoming boundary circles and $b^+$ outgoing boundary circles, it is convenient to interpret this amplitude as a bounded (in fact Hilbert-Schmidt) operator 
\begin{equation}\label{amplitude_as_operator}
\mc{A}_{S_g,\bs{\alpha}} :\mc{H}^{\otimes b^-}\to \mc{H}^{\otimes b^+}
\end{equation} 
defined through the relation
\begin{equation}\label{amplitude_as_operator_exp}
 \cjg \mc{A}_{S_g,\bs{\alpha}}(u_{b^++1}\otimes \dots \otimes u_{b}), \bar{u}_{1}\otimes \dots \otimes \bar{u}_{b^+}\cjd_{\mc{H}^{\otimes b^+ }}:=
\cjg \mc{A}_{S_g,\bs{\alpha}},\bbar{ u_{1}\otimes \dots \otimes u_{b}}\cjd_{\mc{H}^{\otimes b}}
\end{equation}
where, in the right-hand side, $\mc{A}_{S_g,\bs{\alpha}}$ is viewed as an $L^2((H^{-s}(\T))^b,\dd\mu_0^{\otimes b})$ function.
Here we use the convention that $\mc{H}^{\otimes 0}=\C$, so that when $b=b^-$, one has  $\mc{A}_{S_g,\bs{\alpha}} :\mc{H}^{\otimes b}\to \C$. In fact,  \cite[Theorem 4.4]{GKRV21_Segal}  implies the stronger result
$\mc{A}_{S_g,\bs{\alpha}}\in (e^{\delta \rho}\mc{H})^{\otimes b}$, for some small $\delta>0$, if the Seiberg bounds are satisfied. We can then extend \eqref{amplitude_as_operator} as a bounded map 
\begin{equation}\label{amplitude_as_operator2}
\mc{A}_{S_g,\bs{\alpha}} : (e^{-\delta \rho}\mc{H})^{\otimes b^-}\to (e^{\delta \rho}\mc{H})^{\otimes b^+}.
\end{equation} 
In what follows and to avoid multiple notations, we will write $\mc{A}_{S_g,\bs{\alpha}}$ for both the operator 
\eqref{amplitude_as_operator} and the integral kernel (in $\mc{H}^{\otimes b}$) of this operator, and it will be clear from the context 
when this is considered as an operator or an integral kernel. Notice also that the gluing/cutting formulas in 5) of
 Proposition \ref{p:amplitudes_properties} correspond to the composition of operators for a cut $\mc{C}$ such that $\Sigma\setminus \mc{C}$ 
 has two connected components, or a partial trace otherwise.
 
The mapping properties of the amplitudes are stronger, as we explain now.
\begin{lemma}\label{property_pant_amplitude}
Let $S=(\Sigma,J,{\bf x},\bs{\zeta})$ be an admissible Riemann surface with $b^+$ outgoing and $b^-$ incoming boundaries, let $g$ be an admissible metric and let $\bs{\alpha}$ be weights satisfying Seiberg bounds. 
Then, for $S_g=(\Sigma,g,{\bf x},\bs{\zeta})$, $\mc{A}_{S_g,\bs{\alpha}}$ extends as a bounded  operator
\begin{equation}\label{bound_dual}  
\mc{A}_{S_g,\bs{\alpha}}\in \mc{L}(\mc{D}'(\mc{Q})^{\otimes b^-},\mc{D}(\mc{Q})^{\otimes b^+}).
\end{equation}
with the convention that $\mc{D}'(\mc{Q})^{\otimes 0}=\mc{D}(\mc{Q})^{\otimes 0}=\C$.
\end{lemma}
\begin{proof} 
Let us use the parametrisations $\zeta_j:\T\to \pl_j\Sigma$ such that for $j=1,\dots,b^+$ (resp. $j=b^++1,\dots b$), 
they extend as holomorphic maps $\zeta_j:\A_{\delta}\to V_j$ near 
$\pl_j\Sigma$ (resp. $\zeta_j:\A_{1,\delta^{-1}}\to V_j$) for some $\delta<1$. We can  actually  assume that they   do in a slightly bigger annulus $\A_{\delta'}$ (resp. $\A_{1,{\delta'}^{-1}}$).
One can write $\Sigma=\Sigma'\# \bigcup_{j=1}^b V_j$ where $\Sigma'$ is a slightly smaller surface 
and we can assume (taking $\delta$ close enough to $1$) that the metric has the form $g=(\zeta_j)_*(|\dd z|^2/|z|^2)$. Here the parametrisation of 
$\pl_j\Sigma'$ is given by $\zeta_j':=e^{i\theta}\mapsto \zeta_j(\delta^{\sigma_j} e^{i\theta})$ provided $\pl_j\Sigma$ is outgoing ($\sigma_j=1$)/incoming($\sigma_j=-1$).
In terms of amplitudes, using the cutting/gluing formula of Proposition \ref{p:amplitudes_properties}, we have for $u_1,\dots,u_{b^-}\in \mc{H}$
\[\mc{A}_{S_g,\bs{\alpha}}(u_1\otimes\dots \otimes u_{b^-})= \left(\frac{C(\delta)}{\sqrt{2}\pi}\right)^b
(\otimes^{b^+}e^{\log(\delta){\bf H}})(\mc{A}_{\Sigma',g,{\bf x},\bs{\zeta}',\bs{\alpha}}(e^{\log(\delta){\bf H}}u_{1} \otimes \dots \otimes e^{\log(\delta){\bf H}}u_{b^-}))\]
where we used that $\mc{A}_{\mathbb{A}_{1,\delta^{-1}},|\dd z|^2/|z|^2,\bs{\zeta}_{\mathbb{A}}}$ is the integral kernel of the operator 
$C(\delta)e^{\log(\delta)\bH}$ for some constant $C(\delta)>0$ by \cite[Proposition 6.1]{GKRV21_Segal}, where $\bs{\zeta}_{\mathbb{A}}(e^{i\theta})=(\delta^{-1}e^{i\theta},e^{i\theta})$ and, by \cite[Lemma 6.5]{GKRV20_bootstrap}, $e^{-t\bH}:\mc{D}'(\mc{Q})\to \mc{D}(\mc{Q})$ is bounded if $t>0$. 
This implies the claim since $\mc{A}_{\Sigma',g,{\bf x},\bs{\zeta}',\bs{\alpha}}\in \mc{L}(\mc{H}^{\otimes b^-},\mc{H}^{\otimes b^+})$.
\end{proof}

Next, we describe how reverting the orientation of outgoing boundary components modifies the amplitudes. Define the reverting orientation operator by (see \cite[Section 6.6]{GKRV21_Segal})
\[{\bf O}: \mc{H}\to \mc{H}, \quad {\bf O}u(\varphi)=u(\varphi\circ o),\]
where we recall that $o(z)=1/z$ is the inversion. This operator is bounded on $\mc{H}$, and 
it commutes with ${\bf H}$ (\cite[Lemma 6.1]{GKRV21_Segal}). ${\bf O}: \mc{D}(\mc{Q})\to \mc{D}(\mc{Q})$ is also bounded. We extend its action to $\mc{H}^{\otimes b}$ by 
${\bf O}{\bf u}={\bf O}u_1\otimes\dots\otimes {\bf O}u_b$ if ${\bf u}=u_1\otimes \dots \otimes u_b$.
\begin{lemma}\label{changing_orientation_amp}
Let $S_g=(\Sigma,g,{\bf x},\bs{\zeta})$ and $\bs{\alpha}$ as in Lemma \ref{property_pant_amplitude}. Then the following holds  
\[ \cjg \mc{A}_{\Sigma,g,{\bf x},\bs{\zeta},\bs{\alpha}}({\bf u}_-),{\bf u}_+\cjd_{\mc{H}^{\otimes b^+}} = \mc{A}_{\Sigma,g,{\bf x},\bs{\zeta}^{\rm in},\bs{\alpha}}( {\bf O}\bar{{\bf u}}_+ \otimes {\bf u}_-)\] 
for each ${\bf u}_+\in \mc{H}^{\otimes b^+}$ and ${\bf u}_-\in \mc{H}^{\otimes b^-}$.
\end{lemma}
\begin{proof}
Using that the map $\tilde{\varphi}\mapsto \tilde{\varphi}\circ o$ preserves the measure 
$\mu_0$, we can write (for $s>0$)
\[
\begin{split}
\cjg \mc{A}_{\Sigma,g,{\bf x},\bs{\zeta},\bs{\alpha}}({\bf u}_-),{\bf u}_+\cjd_{\mc{H}^{\otimes b^+}} =& \int_{H^{-s}(\T)^b}  \mc{A}_{\Sigma,g,{\bf x},\bs{\zeta},\bs{\alpha}}(\tilde{\bs{\varphi}}){\bf u}_-(\tilde{\bs{\varphi}})\bbar{{\bf u}_+(\tilde{\bs{\varphi}})}\dd\mu_0(\tilde{\bs{\varphi}})\\
=& \int_{H^{-s}(\T)^b}  \mc{A}_{\Sigma,g,{\bf x},\bs{\zeta},\bs{\alpha}}(\tilde{\varphi}_1\circ o,\dots,\tilde{\varphi}_{b^+}\circ o,\tilde{\varphi}_{b^++1},\dots,\tilde{\varphi}_b){\bf u}_-(\tilde{\bs{\varphi}})\bbar{{\bf O}{\bf u}_+(\tilde{\bs{\varphi}})}\dd\mu_0(\tilde{\bs{\varphi}})
\end{split}\]
where $\tilde{\bs{\varphi}}:=(\tilde{\varphi}_1,\dots,\tilde{\varphi}_{b})$ and ${\bf u}_+(\tilde{\bs{\varphi}}):=\prod_{j=1}^{b^+}u_j(\tilde{\varphi}_j)$, ${\bf u}_-(\tilde{\bs{\varphi}}):=\prod_{j=b^++1}^{b}u_j(\tilde{\varphi}_j)$, provided ${\bf u}_+=u_1\otimes \dots \otimes u_{b^+}$ and 
${\bf u}_-=u_{b^++1}\otimes \dots \otimes u_b$. This implies the result 
since composing $\varphi_j$ by $o$ is equivalent to replacing the parametrisation $\zeta_j$ by $\zeta_j\circ o$ in the definition of the Liouville amplitudes in Definition \ref{def:amp}.
\end{proof}

Let us now recall the result \cite[Theorem 11.1]{GKRV21_Segal} which expresses the matrix coefficients of the Liouville amplitudes of complex building blocks on the eigenbasis\footnote{In that reference, the basis that is used is made of the eigenfunctions $\Psi_{Q+{\rm i}p,\nu,\tilde{\nu}}$, but we directly get the announced result using the relation \eqref{def_of_H_Q+ip} in terms of Shapovalov matrices.} $H_{Q+{\rm i}p,\nu,\tilde{\nu}}$ of ${\bf H}$. 
We fix a compatible metric $g_{\rm dozz}$ on the Riemann sphere $(\hat{\C},J_\C)$ given by 
\begin{equation}\label{gdozz}
g_{\rm dozz}:=\max(1,|z|)^{-4}|\dd z|^2.
\end{equation}
\begin{lemma}[\textbf{Matrix coefficients}]\label{matrix_coefficients}
Let $(\mc{P},J,{\bf x},\bzeta)$ be a complex building block of type $(m,b^+,b^-)$ with $b^-$ incoming boundaries $\zeta_{b^++1},\dots,\zeta_{b}$ and $b^+$ outgoing boundaries $\zeta_{1},\dots,\zeta_{b^+}$, and $g$ an admissible metric compatible with $J$ on $\mc{P}$. Let $\bs{\psi}=(\psi_1,\dots,\psi_b)$ be the centered model form for $(\mc{P},J,{\bf x},\bzeta^{\rm in})$ and $\bs{\sigma}=(\sigma_1,\dots,\sigma_b)$ the orientation signs of the boundary components. Let  $\bs{\alpha}=(\alpha_{b+1},\dots,\alpha_3)$ satisfy Seiberg bounds, and ${\bf p}=(p_1,\dots,p_b)\in \R^b$ and denote $\bs{\sigma}{\bf p}=(\sigma_1p_1,\dots,\sigma_bp_b)$. Let
\begin{align}
& \bs{\alpha}({\bf p})=(\alpha_1({\bf p}),\alpha_2({\bf p}),\alpha_3({\bf p})), \quad \alpha_{k}({\bf p})=Q+{\rm i}p_{k} \textrm{ for }k=1,\dots,b, \quad 
\alpha_{k}({\bf p})=\alpha_k \textrm{ for }k=b+1,\dots,3,\label{alpha(p)}
\end{align}

For  $\bs{\nu}=(\nu_1,\dots,\nu_b)\in \mc{T}^b$, $\tilde{\bs{\nu}}=(\tilde{\nu}_1,\dots,\tilde{\nu}_b)\in \mc{T}^b$, one has
\begin{align}
& \big\cjg \mc{A}_{\mc{P},g,{\bf x},\bzeta,\bs{\alpha}},\bigotimes_{k=1}^{b^+} H_{\alpha_k({\bf p}),\nu_{k},\tilde{\nu}_k}\otimes \bigotimes_{k=b^++1}^b\bbar{H_{\alpha_k({\bf p}),\nu_{k},\tilde{\nu}_k}}\big\cjd_{\mc{H}^{\otimes b}}=\label{Matrix_coef_pairofpants} \\
& C(\bs{\psi},g,\bs{\Delta}_{\bs{\alpha}({\bf p})}) C^{\rm DOZZ}(\bs{\alpha}(-\bs{\sigma}{\bf p})) w_{\bs{\psi}}(\bs{\Delta}_{\bs{\alpha}({\bf p})},\bs{\nu})\bbar{w_{\bs{\psi}}(\bs{\Delta}_{\bs{\alpha}({\bf p})},\tilde{\bs{\nu}})}\nonumber
\end{align}
where, for $\tilde{\bs{\alpha}}\in \C^3$ and  $\bs{\Delta}_{\tilde{\bs{\alpha}}}=(\Delta_{\tilde{\alpha}_1},\Delta_{\tilde{\alpha}_2},\Delta_{\tilde{\alpha}_3})$, the constants $w_{\bs{\psi}}(\bs{\Delta}_{\tilde{\bs{\alpha}}},\bs{\nu})$ are such that 
$w_{\bs{\psi}}(\bs{\Delta}_{\tilde{\bs{\alpha}}},\bs{\nu})=1$, if $\nu_k=0$ for each $k=1,\dots,b$. Furthermore,  $w_{\bs{\psi}}(\bs{\Delta}_{\tilde{\bs{\alpha}}},\bs{\nu})$ are of the form
\[
w_{\bs{\psi}}(\bs{\Delta},\bs{\nu})=
\sum_{|\nu'_j|=|\nu_j|}\Big(\prod_{j=1}^b F_{Q+ip_j}^{-1/2}(\nu_j,\nu'_j)\Big)
w^{\rm ward}_{\bs{\psi}}(\bs{\Delta},\bs{\nu}')
\]
with $w^{\rm ward}_{\bs{\psi}}(\bs{\Delta},\bs{\nu}')$ polynomials in $\bs{\Delta}_{\tilde{\bs{\alpha}}}=(\Delta_{\tilde{\alpha}_1},\Delta_{\tilde{\alpha}_2},\Delta_{\tilde{\alpha}_3})$  and the constant $C(\bs{\psi},g,\bs{\Delta}_{\tilde{\bs{\alpha}}})$ is given by 
\begin{equation}\label{constantC(P,g)} 
C(\bs{\psi},g,\bs{\Delta}_{\tilde{\bs{\alpha}}})=C_b2^{2\Delta_{\tilde{\alpha}_2}-2\Delta_{\tilde{\alpha}_1}-2\Delta_{\tilde{\alpha}_3}}\prod_{k=1}^b|\psi_k'(0)|^{2\Delta_{\tilde{\alpha}_k}} e^{c_{\rm L}S_{\rm L}^0(\hat{\C},g_{\rm dozz},\hat{g})-\sum_{k=b+1}^3\Delta_{\tilde{\alpha}_k}\omega(z^{\bullet}_k)}
\end{equation}
for some explicit universal constant $C_b>0$ depending only on $b$. Here, $\hat{g}$ is the filling metric of $g$ on $\hat{\C}$ (recall Definition \ref{LiouvillePant}): it is equal to $g$ on $\mc{P}$,   to $(\psi_k)_*(|\dd z|^2)$ on $\mc{D}_k=\psi_k(\D)$, and $\hat{g}=e^{\omega}g_{{\rm dozz}}$. The constant $C^{\rm DOZZ}(\tilde{\bs{\alpha}})$ is the DOZZ structure constant given by \eqref{DOZZformula}.
\end{lemma}

Although this is not important for our purposes, the constant $C_b$ is given by
\[C_b=\frac{(\sqrt{2}\pi)^{b-1}}{2\sqrt{2}Z_{\D,g_\D}^{b}}  \big({\frac{{\rm v}_{g_{\rm dozz}}(\hat\C)}{{\det}(\Delta_{g_{\rm dozz}})}} \big)^\hf=\frac{(\sqrt{2}\pi)^{b-1}}{2eZ_{\D,g_\D}^{b-2}}\] 
with $Z_{\D,g_\D}= e^{\frac{1}{4}}\times 2^{\frac{1}{12}}\pi^{\frac{1}{4}}e^{\frac{5}{24}+\zeta_R'(-1)}$ where $\zeta_R$ is the Riemann zeta function.

Notice in particular from \eqref{Matrix_coef_pairofpants} that, for $p_k>0$,
\begin{equation}\label{w_as_quotient}
w_{\bs{\psi}}(\bs{\Delta}_{\bs{\alpha}({\bf p})},\bs{\nu})=\frac{ \cjg \mc{A}_{\mc{P},g,{\bf x},\bs{\zeta},\bs{\alpha}},\bigotimes_{k=1}^{b^+} H_{Q+{\rm i}p_{k},\nu_{k},0}\otimes \bigotimes_{k=b^++1}^b\bbar{H_{Q+{\rm i}p_{k},\nu_{k},0}}\cjd_{\mc{H}^{\otimes b}}}{ \cjg \mc{A}_{\mc{P},g,{\bf x},\bs{\zeta},\bs{\alpha}},\bigotimes_{k=1}^{b^+} H_{Q+{\rm i}p_k}\otimes \bigotimes_{k=b^++1}^b\bbar{H_{Q+{\rm i}p_k}}\cjd_{\mc{H}^{\otimes b}}}.
\end{equation}

\subsection{Variations of amplitudes}\label{variation_incoming_amplitude}
Let $S=(\Sigma,J,{\bf x},\bs{\zeta})$ be an admissible Riemann surface with incoming boundary and $\bs{\alpha}$ satisfy the Seiberg bounds. For $g$ an admissible metric,
by \cite[Theorem 4.21]{BGKR1}, the map $\bs{f}\mapsto \mc{A}_{\Sigma^{\bs{f}},g^{\bs{f}}, 
{\bf x},\bs{\zeta}^{\bs{f}},\bs{\alpha}}$ is differentiable on 
$\mc{U}_{\eps}^b$ if $b$ is the number of boundary components, 
 where $\mc{U}_{\eps}^b$, $\Sigma^{\bs{f}}$, $\bs{\zeta}^{\bs{f}}$, $g^{\bs{f}}$ are defined in Section \ref{sec:deformations}. The differential can be written in terms of Virasoro operators. Let us recall this result and we will extend it to the case with outgoing boundaries.

To simplify the notations in what follows, let us denote for $v_j\in {\rm Hol}_\eps(\A)$ and ${\rm v}_j=v_j(z)\pl_z$,
\begin{equation}\label{notationsHvj} 
\begin{split}
&  \bH_{{\rm v}_j}^{(j)}:  \otimes^b \mc{D}(\mc{Q})\to \otimes^b \mc{D}'(\mc{Q}),  \quad   \bH_{{\rm v}_j}^{(j)}(h_1\otimes \dots \otimes h_b):=
h_1\otimes \dots\otimes  {\bH}_{{\rm v}_j} h_j\otimes \dots \otimes h_b.
\end{split}
\end{equation}

\begin{proposition}\label{th:derivative_general}
Let $S=(\Sigma,J,{\bf x},\bs{\zeta})$ be an admissible Riemann surface with $b^+$ outgoing and $b^-$ incoming boundary circles, 
and let $\bs{\alpha}$ satisfy the Seiberg bounds. Let $g$ be an admissible Riemannian metric compatible on $S$ and denote by $S_g=(\Sigma,g,{\bf x},\bs{\zeta})$ the Riemann surface equipped with the metric $g$.
We consider $\mc{U}_{\eps}^b$, $\Sigma^{\bs{f}}$, $\bs{\zeta}^{\bs{f}}$, 
as defined by \eqref{sigma^fzeta^f}, and let $\bs{f}\mapsto g^{\bs{f}}$ be a $C^1$ family of admissible metrics  
on $\Sigma^{\bs{f}}$, equal to $g$ outside a fixed small neighbourhood of the boundary, and near the marked points, and with $g^{{\rm Id}}=g$; for instance it can be  defined by \eqref{defgpsi}. Denote 
\begin{equation}\label{def_Sgf}
S_g\cdot \bs{f}:=(\Sigma^{\bs{f}},g^{\bs{f}},{\bf x},
\bs{\zeta}^{\bs{f}}).
\end{equation}
The map 
\[ \mc{A}_{S_g \cdot ,\bs{\alpha}}:  \bs{f}\mapsto \mc{A}_{S_g\cdot \bs{f},\bs{\alpha}}\in \mc{L}(\mc{D}(\mc{Q})^{\otimes b^-}, (\mc{D}(\mc{Q})^{\otimes b^+})') \]
is $C^1$ and, for all tangent vector ${\bv}=({\rm v}_1,\dots,{\rm v}_b)\in T_{\rm Id}\mc{U}_\eps^b$ 
and ${\bf u}=u_{b^++1}\otimes \dots\otimes u_b$ with $u_j\in \mc{D}(\mc{Q})$, the differential of the amplitude at $\bs{f}=({\rm Id},\dots,{\rm Id})$ is given by  
\begin{equation}\label{derivative_of_amplitude} 
\begin{split} 
D_\bv \mc{A}_{S_g\cdot ,\bs{\alpha}}({\rm Id}){\bf u}={}& 
 \Big(\sum_{j=1}^{b^+}{\bf H}_{{\rm v}_j}^{(j)}\Big)
 \mc{A}_{S_g,\bs{\alpha}}({\bf u})- \mc{A}_{S_g,\bs{\alpha}}
 \Big(\sum_{j=1}^{b^-}{\bf H}^{(j)}_{{\rm v}_{b^++j}}{\bf u}\Big)\\
 & +c_{\rm L}\Big(\sum_{j=1}^b\frac{\sigma_j{\rm Re}(v_{j0})}{12}+ D_\bv \mc{S}_{S_g\cdot}({\rm Id})\Big)\mc{A}_{S_g,\bs{\alpha}}({\bf u})
\end{split}
\end{equation}
where $\sigma_j\in \{\pm 1\}$ is the orientation sign of the j-th boundary component,  $v_j(z)=\sum_{n\in \Z}v_{jn}z^{n+1}$ and $\mc{S}$ is the Weyl anomaly functional 
\begin{equation}\label{Def_of_SX} 
\mc{S}_{S_g\cdot}: \mc{U}_{\eps}^b\to \R , \quad \mc{S}_{S_g\cdot \bs{f}}=S_{\rm L}^0(\Sigma^{\bs{f}},g,g^{\bs{f}}).
\end{equation}
The right-hand side of \eqref{derivative_of_amplitude} makes sense due to \eqref{bound_dual} and the boundedness 
of $\bH_{{\rm v}_j}:\mc{D}(\mc{Q})\to \mc{D}'(\mc{Q})$.
\end{proposition}
\begin{proof} In \cite[Theorem 4.21]{BGKR1}, the result is proved when $b^+=0$, i.e. all boundary circles are incoming. To deal with the outgoing case, we shall use the reverting orientation operator ${\bf O}$. By Lemma \ref{changing_orientation_amp}, and the fact that ${\bf O}:\mc{D}(\mc{Q})\to \mc{D}(\mc{Q})$ is bounded, we deduce the differentiability of $\mc{A}_{S_g\cdot}$ from the purely incoming case. 
Without loss of generality we will assume that $b^+=1$ and only ${\rm v}_1=v_1(z)\pl_z$ is non-zero, thus $\bv=({\rm v}_1,0,\dots,0)$. 
We have, for each ${\bf u}_+\in \mc{H}$ and ${\bf u}_-\in \mc{H}^{\otimes b^-}$,  
\begin{equation}\label{Amp_O}
 \cjg \mc{A}_{S_g\cdot \bs{f},\bs{\alpha}}({\bf u}_-),{\bf u}_+\cjd_{\mc{H}} = \mc{A}_{\Sigma^{\bs{f}},g^{\bs{f}},{\bf x},(\bs{\zeta}^{\bs{f}})^{\rm in},\bs{\alpha}}( {\bf O}\bar{{\bf u}}_+ \otimes {\bf u}_-)\end{equation} 
Let $(\hat{\Sigma},\hat{J})$ be the closed filling Riemann surface from Lemma \ref{FillingSigma} by gluing disks $\mc{D}_1,\dots,\mc{D}_b$ to $\Sigma$,  and 
let $\bs{\psi}^0=(\psi_1^0,\dots,\psi_b^0)$  be the associated filling biholomorphisms mapping $\psi_k^0:\D\to \mc{D}_k\subset \hat{\Sigma}$ for $k=2,\dots,b$ and $\psi_1^0:\D^*\to \mc{D}_1\subset \hat{\Sigma}$.

We consider a variation $f_t(z)=z+tv_1(z)$ in ${\rm BiHol}_\eps(\A)$ written under the form 
$v_1(z)=\sum_{n\in \Z} v_{1n}z^{n+1}\in {\rm Hol}_\eps(\A)$ for $t\in (-t_0,t_0)$ small, and write 
\[\psi_1^t:=\psi_1^0\circ f_t= (\psi_1^0)^{\rm in}\circ (o\circ f_t\circ o)\circ o.\]
We denote $f^*_t:=o\circ f_t\circ o$ which belongs to ${\rm BiHol}_{\eps}(\A)$, and notice that
\[ f^*_t(z)=z+tv_1^*(z)+\mc{O}(t^2), \quad v_1^*(z):=-\sum_{n\in \Z} v_{1n}z^{-n+1}.\] 
 We also let ${\rm v}_1^*:=v_1^*(z)\pl_z$.
In particular, $(\psi_1^t)^{\rm in}=\psi_1^{\rm in}\circ f^*_t $, and we can use \cite[Theorem 4.24]{BGKR1} and \eqref{Amp_O}: denoting $\bs{f}_t=(f_t,{\rm Id},\dots,{\rm Id})$, we have for ${\bf u}_+\in \mc{D}(\mc{Q})$ and ${\bf u}_-\in \mc{D}(\mc{Q})^{\otimes b^-}$
\[ \begin{split}
\pl_t \big\cjg \mc{A}_{S_g\cdot \bs{f}_t,\bs{\alpha}}({\bf u}_-),{\bf u}_+\big\cjd_{\mc{H}}|_{t=0} =& \pl_t\mc{A}_{\Sigma^{\bs{f}_t},g^{\bs{f}_t},{\bf x},(\bs{\zeta}^{\bs{f}_t})^{\rm in},\bs{\alpha}}( {\bf O}\bar{{\bf u}}_+ \otimes {\bf u}_-)|_{t=0}\\
=&-\mc{A}_{\Sigma,g,{\bf x},\bs{\zeta}^{\rm in},\bs{\alpha}}( {\bf H}_{{\rm v}^*_1}{\bf O}\bar{{\bf u}}_+ \otimes {\bf u}_-)\\
& +c_{\rm L}\big(\frac{{\rm Re}(v_{10})}{12}+D_{\bv}\mc{S}_{S_g\cdot}({\rm Id})\big)\mc{A}_{\Sigma,g,{\bf x},\bs{\zeta}^{\rm in},\bs{\alpha}}({\bf O}\bar{{\bf u}}_+ \otimes {\bf u}_-)
\end{split}\]
where ${\bf H}_{{\rm v}_1^*}$ is defined as in \eqref{def_Hv}.
Now, by \cite[Lemma 6.1]{GKRV21_Segal}, if $\bf{C}$ denotes the complex conjugation, we have ${\bf L}_n{\bf O}{\bf C}={\bf O}{\bf C}{\bf L}_n$ and $\tilde{{\bf L}}_n{\bf O}{\bf C}={\bf O}{\bf C}\tilde{{\bf L}}_n$ on $\mc{D}(\mc{Q})$, thus 
\[ {\bf H}_{{\rm v}_1^*}{\bf O}{\bf C}={\bf O}{\bf C}\big(\sum_{n\in \Z} \bbar{v}_{1n}{\bf L}_{-n}+v_{1n}\tilde{{\bf L}}_{-n}\big)=-{\bf O}{\bf C}{\bf H}_{{\rm v}_1}^*.\]
Therefore we obtain (using again Lemma \ref{changing_orientation_amp})
\[\begin{split}
\pl_t \big\cjg \mc{A}_{S_g\cdot \bs{f}_t}({\bf u}_-),{\bf u}_+\big\cjd_{\mc{H}}|_{t=0} =& \big\cjg \mc{A}_{S_g,\bs{\alpha}} ({\bf u}_-),{\bf H}_{{\rm v}_1}^*{\bf u}_+\big\cjd_{\mc{H}}+c_{\rm L}\big(\frac{{\rm Re}(v_{10})}{12}+D_\bv \mc{S}_{S_g\cdot }({\rm Id})\big)\cjg \mc{A}_{S_g,\bs{\alpha}}({\bf u}_-),{\bf u}_+\cjd_{\mc{H}}\\
=&\big\cjg {\bf H}_{{\rm v}_1}\mc{A}_{S_g,\bs{\alpha}}({\bf u}_-),{\bf u}_+\big\cjd_{\mc{H}}
 +c_{\rm L}(\frac{{\rm Re}(v_{10})}{12}+D_\bv \mc{S}_{S_g\cdot}({\rm Id})\big)\big\cjg \mc{A}_{S_g,\bs{\alpha}}({\bf u}_-),{\bf u}_+\big\cjd_{\mc{H}}
\end{split}\]
which concludes the proof.
\end{proof}

\begin{corollary}[\textbf{Ward identity for Segal amplitudes}]\label{th:Ward_amplitude}
Let $S=(\Sigma,J,{\bf x},\bs{\zeta})$ be an admissible Riemann surface with $b^+$ outgoing and $b^-$ incoming boundary circles,
and let $\bs{\alpha}$ satisfy the Seiberg bounds. Let $g$ be an admissible Riemannian metric compatible on $S$ and denote by $S_g=(\Sigma,g,{\bf x},\bs{\zeta})$ the Riemann surface equipped with the metric $g$. Let $(\hat{\Sigma},\hat{J})$ be the closed filling Riemann surface from Lemma \ref{FillingSigma} and let ${\rm v}$ be a holomorphic vector field defined in a neighbourhood of $\Sigma$ contained in $(\hat{\Sigma},\hat{J})$. Set $V:={\rm v}+\bar{{\rm v}}$ and let ${\rm div}_{g}V$ be the divergence of $V$ with respect to $g$. Then the following holds
 \[\begin{split}
0=& 
 \sum_{j=1}^{b^+}{\bf H}_{{\rm v}_j}^{(j)}
 \mc{A}_{S_g,\bs{\alpha}}- \mc{A}_{S_g,\bs{\alpha}}
 \sum_{j=1}^{b^-}{\bf H}^{(j)}_{{\rm v}_{b^++j}}+\sum_{j=1}^m (\Delta_{\alpha_j}({\rm div}_{g}V)(x_j)+V_{\bf x}^{(j)})\mc{A}_{S_g,\bs{\alpha}}\\
&  +c_{\rm L}\Big(\sum_{j=1}^b\frac{\sigma_j{\rm Re}(v_{j0})}{12}-\frac{1}{96\pi}\int_{\Sigma}2K_{g}({\rm div}_{g}V){\rm dv}_{g}\Big)\mc{A}_{S_g,\bs{\alpha}}
  \end{split}\]
 where $v_j\pl_z=\psi_j^*{\rm v}$ and $V_{\bf x}^{(j)}$ denotes the vector field $V$ acting in the j-th variable of a function of $m$ 
 variables $x_1,\dots,x_m$.
\end{corollary}
\begin{proof}
Let $f_t$ be the flow of ${\rm v}$ for small $t$, and define 
\[\Sigma^t=f_t(\Sigma), \quad \bs{\zeta}^t=(f_t\circ \zeta_1,\dots,f_t\circ \zeta_b), \quad {\bf x}^t=(f_{-t}(x_1),\dots,f_{-t}(x_m)).\] 
Choose a $C^1$ family of metrics $g^t$ conformal to $g$ such that  $f_t^{*}g^t=\exp(\omega^t)g$ with $\omega^t=0$ near $\pl \Sigma$ and $g^t=g$ outside a small neighbourhood of $\pl \Sigma$ and $\omega^0=0$. By Proposition \ref{p:amplitudes_properties}, we have 
\[ \mc{A}_{\Sigma^t,g^t,{\bf x},\bs{\zeta}^t,\bs{\alpha}}= \mc{A}_{\Sigma,f_t^*g^t,{\bf x}^t,\bs{\zeta},\bs{\alpha}}.\] 
Using that the left-hand side is differentiable, the right-hand side is as well and we get, by Proposition \ref{th:derivative_general},
\begin{align*}
 &\sum_{j=1}^{b^+}{\bf H}_{{\rm v}_j}^{(j)}
 \mc{A}_{S_g,\bs{\alpha}}- \mc{A}_{S_g,\bs{\alpha}}
 \sum_{j=b^++1}^{b}{\bf H}^{(j)}_{{\rm v}_{j}}
  +c_{\rm L}\Big(\sum_{j=1}^b\frac{\sigma_j{\rm Re}(v_{j0})}{12}+ \pl_tS_{\rm L}^0(\Sigma^t,g,g^t)|_{t=0}\Big)\mc{A}_{S_g,\bs{\alpha}}\\
&= \pl_t \big(\mc{A}_{\Sigma,f_t^*g^t,{\bf x}^t,\bs{\zeta},\bs{\alpha}}\big)|_{t=0}
\end{align*}
We can use the Weyl anomaly formula \eqref{weyl_anomaly_formula_amp} for amplitudes: we get
\[ \pl_t \big(\mc{A}_{\Sigma,f_t^*g^t,{\bf x}^t,\bs{\zeta},\bs{\alpha}}\big)|_{t=0}= \pl_t \big(\mc{A}_{\Sigma,g,{\bf x}^t,\bs{\zeta},\bs{\alpha}}\big)|_{t=0}+(c_{\rm L}\pl_t S_{\rm L}^0(\Sigma,g,f_t^*g^t)|_{t=0}- \sum_{j=1}^m \Delta_{\alpha_j}\dot{\omega}(x_j))\mc{A}_{S_g,\bs{\alpha}}.\]
where $\dot{\omega}=\pl_t\omega^t|_{t=0}$. We notice that
\[S_{\rm L}^0(\Sigma^t,g,g^t)-S_{\rm L}^0(\Sigma,g,f_t^*g^t)=S_{\rm L}^0(\Sigma,f_t^*g,f_t^*g^t)+
S_{\rm L}^0(\Sigma,f_t^*g^t,g)=S_{\rm L}^0(\Sigma,f_t^*g,g)\]
which gives us 
\[\begin{split}
 \pl_t \big(\mc{A}_{\Sigma,g,{\bf x}^t,\bs{\zeta},\bs{\alpha}}\big)|_{t=0}={}& \Big(\sum_{j=1}^{b^+}{\bf H}_{{\rm v}_j}^{(j)}\Big)
 \mc{A}_{S_g,\bs{\alpha}}- \mc{A}_{S_g,\bs{\alpha}}
 \Big(\sum_{j=b^++1}^{b}{\bf H}^{(j)}_{{\rm v}_{j}}\Big)+c_{\rm L}\pl_tS_{{\rm L}}^0(\Sigma,f_t^*g,g)\mc{A}_{S_g,\bs{\alpha}}\\
 & +c_{\rm L}\sum_{j=1}^b\frac{\sigma_j{\rm Re}(v_{j0})}{12}\mc{A}_{S_g,\bs{\alpha}}+ \sum_{j=1}^m \Delta_{\alpha_j}\dot{\omega}(x_j)\mc{A}_{S_g,\bs{\alpha}}.
\end{split}\]
Let $V=2{\rm Re}(v(z)\pl_z)$.  Now we can compute, writing $f_t^*g=\exp(a^t)g$ and $\dot{a}=\pl_{t}a^t|_{t=0}={\rm div}_{g}V$ 
\[ 96\pi \pl_tS_{\rm L}^0(\Sigma,f_t^*g,g)|_{t=0}=-\int_{\Sigma}2K_{g}\dot{a}{\rm dv}_{g}=-\int_{\Sigma}2K_{g}({\rm div}_{g}V){\rm dv}_{g}.\qedhere\]
\end{proof}

\section{Conformal block amplitudes of marked  complex building blocks}\label{definition_conformal_block_amplitudes}
 
In this section, we define the conformal block amplitudes of complex building blocks by using the Segal amplitudes. To put things into context, let us first recall the definition of conformal blocks in the language of Vertex Operator Algebras (VOA). 
We start with a short subsection, somehow quite informal, explaining our construction in the following subsections of conformal blocks on the basic components of complex surfaces that we called complex building blocks. 

\subsection{Conformal blocks in VOA and the relation with Segal amplitudes} 
Fix a closed Riemann surface $(\Sigma,J)$ with marked points ${\bf x}=(x_1,\dots,x_m)$ and local holomorphic coordinates $\psi_1,\dots,\psi_m$ such that $\psi_k:\D\to \mc{D}_k$ is biholomorphic with $\psi_k(0)=x_k$, and the disks $\mc{D}_k$ are all disjoint. To each marked point $x_k$, we attach a module  $\mc{V}_{\alpha_k}$ corresponding to a highest weight representation of the Virasoro algebra $({\bf  L}_n)_{n\in \Z}$ with central charge $c_{\rm L}$ on each $\mc{V}_{\alpha_k}$, with weight $\Delta_{\alpha_k}$. In our case, we consider $\mc{V}_\alpha$ to be the highest weight module generated by $\Psi_{\alpha}$. 
We set $\bs{\alpha}=(\alpha_1,\dots,\alpha_m)$ and $\bs{\psi}=(\psi_1,\dots,\psi_m)$. A conformal block (with weights $\bs{\alpha}$) is a continuous linear functional 
\begin{equation}\label{block_in the_VOA}
\mc{B}_{\bs{\psi},\bs{\alpha}}: \mc{V}_{\alpha_1}\otimes \dots\otimes \mc{V}_{\alpha_m}\to \C
\end{equation}
satisfying the so-called Ward identities: for each meromorphic vector field ${\rm v}$ on $\Sigma$ with poles contained in the set $\{{\bf x}\}$ of marked points, we have 
\begin{equation}\label{WardVOA}
\mc{B}_{\bs{\psi},\bs{\alpha}}({\bf L}_{{\rm v}_1}\otimes {\rm Id}\otimes \dots\otimes {\rm Id}+\dots+{\rm Id}\otimes \dots\otimes {\rm Id}\otimes {\bf L}_{{\rm v}_m})={\rm c}_{\rm L}G_{\bs{\psi}}({\rm v}_1,\dots,{\rm v}_m)\mc{B}_{\bs{\psi},\bs{\alpha}},
\end{equation} 
where ${\rm v}_k=\psi_k^*{\rm v}=-\sum_{n\in \Z}v_{kn}z^{n+1}\pl_z$, $G_{\bs{\psi}}({\rm v}_1,\dots,{\rm v}_m)\in \C$ 
is some  linear functional on the ${\rm v}_k$ depending on the local coordinates $\bs{\psi}$ (this will be described later), and 
\[ {\bf L}_{{\rm v}_k}:=\sum_{n\in \Z}v_{kn}{\bf L}_n.\]
For this definition to make sense, one needs to set up a topology on $\mc{V}_{\alpha_k}$ and consider the domain in $\mc{V}_{\alpha_k}$ on which 
${\bf L}_{{\rm v}_k}$ acts (it is in general an unbounded operator).
In the case of Liouville theory, the conformal weights we consider belong to $(-\infty,Q)\cup (Q+{\rm i}\R_+)$ and satisfy the Seiberg bounds
$\sum_{k=1}^m{\rm Re}(\alpha_k)-Q\chi(\Sigma)>0$. When $\alpha=Q+{\rm i}p\in Q+{\rm i}\R^*$, we have seen in \eqref{Verma_Hilbert} and \eqref{Hilbert_st_verma}  that $\mc{V}_{Q+{\rm i}p}$ has a Hilbert space structure isomorphic to $L^2(\mc{T})$, but  ${\bf L}_{{\rm v}_k}$ are not bounded on $\mc{V}_{Q+{\rm i}p}$ with respect to this topology. We shall rather need to act on vectors in $\mc{V}_{Q+{\rm i}p}$ of the form $\sum_{\nu \in\mc{T}}f(\nu)H_{Q+{\rm i}p,\nu,0}$ with $\nu \mapsto |\nu|^{1/2}f(\nu)\in L^2(\mc{T})$, i.e. on weighted spaces, which is related to the fact that ${\bf L}_{{\rm v}_k}$ are bounded as maps $\mc{D}(\mc{Q})\to \mc{D}'(\mc{Q})$ (recall \eqref{def_Hv}).
When $\alpha<Q$ and not in the Kac table $Q-\frac{\gamma}{2}\N^*-\frac{2}{\gamma}\N^*$, a similar description can be given.

In this section, we consider the case where the underlying closed surface is the Riemann sphere $\hat{\C}$ with $3$ marked points $x_1,x_2,x_3$. 
We can assume that these points are fixed to be $x_k=z^\bullet_k$ for $k=1,2,3$ with $(z^\bullet_1,z^\bullet_2,z^\bullet_3)=(-1,0,1)$ as before. The local coordinate systems $\psi_k:\D\to \psi_k(\D)$ satisfy $\psi_k(0)=z^\bullet_k$, and induce a complex building block 
$(\Sigma,J,\bs{\zeta})\in \mc{P}_{0,3}$ defined by
\[ \Sigma =\hat{\C}\setminus \bigcup_{k=1}^3\psi_k(\D^\circ), \quad J=J_\C, \quad\bs{\zeta}=(\psi_1|_{\T},\psi_2|_{\T},\psi_3|_{\T}),\]
which is already under model form in the sense of Lemma \ref{modelform}. If $\psi_k$ are M\"obius transformation, the term $G_{\bs{\psi}}$ vanishes in \eqref{WardVOA}. For $\bs{\alpha}=\bs{\alpha}({\bf p})=(Q+{\rm i}p_1,Q+{\rm i}p_2,Q+{\rm i}p_3)$ with ${\bf p}=(p_1,p_2,p_3)\in \R_+^3$, 
we can show by using inductively the Ward identity \eqref{WardVOA} 
(with vector fields with poles located only at one marked point) that the space of solutions of \eqref{WardVOA} is 1-dimensional, i.e. for each triplet $\bs{\nu}=(\nu_1,\nu_2,\nu_3)\in \mc{T}^3$ of Young diagrams,
\[ \mc{B}_{\bs{\psi},\bs{\alpha}({\bf p})}(H_{Q+{\rm i}p_1,\nu_1,0}\otimes H_{Q+{\rm i}p_2,\nu_2,0}\otimes H_{Q+{\rm i}p_3,\nu_3,0})=w_{\bs{\psi}}(\bs{\alpha}({\bf p}),\bs{\nu})
\mc{B}_{\bs{\psi},\bs{\alpha}({\bf p})}(\Psi_{Q+{\rm i}p_1}\otimes \Psi_{Q+{\rm i}p_2}\otimes \Psi_{Q+{\rm i}p_3}) \]
for some coefficients $w_{\bs{\psi}}(\bs{\alpha}({\bf p}),\bs{\nu})\in \C$. The main difficulty lies in the apparent intractability of establishing the boundedness of $\mc{B}_{\bs{\psi},\bs{\alpha}({\bf p})}$ on $\bigotimes_{k=1}^3 \mc{V}_{Q+{\rm i}p_k}$ based solely on the algorithm that computes the coefficients $w_{\bs{\psi}}(\bs{\alpha}({\bf p}),\bs{\nu})$. 

When considering higher genus  surfaces or the sphere with $m\geq 4$ marked points, the space of formal solutions  to the Ward identity \eqref{WardVOA} becomes infinite dimensional, and constructing such solutions appears to be highly non-trivial. The classical approach consists in composing conformal blocks associated with three-point spheres by gluing along one of their entries, using a non-degenerate quadratic form on the Verma modules.
However, this composition is well-defined only if one knows that the three-point conformal block $\mc{B}_{\bs{\psi},\bs{\alpha}({\bf p})}$  is bounded.

To address the crucial issue of boundedness  of   conformal blocks, we construct them from the Segal amplitudes 
$\mc{A}_{\Sigma,g,\bs{\zeta}}: \mc{H}^{\otimes 3}\to \C$ by restricting these amplitudes to 
$\bigotimes_{k=1}^3 \mc{V}_{\alpha_k}$, for $g$ an admissible metric. In practice, since $\mc{V}_{\alpha_k}$ is not strictly a subspace of $\mc{H}$ (but rather of 
weighted spaces $e^{-\eps \rho}\mc{H}$ for $\eps>0$),  we instead consider the restriction of $\mc{A}_{\Sigma,g,\bs{\zeta}}$ to the (tensor product of) holomorphic Hilbert space $\mc{V}=\mc{H}^{1,0}$ introduced in \eqref{VQ+ip}. This space can be identified with the direct integral $\int_\oplus \mc{V}_{Q+{\rm i}p}\dd p$ of Verma modules. Via the map $\Pi_{\mc{V}}^*:L^2(\R_+\times \mc{T})\to \mc{V}$ described in Lemma \ref{PiV^*PiV}, the restriction 
$\mc{A}_{\Sigma,g,\bs{\zeta}}|_{\mc{V}^{\otimes 3}}$ can be characterized by   the bounded operator
\[ \mc{A}_{\Sigma,g,\bs{\zeta}}\circ \bigotimes^3 \Pi_{\mc{V}}^* : L^2(\R_+\times \mc{T})^{\otimes 3}\to \C.\]
Applied to a vector of the form $F=f(p_1,p_2,p_3)\delta_{\nu_1}\delta_{\nu_2}\delta_{\nu_3}$, where $f\in L^2(\R_+^3)$ is
compactly supported and $\delta_{\nu}$ denotes the characteristic function of $\{\nu\}$ on $\mc{T}$, this operator yields 
\[  \mc{A}_{\Sigma,g,\bs{\zeta}}(\bigotimes^3 \Pi_{\mc{V}}^*F)=\frac{1}{(2\pi)^{3/2}}\int_{\R_+^3}f(p_1,p_2,p_3)\mc{A}_{\Sigma,g,{\bf x},\bs{\zeta},\bs{\alpha}}(\bigotimes_{k=1}^3 H_{Q+{\rm i}p_k,\nu_k,0})\dd p_1\dd p_2\dd p_3.\]
As a consequence of Lemma \ref{matrix_coefficients}, this expression can be rewritten as
\[\mc{A}_{\Sigma,g,\bs{\zeta}}(\bigotimes^3 \Pi_{\mc{V}}^*F)=\frac{1}{(2\pi)^{3/2}}\int_{\R_+^3}f({\bf p}) w_{\bs{\psi}}(\bs{\alpha}({\bf p}),\bs{\nu}) \mc{A}_{\Sigma,g,\bs{\zeta}}(\bigotimes_{k=1}^3 \Psi_{Q+{\rm i}p_k})\dd {\bf p}.
\]
In particular, this representation allows us to extract the conformal block coefficient $w_{\bs{\psi}}(\bs{\alpha}({\bf p}),\bs{\nu})$, whose growth as $|\nu_k|\to \infty$ governs the boundedness of the conformal block. 

It remains to determine the normalization term 
\[\Gamma({\bf p},\bs{\psi}):=\mc{B}_{\bs{\psi},\bs{\alpha}({\bf p})}(\Psi_{Q+{\rm i}p_1}\otimes \Psi_{Q+{\rm i}p_2}\otimes \Psi_{Q+{\rm i}p_3})\]
as a function of the holomorphic coordinates maps $\bs{\psi}=(\psi_1,\psi_2,\psi_3)$. In order to satisfy the Ward 
identity \eqref{WardVOA}, the function $\Gamma({\bf p},\bs{\psi})$ must, up to an arbitrary multiplicative constant independent of $\psi_k$, be given by 
\begin{equation}\label{Gamma_fct_first_def}
\Gamma({\bf p},\bs{\psi})= \prod_{k=1}^3 (\psi_k'(0))^{\Delta_{Q+{\rm i}p_k}}.
\end{equation}
To make sense of this expression, one must specify the argument of $\psi_k'(0)$. Geometrically, this amounts to working on the universal cover of the space of triples of biholomorphisms $\bs{\psi}$ satisfying $\psi_k(0)=z^\bullet_k$. This universal cover is precisely the Teichm\"uller space
  $\mc{T}_{0,0,3}$ of pairs of pants with boundary parametrisations, as described in Proposition 
\ref{identification_Teichmuller}. More generally, if the model form is not centered and satisfies $(\psi_1(0),\psi_2(0),\psi_3(0))=(z_1,z_2,z_3)$, 
the normalization term is
\[\Gamma({\bf p},\bs{\psi})= (\psi_1(0)-\psi_2(0))^{\Delta_{\alpha_3}-\Delta_{\alpha_2}-\Delta_{\alpha_1}}
(\psi_2(0)-\psi_3(0))^{\Delta_{\alpha_1}-\Delta_{\alpha_2}-\Delta_{\alpha_3}}(\psi_1(0)-\psi_3(0))^{\Delta_{\alpha_2}-\Delta_{\alpha_3}-\Delta_{\alpha_1}} \prod_{k=1}^3 \psi_k'(0)^{\Delta_{Q+{\rm i}p_k}},\]
which reflects the invariance by M\"obius transformations.
It is therefore natural to view the conformal blocks as a family ${\bf p}\mapsto \mc{B}_{\bs{\psi},\bs{\alpha}({\bf p})}$, 
 for ${\bf p}\in \R_+^3$, acting as operators on $L^2(\mc{T})^{\otimes 3}$, which can be identified with the space $\bigotimes_{k=1}^{3}\mc{V}_{Q+{\rm i}p_k}$. Using the properties of $\Pi_{\mc{V}}^*$, one may instead define the family of conformal blocks $\mc{B}_{\bs{\psi}}$   as the operator: 
\[\mc{B}_{\bs{\psi}}f:=(2\pi)^{b/2}\mc{A}_{\Sigma,g,\bs{\zeta}}\Big(\bigotimes^3 \Pi_{\mc{V}}^* ( \Omega f)\Big), \quad \Omega({\bf p},\bs{\nu}):=\frac{\Gamma({\bf p},\bs{\psi})}{\mc{A}_{\Sigma,g,\bs{\zeta}}(\bigotimes_{k=1}^3 \Psi_{Q+{\rm i}p_k})}\]
for $f\in L^2(\R_+\times \mc{T})^{\otimes 3}\simeq L^2(\R^3_+,L^2(\mc{T})^{\otimes 3})$ such that 
$\Omega f\in L^2(\R_+^3,L^2( \mc{T})^{\otimes 3})$,  ${\bf p}=(p_1,p_2,p_3)\in \R_+^3$ and $\bs{\nu}=(\nu_1,\nu_2,\nu_3)\in \mc{T}^3$. This operator has an integral kernel $\mc{B}_{\bs{\psi}}({\bf p},\bs{\nu})=w_{\bs{\psi}}(\bs{\alpha}({\bf p}),\bs{\nu})\Gamma({\bf p},\bs{\psi})$, i.e. we have the following expression
\[ \mc{B}_{\bs{\psi}}f=\int_{\R_+^3}\sum_{\bs{\nu}\in \mc{T}^3}w_{\bs{\psi}}(\bs{\alpha}({\bf p}),\bs{\nu})\Gamma({\bf p},\bs{\psi})f({\bf p},\bs{\nu})\dd {\bf p}.\]
Fixing ${\bf p}=(p_1,p_2,p_3)\in \R_+^3$ and $\bs{\nu}=(\nu_1,\nu_2,\nu_3)$, the conformal block in the sense of \eqref{block_in the_VOA} is simply expressed in terms of this integral kernel as
\[\mc{B}_{\bs{\psi},\bs{\alpha}({\bf p})}(H_{Q+{\rm i}p_1,\nu_1,0}\otimes H_{Q+{\rm i}p_2,\nu_2,0}\otimes H_{Q+{\rm i}p_3,\nu_3,0})=\mc{B}_{\bs{\psi}}({\bf p},\bs{\nu}) \]

More generally, we shall consider conformal blocks associated with decorated marked complex building blocks of type $(m,b^+,b^-)$ (with $m+b^++b^-=3$) as functions on $\tT_{0,m,b^+,b^-}$ taking values in bounded operators from $L^2(I \times \mc{T})^{\otimes b^-}$ to $L^2(I \times \mc{T})^{\otimes b^+}$ for each compact interval $I\subset (0,\infty)$. Incoming and outgoing boundaries correspond, respectively, to viewing the conformal blocks either as elements of the dual space  $(\mc{V}_{Q+{\rm i}p})^*$ (incoming case) or as elements of $\mc{V}_{Q+{\rm i}p}$ (outgoing case). This perspective is convenient for composing conformal blocks when gluing surfaces, in analogy with the Segal axioms for Liouville amplitudes.

\subsection{Complex conformal radius.}\label{sec:complex_conformal_radius}
In this section, we define rigorously the function $\Gamma({\bf p},\bs{\psi})$ introduced above. To this end, we rely on the description of the Teichmüller space   $\mc{T}_{0,0,3}$ given in Proposition \ref{identification_Teichmuller}, and more generally on the description of 
$\tT_{0,m,b^+,b^-}$ for $m+b^++b^-=3$, corresponding to complex building blocks of other types.

Consider the configuration space
\[\mc{O} := \{ {\bf z}=(z_1,z_2,z_3)\in \C^3 \,|\, \forall i\not=j, z_i\not= z_j\}.\]
For complex weights $\bs{\alpha}=(\alpha_1,\alpha_2,\alpha_3)\in\C^3$, we define the real analytic function on $\mc{O}$ 
\begin{equation}\label{Gammaterm} 
\Gamma(\bs{\alpha},{\bf z}):= (z_1-z_2)^{\Delta_{\alpha_3}-\Delta_{\alpha_2}-\Delta_{\alpha_1}}
(z_2-z_3)^{\Delta_{\alpha_1}-\Delta_{\alpha_2}-\Delta_{\alpha_3}}(z_1-z_3)^{\Delta_{\alpha_2}-\Delta_{\alpha_3}-\Delta_{\alpha_1}} \prod_{k=1}^3e^{{\rm i}\Delta_{\alpha_k}\arg M_{\bf z}'(z_k^\bullet)},
\end{equation}
where $M_{\bf z}$ is the unique M\"obius transformation mapping ${\bf z}^\bullet=(z^\bullet_1,z^\bullet_2,z^\bullet_3)=(-1,0,1)$ to ${\bf z}=(z_1,z_2,z_3)\in\mc{O}$. Here $\log(z_i-z_j)$ and $\arg M_{\bf z}'(z_k^\bullet)$ are determined by fixing
\begin{equation}\label{Gammaterm_initial_value}
\Gamma(\bs\alpha,{\bf z}^\bullet):=e^{-\pi{\rm i}(\Delta_{\alpha_1}+\Delta_{\alpha_2}+\Delta_{\alpha_3})}2^{\Delta_{\alpha_2}-\Delta_{\alpha_3}-\Delta_{\alpha_1}}
\end{equation}
and extending continuously along a continuous family $t\in[0,1]\mapsto M_t\in {\rm PSL}_2(\C)$ of M\"obius transformations with $M_0={\rm Id}$, $M_1=M_{{\bf z}}$ such that $M_t({\bf z}^\bullet)\in\mc{O}$ for all $t$. The resulting function is a priori defined on the universal cover of $\mc{O}$, but we claim that it descends to $\mc{O}$:

\begin{lemma}\label{Gammaterm_lemma}
For $\bs{\alpha}\in\C^3$, the function $\Gamma({\bs\alpha},\cdot):\mc{O}\to\C$ is well-defined and M\"obius covariant in the following sense: for ${\bf z}^1,{\bf z}^2\in\mc{O}$, let $M_{{\bf z}^1,{\bf z}^2}$ be the unique M\"obius transformation mapping ${\bf z}^1$ to ${\bf z}^2$, then
\begin{equation}\label{Gammaterm_Moebius}
\Gamma(\bs{\alpha},{\bf z}^2)=\Gamma(\bs{\alpha},{\bf z}^1)\prod_{k=1}^3|M_{{\bf z}^1,{\bf z}^2}'(z_k^1)|^{-\Delta_{\alpha_k}}.
\end{equation}
In particular, we have for ${\bf z}\in\mc{O}$
\begin{equation}\label{Gammaterm_formula}
\Gamma(\bs{\alpha},{\bf z})=e^{-\pi{\rm i}(\Delta_{\alpha_1}+\Delta_{\alpha_2}+\Delta_{\alpha_3})}2^{\Delta_{\alpha_2}-\Delta_{\alpha_3}-\Delta_{\alpha_1}}\prod_{k=1}^3|M_{\bf z}'(z_k^\bullet)|^{-\Delta_{\alpha_k}}.
\end{equation}
\end{lemma}
\begin{proof}
Since the fundamental group of $\mc{O}$ is the pure braid group, it suffices to verify that $\Gamma(\bs{\alpha},{\bf z})$ is invariant under the change of a $z_j$ winding around another $z_k$. For notational convenience, say $j=1$, $k=2$, and consider a continuous family $t\in[0,1]\mapsto M_t\in {\rm PSL}_2(\C)$ with ${\bf z}_t=M_t({\bf z}^\bullet)=(z_{1,t},z_2,z_3)$ where $z_{1,t}=z_2+\varepsilon e^{2\pi{\rm i}t}$ for a sufficiently small $\varepsilon>0$. The monodromy of the first three terms in \eqref{Gammaterm} comes only from $(z_{1,t}-z_2)^{\Delta_{\alpha_3}-\Delta_{\alpha_2}-\Delta_{\alpha_1}}$, giving $e^{2\pi{\rm i}(\Delta_{\alpha_3}-\Delta_{\alpha_2}-\Delta_{\alpha_1})}$. To compute the monodromy of the last product, let $M^1$ be the M\"obius transformation mapping ${\bf z}'=(1,0,\infty)$ to ${\bf z}^\bullet$ and $M^2$ the M\"obius transformation mapping $(z_{1,0},z_2,z_3)$ to ${\bf z}'=(1,0,\infty)$. Clearly $t\mapsto M^2(z_{1,t})$ also winds around $0=M^2(z_2)$ exactly once. Since $M^2\circ M_t\circ M^1$ fixes $0,\infty$ and maps $1$ to $M^2(z_{1,t})$, we have $(M^2\circ M_t\circ M^1)(z)=M^2(z_{1,t})z$. It is easy to check that $(M^2\circ M_t\circ M^1)'(z_k')/M_t'(z_k^\bullet)\in\C\setminus\{0\}$ has zero winding for all $k=1,2,3$, where the derivative at $\infty$ of a M\"obius transformation $M$ fixing $\infty$ is defined by $M'(\infty):=(\iota\circ M\circ \iota)'(0)$, with $\iota:z\mapsto 1/z$. Thus the monodromy is the same for $M_t$ at ${\bf z}^\bullet$ and for $M^2\circ M_t\circ M^1$ at ${\bf z}'$. Now $\prod_{k=1}^3e^{{\rm i}\Delta_{\alpha_k}\arg (M^2\circ M_t\circ M^1)'(z_k')}=e^{{\rm i}(\Delta_{\alpha_1}+\Delta_{\alpha_2}-\Delta_{\alpha_3})\arg M^2(z_{1,t})}$, giving the monodromy $e^{2\pi{\rm i}(\Delta_{\alpha_1}+\Delta_{\alpha_2}-\Delta_{\alpha_3})}$. This cancels out the monodromy of the first three terms. Thus we have proved that $\Gamma_{\bs{\alpha}}$ is well-defined on $\mc{O}$.

The M\"obius covariance follows from the property $(M(z_1)-M(z_2))^2=(z_1-z_2)^2M'(z_1)M'(z_2)$ for any M\"obius transformation $M$ and $z_1,z_2,M(z_1),M(z_2)\in\C$. Indeed, taking $\log$ formally gives \eqref{Gammaterm_Moebius} by a simple calculation. To make this rigorous, we use the following argument. For ${\bf z}^1\in\mc{O}$, let $S_{{\bf z}^1}$ be the set of ${\bf z}^2\in\mc{O}$ such that \eqref{Gammaterm_Moebius} holds. Since both sides are continuous, this set is closed in $\mc{O}$. The previous $\log$ argument is valid for ${\bf z}^2$ close to ${\bf z}^1$, so $S_{\bf z}$ contains a neighbourhood of $\bf z$ for any ${\bf z}\in\mc{O}$. Now we observe that ${\bf z}^2\in S_{{\bf z}^1}$ implies $S_{{\bf z}^2}\subset S_{{\bf z}^1}$ thanks to the cocycle property $|M_{{\bf z}^2,{\bf z}^3}'(z_k^2)|^{-\Delta_{\alpha_k}}|M_{{\bf z}^1,{\bf z}^2}'(z_k^1)|^{-\Delta_{\alpha_k}}=|M_{{\bf z}^1,{\bf z}^3}'(z_k^1)|^{-\Delta_{\alpha_k}}$ for ${\bf z}^1,{\bf z}^2,{\bf z}^3\in\mc{O}$. Thus $S_{{\bf z}^1}$ is also open. Since $\mc{O}$ is clearly connected, we get $S_{{\bf z}^1}=\mc{O}$.
\end{proof}

Next, we give a rigorous definition of the function \eqref{Gamma_fct_first_def}, which needs to be defined on the decorated Teichm\"uller space $\tT_{0,m,b}$, in the case with marked points. Note that in the case $m=0$, $b=3$, we have $\tT_{0,0,3}=\mc{T}_{0,0,3}$ by definition. Recalling Lemma \ref{another_desc_of_Teich2} and using the map $\Lambda$ from \eqref{Lambda}, we define 
\begin{equation}
\Gamma_{\bs{\alpha}}:\R^b\times\tT_{0,m,b}\to\C, \quad ({\bf p},[F])\mapsto\Gamma(\bs{\alpha}({\bf p}),{\bf z}^\bullet)\prod_{k=1}^3 e^{\Lambda_k(F)\Delta_{\alpha_k({\bf p})}}\prod_{k=1}^b 4^{-\Delta_{\alpha_k({\bf p})}},
\label{Gamma_alpha_mb}
\end{equation}
where $\bs{\alpha}=(\alpha_{b+1},\dots,\alpha_3)\in \C^m$ is a set of weights if $b<3$ and $\bs{\alpha}({\bf p})=(\alpha_1({\bf p}),\alpha_2({\bf p}),\alpha_3({\bf p}))$ is as in \eqref{alpha(p)}. The term $4^{-\Delta_{\alpha_k({\bf p})}}$ comes from 
\eqref{normalisationF} with the scaling $z_k^\bullet+z/4$ for the model parametrisation.
 If $b=3$, we omit $\bs{\alpha}$ from the notation. In terms of the model form $\tilde{\Theta}_{m,b}:\tT_{0,m,b}\xrightarrow{\sim}\bigcup_{\eps>0}\tT_{m,b}^\eps$ (Proposition \ref{identification_Teichmuller}), $\Gamma_{\bs\alpha}$ has the form
\begin{equation}\label{Gammamodelform}
\Gamma_{\bs{\alpha}}({\bf p},\bs{\psi},\bs{\theta},\bs{\la})=
\Gamma(\bs{\alpha}({\bf p}),{\bf z}^\bullet)\prod_{j=1}^b |\psi_j'(0)|^{\Delta_{Q+{\rm i}p_j}}e^{{\rm i}\Delta_{Q+{\rm i}p_j}\theta_j}\prod_{j=1}^m e^{\Delta_{\alpha_{b+j}}\la_j}.
\end{equation}
Under this form, this function is naturally defined on a larger domain of $({\bf p},\bs{\psi},\bs{\theta},{\bf x},\bs{\la})\in \R^b \times \tBiHol_\eps(\D)^b\times \R^b\times \C^m\times \C^m$, such that $(\psi_1(0),\dots,\psi_b(0),x_1,\dots,x_m)\in \mc{O}$, $\psi_k(\D)$ are all disjoint
and $x_j\notin \cup_{k=1}^b\psi_k(\D)\cup \{\infty\}$ for each $j$: it is given by the expression
\[\Gamma_{\bs{\alpha}}({\bf p},\bs{\psi},\bs{\theta},{\bf x},\bs{\la}):=
\Gamma(\bs{\alpha}({\bf p}),{\bf z}_{\bs{\psi},{\bf x}})\prod_{j=1}^b |\psi_j'(0)|^{\Delta_{Q+{\rm i}p_j}}e^{{\rm i}\Delta_{Q+{\rm i}p_j}\theta_j}\prod_{j=1}^m e^{\Delta_{\alpha_{b+j}}\la_j}.\]
with ${\bf z}_{\bs{\psi},{\bf x}}:=(\psi_1(0),\ldots,\psi_b(0),x_1\dots,x_m)\in\mc{O}$. By Lemma \ref{Gammaterm_lemma}, it is M\"obius invariant in the sense that
\begin{equation}\label{Gamma_alpha_Moebius}
\Gamma_{\bs\alpha}({\bf p},\bs{\psi},\bs{\theta},{\bf x},\bs{\lambda})=\Gamma_{\bs\alpha}({\bf p},M\circ\bs{\psi},\bs{\theta},M({\bf x}),\bs{\lambda}+\log|M'({\bf x})|)
\end{equation}
for any M\"obius transformation $M$ such that $M\circ\bs{\psi}\in\tBiHol_\eps(\D)^b$ (i.e.\ $M(\psi_k(0))\neq\infty$ for $k=1,\ldots,b$) and $M(x_k)\neq\infty$ for $k=1,\ldots,m$, 
where $M({\bf x}):=(M(x_1),\dots, M(x_m))$ and $\log|M'({\bf x})|:= (\log|M'(x_1)| ,\dots,\log |M'(x_m)|)$. When ${\bf x}=(z^\bullet_{b+1},\dots,z^\bullet_{3})$, we simply use the notation $\Gamma_{\bs{\alpha}}({\bf p},\bs{\psi},\bs{\theta},\bs{\la})$.

We now make a useful remark concerning the function $\Gamma_{\bs{\alpha}}$ when it is considered on a simply connected neighbourhood of $(\bs{\psi}^0,\bs{\theta}^0,\bs{0})$ in $\tT^\eps_{m,b}$.
This map is $C^1$, and its derivative satisfies the following property: for   $\bv=(v_1\pl_z,\dots,v_b\pl_z) \in T_{(\bs{\psi}^0,\bs{\theta}^0)}\mc{T}^\eps_{m,b}$ (recall the convention \eqref{Tangent_space_Teich}) and $\bs{\la}^0=(\la_{1}^0,\dots,\la_m^0)\in \C^m$, ${\bf t}=(t_1,\dots,t_m)\in \C^m\simeq T_{\bs{\lambda}^0}\C^m$,
\begin{equation}\label{dlogGamma} 
\frac{D_{\bv,{\bf t}}\Gamma_{\bs{\alpha}}({\bf p},\bs{\psi}^0,\bs{\theta}^0,\bs{\la}^0)}{\Gamma_{\bs{\alpha}}({\bf p},\bs{\psi}^0,\bs{\theta}^0,\bs{\la}^0)}=
\sum_{j=1}^b \Delta_{Q+{\rm i}p_j} v'_j(0)+\sum_{j=1}^m \Delta_{\alpha_{b+j}} t_j.
\end{equation}

\subsection{Definition of conformal block amplitudes for complex building blocks.}
We start by defining the conformal block, which we call conformal block amplitude, of an element $\tilde{\mc{X}}\in \tT_{0,m,b}$ ($m+b=3$) in the Teichm\"uller space of complex building blocks with incoming boundaries. By Proposition \ref{identification_Teichmuller}, $\tilde{\mc{X}}$ can be represented as an element $\tilde{\Theta}_{m,b}(\tilde{\mc{X}})=(\bs{\psi},\bs{\theta},\bs{\la})\in \tT^\eps_{m,b}$ for some $\eps>0$ using the model form map $\tilde{\Theta}_{m,b}$.

\begin{definition}[\textbf{Conformal block amplitude of marked complex building block with incoming boundaries}] \label{defblockamp_pant}
 Let $\tilde{\mc{X}}\in \tT_{0,m,b}$ and  $\tilde{\Theta}_{m,b}(\tilde{\mc{X}})=(\bs{\psi},\bs{\theta},\bs{\la})\in \tT^\eps_{m,b}$ for some $\eps>0$, and let $g$ be an admissible metric on $\mc{P}=\hat{\C}\setminus \cup_{k=1}^b\psi_k(\D^\circ)$ with boundary parametrisation $\bs{\zeta}=(
 \psi_1|_{\T},\dots,\psi_b|_{\T})$ and marked points ${\bf x}^\bullet=(z^\bullet_{b+1},\dots,z^{\bullet}_{3})$ if $m\geq 1$, in which case we also let $\bs{\alpha}=(\alpha_{b+1},\dots,\alpha_3)\in \C^m$ be weights satisfying the generalised Seiberg bound \eqref{generalisedSB}, and such that 
\begin{equation}\label{cdozznot=0}
C^{\rm DOZZ}(\bs{\alpha}({\bf p}))\neq 0
\quad\text{for every }{\bf p}\in(0,\infty)^b
\end{equation}
where $C^{\rm DOZZ}(\cdot)$ denotes the DOZZ constant \eqref{DOZZformula} and $\bs{\alpha}({\bf p})$ is defined by \eqref{alpha(p)}.
 The \textbf{conformal block amplitude} of $\tilde{\mc{X}}$, or alternatively of $(\bs{\psi},\bs{\theta},\bs{\la})$, denoted both by 
 $\mc{B}_{\tilde{\mc{X}},\bs{\alpha}}$  and $\mc{B}_{\bs{\psi},\bs{\theta},\bs{\la},\bs{\alpha}}$, is defined by 
 \begin{equation}\label{propertyB_pant}
 \begin{split}
&\mc{B}_{\tilde{\mc{X}},\bs{\alpha}}: C_c(\R_+\times \mc{T})^{\otimes b}\to  \C, \qquad \mc{B}_{\tilde{\mc{X}},\bs{\alpha}}{\bf u}=(2\pi)^{b/2}\mc{A}_{\mc{P},g,{\bf x}^\bullet,\bzeta,\bs{\alpha}}\Big (\bigotimes^{b}\Pi_{\mc{V}}^*(
\Omega {\bf u})\Big)
 \end{split}\end{equation}
where $\Pi_{\mc{V}}$ is the operator of \eqref{PiV} and $\Omega({\bf p},\bs{\nu})=\Omega({\bf p})$ is the function 
\[\Omega({\bf p}):=\frac{\Gamma_{\bs{\alpha}}({\bf p},\bs{\psi},\bs{\theta},\bs{\la})}{\mc{A}_{\mc{P},g,{\bf x}^\bullet,\bs{\zeta},\bs{\alpha}}(\bigotimes_{k=1}^b \Psi_{Q+{\rm i}p_k})}.\]
\end{definition}
Notice that the condition \eqref{cdozznot=0} is satisfied if ${\rm Re}(\alpha_k)\leq Q$ and $\alpha_k\not=Q$ for all $k$.
By item 4) of Proposition \ref{p:amplitudes_properties}, notice that 
$\mc{B}_{\bs{\psi},\bs{\theta},\bs{\la},\bs{\alpha}}$ does not depend on the choice of admissible metric $g$.
 \\

\begin{remark}\label{rem:extension} 
Using Lemma \ref{extension_holomorphe_amp}, the conformal block amplitude  
\begin{equation}\label{block_holo_inalpha}
\bs{\alpha}\mapsto \mc{B}_{\tilde{\mc{X}},\bs{\alpha}}
\end{equation}
is holomorphic in the region $\{\bs{\alpha}\in \C^m\,|\,\sum_{\ell=1}^m |{\rm Re}(\alpha_\ell)-Q|<Q \}$.
\end{remark}

For $\eps>0$,  and $I\subset (0,\infty)$ a compact interval, let us define the space $e^{-\eps|\nu|}L^2(I\times \mc{T})$ as the Hilbert space consisting of functions of the form $e^{-\eps|\nu|}u(p,\nu)$ with $u\in L^2(I\times \mc{T})$ and with 
norm defined by $\|e^{-\eps|\nu|}u\|_{e^{-\eps|\nu|}L^2(I\times \mc{T})}:=\|u\|_{L^2(I\times \mc{T})}$.

\begin{lemma}\label{lem:estimate_blocks}
The condition \eqref{cdozznot=0} can be removed in Definition \ref{defblockamp_pant} and the conformal blocks defined for all 
 $\bs{\alpha}\in \C^m$ satisfying the generalised Seiberg bound. If $u_k\in C_c(\R_+\times \mc{T})$ for $k=1,\dots,b$, the conformal block amplitude of $(\bs{\psi},\bs{\theta},\bs{\la})\in \tT^\eps_{m,b}$ satisfies 
\[\mc{B}_{\bs{\psi},\bs{\theta},\bs{\la},\bs{\alpha}}(u_1\otimes \dots \otimes u_b)=\int_{\R_+^b}\sum_{\bs{\nu}\in \mc{T}^b}\mc{B}_{\bs{\psi},\bs{\theta},\bs{\la},\bs{\alpha}}({\bf p},\bs{\nu})\prod_{k=1}^bu_k(p_k,\nu_k) \dd {\bf p}\]
\begin{equation}\label{coef_B_in_terms_of_w}
\textrm{with } \quad\mc{B}_{\bs{\psi},\bs{\theta},\bs{\la},\bs{\alpha}}({\bf p},\bs{\nu}):=w_{\bs{\psi}}(\bs{\Delta}_{\bs{\alpha}({\bf p})},\bs{\nu})\Gamma_{\bs{\alpha}}({\bf p},\bs{\psi},\bs{\theta},\bs{\la})
\end{equation}
where $\bs{\alpha}({\bf p})$ is defined by \eqref{alpha(p)},  $w_{\bs{\psi}}(\bs{\Delta}_{\bs{\alpha}({\bf p})},\bs{\nu})$ are the matrix coefficients in Lemma \ref{matrix_coefficients}. Moreover, there exists $\eps>0$ such that
\begin{equation}\label{bound_bloc_pant_Dozz^4} 
\int_{\R_+^b}
|C^{\rm DOZZ}(\bs{\alpha}({\bf p}))|^2 e^{\eps|{\bf p}|^2}\big(\sum_{\bs{\nu}\in \mc{T}^b}e^{2\eps|\bs{\nu}|}
|\mc{B}_{\bs{\psi},\bs{\theta},\bs{\la},\bs{\alpha}}({\bf p},\bs{\nu})|^2\big)\big(\sum_{\bs{\nu}\in \mc{T}^b}e^{2\eps|\bs{\nu}|}
|\mc{B}_{\bs{\psi},\bs{\theta},\bs{\la},\bbar{\bs{\alpha}}}({\bf p},\bs{\nu})|^2\big) \dd {\bf p}<\infty,
\end{equation}
\begin{equation}\label{bound_bloc_pant_Dozz^2} 
\int_{\R_+^b}|\Gamma_{\bbar{\bs{\alpha}}}({\bf p},\bs{\psi},\bs{\theta},\bs{\la})|^2e^{\eps|{\bf p}|^2}|C^{\rm DOZZ}(\bs{\alpha}({\bf p}))|^2
\sum_{\bs{\nu}\in \mc{T}^b}e^{2\eps|\bs{\nu}|}|\mc{B}_{\bs{\psi},\bs{\theta},\bs{\la},\bs{\alpha}}({\bf p},\bs{\nu})|^2\dd{\bf p}<\infty.
\end{equation}
In particular, for each compact interval $I\subset (0,\infty)$ and $\bs{\alpha}$ such that \eqref{cdozznot=0} holds, $\mc{B}_{\tilde{\mc{X}},\bs{\alpha}}=\mc{B}_{\bs{\psi},\bs{\theta},\bs{\la},\bs{\alpha}}$ extends  as a bounded map
\begin{equation}\label{bloc_decorated_bb} 
\mc{B}_{\tilde{\mc{X}},\bs{\alpha}}: (e^{\eps|\nu|}L^2 (I\times \mc{T}))^{\otimes b}\to \C
\end{equation}
which is holomorphic in $\bs{\alpha}$ in the region where the generalised Seiberg bounds and \eqref{cdozznot=0} hold.
\end{lemma}
\begin{proof}
The fact that condition \eqref{cdozznot=0} can be removed in Definition \ref{defblockamp_pant} follows from Lemma \ref{matrix_coefficients}, which in turn gives \eqref{coef_B_in_terms_of_w}.
For $(\mc{P},J_\C,{\bf x}^\bullet,\bzeta)$ the complex building block associated to $(\bs{\psi},\bs{\theta})$ and $g$ an admissible metric, using Lemma \ref{matrix_coefficients} we have, for $\bs{\alpha}$ real valued satisfying the Seiberg bounds,
\begin{equation}\label{matrix_coef_again}
\begin{aligned}
\cjg \mc{A}_{\mc{P},g,{\bf x}^\bullet,\bzeta,\bs{\alpha}},\otimes_{k=1}^b\bbar{H_{Q+{\rm i}p_k,\nu_k,\tilde{\nu}_k}}\cjd_{\mc{H}^{\otimes b}}
&=
C_b C^{\rm DOZZ}(\bs{\alpha}({\bf p})) e^{c_{\rm L}S_{\rm L}^0(\hat{\C},g_{\rm dozz},\hat{g})-
\sum_{k=b+1}^3\Delta_{\alpha_k}\omega(z_k^\bullet)} \\
&\qquad\qquad \times\, |\Gamma_{\bs{\alpha}}({\bf p},\bs{\psi},\bs{\theta},0)|^2 w_{\bs{\psi}}(\bs{\Delta}_{\bs{\alpha}({\bf p})},\bs{\nu})\bbar{w_{\bs{\psi}}(\bs{\Delta}_{\bs{\alpha}({\bf p})},\tilde{\bs{\nu}})},
\end{aligned}
\end{equation}
where $\hat{g},g_{\rm dozz},\omega$ are defined there. The left hand side extends holomorphically in $\bs{\alpha}$ in the region 
\begin{equation}\label{region_holomorphy} 
\sum_{\ell=1}^m |{\rm Re}(\alpha_\ell)-Q|<Q
\end{equation}
by Lemma \ref{extension_holomorphe_amp}. Using that $w_{\bs{\psi}}(\bs{\Delta}_{\bs{\alpha}({\bf p})},\bs{\nu})$ is polynomial in the weights $\Delta_{\alpha_k}$, by Lemma \ref{matrix_coefficients} (following \cite[Theorem 11.1]{GKRV21_Segal}) and that the DOZZ function extends holomorphically in $\bs{\alpha}$ in \eqref{region_holomorphy}, the right-hand side also extends holomorphically in \eqref{region_holomorphy} and \eqref{matrix_coef_again} becomes there
\begin{equation}\label{factorisation_extended}
\begin{aligned}
\cjg \mc{A}_{\mc{P},g,{\bf x}^\bullet,\bzeta,\bs{\alpha}},\otimes_{k=1}^b\bbar{H_{Q+{\rm i}p_k,\nu_k,\tilde{\nu}_k}}\cjd_{\mc{H}^{\otimes b}}
&=
C_b C^{\rm DOZZ}(\bs{\alpha}({\bf p})) e^{c_{\rm L}S_{\rm L}^0(\hat{\C},g_{\rm dozz},\hat{g})-
\sum_{k=b+1}^3\Delta_{\alpha_k}\omega(z_k^\bullet)} \\
&\qquad\qquad \times\, \Gamma_{\bs{\alpha}}({\bf p},\bs{\psi},\bs{\theta},0)\bbar{\Gamma_{\bbar{\bs{\alpha}}}({\bf p},\bs{\psi},\bs{\theta},0)} w_{\bs{\psi}}(\bs{\Delta}_{\bs{\alpha}({\bf p})},\bs{\nu})\bbar{w_{\bs{\psi}}(\bs{\Delta}_{\bbar{\bs{\alpha}}({\bf p})},\tilde{\bs{\nu}})},
\end{aligned}
\end{equation}
where $\bbar{\bs{\alpha}}({\bf p})$ consists of the conjugation of the $\alpha_k$ coefficients but not the $p_k$.
The formula \eqref{coef_B_in_terms_of_w} follows directly from this. We then obtain
\[|\cjg \mc{A}_{\mc{P},g,{\bf x}^\bullet,\bzeta,\bs{\alpha}},\otimes_{k=1}^b\bbar{H_{Q+{\rm i}p_k,\nu_k,\tilde{\nu}_k}}\cjd_{\mc{H}^{\otimes b}}| = C_1(\bs{\alpha}) |C^{\rm DOZZ}(\bs{\alpha}({\bf p}))|\, |\mc{B}_{\bs{\psi},\bs{\theta},0,\bs{\alpha}}({\bf p},\bs{\nu})|\, |\mc{B}_{\bs{\psi},\bs{\theta},0,\bbar{\bs{\alpha}}}({\bf p},\tilde{\bs{\nu}})|\]
for some constant $C_1(\bs{\alpha})$ depending on $\bs{\alpha}$ but not on ${\bf p}$.
Let $\delta=e^{-\eps}<1$ be close enough to $1$ so that $\psi_j$ are holomorphic in $\delta^{-1}\D$ with disjoint images $\psi_j(\delta^{-1}\D)$ and 
let $\mc{P}^{\delta}$ be the model building block associated to 
\[ \bs{\psi}^\delta:=(\psi_1(\delta^{-1}\cdot),\dots,\psi_{b}(\delta^{-1}\cdot)).\] 
We write $\bs{\zeta}^\delta:=\bs{\psi}^\delta|_{\T}$.
By \cite[Corollary 11.4]{GKRV21_Segal} and \eqref{Gamma_alpha_mb}, we have 
\begin{equation}\label{scaling_w}
w_{\bs{\psi}^\delta}(\bs{\Delta}_{\bs{\alpha}({\bf p})},\bs{\nu})=\prod_{j=1}^b \delta^{-|\nu_j|}w_{\bs{\psi}}(\bs{\Delta}_{\bs{\alpha}({\bf p})},\bs{\nu}), \quad |\Gamma_{\bs{\alpha}}({\bf p},\bs{\psi}^\delta,\bs{\theta},\bs{\la})|^{2}=\prod_{k=1}^b |\delta|^{-(Q^2+p_k^2)/2}|\Gamma_{\bs{\alpha}}({\bf p},\bs{\psi},\bs{\theta},\bs{\la})|^{2}.
\end{equation}
Using Lemma \ref{l:diagonalisation} and $C^{\rm DOZZ}(\bbar{\bs{\beta}})=\bbar{C^{\rm DOZZ}(\bs{\beta})}$ if $\sum_{k=1}^3 {\rm Re}(\beta_k)+Q>0$, for $\bs{\alpha}$ such that \eqref{region_holomorphy} holds, we have 
\begin{equation}\label{upper_bound_block_model}
\begin{split}
\|\mc{A}_{\mc{P}^\delta,g,{\bf x}^\bullet,\bzeta^\delta,\bs{\alpha}}\|^2_{\mc{H}^{\otimes b}}&= \|\mc{A}_{\mc{P}^\delta,g,{\bf x}^\bullet,\bzeta^\delta,\bbar{\bs{\alpha}}}\|^2_{\mc{H}^{\otimes b}}\\
&=
\frac{1}{(2\pi)^b}\int_{\R_+^b}\sum_{\bs{\nu},\tilde{\bs{\nu}}\in \mc{T}^b}|\cjg \mc{A}_{\mc{P}^\delta,g,{\bf x}^\bullet,\bzeta^\delta,\bs{\alpha}},\otimes_{k=1}^b\bbar{H_{Q+{\rm i}p_k,\nu_k,\tilde{\nu}_k}}\cjd_{\mc{H}^{\otimes b}}|^2\dd {\bf p}\\
\end{split}
\end{equation}
and the right-hand side is equal to 
\[\frac{|C_1(\bs{\alpha})|^2\delta^{-bQ^2}}{(2\pi)^b} \int_{\R_+^b}|C^{\rm DOZZ}(\bs{\alpha}({\bf p}))|^2\delta^{-|{\bf p}|^2}\big(\sum_{\bs{\nu}\in \mc{T}^b}
\delta^{-2|\bs{\nu}|}|\mc{B}_{\bs{\psi},\bs{\theta},0,\bs{\alpha}}({\bf p},\bs{\nu})|^2\big)\big(\sum_{\bs{\nu}\in \mc{T}^b}
\delta^{-2|\bs{\nu}|}|\mc{B}_{\bs{\psi},\bs{\theta},0,\bbar{\bs{\alpha}}}({\bf p},\bs{\nu})|^2\big) \dd {\bf p}.\]
This proves \eqref{bound_bloc_pant_Dozz^4} for any $\bs\alpha=(\alpha_1,\dots,\alpha_m)$ satisfying \eqref{region_holomorphy}, since 
$\mc{B}_{\bs{\psi},\bs{\theta},\bs{\la},\bs{\alpha}}({\bf p},\bs{\nu})=e^{\sum_{k=1}^m \la_k \Delta_{\alpha_{b+k}}}\mc{B}_{\bs{\psi},\bs{\theta},0,\bs{\alpha}}({\bf p},\bs{\nu})$. Similarly, 
\[\begin{split}
 \|\mc{A}_{\mc{P}^\delta,g,{\bf x}^\bullet,\bzeta^\delta,\bs{\alpha}}\|^2_{\mc{H}^{\otimes b}}&\geq  \frac{1}{(2\pi)^b}\int_{\R_+^b}\sum_{\bs{\nu}\in \mc{T}^b}|\cjg \mc{A}_{\mc{P}^\delta,g,{\bf x}^\bullet,\bzeta^\delta,\bs{\alpha}},\otimes_{k=1}^b\overline{H_{Q+{\rm i}p_k,\nu_k,0}}\cjd_{\mc{H}^{\otimes b}}|^2\dd {\bf p}
\end{split}\]
and the right-hand side is equal to 
\[\frac{|C_1(\bs{\alpha})|^2\delta^{-bQ^2}}{(2\pi)^b} \int_{\R_+^b}|C^{\rm DOZZ}(\bs{\alpha}({\bf p}))|^2\delta^{-|{\bf p}|^2}|\Gamma_{\bbar{\bs{\alpha}}}({\bf p},\bs{\psi},\bs{\theta},0)|^2\big(\sum_{\bs{\nu}\in \mc{T}^b}\delta^{-2|\bs{\nu}|}
|\mc{B}_{\bs{\psi},\bs{\theta},0,\bs{\alpha}}({\bf p},\bs{\nu})|^2\big)
 \dd {\bf p}.\]
This proves \eqref{bound_bloc_pant_Dozz^2} when $\bs{\la}=0$. Since both $\Gamma_{\bs{\alpha}}({\bf p},\bs{\psi},\bs{\theta},\bs{\la})$ and $\mc{B}_{\bs{\psi},\bs{\theta},\bs{\la},\bs{\alpha}}({\bf p},\bs{\nu})$ differ from their values at $\bs{\la}=0$ by the factor $e^{\sum_{k=b+1}^3\la_k\Delta_{\alpha_{k}}}$, and similarly with $\bs{\alpha}$ replaced by $\bbar{\bs{\alpha}}$, the general case follows.. The holomorphy in $\bs{\alpha}$ follows separately from \eqref{block_holo_inalpha}.
\end{proof}
It is not hard to check that \eqref{bloc_decorated_bb} extends to all $\bs{\alpha}$ satisfying the generalised Seiberg bounds: this can be done 
using Cauchy's formula to write the conformal block at value $\bs{\alpha}$ where $C^{\rm DOZZ}(\bs{\alpha}({\bf p}))=0$ for some ${\bf p}$ 
as an integral on a product of $m$ small circles in $\C^m$, where \eqref{bloc_decorated_bb} holds and  \eqref{cdozznot=0} holds.

The conformal block amplitudes of marked complex building blocks obey the following transformation rules under Dehn twists around the boundary circles and braidings:

\begin{lemma}[\textbf{Action of Dehn twist on conformal block amplitude}]\label{Dehn_Twist_change}
Let $m+b=3$ and $\bs{\alpha}=(\alpha_{b+1},\dots,\alpha_3)\in \C^m$ satisfying  the generalised Seiberg bound. If $T_k$ denotes the Dehn twist around $\pl_k\Sigma_{m,b}^\bullet$ when $k\leq b$ or the Dehn twist around the marked point $z_k^\bullet$ when $k>b$, we have for $\tilde{\mc{X}}\in \tT_{0,m,b}$ and each ${\bf u}\in (e^{\eps|\nu|}L^2 (I\times \mc{T}))^{\otimes b}$ for $\eps>0$ small enough
\[\mc{B}_{\tilde{\mc{X}}\cdot T_k,\bs{\alpha}}{\bf u}= \mc{B}_{\tilde{\mc{X}},\bs{\alpha}}(e^{2\pi {\rm i}\Delta_{\alpha_k({\bf p})}}{\bf u}), \]
where $\alpha_k({\bf p})$ is defined by \eqref{alpha(p)}.
\end{lemma}
\begin{proof} Let $\tilde{\Theta}_{m,b}(\tilde{\mc{X}})=(\bs{\psi},\bs{\theta},\bs{\la})\in \tT^\eps_{m,b}$ for some $\eps>0$.
By 1) of Proposition \ref{identification_Teichmuller}, we have $\tilde{\Theta}_{m,b}(\tilde{\mc{X}}\cdot T_k)=(\bs{\psi},(\bs{\theta},\bs{\la})+2\pi e_k)$, and by definition,
\[ \Gamma_{\bs{\alpha}}({\bf p},\bs{\psi},(\bs{\theta},\bs{\la})+2\pi e_k)= e^{2\pi {\rm i}\Delta_{\alpha_k({\bf p})}}\Gamma_{\bs{\alpha}}({\bf p},\bs{\psi},\bs{\theta},\bs{\la}).\]
Since the matrix coefficients $w_{\bs{\psi}}(\bs{\alpha}({\bf p}),\bs{\nu})$ in Lemma \ref{matrix_coefficients} only depend on $\bs{\psi}$ (thus on $\bs{\theta}$ modulo $2\pi \Z^b$), we obtain the result by using the expression \eqref{coef_B_in_terms_of_w} of the conformal block in Lemma \ref{lem:estimate_blocks}.
\end{proof}

\begin{lemma}[\textbf{Action of braiding on conformal block amplitude}]\label{braiding_change}
Let $m+b=3$ and $\bs{\alpha}=(\alpha_{b+1},\dots,\alpha_3)\in \C^m$ satisfying the generalised  Seiberg bound. Let $B_{jk}$ be the braiding map of \eqref{braiding} with $\{j,k,\ell\}=\{1,2,3\}$, where we assume that either $j<k\leq b$ or $b<j<k$. Recall the notation $\tau$ from 2) of Proposition \ref{identification_Teichmuller}. We have for $\tilde{\mc{X}}\in \tT_{0,m,b}$ and each ${\bf u}\in (e^{\eps|\nu|}L^2 (I\times \mc{T}))^{\otimes b}$ for $\eps>0$ small enough
\begin{gather*}
\mc{B}_{\tilde{\mc{X}}\cdot B_{jk},\bs{\alpha}}{\bf u}=  \mc{B}_{\tilde{\mc{X}},\bs{\alpha}_\tau}(e^{\pi{\rm i}(\Delta_{\alpha_j({\bf p})}+\Delta_{\alpha_k({\bf p})}+(-1)^{\ell+1}\Delta_{\alpha_\ell({\bf p})})}{\bf u}_\tau),
\end{gather*}
where $\bs{\alpha}({\bf p})=(\alpha_1({\bf p}),\alpha_2({\bf p}),\alpha_3({\bf p}))$ is defined by \eqref{alpha(p)}, and $M_{jk}$ is the unique M\"obius transformation with $(M_{jk}(z^\bullet_j),M_{jk}(z^\bullet_k),M_{jk}(z^\bullet_\ell))=(z^\bullet_k,z^\bullet_j,z^\bullet_\ell)$. Here the action of $B_{jk}$ on $\tT_{0,m,b}$ (and similarly on $\mc{T}_{0,m,b}$) is defined as in Proposition \ref{identification_Teichmuller}, i.e. by 
\[[(F,\Sigma,J,{\bf x},\bs{\zeta},{\bf Y})]\cdot B_{jk}:=[(F\circ B_{jk},\Sigma,J,{\bf x}_\tau,\bs{\zeta}_\tau,{\bf Y}_\tau)].\] 
\end{lemma}
\begin{proof}
Again we reindex $\bs{\lambda}=(\lambda_{b+1},\ldots,\lambda_3)$.
We would like to show
\[\mc{B}_{\tilde{\mc{X}}\cdot B_{jk},\bs{\alpha}}({\bf p},\bs{\nu}) = e^{\pi{\rm i}(\Delta_{\alpha_j({\bf p})}+\Delta_{\alpha_k({\bf p})}+(-1)^{\ell+1}\Delta_{\alpha_\ell({\bf p})})} \mc{B}_{\tilde{\mc{X}},\bs{\alpha}_\tau}({\bf p}_\tau,\bs{\nu}_\tau).\]
Let $\tilde{\Theta}_{m,b}(\tilde{\mc{X}})=(\bs{\psi},\bs{\theta},\bs{\lambda})\in \tT^\eps_{m,b}$ for some $\eps>0$. By 2) of Proposition \ref{identification_Teichmuller}, we have 
\[\tilde{\Theta}_{m,b}(\tilde{\mc{X}}\cdot B_{jk})=(M_{jk}\circ\bs{\psi}_\tau,(\bs{\theta}_\tau,\bs{\lambda}_\tau)+\pi(e_j+e_k+(-1)^{\ell+1}e_\ell)).\] 
except when $(b,m,j,k)=(1,2,2,3)$, where \eqref{Braiding_under_Theta23} holds.
Since the matrix coefficients in Lemma \ref{matrix_coefficients} depend only on the complex structure of the complex building block (i.e.\ its class in the moduli space), we have
\[w_{M_{jk}\circ\bs{\psi}_\tau}(\bs{\alpha}({\bf p}),\bs{\nu})= w_{\bs{\psi}}(\bs{\alpha}({\bf p})_\tau,\bs{\nu}_\tau)=w_{\bs{\psi}}(\bs{\alpha}_\tau({\bf p}_\tau),\bs{\nu}_\tau).\]
Assume first that $B_{jk}$ is not a permutation of two marked points.
By definition and \eqref{Gamma_alpha_Moebius}, together with $|M_{12}'(z_3^\bullet)|=1$, we obtain
\[\Gamma_{\bs{\alpha}}({\bf p},M_{jk}\circ\bs{\psi}_\tau,(\bs{\theta}_\tau,\bs{\lambda}_\tau)+\pi(e_j+e_k+(-1)^{\ell+1}e_\ell)) = e^{\pi{\rm i}(\Delta_{\alpha_j({\bf p})}+\Delta_{\alpha_k({\bf p})}+(-1)^{\ell+1}\Delta_{\alpha_\ell({\bf p})})} \Gamma_{\bs{\alpha}}({\bf p},\bs{\psi}_\tau,\bs{\theta}_\tau,\bs{\lambda}_\tau).\]
Writing $\tau$ for the transposition $j\to k$, this gives
\begin{align*}
\Gamma_{\bs{\alpha}}({\bf p},\bs{\psi}_\tau,\bs{\theta}_\tau,\bs{\lambda}_\tau) &= \Gamma(\bs{\alpha}({\bf p}),{\bf z}_\tau^\bullet) \prod_{i=1}^b|\psi_{\tau(i)}'(0)|^{\Delta_{Q+{\rm i}p_i}}e^{{\rm i}\Delta_{Q+{\rm i}p_i}\theta_{\tau(i)}}\prod_{i=b+1}^3e^{\Delta_{\alpha_{i}}\lambda_{\tau(i)}}\\
&=\Gamma(\bs{\alpha}({\bf p}),{\bf z}_\tau^\bullet) \prod_{i=1}^b|\psi_{i}'(0)|^{\Delta_{Q+{\rm i}p_{\tau(i)}}}e^{{\rm i}\Delta_{Q+{\rm i}p_{\tau(i)}}\theta_{i}}\prod_{i=b+1}^3e^{\Delta_{\alpha_{\tau(i)}}\lambda_{i}}\\
&=\frac{\Gamma(\bs{\alpha}({\bf p}),{\bf z}_\tau^\bullet)}{\Gamma(\bs{\alpha}_\tau({\bf p}_\tau),{\bf z}^\bullet)} \Gamma_{\bs{\alpha}_\tau}({\bf p}_\tau,\bs{\psi},\bs{\theta},\bs{\lambda}),
\end{align*}
so it remains to show $\Gamma(\bs{\alpha}_\tau({\bf p}_\tau),{\bf z}^\bullet)=\Gamma(\bs{\alpha}({\bf p}),{\bf z}_\tau^\bullet)$. Using the formula \eqref{Gammaterm_formula}, it is easy to see that, for $\bs{\alpha}$ real, $\frac{\Gamma(\bs{\alpha}({\bf p}),{\bf z}_\tau^\bullet)}{\Gamma(\bs{\alpha}_\tau({\bf p}_\tau),{\bf z}^\bullet)}\in\R_+$, so
it suffices to show $|\Gamma(\bs{\alpha}_\tau({\bf p}_\tau),{\bf z}^\bullet)|=|\Gamma(\bs{\alpha}({\bf p}),{\bf z}_\tau^\bullet)|$, which follows from the definition. This identity then extends holomorphically to  complex weights $\bs{\alpha}$. In the case of braidings permuting marked points (we then have $(j,k)=(2,3)$ and $b=1$), 
\begin{align*}
&\Gamma_{\bs{\alpha}}(p_1,M_{23}\circ\psi_1,\theta_1+\pi,\la_3-\log4+{\rm i}\pi,\la_2+\log4+{\rm i}\pi)\\
&=\Gamma(\bs{\alpha}({\bf p}),{\bf z}^\bullet)e^{{\rm i}\pi(\Delta_{Q+{\rm i}p_1}+\Delta_{\alpha_2}+\Delta_{\alpha_3})}4^{\Delta_{\alpha_3}-\Delta_{\alpha_2}}e^{{\rm i}\theta_1\Delta_{Q+{\rm i}p_1}}|\psi_1'(0)|^{\Delta_{Q+{\rm i}p_1}}e^{\Delta_{\alpha_3}\la_2+\Delta_{\alpha_2}\la_3}\\
&= \Gamma(\bs{\alpha}_\tau({\bf p}_\tau),{\bf z}^\bullet)e^{{\rm i}\pi(\Delta_{Q+{\rm i}p_1}+\Delta_{\alpha_2}+\Delta_{\alpha_3})}e^{{\rm i}\theta_1\Delta_{Q+{\rm i}p_1}}|\psi_1'(0)|^{\Delta_{Q+{\rm i}p_1}}e^{\Delta_{\alpha_3}\la_2+\Delta_{\alpha_2}\la_3}\\
&= e^{{\rm i}\pi(\Delta_{Q+{\rm i}p_1}+\Delta_{\alpha_2}+\Delta_{\alpha_3})}\Gamma_{\bs{\alpha}_\tau}({\bf p}_\tau,\bs{\psi},\bs{\theta},\bs{\lambda}),
\end{align*}
which gives the announced result.
\end{proof}

Next, we define the conformal block amplitude of complex building blocks with outgoing boundaries. 

\begin{definition}[\textbf{Conformal block amplitude of marked complex building block}] \label{defblockamp} Let $m+b=3$
and $\bs{\alpha}=(\alpha_{b+1},\dots,\alpha_3)\in \C^m$ satisfying the generalised Seiberg bound.
Let $b^++b^-=b$ and $\bs{\sigma}=(\sigma_1,\dots,\sigma_b)\in \{\pm1\}^b$ be orientation signs with $\sigma_k=+1$ for $k=1,\ldots,b^+$, $\sigma_k=-1$ for $k=b^++1,\ldots,b$. For $\tilde{\mc{X}}=[(F,\Sigma,J,{\bf x},\bs{\zeta},{\bf Y})]\in \tT_{0,m,b}$, let $\tilde{\mc{X}}^{\bs{\sigma}}=[(F,\Sigma,J,{\bf x},\bs{\zeta}^{\bs{\sigma}},{\bf Y})]\in \tT_{0,m,b^+,b^-}$ be defined by \eqref{decor_teich_to_teich_sigma}.
Let $I$ be a compact interval in $(0,\infty)$.  For $\eps>0$ as in Lemma \ref{lem:estimate_blocks},
the \textbf{conformal block amplitude} of $\tilde{\mc{X}}^{\bs{\sigma}}$ with weights $\bs{\alpha}=(\alpha_{b+1},\dots,\alpha_3)$ is the operator
\[\mc{B}_{\tilde{\mc{X}}^{\bs{\sigma}},\bs{\alpha}}: (e^{\eps|\nu|}L^2(I\times \mc{T}))^{\otimes b^-}\to (e^{-\eps|\nu|}L^2(I\times\mc{T}))^{\otimes b^+}\]
defined as follows: for any $u_1,\dots, u_b\in e^{\eps |\nu|}L^2(I\times \mc{T})$,
 \begin{equation}\label{propertyB} 
 \cjg \mc{B}_{\tilde{\mc{X}}^{\bs{\sigma}},\bs{\alpha}} (u_{b^++1}\otimes \cdots \otimes u_{b}), \bar{u}_{1}\otimes \dots \otimes \bar{u}_{b^+}\cjd_{L^2(I\times \mc{T})^{\otimes b^+}}=  \mc{B}_{\tilde{\mc{X}},\bs{\alpha}} (u_1\otimes \dots \otimes u_b).
\end{equation}
\end{definition}
We notice that for ${\bf u}\in (e^{\eps|\nu|}L^2(I\times\mc{T}))^{\otimes b^-}$, $\mc{B}_{\tilde{\mc{X}}^{\bs{\sigma}},\bs{\alpha}}{\bf u}\in (e^{-\eps|\nu|}L^2(I\times\mc{T}))^{\otimes b^+}$ has the form
\begin{equation}\label{block_amplitude_pant_operator}
\mc{B}_{\tilde{\mc{X}}^{\bs{\sigma}},\bs{\alpha}}{\bf u}({\bf p}^+,\bs{\nu}^+)=(2\pi)^{b/2}\Big(\big(\bigotimes^{b^+}\Pi_{\mc{V}}\big)\circ \mc{A}_{\Sigma,g,{\bf x}^\bullet,\bzeta^{\bs{\sigma}},\bs{\alpha}}\circ \big(\bigotimes^{b^-}\Pi_{\mc{V}}^*\big)(\Omega_{{\bf p}^+}{\bf u})\Big)({\bf p}^+,\bs{\nu}^+),
\end{equation}
where $g$ is an admissible metric on $\Sigma$ and for ${\bf p}^+\in I^{b^+}$, $\Omega_{{\bf p}^+}$ is the function on $I^{b^-}$ defined by 
\[\Omega_{{\bf p}^+}({\bf p}^-):=\frac{\Gamma_{\bs{\alpha}}({\bf p},\bs{\psi},\bs{\theta},\bs{\la})}{\mc{A}_{\Sigma,g,{\bf x}^\bullet,\bs{\zeta},\bs{\alpha}}(\bigotimes_{k=1}^b \Psi_{Q-{\rm i}\sigma_kp_k})},\]
with the notation ${\bf p}=(p_1,\ldots,p_b)$, ${\bf p}^+=(p_1,\ldots,p_{b^+})$, ${\bf p}^-=(p_{b^++1},\ldots,p_b)$.
The integral kernel of $\mc{B}_{\tilde{\mc{X}}^{\bs{\sigma}},\bs{\alpha}}$ is given, for $\bs{\nu}=(\bs{\nu}^+,\bs{\nu}^-)\in \mc{T}^{b^+}\times \mc{T}^{b^-}=\mc{T}^b$, by 
\[\mc{B}_{\tilde{\mc{X}}^{\bs{\sigma}},\bs{\alpha}}({\bf p}^+,\bs{\nu}^+,{\bf p}^-,\bs{\nu}^-)=\mc{B}_{\tilde{\mc{X}},\bs{\alpha}}({\bf p},\bs{\nu}).\]

\section{Variation of conformal block amplitudes of marked complex building blocks}\label{sec:variation_blocks}
In this section, we compute the differential of conformal block amplitudes as functions of the complex building blocks in terms of the action of the (left-moving) Virasoro algebra. We shall assume that $\bs{\alpha}$ are real valued and satisfy the Seiberg bounds.
We first introduce the following notation for $j=1,\ldots,b$ (with $b$ implicit):
\begin{align*} 
& \bs{\ell}_{\v_j}^{(j)}:  \otimes^b h_0^{-1/2}L^2(\R_+\times \mc{T})\to \otimes^b h_0^{1/2}L^2(\R_+\times \mc{T}), \quad  \bs{\ell}_{\v_j}^{(j)}(h_1\otimes \dots \otimes h_b):= h_1\otimes \dots\otimes  \bs{\ell}_{\v_j} h_j\otimes \dots \otimes h_b,\\
& \bL_{\v_j}^{(j)}:  \otimes^b \mc{D}(\mc{Q})^{1,0}\to \otimes^b \mc{D}'(\mc{Q}),  \quad   \bL_{\v_j}^{(j)}(h_1\otimes \dots \otimes h_b):=
h_1\otimes \dots\otimes  {\bf L}_{\v_j} h_j\otimes \dots \otimes h_b,
\end{align*}
where $b$ is a nonnegative integer which will be clear from context, and we recall the definitions \eqref{def_Hv} and \eqref{def:ell_bv}. Here $h_{0}(p,\nu):=h_0(p,\nu,0)=(2\Delta_{Q+{\rm i}p}+|\nu|)$ as before. Also, we shall write below $\v_j=v_j(z)\pl_z$ where we make the distinction that $\v_j$ is a holomorphic vector field and $v_j$ is a complex valued function.
Recall also the actions $\mc{X}\cdot \bs{f}$ and $\tilde{\mc{X}}\cdot \bs{f}$ 
 of $\mc{U}_\eps^b$ on $\mc{T}_{0,m,b^+,b^-}$ and $\tT_{0,m,b^+,b^-}$ defined in \eqref{action_bsf_tildeX}. 
The goal of this section is to prove the following result. 

\begin{proposition}
\label{general_derivative_block_amp}
Let $\tilde{\mc{X}}=[(F,\mc{P},J,{\bf x}, \bs{\zeta},{\bf Y})] \in \tT_{0,m,b}$ with $m+b=3$ and $\tilde{\Theta}_{m,b}(\tilde{\mc{X}})=(\bs{\psi},\bs{\theta},\bs{\la})\in \tT^\eps_{m,b}$ for some $\eps>0$. Let $\bs{\sigma}=(\sigma_1,\dots,\sigma_b)$ be orientation signs with $\sigma_k=+1$ for $k\leq b^+$ and  $\sigma_k=-1$ for $k>b^+$, and 
 let $\bs{\alpha}=(\alpha_{b+1},\dots,\alpha_3)\in \R^m$ be weights satisfying the Seiberg bounds. For each compact interval $I\subset (0,\infty)$,
\[ 
\mc{B}_{\tilde{\mc{X}}^{\bs{\sigma}}\cdot, \bs{\alpha}} :\left\{ \begin{array}{ccl}
\mc{U}_\eps^b & \to & \mc{L}((h_0^{-1/2}L^2(I\times \mc{T}))^{\otimes b^-}, (h_0^{1/2}L^2(I\times \mc{T}))^{\otimes b^+}), \\
 \bs{f} & \mapsto &  \mc{B}_{\tilde{\mc{X}}^{\bs{\sigma}}\cdot \bs{f},\bs{\alpha}}
\end{array}\right. 
\]
is $C^1$ and its differential at  $\bs{f}={\rm Id}:=({\rm Id},\dots,{\rm Id})$ is given as follows:
for all $\bv=(v_1\pl_z,\dots,v_b\pl_z)\in T_{{\rm Id}}\mc{U}_\eps^b$ and ${\bf u}\in (h_0^{-1/2}L^2(I\times \mc{T}))^{\otimes {b^-}}$,
\begin{equation}\label{lastformula}
\begin{split}
D_{\bf v}\mc{B}_{\tilde{\mc{X}}^{\bs{\sigma}}\cdot,\bs{\alpha}} ({\rm Id})({\bf u})=& \sum_{j=1}^{b^+}\bs{\ell}_{\v_j}^{(j)}
\mc{B}_{\tilde{\mc{X}}^{\bs{\sigma}},\bs{\alpha}} ({\bf u})+\sum_{j=1}^{b^+}\Big( \frac{c_{\rm L}}{24\pi {\rm i}}\int_{\T} v_j(z) {\rm S}(\psi_j\circ o)(z) \dd z\Big)\mc{B}_{\tilde{\mc{X}}^{\bs{\sigma}},\bs{\alpha}}({\bf u})\\
 &- \sum_{j=1}^{b^-}\mc{B}_{\tilde{\mc{X}}^{\bs{\sigma}},\bs{\alpha}} (\bs{\ell}_{\v_{b^++j}}^{(j)}{\bf u}) - \sum_{j=b^++1}^b\Big( \frac{c_{\rm L}}{24\pi {\rm i}}\int_{\T} v_j(z) {\rm S}(\psi_j)(z) \dd z\Big)\mc{B}_{\tilde{\mc{X}}^{\bs{\sigma}},\bs{\alpha}}({\bf u}),
\end{split}
\end{equation}
where ${\rm S}(\psi)$ denotes the Schwarzian derivative of $\psi$, and the terms $v_j(z) {\rm S}(\psi_j)(z) \dd z,v_j(z) {\rm S}(\psi_j\circ o)(z) \dd z$ are viewed as holomorphic $1$-forms in a neighbourhood of the unit circle $\T$, which are integrated along $\T$ with its natural orientation.
\end{proposition}

%

We prove this proposition in several steps. We begin with the case where all boundary orientations are incoming and consider holomorphic deformations of the boundary parametrisations.
 \begin{lemma}[\textbf{Deformations in $\mc{U}_\eps^-$, incoming case}]\label{prop:variation_block_amplitude}
Let $\tilde{\mc{X}}=[(F,\mc{P},J,{\bf x},\bs{\zeta},{\bf Y})]\in \tT_{0,m,b}$ and  
$\tilde{\Theta}_{m,b}(\tilde{\mc{X}})=(\bs{\psi},\bs{\theta},\bs{\la})\in \tT^\eps_{m,b}$ for some $\eps>0$ and let $\bs{\alpha}=(\alpha_{b+1},\dots,\alpha_3)\in \R^m$ be weights satisfying the Seiberg bounds. For each compact interval $I\subset (0,\infty)$, the map  
\[
\mc{B}_{\tilde{\mc{X}}\cdot,\bs{\alpha}} :\left\{ \begin{array}{lll}
(\mc{U}_\eps^-)^b & \to & \mc{L}((h_0^{-1/2}L^2(I\times \mc{T}))^{\otimes b}, \C), \\
 \bs{f} & \mapsto & \mc{B}_{\tilde{\mc{X}}\cdot \bs{f},\bs{\alpha}}  
\end{array}\right. 
\]
is $C^1$ with differential at  $\bs{f}={\rm Id}:=({\rm Id},\dots,{\rm Id})$ given by: 
for all $\bv=(v_1\pl_z,\dots,v_b\pl_z)\in (T_{{\rm Id}}\mc{U}_\eps^-)^{b}$ and ${\bf u}\in (h_0^{-1/2}L^2(I\times \mc{T}))^{\otimes b}$,
\begin{equation}\label{derivative_of_amplitudeB1}
\begin{split}
D_{\bf v}\mc{B}_{\tilde{\mc{X}}\cdot,\bs{\alpha}} ({\rm Id}){\bf u}=  -\mc{B}_{\tilde{\mc{X}},\bs{\alpha}}\Big(\sum_{j=1}^b \bs{\ell}_{\v_j}^{(j)}{\bf u}\Big).
\end{split}
\end{equation}
The right-hand side of \eqref{derivative_of_amplitudeB1} makes sense due to Lemma \ref{lem:estimate_blocks} and the boundedness 
of $\bs{\ell}_{\v_j}:h_0^{-1/2}L^2(I\times \mc{T})\to h_0^{1/2}L^2(I\times \mc{T})$.
\end{lemma}
\begin{proof} 
Let $S:=(\mc{P},J,{\bf x},\bs{\zeta})$ and $S_g:=(\mc{P},g,{\bf x},\bs{\zeta})$ for some admissible metric $g$.
Then, recall the expression \eqref{propertyB_pant} for the conformal block amplitude. To prove its differentiability, we analyse each term in its definition.
For $\bs{f}=(f_1,\dots,f_b)$ with $f_k\in \mc{U}_\eps^-$, and denoting 
\begin{equation}\label{bspsicdotf}
\begin{split}
(\bs{\psi},\bs{\theta})\cdot {\bs f}&\phantom{:}= (\bs{\psi}\cdot \bs{f}, \bs{\theta}\cdot \bs{f})\\
&:= ((\psi_1\circ f_1,\dots,\psi_b\circ f_b), (\theta_1+{\rm arg} f_1'(0),\dots,\theta_b+{\rm arg} f_b'(0))),
\end{split}
\end{equation}
the map $\bs{f}\mapsto \Gamma_{\bs{\alpha}}({\bf p},(\bs{\psi},\bs{\theta})\cdot {\bs f},\bs{\la})$ is $C^1$. Here the argument ${\rm arg} f'(0)$ is uniquely defined for $f\in\mc{U}_\eps^-$ for $\mc{U}_\eps^-$ small enough.
Next, for ${\bf p}$ in the compact set $I^b\subset(0,\infty)^b$ the map (recall \eqref{Matrix_coef_pairofpants})
\[\bs{f}\mapsto \mc{A}_{S_g\cdot \bs{f},\bs{\alpha}}(\Psi_{Q+{\rm i}p_1},\dots,\Psi_{Q+{\rm i}p_b})=C(\bs{\psi}\cdot \bs{f},g^{\bs{f}},\bs{\alpha}({\bf p}))C^{\rm DOZZ}(\bs\alpha({\bf p}))\]
is $C^1$, in view of the expression \eqref{constantC(P,g)}, and has norm bounded from below by a positive number for ${\bf p}\in I^b$.
Since $\Pi_{\mc{V}}^*: h_0^{-1/2}L^2(\R_+\times \mc{T})\to \mc{D}(\mc{Q})$ is bounded, it follows that $\bs{f}\mapsto \mc{B}_{\tilde{\mc{X}}\cdot\bs{f},\bs{\alpha}}$ is $C^1$ by Proposition \ref{th:derivative_general}. 

Notice that $\Phi_{\bs{f}}={\rm Id}$ for $\bs{f}\in(\mc{U}_\eps^-)^b$, so the action $\mc{R}_{\bs{\alpha}}$ is trivial in this case, and we can keep $\bs{\lambda}$ constant in what follows. To compute the differential of $\bs{f}\mapsto \mc{B}_{\tilde{\mc{X}}\cdot\bs{f},\bs{\alpha}}$, we can assume that $\bv=(v_1\pl_z,0,\dots,0)$ with $v_1\in {\rm Hol}_\eps(\D)$ 
such that $v_1(0)=0$.
First we consider the case where  $v_1(z)=\mc{O}(z^2)$ at $z=0$. In that case, by \eqref{dlogGamma} (and the similar expression for the conjugate), we see that
\begin{equation}\label{variationGamma^t=0} 
 D_{\bf v} \Gamma_{\bs{\alpha}}({\bf p}, (\bs{\psi},\bs{\theta})\cdot,\bs{\la})|_{\bs{f}={\rm Id}}=0, \quad D_{\bf v} \bbar{\Gamma_{\bs{\alpha}}({\bf p},(\bs{\psi},\bs{\theta})\cdot,\bs{\la})}|_{\bs{f}={\rm Id}}=0.
 \end{equation} 
The operator $\otimes^b\Pi_{\mc{V}}^*$  of Lemma \ref{PiV^*PiV} can also be viewed as mapping  $L^2((\R_+\times \mc{T})^b)\to L^2(H^{-s}(\T)^{b},d\mu^{\otimes b})$ for $s>0$. Hence, for $u_j\in h_0^{-1/2}L^2(I\times \mc{T})$ (extended by $u_j(p,\nu)=0$ when $p\notin I$) and ${\bf u}:=u_1\otimes \dots \otimes u_b$, one can define 
\begin{equation}\label{definition_of_F}
F(\bs{\psi},\bs{\theta},\bs{\la}):=(2\pi)^{\frac{b}{2}}\otimes^b\Pi_{\mc{V}}^* \Big(\frac{\Gamma_{\bs{\alpha}}({\bf p},\bs{\psi},\bs{\theta},\bs{\la})\,  {\bf u}}{C^{\rm DOZZ}(\bs{\alpha}({\bf p}))C(\bs{\psi},g,\bs{\alpha}({\bf p}))}\Big)\in \otimes^b \mc{D}(\mc{Q})^{1,0},
\end{equation}
where $C(\bs{\psi},g,\bs{\alpha}({\bf p}))$ is defined by \eqref{constantC(P,g)}. 
In particular, we have $\mc{B}_{\tilde{\mc{X}},\bs{\alpha}}({\bf u})=\mc{A}_{\mc{P},g,{\bf x},\bs{\zeta},\bs{\alpha}}(F(\bs{\psi},\bs{\theta},\bs{\la}))$ by Lemma \ref{matrix_coefficients}. By Leibniz rule and \eqref{variationGamma^t=0}, we obtain
\begin{equation}\label{deriveeB^t}
 \begin{split}
D_{\bv}\mc{B}_{\tilde{\mc{X}}\cdot, \bs{\alpha}} ({\rm Id}){\bf u}=D_{\bv}\mc{A}_{S_g \cdot ,\bs{\alpha}}({\rm Id})F(\bs{\psi},\bs{\theta},\bs{\la})-
 \mc{A}_{S_g,\bs{\alpha}} \Big(\otimes^b\Pi_{\mc{V}}^* \big( \frac{D_\bv \mc{C}_{\bs{\alpha}({\bf p})}({\rm Id})}{\mc{C}_{\bs{\alpha}({\bf p})}({\rm Id})}\frac{(2\pi)^{\frac{b}{2}}\Gamma_{\bs{\alpha}}({\bf p},\bs{\psi},\bs{\theta},\bs{\la})\,  {\bf u}}{C^{\rm DOZZ}(\bs{\alpha}({\bf p}))\mc{C}_{\bs{\alpha}({\bf p})}({\rm Id})}
 \big)\Big),
  \end{split}\end{equation}
where  $\mc{C}_{\bs{\alpha}({\bf p})}(\bs{f})$ is defined by 
\begin{equation}\label{constantmcC} 
\mc{C}_{\bs{\alpha}({\bf p})}(\bs{f}):=C(\bs{\psi}\cdot {\bs{f}},g^{\bs{f}},\bs{\alpha}({\bf p})).
\end{equation}

The first term in the right-hand side of \eqref{deriveeB^t} can be obtained by Proposition \ref{th:derivative_general}:
\[ \begin{split}
D_{\bv}\mc{A}_{S_g\cdot,\bs{\alpha}}({\rm Id})F(\bs{\psi},\bs{\theta},\bs{\la})&= 
\mc{A}_{S_g,\bs{\alpha}}((-{\bf H}_{\v_1}^{(1)}+ c_{\rm L}D_\bv\mc{S}_{S_g\cdot}({\rm Id}))F(\bs{\psi},\bs{\theta},\bs{\la}))\\
 &=\mc{A}_{S_g,\bs{\alpha}}((-{\bf L}_{\v_1}^{(1)}+ c_{\rm L}D_\bv \mc{S}_{S_g\cdot} ({\rm Id}))F(\bs{\psi},\bs{\theta},\bs{\la})),
\end{split}\]
where we used that $v_1'(0)=0$ and $\bH_{\v_1}^{(1)}F(\bs{\psi},\bs{\theta},\bs{\la})=\bL_{\v_1}^{(1)}F(\bs{\psi},\bs{\theta},\bs{\la})$ since $F(\bs{\psi},\bs{\theta},\bs{\la})\in \otimes^b\mc{D}(\mc{Q})^{1,0}$ and $\tilde{{\bf L}}_nh=0$ for all $n>0$ and $h\in \mc{D}^{1,0}(\mc{Q})$. Therefore \eqref{deriveeB^t} becomes 
\begin{align}
D_{\bv}\mc{B}_{\tilde{\mc{X}}\cdot, \bs{\alpha}} ({\rm Id}){\bf u}&=
\mc{A}_{S_g,\bs{\alpha}}\Big(\big(-{\bf L}_{\v_1}^{(1)}+  c_{\rm L}D_\bv \mc{S}_{S_g\cdot} ({\rm Id})\big)F(\bs{\psi},\bs{\theta},\bs{\la})-\otimes^b\Pi_{\mc{V}}^* \big( \frac{D_\bv \mc{C}_{\bs{\alpha}({\bf p})}({\rm Id})}{\mc{C}_{\bs{\alpha}({\bf p})}({\rm Id})}\frac{(2\pi)^{\frac{b}{2}}\Gamma_{\bs{\alpha}}({\bf p},\bs{\psi},\bs{\theta},\bs{\la})\,  {\bf u}}{C^{\rm DOZZ}(\bs{\alpha}({\bf p}))\mc{C}_{\bs{\alpha}({\bf p})}({\rm Id})}
 \big)\Big)\nonumber\\
&= \mc{B}_{\tilde{\mc{X}},\bs{\alpha}}\big(-\bs{\ell}_{\v_1}^{(1)}
-D_\bv \log\big(e^{-c_{\rm L}\mc{S}_{S_g\cdot}}\mc{C}_{\bs{\alpha}({\bf p})}\big)({\rm Id})\big){\bf u},\label{dtA^{1,0}}
\end{align}
where we used in the second line the identity $\bL_{\v_1}\Pi_{\mc{V}}^*=\Pi_{\mc{V}}^*\bs{\ell}_{\v_1}$.
Next, we are going to show that 
\begin{equation}\label{identitytoproveSL}
D_\bv \log\big(e^{-c_{\rm L}\mc{S}_{S_g\cdot}}\mc{C}_{\bs{\alpha}({\bf p})}\big)({\rm Id})=0.
\end{equation} 
We compute the logarithmic derivative of
$\mc{C}_{\bs{\alpha}({\bf p})}$ directly by testing the
variation formula on primary states. Set
$a:=v_1'(0)$ and let
\[
{\bf u}'({\bf p},\bs{\nu},\widetilde{\bs{\nu}})
:=\chi({\bf p})\prod_{k=1}^b
{\bf 1}_{\{0\}}(\nu_k){\bf 1}_{\{0\}}(\widetilde{\nu}_k),
\qquad
\chi({\bf p}):=\prod_{k=1}^b\chi_k(p_k),
\]
where $\chi_k\in C_c^\infty(I^\circ)$. Define
\[
\hat{\mc{A}}(\bs{f})
:=\mc{A}_{S_g\cdot\bs{f},\bs{\alpha}}
\circ\bigotimes^b\Pi_{\mc{V}\otimes\bar{\mc{V}}}^{-1}.
\]
Since $\bs{\ell}^{(1)}_n{\bf u}'=0$ for all $n>0$, and $\ell_0^{(1)}{\bf u}'=\tilde{\ell}_0^{(1)}{\bf u}'=\Delta_{Q+{\rm i}p_1}{\bf u}'$, we have
\[
\big(\bs{\ell}_{{\rm v}_1}^{(1)}
+\widetilde{\bs{\ell}}_{{\rm v}_1}^{(1)}\big){\bf u}'
=-(a+\bar a)\Delta_{Q+{\rm i}p_1}{\bf u}'.
\]
Thus Proposition \ref{th:derivative_general} gives
\begin{equation}\label{dvhatA} 
D_{\bv}\hat{\mc{A}}({\rm Id}){\bf u}' = \hat{\mc{A}}({\rm Id}) \Big(
(a+\bar a)\Delta_{Q+{\rm i}p_1}-\frac{c_{\rm L}}{12}{\rm Re}(a)
+c_{\rm L}D_{\bv}\mc{S}_{S_g\cdot}({\rm Id})\Big){\bf u}'.
\end{equation}
On the other hand, Lemma \ref{matrix_coefficients} yields
\[\hat{\mc{A}}(\bs{f}){\bf u}'= \frac{1}{(2\pi)^{b/2}}\int_{I^b}\mc{C}_{\bs{\alpha}({\bf p})}(\bs{f})C^{\rm DOZZ}(\bs{\alpha}({\bf p}))
\chi({\bf p})\,\dd{\bf p}.\]
Using the explicit expression \eqref{constantC(P,g)}, we can differentiate this integral: the coefficient and its derivative
depend continuously on $\bs{f}$ and ${\bf p}$, with ${\bf p}$
restricted to a compact set. Comparing the result with \eqref{dvhatA} gives 
\[\begin{aligned}
0=\int_{I^b}\Big[D_{\bv}\mc{C}_{\bs{\alpha}({\bf p})}({\rm Id})-\mc{C}_{\bs{\alpha}({\bf p})}({\rm Id})
\Big((a+\bar a)\Delta_{Q+{\rm i}p_1}-\frac{c_{\rm L}}{12}{\rm Re}(a)
+c_{\rm L}D_{\bv}\mc{S}_{S_g\cdot}({\rm Id})
\Big)\Big]C^{\rm DOZZ}(\bs{\alpha}({\bf p}))\chi({\bf p})\,\dd{\bf p}.
\end{aligned}\]
Since the functions $\chi_k$ are arbitrary, the integrand
without $\chi$ vanishes in the sense of distributions.
Under the real Seiberg bounds conditions on $\bs{\alpha}$,
$C^{\rm DOZZ}(\bs{\alpha}({\bf p}))\neq0$ for
${\bf p}\in(0,\infty)^b$, and
$\mc{C}_{\bs{\alpha}({\bf p})}({\rm Id})>0$.
Dividing by these factors and using continuity in ${\bf p}$,
we obtain, for every ${\bf p}\in I^b$,
\[
D_{\bv}\log\left(
e^{-c_{\rm L}\mc{S}_{S_g\cdot}}
\mc{C}_{\bs{\alpha}({\bf p})}
\right)({\rm Id})
=
(a+\bar a)\Delta_{Q+{\rm i}p_1}
-\frac{c_{\rm L}}{12}{\rm Re}(a).
\]
For $v_1(z)=\mc{O}(z^2)$, this proves
\eqref{identitytoproveSL}. Notice that for $v_1(z)=\lambda_0z$,
this argument gives
\begin{equation}\label{identite_dWdC}
D_\bv \log\big(e^{-c_{\rm L}\mc{S}_{S_g\cdot}}\mc{C}_{\bs{\alpha}({\bf p})}\big)({\rm Id})=(\la_0+\bar{\la}_0)\Delta_{Q+{\rm i}p_1}-\frac{c_{\rm L}}{12}{\rm Re}(\la_0),
\end{equation}
Now, using the same approach as for proving \eqref{dtA^{1,0}} (based on \eqref{derivative_of_amplitude}), we now obtain in this case
\[D_{\bv}\mc{B}_{\tilde{\mc{X}}\cdot, \bs{\alpha}} ({\rm Id}){\bf u} = \mc{B}_{\tilde{\mc{X}},\bs{\alpha}}\big(\la_0\bs{\ell}_{0}^{(1)}+ \bar{\la}_0\tilde{\bs{\ell}}_{0}^{\,(1)}+\la_0\Delta_{Q+{\rm i}p_1}-\frac{c_{\rm L}}{12}{\rm Re}(\la_0)-D_\bv \log\big(e^{-c_{\rm L}\mc{S}_{S_g\cdot}}\mc{C}_{\bs{\alpha}({\bf p})}\big)({\rm Id})\big){\bf u}.\]
The term $\la_0\Delta_{Q+{\rm i}p_1}$ comes from the variation \eqref{variationGamma^t=0} which is not vanishing anymore.
By \cite[Proposition 3.6]{BGKRV} and \cite[Proposition 4.9]{GKRV20_bootstrap} we have $\tilde\bL_0H_{Q+{\rm i}p,\nu,0}=\Delta_{Q+{\rm i}p}H_{Q+{\rm i}p,\nu,0}$, so $\tilde{\bs{\ell}}_0^{(1)}{\bf u}=\Delta_{Q+{\rm i}p}{\bf u}$. Combining with \eqref{identite_dWdC}, we get the desired formula for the variation.
\end{proof}

We obtain, as a consequence of Lemma \ref{prop:variation_block_amplitude}, the differentiability of a general block amplitude under all possible deformations.
\begin{corollary}\label{differentiabiliteblock}
Let $\tilde{\mc{X}}=[(F,\mc{P},J,{\bf x},\bs{\zeta},{\bf Y})]\in \tT_{0,m,b}$ and  
$\tilde{\Theta}_{m,b}(\tilde{\mc{X}})=(\bs{\psi},\bs{\theta},\bs{\la})\in \tT^\eps_{m,b}$ for some $\eps>0$. Let $\bs{\sigma}=(\sigma_1,\dots,\sigma_b)$ be orientation signs with $\sigma_k=+1$ for $k\leq b^+$ and  $\sigma_k=-1$ for $k>b^+$, and let $\bs{\alpha}=(\alpha_{b+1},\dots,\alpha_3)$ be weights satisfying the Seiberg bounds. For each compact interval $I\subset (0,\infty)$, the following  map  is $C^1$:
\begin{equation}\label{blockchi_out} 
\mc{B}_{\tilde{\mc{X}}^{\bs{\sigma}}\cdot,\bs{\alpha}} :\left\{ \begin{array}{lll}
\mc{U}_\eps^b & \to & \mc{L}((h_0^{-1/2}L^2(I\times \mc{T}))^{\otimes b^-},(h_0^{1/2}L^2(I\times \mc{T}))^{\otimes b^+ }), \\
 \bs{f} & \mapsto & \mc{B}_{\tilde{\mc{X}}^{\bs{\sigma}}\cdot \bs{f},\bs{\alpha}}.
\end{array}\right.
\end{equation}
\end{corollary}
\begin{proof} Consider the map 
\begin{equation}\label{model_form_after_sigma} 
\bs{f}\in \mc{U}_\eps^b \mapsto  (\bs{\psi}^{\bs{f}},\bs{\theta}^{\bs{f}},\bs{\lambda}^{\bs{f}})=\tilde{\Theta}_{m,b}((\tilde{\mc{X}}^{\bs{\sigma}}\cdot \bs{f})^{\rm in})
\end{equation}
defined using \eqref{teich_to_teich_in}. It follows from Lemma \ref{PhiC1} that it is a $C^1$ map. 
The result then follows from the definition \eqref{propertyB}.
\end{proof}
Next, as a  consequence of Lemma \ref{prop:variation_block_amplitude}, we compute the variation of a block amplitude of a general building block when the deformations are by elements in $\mc{U}_\eps^-$ for the incoming components and in $\mc{U}_\eps^+$ for the outgoing ones.
\begin{corollary}\label{derivativeblock_cor2}
Let $\tilde{\mc{X}}=[(F,\mc{P},J,{\bf x},\bs{\zeta},{\bf Y})]\in \tT_{0,m,b}$ and  
$\tilde{\Theta}_{m,b}(\tilde{\mc{X}})=(\bs{\psi},\bs{\theta},\bs{\la})\in \tT^\eps_{m,b}$ for some $\eps>0$. Let $\bs{\sigma}=(\sigma_1,\dots,\sigma_b)$ be orientation signs with $\sigma_k=+1$ for $k\leq b^+$ and  $\sigma_k=-1$ for $k>b^+$, and let $\bs{\alpha}=(\alpha_{b+1},\dots,\alpha_3)$ be weights satisfying the Seiberg bounds. For each compact interval $I\subset (0,\infty)$, the map  \eqref{blockchi_out}
has for differential at  ${\rm Id}$ given by:
for all ${\bf u}\in (h_0^{-1/2}L^2(I \times \mc{T}))^{\otimes b^-}$
\begin{equation}\label{derivative_of_amplitudeB2}
D_\bv \mc{B}_{\tilde{\mc{X}}^{\bs{\sigma}}\cdot,\bs{\alpha}} ({\rm Id}){\bf u}=
\Big(\sum_{k=1}^{b^+}\bs{\ell}_{\v_k}^{(k)}\Big)\mc{B}_{\tilde{\mc{X}}^{\bs{\sigma}},\bs{\alpha}} 
 {\bf u} -\mc{B}_{\tilde{\mc{X}}^{\bs{\sigma}},\bs{\alpha}}\Big(\sum_{k=1}^{b^-} 
\bs{\ell}_{\v_{b^++k}}^{(k)}\Big){\bf u}
\end{equation}
where  $\bv=(\v_1,\dots,\v_b)$ with $\v_j \in T_{{\rm Id}}\mc{U}^+_\eps$ for $j=1,\dots,b^+$ and 
$\v_j \in T_{{\rm Id}}\mc{U}^-_\eps$ for $j=b^++1,\dots,b$.
\end{corollary}
\begin{proof} 
The map \eqref{model_form_after_sigma} turns out to be quite explicit in this case when restricted to $(\mc{U}_\eps^+)^{b^+}\times (\mc{U}_\eps^-)^{b^-}$. Indeed, it is given for $\bs{f}=(f_1,\dots,f_b)$ by 
\[ \psi_k^{\bs{f}}= \psi_{k}\circ f_{k} \textrm{ if }k=b^++1,\dots,b, \quad \psi_k^{\bs{f}}= 
\psi_{k}\circ (o\circ f_{k}\circ o)\textrm{ if }k=1,\dots,b^+,
\]
and $\theta_k^{\bs{f}}$ is the argument of $(\psi_k^{\bs{f}})'(0)$ defined uniquely in a neighbourhood of $\theta_k$ if $\mc{U}_\eps^\pm$ are small enough. As in the case of Lemma \ref{prop:variation_block_amplitude}, $\bs{\lambda}^{\bs{f}}$ is constant.
For $k=1,\dots,b^+$, using \eqref{Tangent_U_eps}, we write $\v_k=v_k(z)\pl_z$ with $v_k(z)=-z^{2}v_k^*(o(z))$ for some $v_k^*\in {\rm Hol}_\eps(\D)$ such that $v_k^*(0)=0$.
Denoting $v_k^*(z)=\sum_{n\geq 0}v_{kn}^*z^{n+1}$, we get $v_k(z)=-\sum_{n\geq 0}v_{kn}^*z^{-n+1}$ and then 
\begin{equation}\label{expression_for_ell_vk*}
-\bs{\ell}_{{\rm v}_k^*}^{\top}=\sum_{n\geq 0}v^*_{kn}\bs{\ell}_{-n}=\bs{\ell}_{{\rm v}_k}
\end{equation}
where ${\rm v}_k^*=v_k^*(z)\pl_z$.
By \eqref{propertyB}, we have for $u_1,\dots,u_b\in h_0^{-1/2}L^2(I\times \mc{T})$
\[ \cjg D_\bv \mc{B}_{\tilde{\mc{X}}^{\bs{\sigma}}\cdot ,\bs{\alpha}} ({\rm Id})(u_{b^++1}\otimes \dots \otimes u_b), \bar{u}_{1}\otimes \dots \bar{u}_{b^+}\cjd_{L^2(\R_+\times \mc{T})^{\otimes b^+}} =D_{(\bv_+^*,\bv_-)} \mc{B}_{\tilde{\mc{X}}\cdot,\bs{\alpha}}({\rm Id})(u_1\otimes\dots \otimes u_b),\]
where $\bv_-=({\rm v}_{b^++1},\dots,{\rm v}_b)$ and $\bv^*_+:=({\rm v}_1^*,\dots,{\rm v}_{b^+}^*)$.
We can then use Lemma \ref{prop:variation_block_amplitude} to write the right-hand side in terms of $\bs{\ell}_{{\rm v}_k^*}$ and $\bs{\ell}_{{\rm v}_k}$: for ${\bf u}=u_1\otimes \dots\otimes u_b$,  
\[\begin{split}
D_{(\bv_+^*,\bv_-)} \mc{B}_{\tilde{\mc{X}}\cdot,\bs{\alpha}}({\rm Id}){\bf u}&=-\mc{B}_{\tilde{\mc{X}},\bs{\alpha}}\Big(\sum_{k=1}^{b^+} \bs{\ell}_{{\rm v}_k^*}^{(k)}+ \sum_{k=1}^{b^-} 
\bs{\ell}_{{\rm v}_{b^++k}}^{(b^++k)}\Big){\bf u}\\
&= -\Big\cjg \mc{B}_{\tilde{\mc{X}}^{\bs{\sigma}},\bs{\alpha}}\Big(\sum_{k=1}^{b^-} 
\bs{\ell}_{{\rm v}_{b^++k}}^{(k)}\Big)(u_{b^++1}\otimes\dots \otimes u_b),\overline{u_1\otimes \dots\otimes u_{b^+}}\Big\cjd_{L^2(\R_+\times \mc{T})^{\otimes b^+}}\\
& \phantom{={}} - \Big\cjg \mc{B}_{\tilde{\mc{X}}^{\bs{\sigma}},\bs{\alpha}}(u_{b^++1}\otimes\dots \otimes u_b), \overline{\Big(\sum_{k=1}^{b^+} \bs{\ell}_{{\rm v}_k^*}^{(k)}\Big)(u_1\otimes \dots\otimes u_{b^+})}\Big\cjd_{L^2(\R_+\times \mc{T})^{\otimes b^+}},
\end{split}\]
and the last term can be rewritten, using \eqref{expression_for_ell_vk*}, as 
\[\Big\cjg \Big(\sum_{k=1}^{b^+} \bs{\ell}_{{\rm v}_k}^{(k)}\Big)\mc{B}_{\tilde{\mc{X}}^{\bs{\sigma}},\bs{\alpha}}(u_{b^++1}\otimes\dots \otimes u_b), \overline{u_1\otimes \dots\otimes u_{b^+}}\Big\cjd_{L^2(\R_+\times \mc{T})^{\otimes b^+}}\]
to obtain the desired result.  
\end{proof}

The next step is to compute the variation of block amplitudes with both outgoing and incoming boundaries 
in the direction of deformations in $\mc{U}_\eps^-$. 
\begin{lemma}[\textbf{Deformations in $\mc{U}_\eps^-$, general case}]\label{derivativeoutblock}
Let $\tilde{\mc{X}}=[(F,\mc{P},J,{\bf x},\bs{\zeta},{\bf Y})]\in \tT_{0,m,b}$ and  
$\tilde{\Theta}_{m,b}(\tilde{\mc{X}})=(\bs{\psi},\bs{\theta},\bs{\la})\in \tT^\eps_{m,b}$ for some $\eps>0$. Let $\bs{\sigma}=(\sigma_1,\dots,\sigma_b)$ be orientation signs with $\sigma_k=+1$ for $k\leq b^+$ and  $\sigma_k=-1$ for $k>b^+$, and let $\bs{\alpha}=(\alpha_{b+1},\dots,\alpha_3)$ be weights satisfying the Seiberg bounds. For each  compact interval $I\subset (0,\infty)$, the map  
\[
\mc{B}_{\tilde{\mc{X}}^{\bs{\sigma}}\cdot, \bs{\alpha}} :\left\{ \begin{array}{ccl}
(\mc{U}_\eps^-)^b& \to & \mc{L}((h_0^{-1/2}L^2(I\times \mc{T}))^{\otimes b^-}, (h_0^{1/2}L^2(I\times \mc{T}))^{\otimes b^+}), \\
 \bs{f} & \mapsto & \mc{B}_{\tilde{\mc{X}}^{\bs{\sigma}}\cdot \bs{f},\bs{\alpha}}  
\end{array}\right. 
\]
is $C^1$ with differential at  ${\rm Id}$ given by: 
for all ${\bf u}\in (h_0^{-1/2}L^2(I\times \mc{T}))^{\otimes b^-}$,
\begin{equation}\label{derivative_of_amplitudeB3}
\begin{split}
D_\bv \mc{B}_{\tilde{\mc{X}}^{\bs{\sigma}}\cdot ,\bs{\alpha}} ({\rm Id}){\bf u}&= 
\Big(\sum_{k=1}^{b^+}\bs{\ell}_{\v_k}^{(k)}\Big)\mc{B}_{\tilde{\mc{X}}^{\bs{\sigma}},\bs{\alpha}} 
 {\bf u} -\mc{B}_{\tilde{\mc{X}}^{\bs{\sigma}},\bs{\alpha}}\Big(\sum_{k=1}^{b^-} 
\bs{\ell}_{\v_{b^++k}}^{(k)}\Big){\bf u}\\
 & \phantom{={}} +\sum_{k=1}^{b^+}\Big( \frac{c_{\rm L}}{24\pi {\rm i}} \int_{\T} v_k(z) {\rm S}( \psi_k\circ o)(z) \dd z\Big)\mc{B}_{\tilde{\mc{X}}^{\bs{\sigma}},\bs{\alpha}} {\bf u}
\end{split}
\end{equation}
where  $\bv=(v_1\pl_z,\dots,v_b\pl_z)$ with $v_k\pl_z\in T_{{\rm Id}}\mc{U}_\eps^-$ for all $k$ and  ${\rm S}(\cdot)$ denotes the Schwarzian derivative.
\end{lemma}
\begin{proof} Let $S:=(\mc{P},J,{\bf x},\bs{\zeta}^{\bs{\sigma}})$ and $S_g:=(\mc{P},g,{\bf x},\bs{\zeta}^{\bs{\sigma}})$ for some admissible metric $g$. To simplify the exposition, we shall assume that $b^+=1$, the proof being the same when $b^+>1$.
The case ${\rm v}_1=0$ follows from Lemma \ref{prop:variation_block_amplitude}, so it remains to consider the case where 
${\rm v}_j=0$ for all $j>1$ and ${\rm v}_1\not=0$. For $\bs{f}\in(\mc{U}_\eps^-)^b$, let $(\bs{\psi}^{\bs{f}},\bs{\theta}^{\bs{f}},\bs{\lambda}^{\bs{f}})$ be defined by \eqref{model_form_after_sigma} and $S_g\cdot \bs{f}$ by \eqref{def_Sgf}.
Define also  
\[ \mc{S}_{S_g\cdot \bs{f}}:=S_{\rm L}^0(\mc{P}^{\bs{f}},g,g^{\bs{f}}),\quad \mc{C}_{\bs{\alpha}({\bf p})}(\bs{f}):=C(\bs{\psi}^{\bs{f}},g^{\bs{f}},\bs{\alpha}({\bf p})), \]
where $g^{\bs{f}}$ is a $C^1$ family of admissible metrics on $S_g\cdot\bs{f}$ and $C(\bs{\psi},g,\bs{\alpha}({\bf p}))$ is defined by \eqref{constantC(P,g)}.
We may assume that $\mc{P}=\hat{\C}\setminus \cup_{k=1}^b\psi_k(\D^\circ)$ (with $J=J_\C$) is under model form, with the marked points ${\bf x}^\bullet=(z^{\bullet}_{b+1},\dots,z^{\bullet}_3)$. By \eqref{extension_block_def2}, we have $\bs{\lambda}^{\bs{f}}=\bs{\lambda}+\log\Phi'_{\bs{f}}({\bf x}^\bullet)$ with $\log \Phi'_{\bs{f}}({\bf x}^\bullet):=(\log \Phi_{\bs{f}}'(z^\bullet_{b+1}),\dots,\log \Phi_{\bs{f}}'(z^\bullet_3))$ and $\Phi_{\bs{f}}$ the uniformising map of $S_g\cdot\bs{f}$ defined in Lemma \ref{PhiC1}. Here each $\log \Phi'_{\bs{f}}(z^{\bullet}_k)$ is well-defined since $\Phi'_{\bs{f}}(z^{\bullet}_k)$ belongs to a small neighbourhood of $1$ for $\mc{U}_\eps^-$ small enough. Thus we obtain for ${\bf u}:=u_2\otimes \dots \otimes u_b$ with $u_j\in h_0^{-1/2}L^2(I\times \mc{T})$
\[\mc{B}_{\tilde{\mc{X}}^{\bs{\sigma}}\cdot \bs{f},\bs{\alpha}}{\bf u}= (2\pi)^{b/2}\Pi_{\mc{V}}\mc{A}_{S_g\cdot \bs{f},\bs{\alpha}} 
\bigotimes^{b-1}\Pi_{\mc{V}}^*\Big( \frac{ \Gamma_{\bs{\alpha}}({\bf p},\bs{\psi}^{\bs{f}},\bs{\theta}^{\bs{f}},\bs{\la} +\log \Phi'_{\bs{f}}({\bf x}^\bullet))}{\mc{C}_{\bs{\alpha}({\bf p})}(\bs{f})C^{\rm DOZZ}(\bs{\alpha}(-\bs{\sigma}{\bf p}))}{\bf u}\Big) .\]
Define the function 
\begin{equation}\label{GammaPhi_alpha_p}
\bs{f}\in \mc{U}_\eps^b\mapsto  \Gamma^{\Phi}_{\bs{\alpha},{\bf p}}(\bs{f}):= \Gamma_{\bs{\alpha}}({\bf p},\bs{\psi}^{\bs{f}},\bs{\theta}^{\bs{f}}, \bs{\la}+\log \Phi'_{\bs{f}}({\bf x}^\bullet))
\end{equation}
and observe that (from \eqref{Gamma_alpha_mb} the $\bs{\la}$ term cancels out)
\[ \frac{\pl_t \Big(\Gamma_{\bs{\alpha}}({\bf p},\bs{\psi}^{{\rm Id}+t{\bf v}},\bs{\theta}^{{\rm Id}+t{\bf v}},\bs{\la} +\log \Phi'_{{\rm Id}+t\bv}({\bf x}^\bullet))\Big)\Big|_{t=0}}{\Gamma_{\bs{\alpha}}({\bf p},\bs{\psi},\bs{\theta},\bs{\la})}= \frac{D_\bv \Gamma^{\Phi}_{\bs{\alpha},{\bf p}}({\rm Id})}{\Gamma^{\Phi}_{\bs{\alpha},{\bf p}}({\rm Id})}.\]
We thus obtain by Proposition \ref{th:derivative_general} for ${\rm v}_1=v_1(z)\pl_z$ with $v_1\in {\rm Hol}_\eps(\D)$ and $v_1(0)=0$
\[\begin{split}
D_\bv \mc{B}_{\tilde{\mc{X}}^{\bs{\sigma}}\cdot,\bs{\alpha}}({\rm Id}){\bf u}&= \Pi_{\mc{V}}{\bH}_{{\rm v}_1}\mc{A}_{S_g,\bs{\alpha}}F+
c_{\rm L} \mc{B}_{\tilde{\mc{X}}^{\bs{\sigma}},\bs{\alpha}}\Big(\big(\frac{{\rm Re}(v_1'(0))}{12}+ D_\bv \mc{S}_{S_g\cdot }({\rm Id})\big){\bf u}\Big) \\
&\phantom{={}} + \mc{B}_{\tilde{\mc{X}}^{\bs{\sigma}},\bs{\alpha}} \Big( \frac{D_\bv \Gamma^{\Phi}_{\bs{\alpha},{\bf p}}({\rm Id})}{\Gamma^{\Phi}_{\bs{\alpha},{\bf p}}({\rm Id})}{\bf u}\Big)-\mc{B}_{\tilde{\mc{X}}^{\bs{\sigma}},\bs{\alpha}} \Big( \frac{D_\bv \mc{C}_{\bs{\alpha}({\bf p})}({\rm Id})}{\mc{C}_{\bs{\alpha}({\bf p})}({\rm Id})}{\bf u}\Big),\end{split}\]
where we have set 
\[F:=(2\pi)^{b/2}\otimes^{b-1}\Pi_{\mc{V}}^* \Big(\frac{\Gamma^{\Phi}_{\bs{\alpha},{\bf p}}({\rm Id})\, {\bf u}}{C^{\rm DOZZ}(\bs{\alpha}(-\bs{\sigma}{\bf p}))\mc{C}_{\bs{\alpha}({\bf p})}({\rm Id})}\Big)\in \otimes^{b-1} \mc{D}(\mc{Q})^{1,0}.\]
Let us compute $\Pi_{\mc{V}}\bH_{{\rm v}_1}$. By linearity, it suffices to consider the case ${\rm v}_1 =v_{1n} z^{n+1}\pl_z$ for some $v_{1n}\in \C$ and $n\geq 0$, so that $\bH_{{\rm v}_1}=-v_{1n}{\bf L}_n-\bar{v}_{1n}\tilde{{\bf L}}_n$.
Using the definition \eqref{PiV} of $\Pi_{\mc{V}}$ together with ${\bf L}_n^*={\bf L}_{-n}$ and $\tilde{{\bf L}}_n^*=\tilde{{\bf L}}_{-n}$, we get
\[\begin{split}
(\Pi_{\mc{V}}\bH_{{\rm v}_1}u)(p,\nu)=&\frac{1}{\sqrt{2\pi}}
\cjg \bH_{{\rm v}_1} u, H_{Q+{\rm i}p,\nu,0}\cjd_{\mc{H}}
=-\frac{1}{\sqrt{2\pi}}
\cjg u, (\bar{v}_{1n}{\bf L}_{-n}+v_{1n} \tilde{\bf L}_{-n})H_{Q+{\rm i}p,\nu,0}\cjd_{\mc{H}}\\
=&  -v_{1n}(\bs{\ell}_{n}\circ \Pi_{\mc{V}}u)(p,\nu)-\frac{\bar{v}_{1n}}{\sqrt{2\pi}}\Big\cjg u,\sum_{\tilde{\nu}\in \mc{T}, |\tilde{\nu}|=n}F_{Q+{\rm i}p}^{1/2}(n,\tilde{\nu})H_{Q+{\rm i}p,\nu,\tilde{\nu}}\Big\cjd_{\mc{H}},
\end{split}\]
where $n$ denotes the Young diagrams with the single entry $n$, and we used \eqref{Lnln} and \eqref{matrixln} to extract the first term in the second line. If $n>0$, this gives, using Lemma~\ref{matrix_coefficients}, 
\[ \Pi_{\mc{V}}{\bH}_{{\rm v}_1}\mc{A}_{S_g,\bs{\alpha}}F=\bs{\ell}_{{\rm v}_1} \mc{B}_{\tilde{\mc{X}}^{\bs{\sigma}},\bs{\alpha}}{\bf u}
-\mc{B}_{\tilde{\mc{X}}^{\bs{\sigma}},\bs{\alpha}}(
\bbar{v_{1n} w'_{\bs{\psi}}(\bs{\alpha}({\bf p}),\bs{\nu}_n)}{\bf u})\]
where $\bs{\nu}_n=(n,0,\dots,0)\in \mc{T}^b$, and 
\[w'_{\bs{\psi}}(\bs{\alpha}({\bf p}),\bs{\nu}_n)=\sum_{\tilde{\nu}\in \mc{T}, |\tilde{\nu}|=n}
F_{Q+{\rm i}p}^{1/2}(n,\tilde{\nu})w_{\bs{\psi}}(\bs{\alpha}({\bf p}),(\tilde{\nu},0,\dots,0)).\]
Similarly, if $n=0$ (since $\tilde{{\bf L}}_0\Psi_{Q+{\rm i}p,\nu',0}=\Delta_{Q+{\rm i}p}\Psi_{Q+{\rm i}p,\nu',0}$) 
\[ \Pi_{\mc{V}}{\bH}_{{\rm v}_1}\mc{A}_{S_g,\bs{\alpha}}F=\bs{\ell}_{{\rm v}_1}
\mc{B}_{\tilde{\mc{X}}^{\bs{\sigma}},\bs{\alpha}}{\bf u}
-\bbar{v}_{10}\Delta_{Q+{\rm i}p_1}\mc{B}_{\tilde{\mc{X}}^{\bs{\sigma}},\bs{\alpha}}
{\bf u}.\]
By  the same reasoning as to get \eqref{identite_dWdC}, we obtain (with $\bv=(v_1\pl_z,0,\dots,0)$)
\begin{align*}
& \textrm{ if }n>0,\quad c_{\rm L}\frac{{\rm Re}(v_1'(0))}{12}-D_\bv \log(e^{-c_{\rm L}\mc{S}_{S_g\cdot}}\mc{C}_{\bs{\alpha}({\bf p})})({\rm Id})-\bbar{v_{1n}w'_{\bs{\psi}}(\bs{\alpha}({\bf p}),\bs{\nu}_n)} -v_{1n} w'_{\bs{\psi}}(\bs{\alpha}({\bf p}),\bs{\nu}_n)=0,\\
& \textrm{ if }n=0,\quad c_{\rm L}\frac{{\rm Re}(v_1'(0))}{12}-D_\bv \log(e^{-c_{\rm L}\mc{S}_{S_g\cdot}}\mc{C}_{\bs{\alpha}({\bf p})})({\rm Id}) -(\bbar{v}_{10} +v_{10}) \Delta_{Q+{\rm i}p_1}=0,
\end{align*}
thus we deduce that for a general vector field ${\rm v}_1=v_1(z)\pl_z$ with $v_1(z)=\sum_{n=0}^\infty v_{1n}z^{n+1}\in {\rm Hol}_\eps(\D)$,
\begin{equation}\label{laststep}
D_\bv \mc{B}_{\tilde{\mc{X}}^{\bs{\sigma}}\cdot,\bs{\alpha}}({\rm Id})({\bf u})=\bs{\ell}_{{\rm v}_1} \mc{B}_{\tilde{\mc{X}}^{\bs{\sigma}},\bs{\alpha}}{\bf u}+\mc{B}_{\tilde{\mc{X}}^{\bs{\sigma}},\bs{\alpha}}\Big(\big( \frac{D_\bv \Gamma^{\Phi}_{\bs{\alpha},{\bf p}}}{\Gamma^{\Phi}_{\bs{\alpha},{\bf p}}}({\rm Id})
+\sum_{n=1}^\infty v_{1n} w'_{\bs{\psi}}(\bs{\alpha}({\bf p}),\bs{\nu}_n)+v_{10}\Delta_{Q+{\rm i}p_1}\big){\bf u}\Big).\end{equation}
To prove \eqref{derivative_of_amplitudeB3}, it thus suffices to prove that for each $v_1(z)=\sum_{n\geq 0}v_{1n}z^{n+1}\in {\rm Hol}_{\eps}(\D)$,  
\begin{equation}\label{dGamma_wn}
  \frac{D_\bv \Gamma^{\Phi}_{\bs{\alpha},{\bf p}}}{\Gamma^{\Phi}_{\bs{\alpha},{\bf p}}}({\rm Id})
+\sum_{n=1}^\infty v_{1n} w'_{\bs{\psi}}(\bs{\alpha}({\bf p}),\bs{\nu}_n)+v_{10}\Delta_{Q+{\rm i}p_1}= \frac{c_{\rm L}}{24\pi {\rm i}} \int_{\T} v_1(z) {\rm S}(\psi_1\circ o)(z) \d z,
\end{equation}
where ${\rm S}(\cdot)$ denotes the Schwarzian derivative and $\bs{\psi}=(\psi_1,\dots,\psi_b)$. By linearity, it suffices to consider the case where $v_1$ is a polynomial of $z$.  We first prove it when $v_1(z)=z$. For this case, a direct computation gives 
\[\int_{\T} v_1(z) {\rm S}(\psi_1\circ o)(z) \d z=0, \quad   \frac{D_{({\rm v}_1,0,\dots,0)}\Gamma^{\Phi}_{\bs{\alpha},{\bf p}}}{\Gamma^{\Phi}_{\bs{\alpha},{\bf p}}}({\rm Id})=-\Delta_{Q+{\rm i}p_1}\]
showing \eqref{dGamma_wn}.
We will use both notations 
$DF({\rm Id},0)(\bv , {\bf t})$ and $D_{(\bv , {\bf t})}F({\rm Id},0)$ for the differential at $({\rm Id},0)$ of a $C^1$ function 
$F:(\mc{U}_\eps^-)^b \times \C^m\to \C$ in the direction $(\bv , {\bf t})$.
Let us consider a meromorphic vector field ${\rm a}=a\pl_z$ on $\hat{\C}$ with a  single pole of order $N-1\geq 1$ at $z=z_1^\bullet$ and vanishing at $z_2^\bullet$ and $z_3^\bullet$, and write 
${\rm a}_j=a_j\pl_z:=(\psi_j)^*(a\pl_z)$ for $j=1,\dots,b$. 
Let $a_1(z)=\sum_{n=-N}^\infty a_{1n}z^{n+1}$ denote the Laurent expansion of $a_1$. 
Now, recall the result of \cite[Proposition 11.3]{GKRV21_Segal}: the following identity holds 
\[\begin{split} 
\sum_{n=-N}^{-1} a_{1n}w'_{\bs{\psi}}(\bs{\alpha}({\bf p}),\bs{\nu}_{|n|}) +\Delta_{\alpha_1({\bf p})}a_{10}& = -\frac{c_{\rm L}}{12}{\rm Res}_{z_1^\bullet}(a {\rm S}(\psi_1^{-1}))
 - \Delta_{\alpha_2({\bf p})}a'(z_2^\bullet)-\Delta_{\alpha_3({\bf p})}a'(z_3^\bullet),
\end{split}
\]
where ${\rm S}(\psi_1^{-1})$ is the Schwarzian derivative of $\psi_1^{-1}$. We may rewrite this identity as 
\[
\begin{split}
\sum_{n=-N}^{-1} a_{1n}w_{\bs{\psi}}'(\bs{\alpha}({\bf p}),\bs{\nu}_{|n|})=& -\frac{D_{({\bf a},{\bf t})}\Gamma_{\bs{\alpha}}({\bf p},(\bs{\psi},\bs{\theta})\cdot,\cdot)({\rm Id},0)}{\Gamma_{\bs{\alpha}}({\bf p},\bs{\psi},\bs{\theta},0)} -\Delta_{Q+{\rm i}p_1}a_{10}-\frac{c_{\rm L}}{12}{\rm Res}_{z_1^\bullet}(a{\rm S}(\psi_1^{-1})),
\end{split}
\]
where ${\bf a}=(0,{\rm a}_2,\dots,{\rm a}_b)$ and ${\bf t}=(a'(z^{\bullet}_{b+1}),\dots,a'(z^{\bullet}_3))$. Let ${\rm a}_1^*=a_1^*\pl_z :=o^*(a_1\pl_z)=(\psi_1\circ o)^*(a\pl_z)$ with $a_1^*(z)=-z^2 a_1(1/z)=-\sum_{n=-N}^\infty a_{1n}z^{-n+1}$ and let $(a_1^*)_+:=-\sum_{n=-N}^0a_{1n}z^{-n+1}$. Using that 
\[{\rm S}(\psi_1^{-1})(z)= {\rm S}(o^{-1}\circ\psi_1^{-1})=-(((\psi_1\circ o)^{-1})')^2{\rm S}(\psi_1\circ o)\circ (\psi_1\circ o)^{-1}\] 
and using that $o$ reverses orientation on $\T$, we observe that 
\[{\rm Res}_{z_1^\bullet}(a{\rm S}(\psi_1^{-1})) =\frac{1}{2\pi{\rm i}}\int_{\T} a_1^*(z) {\rm S}(\psi_1\circ o)(z) \d z=\frac{1}{2\pi{\rm i}}\int_{\T} (a_1^*)_+(z) {\rm S}(\psi_1\circ o)(z) \d z.\]
Hence, we obtain 
\begin{equation}\label{secondreduction}
\begin{split}
\sum_{n=-N}^{-1} a_{1n}w'_{\bs{\psi}}(\bs{\alpha}({\bf p}),\bs{\nu}_{|n|})=& -\frac{D_{({\bf a},{\bf t})}\Gamma_{\bs{\alpha}}({\bf p},(\bs{\psi},\bs{\theta})\cdot,\cdot)({\rm Id},0)}{\Gamma_{\bs{\alpha}}({\bf p},\bs{\psi},\bs{\theta},0)} -\Delta_{Q+{\rm i}p_1}a_{10}\\
&-\frac{c_{\rm L}}{24\pi {\rm i}}\int_{\T} (a_1^*)_+(z) {\rm S}(\psi_1\circ o)(z) \dd z.
\end{split}\end{equation}
Now let us consider $D_{{\bf a}_1^*} \Gamma^{\Phi}_{\bs{\alpha},{\bf p}}({\rm Id})$ with ${\bf a}_1^*=({\rm  a}_1^*,0,\dots,0)$. 
If $f_t$ denotes the flow of ${\rm a}$ for small $t$, the complex building block $\mc{P}^t$ obtained by deforming only  the
parametrisation of the first boundary $\psi_1|_{\T}$ to  $(f_t\circ \psi_1\circ o)|_{\T}$ is 
biholomorphic to $f_{-t}(\mc{P}^t)$, where now only the $j=2,\dots,b$ boundary parametrisations are depending on $t$ and are given by $(f_{-t}\circ \psi_j)|_{\T}$. Denote $\bs{f}_t=((\psi_1\circ o)^{-1}\circ f_t\circ \psi_1\circ o,{\rm Id}\dots,{\rm Id})$ and notice that $\pl_t\bs{f}_t|_{t=0}=(a_1^*,0,\dots,0)$. Then 
$\mc{X}^{\bs{\sigma}}\cdot \bs{f}_t$ has for model form (using the notation \eqref{model_form_after_sigma})
\[ (\bs{\psi}^{\bs{f}_t},\bs{\theta}^{\bs{f}_t})=\big((\psi_1,f_{-t}\circ \psi_2,\dots,f_{-t}\circ \psi_b), (\theta_1,\theta_2+{\rm arg}f'_{-t}(z_2^\bullet),\dots,\theta_b+{\rm arg}f'_{-t}(z_b^\bullet))\big),\] 
since the uniformising map $\Phi_{\bs{f}_t}$ of Lemma \ref{PhiC1} is given by 
$\Phi_{\bs{f}_t}=f_{-t}$. Observe that 
\[\begin{gathered}
(\bs{\psi}^{\bs{f}_t},\bs{\theta}^{\bs{f}_t})=(\bs{\psi},\bs{\theta})\cdot {\bf h}_t ,\quad 
{\bf h}_t:=({\rm Id},\psi_2^{-1}\circ f_{-t}\circ \psi_2,\dots,\psi_b^{-1}\circ f_{-t}\circ \psi_b),
\end{gathered}\]
and $\pl_t{\bf h}_t|_{t=0}=(0,-{\rm a}_2,\dots,-{\rm a}_b)=-{\bf a}$.
It follows that 
\[ \begin{split} 
\frac{D_{{\bf a}_1^*} \Gamma^{\Phi}_{\bs{\alpha},{\bf p}}}{\Gamma^{\Phi}_{\bs{\alpha},{\bf p}}}({\rm Id})=& -\frac{D_{{\bf a}}\Gamma_{\bs{\alpha}}({\bf p},(\bs{\psi},\bs{\theta})\cdot)}{\Gamma_{\bs{\alpha}}({\bf p},\bs{\psi},\bs{\theta})}({\rm Id})-\sum_{j=b+1}^3 \Delta_{\alpha_j}a'(z^{\bullet}_j)\\
=& -\frac{D_{({\bf a},{\bf t})}\Gamma_{\bs{\alpha}}({\bf p},(\bs{\psi},\bs{\theta})\cdot,\cdot)}{\Gamma_{\bs{\alpha}}({\bf p},\bs{\psi},\bs{\theta},0)}({\rm Id},0).
\end{split}\]
Combining with \eqref{secondreduction}, we obtain 
\begin{equation}\label{dGamma_wn_case1}
\frac{D_{{\bf a}_1^*} \Gamma^{\Phi}_{\bs{\alpha},{\bf p}}}{\Gamma^{\Phi}_{\bs{\alpha},{\bf p}}}({\rm Id})-\sum_{n=-N}^{-1} a_{1n}w'_{\bs{\psi}}(\bs{\alpha}({\bf p}),\bs{\nu}_{|n|})-\Delta_{Q+{\rm i}p_1}a_{10}=\frac{c_{\rm L}}{24\pi {\rm i}}\int_{\T} (a_1^*)_+(z) {\rm S}(\psi_1\circ o)(z) \d z.
\end{equation}
Observe that for any holomorphic vector field $\hat{{\rm v}}_1=\hat{v}_1\pl_z$ in ${\rm Hol}_\eps(\D^*)$ with $\hat{v}_1(z)=\sum_{n=1}^\infty \hat{v}_{1n}z^{-n+1}$, the flow $\hat{f}_t$ belongs to ${\rm BiHol}_\eps(\D^*)$ and we have, for $\widehat{\bs{f}}_t=(\hat{f}_t,{\rm Id},\dots,{\rm Id})$,
\[ (\bs{\psi}^{\widehat{\bs{f}}_t},\bs{\theta}^{\widehat{\bs{f}}_t})=(\bs{\psi},\bs{\theta})\cdot \widehat{{\bf h}}_t, \quad \widehat{{\bf h}}_t=(o\circ \hat{f}_t\circ o,{\rm Id},\dots,{\rm Id}),\]
and $\hat{{\bf v}}:=\pl_t\widehat{{\bf h}}_t|_{t=0}=(\hat{{\rm v}}_1^*,0,\dots,0)$ with $\hat{{\rm v}}_1^*=\hat{v}_1^*(z)\pl_z$ for 
$\hat{v}_1^*(z)=-\sum_{n=1}^\infty \hat{v}_{1n}z^{n+1}$. Since also $\Phi_{\widehat{\bs{f}}_t}={\rm Id}$, we get 
\[\frac{D_{(\hat{{\rm v}}_1,0,\dots,0)} \Gamma^{\Phi}_{\bs{\alpha},{\bf p}}}{\Gamma^{\Phi}_{\bs{\alpha},{\bf p}}}({\rm Id})=0.\]
Adding this to \eqref{dGamma_wn_case1}, we obtain, for ${\bf v}:=({\rm v}_1,0,\dots,0)$ with 
${\rm v}_1:=({\rm a}_1^*)_+=\sum_{n=0}^Nv_{1n}z^{n+1}\pl_z$ which is polynomial in $z$ with $-a_{1n}=v_{1|n|}$ for $n=-N,\dots,-1$, 
\[\frac{D_{\bv} \Gamma^{\Phi}_{\bs{\alpha},{\bf p}}}{\Gamma^{\Phi}_{\bs{\alpha},{\bf p}}}({\rm Id})+\sum_{n=1}^{N} v_{1n}w_{\bs{\psi}}'(\bs{\alpha}({\bf p}),\bs{\nu}_{|n|})+\Delta_{Q+{\rm i}p_1}v_{10}=\frac{c_{\rm L}}{24\pi {\rm i}}\int_{\T} v_1(z) {\rm S}(\psi_1\circ o)(z) \dd z.\]
We have thus shown that \eqref{dGamma_wn} holds for any 
polynomials $v_1=\sum_{n=0}^N v_{1n}z^{n+1}$ for which there exists a meromorphic vector field ${\rm v}=v\pl_z$ on $\hat{\C}$, vanishing at $z_2^\bullet,z_3^\bullet$, with only pole at $z=z_1^\bullet$ and such that  $\psi_1^*(v\pl_z)=-\sum_{n=0}^N v_{1n}z^{-n+1}\pl_z+\mc{O}(z^2)$ near $z=0$. By induction on the order $N-1$ 
of the pole at $z=z_1^\bullet$ and since $\psi_1(z)=z_1^\bullet+\omega z+\mc{O}(z^2)$ for some $\omega\not=0$, it is direct to see that for 
any $v_{11},\dots,v_{1N}$ with $N\geq 1$, there  exists a meromorphic vector field ${\rm v}=v\pl_z$ on $\hat{\C}$ with poles at $z_1^\bullet$, vanishing at $z_2^\bullet,z_3^\bullet$ such that $\psi_1^*(v\pl_z)=-\sum_{n=1}^N v_{1n}z^{-n+1}\pl_z+cz+\mc{O}(z^2)$ for some $c\in \C$.
Since we already know that \eqref{dGamma_wn} holds for $v_1(z)=cz$, this shows \eqref{dGamma_wn} for all polynomials. Therefore it holds 
 for all $v_1\in {\rm Hol}_\eps(\D)$ vanishing at $z=0$ by density, since polynomials are dense in the relevant holomorphic topology on every slightly smaller disk and both sides of \eqref{dGamma_wn} depend continuously on $v_1$.
 By \eqref{laststep}, the proof of the Lemma is done.
\end{proof}

\begin{remark}\label{generalisation521}
We notice that \eqref{dGamma_wn} also holds for ${\rm v}=v(z)\pl_z$ with $v\in {\rm Hol}_\eps(\A)$. Indeed, to prove this, it suffices by linearity to consider the case
$v(z)=v_{-n}z^{-n+1}$ with $n>0$, and \eqref{dGamma_wn} amounts to prove 
\begin{equation}\label{DGamma=0}
D_{\bv} \Gamma^{\Phi}_{\bs{\alpha},{\bf p}}=0.
\end{equation}
For such $v$ and any small $t\in \R$ (and with the same notations as in the proof of Lemma \ref{derivativeoutblock}), and letting 
\[(\bs{\psi}^t,\bs{\theta}^t,\bs{\la}^t):=\tilde{\Theta}_{m,b}((\tilde{\mc{X}}^{\bs{\sigma}}\cdot ({\rm Id}+tv,
{\rm Id},\dots,{\rm Id}))^{\rm in})
\]
we have, as $t\to 0$,
\[\bs{\psi}^t=\bs{\psi}^0-t(v_{-n}z^{n+1}(\psi_1^0)'(z),0,\dots,0)+\mc{O}(t^2), \quad \bs{\la}^t=\bs{\la}^0+\mc{O}(t^2).\]
It follows directly that \eqref{DGamma=0} holds.
\end{remark}
Finally, we compute the variation of block amplitudes by deformations in $\mc{U}_\eps^+$.
\begin{lemma}\label{derivativeinblock+}
Let $\tilde{\mc{X}}=[(F,\mc{P},J,{\bf x},\bs{\zeta},{\bf Y})]\in \tT_{0,m,b}$ and  
$\tilde{\Theta}_{m,b}(\tilde{\mc{X}})=(\bs{\psi},\bs{\theta},\bs{\la})\in \tT^\eps_{m,b}$ for some $\eps>0$. Let $\bs{\sigma}=(\sigma_1,\dots,\sigma_b)$ be orientation signs with $\sigma_k=+1$ for $k\leq b^+$ and  $\sigma_k=-1$ for $k>b^+$, and let $\bs{\alpha}=(\alpha_{b+1},\dots,\alpha_3)$ be weights satisfying the Seiberg bounds. For each compact interval $I\subset (0,\infty)$, the map  
\[
\mc{B}_{\tilde{\mc{X}}^{\bs{\sigma}}\cdot, \bs{\alpha}} :\left\{ \begin{array}{ccl}
(\mc{U}_\eps^+)^b& \to & \mc{L}((h_0^{-1/2}L^2(I\times \mc{T}))^{\otimes b^-}, (h_0^{1/2}L^2(I\times \mc{T}))^{\otimes b^+}), \\
 \bs{f} & \mapsto & \mc{B}_{\tilde{\mc{X}}^{\bs{\sigma}}\cdot \bs{f},\bs{\alpha}}  
\end{array}\right. 
\]
is $C^1$ with differential at  ${\rm Id}$ given by: for all ${\bf u}\in (h_0^{-1/2}L^2(I\times \mc{T}))^{\otimes b^-}$,
\[\begin{split}
D_\bv \mc{B}_{\tilde{\mc{X}}^{\bs{\sigma}}\cdot ,\bs{\alpha}} ({\rm Id}){\bf u}={}& 
\Big(\sum_{k=1}^{b^+}\bs{\ell}_{{\rm v}_k}^{(k)}\Big)\mc{B}_{\tilde{\mc{X}}^{\bs{\sigma}},\bs{\alpha}}{\bf u}-\mc{B}_{\tilde{\mc{X}}^{\bs{\sigma}},\bs{\alpha}} \Big(\sum_{k=1}^{b^-}
\bs{\ell}_{{\rm v}_{b^++k}}^{(k)}\Big){\bf u}\\
 & - \sum_{k=b^++1}^b\Big( \frac{c_{\rm L}}{24\pi {\rm i}} \int_{\T} v_k(z) {\rm S}(\psi_k)(z) \dd z\Big)\mc{B}_{\tilde{\mc{X}}^{\bs{\sigma}},\bs{\alpha}}{\bf u}
\end{split}\]
where $\bv=(v_1\pl_z,\dots,v_b\pl_z)$ with $v_k\pl_z\in T_{{\rm Id}}\mc{U}_\eps^+$ for all $k$ and  ${\rm S}(\cdot)$ denotes the Schwarzian derivative. 
\end{lemma}
\begin{proof} 
We use the same argument as in Corollary \ref{derivativeblock_cor2}.  By linearity and using Corollary \ref{derivativeblock_cor2} it suffices to consider the case where ${\rm v}_j=0$ for $j\leq b^+$. To simplify the exposition, we assume that $(b^+,b^-)=(0,b)$, i.e. the boundary components are all incoming. In that case $\bs{\sigma}=(-1,\dots,-1)$ 
and $\tilde{\mc{X}}^{\bs{\sigma}}=\tilde{\mc{X}}$. Let $\tilde{\mc{X}}^+:=\tilde{\mc{X}}^{\bs{\sigma}_+}$ where  $\bs{\sigma}_+=(1,\dots,1)$, i.e. $\tilde{\mc{X}}^{+}$ has only outgoing boundary components. 
We have 
\[ D_{\bv}\mc{B}_{\tilde{\mc{X}}\cdot,\bs{\alpha}}({\rm Id}){\bf u}=\pl_t \mc{B}_{\tilde{\mc{X}}\cdot \bs{f}_t,\bs{\alpha}}{\bf u}\big|_{t=0} =
\pl_t\big\cjg \mc{B}_{\tilde{\mc{X}}^+\cdot \bs{f}^*_t,\bs{\alpha}},\bbar{{\bf u}}\big\cjd_{L^2(I\times \mc{T})^{\otimes b}}\big|_{t=0}\]
where $\bs{f}_t(z)=(z+tv_1(z),\dots,z+tv_b(z))$ for small $t$ and 
\[\bs{f}_t^*(z)=\Big(\frac{z}{1+tzv_1(o(z))},\dots,\frac{z}{1+tzv_b(o(z))}\Big)= (z+tv_1^*(z),\dots,z+tv_b^*(z))+\mc{O}(t^2)\]
 with $v_k^*(z)=-z^{2}v_k(1/z)$ which belongs to ${\rm Hol}_\eps(\D)$ and satisfies $v_k^*(0)=0$. We also denote $\bv^*:=({\rm v}_1^*,\dots,{\rm v}_b^*)$ with ${\rm v}_k^*=v_k^*\pl_z$.
We can apply Lemma \ref{derivativeoutblock} to deduce that 
\[\begin{split}
D_{\bv}\mc{B}_{\tilde{\mc{X}}\cdot,\bs{\alpha}}({\rm Id}){\bf u}&=\cjg D_{\bv^*}\mc{B}_{\tilde{\mc{X}}^+\cdot,\bs{\alpha}}({\rm Id}),\bbar{{\bf u}}\cjd_{L^2(I\times \mc{T})^{\otimes b}}\\
&= \Big\cjg\Big(\sum_{k=1}^{b}\bs{\ell}_{{\rm v}^*_k}^{(k)}\Big) \mc{B}_{\tilde{\mc{X}}^+,\bs{\alpha}},\bbar{{\bf u}}\Big\cjd_{L^2(I\times \mc{T})^{\otimes b}}+\sum_{k=1}^{b}\Big( \frac{c_{\rm L}}{24\pi {\rm i}} \int_{\T} v_k^*(z) {\rm S}( \psi_k\circ o)(z) \dd z\Big)\cjg \mc{B}_{\tilde{\mc{X}}^+,\bs{\alpha}},\bbar{{\bf u}}\cjd_{L^2(I\times \mc{T})^{\otimes b}}.
\end{split}
\]
Since $v_k^*(z)=\sum_{n\geq 0}v_{kn}^*z^{n+1}$ and $v_k(z)=-\sum_{n\geq 0}v_{kn}^*z^{-n+1}$, we have 
$-\bs{\ell}_{\v_k^*}^{\top}=\sum_{n\geq 0}v^*_{kn}\bs{\ell}_{-n}=\bs{\ell}_{{\rm v}_k}$. We thus obtain
\[D_{\bv}\mc{B}_{\tilde{\mc{X}}\cdot,\bs{\alpha}}({\rm Id}){\bf u}=-\mc{B}_{\tilde{\mc{X}},\bs{\alpha}}\Big(\sum_{k=1}^{b}\bs{\ell}_{{\rm v}_k}^{(k)}\Big){\bf u}-\sum_{k=1}^{b}\Big( \frac{c_{\rm L}}{24\pi {\rm i}} \int_{\T} v_k(z) {\rm S}( \psi_k)(z) \dd z\Big)\mc{B}_{\tilde{\mc{X}},\bs{\alpha}}{\bf u},\]
where we used a change of variable $z\mapsto o(z)=1/z$ in the last integral.
This completes the proof.
\end{proof}

\begin{proof}[Proof of Proposition \ref{general_derivative_block_amp}]
Now, by linearity, we can gather the results of Lemma \ref{prop:variation_block_amplitude}, Corollary \ref{differentiabiliteblock}, Lemma \ref{derivativeoutblock} and Lemma \ref{derivativeinblock+}: since for $v(z)=\sum_{n\in \Z} v_nz^{n+1} \in {\rm Hol}_\eps(\A)$ (for some $\eps>0$ small) 
we have 
\[\int_{\T} v(z) {\rm S}(\psi_j)(z) \dd z = \int_{\T} v_-(z) {\rm S}(\psi_j)(z) \dd z\]
where $v_-(z)=\sum_{n<-1}v_nz^{n+1}$, we obtain the general variation formula for the conformal block amplitudes of complex building blocks announced in Proposition \ref{general_derivative_block_amp}.
\end{proof}

\section{Conformal blocks on Teichm\"uller space}

In this section, we will construct the conformal blocks of a marked Riemann surface equipped with a marked complex building block decomposition.  This procedure is done in Section 8.2 of \cite{GKRV21_Segal} without the marking but at the cost of only a local definition of the conformal blocks as functions on the deformation space. In the present work, we want to define globally on Teichm\"uller space the conformal blocks as holomorphic sections of line bundles. We use several times implicitly in this section the classical fact (which follows from  Cauchy's formula), that holomorphy of a function $F$ defined on an open set of $\C^n$ with values in a Hilbert space follows from holomorphy in the weak topology together with continuity of $F$ in the strong topology of the Hilbert space.

\subsection{Gluing of conformal blocks of marked complex building blocks}\label{s:gluing_conf_blocks}

In this section, we construct the conformal blocks of marked Riemann surfaces with a marked complex building block decomposition, by composing the conformal block amplitudes of the marked complex building blocks.

Let $\mc{G}$ be a gluing graph in the sense of Definition \ref{def:gluing_graph} with $i_0$ internal edges and $j_0$ internal vertices, where 
each internal vertex ${\rm ve}_j$ is of type
$b_j=3-m_j\in\{1,2,3\}$, i.e.\ it is incident to $m_j$ external marked edges, whose half-edges at ${\rm ve}_j$ are $
(j,b_j+1),\ldots,(j,3).$
We denote by $b:=\sum_{j=1}^{j_0}b_j-2i_0$ the  number of external edges of $\mc{G}$ associated to boundary circles.

Consider the Hilbert space 
\[ \mc{H}_{\mc{T}}:=L^2(\mc{T})=L^2(\mc{T},\mu_{\mc{T}})\]
where $\mu_{\mc{T}}$ is the measure putting mass $1$ on each $\nu\in \mc{T}$. It has 
 orthonormal basis $(\textbf{1}_{\nu})_{\nu \in \mc{T}}$ where $\textbf{1}_{\nu}$ denotes the characteristic function of $\{\nu\}$. 
Denote by $\mc{H}_{\mc{T}}^*$ its dual, which identifies with $\mc{H}_{\mc{T}}$, and let 
$\delta_{\nu}: \mc{H}_{\mc{T}}\to \C$ be defined by $\delta_{\nu}(f):=f(\nu)=\cjg f,{\bf 1}_{\nu}\cjd_{\mc{H}_{\mc T}}$. Then $(\delta_{\nu})_{\nu \in \mc{T}}$ is the dual basis to $(\textbf{1}_{\nu})_{\nu \in \mc{T}}$.  
For a Hilbert space $\mc{H}$ and $b\in\N$, we shall write $\otimes^b\mc{H}=\mc{H}^{\otimes b}$ for the tensor product of $b$ copies of $\mc{H}$, and we shall identify $\otimes^b\mc{H}^*\simeq(\otimes^b\mc{H})^*$. Here, we identify the Hilbertian tensor products $\otimes^b\mc{H}_{\mc{T}}$ and $\otimes^b\mc{H}_{\mc{T}}^*$ with $L^2(\mc{T}^b,\mu_{\mc{T}}^b)$. For $\eps>0$, define 
$e^{-\eps|\nu|}\mc{H}_{\mc{T}}:=\{f: \mc{T}\to \C \,|\, e^{\eps|\nu|}f\in L^2(\mc{T})\}$ and 
$(e^{\eps|\nu|}\mc{H}_{\mc{T}})^*\simeq e^{-\eps|\nu|}\mc{H}_{\mc{T}}$ its dual. An element in $\bigotimes_{j=1}^{j_0}( \bigotimes^{b_j}(e^{\eps|\nu|}\mc{H}_{\mc{T}})^*)$ is identified with a function  $W:(\nu_1,\dots,\nu_{b'})\to W(\nu_1,\dots,\nu_{b'})$ in $e^{-\eps(\sum_{i=1}^{b'} |\nu_i|)}L^2(\mc{T}^{b'})$ where $b'=\sum_{j=1}^{j_0}b_j=b+2i_0$. 
We define the linear operator called \textbf{$\mc{G}$-trace}, associated to the graph $\mc{G}$ as follows: 
\[{\rm Tr}^{\mc{G}}_{\mc{H}_{\mc{T}}}:\bigotimes_{j=1}^{j_0}( \bigotimes^{b_j}(e^{\eps|\nu|}\mc{H}_{\mc{T}})^*)\simeq \underbrace{\bigotimes^{b'}(e^{\eps|\nu|}\mc{H}_{\mc{T}})^*}_{\simeq e^{-\eps(\sum_{i=1}^{b'} |\nu_i|)}L^2(\mc{T}^{b'})}\to \underbrace{\bigotimes^{b}(e^{\eps'|\nu|}\mc{H}_{\mc{T}})^*}_{\simeq e^{-\eps'(\sum_{i=1}^{b} |\nu_i|)}L^2(\mc{T}^{b})}\]
for any $\eps'\in (0,\eps)$ as the composition of trace operators associated to the internal edges of $\mc{G}$, i.e.\ for $W\in \otimes_{j=1}^{j_0}( \otimes^{b_j}(e^{\eps|\nu|}\mc{H}_{\mc{T}})^*)$,
\begin{equation}\label{def_of_TrG}
{\rm Tr}^{\mc{G}}_{\mc{H}_{\mc{T}}}(W):=\Big(\prod_{i=1}^{i_0} {\rm Tr}_{\mc{H}_{\mc{T}}}^{{\rm e}_i}\Big)(W)\in \bigotimes^{b}(e^{\eps'|\nu|}\mc{H}_{\mc{T}})^*\simeq e^{-\eps'(\sum_{i=1}^{b} |\nu_i|)}L^2(\mc{T}^{b}),
\end{equation}
where for an internal edge ${\rm e}=\{(j_1,k_1),(j_2,k_2)\}$ of $\mc{G}$ (we may assume that $j_1\leq j_2$), the ${\rm e}$-trace ${\rm Tr}_{\mc{H}_{\mc{T}}}^{{\rm e}}$ is the linear operator defined by the following tracing operation: 
if $W\in  e^{-\eps(\sum_{i=1}^{b'} |\nu_i|)}L^2(\mc{T}^{b'})$ is viewed as an element $W$ in $e^{-\eps(\sum_{i=1}^{b'} |\nu_i|)}L^2(\mc{T}^{b'})\subset e^{-\eps(\sum_{i=1}^{b'} |\nu_i|)}L^\infty(\mc{T}^{b'})$
\[{\rm Tr}_{\mc{H}_{\mc{T}}}^{{\rm e}}(W):=\sum_{\nu\in \mc{T}}W(\dots,\underbrace{\nu}_{i_1},\dots,\underbrace{\nu}_{i_2},\dots ).\]
where $i_1:=(\sum_{\ell=1}^{j_1-1}b_\ell)+k_1$ and $i_2:=(\sum_{\ell=1}^{j_2-1}b_\ell)+k_2$.
The convergence of the series and the fact that ${\rm Tr}^{\mc{G}}_{\mc{H}_{\mc{T}}}(W)\in e^{-\eps(\sum_{i=1}^{b} |\nu_i|)}L^\infty(\mc{T}^{b})\subset e^{-\eps'(\sum_{i=1}^{b} |\nu_i|)}L^2(\mc{T}^{b})$ for $\eps'\in (0,\eps)$ follow from the fact that, for each $\delta>0$,
\[\sum_{\nu\in \mc{T}} e^{-\delta |\nu|}<\infty\]
by using that $\{\nu\in \mc{T}\,|\, |\nu|=n\}=\mc{O}(e^{C\sqrt{n}})$ for some $C>0$.
This argument also implies that, if $(\eps_1,\dots,\eps_{b'})\in \R^{b'}$ are such that 
\[ \eps_{i_1}+\eps_{i_2}>0, \textrm{ for }i_1=(\sum_{\ell=1}^{j_1-1}b_\ell)+k_1 \textrm{ and }i_2=(\sum_{\ell=1}^{j_2-1}b_\ell)+k_2,\]
the trace operation ${\rm Tr}^{\rm e}_{\mc{H}_{\mc{T}}}$ extends as a continuous map  
\[{\rm Tr}^{\rm e}_{\mc{H}_{\mc{T}}}: e^{-\sum_{i=1}^{b'} \eps_i|\nu_i|}L^2(\mc{T}^{b'})\to  e^{-\sum_{i=1}^{b'-2} \eps_i'|\nu_i|}L^2(\mc{T}^{b'-2})\]
for any $\eps_i'<\hat{\eps}_i$, where 
\[(\hat{\eps}_1,\dots,\hat{\eps}_{b'-2})=(\eps_1,\dots,\eps_{i_1-1},\eps_{i_1+1},\dots,\eps_{i_2-1},\eps_{i_2+1},\dots, \eps_{b'}).\]
Consequently, rewriting the components $(\nu_1,\dots,\nu_{b'})\in \mc{T}^{b'}$ under the form 
$(\nu_{11},\dots,\nu_{1b_1},\dots,\nu_{j_01},\dots,\nu_{j_0b_{j_0}})$, we deduce that if 
\begin{equation}\label{condition_trace}
\eps_{jk}+\eps_{j'k'}>0 \textrm{ for each internal edge }\{(j,k),(j',k')\}
\end{equation}
then the trace 
${\rm Tr}_{\mc{H}_{\mc{T}}}^{\mc{G}}$ extends continuously as a map 
\begin{equation}\label{weightedtr}
{\rm Tr}_{\mc{H}_{\mc{T}}}^{\mc{G}}: e^{-\sum_{j=1}^{j_0}\sum_{k=1}^{b_j} \eps_{jk}|\nu_{jk}|}L^2(\mc{T}^{b'})\to e^{-\sum_{\ell=1}^b\eps'_{j(\ell)k(\ell)}|\nu_{j(\ell)k(\ell)}|}L^2(\mc{T}^{b})
\end{equation}
for each $\eps_{jk}'<\eps_{jk}$,
where ${\rm e}^{\rm b}_\ell=\{(\ell,\circ),(j,k)\}$ (recall the notation of Definition \ref{def:gluing_graph}).
We shall use this continuity property below after integrating the
$\mc G$-contraction against a compactly supported test function in
the internal spectral paramter.


Given a gluing graph $\mc{G}$ (recall Definition \ref{def:gluing_graph}) with $j_0$ internal vertices and $m$ external vertices associated to marked points, for each $j=1,\dots,j_0$, we consider some complex weights $\bs{\alpha}_j=(\alpha_{j1},\dots,\alpha_{jm_j})\in \C^{m_j}$ for each internal vertex (possibly with $m_j=0$). We denote by 
\begin{equation}\label{alpha_to_alpha_j}
\bs{\alpha}=(\alpha_1,\ldots,\alpha_m), \qquad \alpha_i=\alpha_{jr} \quad\textrm{if}\quad {\rm e}_i^{\rm m}
 =\{(i,\bullet),(j,b_j+r)\}, \qquad 1\leq r\leq m_j.
\end{equation}
the collection of all these weights, that we think of being associated to the marked points of the surface obtained by gluing complex building blocks.
The association \eqref{alpha_to_alpha_j} allows us to pass between weights $\bs{\alpha}$ on $\Sigma$ and weights $\bs{\alpha}_j$ on each complex building block $\mc{P}_j$ with respect to any marked complex building block decomposition $\bs{\Xi}$ of $(\Sigma,J,{\bf x},\bs{\zeta},{\bf Y})$ with gluing graph $\mc{G}$. We shall say that $\bs{\alpha}\in\C^m$ satisfies the \textbf{generalised Seiberg bound with respect to $\mc{G}$} if for  each $j$,  ${\rm Re}(\bs{\alpha}_j)$ satisfies 
\begin{equation}\label{e:generalised_Seiberg_bounds_G}
\sum_{k=1}^{m_j}|{\rm Re}(\alpha_{jk})-Q|<Q.
\end{equation}
Notice that this condition depends only on $\mc{G}$.
We observe that if $S=(\Sigma,J,{\bf x})$ is a closed Riemann surface with marked points and if $\bs{\alpha}\in (Q-\eps,Q)^m$ for $\eps>0$ small enough, then for each marked building block decomposition $\bs{\Xi}$ of $S$, the generalised Seiberg bound is satisfied with respect to the gluing graph $\mc{G}$ of $\bs{\Xi}$.
We are now in position to define the conformal blocks.
\begin{lemma}[\textbf{Conformal block on $\BDTT_{g,m,b}$}]\label{block_glued}
Let $\tilde{S}=(\Sigma,J,{\bf x},\bs{\zeta},{\bf Y})$ be a decorated Riemann surface of type $(g,m,0,b)$ equipped with a marking $F$ and a marked complex building block decomposition $\bs{\Xi}\in \tilde{\mc{BD}}(\tilde{S})$ (Definition \ref{def:cbb_marking}). Then $\tilde{\mc{X}}=[(F,\Sigma,J,{\bf x},\bs{\zeta},{\bf Y})]\in \tT_{g,m,b}$ and $\tilde{\mc{X}}_{\bs{\Xi}}=[(F,\Sigma,J,{\bf x},\bs{\zeta},{\bf Y},\bs{\Xi})]\in \BDTT_{g,m,b}$ (Definition \ref{def:Decoratd_teich_DT}). Let $\mc{G}$ be the gluing graph of $\bs{\Xi}$ (Definition \ref{def:gluing_graph}), with $i_0=3g-3+m+b$ internal edges and $j_0=2g-2+m+b$ internal vertices. Recall the model form \eqref{def_MF}
\[\tilde{\mc{MF}}(\tilde{\mc{X}}_{\bs{\Xi}})=(\tilde{\Theta}_{m_1,b_1}(\tilde{\mc{X}}_1),\dots,\tilde{\Theta}_{m_{j_0},b_{j_0}}(\tilde{\mc{X}}_{j_0}))\] 
with $\tilde{\mc{X}}_j\in \tT_{0,m_j,b_j}$ associated to $\Xi_j$ as in \eqref{element_in_subTeich}, where $m_j+b_j=3$, $b_j\geq1$. 
Let $\bs{\alpha}=(\alpha_1,\ldots,\alpha_m)\in\C^m$ satisfying the generalised Seiberg bounds with respect to $\mc{G}$ and \eqref{cdozznot=0}, and denote by 
$\bs{\alpha}_j=(\alpha_{j1},\dots,\alpha_{jm_j})\in \C^{m_j}$ the corresponding weights associated to the $j$-th marked complex building block, as in 
\eqref{alpha_to_alpha_j}.
Let $\mc{B}_{\tilde{\mc{X}}_j,\bs{\alpha}_j}({\bf p}_j)$ be the conformal block amplitude of the decorated marked complex building block $\tilde{\mc{X}}_j$ defined by \eqref{bloc_decorated_bb}, viewed as a function of ${\bf p}_j=(p_{j1},\dots,p_{jb_j})\in \R_+^{b_j}$ with values in $((e^{\eps|\nu|}\mc{H}_{\mc{T}})^*)^{\otimes b_j}$ for some $\eps>0$.
For ${\bf p}=(p_1,\dots,p_{i_0})\in \R_+^{i_0}$ and ${\bf p}'=(p'_1,\dots,p'_b)\in \R_+^{b}$, we denote 
\begin{equation}\label{p_j^G} 
{\bf p}_{j}^{\mc{G}}=(p_{j1},\dots,p_{jb_j}), \quad p_{jk}=p_{j'k'}=p_i \, \textrm{ if } {\rm e}_i=\{(j,k),(j',k')\}, \quad  p_{jk}=p'_i \, \textrm{ if } {\rm e}_i^{\rm b}=\{(i,\circ),(j,k)\}.
\end{equation}
Then, for almost every ${\bf p}\in \R_+^{i_0}$ and ${\bf p}'\in \R_+^{b}$, the $\mc{G}$-trace 
\[  \mc{B}_{\tilde{\mc{X}}_{\bs{\Xi}},\bs{\alpha}}({\bf p},{\bf p}'):={\rm Tr}^{\mc{G}}_{\mc{H}_{\mc{T}}}(\mc{B}_{\tilde{\mc{X}}_1,\bs{\alpha}_1}({\bf p}^{\mc{G}}_1)\otimes\dots \otimes\mc{B}_{\tilde{\mc{X}}_{j_0},\bs{\alpha}_{j_0}}({\bf p}^{\mc{G}}_{j_0}))\in \otimes^{b}(e^{\eps'|\nu|}\mc{H}_{\mc{T}})^*\]
is well-defined for any $\eps'\in (0,\eps)$, and for each compact interval $I\subset (0,\infty)$ it induces a map, called the \textbf{conformal block (amplitude)} of $(\tilde{\mc{X}}_{\bs{\Xi}},\bs{\alpha})$, 
\[ {\bf p}\in I^{i_0}\mapsto \mc{B}_{\tilde{\mc{X}}_{\bs{\Xi}},\bs{\alpha}}({\bf p}) \in ((e^{\eps'|\nu|}L^2(I\times \mc{T}))^*)^{\otimes b},\]
defined by setting, for $u_1,\dots,u_b\in e^{\eps'|\nu|}L^2(I\times \mc{T})$,
\[ \mc{B}_{\tilde{\mc{X}}_{\bs{\Xi}},\bs{\alpha}}({\bf p})(u_1\otimes \dots \otimes u_{b})=\int_{I^{b}} \mc{B}_{\tilde{\mc{X}}_{\bs{\Xi}},\bs{\alpha}}({\bf p},{\bf p}')(u_1(p_1')\otimes \dots \otimes u_{b}(p'_b))\dd{\bf p'}.\]
The conformal block depends only on the class $\tilde{\mc{X}}_{\bs{\Xi}}\in \BDTT_{g,m,b}$, it is not identically zero when $\bs{\alpha}$ is real-valued and it belongs to $L^1(I^{i_0},((e^{\eps'|\nu|}L^2(I\times \mc{T}))^*)^{\otimes b})$  with locally uniform bounds when $\bs\alpha$ ranges over compact
subsets of the open region of weights satisfying the generalised
Seiberg bounds and \eqref{cdozznot=0}. Moreover, it depends holomorphically on $\bs\alpha$ with values in this space, in that region.
If the graph has no self-loops and the weights are real, the conformal block
belongs to the corresponding $L^2$-space. In the exceptional case
of a once-punctured torus, it also belongs to $L^2(I)$ for a real weight $\alpha\in(0,Q)$.
\end{lemma}
\begin{proof} First, we notice that the expression for conformal blocks is invariant under the equivalence in
Definition \ref{def:Decoratd_teich_DT}, since every local decorated
Teichm\"uller point $\tilde{\mc X}_j$ is unchanged. The announced result is essentially a rewriting of the main result of \cite[Sections 7 and 8.2]{GKRV21_Segal}, the only real difference is that we allow complex values $\bs{\alpha}$ here, but the proof works the same once we use Lemma \ref{lem:estimate_blocks} for the estimates of the conformal block amplitudes of  marked complex building blocks.
In that reference, the definition of the conformal block differs from the definition used here, but only through the functions $\Gamma_{\bs{\alpha}_j}({\bf p}_j^{\mc{G}},\tilde{\Theta}_{m_j,b_j}(\tilde{\mc{X}}_j))$ depending on $(\tilde{\mc{X}}_j,\bs{\alpha}_j)$ and smoothly on ${\bf p},{\bf p}'$, thus not affecting the claimed properties. The boundedness in the weighted space $\otimes^{b}(e^{\eps'|\nu|}\mc{H}_{\mc{T}})^*$ follows from Lemma \ref{lem:estimate_blocks}. 
Assume first that the gluing graph contains no self-loops and that
the weights \(\bs\alpha\) are real. For each internal vertex ${\rm ve}_j\in V$, and denoting 
$\mc{B}_{\tilde{\mc{X}}_j,\bs{\alpha}_j}$ for the block amplitude associated to ${\rm ve}_j$, set
\[ F_j({\bf p}_j):= \sum_{\bs\nu_j\in\mc T^{b_j}}e^{2\eps|\bs\nu_j|} \left| \mc B_{\tilde{\mc X}_j,\bs\alpha_j} ({\bf p}_j,\bs\nu_j) \right|^2.
\]
By Lemma~\ref{lem:estimate_blocks} and the fact that the corresponding
DOZZ factor is bounded from below by a strictly positive constant on
$I^{b_j}$, we have $F_j\in L^2(I^{b_j})$.
Weighted Cauchy--Schwarz in the Young-diagram indices associated with
the internal edges gives, for every $0<\eps'<\eps$,
\[ \sum_{\bs\nu \in\mc T^b} e^{2\eps'|\bs\nu|}
 \left| \mc B_{\tilde{\mc X}_{\bs\Xi},\bs\alpha} ({\bf p},{\bf p}',\bs\nu) \right|^2
 \leq C_{\eps,\eps'} \prod_{j=1}^{j_0}F_j({\bf p}_j^{\mc G}).
\]
Since the graph has no self-gluing loops, every internal spectral parameter $p_i$
occurs in two distinct vertex factors. The generalised
Cauchy--Schwarz inequality on the gluing graph therefore yields
\[  \int_{I^{i_0+b}}  \prod_{j=1}^{j_0}F_j({\bf p}_j^{\mc G})\, \dd{\bf p}\,\dd{\bf p}'
 \leq  |I|^{b/2} \prod_{j=1}^{j_0} \|F_j\|_{L^2(I^{b_j})}<\infty.
\]
When the graph has no self-loops and the weights are real, this proves that
\[ \mc{B}_{\tilde{\mc X}_{\bs\Xi},\bs\alpha} \in L^2\Big(I^{i_0}, \big(
   \big(e^{\eps'|\nu|}L^2(I\times\mc T)\big)^* \big)^{\otimes b} \Big).
\]
For complex weights satisfying the generalised Seiberg bounds and \eqref{cdozznot=0}, equation \eqref{bound_bloc_pant_Dozz^2} gives 
$F_{j}({\bf p}_j)  \in L^1(I^{b_j})$. Write ${\bf p}_j=({\bf p}_{j,\mathrm{int}}, {\bf p}_{j,\mathrm{ext}})$ according to the internal and external edges incident to the vertex ${\rm ve}_j$, and set 
\[ A_{j}({\bf p}_{j,\mathrm{int}}) := \int F_{j} ({\bf p}_{j,\mathrm{int}},{\bf p}_{j,\mathrm{ext}}) \,\dd{\bf p}_{j,\mathrm{ext}}. \] 
By the weighted Cauchy--Schwarz inequality in the Young-diagram indices, and integrating in ${\bf p}_{j,{\rm ext}}$, 
\[ \left\| \mc B_{\tilde{\mc X}_{\bs\Xi},\bs\alpha}({\bf p}) \right\|_ {((e^{\eps'|\nu|}L^2(I\times\mc T))^*)^{\otimes b}} \leq C_{\eps,\eps'} \prod_{j=1}^{j_0} A_{j}({\bf p}_{j,\mathrm{int}})^{1/2}. \] 
When computing the $L^2(I^{i_0}, ((e^{\eps'|\nu|}L^2(I\times\mc T))^*)^{\otimes b},\dd {\bf p})$ norm of $\mc{B}_{\tilde{\mc X}_{\bs\Xi},\bs\alpha}$ 
we use that the graph has no self-gluing loops, and thus every internal edge variable $p_i$ of ${\bf p}$ occurs in exactly two factors. The  Cauchy--Schwarz inequality therefore yields (for some constant $C_{\eps,\eps'}$)
\[ 
\| \mc B_{\tilde{\mc X}_{\bs\Xi},\bs\alpha} \|_ {L^1(I^{i_0}; ((e^{\eps'|\nu|}L^2(I\times\mc T))^*)^{\otimes b})} \leq C_{\eps,\eps'} \prod_{j=1}^{j_0} \|A_{j}\|_{L^1}^{1/2} = C_{\eps,\eps'} \prod_{j=1}^{j_0} \|F_{j}\|_{L^1(I^{b_j})}^{1/2} <\infty.  \] 
This proves the claimed $L^1$ property when the graph has no self-gluing loop.

In the case of a gluing graph containing self-gluing loops, say for example if ${\rm e}_1=\{(1,1),(1,2)\}$ with $\mc{X}_1$ of type $(m_1,b_1)=(0,3)$, it is not a priori clear that $\mc{B}_{\tilde{\mc{X}}_1}(p_1,p_1,p_{13})$ makes sense for almost all $(p_1,p_{13})\in I^2$ since $\mc{B}_{\tilde{\mc{X}}_1}(p_1,p_2,p_3)$ is only a priori $L^2(I^3)$ in the variables $(p_1,p_2,p_3)$. 
We apply \cite[Lemma 7.6]{GKRV21_Segal} to the Segal
amplitude $\mc{A}_{S_1}$ (with $S_1$ a representative of $\pi_{\tilde{\mc{T}}\to\mc{T}}(\tilde{\mc{X}}_1)$). This gives a factorisation of  
\begin{equation}\label{factorisation_bloc_ampl}
\mc{A}_{S_1}(\tilde{\varphi}_1,\tilde{\varphi}_2,\tilde{\varphi}_3)=\int_{H^{-s}(\T)} K_1(\tilde{\varphi}_1,\tilde{\varphi})K_2(\tilde{\varphi},\tilde{\varphi}_2,\tilde{\varphi}_3)\dd \mu_0(\tilde{\varphi})
\end{equation}
into a composition of two Hilbert-Schmidt operators. 
Its spectrally transformed amplitude
$\widehat{\mc{A}}_{S_1}=\overline{\Pi_{\mc{V}\otimes \bbar{\mc{V}}}^{\otimes 3}(\overline{\mc{A}_{S_1}})}$  then also factorises and, using (the proof of) \cite[Lemma 7.2]{GKRV21_Segal}, its self-gluing diagonal
\begin{equation}\label{DDiagonalrest}
D(p,\nu,\tilde{\nu},p_3,\nu_3,\tilde{\nu}_3):=\widehat{\mc{A}}_{S_1}(p,p,p_3,\nu,\nu,\nu_3,\tilde{\nu},\tilde{\nu},\tilde{\nu}_3) 
\end{equation}
is canonically defined and
\[ \int_I\sum_{\nu,\tilde\nu\in\mc T} \left\| D(p,\nu,\tilde\nu,\cdot)
 \right\|_{(e^{\eps'(|\nu_3|+|\tilde{\nu}_3|)}L^2(I\times \mc{T}^2))^*}\,\dd p<\infty.
\]
It follows that 
 \[ \int_I\sum_{\nu\in\mc T}\Big(\int_I\sum_{\nu_3\in\mc T}
  e^{2\eps'|\nu_3|}  |D(p,\nu,0,p_3,\nu_3,0)|^2  \,\dd p_3 \Big)^{1/2}\dd p<\infty.
 \]
Then, by the factorisation formula \eqref{factorisation_extended},  one has 
\[D(p,\nu,\tilde{\nu},p_3,\nu_3,\tilde{\nu}_3)= C_{\mc{X}_1,g}(p,p_3) \mc{B}_{\mc{X}_1}(p,p,p_3,\nu,\nu,\nu_3)
 \overline{\mc{B}_{\mc{X}_1}(p,p,p_3,\tilde{\nu},\tilde{\nu},\tilde{\nu}_3)},\]
for some $C_{\mc{X}_1,g}(p,p_3)$ which is the product of the DOZZ constant $C^{\rm DOZZ}(Q+{\rm i}p,Q-{\rm i}p,Q+{\rm i}p_3)$ 
and geometric normalisation factors independent of $(p,p_3)$. Moreover, $\mc{B}_{\mc X_1}(p,p,p_3,0,0,0)=\Gamma_{\mc X_1}(p,p,p_3)$
is nonzero. Hence, using \eqref{cdozznot=0}, on $I^2$, the
function
\[ (p,p_3)\in I^2\mapsto C_{\mc X_1,g}(p,p_3) \overline{\Gamma_{\mc X_1}(p,p,p_3)}
\]
is bounded from below in modulus by a strictly positive constant.
Consequently,
\[ \int_I\sum_{\nu\in\mc T}\left\| \mc{B}_{\mc X_1}(p,p,\cdot,\nu,\nu,\cdot)\right\|_
 {(e^{\eps'|\nu_3|} L^2(I\times\mc T))^*} \,\dd p<\infty.
\]
Thus the series
\[ b_1(p):=\sum_{\nu\in\mc T}\mc{B}_{\mc X_1}(p,p,\cdot,\nu,\nu,\cdot)
\]
converges absolutely in $(e^{\eps'|\nu_3|} L^2(I\times\mc T))^*$ for
almost every $p$, and $b_1\in L^1(I;(e^{\eps'|\nu_3|} L^2(I\times\mc T))^*).$
The same construction is performed at every vertex carrying a
self-loop. Except in the once-punctured torus case considered below,
such a vertex is of type $(0,3)$ and hence carries no marked points.
Combining these bounds with those obtained in the discussion above for vertices with no self-gluing loop, 
a last application of Cauchy-Schwarz along the
remaining internal edges gives the claimed global $L^1$ bound.
We deduce that the conformal block belongs to $L^1(I^{i_0},((e^{\eps'|\nu|}L^2(I\times\mc T))^*)^{\otimes b})$
and depends holomorphically on the weights with values in this space.

If the self-gluing loop vertex has type $(m_1,b_1)=(1,2)$ (then the whole surface is a once-punctured torus) 
and $\alpha$ satisfies the generalised Seiberg bounds with \eqref{cdozznot=0}, the same argument applies. Indeed, if
$D_\alpha(p,\nu,\tilde\nu) :=\widehat{\mc A}_{S_1,\alpha} (p,p,\nu,\nu,\tilde\nu,\tilde\nu),$
then
\[\int_I\sum_{\nu,\tilde\nu\in\mc T}|D_\alpha(p,\nu,\tilde\nu)|\,\dd p<\infty.\]
Restricting to \(\tilde\nu=0\) and using the factorisation \eqref{factorisation_extended} as above gives
\[ \int_I\sum_{\nu\in\mc T}|\mc B_{\mc X_1,\alpha}(p,p,\nu,\nu)|\,\dd p<\infty.
\]
Consequently, the once-punctured torus block is well-defined and
belongs to $L^1(I)$ for complex $\alpha$ satisfying the generalised
Seiberg bound. For real $\alpha\in(0,Q)$, the stronger $L^2(I)$ property follows
from \cite[Theorem~9.2]{GKRV21_Segal}, since the DOZZ constant
appearing in the bootstrap formula is continuous and bounded from
below by a strictly positive constant on $I$.

To conclude the proof, we show that $\mc B_{\tilde{\mc X}_{\bs\Xi},\bs\alpha}$ is not the zero element. Assume first that $b>0$ and that $\bs\alpha$ is real valued and satisfies the Seiberg bounds. Let $S^{\#2}$ be the double of the surface $S$ representing $\tilde{\mc X}$, obtained by gluing $S$ to its mirror image along their $b$ boundary components. We equip $S^{\#2}$ with the doubled complex building block decomposition $\bs\Xi^{\#2}:=\bs\Xi\cup\tau(\bs\Xi)$ and with the marked points ${\bf x}^{\#} = (x_1,\ldots,x_m,\tau(x_1),\ldots,\tau(x_m))$, carrying the weights induced by $\bs\alpha$. Let $g$ be an admissible metric on $S$ and let $g^{\#2}$ be its $\tau$-invariant extension to $S^{\#2}$. By \cite[Theorem 8.4]{GKRV21_Segal}, the bootstrap decomposition associated with $\bs\Xi^{\#2}$ gives 
\[ \begin{split} 
& \big\langle \prod_{k=1}^mV_{\alpha_k}(x_k) \prod_{k=1}^mV_{\alpha_k}(\tau(x_k)) \big\rangle_{S^{\#2},g^{\#2}} \\ 
&\quad = C(g,\bs\alpha) \int_{\R_+^{i_0}\times\R_+^b\times\R_+^{i_0}} \rho_{\bs\alpha} ({\bf p},{\bf p}',{\bf p}'') 
\Big| \sum_{\bs\nu\in\mc T^b} \mc B_{\tilde{\mc X}_{\bs\Xi},\bs\alpha} ({\bf p},{\bf p}',\bs\nu) 
\overline{\mc B_{\tilde{\mc X}_{\bs\Xi},\bs\alpha} ({\bf p}'',{\bf p}',\bs\nu)} \Big|^2 \,\dd{\bf p}\,\dd{\bf p}'\,\dd{\bf p}'',
\end{split}
\] 
where $C(g,\bs\alpha)>0$ and $ \rho_{\bs\alpha}({\bf p},{\bf p}',{\bf p}'')$ is a product of DOZZ constants. The correlation function on the left-hand side is strictly positive by its probabilistic definition. If $\mc B_{\tilde{\mc X}_{\bs\Xi},\bs\alpha}$ were zero as an element of the corresponding space, then 
$\mc B_{\tilde{\mc X}_{\bs\Xi},\bs\alpha} ({\bf p},{\bf p}',\bs\nu)=0$ for almost every $({\bf p},{\bf p}')$ and every $\bs\nu\in\mc T^b$, and the right-hand side would vanish. This is a contradiction. When $b=0$, the same conclusion follows by applying the bootstrap formula directly to the closed surface $S$: 
\[ \Big\cjg\prod_{k=1}^mV_{\alpha_k}(x_k)\Big\cjd_{S,g} = C(g,\bs\alpha) \int_{\R_+^{i_0}} \rho_{\bs\alpha}({\bf p}) \left| \mc B_{\tilde{\mc X}_{\bs\Xi},\bs\alpha}({\bf p}) \right|^2 \,\dd{\bf p}>0. \] 
Thus, in all cases, $\mc B_{\tilde{\mc X}_{\bs\Xi},\bs\alpha}$ is not the zero element.
\end{proof}

Notice that the conformal block of $(\tilde{\mc{X}}_{\bs{\Xi}},\bs{\alpha})$ can be represented by a function 
\[ ({\bf p},{\bf p}',\bs{\nu})\in \R_+^{i_0}\times \R_+^{b}\times \mc{T}^{b} \mapsto  \mc{B}_{\tilde{\mc{X}}_{\bs{\Xi}},\bs{\alpha}}({\bf p},{\bf p}',\bs{\nu})\in\C,\]
and the action of $\mc{B}_{\tilde{\mc{X}}_{\bs{\Xi}},\bs{\alpha}}({\bf p})$ on ${\bf u}=u_1\otimes \dots\otimes u_b\in (e^{\eps'|\nu|}L^2(I\times \mc{T}))^{\otimes b}$ (with $I\subset (0,\infty)$ a compact interval), for almost every ${\bf p}\in I^{i_0}$, is given by 
\[\mc{B}_{\tilde{\mc{X}}_{\bs{\Xi}},\bs{\alpha}}({\bf p}){\bf u}=\int_{I^b}\sum_{\bs{\nu}\in \mc{T}^b}\mc{B}_{\tilde{\mc{X}}_{\bs{\Xi}},\bs{\alpha}}({\bf p},{\bf p}',\bs{\nu})\prod_{k=1}^{b}u_k(p'_k,\nu_k)\dd {\bf p}', \]
where ${\bf p}'=(p_1',\dots,p'_{b})$ and $\bs{\nu}=(\nu_1,\dots,\nu_{b})$.
The parameters ${\bf p}$ are called the \textbf{internal channel momenta} in the spectrum, since they are associated to the cutting curves of the complex building block decomposition and to the Verma modules $\mc{V}_{Q+{\rm i}p_k}\simeq \mc{H}_{\mc{T}}$ for $k=1,\dots,i_0$, 
whereas the parameters ${\bf p'}$ are called the \textbf{external channel momenta} since they are associated to the boundary components of the surface and to the Verma modules $\mc{V}_{Q+{\rm i}p'_k}\simeq \mc{H}_{\mc{T}}$ for $k=1,\dots,b$.

It is important to notice that the conformal block $\mc{B}_{\tilde{\mc{X}}_{\bs{\Xi}},\bs{\alpha}}$ does \textit{not} descend as a function of the Teichm\"uller point $\tilde{\mc{X}}\in \tT_{g,m,b}$ but depends a priori on the marked complex building block decomposition $\bs{\Xi}$ on $\tilde{S}$. 
We shall study this dependence on $\bs{\Xi}$ in what follows.

We can also consider Riemann surfaces with outgoing boundaries as follows. 
If $\tilde{\mc{X}}_{\bs{\Xi}}=[(F,\Sigma,J,{\bf x},\bs{\zeta},{\bf Y},\bs{\Xi})]\in \mc{D}\tT_{g,m,b}$ has type $(g,m,0,b)$ and $\bs{\sigma}=(\sigma_1,\dots,\sigma_b)$ are orientation signs with $\sigma_{k}=+1$ for $k=1,\dots,b^+$ and
$\sigma_{k}=-1$ for $k=b^++1,\dots,b$, we define
$\tilde{\mc{X}}^{\bs{\sigma}}_{\bs{\Xi}}:=[(F,\Sigma,J,{\bf x},\bs{\zeta}^{\bs{\sigma}},{\bf Y},\bs{\Xi})]\in\mc{D}\tT_{g,m,b^+,b^-}$ of type $(g,m,b^+,b^-)$ (generalising Definition \ref{def:Decoratd_teich_DT} in the obvious way), where $\bs{\zeta}^{\bs{\sigma}}$ is as in \eqref{teich_to_teich_sigma}. 
The conformal block of $(\tilde{\mc{X}}^{\bs{\sigma}}_{\bs{\Xi}},\bs{\alpha})$ is then the map (for some $\eps>0$)
\[ {\bf p}\in I^{i_0}\mapsto \mc{B}_{\tilde{\mc{X}}^{\bs{\sigma}}_{\bs{\Xi}},\bs{\alpha}}({\bf p}) \in \mc{L}((e^{\eps|\nu|}L^2(I\times \mc{T}))^{\otimes b^-}, (e^{-\eps|\nu|}L^2(I\times \mc{T}))^{\otimes b^+}),\]
defined, for every compact interval $I\subset (0,\infty)$, by setting for $u_1,\dots,u_b\in e^{\eps|\nu|}L^2(I\times \mc{T})$
\[ \cjg \mc{B}_{\tilde{\mc{X}}^{\bs{\sigma}}_{\bs{\Xi}},\bs{\alpha}}({\bf p})(u_{b^++1}\otimes \dots \otimes u_{b}),\bar{u}_{1}\otimes \dots \otimes \bar{u}_{b^+}\cjd_{L^2(I\times \mc{T})^{\otimes b^+}}=\mc{B}_{\tilde{\mc{X}}_{\bs{\Xi}},\bs{\alpha}}({\bf p})(u_1\otimes \dots \otimes u_b).\]
As before, it is defined for almost every ${\bf p}$ and is $L^1(I^{i_0})$ in the ${\bf p}$ variable.

\subsection{Schwarzian form and Schwarzian cocycle}

In the variation formula of the conformal block amplitude of a marked complex building block (Proposition \ref{general_derivative_block_amp}), we see that the Schwarzian derivatives of the uniformising maps $\psi_j$ are involved in a non-trivial way. This reflects the projectivity of the conformal blocks, which can be described precisely using geometric quantities on the fiber bundle $\mc{DT}_{g,m,b}\to\mc{T}_{g,m,b}$.

To motivate the next definition, we define the following natural $1$-form associated to deformations on $\mc{DT}_{g,m,b}$. Let $\mc{G}$ be a gluing graph and $\mc{X}_{\bs{\Xi}}=[(F,\Sigma,J,{\bf x},\bs{\zeta},\bs{\Xi})]\in \mc{D}\mc{T}_{g,m,b}^{\mc{G}}$ with $\mc{X}=\pi_{\mc{DT}\to\mc{T}}(\mc{X}_{\bs{\Xi}})=[(F,\Sigma,J,{\bf x},\bs{\zeta})]\in\mc{T}_{g,m,b}$. Via the homeomorphism $\mc{MF}^{\rm aug}:\BDT_{g,m,b}^{\mcG}\to\mc{T}_{g,m,b}^{\mcG}\times\piBD_{g,m,b}^{\mcG}$ introduced in \eqref{def_MFhat}, the action of $\mc{U}_\eps^{\mc{G}}:=\prod_{j=1}^{j_0}\mc{U}_\eps^{b_j}$ on $\mc{T}_{g,m,b}^{\mcG}$ from \eqref{action_bsf_tildeX} induces an action on $\BDT_{g,m,b}^{\mcG}$.
We define the complex valued $1$-form $\Theta_{\bs{\Xi}}'$ at $\mc{X}_{\bs{\Xi}}$ by
\[\Theta_{\bs{\Xi}}'(\bs{\v}):= - \sum_{j=1}^{j_0} \sum_{k=1}^{b_j} \frac{1}{2\pi {\rm i}}\oint_{\T} {\rm S}(\psi_{jk})v_{jk} \dd z\]
for $\bs{\v}=((\v_{jk})_{k=1}^{b_j})_{j=1}^{j_0}\in\prod_{j=1}^{j_0}T_{\rm Id}\mc{U}_\eps^{b_j}$, where $\mc{MF}({\bs{\Xi}})=((\psi_{jk})_{k=1}^{b_j},(\theta_{jk})_{k=1}^{b_j})_{j=1}^{j_0}$, i.e. $(\psi_{jk})_{k=1}^{b_j}$ is the model form in $\mc{P}_{m_j,b_j}^\eps$ of the complex building block in $\Sigma$ defined by $\Xi_j$. This is the factor of the projective term in the variation of the conformal block on $\mc{DT}_{g,m,b}^{\mc{G}}$ under a deformation in $\mc{U}_\eps^{\mc{G}}$. We are especially interested in the case where the deformation is along the fiber $\pi_{\mc{DT}\to\mc{T}}^{-1}(\mc{X})$. A sufficient condition for $\bs{\v}=((\v_{jk})_{k=1}^{b_j})_{j=1}^{j_0}$ to generate such a deformation is if
\begin{equation}\label{fiber_tangent_condition}
v_{jk}(z)=-z^2v_{j'k'}(o(z)) \, \textrm{ if } {\rm e}_i=\{(j,k),(j',k')\}, \quad  \v_{jk}=0 \, \textrm{ if } {\rm e}_i^{\rm b}=\{(i,\circ),(j,k)\}.
\end{equation}
The subspace $T_{{\rm Id}}^{\rm vert}\mc{U}_\eps^{\mc{G}}$ of tangent vectors satisfying \eqref{fiber_tangent_condition} is isomorphic to $T_{\rm Id}\mc{U}_\eps^{i_0}$ (recall that $i_0$ is the number of internal edges), but the identification depends on a choice of the orientations of the internal edges. Indeed, if $\vec{\mc{G}}$ is such a choice for $\mc{G}$ with each $\vec{\rm e}_i=((j^-(i),k^-(i)),(j^+(i),k^+(i)))$, then we have an isomorphism
\[\pi^{\vec{\mc{G}}}:T_{{\rm Id}}^{\rm vert}\mc{U}_\eps^{\mc{G}}\to T_{\rm Id}\mc{U}_\eps^{i_0},\quad((\v_{jk})_{k=1}^{b_j})_{j=1}^{j_0}\mapsto(\v_{j^-(i),k^-(i)})_{i=1}^{i_0},\]
and we have
\[\Theta_{\bs{\Xi}}'((\pi^{\vec{\mc{G}}})^{-1}({\rm v}_1,\dots,{\rm v}_{i_0}))=\sum_{i=1}^{i_0} \frac{1}{2\pi {\rm i}}\oint_{\T} ({\rm S}(\psi_{j^+(i)k^+(i)}\circ o)-{\rm S}(\psi_{j^-(i)k^-(i)}))v_i \dd z,\]
where we used the identity
\begin{equation}\label{identity_Schwarzian_inversion}
\oint_{\T} {\rm S}(\psi)v \dd z=-\oint_{\T} {\rm S}(\psi\circ o)v^* \dd z,\quad v^*(z)=-z^2v(o(z)).
\end{equation}
This leads us to the following definition:

\begin{definition}[\textbf{Schwarzian form}]\label{Schwarzian_form}
Let $\mc{X}_{\bs{\Xi}}=[(F,\Sigma,J,{\bf x},\bs{\zeta},\bs{\Xi})]\in \mc{D}\mc{T}_{g,m,b}^{\mc{G}}$ with model form 
$\mc{M}\mc{F}(\bs{\Xi})=((\bs{\psi}_1,\bs{\theta}_1),\dots,(\bs{\psi}_{j_0},\bs{\theta}_{j_0}))$ as defined in \eqref{definition_MFXi}.
We pick an orientation $\vec{\mc{G}}$ of $\mc{G}$ where each internal edge $\vec{\rm e}_i=((j^-(i),k^-(i)),(j^+(i),k^+(i)))$.
For $\eps>0$ sufficiently small, we define the $1$-form, called the \textbf{Schwarzian $1$-form},  
\[ \Theta_{\bs{\Xi}}: T_{{\rm Id}}\mc{U}_\eps^{i_0}\to \C, \quad 
\Theta_{\bs{\Xi}}({\rm v}_1,\dots,{\rm v}_{i_0})=\sum_{i=1}^{i_0} \frac{1}{2\pi {\rm i}}\oint_{\T} ({\rm S}(\psi_{j^+(i)k^+(i)}\circ o)-{\rm S}(\psi_{j^-(i)k^-(i)}))v_i \dd z,\]
where ${\rm v}_i=v_i(z)\pl_z$ and ${\rm S}(\cdot)$ denotes the Schwarzian as before. Notice that $\Theta_{\bs{\Xi}}$ depends only on $\mc{X}_{\bs{\Xi}}$ and implicitly on a fixed orientation $\vec{\mc{G}}$ of $\mc{G}$.
\end{definition}

Let $\bs{\xi}=([\xi_{1}],\dots,[\xi_{i_0}])$ be the maximal family of cutting curves $\mc{C}_i=\xi_i(\T)$ in $\Sigma$
(in the sense of Definition \ref{def:maximal_family_cutting_curves}) induced by $\bs{\Xi}$, where the orientation of $\xi_i$ is determined 
by $\xi_i=\Xi_{j^-(i)}\circ \zeta_{k^{-}(i)}^\bullet$ if $\vec{\rm e}_i=((j^-(i),k^{-}(i)),(j^+(i),k^{+}(i))$. For each $j$, let $\Phi_j:\mc{P}_j\to \hat{\C}\setminus \cup_{k=1}^{b_j}\psi_{jk}(\D^\circ)$ be the model form uniformisation of the complex building block $\mc{P}_j$ defined by $\Xi_j$.
We have $\Phi_{j^-(i)}\circ \xi_i=\psi_{j^{-}(i)k^{-}(i)}|_{\T}$ 
and $\Phi_{j^+(i)}\circ \xi_i=\psi_{j^+(i)k^+(i)}\circ o|_{\T}$. 
The Schwarzian $1$-form can also be written as 
\[ \Theta_{\bs{\Xi}}({\rm v}_1,\dots,{\rm v}_{i_0})=-\sum_{i=1}^{i_0} \frac{1}{2\pi{\rm i}}\oint_{\Phi_{j^+(i)}(\mc{C}_{i})}\big({\rm S}(\Phi_{j^-(i)}\circ \Phi^{-1}_{j^+(i)}) \dd z^2\big)\big((\Phi_{j^+(i)}\circ \xi_i)_*({\rm v}_i)\big),\]
where $\mc{C}_i=\pl_{k^+(i)}\mc{P}_{j^+(i)}=\pl_{k^-(i)}\mc{P}_{j^-(i)}$ is oriented by $\xi_i$ and $\Phi_{j^+(i)}(\mc{C}_{i})$ by $\Phi_{j^+(i)}\circ\xi_i$. 
Indeed, one has 
\[\Phi_{j^+(i)}^{-1}\circ \psi_{j^+(i)k^+(i)}\circ o=\Phi_{j^-(i)}^{-1}\circ\psi_{j^-(i)k^-(i)}=\xi_i\] and thus
\[ \begin{split}
{\rm S}(\Phi_{j^-(i)}\circ \Phi^{-1}_{j^+(i)})&={\rm S}(\psi_{j^-(i)k^-(i)}\circ(\psi_{j^+(i)k^+(i)}\circ o)^{-1})\\
&= (((\psi_{j^+(i)k^+(i)}\circ o)^{-1})')^2({\rm S}(\psi_{j^-(i)k^-(i)})-{\rm S}(\psi_{j^+(i)k^+(i)}\circ o))\circ (\psi_{j^+(i)k^+(i)}\circ o)^{-1},
\end{split}\]
and changing variables via $\psi_{j^+(i)k^+(i)}\circ o=\Phi_{j^+(i)}\circ \xi_i$ in the expression of $\Theta_{\bs{\Xi}}(\bv)$, we get the announced expression.

If we integrate the Schwarzian $1$-form along the fibers of the fiber bundle $\mc{DT}_{g,m,b}\to\mc{T}_{g,m,b}$, we obtain the following quantity:

\begin{definition}[\textbf{Schwarzian cocycle}]\label{d:cocycle}
Let $\mc{G}$ be a gluing graph and $[\bs{\Xi}^\bullet]\in\piBD_{g,m,b}^{\mc{G}}$. Let $\mc{X}_{\bs{\Xi}^1},\mc{X}_{\bs{\Xi}^2}\in\mc{DT}_{g,m,b}^{[\bs{\Xi}^\bullet]}$ with $\pi_{\mc{DT}\to\mc{T}}(\mc{X}_{\bs{\Xi}^1})=\pi_{\mc{DT}\to\mc{T}}(\mc{X}_{\bs{\Xi}^2})=\mc{X}\in\mc{T}_{g,m,b}$. Pick a representative 
$(F,S)$ of $\mc{X}$ with $S=(\Sigma,J,{\bf x},\bs{\zeta})$. Pick representatives $(F,S,\bs{\Xi}^1),(F,S,\bs{\Xi}^2)$ of $\mc{X}_{\bs{\Xi}^1},\mc{X}_{\bs{\Xi}^2}$, respectively, with $\bs{\Xi}^1,\bs{\Xi}^2\in\mc{BD}(S)$. Choose a continuous family $t\in[1,2]\mapsto\bs{\Xi}^t\in\mc{BD}(S)$ of marked complex building block decompositions of $S$ with $\bs{\Xi}^t|_{t=1}=\bs{\Xi}^1$, $\bs{\Xi}^t|_{t=2}=\bs{\Xi}^2$, which is $C^1$ in the sense that the model form 
$\mc{M}\mc{F}(\bs{\Xi}^t)\in\mc{T}_{g,m,b}^{\mc{G}}$ is $C^1$. Again, we pick an orientation $\vec{\mc{G}}$ of $\mc{G}$ where each internal edge $\vec{\rm e}_i=((j^-(i),k^-(i)),(j^+(i),k^+(i)))$. For each $i$ and $t$, let $\xi_i^t:=\Xi_{j^-(i)}^t\circ\zeta_{k^-(i)}^\bullet$ as above, which extends to a small neighbourhood of $\T$. The \textbf{Schwarzian cocycle} is then defined by
\begin{equation}\label{DefUpsilon}
\Upsilon(\mc{X}_{\bs{\Xi}^1},\mc{X}_{\bs{\Xi}^2}):=\Upsilon_{\mc{X}}(\bs{\Xi}^1,\bs{\Xi}^2):=\int_1^2\Theta_{\bs{\Xi}^t}({\bf v}^t)\dd t,
\end{equation}
where ${\bf v}^t=({\rm v}_1^t,\dots,{\rm v}_{i_0}^t)\in T_{\rm Id}\mc{U}_\eps^{i_0}$ with ${\rm v}_i^t:=\pl_s((\xi_i^t)^{-1}\circ \xi_i^{t+s})|_{s=0}$ the vector field associated to the variation of the cutting curve $\xi_i^t$. 
\end{definition}

We shall see below (Corollary \ref{l:cocycle}) that this integral does not depend on the continuous family $t\mapsto\bs{\Xi}^t$. It is easy to see that it does not depend on the orientation $\vec{\mc{G}}$ of $\mc{G}$ using again the identity \eqref{identity_Schwarzian_inversion} (or by noting that the integrand factors through the orientation-independent form $\Theta_{\bs{\Xi}}'$ since the deformation is in $\mc{U}_\eps^{\mc{G}}$).
A change of the decorated marked complex building blocks by diffeomorphisms $h_j$ as in Definition \ref{def:Decoratd_teich_DT} leaves the boundary parametrisations and the local model forms unchanged. The Schwarzian form is therefore invariant under  such changes.


We now observe that the real part of the Schwarzian cocycle can be expressed in terms of Liouville actions. 

\begin{lemma}\label{Re(Upsilon)}
Let $\mc{X}\in \mc{T}_{g,m,b}$ and consider a representative 
$(F,S)$ of $\mc{X}$ with $S=(\Sigma,J,{\bf x},\bs{\zeta})$. Let $\bs{\Xi}^1$ and $\bs{\Xi}^2$ be two marked complex building block decompositions of $S$ in the same isotopy class $[\bs{\Xi}^\bullet]\in \piBD_{g,m,b}$.
For $\ell=1,2$, we consider:
\begin{itemize} 
\item a smooth metric $g_{\Sigma}$ on $\Sigma$ compatible with $J$ and a smooth metric $g_{\hat{\C}}$ on $\hat{\C}$ compatible with $J_{\C}$;
\item a smooth metric $g^\ell$ on $\Sigma$ compatible with $J$ and which is admissible 
on each complex building block $(\mc{P}_j^\ell,J,{\bf x}_j,\bs{\zeta}_j^\ell)$ induced by $\bs{\Xi}^\ell$; 
\item  the filling metric $g_j^\ell$ of
$g^\ell|_{\mc P_j^\ell}$  on $\hat{\C}$ in the sense of Definition \ref{LiouvillePant}. 
\end{itemize}
Then the following holds: 
\[  \frac{1}{6}{\rm Re}(\Upsilon_{\mc{X}}(\bs{\Xi}^1,\bs{\Xi}^2))= \Big(S_{\rm L}^0(\Sigma,g_\Sigma,g^2)- \sum_{j=1}^{j_0}S_{\rm L}^0(\hat{\C},g_{\hat{\C}},g_j^2)\Big) - \Big(S_{\rm L}^0(\Sigma,g_\Sigma,g^1)
- \sum_{j=1}^{j_0}S_{\rm L}^0(\hat{\C},g_{\hat{\C}},g^1_j)\Big).
\]
\end{lemma}
\begin{proof} First, let $g$ and $g'=e^{\omega}g$ be two conformal metrics that are admissible on each complex building block; we have $\omega=0$ near the cutting curves and the boundary. Let 
$W(g):=S_{\rm L}^0(\Sigma,g_\Sigma,g)- \sum_{j=1}^{j_0}S_{\rm L}^0(\hat{\C},g_{\hat{\C}},g_j)$ where $g_j$ is the filling of $g|_{\mc{P}_j}$ (and similarly $g'_j$ the filling of $g'|_{\mc{P}_j}$). 
By the cocycle property of $S_{\rm L}^0$, 
\[W(g')-W(g)=S_{\rm L}^0(\Sigma,g,g')- 
\sum_{j=1}^{j_0}S_{\rm L}^0(\hat{\C},g_j,g_j').\]
Since $g,g'$ are admissible on the complex building blocks, $g_j=g_j'$ near $\pl \Phi_j(\mc{P}_j)$ (with $\Phi_j:\mc{P}_j\to \hat{\C}$ the uniformising map), thus
$S_{\rm L}^0(\hat{\C},g_j,g'_j)=S_{\rm L}^0(\mc{P}_j,g,g')$, which implies that $W(g)=W(g')$. 

Now to prove this result, we consider a $C^1$ curve $t\in [1,2]\mapsto \bs{\Xi}^t$ of marked complex building block decompositions interpolating between $\bs{\Xi}^1$ and $\bs{\Xi}^2$, and we differentiate ${\rm Re}(\Upsilon_{\mc{X}}(\bs{\Xi}^1,\bs{\Xi}^t))$ with  respect to $t$. Denote by $(\bs{\psi}_j^t,\bs{\theta}_j^t)$ the model form of the marked complex building block $\mc{X}_j^t$ induced by $\Xi_j^t$.  
By linearity, it suffices to treat a deformation for which only the cutting curve associated to the first edge varies: we may assume that for ${\rm e}_1=\{(j^-(1),k^-(1)),(j^+(1),k^+(1))\}$, the model forms $\bs{\psi}_j^t$ are constant in $t$ for $j\neq j^\pm(1)$. 
We start with the case $j^+(1)\not=j^-(1).$ 
Let us  write 
$\vec{\rm e}_1=((j^-(1),k^-(1)),(j^+(1),k^+(1)))$, and to simplify the notations, we denote $\psi_+^t=\psi_{j^+(1)k^+(1)}^t$ and $\psi_-^t=\psi_{j^-(1)k^-(1)}^t$, and similarly $\bs{\psi}_+^t=\bs{\psi}_{j^+(1)}^t$ 
and $\bs{\psi}_-^t=\bs{\psi}_{j^-(1)}^t$. The associated cutting curve is $\xi_1^t=\Xi_{j^-(1)}^t\circ \zeta^\bullet_{k^-(1)}$.
One has 
\begin{equation}\label{variation_upsilon_X}
\pl_t \Upsilon_{\mc{X}}(\bs{\Xi}^1,\bs{\Xi}^t)= \frac{1}{2\pi {\rm i}}\oint_{\T}v^t({\rm S}(\psi_+^t\circ o)-
{\rm S}(\psi_-^t))\dd z
\end{equation} 
where $\v^t=\pl_s((\xi_1^{t})^{-1}\circ \xi_1^{t+s})|_{s=0}=v^t(z)\pl_z$ 
and we write\footnote{Here the coefficients $v_n^t$ are the geometric Laurent coefficients,
defined by $v^t(z)=\sum_n v_n^tz^{n+1}$; they are therefore the opposites of the coefficients used in the convention \eqref{def_Hv}.} $v^t(z)=\sum_{n\in \Z}v_n^tz^{n+1}$.
Using \eqref{dGamma_wn}, more precisely its generalisation explained in Remark \ref{generalisation521}, we have (with the notations used there)
\begin{equation}\label{real_Schwarzian_cocycle_proof_identity}
\frac{\pl_t\Gamma^{\Phi}_{\bs{\alpha},{\bf p}}(\bs{\psi}_+^t)}{\Gamma^{\Phi}_{\bs{\alpha},{\bf p}}(\bs{\psi}_+^t)}
+\sum_{n=1}^\infty v_{n}^t w_{\bs{\psi}_+^t}'(\bs{\alpha}({\bf p}),\bs{\nu}_n) +v_{0}^t\Delta_{Q+{\rm i}p_1}= \frac{c_{\rm L}}{24\pi {\rm i}} \oint_{\T} v^t{\rm S}(\psi_{+}^t\circ o)\dd z.
\end{equation}
We choose $g^t$ to be a $C^1$  family of metrics on $\Sigma$ that are admissible on each complex building block $\mc{P}_j^t$, such that 
$g^t$ is equal to $g^1$ outside a small neighbourhood of $\xi_1^t(\T)$, and we write $g_\pm^t:=g^t|_{\mc{P}_\pm^t}$ if $\mc{P}_\pm^t$ is the complex building block associated to $\Xi^t_{j^{\pm}(1)}$. 
We consider the first boundary parametrisation of $\mc{P}_+^t$ to be outgoing and given by 
$\Xi_{j^+(1)}^t\circ \zeta_{k^+(1)}^\bullet \circ o$, 
and all the other boundary parametrisations are incoming (and do not depend on $t$).
Using Lemma \ref{matrix_coefficients} (first line) and Proposition \ref{th:derivative_general} (second line), we have 
 \[ \begin{split}
\sum_{n=1}^{\infty} v_{n}^t w_{\bs{\psi}_+^t}'(\bs{\alpha}({\bf p}),\bs{\nu}_n) +\Delta_{Q+{\rm i}p_1}v_{0}^t=&
 -\frac{ \cjg \frac{1}{2}(\bH_{{\rm v}^t}-{\rm i}\bH_{{\rm i}{\rm v}^t})\mc{A}_{\mc{P}^t_+,g_+^t,{\bf x},\bzeta_+^t,\bs{\alpha}}(\otimes_{k=2}^{b} \Psi_{Q+{\rm i}p_k}),\Psi_{Q+{\rm i}p_1}\cjd_{\mc{H}}}{C(\bs{\psi}_+^t,g_+^t,\bs{\alpha}({\bf p}))C^{\rm DOZZ}(\bs{\alpha}(-\bs{\sigma}{\bf p}))}\\
 =& -\frac{1}{2}\Big( \frac{ \pl_t \cjg \mc{A}_{\mc{P}_+^t,g_+^t,{\bf x},\bzeta_+^t,\bs{\alpha}}(\otimes_{k=2}^{b} \Psi_{Q+{\rm i}p_k}),\Psi_{Q+{\rm i}p_1}\cjd_{\mc{H}}}{C(\bs{\psi}_+^t,g_+^t,\bs{\alpha}({\bf p}))C^{\rm DOZZ}(\bs{\alpha}(-\bs{\sigma}{\bf p}))}-c_{\rm L}\pl_s S_{\rm L}^0(\mc{P}_+^s,g_+^t,g_+^s)|_{s=t}
 \Big)\\
  & +\frac{{\rm i}}{2} \Big(\frac{ \pl_t \cjg \mc{A}_{\mc{P}_{*}^t,g_*^t,{\bf x},\bzeta_{*}^t,\bs{\alpha}}(\otimes_{k=2}^{b} \Psi_{Q+{\rm i}p_k}),\Psi_{Q+{\rm i}p_1}\cjd_{\mc{H}}}{C(\bs{\psi}_{+*}^t,g_*^t,\bs{\alpha}({\bf p}))C^{\rm DOZZ}(\bs{\alpha}(-\bs{\sigma}{\bf p}))}-c_{\rm L}\pl_s S_{\rm L}^0(\mc{P}_*^s,g_*^t,g_*^s)|_{s=t}\Big)+\frac{c_{\rm L}}{24}v_0^t\\
  =& -\frac{1}{2}\Big( \frac{ \pl_tC(\bs{\psi}_+^t,g_+^t,\bs{\alpha}({\bf p}))}{C(\bs{\psi}_+^t,g_+^t,\bs{\alpha}({\bf p}))}-c_{\rm L}\pl_s S_{\rm L}^0(\mc{P}_+^s,g_+^t,g_+^s)|_{s=t}\Big) \\
 & +\frac{{\rm i}}{2} \Big(\frac{ \pl_tC(\bs{\psi}_{+*}^t,g_*^t,\bs{\alpha}({\bf p}))}{C(\bs{\psi}_{+*}^t,g_*^t,\bs{\alpha}({\bf p}))}-c_{\rm L}\pl_{s} S_{\rm L}^0(\mc{P}_*^s,g_*^t,g_*^s)|_{s=t}\Big)+\frac{c_{\rm L}}{24}v_0^t,
 \end{split}\]
where $(\mc{P}^t_*,\boldsymbol{\zeta}_*^t)$ are defined using the flow of ${\rm i}{\rm v}^t$ and $g^t_*$ is admissible on $\mc{P}_*^t$, and $\bs{\psi}_{+*}^t$ is the associated model form.
Combining this with \eqref{real_Schwarzian_cocycle_proof_identity}, we thus get (recall the definition \eqref{GammaPhi_alpha_p})
\[\begin{split}
\frac{c_{\rm L}}{24\pi {\rm i}} \oint_{\T} v^t{\rm S}(\psi_{+}^t\circ o)\dd z=& - \frac{1}{2}\frac{ \pl_tC(\bs{\psi}_+^t,g_+^t,\bs{\alpha}({\bf p}))}{C(\bs{\psi}_+^t,g_+^t,\bs{\alpha}({\bf p}))}+ \frac{c_{\rm L}}{2}\pl_{s}S_{\rm L}^0(\mc{P}_+^s,g_+^t,g_+^s)|_{s=t} +\frac{c_{\rm L}}{24}v_0^t+\frac{\pl_t\Gamma^{\Phi}_{\bs{\alpha},{\bf p}}(\bs{\psi}_+^t)}{\Gamma^{\Phi}_{\bs{\alpha},{\bf p}}(\bs{\psi}_+^t)}\\
& +\frac{{\rm i}}{2}\frac{ \pl_tC(\bs{\psi}_{+*}^t,g_*^t,\bs{\alpha}({\bf p}))}{C(\bs{\psi}_{+*}^t,g_*^t,\bs{\alpha}({\bf p}))}-{\rm i} \frac{c_{\rm L}}{2}\pl_{s}S_{\rm L}^0(\mc{P}_*^s,g_*^t,g_*^s)|_{s=t}.
\end{split}\]
Using \eqref{constantC(P,g)} and taking the real part of this identity, we obtain for $\hat{g}_+^t$ the filling metric of $g_+^t$
\[ {\rm Re}(\frac{1}{2\pi {\rm i}} \int_{\T} v^t{\rm S}(\psi_{+}^t\circ o)\dd z)=6(\pl_{s}S_{\rm L}^0(\mc{P}_+^s,g_+^t,g_+^s)|_{s=t}-\pl_tS_{\rm L}^0(\hat{\C},g_{\hat{\C}},\hat{g}_+^t))+\frac{1}{2}{\rm Re}(v_0^t).\]
Applying the same reasoning to $\bs{\psi}_-^t$, we obtain for $\hat{g}_-^t$ the filling metric of $g_-^t$
\[ {\rm Re}(\frac{1}{2\pi {\rm i}} \int_{\T} v^t{\rm S}(\psi_{-}^t)\dd z)=-6(\pl_{s}S_{\rm L}^0(\mc{P}_-^s,g_-^t,g_-^s)|_{s=t}-\pl_tS_{\rm L}^0(\hat{\C},g_{\hat{\C}},\hat{g}_-^t))+\frac{1}{2}{\rm Re}(v_0^t).\]
Therefore 
\[\begin{split}
\pl_t {\rm Re}(\Upsilon_{\mc{X}}(\bs{\Xi}^1,\bs{\Xi}^t))&=6\big(\pl_{s}S_{\rm L}^0(\mc{P}_+^s,g_+^t,g_+^s)|_{s=t}+\pl_{s}S_{\rm L}^0(\mc{P}_-^s,g_-^t,g_-^s)|_{s=t}-\pl_tS_{\rm L}^0(\hat{\C},g_{\hat{\C}},\hat{g}_+^t)-\pl_tS_{\rm L}^0(\hat{\C},g_{\hat{\C}},\hat{g}_-^t)\big) \\
&= 6\big(\pl_{t}S_{\rm L}^0(\Sigma,g^1,g^t)-\pl_tS_{\rm L}^0(\hat{\C},g_{\hat{\C}},\hat{g}_+^t)-\pl_tS_{\rm L}^0(\hat{\C},g_{\hat{\C}},\hat{g}_-^t)\big).
\end{split}
\]
This proves the differential identity when ${\rm e}_1$ is not a
self-gluing loop, by the invariance property with respect to the choice of $g$ far from the cutting curves, explained in the beginning of the proof.
It remains to consider the case where ${\rm e}_1$ is a self-gluing loop.
In that case $j^-(1)=j^+(1)=j$ and ${\rm e}_1=\{(j,k_-),(j,k_+)\}$, 
$k_-\neq k_+$,
so that the two sides of the cutting curve correspond to two distinct
boundary components of the same complex building block $\mc P_j^t$.
Consequently, $\mc P_+^t$ and $\mc P_-^t$ must not be regarded as two
distinct building blocks. At each fixed $t$, decompose the tangent vector of the deformation of
$\mc P_j^t$ as $\bs{\rm v}_j^t=\bs{\rm v}_{j,+}^t+\bs{\rm v}_{j,-}^t$,
where the two terms denote the contributions associated respectively
with the boundary charts $\psi_+^t\circ o$ and $\psi_-^t$. We keep a single family
of metrics $g_j^t:=g^t|_{\mc{P}_j^t}$ which is admissible simultaneously at the two boundary components,
and denote by $\hat{g}_j^t$ its filling metric.
Proposition~\ref{th:derivative_general} can be applied directly with
the two corresponding boundary vector fields. Since its right-hand
side is linear in these vector fields, the resulting expression is
the sum of the two one-boundary computations above. Hence
\[{\rm Re}\left( \frac{1}{2\pi{\rm i}}
\oint_{\T}v^t\big({\rm S}(\psi_+^t\circ o)-{\rm S}(\psi_-^t)\big)\dd z\right)=
6\,\partial_s\big[S_{\rm L}^0(\mc P_j^s,g_j^t,g_j^s)-S_{\rm L}^0(\hat{\C},g_{\hat{\C}},\hat g_j^s)\big]_{s=t}.
\]
Here the derivative on the right-hand side is the total derivative
of the single building block $\mc P_j^t$, including the variations of
both boundary components. Integrating with respect to $t$ and using the independence with
respect to the choice of admissible metrics proved above concludes
the proof.
\end{proof}

\subsection{Ward identity}

We now give another version of the Ward identity for a complex building block, connecting to the usual definition given in the language of Vertex Operator Algebras.

\begin{proposition}\label{Ward_VOA_pant}
Let $\tilde{\mc{X}}=[(F,\mc{P},J_\C,{\bf x},\bs{\zeta},{\bf Y})]\in \tT_{0,m,b}$ with $m+b=3$, $b\geq1$ where $\tilde{S}=(\mc{P},J_\C,{\bf x},\bs{\zeta},{\bf Y})$ is a decorated complex building block in model form, i.e.  
$\mc{P}=\hat{\C}\setminus \cup_{k=1}^b\psi_k(\D^\circ)$, $\zeta_k=\psi_k|_{\T}$ 
and ${\bf x}=(x_1,\dots,x_m)=(z^\bullet_{b+1},\dots,z^{\bullet}_3)$ for some $\bs{\psi}=(\psi_1,\dots,\psi_b)\in \mc{P}^\eps_{m,b}$, where $\eps>0$. Let $\bs{\sigma}=(\sigma_1,\dots,\sigma_b)$ be orientation signs with $\sigma_k=+1$ for $k\leq b^+$ and  $\sigma_k=-1$ for $k>b^+$, and let $\tilde{\mc{X}}^{\bs{\sigma}}$ be defined by \eqref{decor_teich_to_teich_sigma}. 
Let $\bs{\alpha}=(\alpha_1,\dots,\alpha_m)$ be weights satisfying the Seiberg bounds.
Let ${\rm v}$ be a holomorphic vector field on a neighbourhood of $\mc{P}$ in $\hat{\C}$ and vanishing at the marked points $\bf x$, and set $\v_k:=(\psi_k\circ o)^* {\rm v}$ for $k=1,\dots,b^+$ and $\v_k:=\psi_k^* {\rm v}$ for $k=b^++1,\dots,b$.
Then the following equation holds, as bounded operators in $\mc{L}((h_0^{-1/2}L^2(I\times \mc{T}))^{\otimes b^-}, (h_0^{1/2}L^2(I\times \mc{T}))^{\otimes b^+})$ for each compact interval $I\subset (0,\infty)$:
\begin{align*}
& \big(\sum_{k=1}^{b^+} \bs{\ell}_{\v_k}^{(k)}\big)\mc{B}_{\tilde{\mc{X}}^{\bs{\sigma}},\bs{\alpha}} - \mc{B}_{\tilde{\mc{X}}^{\bs{\sigma}},\bs{\alpha}}\big(\sum_{k=1}^{b^-} \bs{\ell}_{\v_{b^++k}}^{(k)}\big)+\sum_{k=1}^m\Delta_{\alpha_k}v'(x_k)  \mc{B}_{\tilde{\mc{X}}^{\bs{\sigma}},\bs{\alpha}}\\
&= - \sum_{k=1}^{b^+}\Big(\frac{c_{\rm L}}{24\pi {\rm i}}\oint_{\T} v_k(z) {\rm S}(\psi_k\circ o)(z) \dd z\Big)\mc{B}_{\tilde{\mc{X}}^{\bs{\sigma}},\bs{\alpha}}+  \sum_{k=b^++1}^{b}\Big(\frac{c_{\rm L}}{24\pi {\rm i}}\oint_{\T} v_k(z) {\rm S}(\psi_k)(z) \dd z\Big)\mc{B}_{\tilde{\mc{X}}^{\bs{\sigma}},\bs{\alpha}} ,
\end{align*}
where $\v_k=v_k(z)\pl_z$. In particular, if $\psi_k\in {\rm PSL}_2(\C)$ for $k=1,\dots,b$, i.e. if the parametrisations of the boundary induce a complex projective structure on the closed surface $\hat{\mc{P}}$ filling $\mc{P}$ (see Lemma \ref{FillingSigma}), then the Ward identity reads  
\[\big(\sum_{k=1}^{b^+} \bs{\ell}_{\v_k}^{(k)}\big)\mc{B}_{\tilde{\mc{X}}^{\bs{\sigma}},\bs{\alpha}} -  \mc{B}_{\tilde{\mc{X}}^{\bs{\sigma}},\bs{\alpha}}\big(\sum_{k=1}^{b^-}\bs{\ell}_{\v_{b^++k}}^{(k)}\big)+\sum_{k=1}^m\Delta_{\alpha_k}v'(x_k)  \mc{B}_{\tilde{\mc{X}}^{\bs{\sigma}},\bs{\alpha}}=0.\]
\end{proposition}
\begin{proof} We write the model form as $\tilde{\Theta}_{m,b}(\tilde{\mc{X}})=(\bs{\psi},\bs{\theta},\bs{\la})$. 
Let $f_t$ be the flow of ${\rm v}$ defined for small $t$ in a neighbourhood of $\mc{P}$ in  $\hat{\C}$ and 
consider the decorated marked complex building block 
\[ (f_t\circ F,f_t(\mc{P}),J_\C,{\bf x},f_t\circ \bs{\zeta}^{\bs{\sigma}},(f_t)_*{\bf Y})\]
where $(f_t)_*{\bf Y}:=(f_t'(x_1)Y_1,\dots,f_t'(x_m)Y_m)$ and $F:\Sigma^\bullet_{m,b}\to\mc{P}\subset\hat{\C}$ has been extended smoothly to a small 
neighbourhood of the base surface $\Sigma^\bullet_{m,b}$ defined in \eqref{sigma0mb}. Since $f_t$ is holomorphic and isotopic to the identity, we have for small $t$
\[\tilde{\mc{X}}^{\bs{\sigma}}(t):=[(f_t\circ F,f_t(\mc{P}),J_\C,{\bf x},f_t\circ \bs{\zeta}^{\bs{\sigma}},(f_t)_*{\bf Y})]=[(F,\mc{P},J_\C,{\bf x},\bs{\zeta}^{\bs{\sigma}},{\bf Y})]=\tilde{\mc{X}}^{\bs{\sigma}} \, \textrm{ in }\tT_{0,m,b^+,b^-},\] 
and thus 
\[\mc{B}_{\tilde{\mc{X}}^{\bs{\sigma}}(t),\bs{\alpha}}=\mc{B}_{\tilde{\mc{X}}^{\bs{\sigma}},\bs{\alpha}}.\] 
Let $\bs{f}^t=(f^t_1,\dots,f^t_b)$ with 
\[ f^t_k=(\psi_k\circ o)^{-1}\circ f_t\circ (\psi_k\circ o) \textrm{ for }k=1,\dots,b^+, \quad 
f^t_k=\psi_k^{-1}\circ f_t\circ \psi_k \textrm{ for }k=b^++1,\dots,b,\]
so that $\mc{X}^{\bs{\sigma}}(t)=\mc{X}^{\bs{\sigma}} \cdot \bs{f}^t$ where $\mc{X}^{\bs{\sigma}}(t)$ is the projection of $\tilde{\mc{X}}^{\bs{\sigma}}(t)$ on $\mc{T}_{0,m,b^+,b^-}$. We thus get 
\[\begin{split}
\mc{B}_{\tilde{\mc{X}}^{\bs{\sigma}}(t),\bs{\alpha}}=&\, \mc{B}_{\mc{X}^{\bs{\sigma}}(t),\bs{\alpha}}
(\la_1+\log f_t'(x_1),\dots,\la_m+\log f_t'(x_m))\\
=& \,e^{\sum_{k=1}^m\Delta_{\alpha_k}\log f_t'(x_k)}\mc{B}_{\tilde{\mc{X}}^{\bs{\sigma}}\cdot \bs{f}_t,\bs{\alpha}}
\end{split}\]
where $\bs{\la}=(\la_1,\dots,\la_m)$.
By differentiating the identity above at $t=0$ and using Proposition \ref{general_derivative_block_amp}, we obtain the result since the left-hand side is constant in $t$.
\end{proof}

As we discussed in \eqref{ellvQ+ip}, the operators $\bs{\ell}_{\rm v}$ can be represented as continuous families depending on $p\in \R$ of bounded operators $\bs{\ell}_{\rm v}^{Q+{\rm i}p}: \cjg \nu\cjd^{-1/2}L^2(\mc{T})\to  \cjg \nu\cjd^{1/2}L^2(\mc{T})$, where $\cjg \nu\cjd:=(1+|\nu|^2)^{1/2}$ is comparable to $h_0(p)$ for $p$ in compact sets. The Ward identity of Proposition \ref{Ward_VOA_pant} can then be rewritten in the following way.
For a holomorphic vector field ${\rm v}$ as in Proposition \ref{Ward_VOA_pant}, we have for almost all ${\bf p}=(p_1,\dots,p_b)\in\R_+^b$
\begin{equation}\label{ward_p_fixed}
\begin{aligned}
& \big(\sum_{k=1}^{b^+}  {\rm Id}^{\otimes (k-1)}\otimes \bs{\ell}_{\v_k}^{Q+{\rm i}p_k}\otimes {\rm Id}^{\otimes (b^+-k)}\big)\mc{B}_{\tilde{\mc{X}}^{\bs{\sigma}},\bs{\alpha}}({\bf p}) - \mc{B}_{\tilde{\mc{X}}^{\bs{\sigma}},\bs{\alpha}}({\bf p})\big(\sum_{k=1}^{b^-} ({\rm Id}^{\otimes(k-1)}\otimes \bs{\ell}_{\v_{b^++k}}^{Q+{\rm i}p_{b^++k}}\otimes {\rm Id}^{\otimes(b^--k)}\big)\big)\\
&=-\sum_{k=1}^m\Delta_{\alpha_k}v'(x_k)  \mc{B}_{\tilde{\mc{X}}^{\bs{\sigma}},\bs{\alpha}}({\bf p}) - \sum_{k=1}^{b^+}\Big(\frac{c_{\rm L}}{24\pi {\rm i}}\oint_{\T} v_k(z) {\rm S}(\psi_k\circ o)(z) \dd z\Big)\mc{B}_{\tilde{\mc{X}}^{\bs{\sigma}},\bs{\alpha}}({\bf p})\\
& \quad +  \sum_{k=b^++1}^{b}\Big(\frac{c_{\rm L}}{24\pi {\rm i}}\oint_{\T} v_k(z) {\rm S}(\psi_k)(z) \dd z\Big)\mc{B}_{\tilde{\mc{X}}^{\bs{\sigma}},\bs{\alpha}}({\bf p}),
\end{aligned}
\end{equation}
where $\mc{B}_{\tilde{\mc{X}}^{\bs{\sigma}},\bs{\alpha}}({\bf p}): (e^{\eps|\nu|} L^2(\mc{T}))^{\otimes b^-}\to (e^{-\eps|\nu|}L^2(\mc{T}))^{\otimes b^+}$ is the operator defined by 
\[\mc{B}_{\tilde{\mc{X}}^{\bs{\sigma}},\bs{\alpha}}({\bf p})(u_{b^++1}\otimes \dots \otimes u_{b})(\nu_1,\dots,\nu_{b^+})
=\sum_{(\nu_{b^++1},\dots,\nu_b)\in\mc{T}^{b^-}} \mc{B}_{\bs{\psi},\bs{\theta},\bs{\la},\bs{\alpha}}({\bf p},\nu_1,\dots,\nu_b)\prod_{k=b^++1}^{b}u_k(\nu_k)\]
using the definition \eqref{coef_B_in_terms_of_w} (with $(\bs{\psi},\bs{\theta},\bs{\la})$ the model form of $\tilde{\mc{X}}$) 
and $\eps>0$ is chosen small enough so that \eqref{bound_bloc_pant_Dozz^4} holds. The Ward identity \eqref{ward_p_fixed} holds in the sense of 
bounded operators $(\delta^{-|\nu|} L^2(\mc{T}))^{\otimes b^-}\to (\delta^{|\nu|}L^2(\mc{T}))^{\otimes b^+}$ up to increasing slightly $\delta<1$ to absorb the $\cjg \nu\cjd^{1/2}$ weight loss from the mapping property $\bs{\ell}_{\rm v}^{Q+{\rm i}p}: \cjg \nu\cjd^{-1/2}L^2(\mc{T})\to  \cjg \nu\cjd^{1/2}L^2(\mc{T})$.\\

Next, we generalize the Ward identity of Proposition \ref{Ward_VOA_pant} from complex building blocks to general surfaces obtained by gluing the complex building blocks.

\begin{theorem}[\textbf{Ward identities for conformal blocks}]\label{Ward_VOA_general}
Let $\tilde{\mc{X}}_{\bs{\Xi}}=[(F,\Sigma,J,{\bf x},\bs{\zeta},{\bf Y},\bs{\Xi})]\in \BDTT_{g,m,b}$ with model form 
$\tilde{\mc{MF}}(\bs{\Xi})=((\bs{\psi}_1,\bs{\theta}_1,\bs{\la}_1),\dots,(\bs{\psi}_{j_0},\bs{\theta}_{j_0},\bs{\la}_{j_0}))$ as defined in \eqref{definition_MFXi}. Let $\mc{G}$ be the gluing graph of $\bs{\Xi}$ 
with $i_0$ internal edges and $j_0$ internal vertices, and pick an orientation $\vec{\rm e}_i=((j,k),(j',k'))$ for each internal edge.
For each $j=1,\dots,j_0$, let $\bs{\alpha}_j=(\alpha_{j1},\dots,\alpha_{jm_j})\in (0,Q)^{m_j}$ be weights satisfying the Seiberg bound $\sum_{k=1}^{m_j}\alpha_{jk}>(2-b_j)Q$. As in Lemma \ref{block_glued}, we denote 
\[\bs{\alpha}=(\alpha_1,\ldots,\alpha_m),\qquad \alpha_i=\alpha_{jk}\, \textrm{ if }{\rm e}_i^{\rm m}=\{(i,\bullet),(j,b_j+k)\}.\]
We denote for each $i=1,\ldots b$
\[\psi_i=\psi_{jk},\quad {\rm v}_i=(\Xi_{j}\circ \zeta_k^\bullet)^*{\rm v},\quad\textrm{if }{\rm e}_i^{\rm b}=\{(i,\circ),(j,k)\},\]
and for each $i=1,\ldots,i_0$
\[{\rm v}_{{\rm e}_i}=\xi_i^*{\rm v},\quad v_{\mc{P}_j}\pl_z= (\Phi_{\mc{P}_j})_*{\rm v},\quad\textrm{if }\vec{\rm e}_i=((j,k),(j',k')),\]
where $\xi_i=\Xi_{j}\circ\zeta_{k}^\bullet$ as before and $\Phi_{\mc{P}_j}:\mc{P}_j \to \hat{\C}$ is the uniformising map in Lemma \ref{modelform} putting into model form the complex building block $(\mc{P}_j,J)$ associated to $\Xi_j$ as in \eqref{element_in_subTeich}.
Let $\v$ be a holomorphic vector field on a neighbourhood of $\Sigma$ in its filling $(\hat{\Sigma},\hat{J})$ vanishing at the marked points ${\bf x}$.
Then the following identity holds, as bounded operators 
$(h_0^{-1/2}L^2(I\times\mc{T}))^{\otimes b}\to \C$ 
for almost all ${\bf p}\in \R_+^{i_0}$:
\begin{equation}\label{Ward_identity_block_1}
\begin{split}
& -\mc{B}_{\tilde{\mc{X}}_{\bs{\Xi}}, \bs{\alpha}}({{\bf p}})\Big(\sum_{k=1}^{b} \bs{\ell}_{\v_{k}}^{(k)}\Big)+\sum_{j=1}^{j_0}\sum_{k=1}^{m_j}
\Delta_{\alpha_{jk}}v_{\mc{P}_j}'(z^\bullet_{b_j+k}) \mc{B}_{\tilde{\mc{X}}_{\bs{\Xi}}, \bs{\alpha}}({{\bf p}}) =  \Big(\sum_{k=1}^{b}\frac{c_{\rm L}}{24\pi {\rm i}}\oint_{\T} v_k {\rm S}(\psi_k)\dd z -\frac{c_{\rm L}}{12}\Theta_{\bs{\Xi}}({\bf v}_{\rm e})\Big)
\mc{B}_{\tilde{\mc{X}}_{\bs{\Xi}}, \bs{\alpha}}({{\bf p}}),
\end{split}
\end{equation}
where ${\bf v}_{\rm e}=({\rm v}_{{\rm e}_1},\dots,{\rm v}_{{\rm e}_{i_0}})$ and 
${\rm S}(\cdot)$ denotes the Schwarzian derivative as always. This identity can also be viewed as an identity of bounded operators 
$(\delta^{-|\nu|}L^2(\mc{T}))^{\otimes b}\to \C$ for some $\delta\in (0,1)$ and for almost all $({\bf p},{\bf p}')\in \R_+^{i_0+b}$:
\begin{align*}
& -\mc{B}_{\tilde{\mc{X}}_{\bs{\Xi}}, \bs{\alpha}}({\bf p},{\bf p}')\Big(\sum_{k=1}^{b} {\rm Id}^{\otimes(k-1)}\otimes \bs{\ell}_{\v_{k}}^{Q+{\rm i}p'_k}\otimes {\rm Id}^{\otimes(b-k)}\Big) +\sum_{j=1}^{j_0}\sum_{k=1}^{m_j}
\Delta_{\alpha_{jk}}v_{\mc{P}_j}'(z^\bullet_{b_j+k})\mc{B}_{\tilde{\mc{X}}_{\bs{\Xi}}, \bs{\alpha}}({\bf p},{\bf p}') \\ 
&=  \Big(\sum_{k=1}^{b}\frac{c_{\rm L}}{24\pi {\rm i}}\oint_{\T} v_k {\rm S}(\psi_k)\dd z -\frac{c_{\rm L}}{12}\Theta_{\bs{\Xi}}({\bf v}_{\rm e})\Big)
 \mc{B}_{\tilde{\mc{X}}_{\bs{\Xi}}, \bs{\alpha}}({\bf p},{\bf p}').
 \end{align*}
\end{theorem}
\begin{proof}
Proposition \ref{Ward_VOA_pant} can be reformulated in the following way. Let $m+b=3$, $b\geq1$. For $\bs{\v}=(\v_k)_{k=1}^b\in T_{\rm Id}\mc{U}_\eps^b$ and $\mc{B}\in(L^2(\mc{T})^*)^{\otimes b}$, define $\nabla_{\bs{\v}}\mc{B}:=-\sum_{k=1}^b\mc{B}\,\bs{\ell}_{\v_k}^{(k)}$ (if the composition makes sense). Let $\tilde{\mc{X}}=[(F,\mc{P},J,{\bf x},\bs{\zeta},{\bf Y})]\in\tT_{0,m,b}$ with weights $\bs{\alpha}$ satisfying the Seiberg bounds and $\bs{\v}=(\v_k)_{k=1}^b\in T_{\rm Id}\mc{U}_\eps^b$. If there exists a holomorphic vector field $\v$ on $\mc{P}$ such that $\v_k=\zeta_k^*\v$ for all $k$, then
\begin{equation}\label{identity_nablavB}
\nabla_{\bs{\v}}\mc{B}_{\tilde{\mc{X}},\bs{\alpha}}=\left(-\sum_{k=1}^m\Delta_{\alpha_k}v'(x_k)+\sum_{k=1}^b\frac{c_{\rm L}}{24\pi {\rm i}}\oint_{\T} v_k {\rm S}(\psi_k) \dd z\right)\mc{B}_{\tilde{\mc{X}},\bs{\alpha}}.
\end{equation}
Returning to the theorem, let us apply this to each $\bs{\v}_j=(\v_{jk})_{k=1}^{b_j}$ on $\mc{P}_j$ where $\v_{jk}:=(\Xi_j\circ\zeta_k^\bullet)^*\v$. We obtain for almost all $\bf p,p'$ (omitted from the notation for simplicity)
\[\begin{aligned}
& \sum_{j=1}^{j_0}{\rm Tr}^{\mc{G}}_{\mc{H}_{\mc{T}}}(\mc{B}_{\tilde{\mc{X}}_1,\bs{\alpha}_1}\otimes\cdots\otimes\nabla_{\bs{\v}_j}\mc{B}_{\tilde{\mc{X}}_{j},\bs{\alpha}_{j}}\otimes\cdots\otimes\mc{B}_{\tilde{\mc{X}}_{j_0},\bs{\alpha}_{j_0}}) \\
&= \Big(-\sum_{j=1}^{j_0}\sum_{k=1}^{m_j}
\Delta_{\alpha_{jk}}v_{\mc{P}_j}'(z^\bullet_{b_j+k}) +  \sum_{j=1}^{j_0}\sum_{k=1}^{b_j}\frac{c_{\rm L}}{24\pi {\rm i}}\oint_{\T} v_{jk} {\rm S}(\psi_{jk})\dd z
\Big)\mc{B}_{\tilde{\mc{X}}_{\bs{\Xi}}, \bs{\alpha}}.
\end{aligned}\]
Here, a priori, $\nabla_{\bs{\v}_j}\mc{B}_{\tilde{\mc{X}}_{j},\bs{\alpha}_{j}}\in ((h_0^{-1/2}\mc{H}_{\mc{T}})^*)^{\otimes b_j}$, but from \eqref{identity_nablavB} we see that it belongs to $(\mc{H}_{\mc{T}}^*)^{\otimes b_j}$.
By the discussion before Definition \ref{Schwarzian_form}, the sum of the Schwarzian terms is precisely the right-hand side of \eqref{Ward_identity_block_1}. It remains to show that the $\bs{\ell}_{\v_{jk}}$ terms associated to the internal edges vanish after taking the trace. 
In the case of gluing two complex building blocks associated to an internal edge ${\rm e}_i=\{(j,k),(j',k')\}$, assuming $k=k'=1$ for simplicity, we have an identity of the form (with $\ldots$ denoting fixed elements)
\[{\rm Tr}_{\mc{H_T}}^{{\rm e}_i}(\mc{B}_{\tilde{\mc{X}}_j,\bs{\alpha}_j}(\bs{\ell}_{\v_{j1}}^{Q+{\rm i}p_i}\cdot,\ldots)\otimes\mc{B}_{\tilde{\mc{X}}_{j'},\bs{\alpha}_{j'}}(\cdot,\ldots))+{\rm Tr}_{\mc{H_T}}^{{\rm e}_i}(\mc{B}_{\tilde{\mc{X}}_j,\bs{\alpha}_j}(\cdot,\ldots)\otimes\mc{B}_{\tilde{\mc{X}}_{j'},\bs{\alpha}_{j'}}(\bs{\ell}_{\v_{j'1}}^{Q+{\rm i}p_i}\cdot,\ldots))=0.\]
To justify this cancellation, set $f(u):= \mc B_{\tilde{\mc X}_j,\bs\alpha_j}(u,\ldots)$ and $h(u):= \mc B_{\tilde{\mc X}_{j'},\bs\alpha_{j'}}(u,\ldots)$
 and let $h^*\in\mc H_{\mc T}$ be the Riesz representative characterized by 
 $\cjg h^*,u\cjd_{\mc H_{\mc T}} = \overline{h(u)}$. We have $\bbar{h^*(\nu)}=h(\mathbf 1_\nu)$. Define the (a priori unbounded) linear operators 
 \[ B:=\bs\ell_{\v_{j1}}^{Q+{\rm i}p_i}, \qquad Au:=f(u)\,\bbar{h^*} \] 
on $\mc{H}_{\mc{T}}$. 
Since $\bs\ell_{\v_{j'1}}^{Q+{\rm i}p_i} = -B^{\intercal}$,
where the transpose is taken with respect to the canonical basis $(\mathbf 1_\nu)_{\nu\in\mc T}$, the contribution of the edge ${\rm e}_i$ is 
\[ \sum_{\nu\in\mc T} f(B\mathbf 1_\nu)h(\mathbf 1_\nu) - \sum_{\nu\in\mc T} f(\mathbf 1_\nu)h(B^{\intercal}\mathbf 1_\nu)= {\rm Tr}(AB)-{\rm Tr}(BA). \] 
Indeed,  $\cjg AB\mathbf 1_\nu,\mathbf 1_\nu\cjd = f(B\mathbf 1_\nu)h(\mathbf 1_\nu)$ 
while $\cjg BA\mathbf 1_\nu,\mathbf 1_\nu\cjd = f(\mathbf 1_\nu) h(B^{\intercal}\mathbf 1_\nu)$.
Let $E_\eps$ be the diagonal (unbounded) operator defined by $E_\eps\mathbf 1_\nu=e^{\eps|\nu|}\mathbf 1_\nu$. 
By Lemma~\ref{lem:estimate_blocks}, $A_\eps:=E_\eps A E_\eps$ is trace class on $\mc{H}_{\mc{T}}$ for $\eps>0$ small enough, whereas
  $B_\eps:=E_\eps^{-1}BE_\eps^{-1}$ is bounded on $\mc{H}_{\mc{T}}$. 
Using the absolute convergence of the traces below, we directly see that 
\[ {\rm Tr}(AB)={\rm Tr}(A_\eps B_\eps), \qquad {\rm Tr}(BA)={\rm Tr}(B_\eps A_\eps). \] 
 The cyclicity of the trace for composition of trace class with bounded operators  therefore gives \[ {\rm Tr}(AB)-{\rm Tr}(BA) = {\rm Tr}(A_\eps B_\eps)-{\rm Tr}(B_\eps A_\eps) =0. \] This proves the cancellation of the two terms associated with the internal edge ${\rm e}_i$.

In the self-gluing case, we use a similar argument as in the proof of Lemma~\ref{block_glued}, from the
Hilbert--Schmidt factorisation.  Writing $B_{\nu',\nu}:=\cjg B\mathbf{1}_\nu,\mathbf{1}_{\nu'}\cjd_{\mc{H}_{\mc{T}}}$ and
$K_{\nu',\nu}:= \mc B_{\tilde{\mc X}_j,\bs\alpha_j}
(\mathbf 1_{\nu'},\mathbf 1_\nu,\ldots)$ with $B$ as above, the two internal-edge contributions are
\[
\sum_{\nu',\nu\in\mc T}B_{\nu',\nu}K_{\nu',\nu}
\quad\text{and}\quad
-\sum_{\nu',\nu\in\mc T}B_{\nu,\nu'}K_{\nu,\nu'}.
\]
They cancel after exchanging $\nu'$ and $\nu$. The
Hilbert--Schmidt factorisation used to define the self-gluing trace
and the weighted estimates above justify the absolute convergence of
these series.
\end{proof}

We notice that if the biholomorphisms $\psi_{jk}$ appearing in the model forms of the complex building blocks are M\"obius transformations, then their Schwarzian derivatives vanish and the Ward identity becomes simply
\[\begin{split}
& -\mc{B}_{\tilde{\mc{X}}_{\bs{\Xi}}, \bs{\alpha}}({{\bf p}})\Big(\sum_{k=1}^{b} \bs{\ell}_{\v_{k}}^{(k)}\Big)+\sum_{j=1}^{j_0}\sum_{k=1}^{m_j}
\Delta_{\alpha_{jk}}v_{\mc{P}_j}'(z^\bullet_{b_j+k}) \mc{B}_{\tilde{\mc{X}}_{\bs{\Xi}}, \bs{\alpha}}({{\bf p}}) = 0.
 \end{split}\]

As for complex building blocks, we have a version for surfaces with both incoming and outgoing boundaries. Let $\bs{\sigma}=(\sigma_1,\dots,\sigma_b)$ be orientation signs with $\sigma_k=+1$ for $k\leq b^+$ and  $\sigma_k=-1$ for $k>b^+$, and let $\tilde{\mc{X}}^{\bs{\sigma}}$ be defined by \eqref{decor_teich_to_teich_sigma}.
Then we have the Ward identity
\[\begin{split}
& \Big(\sum_{k=1}^{b^+} \bs{\ell}_{\v_k}^{(k)}\Big) \mc{B}_{\tilde{\mc{X}}^{\bs{\sigma}}_{\bs{\Xi}} ,\bs{\alpha}}({{\bf p}}) -\mc{B}_{\tilde{\mc{X}}^{\bs{\sigma}}_{\bs{\Xi}}, \bs{\alpha}}({{\bf p}})\Big(\sum_{k=1}^{b^-} \bs{\ell}_{\v_{b^++k}}^{(k)}\Big)+\sum_{j=1}^{j_0}\sum_{k=1}^{m_j}
\Delta_{\alpha_{jk}}v_{\mc{P}_j}'(z^\bullet_{b_j+k}) \mc{B}_{\tilde{\mc{X}}^{\bs{\sigma}}_{\bs{\Xi}}, \bs{\alpha}}({{\bf p}}) \\
&=  \Big(-\sum_{k=1}^{b^+}\frac{c_{\rm L}}{24\pi {\rm i}}\oint_{\T} v_k {\rm S}(\psi_k\circ o)\dd z  
+\sum_{k=b^++1}^{b}\frac{c_{\rm L}}{24\pi {\rm i}}\oint_{\T} v_k {\rm S}(\psi_k)\dd z -\frac{c_{\rm L}}{12}\Theta_{\bs{\Xi}}({\bf v}_{\rm e})\Big)
\mc{B}_{\tilde{\mc{X}}^{\bs{\sigma}}_{\bs{\Xi}}, \bs{\alpha}}({{\bf p}}).
 \end{split}\]

\subsection{Local dependence of conformal blocks on the cutting curves}
We now show that the conformal block of a decorated marked Riemann surface $(F,\Sigma,J,{\bf x},\bs{\zeta},{\bf Y})$ equipped with a marked complex building block decomposition $\bs{\Xi}$ depends in an explicit way, in  terms of the Schwarzian cocycle, on the cutting curves induced by $\bs{\Xi}$, under an isotopic deformation of the complex building block decomposition $\bs{\Xi}$. The proof is based on the variation formula (Proposition \ref{general_derivative_block_amp}) of the conformal block amplitudes of marked complex building blocks on the decorated Teichm\"uller space $\tT_{0,m,b}$ of genus $0$ with $m$ marked points and $b$ boundary components with $m+b=3$, $b\geq1$.
This final relation between two conformal blocks associated to two isotopic building block decompositions is a key property to define the conformal blocks as sections of a holomorphic line bundle over the Teichm\"uller space in the case of closed Riemann surfaces with marked points.

\begin{proposition}\label{well-defined_blocks}
Let $[\bs{\Xi}^\bullet]\in\pi_0\tilde{\mc{BD}}_{g,m,b}$ and $\tilde{\mc{X}}\in\tT_{g,m,b}$, and consider  
\[\tilde{\mc{X}}_{\bs{\Xi}^1},\tilde{\mc{X}}_{\bs{\Xi}^2}\in\BDTT_{g,m,b}^{[\bs{\Xi}^\bullet]}
\cap\pi_{\mc{D}\tT\to \tT}^{-1}(\tilde{\mc{X}})\] 
and $\bs{\alpha}\in \C^{m}$ weights satisfying the generalised Seiberg bounds with respect to the gluing graph of $[\bs{\Xi}^\bullet]$ and \eqref{cdozznot=0}. Then the conformal blocks satisfy 
\begin{equation}\label{relation_local_dependence}
\mc{B}_{\tilde{\mc{X}}_{\bs{\Xi}^2},\bs{\alpha}}=e^{\frac{c_{\rm L}}{12}
\Upsilon(\mc{X}_{\bs{\Xi}^1},\mc{X}_{\bs{\Xi}^2})}
\mc{B}_{\tilde{\mc{X}}_{\bs{\Xi}^1},\bs{\alpha}},
\end{equation}
where $\mc{X}_{\bs{\Xi}^\ell}=\pi_{\mc{D}\tT\to\mc{DT}}(\tilde{\mc{X}}_{\bs{\Xi}^\ell})$ is the projection of $\tilde{\mc{X}}_{\bs{\Xi}^\ell}$ on $\BDT_{g,m,b}$, for $\ell=1,2$.
\end{proposition}
\begin{proof} It suffices to consider the case $\bs{\alpha}$ real valued, since the result then extends holomorphically for complex $\bs{\alpha}$.
Consider a family $t\in [1,2]\mapsto\bs{\Xi}^t$ of marked complex building block decompositions of a representative Riemann surface $\tilde{S}=(\Sigma,J,{\bf x},\bs{\zeta},{\bf Y})$ for $\tilde{\mc{X}}=[(F,\Sigma,J,{\bf x},\bs{\zeta},{\bf Y})]$, such that it interpolates between $\bs{\Xi}^1$ and $\bs{\Xi}^2$. We denote 
$\tilde{\mc{X}}_{\bs{\Xi}^t}=[(F,\Sigma,J,{\bf x},\bs{\zeta},{\bf Y},\bs{\Xi}^t)]$, $\mc{X}_{\bs{\Xi}^t}=[(F,\Sigma,J,{\bf x},\bs{\zeta},\bs{\Xi}^t)]$, where only $\bs{\Xi}^t$ depends on $t$. Then $\Upsilon(\mc{X}_{\bs{\Xi}^1},\mc{X}_{\bs{\Xi}^2})$ is defined as in Definition \ref{d:cocycle}, and we use the notations from that definition. For each $j$ and $t$, let $\tilde{\mc{X}}_j^t=[(\Xi_j^t,\mc{P}_j^t,J,{\bf x}_j,\bs{\zeta}_j^t,{\bf Y}_j)]$ be the decorated marked complex building block induced by $\Xi_j^t$, with model form $(\bs{\psi}_j^t,\bs{\theta}_j^t,\bs{\la}_j^t)$.
We shall apply Proposition \ref{general_derivative_block_amp} to each $\bs{\v}_j^t=(\v_{jk}^t)_{k=1}^{b_j}$ on $\mc{P}_j^t$ where $\v_{jk}^t:=\pl_s((\Xi_{j}^{t}\circ\zeta_k^\bullet)^{-1}\circ(\Xi_{j}^{t+s}\circ\zeta_k^\bullet))|_{s=0}$.  
Let 
\[\widehat{\mc{B}}_{\tilde{\mc{X}}_{\bs{\Xi}^t},\bs{\alpha}}:=
\mc{B}_{\tilde{\mc{X}}_1^t,\bs{\alpha}_1}\otimes\cdots\otimes\mc{B}_{\tilde{\mc{X}}_{j_0}^t,\bs{\alpha}_{j_0}}\]
and differentiate with respect to $t$ (using the notation $\nabla_{{\bf v}}$ from the proof of Theorem \ref{Ward_VOA_general}):
\[\pl_t\widehat{\mc{B}}_{\tilde{\mc{X}}_{\bs{\Xi}^t},\bs{\alpha}}=
\sum_{j=1}^{j_0}\mc{B}_{\tilde{\mc{X}}_1^t,\bs{\alpha}_1}\otimes\cdots\otimes\nabla_{\bs{\v}_j^t}\mc{B}_{\tilde{\mc{X}}_{j}^t,\bs{\alpha}_{j}}\otimes\cdots\otimes\mc{B}_{\tilde{\mc{X}}_{j_0}^t,\bs{\alpha}_{j_0}} -\Big( \sum_{j=1}^{j_0}\sum_{k=1}^{b_j}\frac{c_{\rm L}}{24\pi {\rm i}}\oint_{\T} v_{jk}^t {\rm S}(\psi_{jk}^t)\dd z
\Big)\widehat{\mc{B}}_{\tilde{\mc{X}}_{\bs{\Xi}^t}, \bs{\alpha}},\]
as an element in the space $\mc{L}(\otimes_{j=1}^{j_0}(h_0^{-1/2}L^2(I\times \mc{T}))^{\otimes b_j},\C)$ where $I\subset (0,\infty)$ is a compact interval. Here the $t$-derivative is continuous in that topology. Let $\chi\in C_c^\infty(I^{i_0})$ and consider the derivative with respect to $t$ of
\begin{equation}\label{intXtr} 
\int_{\R_+^{i_0}}\chi({\bf p}){\rm Tr}^{\mc{G}}_{\mc{H_T}}(\widehat{\mc{B}}_{\tilde{\mc{X}}_{\bs{\Xi}^t},\bs{\alpha}}({\bf p})) \dd {\bf p}.
\end{equation}
We have 
\[ \nabla_{\bs{\v}_j^t}\mc{B}_{\tilde{\mc{X}}_{j}^t,\bs{\alpha}_{j}}({\bf p}_j^{\mc{G}})=-\sum_{k=1}^{b_j} \mc{B}_{\tilde{\mc{X}}_{j}^t,\bs{\alpha}_{j}}({\bf p}_j^{\mc{G}})\circ \bs{\ell}^{(k)}_{{\rm v}_{jk}^t}\]
and there exists $\eps>0$ such that $\eps_{j\ell}=\eps$ for each $\ell\not= k$, 
and for all $\eps_{jk}<0$ (arbitrarily close to $0$), 
\begin{equation}\label{boundptblockblock}
\mc{B}_{\tilde{\mc{X}}_{j}^t,\bs{\alpha}_{j}}({\bf p}_j^{\mc{G}})\circ \bs{\ell}^{(k)}_{{\rm v}_{jk}^t}\in \big(e^{\sum_{\ell=1}^{b_j}\eps_{j\ell}|\nu_{j\ell}|}L^2((I\times \mc{T})^{b_j})\big)^*, \quad \mc{B}_{\tilde{\mc{X}}_{j}^t,\bs{\alpha}_{j}}({\bf p}_j^{\mc{G}})\in \big(e^{\eps\sum_{\ell=1}^{b_j}|\nu_{j\ell}|}L^2((I\times \mc{T})^{b_j})\big)^*
\end{equation}
with uniform norm (in that space) with respect to $t\in [1,2]$. 

Assume that the gluing graph $\mc G$ has no self-gluing loops. For $\chi\in C_c^\infty(I^{i_0})$, introduce the integrated graph contraction 
\[ \mc{C}_{\chi,\mc G}(W_1,\ldots,W_{j_0}) := 
\int_{I^{i_0}} \chi({\bf p})\, {\rm Tr}^{\mc G}_{\mc H_{\mc T}} (W_1\otimes\cdots\otimes W_{j_0})({\bf p}) \,\dd{\bf p}. \] 
Using the boundedness \eqref{weightedtr} on weighted spaces, 
for each term containing $\mc B_{\tilde{\mc X}_j^t,\bs\alpha_j} \circ\bs\ell_{\v_{jk}^t}^{(k)}$, we choose the weights $\eps_{jk}$ so that they satisfy \eqref{condition_trace}. This is possible due to \eqref{boundptblockblock}.
Since every continuous variable $p_i\in I\subset (0,\infty)$ occurs in two distinct vertex factors, the exponentially weighted Cauchy-Schwarz inequality in the Young-diagram indices, followed by the generalised Cauchy-Schwarz inequality in the ${\bf p}=(p_1,\dots,p_{i_0})$ variable, shows that $\mc{C}_{\chi,\mc G}$ is a continuous multilinear map on the corresponding weighted spaces. More precisely, 
\[ \big\| \mc{C}_{\chi,\mc G}(W_1,\ldots,W_{j_0}) \big\|_{E_{\rm ext}} \leq C_{\chi,I} \prod_{j=1}^{j_0}\|W_j\|_{E_j}, \] 
where $E_j$ denotes the relevant weighted space at the $j$-th vertex and 
\[ E_{\rm ext} = \big((e^{\eps'|\nu|} L^2(I\times\mc T))^*\big)^{\otimes b} \]
 is the space associated with the external edges. The local variation formula above shows that the vertex blocks depend $C^1$ on $t$ in these weighted spaces. By the continuity and multilinearity of $\mc{C}_{\chi,\mc G}$, the expression \eqref{intXtr} is therefore $C^1$ for $t\in[1,2]$, and the usual Leibniz rule for continuous multilinear maps gives 
 \begin{align}
& \pl_t \int_{\R_+^{i_0}}\chi({\bf p}){\rm Tr}^{\mc{G}}_{\mc{H_T}}(\widehat{\mc{B}}_{\tilde{\mc{X}}_{\bs{\Xi}^t},\bs{\alpha}}({\bf p})) \dd {\bf p}\nonumber\\
&= \int_{\R_+^{i_0}}\chi({\bf p})(\sum_{j=1}^{j_0}{\rm Tr}^{\mc{G}}_{\mc{H}_{\mc{T}}}(\mc{B}_{\tilde{\mc{X}}_1^t,\bs{\alpha}_1}\otimes\cdots\otimes\nabla_{\bs{\v}_j^t}\mc{B}_{\tilde{\mc{X}}_{j}^t,\bs{\alpha}_{j}}\otimes\cdots\otimes\mc{B}_{\tilde{\mc{X}}_{j_0}^t,\bs{\alpha}_{j_0}})({\bf p}))\dd {\bf p}\label{well-defined_blocks_identity}\\
& \quad - \Big( \sum_{j=1}^{j_0}\sum_{k=1}^{b_j}\frac{c_{\rm L}}{24\pi {\rm i}}\oint_{\T} v_{jk}^t {\rm S}(\psi_{jk}^t)\dd z
\Big)\int_{\R_+^{i_0}}\chi({\bf p})\mc{B}_{\tilde{\mc{X}}_{\bs{\Xi}^t}, \bs{\alpha}}({\bf p})\dd {\bf p},\nonumber
\end{align}
Now notice that $\v_{jk}^t=0$ for an edge ${\rm e}_i^{\rm b}=\{(i,\circ),(j,k)\}$ associated to the boundary $\pl_i\Sigma$, so there are no boundary Schwarzian terms in \eqref{well-defined_blocks_identity}.
By the discussion before Definition \ref{Schwarzian_form}, the (negative) sum of the Schwarzian terms is precisely $\frac{c_{\rm L}}{12}\Theta_{\bs{\Xi}^t}(\bs{\v}^t)$. As in the proof of Theorem \ref{Ward_VOA_general}, the trace terms sum up to zero. Consequently, \[ \partial_t \int_{I^{i_0}} \chi({\bf p}) \mc B_{\tilde{\mc X}_{\bs\Xi^t},\bs\alpha}({\bf p}) \,\dd{\bf p} = \frac{c_{\rm L}}{12} \Theta_{\bs\Xi^t}({\bf v}^t) \int_{I^{i_0}} \chi({\bf p}) \mc B_{\tilde{\mc X}_{\bs\Xi^t},\bs\alpha}({\bf p}) \,\dd{\bf p}. \] 
and this proves the result since the function $\chi$ is arbitrary.

Next, we consider the case with self-gluing edge in the graph $\mc G$, say ${\rm e}_i=\{(j,1),(j,2)\}$. To simplify the exposition, we only deal with the case where it is not the torus with one point, but that case can be dealt with in the same way.
We use here the canonical self-gluing diagonal constructed in the proof of Lemma~\ref{block_glued}, via the 
Hilbert--Schmidt factorisation of the Segal amplitude given by
\eqref{factorisation_bloc_ampl}.
More precisely, its spectrally transformed amplitude $\widehat{\mc{A}}_{S_j}=\overline{\Pi_{\mc{V}\otimes \bbar{\mc{V}}}^{\otimes 3}(\overline{\mc{A}_{S_j}})}$ then also factorises in the form 
\[\widehat{\mc{A}}_{S_j^t}(p_1,p_2,p_3,\nu_1,\nu_2,\nu_3,\tilde{\nu}_1,\tilde{\nu}_2,\tilde{\nu}_3)=\int_{H^{-s}(\T)}K_j^t(p_1,\nu_1,\tilde{\nu}_1,\tilde{\varphi})N_j^t(p_2,p_3,\nu_2,\tilde{\nu}_2,\nu_3,\tilde{\nu}_3,\tilde{\varphi})\dd \mu_0(\tilde{\varphi})\]
and $K_j^t\in  L^2(I\times \mc{T}^2\times H^{-s}(\T))$ and $N_j^t\in  L^2(I^2\times \mc{T}^4\times H^{-s}(\T))$. The auxiliary decomposition used in
\eqref{factorisation_bloc_ampl} may be chosen depending $C^1$ along the
family $t\in[1,2]$. The differentiability estimates for the Segal amplitudes then show that the resulting factors
$K_j^t,N_j^t$, are $C^1$ provided we work on weighted spaces.
Define the continuous bilinear operator (see \cite[Lemma 7.2]{GKRV21_Segal})
\[\begin{gathered}\
\mc{D}: L^2(I\times \mc{T}^2\times H^{-s}(\T))\times L^2(I^2\times \mc{T}^4\times H^{-s}(\T))\to L^1(I\times \mc{T}^2;L^2(I\times \mc{T}^2))\\  
\mc{D}(K,N)(p,\nu,\tilde{\nu};p_3,\nu_3,\tilde{\nu}_3):=\int_{H^{-s}(\T)}K(p,\nu,\tilde{\nu},\tilde{\varphi})N(p,p_3,\nu,\tilde{\nu},\nu_3,\tilde{\nu}_3,\tilde{\varphi})\dd \mu_0(\tilde{\varphi}).
\end{gathered}\]
working with exponential weighted spaces $e^{-\eps|\nu|}L^2(\mc{T})$ as we did before, the same arguments as above show that 
in $L^1(I\times \mc{T}^2;e^{-\eps(|\nu_3|+|\tilde{\nu}_3|)}L^2(I\times \mc{T}^2))$ we have
\[ \partial_t\mc{D}(K_j^t,N_j^t)
 = \mc{D}(\partial_t K_j^t,N_j^t) + \mc{D}(K_j^t,\partial_t N_j^t).
\]
As in the proof of Lemma \ref{block_glued}, by \eqref{factorisation_extended}, after fixing the anti-holomorphic descendants equal to $0$, this diagonal satisfies 
\[\mc{D}(K_j^t,N_j^t)(p,\nu,0,p',\nu',0) = C_j^t(p,p')\, \mc B_{\mc{X}_j^t} (p,p,p',\nu,\nu,\nu'), \] 
where $C_j^t$ is the product of the DOZZ constant, the metric dependent constants and the $\bbar{\Gamma}({\bf p})$ function. For $p,p'\in I$, $C_j^t$, $(C_j^t)^{-1}$ and their $t$-derivatives are uniformly bounded. 
Hence the self-glued block is $C^1$ in $L^1$, and 
\[
\partial_t{\rm Tr}_{\mc{H}_{\mc{T}}}^{{\rm e}_i}\mc{B}_{\mc{X}_j^t}
={\rm Tr}_{\mc H_{\mc{T}}}^{{\rm e}_i}(\partial_t\mc{B}_{\mc{X}_j^t})
={\rm Tr}_{\mc{H}_{\mc{T}}}^{{\rm e}_i}(\nabla_{\bs\v_j^t}\mc{B}_{\mc{X}_j^t})
-\Big(\sum_{k=1}^{b_j}\frac{c_{\rm L}}{24\pi{\rm i}}\oint_{\T}v_{jk}^t{\rm S}(\psi_{jk}^t)\,\dd z
\Big){\rm Tr}_{\mc{H}_{\mc{T}}}^{{\rm e}_i}(\mc{B}_{\mc{X}_j^t})
\]
in $L^1$. After performing this construction at every self-gluing vertex, the remaining contractions are treated by the continuous multilinear argument used above. Thus \eqref{well-defined_blocks_identity} remains valid. We finally check the cancellation of the two Virasoro terms associated with ${\rm e}_i$. Set $B:=\bs\ell_{\v_{j1}^t}^{Q+{\rm i}p_i}$. 
The gluing relation between the two boundary vector fields gives $\bs\ell_{\v_{j2}^t}^{Q+{\rm i}p_i}=-B^{\intercal}.$
Writing 
\[ B_{\nu',\nu}:= \cjg B\mathbf{1}_\nu,\mathbf{1}_{\nu'}\cjd_{\mc H_{\mc T}}, \qquad K_{\nu',\nu}^t:= \mc{B}_{\mc{X}_j^t} (\mathbf{1}_{\nu'},\mathbf 1_\nu,\ldots), \] 
the two contributions of the self-gluing edge are 
\[ \sum_{\nu',\nu\in\mc T}B_{\nu',\nu}K_{\nu',\nu}^t \quad\text{and}\quad -\sum_{\nu',\nu\in\mc T}B_{\nu,\nu'}K_{\nu,\nu'}^t. \] 
They cancel after exchanging $\nu'$ and $\nu$ in the second sum. The Hilbert--Schmidt factorisation and the weighted energy estimates justify the absolute convergence of these sums and of the associated $p,p'$ integrals. Therefore the self-gluing terms cancel as well, and the differential identity obtained in the case without self-gluing remains valid for an arbitrary gluing graph.
\end{proof}

As a corollary, we obtain the following:
\begin{corollary}[\textbf{Cocycle property}]\label{l:cocycle}
Let $[\bs{\Xi}^\bullet]\in\piBD_{g,m,b}$ and $\mc{X}\in\mc{T}_{g,m,b}$. For two points $\mc{X}_{\bs{\Xi}^1},\mc{X}_{\bs{\Xi}^2}\in\mc{DT}_{g,m,b}^{[\bs{\Xi}^\bullet]}\cap\pi_{\mc{DT}\to\mc{T}}^{-1}(\mc{X})$, the Schwarzian cocycle $\Upsilon(\mc{X}_{\bs{\Xi}^1},\mc{X}_{\bs{\Xi}^2})$ is well-defined, i.e.\ it does not depend on the family of marked complex building block decompositions interpolating between $\mc{X}_{\bs{\Xi}^1}$ and $\mc{X}_{\bs{\Xi}^2}$ used in its definition. Moreover, for $\mc{X}_{\bs{\Xi}^1},\mc{X}_{\bs{\Xi}^2},\mc{X}_{\bs{\Xi}^3}\in\mc{DT}_{g,m,b}^{[\bs{\Xi}^\bullet]}\cap\pi_{\mc{DT}\to\mc{T}}^{-1}(\mc{X})$, we have
\begin{equation}\label{cocycle3}
\Upsilon(\mc{X}_{\bs{\Xi}^1},\mc{X}_{\bs{\Xi}^2})+\Upsilon(\mc{X}_{\bs{\Xi}^2},\mc{X}_{\bs{\Xi}^3})=\Upsilon(\mc{X}_{\bs{\Xi}^1},\mc{X}_{\bs{\Xi}^3}).
\end{equation}
\end{corollary}
\begin{proof}
Let $S=(\Sigma,J,{\bf x},\bs{\zeta})$ be a representative of $\mc{X}=[(F,\Sigma,J,{\bf x},\bs{\zeta})]$ and $\mc{X}_{\bs{\Xi}^\ell}=[(F,\Sigma,J,{\bf x},\bs{\zeta},\bs{\Xi}^\ell)]$ for $\ell=1,2$. Let $\tilde{\mc{X}}=[(F,\Sigma,J,{\bf x},\bs{\zeta},{\bf Y})]\in\tT_{g,m,b}$ with ${\bf Y}=(Y_1,\ldots,Y_m)$ where $Y_i=D\Xi_j^1(x_k^\bullet)Y_k^\bullet$ if $x_i=\Xi_j^1(x_k^\bullet)$. Deforming $\bs{\Xi}^2$ near each marked point $x_k^\bullet$ on $\mc{P}_j^\bullet$ (which does not change $\Upsilon$), we may assume that $D\Xi_j^1(x_k^\bullet)Y_k^\bullet=D\Xi_j^2(x_k^\bullet)Y_k^\bullet$ for all $j,k$. Then $\bs{\Xi}^1,\bs{\Xi}^2$ are both marked complex building block decompositions of the decorated surface $\tilde{S}=(\Sigma,J,{\bf x},\bs{\zeta},{\bf Y})$, and $\tilde{\mc{X}}_{\bs{\Xi}^\ell}=[(F,\Sigma,J,{\bf x},\bs{\zeta},{\bf Y},\bs{\Xi}^\ell)]\in \BDTT_{g,m,b}^{[\bs{\Xi}^\bullet]}$ lifts $\mc{X}_{\bs{\Xi}^\ell}$ for $\ell=1,2$, i.e.\ $\pi_{\BDTT\to\BDT}(\tilde{\mc{X}}_{\bs{\Xi}^\ell})=\mc{X}_{\bs{\Xi}^\ell}$.
We fix weights $\bs{\alpha}\in\R^m$ satisfying the Seiberg bounds with respect to the gluing graph of $[\bs{\Xi}^\bullet]$. 
Applying Proposition \ref{well-defined_blocks} and using that the conformal block $\mc{B}_{\tilde{\mc{X}}_{\bs{\Xi}^1}, \bs{\alpha}}$ is not identically zero by Lemma \ref{block_glued}, we obtain that $\exp(\frac{c_{\rm L}}{12}\Upsilon(\mc{X}_{\bs{\Xi}^1},\mc{X}_{\bs{\Xi}^2}))$ is well-defined.
Since $\Upsilon(\mc{X}_{\bs{\Xi}^1},\mc{X}_{\bs{\Xi}^2})$ is independent of $c_{\rm L}$ and $\bs{\alpha}$, and since $c_{\rm L}$ varies in an open interval as $\gamma$ varies in $(0,2)$, it follows that $\Upsilon(\mc{X}_{\bs{\Xi}^1},\mc{X}_{\bs{\Xi}^2})$ is itself well-defined.

Similarly, modifying the $\bs{\Xi}^\ell$ so that there are lifts $\tilde{\mc{X}}_{\bs{\Xi}^\ell}\in \BDTT_{g,m,b}$ of $\mc{X}_{\bs{\Xi}^\ell}\in \BDT_{g,m,b}$ with a common projection $\pi_{\BDTT\to\tT}(\tilde{\mc{X}}_{\bs{\Xi}^\ell})=\tilde{\mc{X}}\in\tT_{g,m,b}$ for $\ell=1,2,3$, and 
applying Proposition \ref{well-defined_blocks} as above to the pairs $(\tilde{\mc{X}}_{\bs{\Xi}^1},\tilde{\mc{X}}_{\bs{\Xi}^2})$, $(\tilde{\mc{X}}_{\bs{\Xi}^2},\tilde{\mc{X}}_{\bs{\Xi}^3})$, $(\tilde{\mc{X}}_{\bs{\Xi}^1},\tilde{\mc{X}}_{\bs{\Xi}^3})$, 
we deduce the desired formula \eqref{cocycle3}.
\end{proof}

\subsection{Schwarzian line bundles}\label{sub:Schwarzian_line} 

The Schwarzian cocycle leads us to define holomorphic line bundles over the Teichm\"uller space $\mc{T}_{g,m,b}$, which are the subject of the current section.

First we define the following action of $\C^m$ on $\mc{D}\tT_{g,m,b}$.
Recall the one-to-one map $\tilde{\mc{MF}}^{\rm aug}$ of \eqref{def_MFtildehat}. Let $\tilde{\mc{X}}_{\bs{\Xi}}\in\mc{D}\tT_{g,m,b}$ with $\tilde{\mc{MF}}^{\rm aug}(\tilde{\mc{X}}_{\bs{\Xi}}) = (((\bs{\psi}_1,\bs{\theta}_1,\bs{\la}_1),\dots, (\bs{\psi}_{j_0},\bs{\theta}_{j_0},\bs{\la}_{j_0})),[\bs{\Xi}^\bullet])$ for some $[\bs{\Xi}^\bullet]\in\pi_0\tilde{\mc{BD}}_{g,m,b}$. For $\bs{t}\in \C^m$, we define $\bs{t}\cdot \tilde{\mc{X}}_{\bs{\Xi}}\in\mc{D}\tT_{g,m,b}$ by
\[ \tilde{\mc{MF}}^{\rm aug}(\bs{t}\cdot \tilde{\mc{X}}_{\bs{\Xi}})=(((\bs{\psi}_1,\bs{\theta}_1,\bs{\la}_1+\bs{t}_1),\dots, (\bs{\psi}_{j_0},\bs{\theta}_{j_0},\bs{\la}_{j_0}+\bs{t}_{j_0})),[\bs{\Xi}^\bullet]), \]
where $\bs{t}_j\in \C^{m_j}$ are the components in $\bs{t}$ associated to the marked points in $\Xi_j(\mc{P}_j^\bullet)$ (i.e.\ $t_{jk}=t_i$ if $\Xi_j(x_k^\bullet)=x_i$). By definition, if the decorated vectors of $\tilde{\mc{X}}_{\bs{\Xi}}$ are ${\bf Y}=(Y_1,\ldots,Y_m)$, then the decorated vectors of $\bs{t}\cdot\tilde{\mc{X}}_{\bs{\Xi}}$ are $\bs{t}\cdot{\bf Y}=(e^{t_1}Y_1,\ldots,e^{t_m}Y_m)$.

Notice that this action preserves the isotopy class $[\bs{\Xi}^\bullet]$ and the underlying point $\mc{X}=\pi_{\BDTT\to\mc{T}}(\tilde{\mc{X}}_{\bs{\Xi}})$ in $\mc{T}_{g,m,b}$, that is,
\[  \pi_{\mc{D}\tT\to \tT}(\{\bs{t}\cdot \tilde{\mc{X}}_{\bs{\Xi}}\,|\, \bs{t}\in \C^m\})\subset \pi^{-1}_{\tT\to \mc{T}}(\mc{X}).
\]

\begin{definition}[\textbf{Schwarzian line bundles}]\label{def:schwarzian_line}
 
1) Let $[\bs{\Xi}^\bullet]\in \piBD_{g,m,b}$ be an isotopy class of marked complex building block decompositions. For $\mc{X}\in\mc{T}_{g,m,b}$, we define the $1$-dimensional complex vector space
\begin{equation}\label{def_line}
\mc{L}_{\mc{X}}^{[\bs{\Xi}^\bullet]}:=\Big\{ G: \mc{D}\mc{T}_{g,m,b}^{[\bs{\Xi}^\bullet]}\cap\pi_{\mc{DT}\to\mc{T}}^{-1}(\mc{X}) 
\to \C\,|\, \forall \mc{X}_{\bs{\Xi}^1}, \mc{X}_{\bs{\Xi}^2},\,\, G(\mc{X}_{\bs{\Xi}^2})=
G(\mc{X}_{\bs{\Xi}^1})e^{\frac{c_{\rm L}}{12}\Upsilon(\mc{X}_{\bs{\Xi}^1},\mc{X}_{\bs{\Xi}^2})}\Big \}.
\end{equation}
The \textbf{Schwarzian line bundle} associated to $[\bs{\Xi}^\bullet]$ is defined by
\[\mc{L}^{[\bs{\Xi}^\bullet]}:=\bigsqcup_{\mc{X}\in \mc{T}_{g,m,b}}\mc{L}_{\mc{X}}^{[\bs{\Xi}^\bullet]}.\]
2) Let $[\bs{\Xi}^\bullet]\in \pi_0\tilde{\mc{BD}}_{g,m,b}$ be an isotopy class of marked complex building block decompositions in the decorated case. For $\bs{\alpha}\in \C^m$,
we define the $1$-dimensional complex vector space
\begin{equation}\label{def_line_alpha}
\mc{L}_{\mc{X}}^{[\bs{\Xi}^\bullet]}(\bs{\alpha}):=\Big\{ G: \mc{D}\tT_{g,m,b}^{[\bs{\Xi}^\bullet]}\cap\pi_{\mc{D}\tT\to\mc{T}}^{-1}(\mc{X}) 
\to \C\,|\, G \textrm{ satisfies properties }(P1) \textrm{ and }(P2)\Big\},
\end{equation}
where $(P1)$ and $(P2)$ are:
\begin{align*} 
& (P1) \qquad \forall \tilde{\mc{X}}_{\bs{\Xi}^1}, \tilde{\mc{X}}_{\bs{\Xi}^2}\text{ such that }\pi_{\BDTT\to\tT}(\tilde{\mc{X}}_{\bs{\Xi}^1})=\pi_{\BDTT\to\tT}(\tilde{\mc{X}}_{\bs{\Xi}^2}),\\
&\hspace{5em} G(\tilde{\mc{X}}_{\bs{\Xi}^2})=
G(\tilde{\mc{X}}_{\bs{\Xi}^1})e^{\frac{c_{\rm L}}{12}\Upsilon(\mc{X}_{\bs{\Xi}^1},\mc{X}_{\bs{\Xi}^2})}, \textrm{ where }\mc{X}_{\bs{\Xi}^\ell}=\pi_{\BDTT\to \BDT}(\tilde{\mc{X}}_{\bs{\Xi}^\ell});\\
& (P2) \qquad  \forall \tilde{\mc{X}}_{\bs{\Xi}}, \, \forall \bs{t}\in \C^m, \quad 
G(\bs{t}\cdot \tilde{\mc{X}}_{\bs{\Xi}})=
G(\tilde{\mc{X}}_{\bs{\Xi}})e^{\sum_{k=1}^m \Delta_{\alpha_k}t_k}.
\end{align*}
The \textbf{decorated Schwarzian line bundle} associated to $[\bs{\Xi}^\bullet]$ with weights $\bs{\alpha}$ is  defined by 
\[\mc{L}^{[\bs{\Xi}^\bullet]}(\bs{\alpha}):=\bigsqcup_{\mc{X}\in \mc{T}_{g,m,b}}\mc{L}_{\mc{X}}^{[\bs{\Xi}^\bullet]}(\bs{\alpha}).\]
\end{definition}

Let us first show that for $\mc{X}$ fixed, $\mc{L}_{\mc{X}}^{[\bs{\Xi}^\bullet]}$ and $\mc{L}_{\mc{X}}^{[\bs{\Xi}^\bullet]}(\bs{\alpha})$ are indeed complex vector spaces of dimension $1$. We claim that for any fixed $\mc{X}_{\bs{\Xi}^0}\in\BDT_{g,m,b}^{[\bs{\Xi}^\bullet]}$ and $\tilde{\mc{X}}_{\bs{\Xi}^0}\in\BDTT_{g,m,b}^{[\bs{\Xi}^\bullet]}$ over $\mc{X}$, we have linear isomorphisms
\[\Psi:G\in\mc{L}_{\mc{X}}^{[\bs{\Xi}^\bullet]}\mapsto G(\mc{X}_{\bs{\Xi}^0})\in\C,\quad\tilde{\Psi}:G\in\mc{L}_{\mc{X}}^{[\bs{\Xi}^\bullet]}(\bs{\alpha})\mapsto G(\tilde{\mc{X}}_{\bs{\Xi}^0})\in\C.\]
By the definition of $\Upsilon$, it is clear that $\Psi$ is injective, and $\tilde{\Psi}$ is similarly injective since for any $\tilde{\mc{X}}_{\bs{\Xi}^1}, \tilde{\mc{X}}_{\bs{\Xi}^2}\in\mc{D}\tT_{g,m,b}^{[\bs{\Xi}^\bullet]}\cap\pi_{\mc{D}\tT\to\mc{T}}^{-1}(\mc{X})$, there exists a unique $\bs{t}\in \C^m$ such that $\pi_{\BDTT\to\tT}(\bs{t}\cdot\tilde{\mc{X}}_{\bs{\Xi}^1})=\pi_{\BDTT\to\tT}(\tilde{\mc{X}}_{\bs{\Xi}^2})$. Conversely, the function 
\[G_{\bs{\Xi}^0}:\mc{D}\mc{T}_{g,m,b}^{[\bs{\Xi}^\bullet]}\cap\pi_{\mc{D}\mc{T}
\to\mc{T}}^{-1}(\mc{X}) 
\to \C,\qquad G_{\bs{\Xi}^0}(\mc{X}_{\bs{\Xi}}):=\exp(\frac{c_{\rm L}}{12}\Upsilon(\mc{X}_{\bs{\Xi}^0},\mc{X}_{\bs{\Xi}}))\]
is a non-zero vector in $\mc{L}_{\mc{X}}^{[\bs{\Xi}^\bullet]}$ and satisfies $\Psi(G_{\bs{\Xi}^0})=1$. Similarly, we define the function
\begin{equation}\label{G_0(alpha)}
G_{\bs{\Xi}^0}(\bs{\alpha}):\mc{D}\tT_{g,m,b}^{[\bs{\Xi}^\bullet]}\cap\pi_{\mc{D}\tT
\to\mc{T}}^{-1}(\mc{X}) 
\to \C,\qquad G_{\bs{\Xi}^0}(\bs{\alpha})(\tilde{\mc{X}}_{\bs{\Xi}}):=\exp(\frac{c_{\rm L}}{12}\Upsilon(\mc{X}_{\bs{\Xi}^0},\mc{X}_{\bs{\Xi}}))e^{\sum_{k=1}^m\Delta_{\alpha_k}t_k},
\end{equation}
where $\mc{X}_{\bs{\Xi}^0},\mc{X}_{\bs{\Xi}}$ are the projections of $\tilde{\mc{X}}_{\bs{\Xi}^0},\tilde{\mc{X}}_{\bs{\Xi}}$ on $\BDT_{g,m,b}$ and $\bs{t}=(t_1,\ldots,t_m)$ is the unique element in $\C^m$ such that $\pi_{\BDTT\to\tT}(\bs{t}\cdot\tilde{\mc{X}}_{\bs{\Xi}^0})=\pi_{\BDTT\to\tT}(\tilde{\mc{X}}_{\bs{\Xi}})$. Then $G_{\bs{\Xi}^0}(\bs{\alpha})\in\mc{L}_{\mc{X}}^{[\bs{\Xi}^\bullet]}(\bs{\alpha})$ and $\tilde{\Psi}(G_{\bs{\Xi}^0}(\bs{\alpha}))=1$.

To avoid technicalities due to the fact that $\mc{T}_{g,m,b}$ has infinite dimension when $b>0$, we 
now focus on the case $b=0$, so that $\mc{T}_{g,m}=\mc{T}_{g,m,0}$ is a finite dimensional Stein manifold with complex dimension $3g-3+m$.
We shall prove that $\mc{L}^{[\bs{\Xi}^\bullet]}$ and $\mc{L}^{[\bs{\Xi}^\bullet]}(\bs{\alpha})$ admit the structure of holomorphic line bundles over the Teichm\"uller space $\mc{T}_{g,m}$. 

\begin{lemma}\label{l:holomorphy_line_bundle}
Let $S^\bullet=(\Sigma^\bullet,J^\bullet,{\bf x}^\bullet)$ be a fixed base surface and $\tilde{S}^\bullet=(\Sigma^\bullet,J^\bullet,{\bf x}^\bullet,{\bf Y}^\bullet)$ the same surface with a fixed decoration ${\bf Y}^\bullet$. Consider the Teichm\"uller space $\mc{T}_{g,m}=\mc{T}(S^\bullet)$ of marked closed Riemann surfaces of genus $g$ with $m$ marked points. Then for any isotopy class $[\bs{\Xi}^\bullet]\in\piBD_{g,m}=\piBD(S^\bullet)$ of marked complex building block decompositions of $S^\bullet$, there is a structure of holomorphic line bundle over $\mc{T}_{g,m}$ on the collection $\mc{L}^{[\bs{\Xi}^\bullet]}:=\bigsqcup_{\mc{X}\in \mc{T}_{g,m}}\mc{L}_{\mc{X}}^{[\bs{\Xi}^\bullet]}$. Similarly, for any isotopy class $[\bs{\Xi}^\bullet]\in\pi_0\tilde{\mc{BD}}_{g,m}=\pi_0\tilde{\mc{BD}}(\tilde{S}^\bullet)$ and any 
$\bs{\alpha}\in \C^m$, there is a structure of holomorphic line bundle over $\mc{T}_{g,m}$ on the collection $\mc{L}^{[\bs{\Xi}^\bullet]}(\bs{\alpha}):=\bigsqcup_{\mc{X}\in \mc{T}_{g,m}}\mc{L}_{\mc{X}}^{[\bs{\Xi}^\bullet]}(\bs{\alpha})$.
\end{lemma}
\begin{proof}
To define the holomorphic structure on $\mc{L}^{[\bs{\Xi}^\bullet]}$, we need to find a covering of $\mc{T}_{g,m}$ by holomorphic charts
$U^n$ (for $n\in \N$),  define local trivialisations $s^n: B^n\times \C\to \mc{L}^{[\bs{\Xi}^\bullet]}|_{U^n}$ so that for $n,n'\in \N$,
$\Phi_{nn'}:=(s^{n'})^{-1}\circ s^n: (p^n)^{-1}(U^n\cap U^{n'})\times \C \to (p^{n'})^{-1}(U^n\cap U^{n'})\times \C$ is holomorphic, where $p^n:B^n\subset \C^{3g-3+m}\to U^n$ are local holomorphic charts.
To show the base point conveniently, we rather use the definition $\mc{L}^{[\bs{\Xi}^\bullet]}:=\{(\mc{X},G):\mc{X}\in\mc{T}_{g,m},G\in\mc{L}_{\mc{X}}^{[\bs{\Xi}^\bullet]}\}$ in this proof. We will use holomorphic charts $U^n$ given by plumbing coordinates at a point $\mc{X}^n$, using a hyperbolic marked building block decomposition $\bs{\Xi}^n$. The plumbing gives a holomorphic map $p^n:B^n\to\mc{T}_{g,m}$, $\bs{\tau}\mapsto\mc{X}^n(\bs{\tau})$, where $B^n$ is a small ball around $0$ in $\C^{3g-3+m}$.
Take
\begin{equation}\label{trivialization} 
s^n(\btau,\lambda):=(\mc{X}^n(\bs{\tau}),\lambda G^n(\btau)),\quad
G^n(\btau):= 
\exp(\tfrac{c_{\rm L}}{12}\Upsilon(\mc{X}^n_{\bs{\Xi}^n}(\bs{\tau}),\cdot))
\end{equation}
as local trivialisations. 
The inverse is given by $(s^n)^{-1}(\mc{X},G)=((p^n)^{-1}(\mc{X}), G(\mc{X}^n_{\bs{\Xi}^n}((p^n)^{-1}(\mc{X}))))$. 
Since $\btau \mapsto p^{n}(\bs{\tau})$ are holomorphic coordinates, $\phi_{nn'}:=(p^{n'})^{-1}\circ p^n$ is holomorphic on $(p^n)^{-1}(U^n\cap U^{n'})$, and 
\[\begin{split} 
\Phi_{nn'}(\btau,\lambda) =\Big(\phi_{nn'}(\btau), \lambda \exp\big(\tfrac{c_{\rm L}}{12}
\Upsilon(\mc{X}^n_{\bs{\Xi}^n}(\bs{\tau}),\mc{X}^{n'}_{\bs{\Xi}^{n'}}(\phi_{nn'}(\btau)))\big)\Big).
\end{split}\]
It remains to prove that the map
\begin{equation}\label{change_of_coordinates}
\btau\mapsto\Upsilon(\mc{X}^n_{\bs{\Xi}^n}(\bs{\tau}),\mc{X}^{n'}_{\bs{\Xi}^{n'}}(\phi_{nn'}(\btau)))
\end{equation}
is holomorphic in $\btau$ on its domain of definition. Denote $\bs{\tau}'=\phi_{nn'}(\btau)$. First, the surface $S^n(\bs{\tau})=(\Sigma,J^n(\bs{\tau}),{\bf x})$ represents the common point $\mc{X}^n(\bs{\tau})=\mc{X}^{n'}(\btau')$. 
We shall show that for each $\bs{\tau}\in (p^n)^{-1}(U^n\cap U^{n'})$ such that $\mc{X}^n(\bs{\tau})=\mc{X}^{n'}(\bs{\tau}')$, there exists
a family $t\in [0,1]\mapsto \bs{\Xi}^t(\bs{\tau})$ of marked complex building block decompositions on  
$S^n(\bs{\tau})$ with $\bs{\Xi}^{t=0}(\bs{\tau})=\bs{\Xi}^n(\bs{\tau})$ and $\bs{\Xi}^{t=1}(\bs{\tau})=\bs{\Xi}^{n'}(\bs{\tau}')$, such that the model form  $\bs{\psi}_j^t(\bs{\tau})$ of $\Xi_j^t(\bs{\tau})$ are $C^1$ in $t$ and holomorphic in $\bs{\tau}$.
By the definition of $\Upsilon$, this proves the holomorphy of \eqref{change_of_coordinates}. Indeed, the Schwarzian integrand is holomorphic in $\bs\tau$ and continuous in $t$ on a fixed annulus, so its integral over $t\in[0,1]$ is holomorphic in $\bs\tau$.

Now, we construct the covering $(U^n)_{n\in \N}$ of Teichm\"uller space and the marked building block decompositions $\bs{\Xi}^t(\bs{\tau})$ as needed.
 We use the universal curve $\pi_{g,m}: \mc{C}_{g,m}\to \mc{T}_{g,m}$, whose fiber at $\mc{X}\in \mc{T}_{g,m}$ is a marked Riemann surface $S^{\mc{X}}$ representing $\mc{X}$. This is a complex manifold with $\pi_{g,m}$ holomorphic. 
Fix $\mc{X}_0\in \mc{T}_{g,m}$. After using the marking diffeomorphisms to identify the
underlying differentiable surfaces with $\Sigma^\bullet$, the
hyperbolic metrics and the closed geodesics in the isotopy classes of
the cutting curves induced by $[\bs{\Xi}^\bullet]$ depend continuously on the point of
Teichm\"uller space near $\mc{X}_0$, in the topology of analytic parametrisations (here we are fixing locally the origins of the constant
speed parametrisations); this is a consequence of the implicit function theorem and the fact that in constant $-1$ curvature, there is an explicit real analytic expression for a collar neighbourhood of a closed geodesic.  Thus, there exist
$\delta \in (0,1)$ and a neighbourhood $W_{\mc{X}_0}$ of $\mc{X}_0$ such that,
for every $\mc{X}\in W_{\mc X_0}$, all the hyperbolic boundary
parametrisations $\xi^{\mc X}_{jk}:=\Xi^{\mc X}_j\circ\zeta_k^\bullet:\mathbb T\to \Sigma^\bullet$
admit holomorphic extensions to the annuli $\A^\circ_{\delta^4,\delta^{-4}}$ with values in $(\Sigma^\bullet,J^{\mc X})$, where $J^{\mc{X}}$ denotes the complex structure of $\mc{X}$. We still denote by $\xi^{\mc X}_{jk}$ these holomorphic extensions, and the images of the closed collars
$\xi^{\mc X}_{jk}(\A_{\delta^3,\delta^{-3}})$ corresponding to distinct cutting curves are
pairwise disjoint. These extensions can be chosen continuously with
respect to $\mc X$ on $W_{\mc X_0}$.

For every $\mc X\in W_{\mc X_0}$, denote by 
\[ p^{\mc X}:B^{\mc X}\to\mc T_{g,m},  \qquad  p^{\mc X}(\bs\tau)=\mc X(\bs\tau),
\]
 the plumbing coordinate map based on the hyperbolic marked building block decomposition
$\bs\Xi^{\mc X}$, with $B^{\mc{X}}\subset \C^{3g-3+m}$ a small ball near $\bs{\tau}=0$.
Shrinking $W_{\mc X_0}$ and the domains $B^{\mc X}$, if necessary, the plumbing construction gives holomorphic collar maps
\[ \widehat{\xi}^{\, \mc X}_{jk}:B^{\mc X}\times\A^\circ_{\delta^3,\delta^{-3}}\to \mc{C}_{g,m},\]
which are smooth up to the boundary of the annulus. For fixed $\bs\tau$, the map $z\mapsto \widehat{\xi}^{\, \mc X}_{jk}(\bs\tau,z)$ is a holomorphic collar embedding in the fibre over $p^{\mc X}(\bs\tau)$. The map is represented by $(\bs\tau,z)\mapsto(\bs\tau,\xi^{\mc X}_{jk}(e^{{\rm i}\tau_{jk}}z))$
and is therefore holomorphic jointly in $(\bs\tau,z)$. The domains $B^{\mc X}$ are chosen so that $e^{{\rm i}\tau_{jk}}\A_{\delta^3,\delta^{-3}}\subset \A^\circ_{\delta^4,\delta^{-4}}$.
Consequently, by continuity and compactness, we may
shrink $W_{\mc X_0}$ once more so that the following holds. If
$\mc X,\mc X'\in W_{\mc X_0}$ and $\mc{Y}=p^{\mc X}(\bs\tau)= p^{\mc{X'}}(\bs\tau')$,
then, after identifying both plumbed surfaces with the fibre over
$\mc Y$, one has
\[ \widehat{\xi}^{\, \mc X'}_{jk}(\bs\tau',\A_{\delta^2,\delta^{-2}})\subset 
 \widehat{\xi}^{\, \mc X}_{jk}(\bs\tau,\A_{\delta^3,\delta^{-3}}),
\]
and the analogous inclusion with $\mc X$ and $\mc X'$ interchanged.
In particular, the transition map
\[ f_{jk}^{\mc X,\mc X'}(\bs\tau)
 :=  \bigl( \widehat{\xi}^{\, \mc X}_{jk}(\bs\tau,\cdot)\bigr)^{-1}
 \circ  \widehat{\xi}^{\, \mc X'}_{jk} \bigl(\phi_{\mc X,\mc X'}(\bs\tau),\cdot\bigr)\]
is well-defined on $\A_{\delta^2,\delta^{-2}}$, where $ \phi_{\mc X,\mc X'}=(p^{\mc X'})^{-1}\circ p^{\mc X}$ is a biholomorphism between the corresponding subsets of the two plumbing parameter domains.
Then $f_{jk}^{\mc X,\mc X'}(\bs\tau)$ is holomorphic in $\bs\tau$. We also choose $W_{\mc X_0}$ and the domains $B^{\mc X}$ sufficiently small that all these transition maps belong, on a slightly smaller fixed annulus, to a prescribed $C^1$-neighbourhood of ${\rm Id}$, whose elements $f$ satisfy that $f_t(\bs{\tau}):=tf(\bs{\tau})+(1-t){\rm Id}$ are biholomorphisms (on some annulus) for all $t\in [0,1]$.
The sets $(W_{\mc{X}_0})_{\mc{X}_0\in \mc{T}_{g,m}}$ form an open covering of $\mc T_{g,m}$, and we next construct a covering for which this local compatibility property holds on every non-empty intersection.
Since $\mc T_{g,m}$ is diffeomorphic to a disk in $\C^{3g-3+m}$, it is direct to see that there exists a countable locally finite refinement
$(U^n)_{n\in\mathbb N}$ which are plumbing neighbourhoods $U^n=\{\mc{X}^n(\bs{\tau}) \,|\, |\bs{\tau}|< \eps_n\}$ of $\mc{X}^n\in \mc{T}_{g,m}$ satisfying  that for each $n$, there is an index $\mc{X}_0^n$ such
that, if $U^n\cap U^{n'}\ne\varnothing$, then $U^n\cup U^{n'} \subset W_{\mc{X}_0^n}$. 
Thus the two centres $\mc X^n,\mc X^{n'}$, as well as every common point $\mc X^n(\bs\tau) =\mc X^{n'}(\bs\tau') \in U^n\cap U^{n'}$,
belong to one common neighbourhood $W_{\mc{X}_0^n}$. Let us use the notation $\widehat{\xi}^{\, n}_{jk}$ instead of 
$\widehat{\xi}^{\, \mc{X}^n}_{jk}$.
It follows  that
\[ f_{jk}^{nn'}(\bs\tau):= \bigl(\widehat{\xi}^{\, n}_{jk}(\bs\tau,\cdot)\bigr)^{-1}
 \circ  \widehat{\xi}^{\, n'}_{jk}\bigl(\phi_{nn'}(\bs\tau),\cdot\bigr)\]
is well-defined on a fixed annulus and is holomorphic in $\bs\tau$. By the choice of the neighbourhoods $W_{\mc{X}^n_0}$, $f_{jk}^{nn'}(\bs\tau)$ is uniformly close to the identity on a slightly larger annulus. Cauchy estimates then imply that it is
close to the identity in $C^1$ on a smaller annulus. Therefore
\[  f_{jk}^{nn',t}(\bs\tau) :=(1-t){\rm Id} +t f_{jk}^{nn'}(\bs\tau), \qquad 0\leq t\leq1,
\]
remains a holomorphic embedding for every $t$, and depends holomorphically on
$\bs\tau$ and $C^1$ in $t$. 
If $\bs{\Xi}^n(\bs{\tau})$ and $\bs{\Xi}^{n'}(\bs{\tau}')$ are the two marked building block decompositions of the Riemann surface 
$S(\bs{\tau})$ representing $\mc X^n(\bs\tau) =\mc X^{n'}(\bs\tau')$, 
we define $\bs{\Xi}^t(\bs{\tau})$ by deforming $\bs{\Xi}^n(\bs{\tau})$ so that for each oriented edge $\vec{\rm e}_{i}=((j,k),(j',k'))$,
\[
 \Xi_j^t(\bs\tau)\circ\zeta_k^\bullet
 =\widehat{\xi}_{jk}^{\, n}(\bs\tau,\cdot)\circ f_{jk}^{nn',t}(\bs\tau)\big|_{\T},
 \qquad
 \Xi_{j'}^t(\bs\tau)\circ\zeta_{k'}^\bullet
 =\widehat{\xi}_{jk}^{\, n}(\bs\tau,\cdot)\circ f_{jk}^{nn',t}(\bs\tau)\circ o\big|_{\T}.
\]
At $t=0$ these are the two boundary parametrisations of $\bs\Xi^n(\bs\tau)$, while the definition of $f_{jk}^{nn'}$ shows that at $t=1$ they are those of $\bs\Xi^{n'}(\bs\tau')$. The curves remain pairwise disjoint and in the same isotopy classes. The isotopy extension theorem therefore allows us to extend these boundary parametrisations to marked building block embeddings, choosing the extensions at $t=0$ and $t=1$ to be $\bs\Xi^n(\bs\tau)$ and $\bs\Xi^{n'}(\bs\tau')$, respectively. Thus, for every $t\in[0,1]$, $\bs\Xi^t(\bs\tau)$ is a marked building block decomposition of the fixed surface $S(\bs\tau)$ and
\[
 \bs\Xi^{t=0}(\bs\tau)=\bs\Xi^n(\bs\tau),
 \qquad
 \bs\Xi^{t=1}(\bs\tau)=\bs\Xi^{n'}(\bs\tau').
\]

The model forms of $\bs{\Xi}^t(\bs{\tau})$ are given as follows. For each $\vec{\rm e}_{i}=((j,k),(j',k'))$, we define $f_{jk}:=f^{nn',t}_{jk}(\bs{\tau})$, $f_{j'k'}:=o\circ f^{nn',t}_{jk}(\bs{\tau})\circ o$ (omitting the dependence on $t,\bs{\tau}$ for simplicity). For each $j$, we have $\bs{f}_j:=(f_{j1},\ldots,f_{jb_j})\in\mc{U}_\eps^{b_j}$ for $\eps>0$ small. Then for each $j$,
\[\bs{\psi}_{j}^t(\bs{\tau})=(\Phi_{\bs{f}_j,\bs\tau}\circ\psi^n_{j1}(\bs{\tau})\circ f_{j1},\ldots,\Phi_{\bs{f}_j,\bs\tau}\circ \psi^n_{jb_j}(\bs{\tau})\circ f_{jb_j}),\]
where $\Phi_{\bs{f}_j,\bs\tau}$ is the uniformising map from Lemma \ref{PhiC1} with base model form $\bs\psi_j^n(\bs\tau)$. Notice that  the model forms can be written as $\psi^n_{jk}(\bs{\tau})\circ f_{jk}(z)=\psi_{jk}^n(0)(e^{{\rm i}\tau_{jk}}f_{jk}(z))$ for $z$ near 
$\T$, so we are in the setting of  Lemma \ref{PhiC1}.
Since $f^{nn',t}_{jk}(\bs{\tau})$ is $C^1$ in $t$ and holomorphic in $\bs{\tau}$, so is $\bs{\psi}_{j}^t(\bs{\tau})$ by Lemma \ref{PhiC1}.

To deal with $\mc{L}^{[\bs{\Xi}^\bullet]}(\bs{\alpha})$, we proceed similarly. We construct $\tilde{\mc{X}}^n_{\bs{\Xi}^n}(\bs{\tau})\in \mc{D}\tT_{g,m}$ as in  Definition \ref{def:plumbing}, which is a lift of $\mc{X}^n_{\bs{\Xi}^n}(\bs{\tau})$, and we use the following local trivialisations of $\mc{L}^{[\bs{\Xi}^\bullet]}(\bs{\alpha})$:
\begin{equation}\label{trivialization2} 
s_{\bs{\alpha}}^n(\btau,\lambda):=(\mc{X}^n(\bs{\tau}),\lambda G_{\bs{\alpha}}^n(\btau)),\quad
G_{\bs{\alpha}}^n(\btau)(\tilde{\mc{X}}_{\bs{\Xi}}):= 
\exp({\tfrac{c_{\rm L}}{12}\Upsilon(\mc{X}^n_{\bs{\Xi}^n}(\bs{\tau}),\mc{X}_{\bs{\Xi}})}) e^{\sum_{k=1}^m\Delta_{\alpha_k}t^{n}_k(\bs{\tau})},
\end{equation}
where $\mc{X}_{\bs{\Xi}}=\pi_{\mc{D}\tT\to \mc{DT}}(\tilde{\mc{X}}_{\bs{\Xi}})$ and $\bs{t}^n(\bs{\tau})$ is defined by $\pi_{\mc{D}\tT\to \tT}(\bs{t}^n(\bs{\tau})\cdot \tilde{\mc{X}}^n_{\bs{\Xi}^n}(\bs{\tau}))=\pi_{\mc{D}\tT\to \tT}(\tilde{\mc{X}}_{\bs{\Xi}})$.
 We have 
\[((s_{\bs{\alpha}}^{n'})^{-1}\circ s_{\bs{\alpha}}^n)(\btau,\lambda) =\Big(\phi_{nn'}(\btau), \lambda \exp({\tfrac{c_{\rm L}}{12}
\Upsilon\big(\mc{X}^n_{\bs{\Xi}^n}(\bs{\tau}),\mc{X}^{n'}_{\bs{\Xi}^{n'}}(\phi_{nn'}(\btau))\big)}) e^{-\sum_{\ell=1}^m\Delta_{\alpha_\ell}t_\ell^{nn'}(\bs{\tau})}\Big),\]
where $\bs{t}^{nn'}(\bs{\tau})$ is defined by $\pi_{\mc{D}\tT\to \tT}(\tilde{\mc{X}}^n_{\bs{\Xi}^n}(\bs{\tau}))=\pi_{\mc{D}\tT\to \tT}(\bs{t}^{nn'}(\bs{\tau})\cdot \tilde{\mc{X}}^{n'}_{\bs{\Xi}^{n'}}(\phi_{nn'}(\btau)))$.
In terms of model forms, for each $n,j$, let $(\bs{\psi}_j^n(\bs{\tau}),\bs{\theta}_j^n(\bs{\tau}),\bs{\la}_j^n)$ be the model form of the $j$-th marked complex building block of $\tilde{\mc{X}}^n_{\bs{\Xi}^n}(\bs{\tau})$. If the $\ell$-th marked point $x_\ell(\bs{\tau})$ of $\mc{X}^n(\bs{\tau})$ corresponds to the $k$-th marked point of the $j$-th marked complex building block, i.e. ${\rm e}^{\rm m}_\ell=\{(\ell,\bullet),(j,b_j+k)\}$ in the gluing graph, then
\[t_\ell^{nn'}(\bs\tau)=\lambda_{jk}^{n}-\lambda_{jk}^{n'}+
\log\bigl(\Phi'_{\bs f_j^{nn',1}(\bs\tau),\bs\tau}(z_{b_j+k}^\bullet)\bigr),\]
where the logarithm is canonically defined since the coefficient inside the log is close to $1$. 
It remains to check the holomorphic dependence of the vector fields entering
the definition of the Schwarzian cocycle. For the oriented internal edge $\vec{\rm e}_i=((j,k),(j',k'))$, set
\[ \xi_i^t(\bs\tau,z) := \widehat{\xi}_{jk}^{\,n}
 \bigl(\bs\tau,f_{jk}^{nn',t}(\bs\tau,z)\bigr).\]
Since \(\widehat{\xi}_{jk}^{\,n}\) does not depend on \(t\), we have
\[
 \bigl(\xi_i^t(\bs\tau)\bigr)^{-1}
 \circ\xi_i^{t+s}(\bs\tau)
 =
 \bigl(f_{jk}^{nn',t}(\bs\tau)\bigr)^{-1}
 \circ f_{jk}^{nn',t+s}(\bs\tau).
\]
Consequently, the local velocity appearing in
Definition~\ref{d:cocycle} is
\[
 v_i^t(\bs\tau,z)
 =
 \frac{\partial_t f_{jk}^{nn',t}(\bs\tau,z)}
      {\partial_z f_{jk}^{nn',t}(\bs\tau,z)}
 =
 \frac{f_{jk}^{nn'}(\bs\tau,z)-z}
      {(1-t)+t\,\partial_zf_{jk}^{nn'}(\bs\tau,z)}.
\]
The denominator does not vanish since $f_{jk}^{nn',t}(\bs\tau,\cdot)$ is a holomorphic embedding.
Therefore, on a slightly smaller fixed annulus, $v_i^t(\bs\tau,z)$ is jointly holomorphic in
$(\bs\tau,z)$ and continuous in $t$, locally uniformly with respect to \(\bs\tau\). Together with the holomorphic dependence of the model maps
$\bs\psi_j^t(\bs\tau)$, this shows that
\[ \frac{1}{2\pi{\rm i}}\oint_{\T} \left({\rm S}(\psi_{j^+(i)k^+(i)}^t(\bs\tau)\circ o) - {\rm S}(\psi_{j^-(i)k^-(i)}^t(\bs\tau)) \right) v_i^t(\bs\tau)\,\dd z
\]
is holomorphic in $\bs\tau$ and continuous in \(t\), locally
uniformly. Its integral with respect to \(t\in[0,1]\) is therefore
holomorphic in \(\bs\tau\). This proves the holomorphy of
\eqref{change_of_coordinates} and that the transition maps are holomorphic in $\bs{\tau}$.
\end{proof}

Notice that since $\mc{T}_{g,m}$ is a contractible Stein manifold, the line bundles $\mc{L}^{[\bs{\Xi}^\bullet]}$ and 
$\mc{L}^{[\bs{\Xi}^\bullet]}(\bs{\alpha})$ are globally trivialisable as holomorphic line bundles.

We recall that the universal curve is the complex manifold $\mc{C}_{g,m}$ which is a fiber bundle over $\mc{T}_{g,m}$, whose fiber at $\mc{X}=[(F,S)]$ is a Riemann surface with marked points $S=(\Sigma,J,{\bf x})$ representing the class $\mc{X}$; we refer to \cite[Section 5]{Wolpert_Inv} for details in the case $m=0$. 
We shall consider below families of smooth Riemannian metrics $g_{\mc{X}}$ on the fiber $\Sigma$ at $\mc{X}$ of $\mc{C}_{g,m}$ which are compatible with the complex structure $J$ on $\Sigma$ and depend smoothly on $\mc{X}\in \mc{T}_{g,m}$ on the bundle $\mc{C}_{g,m}$. 
When $m=0$, a special case is  
the unique hyperbolic metric on $\Sigma$ (with Gauss curvature $-1$) compatible with $J$.

\begin{definition}[\textbf{Hermitian metric on Schwarzian line bundle}]\label{def_hermitian_metric}
Assume that $\bs{\alpha}\in \R^m$ and 
fix a smooth metric $g_{\hat{\C}}$ on $\hat{\C}$ compatible with the canonical complex structure $J_\C$ and choose a family of smooth Riemannian metrics $g_{\mc{X}}$ on the fiber $\Sigma$ at $\mc{X}$ of $\mc{C}_{g,m}$ which are compatible with the complex structure $J$ induced by $\mc{X}$ on $\Sigma$ and depend smoothly on $\mc{X}\in \mc{T}_{g,m}$. Then we set for $G_1,G_2\in \mc{L}_{\mc{X}}^{[\bs{\Xi}^\bullet]}$ 
\begin{equation}\label{metriconL}
\cjg G_1,G_2\cjd_{\mc{L}_{\mc{X}}^{[\bs{\Xi}^\bullet]}}:= G_1(\mc{X}_{\bs{\Xi}})\bbar{G_2(\mc{X}_{\bs{\Xi}})}e^{-c_{\rm L}S_{\rm L}^0(\Sigma,g_{\mc{X}},g_{\bs{\Xi}})+
c_{\rm L}\sum_{j=1}^{j_0}S_{\rm L}^0(\hat{\C},g_{\hat{\C}},\hat{g}_{\Xi_j})},
\end{equation}
where $\mc{X}_{\bs{\Xi}}=[(F,S,\bs{\Xi})]\in \mc{D}\mc{T}^{[\bs{\Xi}^\bullet]}_{g,m}$ with $S=(\Sigma,J,{\bf x})$, $\bs{\Xi}=(\Xi_1,\dots, \Xi_{j_0})\in \mc{B}\mc{D}(S)$ is a point projecting to $\mc{X}\in \mc{T}_{g,m}$,  
$g_{\bs{\Xi}}$ is any compatible Riemannian metric on $S$ which is admissible on each complex 
building block $\mc{P}_j$ induced by $\Xi_j$, and $\hat{g}_{\Xi_j}$ is the filling metric of $g_{\bs{\Xi}}|_{\mc{P}_j}$ on $\hat{\C}$ in the sense of Definition \ref{LiouvillePant}. Similarly, we set for 
$G_1,G_2\in \mc{L}_{\mc{X}}^{[\bs{\Xi}^\bullet]}(\bs{\alpha})$ 
\begin{equation}\label{metriconLalpha}
\cjg G_1,G_2\cjd_{\mc{L}_{\mc{X}}^{[\bs{\Xi}^\bullet]}(\bs{\alpha})}:= G_1\big(\tilde{\mc{X}}_{\bs{\Xi}}\big)\bbar{G_2\big(\tilde{\mc{X}}_{\bs{\Xi}}\big)}e^{-c_{\rm L}S_{\rm L}^0(\Sigma,g_{\mc{X}},g_{\bs{\Xi}})+
c_{\rm L}\sum_{j=1}^{j_0}S_{\rm L}^0(\hat{\C},g_{\hat{\C}},\hat{g}_{\Xi_j})}\prod_{j=1}^{j_0}\prod_{k=1}^{m_j} 
\Big(\frac{\|(\Xi_j)_*Y_k^\bullet\|_{g_{\mc{X}}}}{\|Y_k^\bullet\|_{g_{\hat{\C}}}}\Big)^{-2\Delta_{\alpha_{jk}}},
\end{equation}
where $\tilde{\mc{X}}_{\bs{\Xi}}=[(F,\tilde{S},\bs{\Xi})]\in \mc{D}\tT^{[\bs{\Xi}^\bullet]}_{g,m}$ with $\tilde{S}=(\Sigma,J,{\bf x},{\bf Y})$, $\bs{\Xi}\in\tilde{\mc{BD}}(\tilde{S})$ is a point projecting to $\mc{X}\in \mc{T}_{g,m}$, and $\alpha_{jk}:=\alpha_\ell$ if
${\rm e}^{\rm m}_\ell =\{(\ell,\bullet),(j,b_j+k)\}$,
or equivalently if $\Xi_j(x_k^\bullet)=x_\ell$.
\end{definition}

Using Lemma \ref{Re(Upsilon)}, we observe that $\cjg G_1,G_2\cjd_{\mc{L}_{\mc{X}}^{[\bs{\Xi}^\bullet]}}$ does not depend on the choice of $\bs{\Xi}$ in the isotopy class $[\bs{\Xi}^\bullet]$, nor on the choice of an admissible metric $g_{\bs{\Xi}}$. It only depends on the choices of $g_{\mc{X}}$ and $g_{\hat{\C}}$. The same holds for $\cjg G_1,G_2\cjd_{\mc{L}_{\mc{X}}^{[\Xi^\bullet]}(\bs{\alpha})}$. Indeed, for $\bs{t}=(t_1,\dots,t_m)\in \C^m$, we have
\[G_1(\bs{t}\cdot \tilde{\mc{X}}_{\bs{\Xi}})\bbar{G_2\big(\bs{t}\cdot \tilde{\mc{X}}_{\bs{\Xi}}\big)}=G_1\big(\tilde{\mc{X}}_{\bs{\Xi}}\big)\bbar{G_2\big(\tilde{\mc{X}}_{\bs{\Xi}}\big)}e^{\sum_{\ell=1}^m 2\Delta_{\alpha_\ell}{\rm Re}(t_\ell)}.\]
Now for $\bs{t}\cdot\bs{\Xi}=(\bs{t}\cdot \Xi_j)_{j=1}^{j_0}$, if ${\rm e}^{\rm m}_\ell =\{(\ell,\bullet),(j,b_j+k)\}$;
then $(\bs t\cdot\Xi_j)_*Y_k^\bullet =e^{t_\ell}(\Xi_j)_*Y_k^\bullet$. This shows the invariance of $\cjg G_1,G_2\cjd_{\mc{L}_{\mc{X}}^{[\Xi^\bullet]}(\bs{\alpha})}$ under the action of $\C^m$.

The same arguments can be extended to the case of complex weights:
\begin{remark}\label{rem:general_weights}
If $\bs{\alpha}\in \C^m$, there is a natural pairing between $G_1\in \mc{L}_{\mc{X}}^{[\bs{\Xi}^\bullet]}(\bs{\alpha})$ and $G_2\in \mc{L}_{\mc{X}}^{[\bs{\Xi}^\bullet]}(\bbar{\bs{\alpha}})$ given by 
\begin{equation}\label{pairing_extended}
\cjg G_1,G_2\cjd_{\bs{\alpha}}:= G_1\big(\tilde{\mc{X}}_{\bs{\Xi}}\big)\bbar{G_2\big(\tilde{\mc{X}}_{\bs{\Xi}}\big)}e^{-c_{\rm L}S_{\rm L}^0(\Sigma,g_{\mc{X}},g_{\bs{\Xi}})+
c_{\rm L}\sum_{j=1}^{j_0}S_{\rm L}^0(\hat{\C},g_{\hat{\C}},\hat{g}_{\Xi_j})}\prod_{j=1}^{j_0}\prod_{k=1}^{m_j} 
\Big(\frac{\|(\Xi_j)_*Y_k^\bullet\|_{g_{\mc{X}}}}{\|Y_k^\bullet\|_{g_{\hat{\C}}}}\Big)^{-2\Delta_{\alpha_{jk}}},
\end{equation}
which does not depend on the choice of $\bs{\Xi}$ in the isotopy class $[\bs{\Xi}^\bullet]$, nor on the choice of an admissible metric $g_{\bs{\Xi}}$, but only depends on the choices of $g_{\mc{X}}$ and $g_{\hat{\C}}$. 
\end{remark}

\subsection{Hilbert space of conformal blocks}
We are finally ready to show that the conformal blocks of closed Riemann surfaces with marked points are global holomorphic sections of the Schwarzian line bundle.
\begin{theorem}[\textbf{Conformal blocks on Teichm\"uller space}]\label{th:block_closed}
Let $[\bs{\Xi}^\bullet]\in \pi_0\tilde{\mc{BD}}_{g,m}$ be an isotopy class of decorated marked complex building block decompositions with
$j_0=2g-2+m$ complex building blocks and $i_0=3g-3+m$ cutting curves.
Let $\bs{\alpha}\in\C^m$ be weights satisfying the generalised Seiberg bounds with respect to the gluing graph of $[\bs{\Xi}^\bullet]$ and \eqref{cdozznot=0}. Then the map
\begin{equation}\label{def:block_closed}
\begin{array}{lll}
\mc{T}_{g,m} & \to & L^1_{\rm loc}(\R_+^{i_0})\otimes \mc{L}^{[\bs{\Xi}^\bullet]}(\bs{\alpha}) \\
\mc{X} &
\mapsto & \mc{F}^{[\bs{\Xi}^\bullet]}_{\bs{\alpha}}(\mc{X}):=(({\bf p},\tilde{\mc{X}}_{\bs{\Xi}})\mapsto \mc{B}_{\tilde{\mc{X}}_{\bs{\Xi}},\bs{\alpha}}({\bf p}))
\end{array}
\end{equation}
is a holomorphic section, called the \textbf{conformal block} associated to the isotopy class $[\bs{\Xi}^\bullet]$. 
Here we view elements of $L^1_{\rm loc}(\R_+^{i_0})\otimes \mc{L}^{[\bs{\Xi}^\bullet]}(\bs{\alpha})$ as functions of ${\bf p}\in \R_+^{i_0}$ and $\tilde{\mc{X}}_{\bs{\Xi}}\in \mc{D}\tT_{g,m}^{[\bs{\Xi}^\bullet]}$. We also denote  $\mc{F}^{[\bs{\Xi}^\bullet]}_{\bs{\alpha}}(\tilde{\mc{X}}_{\bs{\Xi}},{\bf p}):=\mc{B}_{\tilde{\mc{X}}_{\bs{\Xi}},\bs{\alpha}}({\bf p})$. 
\end{theorem}
\begin{proof} 
To show that  $\mc{F}^{[\bs{\Xi}^\bullet]}_{\bs{\alpha}}$ is a holomorphic section, we need to check in local plumbing coordinates $\btau$ that  $\mc{F}^{[\bs{\Xi}^\bullet]}_{\bs{\alpha}}(\mc{X}(\bs{\tau}))/s_{\mc{X}(\bs{\tau})}$ is holomorphic if $s$ is a non-vanishing local holomorphic section of $\mc{L}^{[\bs{\Xi}^\bullet]}(\bs{\alpha})$.
Using the notations from the proof of Lemma \ref{l:holomorphy_line_bundle}, let us use the local holomorphic sections $\bs{\tau}\in U^n\mapsto G^n_{\bs{\alpha}}(\bs{\tau})$ in \eqref{trivialization2} associated to a covering $\bigcup_n U_{\mc{X}^n}$ of Teichm\"uller space $\mc{T}_{g,m}$ using plumbing coordinates.
We need to show that for all $n$, the local function 
\[f^n(\bs{\tau},{\bf p}) :=\frac{\mc{F}_{\bs{\alpha}}^{[\bs{\Xi}^\bullet]}(\mc{X}^n(\bs{\tau}),{\bf p})}{G^n_{\bs{\alpha}}(\bs{\tau})}=\mc{B}_{\tilde{\mc{X}}^n_{\bs{\Xi}^n}(\bs{\tau}),\bs{\alpha}}({\bf p}) \] 
is holomorphic in $\bs{\tau}\in U^n$, where we recall that $\tilde{\mc{X}}^n_{\bs{\Xi}^n}(\bs{\tau})\in \mc{D}\tT^{[\bs{\Xi}^\bullet]}_{g,m}$ is a lift of $\mc{X}^n_{\bs{\Xi}^n}(\bs{\tau})$ with decorating vectors that are fixed (independent of $\bs{\tau}$), and $\mc{X}^n_{\bs{\Xi}^n}(\bs{\tau})$ is
obtained by plumbing $\mc{X}^n_{\bs{\Xi}^n}$ with parameter $\bs{\tau}$ as in Definition \ref{def:plumbing}.
Let $(\bs{\psi}_j(\bs{\tau}),\bs{\theta}_j(\bs{\tau}),\bs{\la}_j)$ be the model form of the decorated marked complex building block associated to $\Xi_j(\bs{\tau})$ (where we used that $\bs{\la}_j$ is independent of $\bs{\tau}$), and $\bs{\alpha}_j$ the weights associated to the marked points of $\Xi_j(\bs{\tau})$ induced by $\bs{\alpha}$ via \eqref{alpha_to_alpha_j}.
It is shown in \cite[Section 8.2]{GKRV21_Segal} that the function (recall \eqref{p_j^G}) 
\begin{equation}\label{dependenceGammaq} 
 \btau\mapsto \frac{\mc{B}_{\tilde{\mc{X}}^n_{\bs{\Xi}^n}(\bs{\tau}),\bs{\alpha}}({\bf p})}{\prod_{j=1}^{j_0}\Gamma_{\bs{\alpha}_j}({\bf p}_j^{\mc{G}},\bs{\psi}_j(\bs{\tau}),\bs{\theta}_j(\bs{\tau}),\bs{\la}_j)}
\end{equation}
is holomorphic in $\btau=(\tau_1,\dots,\tau_{i_0})$ for $\bs{\tau}\in U^n$, i.e. such that $\mc{X}^n(\bs{\tau})\in U_{\mc{X}^n}$ 
with $\bs{\tau}$ being close to $0$. Moreover, setting $\tau_{jk}$ as in \eqref{def_taujk} and $p_{jk}$ as in \eqref{p_j^G}, we have 
\[\Gamma_{\bs{\alpha}_j}({\bf p}_j^{\mc{G}},\bs{\psi}_j(\bs{\tau}),\bs{\theta}_j(\bs{\tau}),\bs{\la}_j)=e^{{\rm i}\sum_{k=1}^{b_j}\Delta_{Q+{\rm i}p_{jk}}\tau_{jk}} \Gamma_{\bs{\alpha}_j}({\bf p}_j^{\mc{G}},\bs{\psi}_j(0),\bs{\theta}_j(0),\bs{\la}_j).\]
which is holomorphic in $\bs{\tau}$.  
Fix a compact interval $I\subset(0,\infty)$ and a compact subset $K\Subset U^n$. After possibly shrinking the plumbing chart, the model
maps $\psi_{jk}(\bs\tau)$, for $\bs\tau\in K$, extend holomorphically to a common collar and remain bounded there in the relevant holomorphic
topology. The locally uniform estimates in Lemma~\ref{block_glued} therefore give
\[ \sup_{\bs\tau\in K}
 \|f^n(\bs\tau,\cdot)\|_{L^1(I^{i_0})}<\infty.\]
Here we also use that, for ${\bf p}\in I^{i_0}$ and $\bs\tau\in K$, the factors
$e^{{\rm i}\sum_k\Delta_{Q+{\rm i}p_{jk}}\tau_{jk}}$ are uniformly bounded.
Choose a jointly measurable representative of $(\bs\tau,{\bf p})\mapsto f^n(\bs\tau,{\bf p})$ which is holomorphic
in $\bs\tau$ for almost every ${\bf p}$. For every
$h\in L^\infty(I^{i_0})$, the function
\[ \bs\tau\mapsto \int_{I^{i_0}}f^n(\bs\tau,{\bf p})
 \overline{h({\bf p})}\,\dd{\bf p}\]
is holomorphic by Cauchy's formula (via the preceding locally
uniform $L^1$-bound and the pointwise holomorphy in $\bs\tau$).
Thus $\bs\tau\mapsto f^n(\bs\tau,\cdot)$ is weakly holomorphic and
locally bounded with values in $L^1(I^{i_0})$. It is therefore strongly holomorphic as an $L^1(I^{i_0})$-valued map. 
Letting $I\subset(0,\infty)$ vary proves the asserted $L^1_{\rm loc}(\R_+^{i_0})$-valued holomorphy.
\end{proof}
Notice that in the case of real valued weights $\bs{\alpha}$ and no self-gluing curves, the conformal block is holomorphic with values in $L^2_{\rm loc}(\R_+^{i_0})$.

Another consequence of \cite[Theorem 8.4]{GKRV21_Segal} is that, if $\alpha\in \R^m$ are real weights satisfying the Seiberg bounds with respect to the gluing graph $\mc{G}$, choosing any smooth family of compatible Riemannian metrics $g_{\mc{X}}$ on a representative Riemann surface $S$ of $\mc{X}$ to induce the Hermitian norm on $\mc{L}^{[\bs{\Xi}^\bullet]}(\bs{\alpha})$, we have
\begin{equation}\label{blocksL2} 
\big({\bf p}\mapsto \|\mc{F}^{[\bs{\Xi}^\bullet]}_{\bs{\alpha}}({\bf p})\|_{\mc{L}^{[\bs{\Xi}^\bullet]}(\bs{\alpha})}^2 \, \bs{\rho}^{\mc{G}}(\bs{\alpha},{\bf p})\big) \in L^1(\R_+^{3g-3+m},\dd{\bf p}),
\end{equation}
where $\mc{G}$ denotes the gluing graph of $[\bs{\Xi}^\bullet]$ and 
\begin{equation}\label{rhoDOZZ}
\bs{\rho}^{\mc{G}}(\bs{\alpha},{\bf p}):=\prod_{j=1}^{2g-2+m} C^{\rm DOZZ}(\bs{\alpha}_j(-\bs{\sigma}_j{\bf p}_j^{\mc{G}})).
\end{equation}
Here $\bs{\sigma}_j=(\sigma_{j1},\ldots,\sigma_{jb_j})$ are the orientation signs of the boundary components of the $j$-th complex building block, induced by fixing a choice of orientation $\vec{\mc{G}}$ of $\mc{G}$; more precisely, we set $\sigma_{jk}=+1$, $\sigma_{j'k'}=-1$ for an oriented internal edge $\vec{\rm e}_i=((j,k),(j',k'))$. In fact, $\bs{\rho}^{\mc{G}}(\bs{\alpha},{\bf p})$ does not depend on the choice of orientation of $\mc{G}$. 
Indeed, a direct consequence of \eqref{symmetry_psi_Q+ip} is that the DOZZ constants satisfy for any $\bs{\alpha} \in (-\infty,Q)^{m}$ and 
${\bf p}\in \R^{b}$ with $m+b=3$
\begin{equation}\label{Cdozzscattering} 
C^{\rm DOZZ}(\bs{\alpha}({\bf p}))=C^{\rm DOZZ}(\bs{\alpha}(-\bs{\sigma}{\bf p}))\prod_{i:\sigma_i=1}\mc{S}(Q+{\rm i}p_i),
\end{equation}
where $\mc{S}(Q+{\rm i}p)$ is the scattering coefficient. In particular, we have
\[\bs{\rho}^{\mc{G}}(\bs{\alpha},{\bf p})=\prod_{j=1}^{2g-2+m} C^{\rm DOZZ}(\bs{\alpha}_j({\bf p}_j^{\mc{G}}))\prod_{i=1}^{3g-3+m}\mc{S}(Q-{\rm i}p_i),\]
which is trivially independent of orientations. For real weights $\bs\alpha$, the reflection relation and the fact that
the two ends of every internal edge carry opposite orientation signs
also imply that $\bs\rho^{\mc G}(\bs\alpha,{\bf p})$ is real valued; here we used 
\begin{equation}\label{dozzconjugue}
\bbar{C^{\rm DOZZ}(a_1,a_2,a_3)}=C^{\rm DOZZ}(\bar{a}_1,\bar{a}_2,\bar{a}_3).
\end{equation}
Moreover, the DOZZ factors occurring in its definition are continuous
and non-vanishing on $\R_+^{i_0}$. Hence $\bs\rho^{\mc G}(\bs\alpha,\cdot)$ is a continuous non-vanishing real
function on the connected set $\R_+^{i_0}$, and therefore has constant
sign. This sign is positive. Indeed, by the conformal bootstrap formula
\cite[Theorem~8.4]{GKRV21_Segal}, the corresponding Liouville
correlation function is the integral of
$\bs\rho^{\mc G}(\bs\alpha,{\bf p})$ multiplied by strictly positive
geometric factors and by the non-negative quantity
$|\mc B_{\tilde{\mc X}_{\bs\Xi},\bs\alpha}({\bf p})|^2$.
Since the Liouville correlation function is strictly positive, the
constant sign of $\bs\rho^{\mc G}$ must be positive. Thus $\bs\rho^{\mc G}(\bs\alpha,{\bf p})>0$ for all ${\bf p}\in\R_+^{i_0}$.

\begin{definition}[\textbf{Hilbert space of conformal blocks}]
Let $\mc{G}$ be a gluing graph for closed surfaces of genus $g$ with $m$ marked points, and let $\bs{\alpha}\in\R^m$ be weights satisfying the Seiberg bounds with respect to $\mc{G}$. We define the Hilbert space 
\[\mc{H}^{\mc{G}}_{\bs{\alpha}}:= L^2(\R^{3g-3+m}_+, \bs{\rho}^{\mc{G}}(\bs{\alpha},{\bf p})\dd {\bf p}).
\]
The \textbf{Hilbert bundle of conformal blocks} over $\mc{T}_{g,m}$ associated to an isotopy class $[\bs{\Xi}^\bullet]\in\pi_0\tilde{\mc{BD}}_{g,m}^{\mc{G}}$ is defined by
\[{\rm CB}^{[\bs{\Xi}^\bullet]}_{\boldsymbol{\alpha}}:=\mc{H}^{\mc{G}}_{\bs{\alpha}}\otimes \mc{L}^{[\bs{\Xi}^\bullet]}(\bs{\alpha}).\]
 \end{definition}

By \eqref{blocksL2}, the conformal block $\mc{F}^{[\bs{\Xi}^\bullet]}_{\bs{\alpha}}(\mc{X})$ is a section of the bundle ${\rm CB}_{\boldsymbol{\alpha}}^{[\bs{\Xi}^\bullet]}$. We notice that since $\mc{L}^{[\bs{\Xi}^\bullet]}(\bs{\alpha})$ can be holomorphically trivialised (since $\mc{T}_{g,m}$ is a Stein manifold), ${\rm CB}^{[\bs{\Xi}^\bullet]}_{\boldsymbol{\alpha}}$ is isomorphic to the trivial Hilbert bundle $\mc{H}^{\mc{G}}_{\bs{\alpha}}$  via such a trivialisation.

\begin{lemma}\label{lem:action_on_Xi_action_on_X}
Let $[\bs{\Xi}^\bullet]\in\pi_0\tilde{\mc{BD}}_{g,m}$ and $[h]\in \tilde{{\rm Mod}}_{g,m}$. Then there exists a  non-vanishing holomorphic section $s$ of ${\rm Hom}(\mc{L}^{[h]\cdot[\bs{\Xi}^\bullet]},[h]^*\mc{L}^{[\bs{\Xi}^\bullet]})$ such that 
\[\mc{F}^{[\bs{\Xi}^\bullet]}(\mc{X}\cdot[h])= s(\mc{X})(\mc{F}^{[h]\cdot[\bs{\Xi}^\bullet]}(\mc{X})).\]
\end{lemma}
\begin{proof}

If $[h]\in \tilde{{\rm Mod}}_{g,m}$, we also denote by $[h]$ its image in ${\rm Mod}_{g,m}$. 
For $\mc X=[(F,\Sigma,J,{\bf x})]$, changing the marking $F$ to $F\circ h$ gives a bijection
\[ \Phi_{[h]}:[(F,\Sigma,J,{\bf x},{\bf Y},\bs\Xi)] \mapsto [(F\circ h,\Sigma,J,{\bf x},{\bf Y},\bs\Xi)]
\]
between the decorated decompositions over $\mc X$ in the
component $[h]\cdot[\bs\Xi^\bullet]$ and those over
$\mc{X}\cdot [h]$ in the component $[\bs\Xi^\bullet]$.
Indeed, $[(F\circ h)^{-1}\circ\bs\Xi]=[ h^{-1}]\cdot[F^{-1}\circ\bs\Xi]$. For
$G\in\mc L_{\mc X}^{[h]\cdot[\bs\Xi^\bullet]}(\bs\alpha)$,
consider the element in $[h]^*\mc L^{[\bs\Xi^\bullet]}(\bs\alpha)$ defined by the expression
\[ (\Phi_{[h]}^{-1})^*G
\big([(F\circ h,\Sigma,J,{\bf x},{\bf Y},\bs\Xi)]\big)
:=
G\big([(F,\Sigma,J,{\bf x},{\bf Y},\bs\Xi)]\big).
\]
The change of marking leaves the Schwarzian cocycle unchanged
and commutes with the decoration action. Hence the function
on the left satisfies properties $(P1)$ and $(P2)$ defining
$\mc{L}_{\mc{X}\cdot [h]}^{[\bs\Xi^\bullet]}(\bs\alpha)$.
The inverse construction using $h^{-1}$ shows that we have defined a line-bundle isomorphism $s:=(\Phi_{[h]}^{-1})^*: \mc L^{[h]\cdot[\bs\Xi^\bullet]}(\bs\alpha)
\to
[h]^*\mc L^{[\bs\Xi^\bullet]}(\bs\alpha).$
The map  $\Phi_{[h]}$ intertwines the local plumbing families used
to define the local holomorphic trivialisations of the two line
bundles; thus the line bundle isomorphism $s$ is holomorphic.
The conformal block depends on $(\Sigma,J,{\bf x},{\bf Y},\bs\Xi)$, but not on the global marking $F$. Evaluating on the corresponding decompositions gives
\[
\mc{F}_{\bs\alpha}^{[\bs\Xi^\bullet]}(\mc X\cdot[h])
= s(\mc{X})
\mc{F}_{\bs\alpha}^{[h]\cdot[\bs\Xi^\bullet]}(\mc X).\qedhere
\]
\end{proof}

\begin{proposition}[\textbf{Density of conformal blocks}]\label{prop:density}
For $\bs{\alpha}\in\R^m$ satisfying the Seiberg bounds assosicated to the gluing graph $\mc{G}$, the complex vector space 
\[{\rm span}\{\mc{F}^{[\bs{\Xi}^\bullet]}_{\bs{\alpha}}(\tilde{\mc{X}}_{\bs{\Xi}^{\rm h}})\,|\, \mc{X}\in \mc{T}_{g,m}\}\]
is dense in  $\mc{H}^{\mc{G}}_{\bs{\alpha}}$, where for each $\mc{X}\in \mc{T}_{g,m}$, we choose a lift $\tilde{\mc{X}}\in \tT_{g,m}$ and a hyperbolic decorated marked complex building block decomposition $\bs{\Xi}^{\rm h}$ associated to $\tilde{\mc{X}}$ in the isotopy class $[\bs{\Xi}^\bullet]\in \pi_0\tilde{\mc{BD}}_{g,m}$.
\end{proposition}
\begin{proof}
Let $u\in \mc{H}^{\mc{G}}_{\boldsymbol{\alpha}}$ satisfying 
$\langle\mc{F}^{[\bs{\Xi}^\bullet]}_{\bs{\alpha}}(\tilde{\mc{X}}_{\bs{\Xi}^{\rm h}}),\bar{u}\rangle_{\mc{H}^{\mc{G}}_{\boldsymbol{\alpha}}}=\int_{\R_+^{i_0}}u({\bf p})\mc{F}^{[\bs{\Xi}^\bullet]}_{\bs{\alpha}}(\tilde{\mc{X}}_{\bs{\Xi}^{\rm h}},{\bf p})\bs{\rho}^{\mc{G}}(\bs{\alpha},{\bf p})\dd{\bf p}=0$ for all $\mc{X}\in \mc{T}_{g,m}$.
Fix a point $\mc{X}\in \mc{T}_{g,m}$, a lift $\tilde{\mc{X}}\in \tT_{g,m}$, and 
$\bs{\Xi}^{\rm h}$ a hyperbolic decorated marked complex building block decomposition for $\tilde{\mc{X}}$; in particular it defines a hyperbolic marked complex building block decomposition for $\mc{X}$.
Consider the plumbing coordinates $\bs{\tau}$ associated to $\tilde{\mc{X}}_{\bs{\Xi}^{\rm h}}$; for this we have fixed an orientation of the internal edges of the gluing graph $\mc{G}$ of $[\bs{\Xi}^\bullet]$. As usual, $j_0=2g-2+m$ is the number of internal vertices of $\mc{G}$ and $i_0=3g-3+m$ is the number of internal edges. We use the notations of Definition \ref{def:plumbing}: for $j=1,\dots,j_0$, $\tilde{\mc{X}}_j(\bs{\tau})\in \tT_{0,m_j,b_j}$ denotes the $j$-th decorated marked complex building block of $\tilde{\mc{X}}_{\bs{\Xi}^{\rm h}}(\bs{\tau})$, and $\tilde{\mc{X}}_j:=\tilde{\mc{X}}_j(0)$. We also denote $\tilde{\Theta}_{m_j,b_j}(\tilde{\mc{X}}_j)=(\bs{\psi}_j,\bs{\theta}_j,\bs{\la}_j)$ for the model form of $\tilde{\mc{X}}_j$.
Using \cite[Corollary 11.4]{GKRV21_Segal}, 
the conformal block amplitude of $(\tilde{\mc{X}}_{j}(\bs{\tau}),\bs{\alpha}_j)$ is given by 
\[\mc{B}_{\tilde{\mc{X}}_j(\btau),\bs{\alpha}_j}({\bf p}_j,\bs{\nu}_j)=\prod_{k=1}^{b_j} e^{{\rm i}(\Delta_{Q+{\rm i}p_{jk}}+|\nu_{jk}|)\tau_{jk}}\mc{B}_{\tilde{\mc{X}}_j,\bs{\alpha}_j}({\bf p}_j,\bs{\nu}_j),\]
where $\tau_{jk}\in \C$ are defined in \eqref{def_taujk}, $\bs{\nu}_j=(\nu_{j1},\dots,\nu_{jb_j})\in \mc{T}^{b_j}$, and $\bs{\alpha}_j$ are the weights attached to the marked points of $\tilde{\mc{X}}_j$.
Using Lemma \ref{block_glued}, when 
${\rm Im}(\tau_i)>-2\eps$ for $\eps>0$ small, we deduce that for ${\bf p}=(p_1,\dots,p_{i_0})\in \R_+^{i_0}$, 
\[\mc{F}^{[\bs{\Xi}^\bullet]}_{\bs{\alpha}}(\tilde{\mc{X}}_{\bs{\Xi}^{\rm h}}(\bs{\tau}),{\bf p})=e^{{\rm i}\sum_{i=1}^{i_0}\Delta_{Q+{\rm i}p_i}\tau_i}
\Big(\prod_{j=1}^{j_0}\Gamma_{\bs{\alpha}_j}({\bf p}_j^{\mc{G}},\bs{\psi}_j,\bs{\theta}_j,\bs{\la}_j)+\sum_{|\bs{\nu}|>0}e^{{\rm i}\sum_{i=1}^{i_0}\tau_i|\nu_i|}\prod_{j=1}^{j_0}\mc{B}_{\tilde{\mc{X}}_j,\bs{\alpha}_j}({\bf p}_j^{\mc{G}},\bs{\nu}_j^{\mc{G}})\Big),\]
where $\bs{\nu}=(\nu_1,\dots,\nu_{i_0})\in \mc{T}^{i_0}$ in the sum, $|{\bs \nu}|:=\sum_{i=1}^{i_0}|\nu_i|$ and we use the notation \eqref{p_j^G} for 
${\bf p}^{\mc{G}}_j$ and similarly $\bs{\nu}_j^{\mc{G}}=(\nu_{j1},\dots,\nu_{jb_j})\in \mc{T}^{b_j}$ with 
$\nu_{jk}=\nu_{j'k'}=\nu_i$ when ${\rm e}_i=\{(j,k),(j',k')\}$.
Recall that $4\Delta_{Q+{\rm i}p}=Q^2+p^2$. We let ${\bf r}=(r_1,\dots,r_{i_0})\in \R_+^{i_0}$ with $r_i=p_i^2/4$ and rewrite the conformal block as a function of ${\bf r}$ under the form 
\[\mc{F}^{[\bs{\Xi}^\bullet]}_{\bs{\alpha}}(\tilde{\mc{X}}_{\bs{\Xi}^{\rm h}}(\bs{\tau}),{\bf p}({\bf r}))= e^{{\rm i}\frac{Q^2}{4}\sum_i\tau_i}
e^{{\rm i}\bs{\tau}\cdot {\bf r}}G({\bf r},{\bf q}), \qquad G({\bf r},{\bf q})=\sum_{{\bf n}\in \N^{i_0}}G_{\bf n}({\bf r}){\bf q}^{\bf n},\]
where ${\bf q}=(q_1,\dots,q_{i_0})=(e^{{\rm i}\tau_1},\dots,e^{{\rm i}\tau_{i_0}})$, ${\bf n}=(n_1,\dots,n_{i_0})$ and ${\bf q}^{\bf n}:=\prod_{i=1}^{i_0}q_i^{n_i}$. Define 
\[ d:=i_0,\qquad |{\bf r}|_1:=\sum_{i=1}^d r_i,
\qquad \dd\mu_{\bs\alpha}({\bf r})
:= \bs\rho^{\mc G}(\bs\alpha,{\bf p}({\bf r}))
\prod_{i=1}^d\frac{\dd r_i}{\sqrt{r_i}}.
\]
Notice that this is precisely the measure obtained from
$\bs\rho^{\mc G}(\bs\alpha,{\bf p})\dd{\bf p}$ under the change of
variables $r_i=p_i^2/4$.
By Lemma \ref{block_glued} and the holomorphic dependence of the
conformal block, there exist $\eta>0$ and $C>0$ such that
\[
 \sup_{\boldsymbol\theta\in[0,2\pi]^d}
 \left\|
 e^{\eta|{\bf r}|_1}
 G\big({\bf r},e^{\eta+{\rm i}\theta_1},\ldots,
                    e^{\eta+{\rm i}\theta_d}\big)
 \right\|_{L^2(\mu_{\bs\alpha})}
 \leq C.
\]
Indeed, this corresponds to taking
$\tau_i=\theta_i-{\rm i}\eta$. Hence, by the $L^2(\mu_{\bs\alpha})$-valued
Cauchy formula,
\[ G_{\bf n}({\bf r})= \frac{1}{(2\pi{\rm i})^d}
 \int_{|q_1|=e^\eta}\cdots\int_{|q_d|=e^\eta} G({\bf r},{\bf q})
 \frac{\dd q_1\cdots\dd q_d}{q_1^{n_1+1}\cdots q_d^{n_d+1}},
\]
and
\begin{equation}\label{weighted_bound_Gn}
 \left\|e^{\eta|{\bf r}|_1}G_{\bf n}\right\|_
 {L^2(\mu_{\bs\alpha})}
 \leq Ce^{-\eta|{\bf n}|}.
\end{equation}
Using Proposition \ref{well-defined_blocks} and property $(P2)$ for
the decorations, the conformal block associated with the hyperbolic
decorated marked complex building block decomposition differs from
the one associated with $\bs\Xi^{\rm h}(\bs\tau)$ by a non-zero scalar
depending on $\bs\tau$ but independent of ${\bf p}$. It follows from
the assumption on $u$ that, for $\bs\tau$ in a real neighbourhood
$U$ of $0$ (here we have omitted the non-zero factor
$e^{{\rm i}\frac{Q^2}{4}\sum_i\tau_i}$)
\begin{equation}\label{orthogonality_expanded}
 0= \sum_{{\bf n}\in\N^d} \int_{\R_+^d} u({\bf r})G_{\bf n}({\bf r}) e^{{\rm i}\bs\tau\cdot({\bf r}+{\bf n})} \dd\mu_{\bs\alpha}({\bf r}).
\end{equation}
For ${\bf n}\in\N^d$, let
$T_{\bf n}({\bf r})={\bf r}+{\bf n}$ and define the complex measure
$\nu:=\sum_{{\bf n}\in\N^d}(T_{\bf n})_*\big(u({\bf r})G_{\bf n}({\bf r})\dd\mu_{\bs\alpha}({\bf r})\big).$
This series converges in total variation, since, by \eqref{weighted_bound_Gn},
\[ \sum_{{\bf n}\in\N^d}\|uG_{\bf n}\|_{L^1(\mu_{\bs\alpha})}
 \leq
 \|u\|_{L^2(\mu_{\bs\alpha})}
 \sum_{{\bf n}\in\N^d}
 \|G_{\bf n}\|_{L^2(\mu_{\bs\alpha})}
 <\infty
\]
 Moreover, for every
$0<\eta'<\eta$, the same estimate shows that
\[
 \int_{\R_+^d}e^{\eta'|{\boldsymbol\xi}|_1}
 \dd|\nu|({\boldsymbol\xi})<\infty.
\]
Consequently, the Fourier transform of $\nu$ is holomorphic in a
complex neighbourhood of $\R^d$. Equation \eqref{orthogonality_expanded} says that it vanishes on the open set
$U\subset\R^d$; hence it vanishes identically, and then $\nu=0$.
Writing $\dd\mu_{\bs\alpha}({\bf r})=w_{\bs\alpha}({\bf r})\dd{\bf r}$ and 
$w_{\bs\alpha}({\bf r}) = \bs\rho^{\mc G}(\bs\alpha,{\bf p}({\bf r})) \prod_{i=1}^d r_i^{-1/2}$, 
we therefore obtain, for almost every
${\boldsymbol\xi}\in\R_+^d$,
\begin{equation}\label{triangular_identity_density}
 \sum_{\substack{{\bf n}\in\N^d\\ n_i<\xi_i\;\forall i}}
 u({\boldsymbol\xi}-{\bf n})
 G_{\bf n}({\boldsymbol\xi}-{\bf n})
 w_{\bs\alpha}({\boldsymbol\xi}-{\bf n})=0,
\end{equation}
and the sum in \eqref{triangular_identity_density} is finite. We now argue by induction over ${\bf k}\in\N^d$, ordered by
$|{\bf k}|$. If ${\boldsymbol\xi}\in(0,1)^d$, only ${\bf n}=0$
contributes, and therefore $u({\boldsymbol\xi})G_{\bf0}({\boldsymbol\xi})w_{\bs\alpha}({\boldsymbol\xi})=0$.
Since $G_{\bf0}({\bf r})=\prod_{j=1}^{j_0}\Gamma_{\bs\alpha_j}({\bf p}^{\mc G}_j,\bs\psi_j,\bs\theta_j,\bs\lambda_j)$
does not vanish, and $w_{\bs\alpha}>0$, we have  $u=0$ almost everywhere on $(0,1)^d$.

Assume that $u=0$ almost everywhere on
${\bf k}'+(0,1)^d$ whenever $|{\bf k}'|<N$, and let
${\bf k}\in\N^d$ satisfy $|{\bf k}|=N$. For
${\boldsymbol\xi}\in{\bf k}+(0,1)^d$, every term in
\eqref{triangular_identity_density} indexed by ${\bf n}\not=0$ involves
${\boldsymbol\xi}-{\bf n}\in({\bf k}-{\bf n})+(0,1)^d$, with
$|{\bf k}-{\bf n}|<N$, and therefore vanishes by the induction
hypothesis. Thus
\[
 u({\boldsymbol\xi})G_{\bf0}({\boldsymbol\xi})
 w_{\bs\alpha}({\boldsymbol\xi})=0,
\]
and hence $u=0$ almost everywhere on
${\bf k}+(0,1)^d$. This proves that $u=0$ almost everywhere on
$\R_+^d$, and therefore proves the density.
\end{proof}

\subsection{Bootstrap formula} In this section, we rewrite the conformal bootstrap formula proved in \cite[Theorem 8.4]{GKRV21_Segal} in terms of the conformal blocks which are now defined globally on Teichm\"uller space $\mc{T}_{g,m}$. We fix the metric on the Riemann sphere $(\hat{\C},J_\C)$ to be the metric $g_{\rm dozz}$ defined in \eqref{gdozz}.
   
\begin{theorem}[\textbf{Conformal bootstrap}]\label{conf_boot}
For $\mc{X}\in \mc{T}_{g,m}$, let $S=(\Sigma,J,{\bf x})$ be a Riemann surface representing the class $\mc{X}$ and $g_{S}$ a compatible Riemannian metric on $S$. 
Let $[\bs{\Xi}^\bullet]\in \pi_0\tilde{\mc{BD}}_{g,m}$ with associated gluing graph $\mc{G}$ and let $\bs{\alpha}=(\alpha_1,\dots,\alpha_m)$ be conformal weights satisfying the Seiberg bounds with respect to $\mc{G}$.
Then 
\begin{equation}\label{Segal_boot_rewrite}
\cjg \prod_{k=1}^mV_{\alpha_k}(x_k)\cjd_{\Sigma,g_S}=C_0\|\mc{F}^{[\bs{\Xi}^\bullet]}_{\bs{\alpha}}(\mc{X})\|_{{\rm CB}_{\bs{\alpha}}^{[\bs{\Xi}^\bullet]}}^2=C_0
\int_{\R_+^{i_0}}\bs{\rho}^{\mc{G}}(\bs{\alpha},{\bf p})\|\mc{F}^{[\bs{\Xi}^\bullet]}_{\bs{\alpha}}(\mc{X},{\bf p})\|_{\mc{L}^{[\bs{\Xi}^\bullet]}(\bs{\alpha})}^2 \dd {\bf p}.
\end{equation}
Here $i_0=3g-3+m$ is the number of cutting curves, the constant $C_0$ is given by
\[C_0=\frac{2^{i_0/2}}{(2\pi)^{2i_0-1}}\prod_{j=1}^{j_0}C_{b_j}\]
with $C_{b_j}$ the constant appearing in \eqref{constantC(P,g)}, $j_0=2g-2+m$ is the number of building blocks, and the Hermitian metric on $\mc{L}^{[\bs{\Xi}^\bullet]}(\bs{\alpha})$ is induced by $g_S$ as in Definition \ref{def_hermitian_metric}. 
\end{theorem}

\begin{proof}
Using the language of this paper, the result \cite[Theorem 8.4]{GKRV21_Segal} can be written as follows. Let $\tilde{\mc{X}}_{\bs{\Xi}}\in\BDTT_{g,m}$ be a lift of $\mc{X}$ having decoration $\bf Y$ on $S$ and decomposition $\bs{\Xi}$ into decorated marked complex building blocks, so that $(\Xi_j)_*Y_k^\bullet=Y_{jk}$ if we define $Y_{jk}=Y_\ell$ when $\Xi_j(z_{b_j+k}^\bullet)=x_\ell$. Let $g_{\bs{\Xi}}$ be a metric that is admissible on each complex building block associated to $\bs{\Xi}$. Then the bootstrap formula reads 
\begin{equation}\label{formule_boot_segal}
\cjg \prod_{k=1}^mV_{\alpha_k}(x_k)\cjd_{\Sigma,g_{\bs{\Xi}}}=C_0
\prod_{j=1}^{j_0}e^{c_{\rm L}S_{\rm L}^0(\hat{\C},g_{\rm dozz},\hat{g}_{\Xi_j})-\sum_{k=1}^{m_j}\Delta_{\alpha_{jk}}(\omega_j(z_{b_j+k}^{\bullet})+2\log|(\Phi_j)_*Y_{jk}|)}
\int_{\R_+^{i_0}}\bs{\rho}^{\mc{G}}(\bs{\alpha},{\bf p})|\mc{F}^{[\bs{\Xi}^\bullet]}_{\bs{\alpha}}(\tilde{\mc{X}}_{\bs{\Xi}},{\bf p})|^2 \dd {\bf p},\end{equation}
where we used the notations of Definition \ref{def_hermitian_metric},  and 
$\omega_j$ is defined as in \eqref{constantC(P,g)} by $(\Phi_j)_*g_{\bs{\Xi}}=e^{\omega_j}g_{\rm dozz}$ if $\Phi_j:\mc{P}_j\to \hat{\C}$ is the model form map (Lemma \ref{modelform}) of the complex building block $\mc{P}_j$ associated to $\Xi_j$. Indeed, in \cite[Theorem 8.4]{GKRV21_Segal}, it is proved that
\begin{align*}
 \cjg \prod_{k=1}^mV_{\alpha_k}(x_k)\cjd_{\Sigma,g_{\bs{\Xi}}}=\frac{2^{i_0/2}}{(2\pi)^{2i_0-1}}
\int_{\R_+^{i_0}}\prod_{j=1}^{j_0}C(\bs{\psi}_j,g_{\bs{\Xi}},\bs{\Delta}_{\bs{\alpha}_j({\bf p})})\bs{\rho}^{\mc{G}}(\bs{\alpha},{\bf p})|\tilde{\mc{F}}_{{\bf p}}(\bs{\alpha})|^2 \dd {\bf p}.
\end{align*}
Here $\tilde{\mc{F}}_{{\bf p}}(\bs{\alpha})$ is the 
normalised conformal block defined exactly as in Lemma \ref{block_glued}, but with $\mc{B}_{\tilde{\mc{X}}_j,\bs{\alpha}_j}$ replaced by the matrix coefficients $w_{\bs{\psi}_j}(\bs{\alpha}_j({\bf p}))$ of Lemma \ref{matrix_coefficients}, where $\bs{\psi}_j$ is the model form (without the marking parameter $\bs{\theta}_j$) of $\mc{P}_j$ and $\bs{\alpha}_j({\bf p})$ is the collection of weights $\alpha_{jk}$ of the marked points $x_{jk}$ in $\mc{P}_j$ and of the internal channels $Q+{\rm i}p_{jk}$ corresponding to internal edges of the gluing graph. The constant $C(\mc{P}_j,g,\bs{\Delta}_{\bs{\alpha}_j({\bf p})})$ are given by \eqref{constantC(P,g)}.
Taking into account the definition of conformal blocks $\mc{B}_{\tilde{\mc{X}}_j,\bs{\alpha}_j}$ in terms of $w_{\bs{\psi}_j}(\bs{\alpha}_j({\bf p}))$ (see Lemma \ref{matrix_coefficients}), and adding the contributions of the decorating vectors gives \eqref{formule_boot_segal}.
Using \eqref{weyl_anomaly_formula_amp}, we get 
\[\begin{split}
\cjg \prod_{k=1}^mV_{\alpha_k}(x_k)\cjd_{\Sigma,g_S}={}& C_0e^{c_{\rm L}S_{\rm L}^0(\Sigma,g_{\bs{\Xi}},g_S)-\sum_{\ell=1}^m\Delta_{\alpha_\ell}(\omega(x_\ell)+2\log|(\Phi_j)_*Y_{\ell}|)}
\prod_{j=1}^{j_0}e^{c_{\rm L}S_{\rm L}^0(\hat{\C},g_{\rm dozz},\hat{g}_{\Xi_j})-\sum_{k=1}^{m_j}\Delta_{\alpha_{jk}}\omega_j(z^{\bullet}_{b_j+k})}\\
& \times 
\int_{\R_+^{i_0}}\bs{\rho}^{\mc{G}}(\bs{\alpha},{\bf p})|\mc{F}^{[\bs{\Xi}^\bullet]}_{\bs{\alpha}}(\tilde{\mc{X}}_{\bs{\Xi}},{\bf p})|^2 \dd {\bf p},
\end{split}\]
where $g_S=e^{\omega}g_{\bs{\Xi}}$.  For each $\ell$, let $j=j(\ell)$ and $k=k(\ell)$ be determined by $\Xi_j(z_{b_j+k}^\bullet)=x_\ell$
Then we have  (below we use the natural Hermitian metrics on the complexified bundles induced by $g_{\rm S}, g_{\rm dozz}$)
\[\begin{split}
\frac{\|Y_\ell\|^2_{g_S}}{\|Y_k^\bullet\|^2_{g_{\rm dozz}}}=& 
\frac{\|Y_k^\bullet\|^2_{(\Phi_j)_*g_S}}{\|Y_k^\bullet\|^2_{g_{\rm dozz}}}e^{2\log|(\Phi_j)_*Y_\ell|}
=\frac{\|Y_k^\bullet\|^2_{(\Phi_j)_*g_{\bs{\Xi}}}}{\|Y_k^\bullet\|^2_{g_{\rm dozz}}}e^{\omega(x_\ell)+2\log|(\Phi_j)_*Y_\ell|}
\\
=&e^{\omega_j(z_{b_j+k}^\bullet)+\omega(x_\ell)+2\log|(\Phi_j)_*Y_\ell|}.
\end{split}\]
Together with the Weyl anomaly formula, this identifies the
multiplicative factor in \eqref{formule_boot_segal} with the Hermitian
metric defined in Definition \ref{def_hermitian_metric}, and proves
\eqref{Segal_boot_rewrite}.
\end{proof}

Using now the estimates \eqref{bound_bloc_pant_Dozz^2} together with the factorisation expression \eqref{factorisation_extended}, the proof of Theorem \ref{conf_boot} contained in \cite{GKRV21_Segal} also applies directly to the case of $\bs{\alpha}\in \C^m$ satisfying the generalised Seiberg bounds with respect to the gluing graph. 
\begin{lemma}\label{bootstrap_complex_alpha}
Let $C(\alpha_1,\dots,\alpha_m)$ be the analytic extension of $\cjg \prod_{k=1}^mV_{\alpha_k}(x_k)\cjd_{\Sigma,g_S}$ to the open set of weights $\bs{\alpha}\in \C^m$  satisfying the generalised Seiberg bound with respect to the gluing graph $\mc{G}$ of $[\bs{\Xi}^\bullet]$. Then, we have  
\begin{equation}\label{Segal_boot_complex}
C(\alpha_1,\dots,\alpha_m)=C_0
\int_{\R_+^{i_0}}\bs{\rho}^{\mc{G}}(\bs{\alpha},{\bf p})\cjg\mc{F}^{[\bs{\Xi}^\bullet]}_{\bs{\alpha}}(\mc{X},{\bf p}),\mc{F}^{[\bs{\Xi}^\bullet]}_{\bbar{\bs{\alpha}}}(\mc{X},{\bf p})\cjd_{\bs{\alpha}} \dd {\bf p}.
\end{equation}
Here $i_0$ and $C_0$ are as in \eqref{Segal_boot_rewrite} and $\cjg\cdot,\cdot\cjd_{\bs{\alpha}}$ is the pairing defined in \eqref{pairing_extended}.
\end{lemma}
\begin{proof}
Notice that the $L^1$-integrability of the individual conformal
blocks does not by itself imply the integrability of their product.
We therefore apply the gluing argument first to the full Segal
amplitudes. By \eqref{factorisation_extended}, the spectrally
transformed amplitude of the $j$-th complex building block has the
form
\[
 \widehat{\mc A}_{j,\bs\alpha_j}
 =
 C_j(\bs\alpha_j,{\bf p}_j)
 C^{\rm DOZZ}(\bs\alpha_j({\bf p}_j))
 \mc B_{j,\bs\alpha_j}({\bf p}_j)
 \otimes
 \overline{\mc B_{j,\overline{\bs\alpha_j}}({\bf p}_j)}.
\]
The estimate \eqref{bound_bloc_pant_Dozz^4} shows that this full
coefficient is square-integrable, with the exponential weights needed
for the gluing. Hence the contraction of the full amplitudes along
the graph $\mc G$ is integrable. For self-gluing edges, the diagonal
is defined using the Hilbert--Schmidt factorisation and
\cite[Lemmas~7.2 and~7.6]{GKRV21_Segal}.

After performing this integrable contraction,
\eqref{factorisation_extended} identifies its integrand with
\[
 \bs\rho^{\mc G}(\bs\alpha,{\bf p})
 \left\langle
 \mc F^{[\bs\Xi^\bullet]}_{\bs\alpha}(\mc X,{\bf p}),
 \mc F^{[\bs\Xi^\bullet]}_{\overline{\bs\alpha}}
       (\mc X,{\bf p})
 \right\rangle_{\bs\alpha}.
\]
Consequently this function belongs to $L^1(\R_+^{i_0})$, and the integral \eqref{Segal_boot_complex} is absolutely
convergent.
\end{proof}

\section{Projective representation of the mapping class group on the space of conformal blocks}\label{sec:projective_representation}

We fix a base closed Riemann surface $(\Sigma^\bullet,J^\bullet)$ of genus $g$ with marked points ${\bf x}^\bullet=(x_1^\bullet,\dots,x_m^\bullet)$ and decorating vectors ${\bf Y}^\bullet=(Y_1^\bullet,\dots,Y_m^\bullet)$. We assume that $2g-2+m>0$. Recall the mapping class groups
\[\begin{aligned}
{\rm Mod}_{g,m}&={\rm Diff}(\Sigma^\bullet,\{{\bf x}^\bullet\})/{\rm Diff}_0(\Sigma^\bullet,\{{\bf x}^\bullet\}),\\
\tilde{{\rm Mod}}_{g,m}&={\rm Diff}(\Sigma^\bullet,\{{\bf x}^\bullet\};{\bf Y}^\bullet)/{\rm Diff}_0(\Sigma^\bullet,\{{\bf x}^\bullet\};{\bf Y}^\bullet).
\end{aligned}\]

\subsection{Local study of the A-move: fusion kernel on an $\R$-symmetric four-holed sphere}\label{sec:local_study}
This subsection describes the local transition kernels for the A-move (of the first type)  in explicit geometric models with anti-holomorphic involutions using the boundary bootstrap proved in the paper \cite{GRW2}.

The transition operator for an A-move is quite involved. Its construction relies on the probabilistic proof of the boundary bootstrap formula for Liouville CFT in \cite{GRW2}. The deduction of the transition operator, also called the \textbf{fusion kernel}, from the boundary bootstrap identity follows the Fourier transform argument of Ghosal, R\'emy, X.~Sun, and Y.~Sun \cite{GRSSbloc}; see also 
a new version including B.~Wu \cite{GRSSWBloc}.

Recall the notion of $\R$-symmetric maps from Section \ref{sub:elementary_moves} in the definition of the A-move, and recall that $\hat{\H}:=\H\cup\{\infty\}$ with $\H:=\{{\rm Im}\,z\geq0\}$, $\hat{\R}:=\R\cup\{\infty\}$ and $\tau(z):=\bar{z}$.

For $m_{\rm A}+b_{\rm A}=4$, consider a decorated Riemann surface of type $(0,m_{\rm A},0,b_{\rm A})$ of the form
\[ \tilde{S}_{\rm A}=(\Sigma_{\rm A},J_\C,{\bf x}_{\rm A}, \bs{\zeta}_{\rm A},{\bf Y}_{\rm A})\]
where $\Sigma_{\rm A}=\hat{\C}\setminus \cup_{k=1}^{b_{\rm A}}\mc{D}_k^\circ$ with 
$\mc{D}_k\subset \C$ disjoint closed topological disks satisfying $\tau(\mc{D}_k)=\mc{D}_k$, $\bs{\zeta}_{\rm A}=(\zeta_1,\ldots,\zeta_{b_{\rm A}})$ with each $\zeta_k:\T\to \pl \mc{D}_k$ an $\R$-symmetric real analytic parametrisation (necessarily with incoming orientation), ${\bf x}_{\rm A}=(x_1,\ldots,x_{m_{\rm A}})$ with each $x_k\in\R$, and finally ${\bf Y}_{\rm A}=(Y_1,\ldots,Y_{m_{\rm A}})$ with each $Y_k=t_k\pl_z$ for some $t_k>0$.
We define $\mc{O}_1,\ldots,\mc{O}_4$ so that 
$\{\mc{O}_1,\dots,\mc{O}_4\}=\{\mc{D}_1,\dots,\mc{D}_{b_{\rm A}},\{x_1\},\dots,\{x_{m_{\rm A}}\}\}$,
positively ordered along $\hat\R$, i.e.\ $y<y'$ for $y\in \mc{O}_k\cap \R$ and $y'\in \mc{O}_{k'}\cap \R$ with $k<k'$.

First let $\xi:\T\to \mc{C}:=\xi(\T)\subset\C\setminus\cup_{k=1}^4\mc{O}_k$ be an $\R$-symmetric real analytic curve such that it encloses a bounded disk $\mc{D}\subset\C$ with $\mc{O}_1\cup \mc{O}_2\subset \mc{D}^\circ$, $\mc{O}_3\cup \mc{O}_4\subset\hat{\C}\setminus\mc{D}$.  
It cuts $\tilde{S}_{\rm A}$ into two decorated complex building blocks of types $(m_1,b_1)$ and $(m_2,b_2)$ with $m_1+m_2=m_{\rm A}$, $b_1+b_2=b_{\rm A}+2$. We denote them by
\[ (\mc{P}_1=\Sigma_{\rm A}\setminus \mc{D}^\circ ,J_\C, {\bf x}_1,\bs{\zeta}_1,{\bf Y}_1)\quad\text{and}\quad
(\mc{P}_2=\Sigma_{\rm A}\cap \mc{D},J_\C, {\bf x}_2,\bs{\zeta}_2,{\bf Y}_2)\]
where for $\ell=1,2$, ${\bf x}_\ell=(x_{\ell1},\dots,x_{\ell m_\ell})$ are the marked points in $\mc{P}_\ell\cap\{{\bf x}_{\rm A}\}$, ${\bf Y}_\ell=(Y_{\ell1},\dots,Y_{\ell m_\ell})$ are the associated decorating vectors, and $\bs{\zeta}_\ell=(\zeta_{\ell1},\dots,\zeta_{\ell b_\ell})$ are the incoming boundary parametrisations induced by $\bs{\zeta}_{\rm A}$ and $\xi$.
Here we choose the unique orderings so that $\zeta_{\ell1}(\T),\dots,\zeta_{\ell b_\ell}(\T),x_{\ell1},\dots,x_{\ell m_\ell}$ is positively ordered along $\hat\R$. Suppose
$\zeta_{1k_1}=\xi$, $\zeta_{2k_2}=\xi\circ o$ for some $1\leq k_1\leq b_1$ and $1\leq k_2\leq b_2$. We take smooth markings $\Xi_\ell^1:\Sigma_{m_\ell,b_\ell}^\bullet\to\mc{P}_\ell$ ($\ell=1,2$) so that $\Xi_1^1$ is $\R$-symmetric (and real analytic near $\cup_{k=1}^4\pl\mc{O}_k$ and $\mc{C}$) and $\Xi_2^1$ satisfies the following: there is an 
$\R$-symmetric diffeomorphism $\Omega:\hat{\C}\to \hat{\C}$, real analytic near $\cup_{k=1}^4\pl\mc{O}_k$ and $\mc{C}$, such that $\Xi_2^1=\Omega\circ T_{k_2,\pi}$ (recall \eqref{Tkalpha}). 
We obtain
\[\tilde{\mc{X}}_\ell:=[(\Xi_\ell^1,\mc{P}_\ell,J_{\C},{\bf x}_\ell,\bs{\zeta}_\ell,{\bf Y}_\ell)]\in\tT_{0,m_\ell,b_\ell}.\]
It has model form $\tilde{\mc{MF}}(\tilde{\mc{X}}_\ell)=(\bs{\psi}_\ell,\bs{\theta}_\ell,\bs{\la}_\ell)$ where $\bs{\psi}_\ell=(\psi_{\ell1},\ldots,\psi_{\ell b_\ell})$ with each $\psi_{\ell k}$ $\R$-symmetric except that $\psi_{2k_2}$ satisfies $\psi_{2k_2}(\bar{z})=\overline{\psi_{2k_2}(z)}$ and is orientation reversing on $\hat\R$, 
$\bs{\theta}_1=0$, $\bs{\theta}_2=\pi e_{k_2}\in\R^{b_2}$ (where $e_i=(\delta_{ik})_{k=1}^{b_2}$), and 
$\bs{\la}_\ell\in\R^{m_\ell}$. 
The corresponding lifted marking parameter is
$\theta_{2k_2}=\pi$, due to the half-twist $T_{k_2,\pi}$.
Similarly, let $\xi^*:\T\to \mc{C}^*:=\xi^*(\T)\subset\C\setminus\cup_{k=1}^4\mc{O}_k$ be an $\R$-symmetric real analytic curve such that it intersects $\xi(\T)$ transversally at exactly two points and encloses a bounded disk $\mc{D}^*\subset\C$ with $\mc{O}_2\cup \mc{O}_3\subset (\mc{D}^*)^\circ$, $\mc{O}_1\cup \mc{O}_4\subset\hat{\C}\setminus\mc{D}^*$. 
It cuts $\tilde{S}_{\rm A}$ into two decorated complex building blocks of types $(m_1^*,b_1^*)$ and $(m_2^*,b_2^*)$ with $m_1^*+m_2^*=m_{\rm A}$, $b_1^*+b_2^*=b_{\rm A}+2$, denoted by
\[ (\mc{P}_1^*=\Sigma_{\rm A}\setminus (\mc{D}^*)^\circ ,J_\C, {\bf x}_1^*,\bs{\zeta}_1^*,{\bf Y}_1^*)\quad\text{and}\quad
(\mc{P}_2^*=\Sigma_{\rm A}\cap \mc{D}^*,J_\C, {\bf x}_2^*,\bs{\zeta}_2^*,{\bf Y}_2^*).\]
Again we choose the unique orderings of each ${\bf x}_\ell^*$ and $\bs{\zeta}_{\ell}^*$ so that $\zeta_{\ell1}^*(\T),\dots,\zeta_{\ell b_\ell^*}^*(\T),x_{\ell1}^*,\dots,x_{\ell m_\ell^*}^*$ is positively ordered along $\hat\R$.
Suppose $\zeta_{1k_1^*}^*=\xi^*$ and $\zeta_{2k_2^*}^*=\xi^*\circ o$ for some $1\leq k_1^*\leq b_1^*$ and $1\leq k_2^*\leq b_2^*$.
We take smooth markings $\Xi_\ell^2:\Sigma_{m_\ell^*,b_\ell^*}^\bullet\to\mc{P}_\ell^*$ for $\ell=1,2$, with $\Xi_1^2$ $\R$-symmetric and $\Xi_2^2=\Omega^*\circ T_{k_2^*,\pi}$ 
with $\Omega^*$ $\R$-symmetric, both being real analytic near $\cup_{k=1}^4\pl\mc{O}_k$ and near the cutting curve $\mc{C}^*$. We obtain
\[\tilde{\mc{X}}_\ell^*:=[(\Xi_\ell^2,\mc{P}_\ell^*,J_{\C},{\bf x}_\ell^*,\bs{\zeta}_\ell^*,{\bf Y}_\ell^*)]\in\tT_{0,m_\ell^*,b_\ell^*},\]
with model form $\tilde{\mc{MF}}(\tilde{\mc{X}}_\ell^*)=(\bs{\psi}_\ell^*,\bs{\theta}_\ell^*,\bs{\la}_\ell^*)$ where $\bs{\psi}_\ell^*=(\psi_{\ell1}^*,\ldots,\psi_{\ell b_\ell^*}^*)$ with each $\psi_{\ell k}^*$ $\R$-symmetric except that $\psi_{2k_2^*}^*$ satisfies $\psi_{2k_2^*}^*(\bar{z})=\overline{\psi_{2k_2^*}^*(z)}$ and is orientation reversing on $\hat\R$, $\bs{\theta}_1^*=0$, $\bs{\theta}^*_2=\pi e_{k_2^*}\in \R^{b_2^*}$, and $\bs{\la}_\ell^*\in\R^{m_\ell^*}$. 

\begin{remark}\label{rem:relation_A_move}
The construction here parallels that of the A-move in Section \ref{sub:elementary_moves}. However, while Section \ref{sub:elementary_moves} describes the topological definition of the A-move on the surface $\Sigma_{\rm A}^\bullet$ obtained by gluing canonical complex building blocks $\mc{P}_j^\bullet,\mc{P}_{j'}^\bullet$, here we are describing the A-move happening on the Riemann subsurface which is the union of two marked complex building blocks $\mc{P}_j,\mc{P}_{j'}\subset\Sigma$, after embedding it holomorphically into $\hat\C$ as the surface $\Sigma_{\rm A}$. Such an embedding is possible if we assume that the model forms of $\mc{P}_j,\mc{P}_{j'}$ satisfy the properties of $\tilde{\mc{MF}}(\tilde{\mc{X}}_\ell)$ above, by a variant of Lemma \ref{FillingSigma}. Then we can choose an $\R$-symmetric marking $\Omega_{\rm A}:\Sigma_{\rm A}^\bullet\to\Sigma_{\rm A}$, holomorphic near $\cup_{k=1}^4\pl\mc{O}_k^\bullet$ and $\xi^{\rm s}(\T)$, such that $\Omega_{\rm A}\circ\xi^{\rm s}=\xi$ and $\Omega_{\rm A}\circ\xi^{\rm t}$ is isotopic to $\xi^*$. Notice that $\Omega_{\rm A}$ is holomorphic near $\hat\xi^{\rm t}(\T)$ by definition. Then the move $(\Xi_1^1,\Xi_2^1)\to(\Xi_1^2,\Xi_2^2)$ (keeping the other markings fixed) on $\Sigma$ matches exactly the definition of the A-move of the first type in the sense of Section \ref{sub:elementary_moves}, up to isotopy. A similar remark applies to the S-move for the next subsection.
\end{remark}

Let $\bs{\alpha}=(\alpha_1,\ldots,\alpha_{m_{\rm A}})\in\R^{m_{\rm A}}$ be weights attached to ${\bf x}_{\rm A}$ and satisfying the Seiberg bounds with respect to the two cuts, and let ${\bf p}'=(p_1,\dots,p_{b_{\rm A}})\in \R_+^{b_{\rm A}}$ be the spectral parameters attached to $\bs{\zeta}_{\rm A}$. For $\ell=1,2$, we define $\bs{\alpha}_\ell=(\alpha_{\ell1},\ldots,\alpha_{\ell m_\ell})$ where $\alpha_{\ell k}=\alpha_j$ if $x_{\ell k}=x_j$, and ${\bf p}_\ell'=(p_{\ell k})_{k\in\{1,\ldots,b_\ell\}\setminus\{k_\ell\}}$ where $p_{\ell k}=p_j$ if $\zeta_{\ell k}=\zeta_j$. We define $\bs{\alpha}_\ell^*,{{\bf p}_\ell^*}'$ analogously. 

Given $\bs{\alpha}$ and ${\bf p}'$ as above, we attach to $\mc{O}_1,\ldots,\mc{O}_4$ the parameters $\bs{\beta}=(\beta_1,\dots,\beta_4)$, and for $j=1,2$ and $k=1,2,3$, we attach the parameter $\beta_{jk}$ to the $k$-th object of $\mc P_j$ (boundary component $\pl_k\mc P_j$ or marked point $\Xi_j(z_k^\bullet)=x_{j(k-b_j)}$): these are defined by
\begin{equation}\label{def_beta_k} 
\beta_{k}:= \begin{cases}
Q + \mathrm{i}p_\ell & \text{if } \mathcal{O}_k = \mathcal{D}_\ell, \\
\alpha_\ell         & \text{if } \mathcal{O}_k = \{x_\ell\},
\end{cases} \qquad 
\beta_{jk}:=
\begin{cases}
Q+{\rm i}p, & k=k_j,\\
Q+{\rm i}p_\ell,&1\leq k\leq b_j, k\not=k_j,\text{ and }\zeta_{jk}=\zeta_\ell,\\
\alpha_\ell,&k'=(k-b_j)>0\text{ and }x_{jk'}=x_\ell,
\end{cases}
\end{equation}
We define $\beta_{jk}^*$ analogously for $\mc P_j^*$ with $p^*$ for $k=k_j^*$.  Thus
$\beta_{1k}$ for $k\not=k_1$ are the weights carried by $\mc O_3,\mc O_4$,
and $\beta_{2k}$ for $k\not=k_2$ are those carried by $\mc O_1,\mc O_2$; in the second channel
the corresponding pairs carry respectively $\mc O_4,\mc O_1$ and
$\mc O_2,\mc O_3$. 

For an internal channel parameter $p\in \R_+$ attached to $\xi$, we define the \textbf{s-channel conformal block} of $\tilde{S}_{\rm A}$ to be
\[ \mc{B}^1_{\bs{\beta}}(p):={\rm Tr}_{k_1k_2}(\mc{B}_{\tilde{\mc{X}}_1,\bs{\alpha}_1}(p,{\bf p}_1')\otimes\mc{B}_{\tilde{\mc{X}}_2,\bs{\alpha}_2}(p,{\bf p}_2')),\]
where ${\rm Tr}_{k_1k_2}$ denotes tracing on the $k_1$-th tensor component of 
$\mc{B}_{\tilde{\mc{X}}_1,\bs{\alpha}_1}(p,{\bf p}_1')$ with the $k_2$-th tensor component of 
$\mc{B}_{\tilde{\mc{X}}_2,\bs{\alpha}_2}(p,{\bf p}_2')$ in the Young diagrams $\nu\in \mc{T}$. Similarly, for an internal channel parameter $p^*\in \R_+$ attached to $\xi^*$, we define the \textbf{t-channel conformal block} of $\tilde{S}_{\rm A}$ to be
\[ \mc{B}^2_{\bs{\beta}}(p^*):={\rm Tr}_{k_1^*k_2^*}(\mc{B}_{\tilde{\mc{X}}_1^*,\bs{\alpha}_1^*}(p^*,{{\bf p}_1^*}')\otimes\mc{B}_{\tilde{\mc{X}}_2^*,\bs{\alpha}_2^*}(p^*,{{\bf p}_2^*}')).\]
We shall describe how these two conformal blocks are related.

Applying \cite[Corollary 11.3]{GRW2} with the parameters $(\mu_1^{\rm b},\dots,\mu_4^{\rm b})=(0,\mu,0,1)$ for any fixed $\mu>0$, we deduce that there exists a constant $h>0$ independent of $\mu$ and ${\bf p}'$ (but depending on  $\bs{\alpha}$) such that 
\begin{equation}\label{formule_bound_boot}
\begin{split}
\int_{\R}H_{0,\mu,0}^{\beta_{1},\beta_{2},Q-{\rm i}p}H_{0,1,0}^{\beta_{3},\beta_{4},Q+{\rm i}p}\tilde{\mc F}^{\mc G}_{p}(\boldsymbol\beta,\bs\nu)\dd p 
=h\int_{\R}H_{1,0,\mu}^{\beta_{4},\beta_{1},Q+{\rm i}p^*}H_{\mu,0,1}^{\beta_{2},\beta_{3},Q-{\rm i}p^*}
\tilde{\mc F}^{\mc G^*}_{p^*}(\boldsymbol\beta,\bs\nu)\dd p^*
\end{split}
\end{equation}
where $H_{\mu_1,\mu_2,\mu_3}^{a_1,a_2,a_3}$ is the structure constant for the boundary Liouville theory, $\tilde{\mc F}^{\mc G}_{p}(\boldsymbol\beta,\bs\nu)$ and $\tilde{\mc F}^{\mc G^*}_{p}(\boldsymbol\beta,\bs\nu)$ are ``normalised conformal blocks''  associated to  the gluing graphs $\mc G$ and $\mc G^*$ relative with
the cuts $\mc C$ and $\mc C^*$.
More precisely, $H_{\mu_1,\mu_2,\mu_3}^{a_1,a_2,a_3}$ is the correlation function of the disk with three particular boundary points and weights $a_1,a_2,a_3$ with boundary cosmological constants $\mu_1,\mu_2,\mu_3$. Its special-function formula was proposed by Ponsot-Teschner (see \cite{PT_boundary_3pt} and proved by \cite{ang2024derivationstructureconstantsboundary}).  The normalised conformal blocks $\tilde{\mc F}^{\mc G}_{p}(\boldsymbol\beta, \bs{\nu})$ are normalised by the condition that (with $(\eps_1,\eps_2,\eps_3)=(-1,1,-1)$)
$\tilde{\mc F}^{\mc G}_{p}(\boldsymbol\beta, \emptyset)$ and $\tilde{\mc F}^{\mc{G}^*}_{p}(\boldsymbol\beta,\emptyset)$ are both a composition of two normalised block amplitudes associated to the two complex building blocks, where the normalised block amplitude of a complex building block is defined by the Ward identity  and the condition that it is equal to $1$ on the empty Young diagram, i.e. its actions on highest weight vectors is $1$. These normalised block amplitudes are functions on the moduli space $\mc{M}_{0,b_j,m_j}$ of building blocks. In particular
the ordering of the boundary parametrisation of the two complex building blocks does not play any role in the values of $\tilde{\mc F}^{\mc G}_{p}(\boldsymbol\beta, \bs{\nu})$ and $\tilde{\mc F}^{\mc{G}^*}_{p}(\boldsymbol\beta,\bs{\nu})$.
The boundary bootstrap relation \eqref{formule_bound_boot} is then always the same whatever the values of $k_1,k_2,k_1^*,k_2^*$.
Using \eqref{Gammamodelform}, they are related to the conformal blocks  $\mc{B}^1_{\bs\beta}(p,\bs\nu)$ and $\mc{B}^2_{\bs\beta}(p^*,\bs\nu)$ by the expressions
\[
\begin{split}
\mc{B}^1_{\bs\beta}(p,\bs\nu)
={}&C^1_{\bs{\alpha}}(\bs{\la})e^{-{\rm i}\pi s(p,\bs\beta)}
\prod_{j=1}^2
\Big(2^{\sum_{k=1}^3\eps_k\Delta_{\beta_{jk}}}
\prod_{k=1}^{b_j}|\psi_{jk}'(0)|^{\Delta_{\beta_{jk}}}\Big)
\tilde{\mc F}^{\mc G}_p(\boldsymbol\beta,\bs\nu),\\
\mc B^2_{\bs\beta}(p^*,\bs\nu)
={}&C^2_{\bs{\alpha}}(\bs{\la}^*)e^{-{\rm i}\pi s(p^*,\bs\beta)}
\prod_{j=1}^2
\Big(2^{\sum_{k=1}^3\eps_k\Delta_{\beta_{jk}^*}}
\prod_{k=1}^{b_j^*}|(\psi_{jk}^*)'(0)|^{\Delta_{\beta^*_{jk}}}\Big)
\tilde{\mc F}^{\mc G^*}_{p^*}(\boldsymbol\beta,\bs\nu),
\end{split}
\]
where $C^1_{\bs{\alpha}}(\bs{\la})$ (resp. $C^2_{\bs{\alpha}}(\bs{\la}^*)$) are constants depending only on $\bs{\alpha}$ and on 
$\bs{\la}=(\bs{\la}_1,\bs{\la}_2)$ (resp. $\bs{\la}^*=(\bs{\la}^*_1,\bs{\la}^*_2)$), and $s(p,\bs{\beta})=\Delta_{Q+{\rm i}p}+\sum_{k=1}^4\Delta_{\beta_k}$.
Indeed, the two factors $2^{(\eps_{k_1}+\eps_{k_2})\Delta_{Q+{\rm i}p}}$ coming from
\eqref{Gammaterm_initial_value} combine with
$|\psi_{1k_1}'(0)\psi_{2k_2}'(0)|^{\Delta_{Q+{\rm i}p}}$ to give the factor
$|2^{\eps_{k_1}+\eps_{k_2}}\psi_{1k_1}'(0)\psi_{2k_2}'(0)|^{\Delta_{Q+{\rm i}p}}$.
Furthermore, the product of the two phases in
\eqref{Gammaterm_initial_value} is
$e^{-{\rm i}\pi(2\Delta_{Q+{\rm i}p}+\sum_k\Delta_{\beta_k})}$,
whereas $\theta_{2k_2}=\pi$ contributes
$e^{{\rm i}\pi\Delta_{Q+{\rm i}p}}$.  Their product is therefore
$e^{-{\rm i}\pi s(p,\bs\beta)}$.  The same computation applies to the
second channel.

Notice that by taking $(\mu_1,\dots,\mu_4)=(0,1,0,\mu)$ in \eqref{formule_bound_boot} for any fixed $\mu>0$, we obtain 
\begin{equation}\label{formule_bound_boot1'}
\begin{split}
\int_{\R}H_{0,1,0}^{\beta_{1},\beta_{2},Q-{\rm i}p}H_{0,\mu,0}^{\beta_{3},\beta_{4},Q+{\rm i}p}e^{{\rm i}\pi s(p,\bs{\beta})}\mc{B}^1_{\bs{\beta}}(p)\dd p  
&=h\int_{\R_+}H_{\mu,0,1}^{\beta_{4},\beta_{1},Q+{\rm i}p^*}H_{1,0,\mu}^{\beta_{2},\beta_{3},Q-{\rm i}p^*}e^{{\rm i}\pi s(p^*,\bs{\beta})}\mc{B}^2_{\bs{\beta}}(p^*)\dd p^*\\
&= \frac{h}{2}\int_{\R}H_{\mu,0,1}^{\beta_{4},\beta_{1},Q+{\rm i}p^*}H_{1,0,\mu}^{\beta_{2},\beta_{3},Q-{\rm i}p^*}e^{{\rm i}\pi s(p^*,\bs{\beta})}\mc{B}^2_{\bs{\beta}}(p^*)\dd p^*
\end{split}
\end{equation}
Similarly, taking 
$(\mu_1,\dots,\mu_4)=(1,0,\mu,0)$ or $(\mu_1,\dots,\mu_4)=(\mu,0,1,0)$ for any fixed $\mu>0$ in \cite[Corollary 11.3]{GRW2}, we obtain for some $h'>0$ 
\begin{equation}\label{formule_bound_boot2}
\begin{gathered}
\int_{\R_+}H_{1,0,\mu}^{\beta_{1},\beta_{2},Q-{\rm i}p}H_{\mu,0,1}^{\beta_{3},\beta_{4},Q+{\rm i}p}e^{{\rm i}\pi s(p,\bs{\beta})}\mc{B}^1_{\bs{\beta}}(p)\dd p  =h'\int_{\R}H_{0,1,0}^{\beta_{4},\beta_{1},Q+{\rm i}p^*}H_{0,\mu,0}^{\beta_{2},\beta_{3},Q-{\rm i}p^*}e^{{\rm i}\pi s(p^*,\bs{\beta})}\mc{B}^2_{\bs{\beta}}(p^*)\dd p^*,\\
\int_{\R_+}H_{\mu,0,1}^{\beta_{1},\beta_{2},Q-{\rm i}p}H_{1,0,\mu}^{\beta_{3},\beta_{4},Q+{\rm i}p}e^{{\rm i}\pi s(p,\bs{\beta})}\mc{B}^1_{\bs{\beta}}(p)\dd p  =h'\int_{\R}H_{0,\mu,0}^{\beta_{4},\beta_{1},Q+{\rm i}p^*}H_{0,1,0}^{\beta_{2},\beta_{3},Q-{\rm i}p^*}e^{{\rm i}\pi s(p^*,\bs{\beta})}\mc{B}^2_{\bs{\beta}}(p^*)\dd p^*.
\end{gathered}
\end{equation}
Here and below, we remind the reader that $\bs{\beta}$ is a function of $\bs{\alpha}$ and ${\bf p}'$.
\begin{lemma}[\cite{GRSSbloc}]\label{GRSSW}
Assume that, for $\bs{\beta}=(\beta_1,\dots,\beta_4)$ defined by \eqref{def_beta_k}, we have
\begin{equation}\label{Seiberg_beta}
{\rm Re}(\beta_1+\beta_4)>Q, \quad {\rm Re}(\beta_2+\beta_3)>Q, \quad {\rm Re}(\beta_1+\beta_2)>Q,\quad 
{\rm Re}(\beta_3+\beta_4)>Q.
\end{equation}
With the same notations and conventions as above, there exists a kernel $K^{{\rm A}}_{\bs{\beta}}(p,p^*)$ which is continuous with respect to $(p,p^*,{\bf p}')$, given for $(p,p^*)\in \R^2$ by 
\begin{equation}\label{A_transform_2to1}
K^{{\rm A}}_{\bs{\beta}}(p,p^*):=\frac{1}{2\pi  C_1^1(p,\bs{\beta})}\int_{\R}e^{-{\rm i}px}e^{(-Q+\beta_1+\beta_2)x}C_{e^{\gamma x}}^2(p^*,\bs{\beta})\dd x,
\end{equation}
where  
\[C_\mu^1(p,\bs{\beta}):=e^{{\rm i}\pi s(p,\bs{\beta})}H_{0,\mu,0}^{\beta_{1},\beta_{2},Q-{\rm i}p}H_{0,1,0}^{\beta_{3},\beta_{4},Q+{\rm i}p}, \quad  
C_\mu^2(p^*,\bs{\beta}):=e^{{\rm i}\pi s(p^*,\bs{\beta})}H_{\mu,0,1}^{\beta_{2},\beta_{3},Q-{\rm i}p^*}H_{1,0,\mu}^{\beta_{4},\beta_{1},Q+{\rm i}p^*},\]
and satisfying the following. Let $K^{{\rm A}}_{\bs{\beta}}:C_c^\infty(\R_+)\to C^0(\R_+)$ 
be the operator $(K^{{\rm A}}_{\bs{\beta}}\varphi)(p):=\int_{\R_+}K^{{\rm A}}_{\bs{\beta}}(p,p^*)\varphi(p^*)\dd p^*$. 
Then there is a constant $h>0$ independent of ${\bf p}'$ such that, in the weak sense,
\begin{equation}\label{K2to1}
\mc{B}^1_{\bs{\beta}}=hK^{\rm A}_{\bs{\beta}}\mc{B}_{\bs{\beta}}^2.
\end{equation}
Moreover, $K^{{\rm A}}_{\bs{\beta}}(p,p^*)$ is invariant under $p\to -p$ and under $p^*\to -p^*$ and we also have 
\begin{equation}\label{formuleKbetabis}
K^{\rm A}_{\bs{\beta}}(p,p^*)=K^{\rm A}_{\beta_3,\beta_4,\beta_1,\beta_2}(p,p^*).
\end{equation}
Conversely, if $\bs{\beta}':=(\beta_2,\beta_3,\beta_4,\beta_1)$, there is a constant $h'>0$ independent of ${\bf p}'$ such that
\begin{equation}\label{K1to2}
\mc{B}^2_{\bs{\beta}}=h'K^{\rm A}_{\bs{\beta'}}\mc{B}_{\bs{\beta}}^1,
\end{equation}
and  $K^{\rm A}_{\beta_2,\beta_3,\beta_4,\beta_1}(p,p^*)=
K^{\rm A}_{\beta_4,\beta_1,\beta_2,\beta_3}(p,p^*)$.
\end{lemma}
\begin{proof} We recall (and simplify slightly) the argument in \cite{GRSSbloc} for completeness.
By \cite{GRSSbloc}, we have 
\[ \begin{gathered}
H_{0,\mu,0}^{\beta_{1},\beta_{2},Q-{\rm i}p}=\mu^{\frac{Q-\beta_{1}-\beta_{2}+{\rm i}p}{\gamma}}H_{0,1,0}^{\beta_{1},\beta_{2},Q-{\rm i}p}, \quad 
|H_{\mu,0,1}^{\beta_{2},\beta_{3},Q-{\rm i}p}|+|H_{1,0,\mu}^{\beta_{4},\beta_{1},Q+{\rm i}p}|=\mc{O}(1) \textrm{ as }\mu\to 0,\\
H_{\mu,0,1}^{\beta_{2},\beta_{3},Q-{\rm i}p}=\mc{O}(\mu^{\frac{Q-{\rm Re}(\beta_{2}+\beta_{3})}{\gamma}}) \textrm{ as }\mu\to \infty, \quad H_{1,0,\mu}^{\beta_{4},\beta_{1},Q+{\rm i}p}=\mc{O}(\mu^{\frac{Q-{\rm Re}(\beta_{4}+\beta_{1})}{\gamma}}) \textrm{ as }\mu\to \infty.
\end{gathered} \]
Combining this with \eqref{formule_bound_boot}, we obtain 
\begin{equation}\label{identity_bound_bootstrap_fourier} 
\int_{\R} \mu^{\frac{Q-\beta_1-\beta_2+{\rm i}p}{\gamma}}C_{1}^1(p,\bs{\beta})\mc{B}_{\bs{\beta}}^1(p)\dd p= h\int_{\R_+}C_{\mu}^2(p^*,\bs{\beta})\mc{B}_{\bs{\beta}}^2(p^*)\dd p^*.
\end{equation} 
Set $x:=\gamma^{-1}\log\mu$. After multiplying the identity by $e^{(-Q+\beta_1+\beta_2)x}$, the left-hand side is the Fourier transform in $x$ of an $L^q$ function of $p$ for every $q\in[1,2]$; it is therefore continuous and belongs to $L^2$ as a function of $x$.
Notice that for $p^*,{\bf p}'$ in a compact set, we have uniformly
\[e^{(-Q+\beta_1+\beta_2)x}|C_{e^{\gamma x}}^2(p^*,\bs{\beta})|=\left\{\begin{array}{ll} 
\mc{O}(e^{-({\rm Re}(\beta_3+\beta_4)-Q)x}) & \textrm{ as }x\to +\infty\\
\mc{O}(e^{({\rm Re}(\beta_1+\beta_2)-Q)x}) & \textrm{ as }x\to -\infty
\end{array}\right..\]
In particular this function is $L^1\cap L^2$ in the $x$ variable, locally uniformly in $(p^*,{\bf p}')$.
Consequently, we may invert the Fourier transform in the sense of tempered distributions in \eqref{identity_bound_bootstrap_fourier}.
Let  $g(p):=C_1^1(p,\bs\beta)\mc B^1_{\bs\beta}(p)$  and 
\[R(x):=\int_{\R}e^{(-Q+\beta_1+\beta_2)x}C^2_{e^{\gamma x}}(p^*,\bs\beta)
\mc B^2_{\bs\beta}(p^*)\,\dd p^*.\]
The latter integral is well-defined for every $x$ by \eqref{identity_bound_bootstrap_fourier}. We use the Fourier convention
$\hat{f}(p)=\int_{\R}e^{-{\rm i}px}f(x)\,\dd x$. Since $g\in L^1(\R)$ we have $R\in C^0(\R)\cap L^\infty(\R)\subset\mc{S}'(\R)$.
Applying the Fourier transform in $\mc{S}'(\R)$, we obtain
$g=\frac{h}{2\pi}\widehat R$.
We therefore define $K^{\rm A}_{\bs\beta}\mc B^2_{\bs\beta}$ as a distribution on $\R$ by
\[
\left\langle K^{\rm A}_{\bs\beta}\mc B^2_{\bs\beta},\varphi\right\rangle:=
\frac{1}{2\pi} \big\cjg\widehat R, \frac{\varphi}{C_1^1(\cdot,\bs\beta)}\big\cjd=
\frac{1}{2\pi}\int_{\R}R(x)
\widehat{\Big(\frac{\varphi}{C_1^1(\cdot,\bs\beta)}\Big)}(x)\,\dd x, \qquad \varphi\in C_c^\infty(\R).
\]
This is well-defined since \(R\) is bounded and the last Fourier
transform is a Schwartz function. We thus obtain $\mc B^1_{\bs\beta}=hK^{\rm A}_{\bs\beta}\mc B^2_{\bs\beta}$ in $\mc{D}'(\R)$.
For each fixed \(p^*\), the estimates above allow us to define
\[
K^{\rm A}_{\bs\beta}(p,p^*)
:=
\frac{1}{2\pi C_1^1(p,\bs\beta)}
\int_{\R}e^{-{\rm i}px}e^{(-Q+\beta_1+\beta_2)x}
C^2_{e^{\gamma x}}(p^*,\bs\beta)\,\dd x .
\]
If $u\in C_c^\infty(\R_+)$, local uniformity in $p^*$ permits
Fubini's theorem and the distribution $K^{\rm A}_{\bs\beta}u$ is
represented by
\[(K^{\rm A}_{\bs\beta}u)(p)=\int_{\R_+}K^{\rm A}_{\bs\beta}(p,p^*)u(p^*)\,\dd p^*.
\]
This in particular gives \eqref{A_transform_2to1}.  Applying the same argument to
\eqref{formule_bound_boot1'} and \eqref{formule_bound_boot2} gives,
respectively, \eqref{formuleKbetabis}, \eqref{K1to2}, and the stated
cyclically permuted expressions for the inverse kernel. 
\end{proof}

It is proved in \cite{GRSSbloc} that $K^{\rm A}_{\bs{\beta}}(\cdot,\cdot)$ can be rewritten in terms of the Ponsot-Teschner fusion 
kernel, found in physics by Ponsot-Teschner \cite{PonsotTeschner1999,PonsotTeschner2001}.
We first remove the harmless phases of the local DOZZ densities. Choose a continuous function $s:\R^*\to\mathbb S^1$ such that
$s(p)^2=\mc{S}(Q+{\rm i}p)$, and $s(-p)=s(p)^{-1}=\bbar{s(p)}$,
where $\mc{S}(Q+{\rm i}p)$ is the scattering coefficient (recall \eqref{symmetry_psi_Q+ip}). Using \eqref{Cdozzscattering} and \eqref{dozzconjugue},  for $\boldsymbol\eta=(\eta_1,\eta_2,\eta_3)\in((-\infty,Q)\cup(Q+{\rm i}\R^*))^3$ and 
we see that
\[ C^{\rm DOZZ}(\boldsymbol\eta)\prod_{\substack{1\leq k\leq3\\
                    \eta_k=Q+{\rm i}p_k}}s(p_k)^{-1}\in\R. \]
Using the explicit DOZZ formula, the generalised Seiberg conditions
and the non-vanishing assumption, the sign of this quantity can be
chosen to be positive, as explained below \eqref{dozzconjugue}.

For
\(\bs\beta\in((-\infty,Q)\cup(Q+{\rm i}\R^*))^4\), set
\[
\chi(\bs\beta)
 :=\prod_{\substack{1\leq k\leq4\\
                    \beta_k=Q+{\rm i}p_k'}}
 s(p_k').
\]
The phases corresponding to the internal weights \(Q+{\rm i}p\) and
\(Q-{\rm i}p\) cancel. Hence there exist positive functions
\(w^1_{\bs\beta},w^2_{\bs\beta}:\R_+\to\R_+^*\) such that
\begin{equation}\label{densitywbeta}
\begin{gathered}
\rho^1_{\bs\beta}(p) :=C^{\rm DOZZ}(\beta_1,\beta_2,Q+{\rm i}p)C^{\rm DOZZ}(\beta_3,\beta_4,Q-{\rm i}p) =\chi(\bs\beta)w^1_{\bs\beta}(p),\\
\rho^2_{\bs\beta}(p) :=C^{\rm DOZZ}(\beta_2,\beta_3,Q+{\rm i}p)C^{\rm DOZZ}(\beta_1,\beta_4,Q-{\rm i}p) =\chi(\bs\beta)w^2_{\bs\beta}(p).
\end{gathered}
\end{equation}
The phase $\chi(\bs\beta)$, which depends only on the exterior
weights, is the same in the two channels. In particular,
$w^\ell_{\bs\beta}=\rho^\ell_{\bs\beta}$ for $\ell=1,2$ when all the $\beta_k$ are real.

We prove in Proposition \ref{prop:unitarity} that $K^{\rm A}_{\bs{\beta}}$ is unitary with respect to a particular $L^2$-product for $\bs{\beta}$ satisfying \eqref{Seiberg_beta}. Using the density result of Lemma \ref{orthogonality}, this implies in particular the following relations:
\begin{corollary}\label{relation_fusions_kernels} 
Let $\bs\beta\in((-\infty,Q)\cup(Q+{\rm i}\R^*))^4$ satisfy the generalised Seiberg bounds \eqref{Seiberg_beta}, and let $w^1_{\bs\beta},w^2_{\bs\beta}$ 
be the positive densities defined by \eqref{densitywbeta}. After multiplying $K^{\rm A}_{\bs\beta}$ by a constant phase if necessary, the operator 
\[K^{\rm A}_{\bs\beta}: L^2(\R_+,w^2_{\bs\beta}(p)\dd p) \to L^2(\R_+,w^1_{\bs\beta}(p)\dd p)\] is unitary. 
Moreover, we have \[ K^{\rm A}_{\beta_1,\beta_2,\beta_3,\beta_4} = K^{\rm A}_{\beta_3,\beta_4,\beta_1,\beta_2}, \] and \[ \big(K^{\rm A}_{\bs\beta}\big)^* = \big(K^{\rm A}_{\bs\beta}\big)^{-1} = K^{\rm A}_{\beta_2,\beta_3,\beta_4,\beta_1} = K^{\rm A}_{\beta_4,\beta_1,\beta_2,\beta_3}, \] where the adjoint is taken with respect to the positive weighted Hilbert space structures above.
\end{corollary}

\subsection{Local study of the S-move: fusion kernel on an $\R$-symmetric one-holed torus}\label{sec:local_studyS}
For $T>0$ and $\tau={\rm i}T$, let us consider the torus
\begin{equation}\label{Ntau} 
N(\tau):=  \{z\in \C \,|\, {\rm Im}(z)\in [-T/2,T/2], {\rm Re}(z)\in [0,1]\}/ (\Z+\tau \Z).
\end{equation}
For notational simplicity, we shall use complex numbers $z\in\C$ to represent points in $N(\tau)$. Notice that $z\mapsto \bar{z}$ is well-defined on $N(\tau)$.
For $m_{\rm S}+b_{\rm S}=1$, consider a decorated Riemann surface of type $(1,m_{\rm S},b_{\rm S})$ of the form
\[ \tilde{S}_{\rm S}=(\Sigma_{\rm S},J_\C, x_{\rm S}, \zeta_{\rm S}, Y_{\rm S})\]
where $\Sigma_{\rm S}= N(\tau)$ and $x_{\rm S}=0$ and $Y_{\rm S}=t\pl_z$ for some $t>0$ (with no $\zeta_{\rm S}$) if $m_{\rm S}=1$,  
and $\Sigma_{\rm S}= N(\tau)\setminus D(0,q_1)$ for some $q_1\in (0,\min(1/4,T/2))$ and 
$\zeta_{\rm S}(e^{{\rm i}\theta})=q_1e^{{\rm i}\theta}$ (with no $x_{\rm S},Y_{\rm S}$) if $b_{\rm S}=1$. Denote $\mc{O}=D(0,q_1)$ if $b_{\rm S}=1$ or $\mc{O}=\{x_{\rm S}\}$ if $m_{\rm S}=1$. 

First let $\xi:\T\to \mc{C}:=\xi(\T)\subset N(\tau)\setminus \mc{O}$ be the real analytic curve defined by 
$\xi(e^{{\rm i}\theta})=\theta/(2\pi)+\tau/2$, which cuts $\Sigma_{\rm S}$ and produces a complex building block $\mc{P}$ 
of type $(m_{\rm S},b^+=1,b^-=b_{\rm S}+1)$, which can be represented as 
$(([0,1]\times [-T/2,T/2])\setminus D(0,q_1))/\cjg z\mapsto z+1\cjd$ if $b_{\rm S}=1$, and as
$([0,1]\times [-T/2,T/2])/\cjg z\mapsto z+1\cjd$ with marked point $x_{\rm S}=0$ if $m_{\rm S}=1$,
with boundary parametrisations given by $\bs{\zeta}=(\zeta_1,\zeta_2,\zeta_{3})$ if $b_{\rm S}=1$ or $\bs{\zeta}=(\zeta_1,\zeta_2)$ if $m_{\rm S}=1$, with $\zeta_1(e^{{\rm i}\theta})=\theta/(2\pi)-\tau/2$  mod $1$ being  outgoing and  $\zeta_2(e^{{\rm i}\theta})=\theta/(2\pi)+\tau/2$  mod $1$ being incoming, and $\zeta_3=\zeta_{\rm S}$ being incoming if $b_{\rm S}=1$. 
Denote by $\pi_{\mc{P}}:\mc{P}\to \Sigma_{\rm S}$  the canonical projection of the cut torus. 
Although it is convenient to describe the cut torus using the mixed convention in which $\zeta_1$ is outgoing and $\zeta_2$ is incoming,
the marked complex building block decompositions of Definition \ref{def:cbb_marking} are defined using incoming boundary parametrisations only. We therefore set
$\zeta^{\rm in}_1:=\zeta_1\circ o$ and $\zeta_2^{\rm in}:=\zeta_2$ and $\bs{\zeta}^{\rm in}$ the collection of incoming parametrisations. Thus $\zeta_1^{\rm in}$ and $\zeta_2^{\rm in}$ are both incoming, and the mixed gluing relation $\zeta_1(z)\sim\zeta_2(z)$ becomes $\zeta^{\rm in}_1(z)\sim\zeta_2^{\rm in}(o(z))$
which is precisely the gluing convention used in Definition \ref{def:cbb_marking}. In all model form and conformal block computations below, the first boundary
parametrisation is understood to be $\zeta_1^{\rm in}$. 

Denote by $b_{\rm S}':=b_{\rm S}+2$ the number of boundary components of $\mc{P}$. Notice that  the 
map $z\mapsto \bar{z}$ induces an anti-holomorphic involution $\iota_{\mc{P}}:\mc P\to \mc{P}$ satisfying $\pi_{\mc{P}}\circ\iota_{\mc{P}}=c\circ\pi_{\mc{P}}$, where $c:z\mapsto\bar z$ on $\Sigma_{\rm S}$, 
$\iota_{\mc{P}}\circ \zeta_1=\zeta_2$ and $\iota_{\mc{P}}\circ \zeta_{\rm S}=\zeta_{\rm S}\circ o$ if $b_{\rm S}=1$. 
An explicit non-centered model form of 
$(\mc P, J_\C,x_{\rm S},\bs{\zeta},Y_{\rm S})$ is obtained from the uniformising map
\[
\Phi(z):= {\rm i}\frac{e^{2\pi {\rm i}z}-1}{e^{2\pi {\rm i}z}+1}.
\]

The biholomorphisms associated with the two incoming parametrisations
$\zeta^{\rm in}_1=\zeta_1\circ o$ and $\zeta_2^{\rm in}=\zeta_2$ are
\[
\chi_1(z):={\rm i}\frac{1-e^{-\pi T}z}{1+e^{-\pi T}z},\qquad
\chi_2(z):=-{\rm i}\frac{1-e^{-\pi T}z}{1+e^{-\pi T}z}.
\]
They satisfy $\chi_1(0)={\rm i}$, $\chi_2(0)=-{\rm i}$ and
$\chi_2(z)=\overline{\chi_1(\bar z)}$. If $b_{\rm S}=1$, we also set
$\chi_3(z):=\Phi(q_1z),$
so that $\chi_3(0)=0$ and $\chi_3(z)=\overline{\chi_3(\bar z)}$. The map
$\Phi:\mc P\to\hat\C\setminus\cup_{k=1}^{b'_{\rm S}}\chi_k(\D)$ satisfies
$\Phi\circ\iota_{\mc{P}}=\overline{\Phi}$ and, if $m_{\rm S}=1$, $\Phi(x_{\rm S})=0$.
We compute that  ${\rm arg}(\chi'_1(0))=-\pi/2$, ${\rm arg}(\chi'_2(0))=\pi/2$ and $\Phi'(0)=-\pi$, i.e. $D\Phi(x_{\rm S})Y_{\rm S}=-\pi t\pl_z$.
The uniformising map 
$\Phi_{\rm S}:\mc P\to \hat{\C} \setminus \cup_{k=1}^{b'_{\rm S}}\psi_k(\D)$ that puts $\mc{P}$ into centered model form is 
$\Phi_{\rm S}=M\circ \Phi$ where $M(z)=(z+{\rm i})/({\rm i}-3z)$ is the M\"obius transformation that maps $({\rm i},-{\rm i},0)$ onto $(-1,0,1)$. 
We have  $\Phi_{\rm S}\circ \zeta_1\circ o=\psi_1|_{\T}$, $\Phi_{\rm S}\circ \zeta_{2}=\psi_2|_{\T}$ and, if  $b_{\rm S}=1$, 
$\Phi_{\rm S}\circ \zeta_{\rm S}=\psi_3|_{\T}$, where $\bs{\psi}=(\psi_1,\dots,\psi_{b'_{\rm S}})$ is the
(unmarked) model form, with each
$\psi_k\in\tBiHol_{\eps}(\D)$
for some $\eps>0$, so that
$\psi_k:\D\to\hat{\C}$ is a holomorphic embedding satisfying
$\psi_k(0)=z_k^\bullet$ whose image $\psi_k(\D)$ is a disk in
$\hat{\C}$ which is not necessarily contained in the affine
chart $\C$.

We compute  that 
$(M'({\rm i}),M'(-{\rm i}),M'(0))=(-{\rm i},-{\rm i}/4,-4{\rm i})$, whose arguments are $-\pi/2$ for the three points.

Let $H:\hat{\C}\to \hat{\C}$  be a diffeomorphism such that $H(z^\bullet_k+z/4)=\psi_k(z)$ for $z\in\D$ and $k=1,\dots,b'_{\rm S}$. If $m_{\rm S}=1$, we also require that
$H(z_3^\bullet+z)=z_3^\bullet+4\pi {\rm i}tz$ near $z=0$, where the coefficient of $z$ is determined by $4\pi {\rm i}t\pl_z=D\Phi_{\rm S}(x_{\rm S})Y_{\rm S}$. There is a choice to be made of the isotopy class of $H$ and we make this choice so that $\Lambda(H)$ (recall \eqref{Lambda}) satisfies
\[ \begin{split}
{\rm Im}(\Lambda(H))&=( {\rm arg}(\chi'_1(0))+{\rm arg}(M'({\rm i})),{\rm arg}(\chi'_2(0))+{\rm arg}(M'(-{\rm i})),{\rm arg}(\Phi'(0))+{\rm arg}(M'(0))),\\
&=(-\pi,0,\pi/2).
\end{split}\]
Here we use the lifted arguments ${\rm arg}(\chi'_1(0))=-\pi/2$, ${\rm arg}(\chi'_2(0))=\pi/2$, ${\rm arg}(\Phi'(0))=\pi$ and ${\rm arg}(M'(z))=-\pi/2$ for $z\in\{{\rm i},-{\rm i},0\}$.
In particular, we have $H:\Sigma^\bullet_{m_{\rm S},b'_{\rm S}}\to \Phi_{\rm S}(\mc P)$ and 
$H\circ \zeta^\bullet_k=\psi_k|_\T$ for all $k$, and we set $\Xi:= \Phi_{\rm S}^{-1}\circ H$. The decorated marked complex building block (we shall simply write $\mc{X}$ when $m_{\rm S}=0$)
\[\tilde{\mc X}=[(\Xi,\mc P,J_\C,x_{\rm S},\bs\zeta^{\rm in},Y_{\rm S})]\] 
has the model form 
$(\bs{\psi},\bs{\theta},\la)$ given by
\[
\bs{\psi}=(M\circ \chi_1,\dots,M\circ \chi_{b'_{\rm S}}),\qquad
\begin{cases}
\bs{\theta}=(-\pi,0),\quad \lambda=\log(4\pi t)+{\rm i}\pi/2,&m_{\rm S}=1,\\
\bs{\theta}=(-\pi,0,\pi/2),&b_{\rm S}=1.
\end{cases}
\]

The map $L:z\mapsto e^{\frac{2\pi}{T}z}$ satisfies 
$L(z+{\rm i}T)=L(z)$, $L(z+1)=\gamma_T(L(z))$ where 
$\gamma_T(w):=e^{\frac{2\pi}{T}}w,$
and therefore induces a biholomorphism
\[
N(\tau)\simeq \C^*/\langle\gamma_T\rangle, \qquad [z]\mapsto[L(z)],
\]
where $[\cdot]$ denotes the respective equivalence classes.
The annulus $\{e^{-2\pi/T}\leq |w|\leq1\}$ is a fundamental domain for this quotient; if $b_{\rm S}=1$, the image of $\mc O$ is removed. Moreover, $L(\bar z)=\overline{L(z)}$.
We consider the second cut of the torus along the curve $\mc{C}^*$ parametrised by $\xi^*:\T\to N(\tau)$ given by $\xi^*(e^{{\rm i}\theta})=1/2+{\rm i}T\theta/(2\pi)$, which by $L$ becomes the circle 
$L(\xi^*(e^{{\rm i}\theta}))=e^{\pi/T+{\rm i}\theta}$. Let $\mc P^*$ be the complex building block obtained from cutting $\Sigma_{\rm S}$ along $\mc{C}^*$; it has the same type $(m_{\rm S},1,b_{\rm S}+1)$ as $\mc{P}$. To represent $\mc P^*$ it is convenient to use the fundamental domain $[-1/2,1/2]\times [-T/2,T/2]$ for $N(\tau)$ and write 
$\mc P^*=\big(([-1/2,1/2]\times[-T/2,T/2])\setminus D(0,q_1)\big)/\cjg z\mapsto z+\tau\cjd$ if $b_{\rm S}=1$, and
$\mc P^*=([-1/2,1/2]\times[-T/2,T/2])/\cjg z\mapsto z+\tau\cjd$ with marked point $x_{\rm S}=0$ if $m_{\rm S}=1$.
We use the parametrisations given by $\bs{\zeta}^*=(\zeta^*_1,\zeta^*_2,\zeta^*_{3})$  if $b_{\rm S}=1$ or $\bs{\zeta}^*=(\zeta^*_1,\zeta^*_2)$ if $m_{\rm S}=1$,  with $\zeta^*_1\color{black}(e^{{\rm i}\theta})= 1/2+{\rm i}T\theta/2\pi$ mod $\tau$ being outgoing and $\zeta^*_2(e^{{\rm i}\theta})= -1/2+{\rm i}T\theta/2\pi$ mod $\tau$ being incoming, and $\zeta_3^*=\zeta_{\rm S}$ being incoming if $b_{\rm S}=1$.
Denote by  $\pi_{\mc{P}^*}:\mc P^*\to \Sigma_{\rm S}$ the canonical projection. As before, we set $\zeta_1^{*,{\rm in}}:=\zeta_1^*\circ o$
and  $\zeta_2^{*,\rm in}:=\zeta_2^*$, and $\bs{\zeta}^{*,{\rm in}}$ the collection of incoming parametrisations.
The map $z\mapsto\bar{z}$ induces an anti-holomorphic involution $\iota_{\mc{P}^*}:\mc P^*\to\mc P^*$ satisfying
$\pi_{\mc{P}^*}\circ\iota_{\mc{P}^*}=c\circ \pi_{\mc{P}^*}$ and  $\iota_{\mc{P}^*}\circ\zeta_k^*=\zeta_k^*\circ o$ for $k=1,2,$
and $\iota_{\mc{P}^*}\circ\zeta_{\rm S}=\zeta_{\rm S}\circ o$ if $b_{\rm S}=1$.
The (unmarked) centered model form $(\psi^*_1,\dots,\psi_{b'_{\rm S}}^*)$ of 
$(\mc P^*,J_\C,x_{\rm S},\bs{\zeta}^*,Y_{\rm S})$ is obtained by applying the $\R$-symmetric uniformising map $\Phi_{\rm S}^*$ where 
\[\Phi_{\rm S}^*(z)=\frac{L(z)}{2-L(z)}, \quad \psi^*_1(z)=\frac{1}{2e^{-\frac{\pi}{T}}z-1}, \quad \psi^*_2(z)=\frac{e^{-\frac{\pi}{T}}z}{2-e^{-\frac{\pi}{T}}z},\]
which satisfy $(\psi^*_1)'(0)=-2e^{-\pi/T}<0$ and $(\psi^*_2)'(0)=e^{-\pi/T}/2>0$. If $b_{\rm S}=1$, then
\[
\psi_3^*(z)=\frac{e^{2\pi q_1z/T}}{2-e^{2\pi q_1z/T}}, \qquad
(\psi_3^*)'(0)=\frac{4\pi q_1}{T}>0.
\]
If $m_{\rm S}=1$, then $\Phi_{\rm S}^*(x_{\rm S})=z_3^\bullet=1$ and
$D\Phi_{\rm S}^*(x_{\rm S})Y_{\rm S}=\frac{4\pi t}{T}\pl_z$. 
We choose a diffeomorphism $H^*:\hat{\C}\to\hat{\C}$ 
such that $H^*(z^\bullet_k+z/4)=\psi_k^*(z)$ for $z\in\D$ and $k=1,\dots,b'_{\rm S}$. If $m_{\rm S}=1$, we also require that
$H^*(z_3^\bullet+z)=z_3^\bullet+\frac{4\pi t}{T}z$
near $z=0$. We choose $H^*$ such that $\Lambda(H^*)$ satisfies ${\rm Im}(\Lambda(H^*))=(-\pi,0,0)$.
We then set (we shall simply write $\mc X^*$ when $m_{\rm S}=0$)
\begin{equation}\label{Xi^*Smove}
\Xi^*:=(\Phi_{\rm S}^*)^{-1}\circ H^*.
\end{equation} The decorated marked complex building block
\[\tilde{\mc X}^*=[(\Xi^*,\mc P^*,J_\C,x_{\rm S},\bs\zeta^{*,{\rm in}},Y_{\rm S})]\] 
has centered model form $(\bs\psi^*,\bs\theta^*,\lambda^*)$ where
\[
\begin{cases}
\bs\theta^*=(-\pi,0),\quad \lambda^*=\log(4\pi t/T),&m_{\rm S}=1,\\
\bs\theta^*=(-\pi,0,0),&b_{\rm S}=1.
\end{cases}
\]
In both decompositions, the model maps $\psi_k$ and $\psi_k^*$ are
elements of $\tBiHol_\eps(\D)$ and are therefore understood as
holomorphic embeddings into $\hat{\C}$. Their expressions in the
affine chart $\C$ may have poles; their images are disks in
$\hat{\C}$ and need not be contained in $\C$.
For later use, we record explicitly the derivatives of the centered
model maps.
Since $\chi_1'(0)=-2{\rm i}e^{-\pi T}$, $\chi_2'(0)=2{\rm i}e^{-\pi T}$,  $M'({\rm i})=-{\rm i}$, $M'(-{\rm i})=-\frac{{\rm i}}4$, $M'(0)=-4{\rm i}$,
we obtain
\[\psi_1'(0)=-2e^{-\pi T},\qquad \psi_2'(0)=\frac{e^{-\pi T}}{2}, \quad\textrm{and if }b_{\rm S}=1, \quad \psi_3'(0)=4\pi{\rm i}q_1.\]
For the second decomposition, direct differentiation gives
\[
(\psi_1^*)'(0)=-2e^{-\pi/T},\qquad (\psi_2^*)'(0)=\frac{e^{-\pi/T}}{2}, \quad\textrm{and if }b_{\rm S}=1,\quad (\psi_3^*)'(0)=\frac{4\pi q_1}{T}.
\]
If $m_{\rm S}=1$, the corresponding decoration parameters are
\[\lambda=\log(4\pi t)+\frac{{\rm i}\pi}{2}, \qquad \lambda^*=\log\left(\frac{4\pi t}{T}\right).\]
In particular,
\begin{equation}\label{derivative_products_model_S}
|\psi_1'(0)\psi_2'(0)|=e^{-2\pi T}=q,
\qquad
|(\psi_1^*)'(0)(\psi_2^*)'(0)|
=e^{-2\pi/T}=q^*.
\end{equation}
On the other hand,
\begin{equation}\label{third_leg_scaling_S}
\textrm{ if }b_{\rm S}=1, \quad |(\psi_3^*)'(0)|=T^{-1}|\psi_3'(0)|,\qquad \textrm{ if }m_{\rm S}=1,\quad \lambda^*-\lambda
=
-\log T-\frac{{\rm i}\pi}{2}.
\end{equation}
Thus even if the third geometric object (corresponding to $\mc{O}$) is unchanged by the S-move, its centered model form coordinate is not.

For an internal spectral parameter $p\in \R_+$ attached to $\xi$, and for $\alpha \in(0,Q)$ in the case $m_{\rm S}=1$, or $p'\in \R_+$ in the case $b_{\rm S}=1$, we attach to $\mc{O}$ the parameter $\beta$ given by
\begin{equation}\label{def_beta_tore} 
\beta:= \begin{cases}
Q + \mathrm{i}p', & \text{if } b_{\rm S}=1, \\
\alpha,          & \text{if } m_{\rm S}=1.
\end{cases} \end{equation}
We define the first channel conformal block of $\tilde{S}_{\rm S}$ (associated to the cut $\mc{C}$ via $\Xi$) to be
\[ \mc{B}^1_{\beta}(p):={\rm Tr}(\mc{B}_{\tilde{\mc X},\alpha}(p,p)) \textrm{ if }m_{\rm S}=1,\qquad 
 \mc{B}^1_\beta(p):={\rm Tr}(\mc{B}_{\mc X}(p,p,p'))  \textrm{ if }b_{\rm S}=1,\]
where ${\rm Tr}$ denotes tracing on the first and second components, i.e.\ 
${\rm Tr}(B):=\sum_{\nu\in\mc{T}}B(\nu,\nu,\ldots)$. 
Similarly, the second channel conformal block of $\tilde{S}_{\rm S}$ (associated to the cut $\mc{C}^*$ via $\Xi^*$) is defined by
\[ \mc{B}^2_{\beta}(p^*):={\rm Tr}(\mc{B}_{\tilde{\mc X}^*,\alpha}(p^*,p^*))  \textrm{ if }m_{\rm S}=1, \qquad 
  \mc{B}^2_\beta(p^*):={\rm Tr}(\mc{B}_{\mc X^*}(p^*,p^*,p'))  \textrm{ if }b_{\rm S}=1.\]
Applying \cite[Corollary 11.3]{GRW2} with the parameters $(\mu_1^{\rm b},\mu_2^{\rm b})=(\mu_1,\mu_2)$ positive (and with the bulk cosmological constant being $0$ in that work), there exists a constant $h>0$ independent of $\mu_1,\mu_2$ and $p'$ (but depending on $\alpha$ if $m_{\rm S}=1$) such that 
\begin{equation}\label{formule_bound_boot_tore}
\begin{split}
\int_{\R_+}\widetilde{\rm G}_{0,\mu_1}(\beta,Q+{\rm i}p)\,
\mu_2\partial_{\mu_2}{\rm U}_{0,\mu_2}(Q-{\rm i}p) e^{{\rm i}\pi s(p,\beta)}\mc{B}^1_\beta (p)\dd p  
=h\int_{\R_+}\mu_2\partial_{\mu_2}
H_{\mu_2,\mu_1,\mu_1}^{Q+{\rm i}p,\beta,Q-{\rm i}p} e^{{\rm i}\pi s'(p,\beta)}\mc{B}^2_\beta (p)\dd p
\end{split}
\end{equation}
where $\widetilde{\rm G}_{0,\mu_1}(\beta,Q+{\rm i}p)$, ${\rm U}_{0,\mu_2}(Q-{\rm i}p)$ are structure constants for the boundary Liouville theory, which enjoy explicit expressions in terms of special functions (\cite{ang2024derivationstructureconstantsboundary}), and
\[ s(p,\beta)= 3\Delta_{Q+{\rm i}p}+\frac{1}{2}\Delta_{\beta}, \quad s'(p,\beta)= 3\Delta_{Q+{\rm i}p}+\Delta_{\beta}.\]
In \eqref{formule_bound_boot_tore}, the integrands are $L^1$ in the $p$ variable.
 We shall not need the explicit expressions of the structure constants, but rather their 
dependence on $\mu_1,\mu_2$.
In \cite[Proposition 11.11]{GRW2}, the identity \eqref{formule_bound_boot_tore} is written in terms of normalised conformal blocks 
$\tilde{\mc{F}}_{p}^{\mc{G}}(Q+{\rm i}p',\nu)$ (or $\tilde{\mc{F}}_{p}^{\mc{G}}(\alpha)$ in the case $m_{\rm S}=1$) 
and $\tilde{\mc{F}}_{p}^{\mc{G}^*}(Q+{\rm i}p',\nu)$ associated to the gluing graphs $\mc{G}$ and $\mc{G}^*$ for the cuts $\mc{C}$ and $\mc{C}^*$. 
By \eqref{Gammaterm_initial_value}, the initial value of the $\Gamma$ factor associated with the three weights
$(Q+{\rm i}p,Q-{\rm i}p,\beta)$ is
\[\Gamma\big((Q+{\rm i}p,Q-{\rm i}p,\beta),{\bf z}^{\bullet}\big)=2^{-\Delta_\beta}e^{-{\rm i}\pi(2\Delta_{Q+{\rm i}p}+\Delta_\beta)}.
\]
Assume first that $b_{\rm S}=1$. With the lifts of the marking parameters used above, namely $(\theta_1,\theta_2,\theta_3)=(-\pi,0,\frac{\pi}{2})$,
we obtain
\begin{equation}\label{S_block_model_boundary_1}
\begin{aligned}
\mc B^1_\beta(p,\bs\nu)={}& 2^{-\Delta_\beta}e^{-{\rm i}\pi(3\Delta_{Q+{\rm i}p}+\Delta_\beta/2)} |\psi_1'(0)\psi_2'(0)|^{\Delta_{Q+{\rm i}p}} |\psi_3'(0)|^{\Delta_\beta} \widetilde{\mc F}^{\mc G}_{p}(\beta,\bs\nu).
\end{aligned}
\end{equation}
For the vertical cut, the corresponding lifted parameters  are $(\theta_1^*,\theta_2^*,\theta_3^*)=(-\pi,0,0)$,
and hence
\begin{equation}\label{S_block_model_boundary_2}
\begin{aligned}
\mc B^2_\beta(p^*,\bs\nu)
={}&2^{-\Delta_{\beta}}e^{-{\rm i}\pi(3\Delta_{Q+{\rm i}p^*}+\Delta_\beta)}|(\psi_1^*)'(0)(\psi_2^*)'(0)|^{\Delta_{Q+{\rm i}p^*}}|(\psi_3^*)'(0)|^{\Delta_\beta} \widetilde{\mc F}^{\mc G^*}_{p^*}(\beta,\bs\nu).
\end{aligned}
\end{equation}
Assume now that $m_{\rm S}=1$. Using the decoration parameters
$\lambda$ and $\lambda^*$ defined above, we obtain
\begin{equation}\label{S_block_model_marked_1}
\begin{aligned}
\mc B^1_\beta(p)={}& 2^{-\Delta_\beta}e^{-{\rm i}\pi(3\Delta_{Q+{\rm i}p}+\Delta_\beta)} |\psi_1'(0)\psi_2'(0)|^{\Delta_{Q+{\rm i}p}}e^{\Delta_\beta \la}
\widetilde{\mc F}^{\mc G}_{p}(\beta),
\end{aligned}
\end{equation}
and
\begin{equation}\label{S_block_model_marked_2}
\begin{aligned}
\mc{B}^2_\beta(p^*) ={}&2^{-\Delta_\beta}e^{-{\rm i}\pi(3\Delta_{Q+{\rm i}p^*}+\Delta_\beta)}
|(\psi_1^*)'(0)(\psi_2^*)'(0)|^{\Delta_{Q+{\rm i}p^*}} e^{\Delta_\beta \lambda^*} \widetilde{\mc F}^{\mc G^*}_{p^*}(\beta).
\end{aligned}
\end{equation}

The existence of a fusion kernel is deduced from \eqref{formule_bound_boot_tore},  as in the proof of the A-move kernel. 
The proof appeared in \cite{GRSSbloc}. We record the precise consequence that will be used below.
\begin{lemma}[\cite{GRSSbloc}]\label{ass:local-S-transform}
Let $\beta\in(0,Q)\cup(Q+{\rm i}\R^*)$. With the same notations and
conventions as above, there exists a kernel
$K^{{\rm S}}_\beta(p,p^*)$ which is continuous with respect to
$(p,p^*)\in\R_+^2$, given by
\begin{equation}\label{K_S_2to1}
K^{{\rm S}}_\beta(p,p^*)
:=
\frac{1}{2\pi D^1_1(p,\beta)}
\widehat D^2_\beta(p,p^*),
\end{equation}
where $p\mapsto \widehat D^2_\beta(p,p^*)$ is the Fourier transform in the sense of tempered distributions of $x\mapsto D^2_x(p^*,\beta)$, with 
\[D^1_1(p,\beta):=\frac{2{\rm i}p}{\gamma}\widetilde{\rm G}_{0,1}(\beta,Q+{\rm i}p){\rm U}_{0,1}(Q-{\rm i}p)e^{{\rm i}\pi s(p,\beta)},
\quad  D^2_x(p^*,\beta):= -\sqrt{2\pi}\frac{2}{\gamma} \partial_x H_{e^{\gamma x/2},1,1}^{Q+{\rm i}p^*,\beta,Q-{\rm i}p^*}e^{{\rm i}\pi s'(p^*,\beta)}.
\]
Let $K^{{\rm S}}_\beta:C_c^\infty(\R_+)\to C^0(\R_+)$ be the operator
$(K^{{\rm S}}_\beta\varphi)(p):=\int_{\R_+}K^{{\rm S}}_\beta(p,p^*)\varphi(p^*)\dd p^*.$
Then there exists a non-zero constant $\kappa$, independent of $p'$ when $b_{\rm S}=1$, such
that, in the weak sense,
\begin{equation}\label{S_transform_2to1}
\mc B^1_\beta=\kappa K^{{\rm S}}_\beta\mc B^2_\beta.
\end{equation}
More precisely, if $h>0$ is the constant occurring in
\eqref{formule_bound_boot_tore}, then
\begin{equation}\label{kappa_S}
\kappa=-\frac{h}{\sqrt{2\pi}}.
\end{equation}
\end{lemma}
\begin{proof}
We use \eqref{formule_bound_boot_tore} with
$\mu_1=1$ and $\mu_2=e^{\gamma x/2}$, for $x\in\R$, together with the
scaling relation
\[
{\rm U}_{0,\mu_2}(Q-{\rm i}p)
=
\mu_2^{2{\rm i}p/\gamma}{\rm U}_{0,1}(Q-{\rm i}p).
\]
Since $\mu_2\partial_{\mu_2}=\frac{2}{\gamma}\partial_x$,  \eqref{formule_bound_boot_tore} becomes
\begin{equation}\label{bootstrap_S_as_distribution}
\int_{\R_+}e^{{\rm i}px}D^1_1(p,\beta) \mc B^1_\beta(p)\dd p
= -\frac{h}{\sqrt{2\pi}}\int_{\R_+}D^2_x(p^*,\beta) \mc B^2_\beta(p^*)\dd p^*.
\end{equation}
The left-hand side is the inverse Fourier transform of an $L^1$
function of $p$ and therefore defines a tempered distribution in $x$.
Consequently, \eqref{bootstrap_S_as_distribution} may be Fourier
transformed in $\mc S'(\R)$.

We now check that the Fourier transform of $D^2_x$ is represented by
a continuous function. By the estimates in the proof of
\cite[Lemma 5.9]{GRSSbloc}, locally uniformly for $p^*$ in compact
subsets of $\R_+$,
\[
\left|
H_{\mu,1,1}^{Q+{\rm i}p^*,\beta,Q-{\rm i}p^*}
\right|
\leq C(p^*)
\begin{cases}
1,&0<\mu\leq1,\\
\mu^{-{\rm Re}(\beta)/\gamma},&\mu\geq1.
\end{cases}
\]
Moreover, by \cite{ang2024derivationstructureconstantsboundary}, the map $\mu\mapsto H_{\mu,1,1}^{Q+{\rm i}p^*,\beta,Q-{\rm i}p^*}$
is $C^1$ at $\mu=0$, locally uniformly for $p^*$ in compact subsets
of $\R_+$. Since $\mu=e^{\gamma x/2}$, this gives
\[
\big|\partial_x H_{e^{\gamma x/2},1,1}^{Q+{\rm i}p^*,\beta,Q-{\rm i}p^*} \big| \leq C(p^*)e^{\gamma x/2}, \qquad x\leq0.
\]
Let $\chi\in C^\infty(\R)$ be equal to $1$ on $(-\infty,0]$ and to $0$ on $[1,\infty)$, and set $H_x(p^*,\beta):= H_{e^{\gamma x/2},1,1}^{Q+{\rm i}p^*,\beta,Q-{\rm i}p^*}$.
We decompose
\[\partial_xH_x = \partial_x(\chi H_x) + \partial_x((1-\chi)H_x) =:F_1(x,p^*)+\partial_xF_2(x,p^*).
\]
The preceding estimates imply that
\[
F_1(\cdot,p^*),F_2(\cdot,p^*)\in L^1(\R),
\]
locally uniformly for $p^*$ in compact subsets of $\R_+$. Therefore,
in the sense of tempered distributions,
\begin{equation}\label{Fourier_D2}
\widehat D^2_\beta(p,p^*)=-\sqrt{2\pi}\frac{2}{\gamma}e^{{\rm i}\pi s'(p^*,\beta)} \left(\int_{\R}e^{-{\rm i}px}F_1(x,p^*)\dd x
+{\rm i}p\int_{\R}e^{-{\rm i}px}F_2(x,p^*)\dd x\right).
\end{equation}
In particular, $\widehat D^2_\beta(p,p^*)$ is continuous in
$(p,p^*)\in\R_+^2$. 
Taking the Fourier transform of
\eqref{bootstrap_S_as_distribution} gives, weakly in the $p$ variable,
\[\mc{B}^1_\beta(p)=-\frac{h}{\sqrt{2\pi}}\int_{\R_+}\frac{\widehat D^2_\beta(p,p^*)}{2\pi D^1_1(p,\beta)}
\mc{B}^2_\beta(p^*)\dd p^*.
\]
Here we use that $D^1_1(p,\beta)\neq0$ for $p>0$, which follows from
the explicit expressions of the boundary structure constants. 
\end{proof}
As for the A-move, the weak identity is sufficient to define the
operator on the span of the torus blocks; its extension to the full
Hilbert space follows from Lemmas \ref{orthogonality_torus} and
\ref{density_torus}.

We first specify the positive measure used below. Set
\[
\rho_\beta(p):=
C^{\rm DOZZ}(\beta,Q+{\rm i}p,Q-{\rm i}p).
\]
If $\beta\in(0,Q)$, the explicit formula \eqref{DOZZformula} shows that
$\rho_\beta(p)>0$ for $p>0$. If $\beta=Q+{\rm i}r$, the conjugation
and reflection relations for the DOZZ constant give (with $\mc{S}(\beta)=\mc{S}(Q+{\rm i}r)$ the scattering coefficient)
\[\overline{\rho_\beta(p)}=
C^{\rm DOZZ}(2Q-\beta,Q+{\rm i}p,Q-{\rm i}p)= \mc S(\beta)^{-1}\rho_\beta(p).
\]
Since $|\mc S(\beta)|=1$, we may choose
$\vartheta_\beta\in\R$ such that $e^{2{\rm i}\vartheta_\beta}=\mc S(\beta)$.
It follows that $e^{-{\rm i}\vartheta_\beta}\rho_\beta(p)\in\R.$
Moreover, $\rho_\beta$ is continuous and does not vanish on $\R_+$.
Its sign is therefore constant, and we choose the square root
$e^{{\rm i}\vartheta_\beta}$ so that
\begin{equation}\label{positive_torus_density}
w_\beta(p):=
e^{-{\rm i}\vartheta_\beta}\rho_\beta(p)>0,
\qquad p>0.
\end{equation}
For $\beta\in(0,Q)$, we take $\vartheta_\beta=0$, so that
$w_\beta=\rho_\beta$.
Using Proposition \ref{prop:unitarity_S}, we deduce the following. 
\begin{corollary}\label{cor:inverse_S_kernel}
For $\beta\in(0,Q)\cup(Q+{\rm i}\R^*)$, the operator
\[K^{\rm S}_\beta: L^2(\R_+,w_\beta(p^*)\dd p^*) \to L^2(\R_+,w_\beta(p)\dd p)
\]
is unitary. Its inverse $(K^{\rm S}_\beta)^{-1}=(K^{\rm S}_\beta)^*$ has, with respect to the Lebesgue measure, the kernel
\begin{equation}\label{inverse_S_kernel}
(K^{\rm S}_\beta)^{-1}(p^*,p)= \frac{w_\beta(p)}{w_\beta(p^*)} \overline{K^{\rm S}_\beta(p,p^*)}.
\end{equation}
Moreover, if $\kappa$ is the constant in \eqref{S_transform_2to1}, then
\begin{equation}\label{S_transform_1to2}
\mc B^2_\beta =\kappa^{-1}(K^{\rm S}_\beta)^{-1} \mc B^1_\beta .
\end{equation}
\end{corollary}

\subsection{Transition operators for conformal blocks}\label{sec:transition_operator}
We return to the general case of conformal blocks on $\mc{T}_{g,m}$. In this section, we describe the transition operator mapping the conformal block associated to one decorated marked complex building block decomposition to the conformal block associated to another non-isotopic decomposition.
Notice that since $\mc{T}_{g,m}$ is a contractible Stein manifold, the line bundle ${\rm Hom}(\mc{L}^{[\bs{\Xi}_1^\bullet]}(\bs{\alpha}),\mc{L}^{[\bs{\Xi}_2^\bullet]}(\bs{\alpha}))$ is always holomorphically trivialisable by the Oka-Grauert principle, so it admits a nowhere vanishing holomorphic section. 
\begin{theorem}[\textbf{Projective unitary representation of mapping class group}]
\label{th:kernelMS}$ $\\
1) Let $[\bs{\Xi}_1^\bullet]\in \pi_0\tilde{\mc{BD}}_{g,m}^{\mcG_1}$ and $[\bs{\Xi}_2^\bullet]\in\pi_0\tilde{\mc{BD}}_{g,m}^{\mcG_2}$ be two isotopy classes with possibly different gluing graphs $\mcG_1,\mcG_2$, and let $\bs{\alpha}$ be weights satisfying the Seiberg bounds with respect to both $\mcG_1$ and $\mcG_2$. Then there exists a unitary operator 
\[K_{\bs{\alpha}}^{[\bs{\Xi}_1^\bullet],[\bs{\Xi}_2^\bullet]}:\mc{H}_{\bs{\alpha}}^{\mcG_1}\to\mc{H}_{\bs{\alpha}}^{\mcG_2}\]
and a nowhere vanishing holomorphic section $s$ of ${\rm Hom}(\mc{L}^{[\bs{\Xi}_1^\bullet]}(\bs{\alpha}),\mc{L}^{[\bs{\Xi}_2^\bullet]}(\bs{\alpha}))
\simeq\mc{L}^{[\bs{\Xi}_2^\bullet]}(\bs{\alpha})\otimes(\mc{L}^{[\bs{\Xi}_1^\bullet]}(\bs{\alpha}))^*$ such that for all $\mc{X}\in \mc{T}_{g,m}$,
\[\mc{F}_{\bs{\alpha}}^{[\bs{\Xi}_2^\bullet]}(\mc{X})=(K_{\bs{\alpha}}^{[\bs{\Xi}_1^\bullet],[\bs{\Xi}_2^\bullet]}\otimes s(\mc{X}))(\mc{F}_{\bs{\alpha}}^{[\bs{\Xi}_1^\bullet]}(\mc{X})).\]
Moreover, such an operator $K_{\bs{\alpha}}^{[\bs{\Xi}_1^\bullet],[\bs{\Xi}_2^\bullet]}$ is unique up to multiplication by a complex constant of modulus $1$.\\
2) Fix a gluing graph $\mc{G}$ and an isotopy class $[\bs{\Xi}^\bullet]\in \pi_0\tilde{\mc{BD}}_{g,m}^{\mcG}$, and let $\bs{\alpha}$ be weights satisfying the Seiberg bounds with respect to $\mc{G}$. Then the map 
\[  U^{\mc{G}}: [h]\in \tilde{{\rm Mod}}_{g,m} \mapsto K_{\bs{\alpha}}^{[\bs{\Xi}^\bullet],[h^{-1}]\cdot[\bs{\Xi}^\bullet]} \in \mc{L}(\mc{H}_{\bs{\alpha}}^{\mc{G}}) \]
is a projective unitary representation of the mapping class group $\tilde{{\rm Mod}}_{g,m}$ on the Hilbert space $\mc{H}_{\bs{\alpha}}^{\mc{G}}$, which descends to a projective unitary representation of ${\rm Mod}_{g,m}$ on $\mc{H}_{\bs{\alpha}}^{\mc{G}}$.
\end{theorem}

\begin{proof}
We shall prove the theorem assuming the existence of the operator $K_{\bs{\alpha}}^{[\bs{\Xi}_1^\bullet],[\bs{\Xi}_2^\bullet]}$ in 1); the rest of the section will be devoted to proving the existence.

First we prove the uniqueness in 1). Let $(K_1,s_1)$ and $(K_2,s_2)$ be two such pairs, so that for all $\mc{X}\in \mc{T}_{g,m}$,
\[\mc{F}_{\bs{\alpha}}^{[\bs{\Xi}_2^\bullet]}(\mc{X})=(K_1\otimes s_1(\mc{X}))(\mc{F}_{\bs{\alpha}}^{[\bs{\Xi}_1^\bullet]}(\mc{X}))=(K_2\otimes s_2(\mc{X}))(\mc{F}_{\bs{\alpha}}^{[\bs{\Xi}_1^\bullet]}(\mc{X})).\]
Since $s_1,s_2$ are nowhere vanishing sections of the same line bundle, we have $s_2=h s_1$ for some nowhere vanishing holomorphic function $h$ on $\mc{T}_{g,m}$. Thus 
\[K_1((s_1\mc{F}_{\bs{\alpha}}^{[\bs{\Xi}_1^\bullet]})(\mc{X}))=h(\mc{X})K_2((s_1\mc{F}_{\bs{\alpha}}^{[\bs{\Xi}_1^\bullet]})(\mc{X}))\in\mc{H}^{\mc{G}_2}_{\bs{\alpha}}\otimes\mc{L}^{[\bs{\Xi}_2^\bullet]}_{\mc{X}}(\bs{\alpha}).\]
Since $K_1,K_2$ are unitary, taking the norm in $\mc{H}^{\mc{G}_2}_{\bs{\alpha}}$ shows $|h(\mc{X})|\equiv1$, so $h$ is necessarily a constant of modulus $1$ by the open mapping theorem. Now by Proposition \ref{prop:density}, the span 
\[{\rm span}\big\{(s_1\mc{F}_{\bs{\alpha}}^{[\bs{\Xi}_1^\bullet]})(\mc{X})/s_{\mc{X}}\,|\,\mc{X}\in\mc{T}_{g,m}\big\}\] 
(where $s_{\mc{X}}$ is any non-zero element in $\mc{L}_{\mc{X}}^{[\bs{\Xi}_2^\bullet]}(\bs{\alpha})$) is dense in $\mc{H}^{\mc{G}_1}_{\bs{\alpha}}$, so $K_1=hK_2$ on $\mc{H}^{\mc{G}_1}_{\bs{\alpha}}$.

Next we prove 2), assuming 1). 
For $[h]\in\tilde{\rm Mod}_{g,m}$, set $\rho^{\mc{G}}_{[h]}:=K_{\bs\alpha}^{[\bs{\Xi}^\bullet],[h]\cdot [\bs{\Xi}^\bullet]}$.
By Lemma \ref{lem:action_on_Xi_action_on_X} and 1), there is a
nowhere vanishing holomorphic section $s_{[h]}\in{\rm Hom}(\mc{L}^{[\bs{\Xi}^\bullet]}(\bs\alpha),
[h]^*\mc{L}^{[\bs{\Xi}^\bullet]}(\bs\alpha))$
such that for all $\mc{X}\in \mc{T}_{g,m}$,
\[\mc{F}_{\bs\alpha}^{[\bs{\Xi}^\bullet]}(\mc{X}\cdot[h])
=(\rho^{\mc{G}}_{[h]}\otimes s_{[h]}(\mc{X}))
\mc{F}_{\bs\alpha}^{[\bs{\Xi}^\bullet]}(\mc X).
\]
Applying this identity successively to $[h_1]$ and $[h_2]$ gives
\[
\mc{F}_{\bs\alpha}^{[\bs{\Xi}^\bullet]}(\mc{X}\cdot[h_1]\cdot[h_2])
=\left(\rho^{\mc{G}}_{[h_2]}\rho^{\mc{G}}_{[h_1]}\otimes
([h_1]^*s_{[h_2]})s_{[h_1]}
\right)
\mc F_{\bs\alpha}^{[\bs{\Xi}^\bullet]}(\mc X).
\]
Using the uniqueness part of 1) for $[h_1h_2]$, we obtain
\[\rho^{\mc{G}}_{[h_1h_2]}=c_{[h_1],[h_2]}\rho^{\mc{G}}_{[h_2]}\rho^{\mc{G}}_{[h_1]}\]
for some complex constant $c_{[h_1],[h_2]}$ of modulus $1$.
Thus $[h]\in\tilde{\rm Mod}_{g,m}\mapsto U^{\mc{G}}([h])=\rho^{\mc{G}}_{[h^{-1}]}$ is a homomorphism into $U(\mc H_{\bs\alpha}^{\mc G})/U(1)$ where $U(\mc H_{\bs\alpha}^{\mc G})\subset\mc{L}(\mc H_{\bs\alpha}^{\mc G})$ is the group of unitary operators.
Now the kernel of the projection $\tilde{\rm Mod}_{g,m}\to{\rm Mod}_{g,m}$ (recall Section \ref{sub:Teichmuller}) is generated by Dehn twists around the marked points. It will follow from the discussion below that the transition operator of such a Dehn twist is a constant multiple of ${\rm Id}$ (see Lemma \ref{lem:T-move_bloc} and below). Hence this homomorphism
descends to ${\rm Mod}_{g,m}$.
\end{proof}




To prove the existence part of 1) of Theorem \ref{th:kernelMS}, we begin with the following two observations.

First, given $[\bs{\Xi}_\ell^\bullet]\in \pi_0\tilde{\mc{BD}}_{g,m}^{\mcG_\ell}$ for $\ell=1,2,3$, if existence is proved for the pairs $[\bs{\Xi}_1^\bullet],[\bs{\Xi}_2^\bullet]$ and $[\bs{\Xi}_2^\bullet],[\bs{\Xi}_3^\bullet]$ with unitary operators $K_{\bs{\alpha}}^{[\bs{\Xi}_1^\bullet],[\bs{\Xi}_2^\bullet]}:\mc{H}_{\bs{\alpha}}^{\mc{G}_1}\to\mc{H}_{\bs{\alpha}}^{\mc{G}_2}$, $K_{\bs{\alpha}}^{[\bs{\Xi}_2^\bullet],[\bs{\Xi}_3^\bullet]}:\mc{H}_{\bs{\alpha}}^{\mc{G}_2}\to\mc{H}_{\bs{\alpha}}^{\mc{G}_3}$ and holomorphic sections $s_1$ of ${\rm Hom}(\mc{L}^{[\bs{\Xi}_1^\bullet]}(\bs{\alpha}),\mc{L}^{[\bs{\Xi}_2^\bullet]}(\bs{\alpha}))$ and $s_2$ of ${\rm Hom}(\mc{L}^{[\bs{\Xi}_2^\bullet]}(\bs{\alpha}),\mc{L}^{[\bs{\Xi}_3^\bullet]}(\bs{\alpha}))$, then taking $K_{\bs{\alpha}}^{[\bs{\Xi}_1^\bullet],[\bs{\Xi}_3^\bullet]}:=K_{\bs{\alpha}}^{[\bs{\Xi}_2^\bullet],[\bs{\Xi}_3^\bullet]}K_{\bs{\alpha}}^{[\bs{\Xi}_1^\bullet],[\bs{\Xi}_2^\bullet]}$ and $s:=s_2s_1$ shows existence for the pair $[\bs{\Xi}_1^\bullet],[\bs{\Xi}_3^\bullet]$.

Second, for fixed $[\bs{\Xi}_1^\bullet],[\bs{\Xi}_2^\bullet]$, it suffices to prove existence for $\bs{\alpha}$ in a non-empty open set in $(-\infty,Q)^m$. Indeed, for a gluing graph $\mc{G}$, let us denote by $\mc{A}^{\mc{G}}\subset\C^m$ the open set of complex weights $\bs{\alpha}$ satisfying the generalised Seiberg bounds with respect to $\mc{G}$, i.e.\ \eqref{generalisedSB} holds on each complex building block with respect to $\mc{G}$.
For $[\bs{\Xi}^\bullet]\in\pi_0\tilde{\mc{BD}}_{g,m}^{\mc{G}}$, it is clear that $\mc{AL}^{[\bs{\Xi}^\bullet]}:=\bigcup_{\bs{\alpha}\in\mc{A}^\mc{G}}\mc{L}^{[\bs{\Xi}^\bullet]}(\bs{\alpha})$ has the structure of a holomorphic line bundle over the product manifold $\mc{A}^{\mc{G}}\times\mc{T}_{g,m}$, and $(\bs{\alpha},\mc{X})\mapsto\mc{F}_{\bs{\alpha}}^{[\bs{\Xi}^\bullet]}(\mc{X})$ is a section of $\mc{AL}^{[\bs{\Xi}^\bullet]}\otimes L_{\rm loc}^1(\R_+^{i_0})$.
For $\ell=1,2$, since $\mc{A}^{\mc{G}_\ell}$ is convex, by Steinness, we can choose a global nowhere vanishing holomorphic section $s_\ell$ of $\mc{AL}^{[\bs{\Xi}_\ell^\bullet]}$ on $\mc{A}^{\mc{G}_\ell}\times\mc{T}_{g,m}$, so that there exists a map $F_\ell:\mc{A}^{\mc{G}_\ell}\times\mc{T}_{g,m}\to L_{\rm loc}^1(\R_+^{i_0})$ such that
\[\mc{F}_{\bs{\alpha}}^{[\bs{\Xi}_\ell^\bullet]}(\mc{X})=F_\ell(\bs{\alpha},\mc{X})s_\ell(\bs{\alpha},\mc{X}).\]
We have $F_\ell(\bs{\alpha},\mc{X})\in\mc{H}_{\bs{\alpha}}^{\mc{G}_\ell}$ for $\bs{\alpha}\in\mc{A}^{\mc{G}_\ell}\cap(-\infty,Q)^m$, and it is holomorphic in $\mc{X}$ for fixed $\bs{\alpha}$ with values in $\mc{H}_{\bs{\alpha}}^{\mc{G}_\ell}$.
Suppose we have proved existence for $\bs{\alpha}$ in some open set $A\subset\mc{A}^{\mc{G}_1}\cap\mc{A}^{\mc{G}_2}\cap(-\infty,Q)^m$. Then there exist a unitary operator $K_{\bs{\alpha}}:\mc{H}_{\bs{\alpha}}^{\mc{G}_1}\to\mc{H}_{\bs{\alpha}}^{\mc{G}_2}$ for each $\bs{\alpha}\in A$ and a nowhere vanishing function $h:A\times\mc{T}_{g,m}\to\C$ such that for all $\bs{\alpha}\in A$ and $\mc{X}\in\mc{T}_{g,m}$,
\[F_2(\bs{\alpha},\mc{X})=h(\bs{\alpha},\mc{X})K_{\bs{\alpha}}F_1(\bs{\alpha},\mc{X}).\]
Notice that $h(\bs{\alpha},\mc{X})$ is necessarily holomorphic in $\mc{X}$.
Consider the positive function on $(\mc{A}^{\mc{G}_1}\cap\mc{A}^{\mc{G}_2}\cap(-\infty,Q)^m)\times\mc{T}_{g,m}$ given by
\[u(\bs{\alpha},\mc{X}):=\frac{\|F_2(\bs{\alpha},\mc{X})\|_{\mc{H}_{\bs{\alpha}}^{\mc{G}_2}}^2}{\|F_1(\bs{\alpha},\mc{X})\|_{\mc{H}_{\bs{\alpha}}^{\mc{G}_1}}^2}=\frac{\|s_1(\bs{\alpha},\mc{X})\|_{\mc{L}_{\mc{X}}^{[\bs{\Xi}_1^\bullet]}(\bs{\alpha})}^2}{\|s_2(\bs{\alpha},\mc{X})\|_{\mc{L}_{\mc{X}}^{[\bs{\Xi}_2^\bullet]}(\bs{\alpha})}^2}.\]
Here we used the conformal bootstrap formula (Theorem \ref{conf_boot}) to get for $\ell=1,2$
\[\langle\prod_{k=1}^m V_{\alpha_k}(x_k)\rangle_{\Sigma,g_S}=C_0\|\mc{F}_{\bs{\alpha}}^{[\bs{\Xi}_\ell^\bullet]}(\mc{X})\|_{{\rm CB}_{\bs{\alpha}}^{[\bs{\Xi}_\ell^\bullet]}}^2=C_0\|F_\ell(\bs{\alpha},\mc{X})\|_{\mc{H}_{\bs{\alpha}}^{\mc{G}_\ell}}^2\|s_\ell(\bs{\alpha},\mc{X})\|_{\mc{L}_{\mc{X}}^{[\bs{\Xi}_\ell^\bullet]}(\bs{\alpha})}^2,\]
where $S=(\Sigma,J,{\bf x})$ is a representative of $\mc{X}=[(F,S)]$ with a compatible metric $g_S$ used for defining $\|\!\cdot\!\|_{\mc{L}_{\mc{X}}^{[\bs{\Xi}_\ell^\bullet]}(\bs{\alpha})}$. By the definition \eqref{metriconLalpha}, it is clear that $\bs{\alpha}\mapsto u(\bs{\alpha},\cdot)$ is real analytic with values in $C(\mc{T}_{g,m})$.
On $A\times\mc{T}_{g,m}$, we have $u(\bs{\alpha},\mc{X})=|h(\bs{\alpha},\mc{X})|^2$, so $\log u(\bs{\alpha},\cdot)$ is pluriharmonic, i.e.\ $\pl_{\mc{X}}\bar{\pl}_{\mc{X}}\log u(\bs{\alpha},\cdot)=0$ for $\bs{\alpha}\in A$. Now since this is real analytic in $\bs{\alpha}$, we get $\pl_{\mc{X}}\bar{\pl}_{\mc{X}}\log u(\bs{\alpha},\cdot)=0$ (in the distributional sense) for all $\bs{\alpha}\in\mc{A}^{\mc{G}_1}\cap\mc{A}^{\mc{G}_2}\cap(-\infty,Q)^m$ by uniqueness of analytic continuation. Since $\mc{T}_{g,m}$ is contractible, for all $\bs{\alpha}\in\mc{A}^{\mc{G}_1}\cap\mc{A}^{\mc{G}_2}\cap(-\infty,Q)^m$, there exists a holomorphic function $h_{\bs{\alpha}}:\mc{T}_{g,m}\to\C$ such that $u(\bs{\alpha},\cdot)=|h_{\bs{\alpha}}|^2$, i.e.\ $\|F_2(\bs{\alpha},\mc{X})\|_{\mc{H}_{\bs{\alpha}}^{\mc{G}_2}}^2=\|h_{\bs{\alpha}}(\mc{X})F_1(\bs{\alpha},\mc{X})\|_{\mc{H}_{\bs{\alpha}}^{\mc{G}_1}}^2$. For fixed $\bs{\alpha}$, consider the function on $\mc{T}_{g,m}\times\mc{T}_{g,m}$ given by
\[f_{\bs{\alpha}}(\mc{X},\mc{X}'):=\langle F_2(\bs{\alpha},\mc{X}),F_2(\bs{\alpha},\mc{X}')\rangle_{\mc{H}_{\bs{\alpha}}^{\mc{G}_2}}-\langle h_{\bs{\alpha}}(\mc{X})F_1(\bs{\alpha},\mc{X}),h_{\bs{\alpha}}(\mc{X}')F_1(\bs{\alpha},\mc{X}')\rangle_{\mc{H}_{\bs{\alpha}}^{\mc{G}_1}}.\]
Now $f_{\bs{\alpha}}(\mc{X},\mc{X}')$ is holomorphic in $\mc{X}$, anti-holomorphic in $\mc{X}'$, and it vanishes on the diagonal $\{\mc{X}=\mc{X}'\}$. Thus $f_{\bs{\alpha}}\equiv0$, i.e.\ the map $h_{\bs{\alpha}}(\mc{X})F_1(\bs{\alpha},\mc{X})\in\mc{H}_{\bs{\alpha}}^{\mc{G}_1}\mapsto F_2(\mc{X})\in\mc{H}_{\bs{\alpha}}^{\mc{G}_2}$ extends to a unitary operator $K_{\bs{\alpha}}':\mc{H}_{\bs{\alpha}}^{\mc{G}_1}\to\mc{H}_{\bs{\alpha}}^{\mc{G}_2}$ for all $\bs{\alpha}\in\mc{A}^{\mc{G}_1}\cap\mc{A}^{\mc{G}_2}\cap(-\infty,Q)^m$, where we used the density of the conformal blocks (Proposition \ref{prop:density}). To recapitulate, we now have for $\bs{\alpha}\in\mc{A}^{\mc{G}_1}\cap\mc{A}^{\mc{G}_2}\cap(-\infty,Q)^m$ and $\mc{X}\in\mc{T}_{g,m}$
\[F_2(\bs{\alpha},\mc{X})=h_{\bs{\alpha}}(\mc{X})K_{\bs{\alpha}}'F_1(\bs{\alpha},\mc{X}).\]
Then $\mc{F}_{\bs{\alpha}}^{[\bs{\Xi}_2^\bullet]}(\mc{X})=(K_{\bs{\alpha}}'\otimes (h_{\bs{\alpha}}s_2s_1^{-1})(\mc{X}))\mc{F}_{\bs{\alpha}}^{[\bs{\Xi}_1^\bullet]}(\mc{X})$, as desired.

Now we claim that it suffices to prove the case where $\bs{\Xi}_1^\bullet\to\bs{\Xi}_2^\bullet$ is one of the elementary moves described in Section \ref{sub:elementary_moves}, namely the A-, S-, B- and T-moves. Notice that if we take $\bs{\alpha}$ such that each $\alpha_j\in(2Q/3,Q)$, then $\bs{\alpha}$ satisfies the Seiberg bounds with respect to all possible gluing graphs. Therefore, using (the decorated version of) Proposition \ref{prop_elementary_moves} and the first observation above, we have proved the theorem for a general pair $[\bs{\Xi}_1^\bullet],[\bs{\Xi}_2^\bullet]$ and $\bs{\alpha}$ in the open set $(2Q/3,Q)^m\subset \mc{A}^{\mc{G}_1}\cap\mc{A}^{\mc{G}_2}\cap(-\infty,Q)^m$. Thus it holds for all $\bs{\alpha}\in\mc{A}^{\mc{G}_1}\cap\mc{A}^{\mc{G}_2}\cap(-\infty,Q)^m$, by the second observation.

For the rest of the section, we shall exhibit a unitary transition operator for each of the A-, S-, B- and T-moves. \\


\noindent $\bullet$  \textbf{T-move.}
Let $[\bs{\Xi}^\bullet]\in \pi_0\tilde{\mc{BD}}_{g,m}^{\mc{G}}$ with gluing graph $\mc{G}$ and a representative $\bs{\Xi}^\bullet\in\tilde{\mc{BD}}(\tilde{S}^\bullet)$. Consider a T-move given by a Dehn twist $T=T_{k,2\pi}$ around the $k$-th boundary component $\pl_k \mc{P}^\bullet_j$ of the $j$-th canonical complex building block $\mc{P}_j^\bullet=\Sigma^\bullet_{m_j,b_j}$. For any decorated marked complex building block decomposition $\bs{\Xi}$ of a decorated surface $\tilde{S}$, we denote for $\ell\in\Z$ the decomposition $\bs{\Xi}\cdot T^\ell:=(\Xi_1,\dots,\Xi_j\circ T^\ell,\dots,\Xi_{j_0})$ of $\tilde{S}$. In particular, we have $[\bs{\Xi}^\bullet\cdot T]\in \pi_0\tilde{\mc{BD}}_{g,m}$ and $[\bs{\Xi}^\bullet]\to[\bs{\Xi}^\bullet\cdot T]$ is the T-move in question. Notice that this does not change the gluing graph, nor the Hilbert space $\mc{H}^{\mc{G}}_{\bs{\alpha}}$.
There is a natural bundle isomorphism
\[\mc{I}_{T} : \mc{L}^{[\bs{\Xi}^\bullet]}\to \mc{L}^{[\bs{\Xi}^\bullet\cdot T]},\quad\mc{I}_{T}(G)(\tilde{\mc{X}}_{\bs{\Xi}}):=G(\tilde{\mc{X}}_{\bs{\Xi}\cdot T^{-1}}).\]
Using Proposition \ref{identification_Teichmuller}, it is easy to see that the bijection $\tilde{\mc{X}}_{\bs{\Xi}}\mapsto\tilde{\mc{X}}_{\bs{\Xi}\cdot T}$ of $\BDTT_{g,m}$ commutes with plumbing in the sense that if we write $(\tilde{\mc{X}}_{\bs{\Xi}})(\bs{\tau})=\tilde{\mc{X}}(\bs{\tau})_{\bs{\Xi}(\bs{\tau})}$ with $\bs{\Xi}(\bs{\tau})$ the associated decomposition of $\tilde{\mc{X}}(\bs{\tau})=\pi_{\mc{D\tilde{T}}\to\tT}((\tilde{\mc{X}}_{\bs{\Xi}})(\bs{\tau}))$, then $(\tilde{\mc{X}}_{\bs{\Xi}\cdot T})(\bs{\tau})=\tilde{\mc{X}}(\bs{\tau})_{\bs{\Xi}(\bs{\tau})\cdot T}$. Thus $\mc{I}_{T}$ is holomorphic.
By Lemma \ref{Dehn_Twist_change}, we obtain:
\begin{lemma}[\textbf{Dehn twist kernel}]\label{lem:T-move_bloc}
Using the notations above, we have for all $\mc{X}\in \mc{T}_{g,m}$
\[\mc{F}_{\bs{\alpha}}^{[\bs{\Xi}^\bullet\cdot T]}(\mc{X},{\bf p})=e^{2\pi {\rm i}\Delta_{Q+{\rm i}p_\ell}}\mc{I}_T\mc{F}_{\bs{\alpha}}^{[\bs{\Xi}^\bullet]}(\mc{X},{\bf p}),\]
where ${\rm e}_\ell=\{(j,k),(j',k')\}$ is the internal edge containing $(j,k)$ in the gluing graph $\mc{G}$ of $[\bs{\Xi}^\bullet]$.
\end{lemma}
In particular, the transition operator for the T-move is 
\[K_{\bs{\alpha}}^{[\bs{\Xi}^\bullet],[\bs{\Xi}^\bullet\cdot T]}{\bf u}({\bf p})=e^{2\pi {\rm i}\Delta_{Q+{\rm i}p_\ell}}{\bf u}({\bf p}),\]
which is unitary on $\mc{H}^{\mc{G}}_{\bs{\alpha}}$.

By the same token, for a T-move given by the Dehn twist $T$ around a marked point $x_\ell^\bullet$ of $\tilde{S}^\bullet$, we have the same holomorphic bundle isomorphism $\mc{I}_{T} : \mc{L}^{[\bs{\Xi}^\bullet]}\to \mc{L}^{[\bs{\Xi}^\bullet\cdot T]}$, and the transition operator is
\[K_{\bs{\alpha}}^{[\bs{\Xi}^\bullet],[\bs{\Xi}^\bullet\cdot T]}{\bf u}({\bf p})=e^{2\pi {\rm i}\Delta_{\alpha_\ell}}{\bf u}({\bf p}).\]
In other words, $K_{\bs{\alpha}}^{[\bs{\Xi}^\bullet],[\bs{\Xi}^\bullet\cdot T]}=e^{2\pi {\rm i}\Delta_{\alpha_\ell}}{\rm Id}$ in this case.\\

\noindent $\bullet$  \textbf{B-move.}
Let $[\bs{\Xi}^\bullet]\in \pi_0\tilde{\mc{BD}}_{g,m}^{\mc{G}}$ with gluing graph $\mc{G}$ and a representative $\bs{\Xi}^\bullet\in\tilde{\mc{BD}}(\tilde{S}^\bullet)$. Consider a B-move $B=B_{k_1k_2}$ on the $j$-th model building block $\mc{P}_j^\bullet=\Sigma^\bullet_{m_j,b_j}$, where either $1\leq k_1<k_2\leq b_j$ (two boundary components) or $b_j<k_1<k_2\leq3$ (two marked points). Let $k_3$ be determined by $\{k_1,k_2,k_3\}=\{1,2,3\}$. For each exterior object of $\Xi_j(\mc{P}_j^\bullet)\subset\Sigma^\bullet$, we set
\[
a_{jk}({\bf p})=\begin{cases}
Q+{\rm i}p_\ell,&\text{if an internal edge ${\rm e}_\ell$ contains $(j,k)$},\\
\alpha_\ell,&\text{if the external edge ${\rm e}^{\rm m}_\ell$ attached to $x_\ell^\bullet$ contains $(j,k)$}.
\end{cases}
\]
As above, for any decorated marked complex building block decomposition $\bs{\Xi}$ of a decorated surface $\tilde{S}$, we denote $\bs{\Xi}\cdot B:=(\Xi_1,\dots,\Xi_j\circ B,\dots,\Xi_{j_0})$ and similarly $\bs{\Xi}\cdot B^{-1}$. In particular, we have $[\bs{\Xi}^\bullet\cdot B]\in \pi_0\tilde{\mc{BD}}_{g,m}$ and $[\bs{\Xi}^\bullet]\to[\bs{\Xi}^\bullet\cdot B]$ is the B-move in question. Notice that the gluing graph $\mc{G}'$ of $[\bs{\Xi}^\bullet\cdot B]$ is generally distinct from $\mc{G}$ of $[\bs{\Xi}^\bullet]$. However, since $C^{\rm DOZZ}(a_1,a_2,a_3)=C^{\rm DOZZ}(a_{\tau(1)},a_{\tau(2)},a_{\tau(3)})$ for any permutation $\tau$ of $\{1,2,3\}$, we have $\rho^{\mc{G}'}(\bs{\alpha},{\bf p})=\rho^{\mc{G}}(\bs{\alpha},{\bf p})$, so that $\mc{H}^{\mc{G}'}_{\bs{\alpha}}=\mc{H}^{\mc{G}}_{\bs{\alpha}}$. There is a natural bundle isomorphism
\[\mc{I}_{B} : \mc{L}^{[\bs{\Xi}^\bullet]}\to \mc{L}^{[\bs{\Xi}^\bullet\cdot B]},\quad \mc{I}_{B}(G)(\tilde{\mc{X}}_{\bs{\Xi}}):=G(\tilde{\mc{X}}_{\bs{\Xi}\cdot B^{-1}}),\]
which is holomorphic since the bijection $\tilde{\mc{X}}_{\bs{\Xi}}\mapsto\tilde{\mc{X}}_{\bs{\Xi}\cdot B}$ of $\BDTT_{g,m}$ commutes with plumbing, again by Proposition \ref{identification_Teichmuller}. By Lemma \ref{braiding_change}, we obtain:
\begin{lemma}[\textbf{Braiding kernel}] \label{lem:B-move_bloc}
Using the notations above, we have for all $\mc{X}\in \mc{T}_{g,m}$
\[\mc{F}_{\bs{\alpha}}^{[\bs{\Xi}^\bullet\cdot B]}(\mc{X},{\bf p})=e^{\pi{\rm i}
(\Delta_{a_{jk_1}({\bf p})}+\Delta_{a_{jk_2}({\bf p})}+(-1)^{k_3+1}\Delta_{a_{jk_3}({\bf p})})}\mc{I}_B \mc{F}_{\bs{\alpha}}^{[\bs{\Xi}^\bullet]}(\mc{X},{\bf p}).\]
\end{lemma}
Thus the transition operator for the B-move is
\[K_{\bs{\alpha}}^{[\bs{\Xi}^\bullet],[\bs{\Xi}^\bullet\cdot B]}{\bf u}({\bf p})=e^{\pi{\rm i}
(\Delta_{a_{jk_1}({\bf p})}+\Delta_{a_{jk_2}({\bf p})}+(-1)^{k_3+1}\Delta_{a_{jk_3}({\bf p})})}{\bf u}({\bf p}),\]
which is unitary as an operator $\mc{H}^{\mc{G}}_{\bs{\alpha}}\to\mc{H}^{\mc{G}'}_{\bs{\alpha}}=\mc{H}^{\mc{G}}_{\bs{\alpha}}$.\\

\noindent $\bullet$ \textbf{A-move}.
Let $[\bs{\Xi}_1^\bullet]\in \pi_0\tilde{\mc{BD}}_{g,m}^{\mc{G}_1}$ and $[\bs{\Xi}_2^\bullet]\in \pi_0\tilde{\mc{BD}}_{g,m}^{\mc{G}_2}$ where $\bs{\Xi}_2^\bullet$ is obtained from $\bs{\Xi}_1^\bullet$ by an A-move of the first type (which suffices by Lemma \ref{lem:A-second-type}). As always, assume that the relevant internal edge of $\mc{G}_1$ is ${\rm e}_1=\{(1,k_1),(2,k_2)\}$. Let ${\bf p}'=(p_2,\ldots,p_{3g-3+m})\in\R_+^{3g-4+m}$ denote spectral parameters associated to the internal edges ${\rm e}_i$ for $i\geq2$. Recall the notations $\mc{O}_1^\bullet,\mc{O}_2^\bullet,\mc{O}_3^\bullet,\mc{O}_4^\bullet$ from Section \ref{sub:elementary_moves} for  marked points and filling disks of $\Sigma_{\rm A}^\bullet$.
We define $\bs{\beta}:=(\beta_1,\beta_2,\beta_3,\beta_4)\in((-\infty,Q)\cup(Q+{\rm i}\R_+))^4$ where $\beta_k:=Q+{\rm i}p_j$ if $\mc{O}_k^\bullet=D_i$ is a disk and $\pl D_i$ corresponds to the internal edge ${\rm e}_j$ ($j\geq2$), or $\beta_k:=\alpha_j$ if 
$\mc{O}^\bullet_k=\{x_j^\bullet\}$ is a marked point of $\Sigma^\bullet_{\rm A}$. The Seiberg bounds say
\[
\Re(\beta_1+\beta_2)>Q,\quad
\Re(\beta_3+\beta_4)>Q,\quad
\Re(\beta_1+\beta_4)>Q,\quad
\Re(\beta_2+\beta_3)>Q.
\]
For $p,p^*\in\R_+$, we write ${\bf p}:=(p,{\bf p}')$, ${\bf p}^*:=(p^*,{\bf p}')$. We define
\begin{align*}
\rho_{\bs{\beta}}(p)&:=C^{\rm DOZZ}(\beta_1,\beta_2,Q+{\rm i}p)C^{\rm DOZZ}(\beta_3,\beta_4,Q-{\rm i}p),\\
\rho_{\bs{\beta}}^*(p^*)&:=C^{\rm DOZZ}(\beta_1,\beta_4,Q+{\rm i}p^*)C^{\rm DOZZ}(\beta_2,\beta_3,Q-{\rm i}p^*).
\end{align*}
Then we have
\[\rho^{\mc{G}_1}(\bs{\alpha},{\bf p})=\rho_{\bs{\beta}}(p)\hat{\rho}(\bs{\alpha},{\bf p}'),\qquad
\rho^{\mc{G}_2}(\bs{\alpha},{\bf p}^*)=\rho_{\bs{\beta}}^*(p^*)\hat{\rho}(\bs{\alpha},{\bf p}'),\]
where $\hat{\rho}(\bs{\alpha},{\bf p}')$ is independent of $p,p^*$. 
Define the unitary operator 
\[K_{\bs{\alpha}}^{[\bs{\Xi}_1^\bullet],[\bs{\Xi}_2^\bullet]}:
L^2(\R_+^{3g-3+m},\rho^{\mc{G}_1}(\bs{\alpha},{\bf p})\dd{\bf p})\to L^2(\R_+^{3g-3+m},\rho^{\mc{G}_2}(\bs{\alpha},{\bf p}^*)\dd{\bf p}^*)
\] 
given by (recall \eqref{A_transform_2to1})
\begin{equation}\label{expression_kernel_A_move}
(K_{\bs{\alpha}}^{[\bs{\Xi}_1^\bullet],[\bs{\Xi}_2^\bullet]}u)(p^*,{\bf p}')
:=\int_0^\infty K^{\rm A}_{\beta_2,\beta_3,\beta_4,\beta_1}(p^*,p)u(p,{\bf p}')
\dd p.\end{equation}
Its unitarity follows from Proposition \ref{prop:unitarity}. 
We now prove the following:
\begin{proposition}[\textbf{A-move kernel}]\label{prop:fusion_A_move}
Let $[\bs{\Xi}_1^\bullet]\in \pi_0\tilde{\mc{BD}}_{g,m}^{\mc{G}_1}$ and $[\bs{\Xi}_2^\bullet]\in\pi_0\tilde{\mc{BD}}_{g,m}^{\mc{G}_{2}}$ be two isotopy classes such that $\bs{\Xi}_1^\bullet\to\bs{\Xi}_2^\bullet$ is an A-move of the first type, and let $\bs{\alpha}$ be weights satisfying the Seiberg bounds with respect to both gluing graphs $\mcG_1$ and $\mcG_2$. Then there exists a nowhere vanishing holomorphic section $s$ of ${\rm Hom}(\mc{L}^{[\bs{\Xi}_1^\bullet]}(\bs{\alpha}),\mc{L}^{[\bs{\Xi}_2^\bullet]}(\bs{\alpha}))
\simeq\mc{L}^{[\bs{\Xi}_2^\bullet]}(\bs{\alpha})\otimes(\mc{L}^{[\bs{\Xi}_1^\bullet]}(\bs{\alpha}))^*$ such that for all $\mc{X}\in \mc{T}_{g,m}$,
\[\mc{F}_{\bs{\alpha}}^{[\bs{\Xi}_2^\bullet]}(\mc{X})=(K_{\bs{\alpha}}^{[\bs{\Xi}_1^\bullet],[\bs{\Xi}_2^\bullet]}\otimes s(\mc{X}))(\mc{F}_{\bs{\alpha}}^{[\bs{\Xi}_1^\bullet]}(\mc{X})).\]
Moreover, such an operator $K_{\bs{\alpha}}^{[\bs{\Xi}_1^\bullet],[\bs{\Xi}_2^\bullet]}$ is unique up to multiplication by a complex constant of modulus $1$.
\end{proposition}
\begin{proof}
It suffices to prove existence. We define the fiber product
\[\BDTT_{g,m}^{[\bs{\Xi}_1^\bullet]}\times_{\mc{T}_{g,m}}\BDTT_{g,m}^{[\bs{\Xi}_2^\bullet]}:=\{(\tilde{\mc{X}}_{\bs{\Xi}^1},\tilde{\mc{X}}_{\bs{\Xi}^2})\in\BDTT_{g,m}^{[\bs{\Xi}_1^\bullet]}\times\BDTT_{g,m}^{[\bs{\Xi}_2^\bullet]} \,|\, \pi_{\BDTT\to\mc{T}}(\tilde{\mc{X}}_{\bs{\Xi}^1})=\pi_{\BDTT\to\mc{T}}(\tilde{\mc{X}}_{\bs{\Xi}^2})\in\mc{T}_{g,m}\}\]
with the associated canonical projection $\pi:\BDTT_{g,m}^{[\bs{\Xi}_1^\bullet]}\times_{\mc{T}_{g,m}}\BDTT_{g,m}^{[\bs{\Xi}_2^\bullet]}\to\mc{T}_{g,m}$. Consider
\[\mc{S}_{\rm A}:=\{(\tilde{\mc{X}}_{\bs{\Xi}^1},\tilde{\mc{X}}_{\bs{\Xi}^2})\in\BDTT_{g,m}^{[\bs{\Xi}_1^\bullet]}\times_{\mc{T}_{g,m}}\BDTT_{g,m}^{[\bs{\Xi}_2^\bullet]} \,|\, \mc{B}_{\tilde{\mc{X}}_{\bs{\Xi}^2},\bs{\alpha}}=\lambda K_{\bs{\alpha}}^{[\bs{\Xi}_1^\bullet],[\bs{\Xi}_2^\bullet]}\mc{B}_{\tilde{\mc{X}}_{\bs{\Xi}^1},\bs{\alpha}}\text{ for some }\lambda\in\C\}.\]
We claim that $\pi(\mc{S}_{\rm A})\subset\mc{T}_{g,m}$ contains a maximal totally real submanifold, i.e.\ a real submanifold $\mc{M}$ of dimension $\dim_\R\mc{M}=\dim_\C\mc{T}_{g,m}=3g-3+m$ such that $T_{\mc{X}}\mc{M}\cap \mc{J}(T_{\mc{X}}\mc{M})=\{0\}$ for all $\mc{X}\in\mc{M}$, where $\mc{J}$ is the complex structure on $\mc{T}_{g,m}$.
Let us first explain why this proves the result, which is a standard argument. By definition, for all $\mc{X}\in\pi(\mc{S}_{\rm A})$, we have $\mc{F}_{\bs{\alpha}}^{[\bs{\Xi}^\bullet_2]}(\mc{X})=(K_{\bs{\alpha}}^{[\bs{\Xi}_1^\bullet],[\bs{\Xi}_2^\bullet]}\otimes s_\mc{X})\mc{F}_{\bs{\alpha}}^{[\bs{\Xi}^\bullet_1]}(\mc{X})$ for some $s_{\mc{X}}\in{\rm Hom}(\mc{L}^{[\bs{\Xi}_1^\bullet]}(\bs{\alpha}),\mc{L}^{[\bs{\Xi}_2^\bullet]}(\bs{\alpha}))_{\mc{X}}={\rm Hom}(\mc{L}_{\mc{X}}^{[\bs{\Xi}_1^\bullet]}(\bs{\alpha}),\mc{L}_{\mc{X}}^{[\bs{\Xi}_2^\bullet]}(\bs{\alpha}))$. We shall show that this holds for all $\mc{X}\in\mc{T}_{g,m}$; then the section $\mc{X}\mapsto s_{\mc{X}}$ is automatically holomorphic since $\mc{F}_{\bs{\alpha}}^{[\bs{\Xi}^\bullet_1]}(\mc{X}),\mc{F}_{\bs{\alpha}}^{[\bs{\Xi}^\bullet_2]}(\mc{X})$ are both holomorphic and nowhere vanishing. Suppose this is false for some $\mc{X}\in\mc{T}_{g,m}$. Let $s_1,s_2$ be global nowhere vanishing holomorphic sections of $\mc{L}^{[\bs{\Xi}_1^\bullet]}(\bs{\alpha})^*,\mc{L}^{[\bs{\Xi}_2^\bullet]}(\bs{\alpha})^*$, respectively. Consider
\[f_1:=(K_{\bs{\alpha}}^{[\bs{\Xi}_1^\bullet],[\bs{\Xi}_2^\bullet]}\otimes s_1)(\mc{F}_{\bs{\alpha}}^{[\bs{\Xi}^\bullet_1]}),\quad f_2:=s_2(\mc{F}_{\bs{\alpha}}^{[\bs{\Xi}^\bullet_2]}),\]
which are nowhere vanishing holomorphic maps $\mc{T}_{g,m}\to\mc{H}_{\bs{\alpha}}^{\mc{G}_2}$. The assumption is equivalent to saying that $f_1(\mc{X})$ and $f_2(\mc{X})$ are not collinear. Then there exist linear functionals $\lambda,\mu\in(\mc{H}_{\bs{\alpha}}^{\mc{G}_2})^*$ such that $\lambda(f_1(\mc{X}))=\mu(f_2(\mc{X}))=1$, $\lambda(f_2(\mc{X}))=\mu(f_1(\mc{X}))=0$. Consider the holomorphic function $D_{\lambda,\mu}:\mc{T}_{g,m}\to\C$ defined by
\[D_{\lambda,\mu}:=\lambda(f_1)\mu(f_2)-\lambda(f_2)\mu(f_1).\]
By assumption, $D_{\lambda,\mu}$ vanishes on $\pi(\mc{S}_{\rm A})\supset\mc{M}$, thus $D_{\lambda,\mu}\equiv0$ on $\mc{T}_{g,m}$. This contradicts $D_{\lambda,\mu}(\mc{X})=1$.

We turn to the proof of the claim. We call an element $\tilde{\mc{X}}_{\bs{\Xi}^1}\in\BDTT_{g,m}^{[\bs{\Xi}_1^\bullet]}$ \textit{good} if the model forms of the marked complex building blocks $\tilde{\mc{X}}_1^1,\tilde{\mc{X}}_2^1$ associated to $\Xi_1^1,\Xi_2^1$ satisfy $\tilde{\mc{MF}}(\tilde{\mc{X}}_\ell^1)=(\bs{\psi}_\ell,\bs{\theta}_\ell,\bs{\la}_\ell)$ ($\ell=1,2$) where $\bs{\psi}_\ell=(\psi_{\ell1},\ldots,\psi_{\ell b_\ell})$ with each $\psi_{\ell k}$ $\R$-symmetric except that $\psi_{2k_2}$ satisfies $\psi_{2k_2}(\bar{z})=\overline{\psi_{2k_2}(z)}$ and is orientation reversing on $\hat\R$, $\bs{\theta}_1=0$, $\bs{\theta}_2=\pi e_{k_2}\in\R^{b_2}$, and 
$\bs{\la}_\ell\in\R^{m_\ell}$. For such an element $\tilde{\mc{X}}_{\bs{\Xi}^1}$, embedding the union of the images of $\Xi_1^1,\Xi_2^1$ into $\hat\C$ and applying the A-move as described in Section \ref{sec:local_study}, we obtain an element $\tilde{\mc{X}}_{\bs{\Xi}^2}\in\BDTT_{g,m}^{[\bs{\Xi}_2^\bullet]}$ such that $\bs{\Xi}^1\to\bs{\Xi}^2$ is an A-move of the first type. By Remark \ref{rem:relation_A_move} and Lemma \ref{GRSSW}, we have $(\tilde{\mc{X}}_{\bs{\Xi}^1},\tilde{\mc{X}}_{\bs{\Xi}^2})\in\mc{S}_{\rm A}$. Thus it suffices to show that the projection of the good elements in $\BDTT_{g,m}^{[\bs{\Xi}_1^\bullet]}$ on $\mc{T}_{g,m}$ contains a maximal totally real submanifold. Fix any good element $\tilde{\mc{X}}_{\bs{\Xi}^1}$. Notice that plumbing with purely imaginary parameters preserves goodness. Then for $N>0$ sufficiently large, $\{\mc{X}(\bs{\tau})=\pi_{\mc{D}\tT\to \mc{T}}(\tilde{\mc{X}}_{\bs{\Xi}^1}(\bs{\tau}))\,|\, \bs{\tau}\in U_{N}\cap ({\rm i}\R)^{3g-3+m}\}$ (recall \eqref{U_N}) is a maximal totally real submanifold of $\mc{T}_{g,m}$, since Lemma \ref{lem:plumbing} shows that $\bs{\tau}$ are holomorphic coordinates on $U_N$ for $\mc{T}_{g,m}$.
\end{proof}

\noindent $\bullet$ \textbf{S-move}.
Let $[\bs{\Xi}_1^\bullet]\in \pi_0\tilde{\mc{BD}}_{g,m}^{\mc{G}_1}$ and
$[\bs{\Xi}_2^\bullet]\in \pi_0\tilde{\mc{BD}}_{g,m}^{\mc{G}_2}$ where $\bs{\Xi}_2^\bullet$ is obtained from $\bs{\Xi}_1^\bullet$ by an
S-move. We use the choices of curve and marking of Section \ref{sub:elementary_moves} corresponding to the horizontal and vertical
cuts in the local S-move model of Section \ref{sec:local_studyS}. As for the A-move, any additional twists or boundary relabellings are
handled by composing with the T- and B-moves already treated.

For notational convenience, assume that the internal edge associated to the S-move is ${\rm e}_1=\{(1,1),(1,2)\}$. Let $p$ and $p^*$ denote its spectral parameters before and after the move, respectively, and denote by ${\bf p}'=(p_2,\ldots,p_{3g-3+m})\in\R_+^{3g-4+m}$ the spectral parameters associated to the internal edges ${\rm e}_i$ for $i\geq2$. The third object of the first complex building block is either a marked point $x_\ell$, in which case the whole surface is a once-punctured torus, or a boundary component corresponding to an unchanged internal edge ${\rm e}_i$ for some $i\geq 2$. As in \eqref{def_beta_tore}, we set
\[\beta:=\beta({\bf p}'):=
\begin{cases}
\alpha_\ell,&\text{in the marked-point case},\\
Q+{\rm i}p_i,&\text{in the boundary case}.
\end{cases}
\]
The weights are real and satisfy the Seiberg bounds for the gluing graphs; in particular, $\alpha_\ell\in(0,Q)$ in the first case.
In the second case, we orient ${\rm e}_i$ towards the first vertex, so that the weight contributed there to the DOZZ factor is
$Q+{\rm i}p_i$. We orient the self-edge by $\vec{\rm e}_1=((1,1),(1,2))$ and leave all other orientations arbitrary.

Recall the densities $\rho_\beta$ and $w_\beta$ and the phase $\vartheta_\beta$ of \eqref{positive_torus_density}. The global
densities factor as
\[\begin{split}
\rho^{\mc G_1}(\bs\alpha,p,{\bf p}')
 &=\rho_{\beta}(p)\hat\rho(\bs\alpha,{\bf p}')
  =w_{\beta}(p)\hat w(\bs\alpha,{\bf p}'),\\
\rho^{\mc G_2}(\bs\alpha,p^*,{\bf p}')
 &=\rho_{\beta}(p^*)\hat\rho(\bs\alpha,{\bf p}')
  =w_{\beta}(p^*)\hat w(\bs\alpha,{\bf p}'),
\end{split}
\]
where $\hat\rho$ is the common product of the remaining vertex
factors and $\hat w(\bs\alpha,{\bf p}') :=e^{{\rm i}\vartheta_{\beta}}
   \hat\rho(\bs\alpha,{\bf p}')>0.$
The last inequality follows from positivity of the global density
and of $w_{\beta}$.
In the once-punctured torus case, ${\bf p}'$ is absent and
$\hat w=\hat\rho=1$. We define
\begin{equation}\label{expression_kernel_S_move}
\begin{split}
(K_{\bs\alpha}^{[\bs\Xi_1^\bullet],[\bs\Xi_2^\bullet]}u)
 (p^*,{\bf p}') &:=(K^{\rm S}_{\beta})^{-1} \big(u(\cdot,{\bf p}')\big)(p^*)=\int_0^\infty  \frac{w_{\beta}(p)}{w_{\beta}(p^*)}
 \overline{K^{\rm S}_{\beta}(p,p^*)} u(p,{\bf p}')\dd p.
\end{split}
\end{equation}
The integral is understood through the Hilbert space extension in
Corollary \ref{cor:inverse_S_kernel}.  Applying that
corollary for almost every ${\bf p}'$ and using the preceding
factorisation shows that
\[
K_{\bs\alpha}^{[\bs\Xi_1^\bullet],[\bs\Xi_2^\bullet]}:
\mc H_{\bs\alpha}^{\mc G_1}\to
\mc H_{\bs\alpha}^{\mc G_2}
\]
is unitary. We keep the constant $\kappa^{-1}$ of
\eqref{S_transform_1to2} in the scalar transition factor below.

\begin{proposition}[\textbf{S-move kernel}]\label{prop:modular_S_move}
Let $[\bs{\Xi}_1^\bullet]\in \pi_0\tilde{\mc{BD}}_{g,m}^{\mc{G}_1}$ and $[\bs{\Xi}_2^\bullet]\in\pi_0\tilde{\mc{BD}}_{g,m}^{\mc{G}_{2}}$ be two isotopy classes such that $\bs{\Xi}_1^\bullet\to\bs{\Xi}_2^\bullet$ is an S-move, and let $\bs{\alpha}$ be weights satisfying the Seiberg bounds with respect to both gluing graphs $\mcG_1$ and $\mcG_2$. Then there exists a nowhere vanishing holomorphic section $s$ of ${\rm Hom}(\mc{L}^{[\bs{\Xi}_1^\bullet]}(\bs{\alpha}),\mc{L}^{[\bs{\Xi}_2^\bullet]}(\bs{\alpha}))
\simeq\mc{L}^{[\bs{\Xi}_2^\bullet]}(\bs{\alpha})\otimes(\mc{L}^{[\bs{\Xi}_1^\bullet]}(\bs{\alpha}))^*$ such that for all $\mc{X}\in \mc{T}_{g,m}$,
\[\mc{F}_{\bs{\alpha}}^{[\bs{\Xi}_2^\bullet]}(\mc{X})=(K_{\bs{\alpha}}^{[\bs{\Xi}_1^\bullet],[\bs{\Xi}_2^\bullet]}\otimes s(\mc{X}))(\mc{F}_{\bs{\alpha}}^{[\bs{\Xi}_1^\bullet]}(\mc{X})).\]
Moreover, such an operator $K_{\bs{\alpha}}^{[\bs{\Xi}_1^\bullet],[\bs{\Xi}_2^\bullet]}$ is unique up to multiplication by a complex constant of modulus $1$.
\end{proposition}
\begin{proof}
As in the proof of Proposition \ref{prop:fusion_A_move}, consider the fiber product $\pi:\BDTT_{g,m}^{[\bs{\Xi}_1^\bullet]}\times_{\mc{T}_{g,m}}\BDTT_{g,m}^{[\bs{\Xi}_2^\bullet]}\to\mc{T}_{g,m}$ and
\[
\mc S_{\rm S}:=\{(\tilde{\mc X}_{\bs\Xi^1},\tilde{\mc X}_{\bs\Xi^2})\in\BDTT_{g,m}^{[\bs{\Xi}_1^\bullet]}\times_{\mc{T}_{g,m}}\BDTT_{g,m}^{[\bs{\Xi}_2^\bullet]}\,|\,
\mc B_{\tilde{\mc X}_{\bs\Xi^2},\bs\alpha}
=\lambda K_{\bs\alpha}^{[\bs\Xi_1^\bullet],[\bs\Xi_2^\bullet]}
\mc B_{\tilde{\mc X}_{\bs\Xi^1},\bs\alpha}
\text{ for some }\lambda\in\C\}.
\]
By the same argument, it suffices to show that $\pi(\mc S_{\rm S})\subset\mc{T}_{g,m}$ contains a maximal
totally real submanifold.

Choose a point
\[\tilde{\mc X}_{\bs\Xi^1}=[(F,\Sigma,J,{\bf x},{\bf Y},\bs\Xi^1)]\in\BDTT_{g,m}^{[\bs\Xi_1^\bullet]}
\]
whose first complex building block is the horizontal cut model $\mc P$ of
Section \ref{sec:local_studyS}, with $T=T_0>0$. In the boundary case,
fix $0<q_1<\min(1/4,T_0/2)$; in the marked point case, fix
$Y_{\rm S}=t\pl_z$ with $t>0$. We use the incoming parametrisations and the lifted marking parameters of that section.
The other complex building blocks are arbitrary. The augmented model form map \eqref{def_MFtildehat} permits this choice in the prescribed class.
As in the proof of Proposition \ref{prop:fusion_A_move}, for $N>0$ sufficiently large, $\{\mc{X}(\bs{\tau})=\pi_{\mc{D}\tT\to \mc{T}}(\tilde{\mc{X}}_{\bs{\Xi}^1}(\bs{\tau}))\,|\, \bs{\tau}\in U_{N}\cap ({\rm i}\R)^{3g-3+m}\}$ is a maximal totally real submanifold of $\mc{T}_{g,m}$.
Thus it suffices to show that the picture of the S-move in Section \ref{sec:local_studyS} is preserved by plumbing by purely imaginary parameters. More precisely, for each $\bs{\tau}\in  U_N\cap ({\rm i}\R)^{3g-3+m}$, we will construct, using an S-move, an element $\tilde{\mc{X}}_{\bs{\Xi}^2_{\bs{\tau}}}\in\BDTT_{g,m}^{[\bs{\Xi}_2^\bullet]}$ such that $(\tilde{\mc{X}}_{\bs{\Xi}^1}(\bs{\tau}),\tilde{\mc{X}}_{\bs{\Xi}^2_{\bs{\tau}}})\in \mc{S}_{\rm S}$.


Let $\bs\tau\in U_N\cap({\rm i}\R)^{3g-3+m}$ and write
$\eta_1={\rm Im}(\tau_1)$. With our orientation convention above,
$\tau_{11}=\tau_1$, $\tau_{12}=0$, and
$\tau_{13}=0$ in the boundary case. Thus only $\psi_1$ changes on the first complex building
block, being replaced by $\psi_1(e^{-\eta_1}\cdot)$.
In the coordinate $z$ of the local model (as in \eqref{Ntau}), this changes
the lower boundary from ${\rm Im}(z)=-T_0/2$ to
${\rm Im}(z)=-T_0/2-\eta_1/(2\pi)$, while the upper boundary
remains ${\rm Im}(z)=T_0/2$. Self-gluing therefore gives
\[
\Sigma_{\rm S}(\bs\tau)=
\begin{cases}
N({\rm i}T(\bs\tau))\setminus D(0,q_1)^\circ,&b_{\rm S}=1,\\
N({\rm i}T(\bs\tau))\text{ with marked point }0,&m_{\rm S}=1,
\end{cases}
\qquad
T(\bs\tau):=T_0+\frac{\eta_1}{2\pi}.
\]
The third parametrisation $\zeta_{\rm S}$, or the decoration
$t\pl_z$, is unchanged. Notice that the old seam is represented by
${\rm Im}(z)=T_0/2$, not by ${\rm Im}(z)=T(\bs\tau)/2$.

We move this seam by an isotopy fixing the boundary or marked point to the
horizontal curve parametrised by
\[
\xi_{\bs\tau}(e^{{\rm i}\theta})
 =\frac{\theta}{2\pi}+\frac{{\rm i}T(\bs\tau)}2.
\]
Such an isotopy exists since $q_1<T_0/2\leq T(\bs\tau)/2$ in
the boundary case, and the moving curves avoid $0$ in the marked point
case. Transporting the marking continuously, with no additional
twist, gives a decomposition $\bs\Xi^{1}_{\bs\tau}$
in the same decorated isotopy class as $\bs\Xi^1(\bs\tau)$.
Its first complex building block is now precisely the horizontal cut model
of Section \ref{sec:local_studyS} at $T=T(\bs\tau)$.
By Proposition \ref{well-defined_blocks},
\begin{equation}\label{S_horizontal_recentring}
\mc B_{\tilde{\mc X}_{\bs\Xi^{\,1}_{\bs\tau}},\bs\alpha}
=a_{\rm S}(\bs\tau,\bs\alpha)
 \mc B_{\tilde{\mc X}_{\bs\Xi^1}(\bs\tau),\bs\alpha},
\qquad a_{\rm S}(\bs\tau,\bs\alpha)\in\C^*,
\end{equation}
where $a_{\rm S}$ is independent of all spectral parameters.

Now cut $\Sigma_{\rm S}(\bs\tau)$ instead along
\[
\xi^*_{\bs\tau}(e^{{\rm i}\theta})
 =\frac12+\frac{{\rm i}T(\bs\tau)\theta}{2\pi}.
\]
The two curves $\xi_{\bs{\tau}}(\T)$ and $\xi^*_{\bs\tau}(\T)$ intersect transversely at exactly one point. We use the
incoming parametrisations and the marking $\Xi^*$ of the local model \eqref{Xi^*Smove} to define
$\bs\Xi^{2}_{\bs\tau}$, leaving all other marked complex building blocks unchanged. These choices are made continuously from the fixed horizontal or vertical models in their
prescribed relative isotopy classes. If necessary, a further decorated isotopy supported in the torus matches the chosen
intersection point and parametrisations in the definition of the S-move in Section \ref{sub:elementary_moves}. We still denote the resulting
decomposition by $\bs\Xi^2_{\bs\tau}$, and notice, by Proposition \ref{well-defined_blocks}, that the conformal block $\mc B_{\tilde{\mc X}_{\bs\Xi^2_{\bs\tau}},\bs\alpha}$ has possibly been multiplied by a constant $a^*_{\rm S}(\bs\tau,\bs\alpha)\in\C^*$ which is independent of all spectral parameters
and equals $1$ when no further isotopy is needed. We also have $F(\bs\tau)^{-1}\circ\bs\Xi^2_{\bs\tau} \in[\bs\Xi_2^\bullet].$
Both decompositions represent the same decorated marked surface
$\tilde S(\bs\tau)$. 

We apply \eqref{S_transform_1to2} to the local torus. In the marked point
case, this gives directly the required identity, with
$\beta=\alpha_\ell$. In the boundary case, it gives, for every
external descendant $\nu\in\mc T$ and almost every $p^*>0$,
\[
\mc B^2_{\beta}(p^*,\nu)=\kappa^{-1} (K^{\rm S}_{\beta})^{-1}
       \big(\mc B^1_{\beta}(\cdot,\nu)\big)(p^*).
\]
Here $\kappa$ is independent of $p^*$ and $\nu$.
We tensor this identity with the conformal blocks of the unchanged other marked complex building blocks  and contract the
remaining edges. This step can be made in the Hilbert
topology as in the A-move argument, from local spectral holomorphy and the locally
uniform Hilbert norm bounds supplied by the bootstrap formula.
Combining this identity with \eqref{S_horizontal_recentring}, we obtain
\[
\mc B_{\tilde{\mc X}_{\bs\Xi^2_{\bs\tau}},\bs\alpha}
=h_{\rm S}(\bs\tau,\bs\alpha)
 K_{\bs\alpha}^{[\bs\Xi_1^\bullet],[\bs\Xi_2^\bullet]}
 \mc B_{\tilde{\mc X}_{\bs\Xi^1}(\bs\tau),\bs\alpha},
\qquad
h_{\rm S}(\bs\tau,\bs\alpha):=\kappa^{-1}a_{\rm S}(\bs\tau,\bs\alpha)a^*_{\rm S}(\bs\tau,\bs\alpha)\in\C^*,
\]
where scalar function $h_{\rm S}$ is independent of $p,p^*$ and ${\bf p}'$.
Hence $(\tilde{\mc{X}}_{\bs{\Xi}^1}(\bs{\tau}),\tilde{\mc{X}}_{\bs{\Xi}^2_{\bs{\tau}}})\in \mc{S}_{\rm S}$.
\end{proof}


\appendix 

\section{The DOZZ three-point function}\label{sec:DOZZ}
The DOZZ function is given by
\begin{equation}\label{DOZZformula}
C_{\gamma,\mu}^{{\rm DOZZ}}(\alpha_1,\alpha_2,\alpha_3)=\frac{ \Big(\pi \mu\frac{\Gamma(\gamma^2/4)}{\Gamma(1-\gamma^2/4)}(\frac{\gamma}{2})^{2-\frac{\gamma^2}{2}}\Big)^{\frac{2Q-\sum_{k=1}^3\alpha_k}{\gamma}} \Upsilon'_{\gamma/2}(0)\prod_{k=1}^3\Upsilon_{\gamma/2}(\alpha_k)}
{\Upsilon_{\gamma/2}(\frac{\alpha_1+\alpha_2+\alpha_3-2Q}{2}) 
\Upsilon_{\gamma/2}(\frac{\alpha_1+\alpha_2-\alpha_3}{2}) \Upsilon_{\gamma/2}(\frac{\alpha_3-\alpha_1+\alpha_2}{2}) \Upsilon_{\gamma/2}(\frac{\alpha_3-\alpha_2+\alpha_1}{2})}.
\end{equation}
The function $\Upsilon_{\gamma/2}$ is defined by $\Upsilon_{\gamma/2}=(\Gamma_{\gamma/2}(z)\Gamma_{\gamma/2}(Q-z))^{-1}$. It is holomorphic with zeros at $-n\gamma/2-m2/\gamma$ and $Q+n\gamma/2+m2/\gamma$ for $n,m\in \N$ (including $n,m=0$).
It follows from the expression above that $C^{\rm DOZZ}(\alpha_1,\alpha_2,\alpha_3)\not=0$ if $\alpha_k\in (-\infty,Q)\cup(Q+{\rm i}\R^*)$ for all $k$ and 
$\sum_{k=1}^3{\rm Re}(\alpha_k)>2Q$.

\section{Unitarity of the fusion kernel for the A-move}
Consider $\hat{\C}$ with marked points $(x_1,x_2,x_3,x_4)=(0,z,t,\infty)$ for $t>1$ close to $1$, where $z\in \Omega$ is defined by 
\begin{equation}\label{def_domain_omega}
 \Omega:=\{z\in\C\,|\, |z|<1\text{ and }|1-z|<1\}.
\end{equation}
We take $g_{\rm dozz}$ as a Riemannian metric and take $\T$, with its canonical parametrisation, as a cut of the sphere. We start with $\bs{\beta}=(\beta_1,\dots,\beta_4)$ real valued satisfying Seiberg bounds relative to the cut around the unit circle $\T$. Using \cite[Section 9.2]{GKRV21_Segal},
the bootstrap formula for the $4$-point function reads, for $z\in \Omega$,
\[ \cjg \prod_{k=1}^4V_{\beta_k}(x_k)\cjd_{\hat{\C},g_{\rm dozz}}=\frac{1}{2\pi}
(\frac{Z_{\D}}{2})^2 \int_{\R_+} t^{2(\Delta_{\beta_3}+\Delta_{\beta_4}-\Delta_{\beta_1}-\Delta_{\beta_2})}\rho^1_{\bs{\beta}}(p)|\mc{F}^{1}_{\bs{\beta}}(p,z/t)|^2\dd p,
\]
where $\rho^1_{\bs{\beta}}(p)=C^{\rm DOZZ}(\beta_1,\beta_2,Q+{\rm i}p)C^{\rm DOZZ}(\beta_3,\beta_4,Q-{\rm i}p)$ and $\mc{F}^{1}_{\bs{\beta}}(p,z)$ is the holomorphic function defined on the universal cover of the annulus $\{0<|z|<1\}$ by (the series converge for almost all $p>0$)
\[\mc{F}^{1}_{\bs{\beta}}(p,z)=z^{\Delta_{Q+{\rm i}p}-\Delta_{\beta_1}-\Delta_{\beta_2}}\sum_{\nu\in \mc{T}}\omega_{\D}(\Delta_{\beta_1},\Delta_{\beta_2},\Delta_{Q+{\rm i}p},\nu)\omega_{\D}(\Delta_{\beta_3},\Delta_{\beta_4},\Delta_{Q-{\rm i}p},\nu)z^{|\nu|},\]
where the coefficients $\omega_{\D}(\cdot,\cdot,\cdot,\nu)$ are obtained from the Ward identity and defined in \cite[Proposition 7.1]{GKRV20_bootstrap}; here $\omega_{\D}(\cdot,\cdot,\cdot,0)=1$. 
Here $Z_{\D}$ is some explicit universal constant (the determinant of the Laplacian on the Dirichlet disk), and we refer in particular to \cite[Equation (9.3)]{GKRV21_Segal}. The limit $t\to 1$ exists (see \cite{GKRV20_bootstrap}) and all terms converge at $t=1$; we therefore assume $t=1$. 
The function $\mc{F}^{1}_{\bs{\beta}}(p,z)$ is the conformal block in the first channel, in some particular coordinates. By the construction of the coefficients $\omega_{\D}$ (see \cite{GKRV20_bootstrap}), it is a real valued function when $z$ is real and all $\Delta_{\beta_k}$ are real (thus in particular when $\beta_k\in (-\infty,Q)\cup (Q+{\rm i}\R^*)$). On the simply connected domain $\Omega$, we choose the branch of $\log z$ which
is real on $(0,1)$. Since all the coefficients of the block are real functions
of the real conformal dimensions, this choice gives, for almost every $p>0$,
\begin{equation}\label{real_block}
\mc{F}^{1}_{\bs{\beta}}(p,\bar{z})=\bbar{\mc{F}^{1}_{\bs{\beta}}(p,z)}.
\end{equation}
The correlation function $C_z(\beta_1,\dots,\beta_4)=\cjg \prod_{k=1}^4V_{\beta_k}(x_k)\cjd_{\hat{\C},g_{\rm dozz}}$ can be written as 
\[C_z(\beta_1,\dots,\beta_4)=\mc{A}_{\Sigma,g_{\rm dozz},\bs{\zeta}}(\Psi_{\beta_1},\Psi_{\beta_2},\Psi_{\beta_3},\Psi_{\beta_4})\]
for some amplitude $\mc{A}$ of a surface which is a sphere with four disks removed (and $\bs{\zeta}$ the boundary parametrisations), and it extends holomorphically, by Lemma~\ref{extension_holomorphe_amp}, to the open set
$
 \mc U_4=\{\bs\beta\in\C^4\,;\,
 \sum_{j=1}^4|\operatorname{Re}(\beta_j)-Q|<2Q\}.
$
In particular, this domain contains the weights such that $\operatorname{Re}(\beta_j)\leq Q$ and
$\sum_{j=1}^4\operatorname{Re}(\beta_j)>2Q$. The amplitude can be cut into two pairs of pants and the proof of the bootstrap in \cite{GKRV21_Segal} applies to show that for $\beta_k\in (-\infty,Q)\cup (Q+{\rm i}\R^*)$ satisfying ${\rm Re}(\beta_1+\beta_2)>Q$ and ${\rm Re}(\beta_3+\beta_4)>Q$
\[ C_z(\beta_1,\dots,\beta_4)=\frac{1}{2\pi}
(\frac{Z_{\D}}{2})^2 \int_{\R_+} \rho^1_{\bs{\beta}}(p)\big|\mc{F}^{1}_{\bs{\beta}}(p,z)\big|^2\dd p.
\]

Next, for $f(z)=t-z$, a direct computation gives $S_{\rm L}^0(\hat{\C},g_{\rm dozz},f^*g_{\rm dozz})=0$. Therefore, for $\beta_k$ real valued satisfying the Seiberg bounds associated to the cut around $\T$,
\[C_z(\beta_1,\dots,\beta_4)=\lim_{t\to 1}\cjg V_{\beta_3}(0)V_{\beta_2}(t-z)V_{\beta_1}(t)V_{\beta_4}(\infty)\cjd_{\hat{\C},g_{\rm dozz}}.\]
Applying the bootstrap formula with the cut at $\T$ as above and letting $t\to 1$, the same method shows that, for $\beta_k\in  (-\infty,Q)\cup (Q+{\rm i}\R^*)$ satisfying ${\rm Re}(\beta_1+\beta_4)>Q$ and ${\rm Re}(\beta_2+\beta_3)>Q$,
\[ C_z(\beta_1,\dots,\beta_4)=\cjg \prod_{k=1}^4V_{\beta_k}(x_k)\cjd_{\hat{\C},g_{\rm dozz}}=\frac{1}{2\pi}
(\frac{Z_{\D}}{2})^2 \int_{\R_+} \rho^2_{\bs{\beta}}(p)|\mc{F}^{2}_{\bs{\beta}}(p,1-z)|^2\dd p,
\]
where $\rho^2_{\bs{\beta}}(p)=C^{\rm DOZZ}(\beta_2,\beta_3,Q+{\rm i}p)C^{\rm DOZZ}(\beta_1,\beta_4,Q-{\rm i}p)$ and $\mc{F}^{2}_{\bs{\beta}}(p,z)=\mc{F}^{1}_{\beta_3,\beta_2,\beta_1,\beta_4}(p,z)$. This shows that
\begin{equation}\label{isometric_relation}
\int_{\R_+} \rho_{\bs{\beta}}^1(p)|\mc{F}^{1}_{\bs{\beta}}(p,z)|^2\dd p=\int_{\R_+} \rho_{\bs{\beta}}^2(p)|\mc{F}^{2}_{\bs{\beta}}(p,1-z)|^2\dd p
\end{equation}
for $z\in \Omega$.
We claim the following orthogonality result:
\begin{lemma}\label{orthogonality}
Let $w_{\bs{\beta}}^1$ and $w_{\bs{\beta}}^2$ defined by \eqref{densitywbeta}, for $\bs{\beta}=(\beta_1,\dots,\beta_4)\in ((-\infty,Q)\cup (Q+{\rm i}\R^*))^4$ such that 
\[{\rm Re}(\beta_1+\beta_4)>Q, \quad {\rm Re}(\beta_2+\beta_3)>Q, \quad {\rm Re}(\beta_1+\beta_2)>Q,\quad 
{\rm Re}(\beta_3+\beta_4)>Q.\]
For any $z,z'\in (0,1)$, we have 
\begin{equation}\label{orthogonality_reel}
\int_{\R_+}w^1_{\bs{\beta}}(p)\mc{F}^{1}_{\bs{\beta}}(p,z)\bbar{\mc{F}^{1}_{\bs{\beta}}(p,z')}\dd p=\int_{\R_+}w^2_{\bs{\beta}}(p)\mc{F}^{2}_{\bs{\beta}}(p,1-z)\bbar{\mc{F}^{2}_{\bs{\beta}}(p,1-z')}\dd p.
\end{equation}
\end{lemma}
\begin{proof}
Consider the map 
\[H^i:(z,z')\in \Omega \times \Omega \mapsto \int_{\R_+}w_{\bs{\beta}}^i(p)\mc{F}^{i}_{\bs{\beta}}(p,z)\mc{F}^{i}_{\bs{\beta}}(p,z')\dd p,\]
where $\Omega$ is defined by \eqref{def_domain_omega}. 
The locally uniform $L^2(\R_+,w^i_{\bs\beta}(p)\dd p)$ estimates on compact
subsets of $\Omega$, together with Cauchy--Schwarz, show that $H^i$ is
holomorphic on $\Omega^2$. The function
\[
 (z,z')\mapsto H^1(z,z')-H^2(1-z,1-z')
\]
vanishes on $\{(z,\bar z);z\in\Omega\}$ by
\eqref{isometric_relation} and \eqref{real_block}. This set is a maximal totally
real submanifold of $\Omega^2$; hence the holomorphic uniqueness theorem gives
$H^1(z,z')=H^2(1-z,1-z')$ on $\Omega^2$. This shows \eqref{orthogonality_reel} by using again \eqref{real_block}.
\end{proof}

Next, we prove a result very similar to Proposition \ref{prop:density}.
\begin{lemma}\label{density_on_R}
Let $0<\la_1<\la_2<1$. For $\bs{\beta}$ as in Lemma \ref{orthogonality}, 
the set ${\rm span}\{ \mc{F}_{\bs{\beta}}^i(\cdot,z)\in L^2(\R_+,w_{\bs{\beta}}^i(p)\dd p) \,|\, z\in (\la_1,\la_2)\}$ is dense in $L^2(\R_+,w_{\bs{\beta}}^i(p)\dd p)$.
\end{lemma}
\begin{proof}
We prove the result for $i=1$, the case $i=2$ being the
same. Let $u\in L^2(\R_+,w_{\bs{\beta}}^1(p)\dd p)$ be such that
\[\int_0^\infty \mc{F}_{\bs{\beta}}^1(p,z)\bbar{u}(p) w_{\bs{\beta}}^1(p)\dd p=0\]
for all $z\in(\la_1,\la_2)$. We use the variable $r=p^2/4$, so that
$\Delta_{Q+{\rm i}p}=Q^2/4+r$, and set $\tau=-\log z$. Reorganising
the Young diagrams according to their level $n\in\N$, we write
\[ \mc{F}_{\bs{\beta}}^1(p,z) = e^{-(r+a)\tau} \sum_{n=0}^\infty
w_{n,\bs{\beta}}(r)e^{-n\tau}, \textrm{ where }  a:=\frac{Q^2}{4}-\Delta_{\beta_1}-\Delta_{\beta_2}\in\R, \qquad w_{0,\bs{\beta}}(r)=1.
\]
It follows that
\[ 0=\int_0^\infty e^{-r\tau} \Big(\sum_{n=0}^\infty w_{n,\bs{\beta}}(r)e^{-n\tau}\Big)U(r)\dd r,
\]
where  $U(r):= w_{\bs{\beta}}^1(2\sqrt r)\bbar{u}(2\sqrt r)/\sqrt{r}$.
Equivalently, $\sum_{n=0}^\infty \mc{L}(w_{n,\bs{\beta}}U)(\tau)e^{-n\tau}=0.$
The locally uniform $L^2$ estimates for the conformal blocks show
that this identity extends holomorphically to
${\rm Re}(\tau)>0$. Hence, for every $\tau>0$ and $s\in\R$, (here $\mc{L}$ denotes the Laplace transform)
\[
\sum_{n=0}^\infty
\mc{L}(w_{n,\bs{\beta}}U)(\tau-{\rm i}s)
e^{-n(\tau-{\rm i}s)}=0.
\]
Proceeding as in the proof of Proposition~\ref{prop:density}, we take
the Fourier transform in $s$ in the sense of tempered distributions.
The convergence of the series in $\mc{S}'(\R_s)$ follows from the
same estimates, and we obtain
\[\sum_{n=0}^\infty w_{n,\bs{\beta}}(\xi-n)U(\xi-n) \mathbf{1}_{\{\xi>n\}}e^{-\tau\xi}=0 \quad  \textrm{ in } \mc{D}'(\R_\xi).\]
For $0<\xi<1$, only the term $n=0$ occurs. Since
$w_{0,\bs{\beta}}=1$, we obtain $U(\xi)=0$ on $(0,1)$.
Assume inductively that $U=0$ on $(0,N)$. If
$\xi\in(N,N+1)$, every term with $n\geq1$ in the series vanishes since
$0<\xi-n<N$ whenever $\xi>n$, and the term $n=0$ gives
$U(\xi)=0$. Thus $U=0$ on $\R_+$.
Since $w_{\bs{\beta}}^1(p)>0$ for almost every $p>0$, this implies
$u=0$ almost everywhere, proving the density. 
\end{proof}

\begin{proposition}[\textbf{Unitarity of the fusion kernel}]\label{prop:unitarity}
Let $\bs{\beta}=(\beta_1,\dots,\beta_4)\in ((-\infty,Q)\cup (Q+{\rm i}\R^*))^4$ satisfying 
\[{\rm Re}(\beta_1+\beta_4)>Q, \quad {\rm Re}(\beta_2+\beta_3)>Q, \quad {\rm Re}(\beta_1+\beta_2)>Q,\quad 
{\rm Re}(\beta_3+\beta_4)>Q.\]
The operator $K^{\rm A}_{\bs{\beta}}: L^2(\R_+,w^{2}_{\bs{\beta}}(p)\dd p)\to L^2(\R_+,w^{1}_{\bs{\beta}}(p)\dd p)$ is unitary.
\end{proposition}
\begin{proof} 
First, by \cite[Corollary 11.4]{GRW2} (see the identity (11.9) there), we have for $z \in(\la,1)$ for some $\la>0$
\[\begin{split}
\int_{\R}H_{0,1,0}^{\beta_{1},\beta_{2},Q-{\rm i}p}H_{0,\mu,0}^{\beta_{3},\beta_{4},Q+{\rm i}p}\mc{F}^1_{\bs{\beta}}(p,z)\dd p  
=&\int_{\R_+}H_{\mu,0,1}^{\beta_{4},\beta_{1},Q+{\rm i}p}H_{1,0,\mu}^{\beta_{2},\beta_{3},Q-{\rm i}p}\mc{F}^2_{\bs{\beta}}(p,1-z)\dd p.
\end{split}\]
Thus, using the proof of Lemma \ref{GRSSW}, we get for all $z\in (\la,1)$
\[\tilde{K}^{\rm A}_{\bs\beta}\mc{F}^2_{\bs\beta}(\cdot,1-z)= \mc{F}^1_{\bs\beta}(\cdot,z),
\qquad z\in(\lambda_1,\lambda_2),\]
in the distributional sense, where (below, $e^{i\pi s(p,\bs{\beta})}$ denotes the multiplication operator by this complex number of modulus $1$)
\[\tilde{K}^{\rm A}_{\bs\beta}= e^{i\pi s(p,\bs{\beta})}\circ K^{\rm A}_{\bs\beta}\circ e^{-i\pi s(p,\bs{\beta})}, \quad s(p,\bs{\beta})=\Delta_{Q+{\rm i}p}+\sum_{k=1}^4\Delta_{\beta_k}.\]
Here notice that with our normalisation of the block, the contstant $h$ of Lemma \ref{GRSSW} is $h=1$.
Choose $\lambda<\lambda_1<\lambda_2<1$, where $\lambda$ is as in
Lemma~\ref{GRSSW}, and set $\mc{E}_2:={\rm span}\left\{\mc F^2_{\bs\beta}(\cdot,1-z)\,;\,z\in(\lambda_1,\lambda_2)
\right\}$. Thus we define
$\tilde{K}^{\rm A}_{\bs\beta}$ on $\mc E_2$ by
\[\tilde{K}^{\rm A}_{\bs\beta}\Big(\sum_{n=1}^N u_n \mc F^2_{\bs\beta}(\cdot,1-z_n)\Big):=\sum_{n=1}^N u_n \mc F^1_{\bs\beta}(\cdot,z_n).
\]
Lemma~\ref{orthogonality} gives, for any $u^2,v^2\in\mc E_2$,
\[\int_{\R_+} (\tilde{K}^{\rm A}_{\bs\beta}u^2)(p)\overline{(\tilde{K}^{\rm A}_{\bs\beta}v^2)(p)} w^1_{\bs\beta}(p)\dd p
=\int_{\R_+}u^2(p)\overline{v^2(p)}w^2_{\bs\beta}(p)\dd p.\]
In particular, this definition does not depend on the representation
of an element of $\mc{E}_2$, and $\tilde{K}^{\rm A}_{\bs\beta}$ is an isometry
on $\mc{E}_2$. By Lemma~\ref{density_on_R}, $\mc E_2$ is dense in
$L^2(\R_+,w^2_{\bs\beta}(p)\dd p)$. Hence
$\tilde{K}^{\rm A}_{\bs\beta}$, and in turn $K^{\rm A}_{\bs\beta}$ extends uniquely to an isometry on this space.
Its range contains
\[
{\rm span}\left\{
\mc F^1_{\bs\beta}(\cdot,z)\,;\,
z\in(\lambda_1,\lambda_2)
\right\},
\]
which is dense in $L^2(\R_+,w^1_{\bs\beta}(p)\dd p)$ by the same
lemma. Since the range of an isometry is closed, it is the whole
target space. Therefore $K^{\rm A}_{\bs\beta}$ is unitary.
\end{proof}

\section{Unitarity of the modular kernel for the S-move}
We use directly the torus blocks occurring in \cite[Theorems 11.8 and
11.9]{GRW2}. For ${\rm Re}(T)>0$, set $q=e^{-2\pi T}$ and $q^*=e^{-2\pi/T}$ 
and define, for $\beta\in (0,Q)\cup (Q+{\rm i}\R^*)$, 
\begin{align}
 \mc F^1_\beta(T,p)
 &:=q^{(6p^2-1)/24}\sum_{\nu \in \mc{T}} w_{\A}(\Delta_{Q+{\rm i}p},\Delta_{\beta},\Delta_{Q+{\rm i}p})q^{|\nu|},
 \label{torus_block_channel_1}\\
 \mc F^2_\beta(T,p) &:=T^{-\Delta_\beta}(q^*)^{(6p^2-1)/24}
\sum_{\nu\in\mc T} w_{\A}(\Delta_{Q+{\rm i}p},\Delta_\beta, \Delta_{Q+{\rm i}p})(q^*)^{|\nu|},
 \label{torus_block_channel_2}
\end{align}
where $T^{-\Delta_\beta}=e^{-\Delta_\beta\log T}$ and $\log T$ is the holomorphic branch on $\{{\rm Re}(T)>0\}$ which is real for
$T>0$. In particular, the blocks are real valued when $T>0$.
We write $\rho_\beta(p)=e^{{\rm i}\vartheta(\beta)}w_\beta(p)$ with $w_\beta(p)>0$ where 
\begin{equation}\label{sec7:rho-torus}
 \rho_{\beta}(p) :=C^{\rm DOZZ}(\beta,Q+{\rm i}p,Q-{\rm i}p)
\end{equation}
and $\vartheta(\beta)$ is independent of $p$.
We prove the two analogues of Lemmas~\ref{orthogonality} and \ref{density_on_R}.

\begin{lemma}\label{orthogonality_torus}
Let $\beta\in(0,Q)\cup(Q+{\rm i}\R^*)$. For all $T,T'>0$, one has
\begin{equation}\label{orthogonality_torus_relation}
\int_{\R_+}w_{\beta}(p) \mc F^1_\beta(T,p)\bbar{\mc F^1_\beta(T',p)}\dd p=
 \int_{\R_+}w_{\beta}(p)\mc F^2_\beta(T,p)\bbar{\mc F^2_\beta(T',p)}\dd p.
\end{equation}
\end{lemma}
\begin{proof}
We first take $\beta\in(0,Q)$ and let
$T\in\C$ with ${\rm Re}(T)>0$. Applying
\cite[Theorem 9.1]{GKRV21_Segal} to the one-point function on the
torus $N({\rm i}T)$, with the horizontal cut and the flat metric
$|\dd z|^2$, gives
\[
\cjg V_\beta(0)\cjd_{N({\rm i}T),|\dd z|^2}
=
C\int_{\R_+}w_\beta(p)
|\mc F^1_\beta(T,p)|^2\dd p,
\]
where $C$ is independent of $T$. The map $z\mapsto z/({\rm i}T)$ descends to a biholomorphism from
$N({\rm i}T)$ to $N({\rm i}/T)$ and maps the horizontal cut to the
vertical one. By diffeomorphism invariance and the Weyl anomaly,
\[
\cjg V_\beta(0)\cjd_{N({\rm i}T),|\dd z|^2}=\cjg V_\beta(0)\cjd_{N({\rm i}/T),|T|^2|\dd z|^2}=|T|^{-2\Delta_\beta}
\cjg V_\beta(0)\cjd_{N({\rm i}/T),|\dd z|^2}.
\]
Applying again \cite[Theorem 9.1]{GKRV21_Segal}, this time to $\cjg V_\beta(0)\cjd_{N({\rm i}/T),|\dd z|^2}$, we deduce that
\[
\cjg V_\beta(0)\cjd_{N({\rm i}T),|\dd z|^2}
=
C\int_{\R_+}w_\beta(p)
|\mc F^2_\beta(T,p)|^2\dd p.
\]
Consequently,
\begin{equation}\label{isometric_relation_torus}
\int_{\R_+}w_\beta(p)|\mc F^1_\beta(T,p)|^2\dd p
=
\int_{\R_+}w_\beta(p)|\mc F^2_\beta(T,p)|^2\dd p
\end{equation}
for ${\rm Re}(T)>0$. For $\beta\in Q+{\rm i}\R^*$, the same identity is first obtained with
$\rho_\beta$ from Lemma~\ref{bootstrap_complex_alpha}, using the
polarized expression
\[
\mc F^i_\beta(T,p)
\overline{\mc F^i_{\bar\beta}(T,p)}.
\]
Since $\Delta_{\bar\beta}=\Delta_\beta$, one has
$\mc F^i_{\bar\beta}=\mc F^i_\beta$. Multiplying both channel
identities by the common phase
$e^{-{\rm i}\vartheta(\beta)}$ gives
\eqref{isometric_relation_torus} with the positive density
$w_\beta$. For $i=1,2$, set
\[ H^i(T,T') := \int_{\R_+}w_\beta(p) \mc F^i_\beta(T,p)\mc F^i_\beta(T',p)\dd p, \qquad {\rm Re}(T),{\rm Re}(T')>0.
\]
The locally uniform $L^2(\R_+,w_\beta(p)\dd p)$ estimates on compact
subsets of the right half-plane, together with Cauchy--Schwarz, show
that $H^i$ is holomorphic in $(T,T')$. Since $\Delta_\beta$ is real, all coefficients of the blocks are
real and our choice of the branch of $\log T$ gives $\mc{F}^i_\beta(\bar{T},p)=\overline{\mc{F}^i_\beta(T,p)}$.
It follows from \eqref{isometric_relation_torus} that $H^1(T,\bar T)=H^2(T,\bar T)$
for every $T$ with ${\rm Re}(T)>0$. The set $\{(T,\bar T)\,;\ {\rm Re}(T)>0\}$ is a maximal totally real submanifold of the product of two right
half-planes. Hence, by holomorphic continuation, $H^1(T,T')=H^2(T,T')$ whenever ${\rm Re}(T),{\rm Re}(T')>0$. Restricting to $T,T'>0$,
and using again the reality of the blocks, proves
\eqref{orthogonality_torus_relation}.
\end{proof}

\begin{lemma}\label{density_torus}
Let $0<T_1<T_2$. For $i=1,2$, the vector space
\[
 {\rm span}\{\mc F^i_\beta(T,\cdot)\,|\,T\in(T_1,T_2)\}
\]
is dense in
$L^2(\R_+,w_{\beta}(p)\dd p)$.
\end{lemma}

\begin{proof}
We prove the statement for $i=1$. We write the conformal block in the
form
\[\mc{F}^1_\beta(T,p)= q^{(6p^2-1)/24} \sum_{n\geq0}w_{n,\beta}(p)q^n, \qquad w_{0,\beta}(p)=1,\]
where $q=e^{-2\pi T}$ and $w_{n,\beta}(p)\in\R$. Suppose that $u\in L^2(\R_+,w_\beta(p)\dd p)$ is orthogonal to
$\mc{F}^1_\beta(T,\cdot)$ for every $T\in(T_1,T_2)$. The locally
uniform $L^2$ estimates for the conformal blocks show that
\[T\mapsto \int_{\R_+}\mc F^1_\beta(T,p)\bbar{u(p)}
w_\beta(p)\dd p\]
is holomorphic on $\{{\rm Re}(T)>0\}$ (and continuous up to ${\rm i}\R$). It therefore vanishes on this
half-plane. Set $r=p^2/2$. After removing the nonzero factor
$e^{\pi T/12}$, which is independent of $p$, the identity becomes
\[\sum_{n\geq0}\mc L\big(\tilde{w}_{n,\beta}U\big)(\pi T)e^{-2\pi nT}=0,
\]
where $\tilde{w}_{n,\beta}(r):=w_{n,\beta}(\sqrt{2r})$ and  $U(r):=w_\beta(\sqrt{2r})\bbar{u(\sqrt{2r})}/\sqrt{2r}$.
In particular, $\widetilde w_{0,\beta}=1$.
We fix $t>0$ and write $T=t+{\rm i}s$. Taking the Fourier transform in
$s$ in the sense of tempered distributions, and rescaling the Fourier
variable by $\pi$, gives
\[
\sum_{n\geq0}
\widetilde w_{n,\beta}(\xi-2n)
U(\xi-2n)\mathbf 1_{\{\xi>2n\}}
e^{-\pi t\xi}=0
\]
in $\mc D'(\R_\xi)$. The convergence of the series in
$\mc S'(\R_s)$ follows from the same locally uniform estimates as
above.
For $0<\xi<2$, only the term $n=0$ occurs. Since
$\widetilde w_{0,\beta}=1$, we obtain $U=0$ on $(0,2)$. Assume
inductively that $U=0$ on $(0,2k)$. If
$\xi\in(2k,2k+2)$, every term with $n\geq1$ vanishes, since
$0<\xi-2n<2k$ whenever $\xi>2n$, and the term $n=0$ gives
$U(\xi)=0$. Thus $U=0$ on $\R_+$. Since $w_\beta(p)>0$ for almost
every $p>0$, it follows that $u=0$ almost everywhere.

For $i=2$, we use $\mc F^2_\beta(T,p)=T^{-\Delta_\beta}\mc F^1_\beta(1/T,p)$.
The factor $T^{-\Delta_\beta}$ is nonzero and independent of $p$,
and $1/T$ ranges over the open interval
$(1/T_2,1/T_1)$. The result therefore follows from the case $i=1$.
\end{proof}

\begin{proposition}[\textbf{Unitarity of the fusion kernel for the S-move}]
\label{prop:unitarity_S}
Let $\beta\in(0,Q)\cup(Q+{\rm i}\R^*)$. The operator
\[ K^{{\rm S}}_\beta:  L^2(\R_+,w_{\beta}(p)\dd p)
 \to  L^2(\R_+,w_{\beta}(p^*)\dd p^*)
\]
defined from \eqref{K_S_2to1} is unitary, i.e. $(K^{\rm S}_\beta)^*=(K^{{\rm S}}_\beta)^{-1}$.
\end{proposition}

\begin{proof}
By \cite[Corollary 11.9]{GRW2}, we have for all sufficiently large $T>0$ 
\[\begin{split}
\int_{\R}\widetilde{\rm G}_{0,\mu_1}(\beta,Q+{\rm i}p)\,
\partial_{\mu_2}{\rm U}_{0,\mu_2}(Q-{\rm i}p)  \mc{F}^1_{\beta}(T,p)\dd p  
=&\sqrt{2\pi}\int_{\R_+}\partial_{\mu_2}
H_{\mu_2,\mu_1,\mu_1}^{Q+{\rm i}p,\beta,Q-{\rm i}p}\mc{F}^2_\beta (T,p)\dd p.
\end{split}\]
By the proof of Lemma \ref{ass:local-S-transform} and the definition of $K^{\rm S}_{\beta}(p,p')$, we get for all $T>0$
\[\mc{F}^1_{\beta}(T,p)=e^{{\rm i}\pi s(p,\beta)}\int_{\R_+} K^{\rm S}_{\beta}(p,p')e^{-{\rm i}\pi s'(p',\beta)}\mc{F}^2_{\beta}(T,p')\dd p'.\]
The proof of the unitarity is then the same as that of Proposition \ref{prop:unitarity} for the A-move, using Lemma \ref{density_torus} and Lemma \ref{orthogonality_torus} in place of Lemma \ref{density_on_R} and Lemma \ref{orthogonality}.
\end{proof}

\bibliographystyle{alpha}
\bibliography{virasoro}

\end{document}